# INTENSIONAL CYBERFORENSICS

Serguei A. Mokhov

A Thesis
in
The Department
of
Computer Science and Software Engineering

Presented in Partial Fulfillment of the Requirements
For the Degree of Doctor of Philosophy (Computer Science)
Concordia University
Montréal, Québec, Canada

September 2013
© Serguei A. Mokhov, 2013

## CONCORDIA UNIVERSITY
## SCHOOL OF GRADUATE STUDIES

This is to certify that the thesis prepared

By:  Serguei A. Mokhov

Entitled: Intensional Cyberforensics

and submitted in partial fulfillment of the requirements for the degree of

Doctor of Philosophy (Computer Science)

complies with the regulations of the University and meets the accepted standards with respect to originality and quality.

Signed by the final examining committee:

| | |
|---|---|
| Dr. Deborah Dysart-Gale | Chair |
| Dr. Weichang Du | External Examiner |
| Dr. Amr Youssef | External to Program |
| Dr. Peter Grogono | Examiner |
| Dr. Terry Fancott | Examiner |
| Dr. Joey Paquet and Dr. Mourad Debbabi | Thesis Supervisor |

Approved by

_______________________________
Chair of Department or Graduate Program Director

September 2013

_______________________________
Dean of Faculty

# Abstract

Intensional Cyberforensics

Serguei A. Mokhov, Ph.D.
Concordia University, 2013

This work focuses on the application of intensional logic to cyberforensic analysis and its benefits and difficulties are compared with the finite-state-automata approach. This work extends the use of the intensional programming paradigm to the modeling and implementation of a cyberforensics investigation process with backtracing of event reconstruction, in which evidence is modeled by multidimensional hierarchical contexts, and proofs or disproofs of claims are undertaken in an eductive manner of evaluation. This approach is a practical, context-aware improvement over the finite state automata (FSA) approach we have seen in previous work. As a base implementation language model, we use in this approach a new dialect of the Lucid programming language, called Forensic Lucid, and we focus on defining hierarchical contexts based on intensional logic for the distributed evaluation of cyberforensic expressions. We also augment the work with credibility factors surrounding digital evidence and witness accounts, which have not been previously modeled.

The Forensic Lucid programming language, used for this intensional cyberforensic analysis, formally presented through its syntax and operational semantics. In large part, the language is based on its predecessor and codecessor Lucid dialects, such as GIPL, Indexical Lucid, Lucx, Objective Lucid, MARFL and JOOIP bound by the underlying intensional programming paradigm.

# Acknowledgments

*To my dear wife, Miao Song, and our little family, who make life worthwhile and meaningful.*

*This thesis is also dedicated to those who made this thesis itself worthwhile[1].*

---

[1] ... and are acknowledged in detail in Section 10.5, page 290

# Contents

# List of Figures

# List of Tables

# List of Algorithms and Listings

# Chapter 1

# Introduction

*Cyberforensics*, or *digital forensics*, is the overarching term for digital crime investigations concerning (among other aspects) digital data recorded on various computing devices and services as evidence for the purposes of analysis, event reconstruction, attribution, and eventual prosecution or exoneration of suspects. Unlike in traditional forensics, cyberforensic processes (see Chapter 2) face a number of extra multifaceted challenges for their methods and results to be usable in a court of law and in information security practice. This thesis presents an endeavor that extends the use of the intensional programming paradigm [131, 350] and the science behind it to formally model and implement a cyberforensics investigation process with backtracing of event reconstruction, formalizing and modeling the evidence as multidimensional hierarchical contexts, and proving or disproving the claims in an intensional manner of expression and eductive evaluation [305, 307] based on Kripke's possible-world semantics [508]. Our proposed solution (Section 1.4) is designed to be a practical, context-aware improvement on the raw finite-state-automata (FSA/LISP) approach we have seen in [83, 135, 136, 137]. The core contribution of this thesis, FORENSIC LUCID, is in Chapter 7, with its most advanced application described in Section 9.5, page 257. What follows are executive summary details of the overall research and approach.

We apply intensional logic to automated cyberforensic analysis, reasoning, and event reconstruction. We explore the benefits and difficulties in comparing the work with the aforementioned finite-state automata/COMMON LISP approach. Intensional logic, as a multidimensional generalization of temporal logic, makes the evaluation of regular mathematical and logic expressions context-aware, where the context of evaluation (an arbitrary part of interest in the "possible world") is a first-class value and can be manipulated in logical expressions via context operators. The fundamental context consists of dimension-value pairs.

To help the study, the foundation of the new FORENSIC LUCID language is presented along with its forensic context operators to navigate evidential statements and crime scene specifications [305]. Our approach additionally introduces the notion of credibility and trustworthiness of evidence and witnesses, which FSA/LISP lacked—via the Dempster–Shafer theory of mathematical evidence to address the problem that some evidence or witness accounts may not be 100% credible, potentially affecting the overall reasoning about a case.

The FORENSIC LUCID compiler and run-time system are being designed within the General Intensional Programming System (GIPSY) [161, 264, 362, 370]. As an added benefit arising from the use of LUCID and its GIPSY implementation, the cyberforensics investigation cases may be conducted using a *scalable distributed/parallel intensional approach* for event reconstruction computation. This work is a natural evolution of and refinement of related works written or cowritten by the author [267, 269, 276, 300, 304, 305, 307, 312, 321].

LUCID and intensional logic are good candidates, proven over time, for knowledge representation and computable context-aware evaluation [171]. The FORENSIC LUCID instance presented in this thesis is arguably the first formally specified and developed language allowing scalable encoding of forensic knowledge of a case, the building of the case's operational description, and the evaluation of claims against them with event reconstruction locally or on a cluster for large volumes of digital evidence. The first ideas of FORENSIC LUCID appeared in [267] in 2007.

As a result, this thesis can be said to be a cross-disciplinary work that touches the breadth of the research in digital forensics, formal methods, intensional logic and programming, software engineering, distributed and parallel computing, soft computing and expert systems, automated reasoning, law, mathematics, pattern recognition and data mining, and graphical visualization. Thus, armed with all these tools and techniques we approach the digital investigation process.

In this chapter subsequently we gradually provide in increasing level of detail, after a more comprehensive overview (Section 1.1) of the research along with the principal motivations for it, we detail a set of problems and gaps in Section 1.3 requiring addressing and the proposed solutions to them in Section 1.4. Further we define the scope of this thesis in Section 1.5 and the overall summary of the research and contributions as well as organization of the remainder of this thesis in Section 1.6.

## 1.1 Overview

FORENSIC LUCID, a functional-intensional forensic case programming/specification language is at the core of the *Intensional Cyberforensics* project. It has undergone extensive design and development including its syntax, semantics, the corresponding compiler, run-time environment, and interactive development environments [266, 267] provided by the General Intensional Programming System (GIPSY) [463]. This work further extends our previous developments in the related work [263, 266, 267, 300, 304].

FORENSIC LUCID, serving as the base declarative specification language model that we use in this approach, is in fact a new dialect of the LUCID intensional-logic-based programming language [24, 25, 26, 27, 509]. As a part of this thesis, we define hierarchical contexts in FORENSIC LUCID (Section 7.2.2) based on intensional logic for the evaluation of cyberforensic expressions, first for modeling example case investigations from the related FSA/LISP work in order to do comparative studies of the old and new approaches [307]. The cases involved disputes between parties surrounding some kind of computer-related equipment. In particular, in this work we model the blackmail and ACME (a fictitious company name) "printing case incident" and make their specification in FORENSIC LUCID for follow-up cyberforensic analysis and event reconstruction. Our approach is based on the said cases, modeled by encoding concepts such as evidence and the related witness accounts as an evidential statement context in a FORENSIC LUCID "program". The evidential statement is an input to the transition function that models the possible "feed-forward" deductions in the case. We then invoke the transition function (actually its reverse, "backtracking") with the evidential statement context, to see if the evidence we encoded agrees with one's claims and then attempt to reconstruct the sequence of events that may explain the claim or disprove it [312]. Following the simple cases, we model more realistic cases and place some of the resulting practical artifacts to work in the actual network and system operations (Section 9.5). Additionally, in the ongoing theoretical and practical work, FORENSIC LUCID is augmented with the Dempster–Shafer theory of mathematical evidence to include credibility factors and similar concepts that are lacking in Gladyshev's model [310]. Specifically, this thesis further refines the theoretical structure and formal model of the observation tuple with credibility weight and other factors for cyberforensic analysis and event reconstruction [310] by extending the first iteration of FORENSIC LUCID that was originally following Gladyshev's approach [135, 136, 137] to only formalize the evidence and the case in question without taking into account witness credibility [300, 304, 310].

As an intensional dialect, FORENSIC LUCID's toolset is practically being designed within the General Intensional Programming System (GIPSY) and the probabilistic model-checking tool PRISM as a backend to compile the FORENSIC LUCID model into PRISM code syntax and check the compiled case model with the PRISM tool at run-time [310]. As briefly mentioned earlier, GIPSY is the middleware platform in this study that performs the FORENSIC LUCID computations. The GIPSY project is an ongoing effort aiming at providing a flexible platform for the investigation of the intensional programming model as realized by the latest instances of the LUCID programming language [24, 25, 26, 27, 509] (a multidimensional context-aware language whose semantics is based on possible-worlds semantics [218, 219]). GIPSY provides an integrated framework for compiling programs written in theoretically all variants of LUCID, and even any language of intensional nature that can be translated into some kind of "generic Lucid" [302] (e.g., in our case GIPL [361, 365, 473, 513]).

## 1.2 Motivation and Applications

### Motivations

A formal approach to cyberforensic analysis is necessary for the artifacts produced to be a credible tool to use in law enforcement and to be viable in courts if challenged. Pavel Gladyshev in his PhD thesis [135] dedicated the entire Chapter 4 to that effect. He subsequently provided formalisms in event reconstruction in digital investigations. Likewise, in order for FORENSIC LUCID (and its surrounding theory and practice) to be a credible tool to use in a court of law (including the implementation of relevant tools for the argumentation), the language ought to have a solid scientific basis, a part of which is formalizing the semantics of the language and proving correctness of the programs written in it.

We move one step further motivated by the fact that truth and credibility can be fuzzy (that is with elements of uncertainty or taintedness of evidence and witness accounts) and not just being completely true or false. Thus, it is natural to want to be able to represent such knowledge, and reason with its presence to discount low credibility claims and give higher-credibility claims higher weight.

The concrete realization of the formal approach also has to be usable by a wider investigative audience, who should be able to represent and visualize the voluminous case knowledge and reason about it efficiently. Thus, it is imperative to have usable scripting and visual aid tools to compose the case and import the digital evidence by human investigators. Additionally, the knowledge representation, case building and management should be friendlier to human investigators and take contextual meaning into the account. The subsequent case evaluation should be scalable and efficient at the same time, given the likely possibility of the need to process a large amount of digital evidential data.

Furthermore, a concrete operational need exists to automate the reasoning about truly offending or false-positive cases in the actual operational environment on a Faculty network to reduce the burden of the very few human analysts manually doing the related investigations while attending to many other duties (cf. Section 9.5).

## Applications

Due to the inherent interdisciplinary nature of this research (cf. page 2), its possible applications and implications can go significantly further beyond the very specific computer and network security investigations mentioned in this thesis and even beyond the cybercrime domain itself.

One possible application of the theoretical framework and formal model of the observation tuple with credibility weight and other factors for cyberforensic analysis is intrusion-detection system (IDS) data and their corresponding event reconstruction [310]. This work may also help with further generalization of the testing methodology of IDSs [355] themselves [310]. In particular, encoding and modeling large volumes of network and other data related to intrusion detection is an important step in incident analysis and response. The data are formalized in FORENSIC LUCID for the purposes of event correlation and reconstruction along with trustworthiness factors (e.g., the likelihood of logs being altered by an intruder) in a common specification of the evidential statement context and a digital crime scene. A possible goal here is to able to collect the intrusion-related evidence as the FORENSIC LUCID's evidential statement from diverse sources like Snort [256, 398, 438], netflows, pcap's data, etc., to do the follow-up investigation and event reconstruction. Another goal is to either be interactive with an investigator present, or fully automated in an autonomous IDS with self-forensics [321] capability [310].

The proposed practical approach in the cyberforensics field can also be used in a normal investigation process involving crimes not necessarily associated with information technology. Combined with an expert system (e.g., implemented in CLIPS [400]), the approach can also be used in training new staff in investigation techniques to help to prevent incomplete analysis that conventional ad-hoc techniques are prone to.

Venturing completely outside of crime investigation (digital or not), the FORENSIC LUCID approach was proposed to be adapted to applications beyond its intended purpose described in this thesis. On one hand, such a cross-disciplinary area of application is archeology for event reconstruction in historical context of the evidence one digs up from the archaeological sites trying to describe and arrange mentally all the evidence and describe what may have happened long ago given all the finds and validate various hypotheses put forward in that context. Extrapolating from that, any application where event reconstruction is needed in the presence of uncertainty, can be approached with FORENSIC LUCID. Another application area comes from even further away—the artistic side and the entertainment industry, such as validating logical consistency of scripts, plots, and strategies for feature films and games of relative complexity (especially if such works have themselves to do with investigations!).

## 1.3 Problem Statement and Gap Analysis

The arguably very first formal approach for evidential statement consistency verification and event reconstruction in cyberforensic investigative analysis appeared in the previously mentioned works [135, 136, 137] by Gladyshev *et al*. That approach (described and recited in detail in Chapter 2) relies on the finite-state automata (FSA) and their transformation and operation to model evidence, witnesses, stories told by witnesses, and their possible evaluation for the purposes of claim validation and event reconstruction (other formalisms were studied in [21, 22], see Section 2.2). There the authors provide a COMMON LISP sample implementation. The examples the works present are the initial use-cases for the proposed technique in this thesis—*Blackmail Investigation* (Section 2.2.5.2) and *ACME Printing* (Section 2.2.5.1) cases. Their approach, however, is unduly complex to use and to understand for non-theoretical-computer science or equivalently minded investigators. While the formalization and implementation of the FSA/LISP approach was very valuable to the community, it is not as elegant as it could have been [83] nor it is very usable by the majority of the less-formal-tech-savvy investigators.

At the origins of *Intensional Programming* (IP) is the LUCID functional intensional programming language, dating back to 1974 [27]. After more than 35 years of development, history has proven that it is a programming paradigm whose languages are diversified and are in constant evolution. However, *Intensional Programming* [131, 350] is an off-main-stream programming paradigm [526] whose concrete applicability still needs to be proven in order to be widely accepted. A lot of intensional dialects have been spawned from the 35+-year-old LUCID [5, 24, 25, 26, 27, 93, 122, 126, 361, 364, 509, 514, 539]. LUCID (see Section 4.1) itself was originally invented with a goal for program correctness verification [23, 25, 26, 244, 533].

Overall, there are a number of unaddressed problems and gaps with theories, techniques, and technologies used, which we summarize below and then elaborate in some detail on the specific points.

1. The FSA/LISP approach to formal cyberforensic analysis:

    (a) does not scale humanly (comprehension, scalability) and computationally (usability)
    (b) has no credibility factors to annotate evidence and testimonies (probabilities and PRISM are subsequently needed)
    (c) has no visualization of the knowledge base and the case management
    (d) requires more realistic cases to test in the actual operational work
    (e) no automatic encoders of the evidence into COMMON LISP or FSA

2. The LUCID language:

    (a) needs hierarchical contexts to represent nested/hierarchical knowledge as streamed arguments and results
    (b) requires object members access and other APIs for arbitrary nesting
    (c) needs the use of formal methods and verification
    (d) needs better usability of the LUCID development and run-time tools for a wider community
    (e) requires an update to the theoretical foundations:
        i. needs a common format for knowledge representation and reasoning
        ii. needs a augmented theoretical logical framework (HOIL and HOIFL, see Section 1.6.1.1, page 13) support for reasoning

3. The GIPSY system:

    (a) needs augmentation to support FORENSIC LUCID
    (b) requires cross-language data types
    (c) needs a compiler for FORENSIC LUCID
    (d) the evaluation engine/run-time systems needs:
        i. multi-tier architecture to address scalability better

ii. refactoring the framework for optimized problem-specific handling

iii. reasoning engines for backtracking and probabilities

(e) configuration management and automation

(f) requires development/configuration environment with graphical support for the better usability

## 1.4 Proposed Solution

This work focuses on the refinement of application of the intensional logic to cyberforensic analysis and its benefits are compared to the pure finite-state automata approach [307]. At the same time we increase the visibility of the Intensional Programming paradigm within the formal methods and cyberforensic analysis communities as the right tool to use in such an approach.

The large part of the solution is the creation of the introduced FORENSIC LUCID dialect of LUCID to foster the research on the *intensional cyberforensics* [300] (see Section 1.6.1.2). To summarize in a few words, this thesis presents *multidimensional context-oriented cyberforensic specification and analysis.* In a large part, FORENSIC LUCID is a union of the syntax and operational semantics inference rules from the comprising intensional languages with its own forensic extensions based on the cited finite-state automata approach [135, 136, 137]. In order to be a credible tool to use, for example, in court, to implement relevant tools for the argumentation, the language ought to have a solid scientific base, a part of which is a complete formalization of the syntax and semantics of the language [304].

Thus, the approach this thesis follows is tailored to addressing a decent subset of problems and gaps outlined in the previous section. Specifically:

1. Addressing the FSA/LISP approach gaps.

   The goal of this work is to lay a foundation to lead to a solution that remedies these two major drawbacks of the FSA approach by introducing credibility factors and improving usability. Additionally, we benefit from the parallel demand-driven context-aware evaluation in terms of the implementing system, which the original COMMON LISP-based implementation [137] approach lacks [305].

   As to the test cases, the two actual illustratory examples Gladyshev's works presented are the mentioned first use-cases for the proposed technique in this thesis—the ACME printer and blackmail case investigations (see Section 2.2.5.1 and Section 2.2.5.2 respectively). These works [136, 137] detail the corresponding formalization using the FSA and the proof-of-concept (PoC) COMMON LISP implementation for the printer case [137]. We first aim at the same cases to model and implement them using the new approach, which paves a way to be more friendly and usable in the actual investigator's work and serve as a basis to further development in the areas of forensic computing and intensional programming using FORENSIC LUCID [307, 312] (see Section 9.3 and Section 9.4 respectively). We then move onto more realistic cases, such as, e.g., MAC spoofer report investigations (see Section 9.5).

Thus, a LUCID approach is a major solution to these problems.

2. Addressing the LUCID gaps.

The LUCID family of languages thrived around intensional logic that makes the notion of context explicit and central, and recently, a first class value [473, 513, 515] that can be passed around as function parameters or as return values and have a set of operators defined upon. We greatly draw on this notion by formalizing our evidence and the stories as a contextual specification of the incident to be tested for consistency against the incident model specification. In our specification model, we require more than just atomic context values—we need a higher-order context hierarchy to specify different level of detail of the incident and being able to navigate into the "depth" of such a context. Luckily, such a proposition has already been made [272] and needs some modifications to the expressions of the cyberforensic context.

In terms of syntax and semantics for FORENSIC LUCID we benefit in large part, as the language is based on its predecessor and codecessor Lucid dialects, such as GIPL, INDEXICAL LUCID, LUCX, OBJECTIVE LUCID, and JOOIP bound by the higher-order intensional logic (HOIL) that is behind them. This work continues to formally specify the operational semantics of the FORENSIC LUCID language extending the previous related work [300, 304].

We further define the specification of hierarchical evidential context expressions and the operators on them when modeling the examples while illustrating related fundamental concepts, operators, and application of context-oriented case modeling and evaluation. COMMON LISP, unlike LUCID, entirely lacks contexts built into its logic, syntax, and semantics, thereby making the implementation of the cases more clumsy and inefficient (i.e., highly sequential) [137]. Our GIPSY system [191, 264, 362, 366, 370] offers a distributed demand-driven evaluation of Lucid programs in a more efficient way and is more general than the LISP's compiler and run-time environment [307, 312].

Thus, HOIL, GIPSY are the solutions here.

3. Addressing the GIPSY gaps.

   To enhance the scalability of GIPSY a number of APIs needed to be updated, redesigned, or designed from scratch to accommodate new middleware technologies, compilers, and the run-time evaluation system. We designed components and amended existing components of the distributed evaluation engine within the General Intensional Programming System (GIPSY) enabling measurements of a wide array of run-time parameters for the purposes of scalability studies, its configuration management, scheduling, and administration with the General Manager Tier (Section 6.2.2.1) component of the GIPSY's multi-tier architecture [161]. We made advances in the software engineering design and implementation of the multi-tier run-time system for the General Intensional Programming System (GIPSY) by further unifying the distributed technologies used to implement the Demand Migration Framework (DMF, Section 6.1.3) in order to streamline distributed execution of hybrid intensional-imperative programs using JAVA [161]. The bulk of multi-tier implementation and scalability studies following the redesign of the APIs were carried out by Han [160] and Ji [191], and graphical configuration management by Rabah *et al.* [393].

   The compiler API support has been further enhanced to allow JOOIP (Section 4.3.2.3) compilation, the semantics and PoC implementation of which was subsequently carried out by Wu [528].

   The author Mokhov supported the design and unification of these APIs and the integration effort within GIPSY following up with his own extensions to support multiple GEE (Section 6.2.2) backends and a compiler to support FORENSIC LUCID, including the complete GIPSY Type System and Theory (see Appendix B). All this work resulted in a number of contributions to GIPSY and its frameworks detailed in Chapter 8.

   As a part of the proposed *near-future work* to improve scalability of information management and presentation and with the ongoing advances in the interaction design [404], the visualization project was proposed to further enhance usability of the discussed language and system and the related tools. Lucid programs are dataflow programs and can be visually represented as data flow graphs (DFGs) and composed visually. FORENSIC LUCID includes the encoding of the evidence (representing the context of evaluation) and the crime scene modeling in order to validate claims against the model and perform event reconstruction, potentially within large swaths of digital evidence. To aid investigators to model the scene and evaluate it, instead of typing a FORENSIC LUCID program, we propose to expand the design and implementation of the Lucid DFG programming onto FORENSIC LUCID case modeling and specification to enhance the usability of the language and the system and its behavior in 3D. We briefly discuss the related work on visual programming and DFG modeling in an attempt to define and select one approach or a composition of approaches for FORENSIC LUCID based on various criteria such as previous implementation, wide use, formal backing in terms of semantics and translation [311] (see Chapter E). Reaching out into a different disciplinary areas of specific interest here—a recent novel concept of documentary knowledge visual representation in illimitable space introduced by Song in [437]. That work may scale when properly re-engineered and enhanced to act as an interactive 3D window into the evidential knowledge grouped into the semantically linked "bubbles" visually representing the documented evidence and by moving such a contextual window, or rather, navigating within theoretically illimitable space and investigator can sort out and re-organize the knowledge items as needed prior launching the reasoning computation. The interaction aspect would be of a particular usefulness to open up the documented case knowledge and link the relevant witness accounts. This is a proposed solution to the large scale visualization problem of large volumes of "scrollable" evidence that does not need to be all visualized at one, but be like in a storage depot. (However, the actual realization of this solution is deferred to later time.)

What follows are the details of our solution along with the related work [274, 313].

## 1.5 Scope

Out of the problems and gaps detailed in the previous sections, we summarize the more concrete objectives within the actual scope of this thesis in Section 1.5.1 and the items that are outside the scope of the thesis and more destined for the immediate future work in Section 1.5.2.

### 1.5.1 Thesis Objectives

The primary objectives are to design the introduced FORENSIC LUCID incident specification/scripting language in terms of formal syntax and semantics and show their use through several case studies. Specifically, in the point form the objectives are to produce:

1. FORENSIC LUCID (Chapter 7)

    (a) FORENSIC LUCID syntax (Section 7.3)
    (b) FORENSIC LUCID operational semantics (Section 7.4)
    (c) Hierarchical higher-order context specification (Section 7.2.2)
    (d) Operators and "transition functions" (Section 7.3.4)
    (e) Observation specification with credibility (Section 7.4.2)

2. FORENSIC LUCID parser (Section 8.1)

3. Run-time evaluation environment (engine) design (Section 8.2)

4. Example cases (see Chapter 9)

### 1.5.2 Out of Scope

- Large scale visualization of the case and systems
- IDS integration and inter-operation
- Other future work items detailed in Section 10.4

## 1.6 Summary

To summarize, we believe and show that the intensional approach with a LUCID-based dialect to the problem is an asset in the fields of cyberforensics and intensional logic and programming as it is promising to be more practical and usable than the plain FSA/LISP in the end [305, 307, 312]. Since LUCID was originally designed and used to prove correctness of programs [24, 25, 26, 509], and is based on the temporal logic functional and data-flow languages, we can relatively easily adopt its computational machinery to backtracking in proving or disproving the evidential statements and claims in the investigation process as simply an evaluation of a forensic expression that either translates to sets of *true* or *false* given all the facts in the formally specified context [312] providing a set of event reconstruction traces (*backtraces*). For that we defined the novel LUCID dialect with a new set of primitives predefined for forensic tasks. Unlike the LISP-based system implementing the finite state automata, we still retain the flexibility of parallel evaluation of several claims or several components of one claim at the same time by relying on the GIPSY's demand-driven general eduction engine (GEE) whose backend is powered by various distributed systems technologies such as the DMS [242, 383, 384, 385, 498, 499, 501] and multi-tier architecture [160, 161, 191, 362]. We also retain the generality of the approach [305, 307, 312].

### 1.6.1 Science and Technology

Here we summarize the science and technology aspects behind this work that we use and rely on as tools and techniques that support the multifaceted nature of this thesis in a variety of ways. The detailed description of these aspects follows in the background chapters in Part I. The specific scientific contributions and the related work done by others and some by the author come from the *Intensional Logic* and its extensions for formalization (Section 1.6.1.1), cyberforensic analysis (Section 1.6.1.2), and GIPSY (Section 1.6.1.4).

#### 1.6.1.1 Intensional Logic and Soft Computing

From the logic perspective, it was shown one can model computations (a computation is also a basic formal unit in the finite state machines in [136, 137]) as logic [222]. When armed with contexts [508] as first-class values and a demand-driven run-time model adopted in the implementation of the LUCID-family of languages [362, 365, 370, 379, 396, 473, 515] that constrains the scope of evaluation in a given set of dimensions, we come to the intensional logic and the corresponding programming artifact. In a nutshell, we model our forensic computation unit in intensional logic and implement it in practice within an intensional programming platform—the General Intensional Programming System (GIPSY) [264, 362, 366, 370]. We project a lot of potential for the results of this work to be successful, beneficial, and usable for cyberforensics investigation as well as simulation and intensional programming communities [307, 312].

From the intensional logic we move up to the concept of *higher-order intensional logic* (HOIL) [301, 302] since FORENSIC LUCID's constructs are that of a function al language. To accommodate the notion of credibility in our formalism [274, 313], we then move to something we define as the *higher-order intensional fuzzy logic* (HOIFL), which is HOIL+Dempster–Shafer mathematical evidence theory since we are dealing with possibly inexact reasoning present in intelligent soft computing systems [204]. For the latter we use the mentioned PRISM tool [467] to check our models. This affects the stated earlier objectives 1a, 1b, and 1e.

For in-depth background on intensional logic and Dempster–Shafer theory see Section 3.2 and Section 3.3.2 respectively.

#### 1.6.1.2 Cyberforensic Analysis

Cyberforensic analysis has to do with automated or semi-automated processing of, and reasoning about, digital evidence, witness accounts, and other details from cybercrime incidents (involving computers, but not limited to them). Analysis is one of the phases in cybercrime investigation (while the other phases focus on evidence collection, preservation, chain of custody, information extraction that precede the analysis) [84]. The phases that follow the analysis are formulation of a report and potential prosecution or exoneration, typically involving expert witnesses [83, 300, 311].

There are quite a few techniques, tools (hardware and software), and methodologies that have been developed for the mentioned phases of the cybercrime investigation. A lot of attention has been paid to tool development for evidence collection and preservation; a few tools have been developed to aid data "browsing" on the confiscated storage media, log files (e.g., Splunk [439]), memory, and so on. Fewer tools have been developed for case analysis of the data (e.g., Sleuthkit [61]), and the existing commercial packages (e.g., EnCase [54, 149] or FTK [2]) are very expensive. Even less so there are case management, event modeling, and event reconstruction, especially with a solid formal theoretical base [300, 311]. More in-depth background on cyberforensics is found in Chapter 2.

FORENSIC LUCID's design and implementation, its theoretical base are being established in this work (see Chapter 7). In the cyberforensic analysis, FORENSIC LUCID is designed to be able to express in a program form the encoding of the evidence, witness stories, and evidential statements, that can be tested against claims to see if there is a possible sequence or multiple sequences of events that explain a given "story". As with the Gladyshev's approach, it is designed to aid investigators to avoid ad-hoc conclusions and have them look at the possible explanations the FORENSIC LUCID program "execution" would yield and refine the investigation, as was Gladyshev has previously shown [135, 136, 137] where hypothetical investigators failed to analyze all the "stories" and their plausibility before drawing conclusions in the cases under investigation [300, 311]. This works improves on Gladyshev's work by being more usable, scalable, expressive, and concise, while dealing with credibility and incorporating the good ideas from Gladyshev.

#### 1.6.1.3 Context-Orientation

Context-oriented computing and reasoning emerged as a domain with a lot of research going into it from various aspects as context is something that provides us with meaning. This includes mobile computing, semantic web [535] and related technologies for natural language understanding [413, 414, 522], intelligent agent computing and distributed systems [42, 414, 454], description logic and ontologies [538, 538], web services and service-oriented architectures [66, 103, 432, 534], human-computer interaction [102], security [538] and environment for requirements engineering in software engineering [254, 401, 413, 482, 522], data mining and pattern recognition [63, 413] among others. Context also helps to deal with an uncertainty sometimes humans have to deal with in an unknown situation using some form of a dialectic approach based on incomplete information [50].

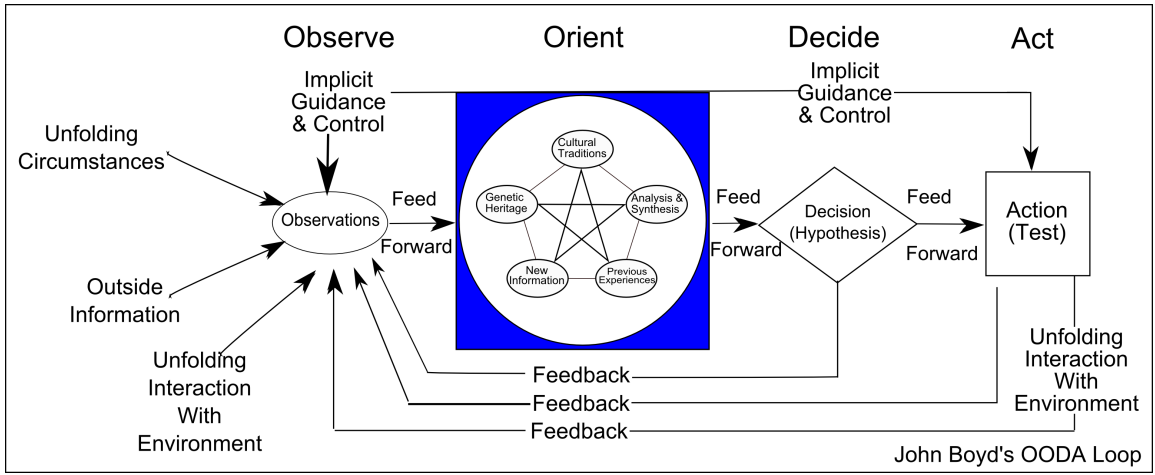

Figure 1: Boyd's OODA loop [51]

Context specification and evaluation is also at the core of the Isabelle/HOL interactive theorem proving framework within its ML implementation [518].

Context helps making decisions about a situation or investigation pertinent to the environment where the situation is taking place. Boyd came up with the notion of the *Observe-Orient-Decide-Act* (OODA) loop (see Figure 1[1]) when describing fighter pilots in a combat situation [51] learning the context of their environment, reading out flight parameters, receiving information from the mission control about the known information about the enemy to plan for the battle style, and then, when engaging, the process continues, albeit, much more rapidly. *Observations* give the evidential parameters as an input of raw observed properties, that formulate the initial context of the situation. The *Orient* step can be termed as the set of context operations, the most important part of the loop, where contextual knowledge is applied to filter out less relevant data about the environment, enemy, and own tools, through the available constraints before arriving at a *Decision* of how to *Act* next to alter the environment and situation further to own advantage and respond to the changes introduced by the enemy for each iteration via the feedback loops. Boyd's military approach of the OODA loop was further expanded to a more complex multi-perspective view of the business world [477] for decision making. (In both cases the approach was to try to run through own loop faster than the opponent's to gain advantage.)

---

[1] The image is by P. E. Moran, reproduction at the CC BY 3.0 license of the Boyd's OODA loop is sourced from the Wikipedia's `http://en.wikipedia.org/wiki/OODA_loop` page.

Context-orientation in security and investigation very similarly go through a similar process, but on a different scale. Context navigation is done with appropriate operators from the existing evidence before arriving at a conclusion, and a possible enforcement action (e.g., enabling firewall restriction, bringing network port down, or intensive follow-up investigation), or a production of a new context to explore further possibilities, options, and hypotheses. This approach in part is also taken by, e.g., Rapid7, the makers of the Metasploit penetration testing suite and their related systems.

A number of approaches have been created to represent contextual knowledge and reason about it. The common mainstream popular way appears to be with ontologies and Web Ontology Language (OWL) [151] and using description logics [521, Chapters 10, 19]. Physically, it is typically a kind of XML-based specification. Aspect-oriented programming (AOP) in away also introduced the notion of context (see Section 8.2.1, page 213). Intensional logic (see Section 1.6.1.1, page 13) has been built around the notion of context nearly from the start and can encompass all the mentioned concepts, logics, and it has a nice executable formalism to go along. We stick with the intensional approach as arguably the oldest sound approach in the scene with solid foundations for formal representation of contextual knowledge and reasoning about it in one concept, instantiated in LUCID. For further in-depth discussion please refer to Section 6.1.4 and Section 3.2.

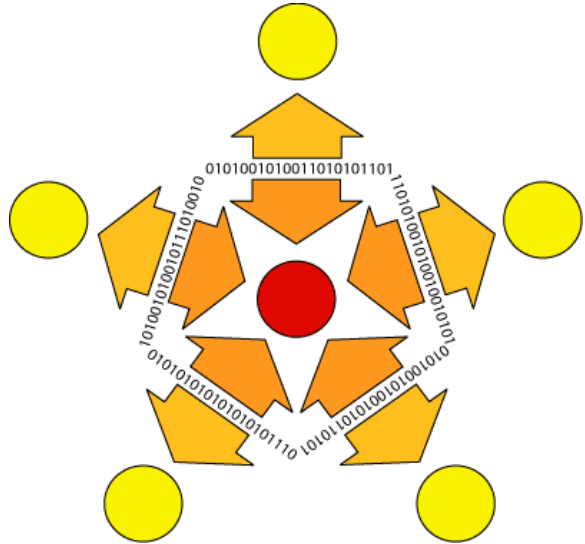

Figure 2: The GIPSY logo representing the distributed nature of GIPSY

#### 1.6.1.4 The General Intensional Programming System (GIPSY)

The General Intensional Programming System (GIPSY) has been built around the mentioned Lucid family of intensional programming languages that rely on the higher-order intensional logic (HOIL) to provide context-oriented multidimensional reasoning of intensional expressions. HOIL combines functional programming with various intensional logics (Section 1.6.1.1) to allow explicit context expressions to be evaluated as first-class values that can be passed as parameters to functions and return as results with an appropriate set of operators defined on contexts. GIPSY's frameworks are implemented in Java as a collection of replaceable components for the compilers of various Lucid dialects and the demand-driven eductive evaluation engine that can run distributively (Figure 2[2]). GIPSY provides support for hybrid programming models that couple intensional and imperative languages for a variety of needs. Explicit context expressions limit the scope of evaluation of mathematical expressions (effectively a Lucid program is a mathematics or physics expression constrained by the context) in tensor physics, regular mathematics in multiple dimensions, etc., and for cyberforensic reasoning as one of the specific use-cases of interest of this thesis. In return, some of this thesis' work also provides GIPSY with more application scenarios to prove its applicability to solve different kinds of problems. Thus, GIPSY is a support testbed for HOIL-based languages some of which enable such reasoning, as in formal cyberforensic case analysis with event reconstruction. In this thesis we discuss in detail the GIPSY architecture, its evaluation engine and example use-cases [302] in Chapter 6 and Chapter 8 respectively.

---

[2] Paquet, 2005

## 1.6.2 Research Approach Overview

As it is becoming evident, the research approach presented in this thesis is multifaceted drawing on a number of aspects. These are visually presented in Figure 3 summarizing the overall research. This figure is drawn up from an inspiration of the multi-tier architecture of GIPSY depicted in Figure 2, with the faces augmented and extend from the logotype of the book by Jordan and Alaghband [196], and lately in part on the context-*orient*ation in the OODA loop in Figure 1. The Figure 3 depicts at the center the core contribution of this thesis—FORENSIC LUCID surrounded by all the supporting tools, techniques, and methodologies from logic programming, to algorithms, to programming languages, to the compile- and run-time middleware systems architecture, and to the data sources used for case evaluations.

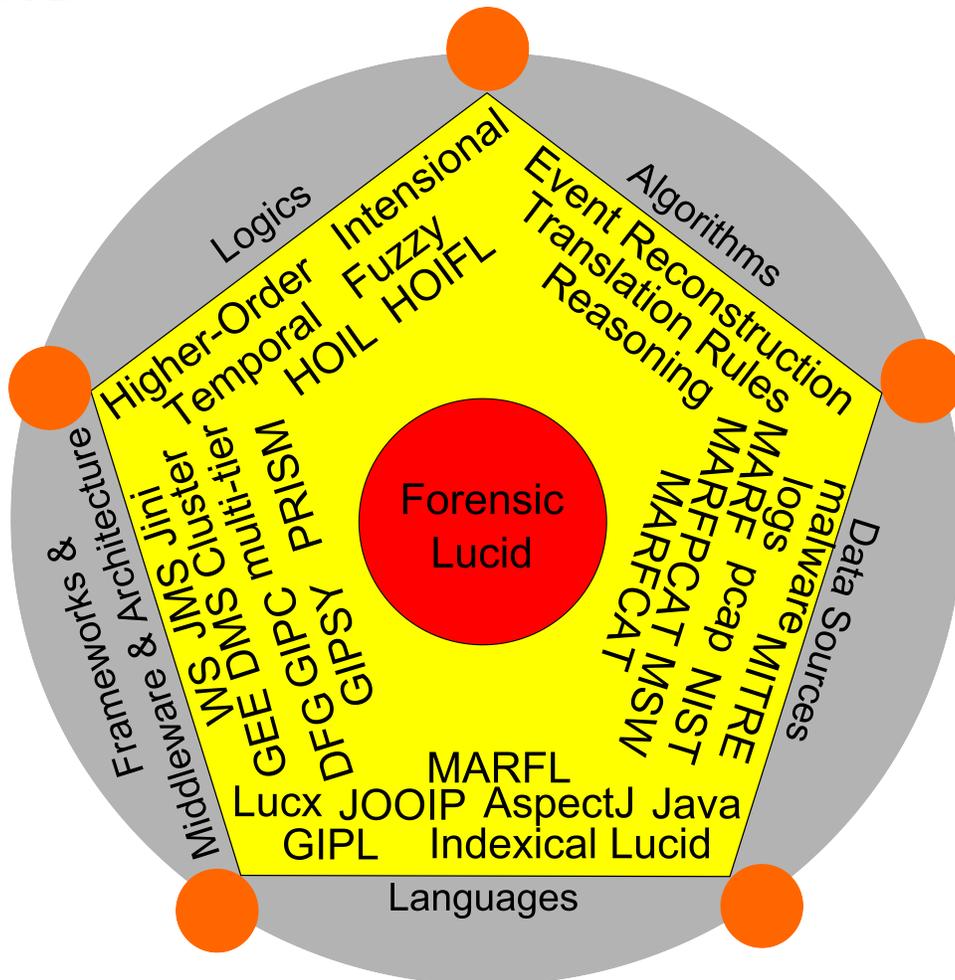

Figure 3: Overall multifaceted research summary

### 1.6.3 Thesis Organization

Due to the multifaceted nature of this thesis, it's organized in parts covering the different facets to make it easier for the reader to navigate and read the relevant parts of the material easier, including the chapter, page, section, figure, etc. hyperlinks provided in the electronic version. A lot of it is dedicated to the background work primarily by others and the author himself to provide the necessary setting, but the background chapters are not required to be read in a strict sequence and the readers may pick and choose the chapters of interest to read based on a particular topic or discussion found in this chapter or in the core methodology part. Some chapters are short that were not fitting into any other chapters, while other chapters are significantly more comprehensive depending on its material's importance for this thesis.

We begin by reviewing the relevant background knowledge (Part I) on cyberforensics (Chapter 2), mathematical and logic foundations (Chapter 3) where we review the notion of intensional logic and programming for the unaware reader, the LUCID programming language (Chapter 4), data mining and pattern recognition aspects (Chapter 5), the General Intensional Programming System (Chapter 6). We subsequently present the refined syntax and the semantics specification of the FORENSIC LUCID language properly attributing the inherited language constructs and rules, and the new extensions followed by the supporting components and systems design and implementation in Part II. We then proceed with the evaluation of the approach using a number of case studies followed by concluding remarks in Part III. After the Bibliography, appendices provide additional details about the theory, design, and implementation aspects supporting the core concepts presented in the main chapters.

### 1.6.4 Publications and Contributions

Here are several works that are directly related to or supporting this research as far as publication aspect concerned. This section lists the select related publications on FORENSIC LUCID (Section 1.6.4.1); GIPSY (Section 1.6.4.2); other aspects related to data mining, pattern recognition, networking and security (Section 1.6.4.3); and self-forensics (Section 1.6.4.4). These contributions provides some initial solutions addressing some of the earlier states gaps and thesis objectives.

#### 1.6.4.1 FORENSIC LUCID

- S. A. Mokhov, J. Paquet, and M. Debbabi. Reasoning about a simulated printer case investigation with Forensic Lucid. In P. Gladyshev and M. K. Rogers, editors, *Proceedings of ICDF2C'11*, number 0088 in LNICST, pages 282–296. Springer, Oct. 2011. Submitted in 2011, appeared in 2012; online at http://arxiv.org/abs/0906.5181

- S. A. Mokhov, J. Paquet, and M. Debbabi. Towards automated deduction in blackmail case analysis with Forensic Lucid. In J. S. Gauthier, editor, *Proceedings of the Huntsville Simulation Conference (HSC'09)*, pages 326–333. SCS, Oct. 2009. Online at http://arxiv.org/abs/0906.0049

- S. A. Mokhov. Encoding forensic multimedia evidence from MARF applications as Forensic Lucid expressions. In T. Sobh, K. Elleithy, and A. Mahmood, editors, *Novel Algorithms and Techniques in Telecommunications and Networking, proceedings of CISSE'08*, pages 413–416, University of Bridgeport, CT, USA, Dec. 2008. Springer. Printed in January 2010

- S. A. Mokhov, J. Paquet, and M. Debbabi. Formally specifying operational semantics and language constructs of Forensic Lucid. In O. Göbel, S. Frings, D. Günther, J. Nedon, and D. Schadt, editors, *Proceedings of the IT Incident Management and IT Forensics (IMF'08)*, LNI140, pages 197–216. GI, Sept. 2008. Online at http://subs.emis.de/LNI/Proceedings/Proceedings140/gi-proc-140-014.pdf

- S. A. Mokhov, J. Paquet, and M. Debbabi. On the need for data flow graph visualization of Forensic Lucid programs and forensic evidence, and their evaluation by GIPSY. In *Proceedings of the Ninth Annual International Conference on Privacy, Security and Trust (PST), 2011*, pages 120–123. IEEE Computer Society, July 2011. Short paper; full version online at http://arxiv.org/abs/1009.5423

- S. A. Mokhov, J. Paquet, and M. Debbabi. Towards automatic deduction and event reconstruction using Forensic Lucid and probabilities to encode the IDS evidence. In S. Jha, R. Sommer, and C. Kreibich, editors, *Proceedings of RAID'10*, LNCS 6307, pages 508–509. Springer, Sept. 2010

- S. A. Mokhov. Enhancing the formal cyberforensic approach with observation modeling with credibility factors and mathematical theory of evidence. [online], also in *;login: vol. 34, no. 6, p. 101*, Dec. 2009. Presented at WIPS at USENIX Security'09, http://www.usenix.org/events/sec09/wips.html

#### 1.6.4.2 GIPSY

- S. A. Mokhov and J. Paquet. Using the General Intensional Programming System (GIPSY) for evaluation of higher-order intensional logic (HOIL) expressions. In *Proceedings of SERA 2010*, pages 101–109. IEEE Computer Society, May 2010. Online at http://arxiv.org/abs/0906.3911

- S. A. Mokhov and J. Paquet. A type system for higher-order intensional logic support for variable bindings in hybrid intensional-imperative programs in GIPSY. In T. Matsuo, N. Ishii, and R. Lee, editors, *9th IEEE/ACIS International Conference on Computer and Information Science, IEEE/ACIS ICIS 2010*, pages 921–928. IEEE Computer Society, May 2010. Presented at SERA 2010; online at http://arxiv.org/abs/0906.3919

- S. A. Mokhov, J. Paquet, and X. Tong. A type system for hybrid intensional-imperative programming support in GIPSY. In *Proceedings of C3S2E'09*, pages 101–107, New York, NY, USA, May 2009. ACM

- B. Han, S. A. Mokhov, and J. Paquet. Advances in the design and implementation of a multi-tier architecture in the GIPSY environment with Java. In *Proceedings of SERA 2010*, pages 259–266. IEEE Computer Society, 2010. Online at http://arxiv.org/abs/0906.4837

- A. Wu, J. Paquet, and S. A. Mokhov. Object-oriented intensional programming: Intensional Java/Lucid classes. In *Proceedings of SERA 2010*, pages 158–167. IEEE Computer Society, 2010. Online at: http://arxiv.org/abs/0909.0764

- J. Paquet, S. A. Mokhov, and X. Tong. Design and implementation of context calculus in the GIPSY environment. In *Proceedings of the 32nd Annual IEEE International Computer Software and Applications Conference (COMPSAC)*, pages 1278–1283, Turku, Finland, July 2008. IEEE Computer Society

#### 1.6.4.3 Data Mining, Pattern Recognition, and Security

- A. Boukhtouta, N.-E. Lakhdari, S. A. Mokhov, and M. Debbabi. Towards fingerprinting malicious traffic. In *Proceedings of ANT'13*, volume 19, pages 548–555. Elsevier, June 2013

- E. Vassev and S. A. Mokhov. Developing autonomic properties for distributed pattern-recognition systems with ASSL: A Distributed MARF case study. *LNCS Transactions on Computational Science, Special Issue on Advances in Autonomic Computing: Formal Engineering Methods for Nature-Inspired Computing Systems*, XV(7050):130–157, 2012. Accepted in 2010; appeared February 2012

- S. A. Mokhov, J. Paquet, M. Debbabi, and Y. Sun. MARFCAT: Transitioning to binary and larger data sets of SATE IV. [online], May 2012. Submitted for publication to JSS; online at http://arxiv.org/abs/1207.3718

- S. A. Mokhov. The use of machine learning with signal- and NLP processing of source code to fingerprint, detect, and classify vulnerabilities and weaknesses with MARFCAT. Technical Report NIST SP 500-283, NIST, Oct. 2011. Report: http://www.nist.gov/manuscript-publication-search.cfm?pub_id=909407, online e-print at http://arxiv.org/abs/1010.2511

- S. A. Mokhov and M. Debbabi. File type analysis using signal processing techniques and machine learning vs. `file` unix utility for forensic analysis. In O. Goebel, S. Frings, D. Guenther, J. Nedon, and D. Schadt, editors, *Proceedings of the IT Incident Management and IT Forensics (IMF'08)*, LNI140, pages 73–85. GI, Sept. 2008

- S. A. Mokhov. Towards syntax and semantics of hierarchical contexts in multimedia processing applications using MARFL. In *Proceedings of the 32nd Annual IEEE International Computer Software and Applications Conference (COMPSAC)*, pages 1288–1294, Turku, Finland, July 2008. IEEE Computer Society

- M. J. Assels, D. Echtner, M. Spanner, S. A. Mokhov, F. Carrière, and M. Taveroff. Multifaceted faculty network design and management: Practice and experience. In B. C. Desai, A. Abran, and S. Mudur, editors, *Proceedings of $C^3S^2E$'11*, pages 151–155, New York, USA, May 2010–2011. ACM. Short paper; full version online at http://www.arxiv.org/abs/1103.5433

- S. A. Mokhov. Towards security hardening of scientific distributed demand-driven and pipelined computing systems. In *Proceedings of the 7th International Symposium on Parallel and Distributed Computing (ISPDC'08)*, pages 375–382. IEEE Computer Society, July 2008

#### 1.6.4.4 Self-Forensics

- S. A. Mokhov, E. Vassev, J. Paquet, and M. Debbabi. Towards a self-forensics property in the ASSL toolset. In *Proceedings of C3S2E'10*, pages 108–113. ACM, May 2010

- S. A. Mokhov. The role of self-forensics modeling for vehicle crash investigations and event reconstruction simulation. In J. S. Gauthier, editor, *Proceedings of the Huntsville Simulation Conference (HSC'09)*, pages 342–349. SCS, Oct. 2009. Online at http://arxiv.org/abs/0905.2449

- S. A. Mokhov and E. Vassev. Self-forensics through case studies of small to medium software systems. In *Proceedings of IMF'09*, pages 128–141. IEEE Computer Society, Sept. 2009

- S. A. Mokhov. Towards improving validation, verification, crash investigations, and event reconstruction of flight-critical systems with self-forensics. [online], June 2009. A white paper submitted in response to NASA's RFI NNH09ZEA001L, http://arxiv.org/abs/0906.1845, mentioned in http://ntrs.nasa.gov/archive/nasa/casi.ntrs.nasa.gov/20100025593_2010028056.pdf

- S. A. Mokhov, J. Paquet, and M. Debbabi. Towards formal requirements specification of self-forensics for autonomous systems. Submitted for review to J. Req. Eng., 2009–2013

# Part I

# Background

# Chapter 2

# Cyberforensics

In this chapter we review the background work in the domain of digital crime investigations. The bulk of the related literature pertinent to this thesis has been summarized here in detail for the readers who want to remind themselves of the core literature that is used in this research (most prominently the FSA formal approach to digital investigations and other formal approaches in cyberforensics). The major literature cited is the one that is either the most fundamental and inspiring work for this thesis and/or of some supplemental relevance and support of this work as this thesis was gradually developed over the years. It is not meant to be a comprehensive survey of the "cutting edge" of the state of the art on the matter. More specifically we review the general aspects of *Forensic Computing* in Section 2.1, the core formal FSA approach by Gladyshev to evidence representation and event reconstruction with examples in Section 2.2, some more recent related approaches are discussed in Section 2.3, and concluding remarks are in Section 2.4.

## 2.1 Forensic Computing

> *Gathering and analyzing data in a manner as free from distortion or bias as possible to reconstruct data or what happened in the past on a system [or a network]* —Dan Farmer, Wietse Venema (1999) [252]

Many ideas in this work come from computer forensics and forensic computing [138, 139, 525]. Both have traditionally been associated with computer crime investigations (cf., Section 1.6.1.2) to seize the evidence, live or "dead", memory contents, disk contents, log files off any storage and computing devices, followed by, among other things, information extraction and analysis [308, 322]. There is a wide scope of research done in forensic computing, including formal and practical approaches, methodologies, and the use associated tools [21, 22, 85, 136, 137, 152, 357] (see Section 2.1.3) [322] and the annual *Digital Forensic Research Workshop (DFRWS)* was established [357] to track progress in the field along with other venues, such as IMF [138, 139], ADFSL, etc.

Digital forensics is also a prominently growing art in court of law with practitioners including attorneys, such as Craig Ball [37], who become more proficient in the digital investigation and the tools, but who are not as widely available as attorneys for other law disciplines because due to very detailed technical aspects of the digital investigation practice that is usually restricted to technical expert witnesses found in law enforcement and similar agencies.

Forensic computing relies mostly on similar approaches in the evidence gathering and usage as the traditional way, summarized below by Lee [231] (where "substance" in item 8 in our case are digital data):

> *Following are the objectives of utilization of forensic evidence found at a crime scene in any [investigation];*
>
>   1. *Information on the corpus delicti;*
>   2. *Information on the modus operandi;*
>   3. *Linkage of persons to other persons, objects, or scenes;*
>   4. *Linkage of evidence to persons, objects or locations;*
>   5. *Determining or eliminating the events and actions that occurred;*
>   6. *Disproving or supporting witness statements or testimony;*
>   7. *Identification or elimination of a suspect;*
>   8. *Identification of unknown substance;*
>   9. *Reconstruction of a crime;*
>   10. *Providing investigative leads.*
>
> *H.C. Lee [231]*

Forensic computing is broadly separated into two aspects: "dead" analysis and "live" analysis. The former is more traditional (working on data off a disk or log files some time after the incident occurred and the computing equipment was powered off) and the latter emerged and gained popularity to capture any live volatile data that may provide more insight on the incident while it is still available (such as memory content that may have unencrypted data and other live data structures, process list, current network connections and traffic flows, current open files, etc., possibly gathering the live data while the incident is unfolding) [59, 252, 373]. *Live forensics* according to McDougal [252] often actually constitutes Incident Response [247] while volatile evidence is "hot" and may still be available. Live forensics is also common in active "honeypots" deployed [172, 464] as a honeypot to lure in attackers to weakened virtual hosts networks and observe their activity to learn and know what and how they attack; the most prominent example is the HoneyNet project [172]. Despite its growing attractiveness, live forensics has also been criticized for potential risks that it may introduce [59], e.g., when an attacker deliberately tries to actively circumvent or feed/leave bogus and noisy data around in disguise for real data making live data findings sometimes less reliable and untrustworthy as evidence (unless the attacker is not in a position to poison the real data or prevent the live analysis tools running on the system under investigation [59]). In practice both live and dead analyses are often used together (e.g., as done by one of our operational case studies in Section 9.5). A combination of live forensics and dead analysis in autonomic environment is termed as *self-forensic computing* (detailed in Appendix D.2.1).

### 2.1.1 Tracing for Event Reconstruction

A lot of aspects in forensic computing have to do with tracing of states, activity, and events via (sys)logging. As a part of the related work on what we call *self-forensics* (see Appendix D for an in-depth discussion) there has been some preliminary work done in the past that was not identified as belonging to that notion. For example, a state-tracing Linux kernel [150] exhibits some elements of self-forensics by tracing its own state for the purposes of forensic analysis [322]. ASPECTJ [29]'s (see Section 8.2.1, page 213 for in-depth discussion) tracing capability of Java programs provides a way of tracing by the observing aspects collecting the forensic data (e.g., data structures content) during the forward tracing of the normal execution flow. (From the self-forensics standpoint this is useful given a large number of web-based applications and distributed systems deployed today are written in JAVA, the use of ASPECTJ, which isn't coupled with the applications' source code and can observe application objects independently, provides a good basis for a self-forensics middleware [322].)

### 2.1.2 Computer Anti-Forensics

There are opposing forces to the forensic analysis (*computer forensics*—CF) and investigation that stand in the way and are also a part of (anti-)forensic computing (*computer anti-forensics*—CAF). Such forces aim at attacking forensic tools or techniques.

As an example, Dahbur and Mohammad introduced [76] the *anti-forensics challenge*s in a brief survey of tools and techniques in the field of anti-forensics together with their classification [288] (excluding network anti-forensics aspects) [288]. In doing so, the authors took the white-hat position on the problem to combat anti-forensics (i.e., the implied "anti-anti-forensics") for successful investigation. They lay a foundation for the terminology of the anti- side [288]. Then follows the problem space definition as a context in the geopolitical spread of data usage on the Internet worldwide [471] including the definition of the relevant and introductory terminology and literature [288]. That allows the authors to raise several sets of classification categories from the surveyed literature based on attack targets, (non-)traditional techniques, functionality, and distinguishing between anti-forensics (forensic analysis prevention by data hiding via cryptographic or steganographic tools) and counter-forensics (direct attack on forensic tools) [288]. They examine the problem from the point of view of constraints found in CF: temporal, financial, and other environmental aspects. This prompts to explore more deeply the challenges posed by CAF to the investigator by describing the evolution of the privacy technologies available to users, encryption, compression bombs, cloud computing, steganography, etc.—overall the nature of the digital evidence and where CAF attacks may come from in the attempt to delay or halt the investigation [288].

On a related topic, Dahbur and Mohammad overlooked some of the related work that can help addressing some of the challenges posed by CAF, specifically on spectral file type analysis via machine learning instead of magic signatures, which can detect file types with relatively good precision when the magic numbers are altered ([290], Section 5.3, page 107).

They also overlook standard Unix utilities that have existed for ages and that can change timestamps, for example, `touch` [409] and what possible attacks on such tools may be [288]. The machine learning approach (Chapter 5) has been shown to work well as well for network forensics quite reliably [49] that in part can address the CAF problem at the network layer.

## 2.1.3 Forensic Tools, Techniques, and Methodologies

> *Dan Farmer and Wietse Venema are generally credited (1999) with the creation of computer forensics as we know it today. They are also the author of one of the [first] freeware tools for doing forensics named The Coroner's Toolkit (TCT). While this tool suit has generally been expanded and enhanced by many others, it certainly is the basis of modern computer forensics at least within the \*NIX world. —Monty McDougal [252]*

This section briefly summarizes the tools and techniques surveyed during this work. The reader is assumed to have some familiarity with or exposure to the languages, tools, techniques, and the concepts listed here. If it is not the case, please refer to the references cited on the subjects in question.

Digital forensic investigation tools have to do with various computing aspects, such as most commonly memory, log, and disk analysis in search for all kinds of evidence, cataloging it properly, and presenting the results. Other aspects in digital investigation touch various operating systems, especially after a compromise due to bad security controls or a misuse [19, 434], including in virtual machines security [356] since virtualization and cloud computing are gaining a lot of prominence lately. Related forensic computing tools also have to do with malware analysis [185, 235, 433] and fingerprinting of malware and its traffic [49, 289, 314]. These activities are becoming a norm for certified computer and network security professionals [54, 97, 444].

There are several forensic toolkits in the open-source, academic, and commercial worlds. Dead/live analysis tools include Sleuthkit along with the chain of custody support, Helix [374] with common tools for most OSes such as TCT [461], Sleuthkit and its Autopsy Browser [60, 61], Windows Forensics Toolkit (WFT) [252], EnCase [149], FTK [2], and others.

We project to have our work included into one or more of those either as a plug-in if the host environment is in JAVA like Forensic Toolkit as JPF Plug-ins [21, 22, 85], Ftklipse [225, 226, 227], or others [249], or as a standalone tool by itself for inclusion into more general forensic toolsets for binary data analysis extracted from files for preliminary classification file data alongside `stegdetect` [386] and many others. We are considering to make it work with the Linux Sleuthkit [60, 61] and commercial tools, such as FTK [2], EnCase [54, 149] and inclusion in the relevant Linux distributions as a part of our future work [290] (see Section 10.4).

### 2.1.4 Formal Cyberforensic Analysis

An increasingly important aspect of forensic computing has to do with formalisms recently developed or being developed to the process of conceiving and applying scientific methodology to cyberforensic investigations. This aspect is a primary interest and pursuit of this thesis. The first more prominent work in that direction discussed is that of Gladyshev *et al.* [83, 136, 137], followed by that of Debbabi *et al.* from the Computer Security Laboratory, Concordia University, from [21, 22, 85] (subsequent discussed in Section 2.3), and more specifically of the author Mokhov *et al.* [269, 274, 300, 304, 307, 310], which comprise the contributions of this thesis. Thus, this chapter's primary remaining focus is on such a formal methodology detailed further in Section 2.2 that serves the as one of the core founding principles of the FORENSIC LUCID foundations described in Chapter 7.

## 2.2 Gladyshev's Formal Approach to Cyberforensic Investigation and Event Reconstruction

For the readers to further understand better the methodology and contribution of this thesis (presented further in Chapter 7), we review in depth the details of the Gladyshev's solution to forensic formalization here along with the two example cases. These formalisms of the FSA approach are recited from the related works of Gladyshev *et al.* [135, 136, 137].

### 2.2.1 Overview

This section reviews in-depth the earlier (Section 1.6.1.2) mentioned first formal approach to cyberforensic analysis that appeared in the works [135, 136, 137] by Gladyshev *et al*. To recapitulate, Gladyshev [135] relies on finite state automata (FSA) and their transformation and operation to model evidence and witnesses accounts, and the subsequent evaluation of their possible explanations (meanings). These are specifically recited in Section 2.2.2 (FSA and terminology definitions), Section 2.2.3 (approach to backtracing for event reconstruction), and Section 2.2.4 (evidence formalization, the core Gladyshev's contribution [135]). Additionally, his works present two example use-cases: the ACME Printing Case and Blackmail Case investigations, reviewed subsequently in Section 2.2.5.

### 2.2.2 Definitions

In Gladyshev's formalization [135], a *finite state machine* is a sequence of four elements $T = (Q, I, \psi, q)$, where:

- $Q$ is a finite set of all possible states

- $I$ is a finite set of all possible events

- $\psi : I \times Q \to Q$ is a transition function that determines the next state for every possible combination of each state and event

- $q \in Q$ is the current system state

A *transition* is the process of state change [83, 135]. Transitions are instantaneous (no duration). A finite computation is a non-empty finite sequence of steps $c = (c_0, c_1, \ldots, c_{|c|-1})$ where each step is a pair $c_j = (c_j^l, c_j^q)$, in which $c_j^l \in I$ is an event, $c_j^q \in Q$ is a state, and any two steps $c_k$ and $c_{k-1}$ are related via a transition function: $\forall k$, such that $1 \leq k < |c|$, $c_k^q = \psi(c_{k-1}^l, c_{k-1}^q)$. The set of all finite computations of the finite state machine $T$ is denoted $C_T$ [83, 135].

A *run* is a sequence of computations $r \in (C_T)^{|r|}$, such that if $r$ is non-empty, its first element is a computation $r_0 \in C_T$, and for all $1 \leq i < |r|, r_i = \psi(r_{i-1})$, where function $\psi$ discards the first element of the given computation [83, 135]. For two computations $x \in C_T$ and $y \in C_T$, a relationship $y = \psi(x)$ exists, if and only if $x = x_0.y$. The *set of all runs* of the finite state machine $T$ is denoted $R_T$. The *run of computation $c$* is a run, whose first computation is $c$ [83, 135]. Any run $r$ is completely determined by its length and its first computation: a computation is a sequence, a run is a computation plus length; the run defines a progress within a computation (see Figure 4). Gladyshev uses runs as an *explanation model* of *an observation* [83, 135].

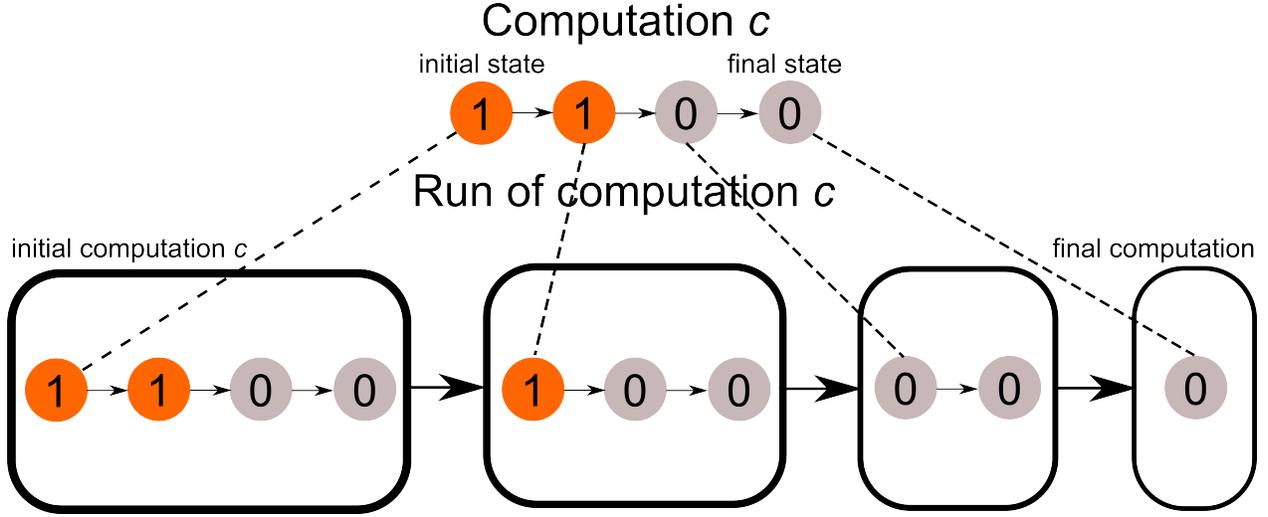

Figure 4: Run of a computation $c$ [135]

A *partitioned run* is a finite sequence of runs $pr \in (R_T)^{|pr|}$, such that concatenation of its elements in the order of listing is also a run [83, 135]:

$$(pr_0.pr_1.pr_2.\cdots.pr_{|pr|-1}) \in R_T \qquad (2.2.2.1)$$

A *set of all partitioned runs* is denoted $PR_T$ [83, 135]. A *partitioning* of run $r \in R_T$ is a partitioned run denoted $pr_r$, such that concatenation of its elements produces $r$:

$$(pr_{r0}.pr_{r1}.pr_{r2}.\cdots.pr_{r|pr|-1}) = r \qquad (2.2.2.2)$$

A partitioned run is used as an explanation model of *an observation sequence* [83, 135]. The condition on the concatenation is to reflect the fact that observations happened after each other, without gaps in time. A partitioned run is determined by a sequence of pairs where the first element of a pair is a computation and the second element is a length [83, 135].

### 2.2.3 Back-tracing

Further, Gladyshev defines the inverse of $\psi$—a function $\psi^{-1}: C_T \to 2^{C_T}$. For any computation $y \in C_T$, it identifies a subset of computations, whose tails are $y : \forall x \in \psi^{-1}(y), y = \psi(x)$. In other words, $\psi^{-1}$ back-traces the given computation [83, 135]. Subsequently, a modified function $\Psi^{-1}: 2^{C_T} \to 2^{C_T}$ is defined as: for $Y \subseteq C_T, \Psi^{-1}(Y) = \bigcup^{\forall y \in Y} \psi^{-1}(y)$. This definition, in addition to uniformity, allows for easier implementation [83, 135]. Both inverse functions are illustrated in Figure 5.

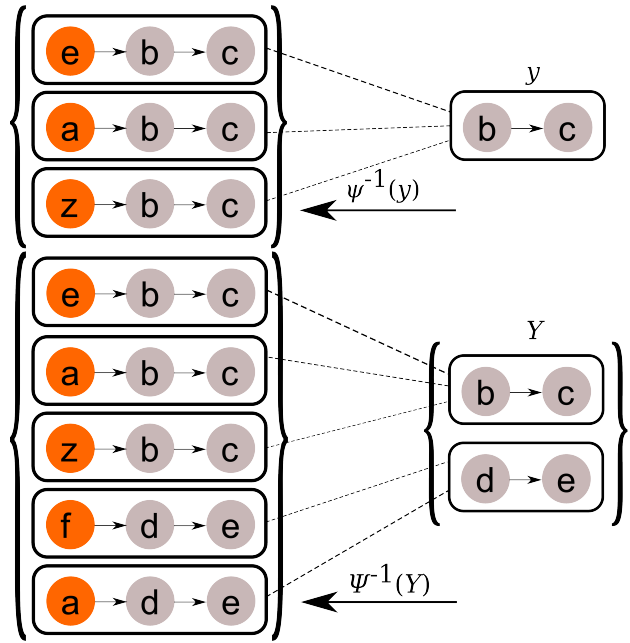

Figure 5: Backtracking example in $\psi^{-1}(y)$ and $\Psi^{-1}(Y)$ [135]

### 2.2.4 Formalization of Evidence

Gladyshev formalizes the evidence (observations) from the observed event flow understanding and storytelling [135]. Every piece of evidence tells its own "story" of the incident. The goal of event reconstruction can be seen as a combination of stories told by witnesses and by various pieces of evidence to make the description of the incident as precise as possible [83, 135]. An *observation* is a statement that system behavior exhibited some property $P$ continuously for some time. Formally, it is defined as a triple $o = (P, min, opt)$, where $P$ is the set of all computations in $T$ that possess the observed property, *min* and *opt* are non-negative integers that specify the duration of an observation [83, 135]. An *explanation* of observation $o$ is a run $r \in R_T$ such that every element of the run $r$ possesses the observed property: for all $0 \leq i < |r|$, $r_i \in P$, and the length of the run $r$ satisfies *min* and *opt*: $min \leq |r| \leq (min+opt)$. The meaning of observation $o$ is the set $R_o \subseteq R_T$ of all runs that explain $o$ [83, 135].

The duration is modeled as lengths of computations. A run explains an observation if the length of the run satisfies the *min* and *opt* requirements, and, each element of the run $r_i$ should possess the property $P$, i.e., the computation $r_i \in P$ [83, 135].

### 2.2.4.1 Types of Observations

Gladyshev divides observations into several types [83, 135]:

- A *fixed-length observation* is an observation of the form of $(P, x, 0)$. Any run explaining it has a length of $x$.

- A *zero-observation* is an observation of the form of $(P, 0, 0)$. The only run explaining it is an empty sequence.

- A *no-observation* is an observation $\$ = (C_T, 0, infinitum)$ that puts no restrictions on computations. The $infinitum$ is an integer constant that is greater than the length of any computation that may have happened (all runs satisfy this observation).

- A *generic observation* is an observation with variable length, i.e., $y > 0$ in $(P, x, y)$. See Section 2.2.4.5 for more details.

### 2.2.4.2 Observation Sequence

An *observation sequence*, per Gladyshev [135], is a non-empty sequence of observations listed in chronological order:

$$os = (observation_A, observation_B, observation_C, \ldots) \qquad (2.2.4.1)$$

An *observation sequence* represents an uninterrupted "eyewitness" (without gaps) story. The next observation in the sequence begins immediately when the previous observation finishes [83, 135]. Gaps in the story are represented by *no-observations* [83, 135], i.e.:

$$\$ = (C_T, 0, infinitum)$$

An *explanation of observation sequence os* is a partitioned run $pr$ such that the length of $pr$ is equal to the length of $os$: $|pr| = |os|$, and each element of $pr$ explains the corresponding observation of $os$ [83, 135]:

$$\forall i : 0 \leq i \leq |os|, pr_i \in R_{os_i} \qquad (2.2.4.2)$$

$$os = (\quad o_1 \quad . \quad o_2 \quad . \quad \cdots \quad . \quad o_n \quad)$$
$$pr = (\quad pr_1 \quad . \quad pr_2 \quad . \quad \cdots \quad . \quad pr_n \quad)$$

where each $pr_i$ is a run that is an explanation of $o_i$ [83, 135]. The *meaning of observation sequence os* is a set $PR_{os} \subseteq 2^{(R_T)^{|os|}}$ of all partitioned runs that explain $os$. A run $r$ satisfies an observation sequence $os$ if and only if there exists a partitioning of $r$ that explains $os$ [83, 135]. There may be more than one partitioning of $r$ that explains $os$ as shown in Figure 6. A computation $c$ satisfies an observation sequence $os$ if and only if there is a run $r$ that satisfies $os$ and $r_0 = c$ [83, 135].

#### 2.2.4.3 Evidential Statement

An *evidential statement* is a non-empty sequence of observation sequences:

$$es = (os_A, os_B, os_C, \ldots) \qquad (2.2.4.3)$$

where ordering of the observation sequences is not important [83, 135]. Each observation sequence is a version of the story. Each principal (i.e., witness) will have their own version (i.e., observation sequence) [83, 135]. An *explanation of an evidential statement es* is a

$$os = ((P_1, 1, infinitum), (P_2, 1, infinitum))$$

$$P_1 = \{\boxed{S_1}, \boxed{S_2}, \boxed{S_3}\}$$
$$P_2 = \{\boxed{S_3}, \boxed{S_4}, \boxed{S_5}\}$$

$$r = \boxed{S_1} \rightarrow \boxed{S_2} \rightarrow \boxed{S_3} \rightarrow \boxed{S_4} \rightarrow \boxed{S_5}$$

$$pr_{A1} = \boxed{S_1} \rightarrow \boxed{S_2} \rightarrow \boxed{S_3}$$
$$pr_{A2} = \boxed{S_4} \rightarrow \boxed{S_5}$$
$$pr_{B1} = \boxed{S_1} \rightarrow \boxed{S_2}$$
$$pr_{B2} = \boxed{S_3} \rightarrow \boxed{S_4} \rightarrow \boxed{S_5}$$

$$explanation_A = (\boxed{S_1} \rightarrow \boxed{S_2} \rightarrow \boxed{S_3}, \boxed{S_4} \rightarrow \boxed{S_5})$$
$$explanation_B = (\boxed{S_1} \rightarrow \boxed{S_2}, \boxed{S_3} \rightarrow \boxed{S_4} \rightarrow \boxed{S_5})$$

where $r$ is a run, $pr_i$ are various partitions of $r$ (which are also runs), and each explanation is a partitioned run.

Figure 6: Explanations as partitioned runs of an observation sequence [135]

sequence of partitioned runs $spr$, such that all elements of $spr$ are partitionings of the same run:

$$spr_{0,0} \cdot spr_{0,1} \cdot \ldots \cdot spr_{0,|spr_0|-1} =$$
$$spr_{1,0} \cdot spr_{1,1} \cdot \ldots \cdot spr_{1,|spr_1|-1} =$$
$$\ldots$$
$$spr_{|es|-1,0} \cdot spr_{|es|-1,1} \cdot \ldots \cdot spr_{|es|-1,|spr_{|es|-1}|-1} = r$$

In other words, it should be a story that explains all the versions and each element of $spr$ explains the corresponding observation sequence in $es$, more specifically: for all $0 \leq i < |es|, spr_i \in PR_{es_i}$ [83, 135].

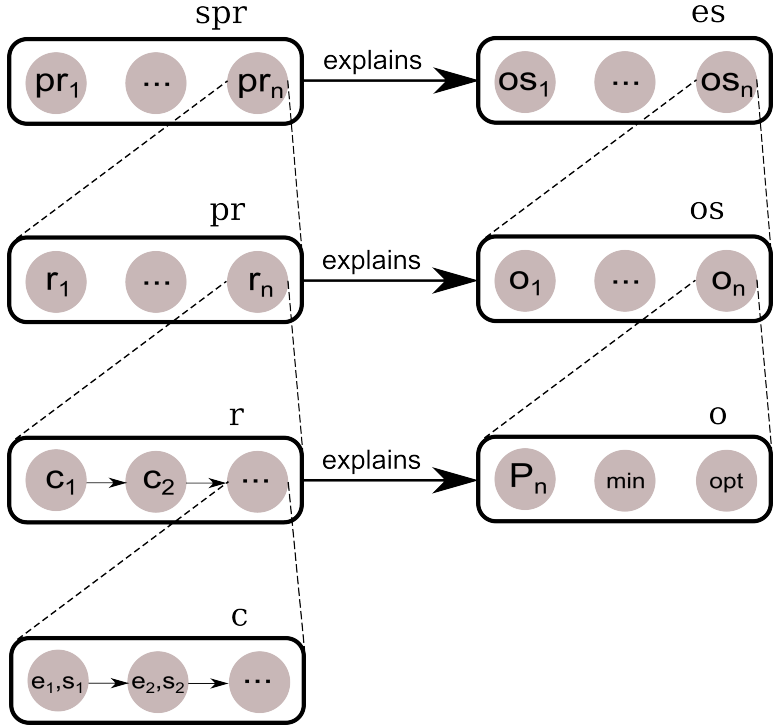
Figure 7: Meaning and explanation hierarchy [135]

The *meaning of evidential statement es* is a set of all sequences of partitioned runs $SPR_{es} \subseteq (PR_{es_0} \times PR_{es_1} \times \ldots \times PR_{es_{|es|-1}})$ that explains *es* [83, 135]. An evidential statement is *inconsistent* if it has an empty set of explanations. The summary of all explanations is in Figure 7 [83, 135].

#### 2.2.4.4 Event Reconstruction Algorithm

Gladyshev's defined *event reconstruction algorithm* (ERA) involves computing the meaning of the given evidential statement with respect to the given state machine [83, 135]. There are two types of observation sequences: *fixed-length observation sequences* and *generic observation sequences* (described in the next section). The meanings of individual observation sequences are *combined* in order to yield the meaning of the evidential statement [83, 135].

**2.2.4.4.1 Fixed-Length Observation Sequences.** Here we review Gladyshev's way to compute the meaning of fixed-length observation sequences [135]. The function $\Psi^{-1}$ provides basic operations to automate back-tracing and proceeds as follows. First, take the set of all computations $C_T$ as the starting point and iteratively back-trace it into the past using

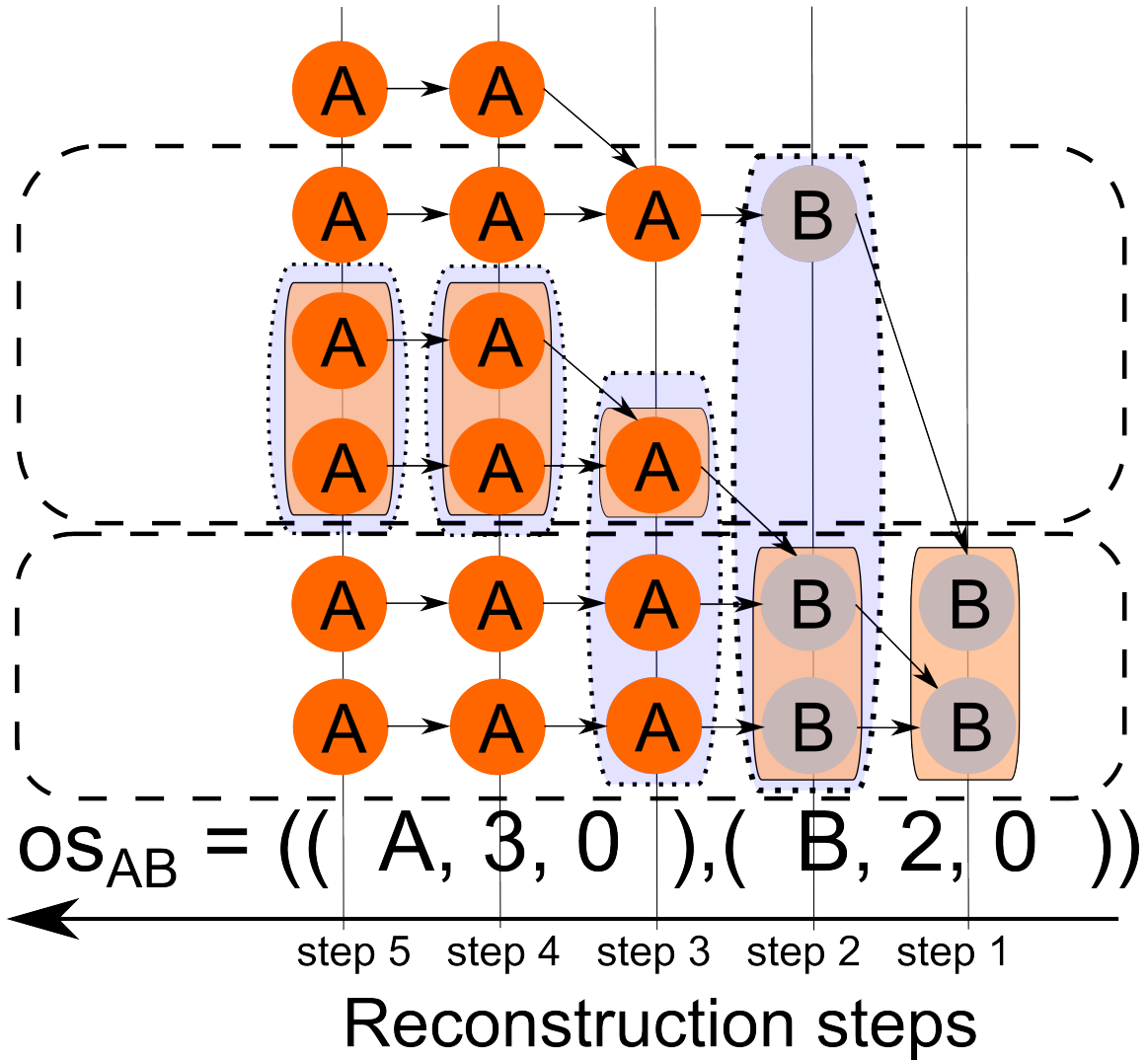

Figure 8: Fixed-length event reconstruction [135]

$\Psi^{-1}$ [83, 135]. At each step, computations that do not possess observed property $P$ are discarded. This is achieved by intersecting the set of back-tracings with the set of computations that possess the property observed at the current step [83, 135]. The result of the intersection is then used as an input for the next invocation of $\Psi^{-1}$. The process continues until either all observations are explained, or the set of computations becomes empty. An illustration of the fixed-length event reconstruction is in Figure 8 [83, 135].

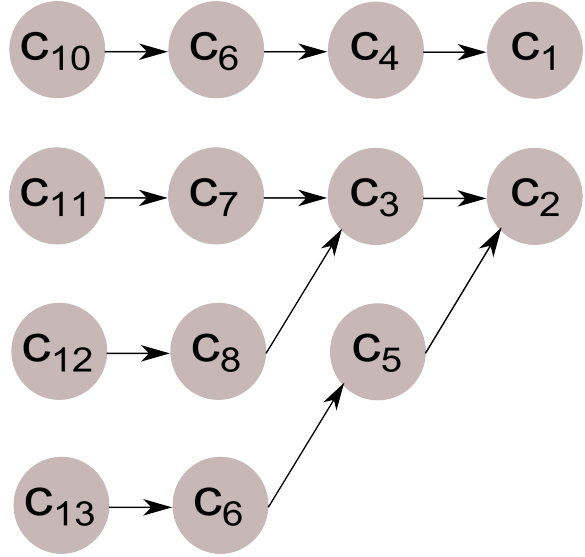

Figure 9: Example computation event sequences

**2.2.4.4.2 Example.** As an example, Gladyshev calculates the meaning of the following fixed-length observation sequence [83, 135]:

$$os_{AB} = (A, 3, 0)(B, 2, 0)$$

Knowing that:

$$A = \{c_6, c_8, c_{10}, c_{12}\}$$
$$B = \{c_1, c_2, c_3, c_4\}$$
$$C_T = \{c_1, \ldots, c_{13}\}$$

such that the relationship of the events is as shown in Figure 9. There is an arrow from $c_x$ to $c_y$ if $\psi(c_x) = c_y$ [83, 135].

**2.2.4.4.3 Meaning Computation.** The observation $os_{AB}$ is equivalent to a sequence $AAABB$. The meaning of $os_{AB}$ can then be computed as follows [83, 135]:

- 2B intersections:

$$B \cap C_T = B$$
$$\Psi^{-1}(B \cap C_T) = \Psi^{-1}(B) = \{c_4, c_3, c_5, c_7, c_8, c_6\}$$
$$B \cap \Psi^{-1}(B \cap C_T) = \{c_3, c_4\}$$
$$\Psi^{-1}(B \cap \Psi^{-1}(B \cap C_T)) = \{c_6, c_7, c_8\}$$

- 3A intersections:

$$A \cap (\Psi^{-1}(B \cap \Psi^{-1}(B \cap C_T))) = \{c_6, c_8\}$$
$$\Psi^{-1}(A \cap (\Psi^{-1}(B \cap \Psi^{-1}(B \cap C_T)))) = \{c_{10}, c_{12}\}$$
$$A \cap (\Psi^{-1}(A \cap (\Psi^{-1}(B \cap \Psi^{-1}(B \cap C_T))))) = \{c_{10}, c_{12}\}$$
$$\Psi^{-1}(A \cap (\Psi^{-1}(A \cap (\Psi^{-1}(B \cap \Psi^{-1}(B \cap C_T)))))) = \{c_{10}, c_{12}\}$$
$$A \cap (\Psi^{-1}(A \cap (\Psi^{-1}(A \cap (\Psi^{-1}(B \cap \Psi^{-1}(B \cap C_T))))))) = \{c_{10}, c_{12}\}$$

**2.2.4.4.4 Map of Partitioned Runs.** To calculate the meaning of an observation sequence Gladyshev [135] uses a set of partitioned runs. A *map of partitioned runs* (MPR) is a representation for a set of partitioned runs. It is a pair $pm = (len, C)$ where $C$ is the set of computations, and *len* is a sequence of their lengths. An MPR could represent the set of all partitioned runs whose initial computations are in $C$, and whose partitions have lengths: $len_0, len_1, \ldots, len_{|len|-1}$ [83, 135]. Gladyshev states that the meaning of a *fixed-length* observation sequence can be expressed by a single MPR. Let $C = \{c_1, c_2, \ldots, c_n\}$ and $len = \langle l_1, l_2, \ldots, l_m \rangle$. Let $MPR = (C, len)$, then its possible illustration is in Figure 10 [83, 135].

### 2.2.4.5 Generic Observation Sequences

For a *generic observation*, whose $opt \neq 0$, the length of the explaining run is not fixed, but is bounded between $min$ and $min + opt$ [83, 135] (Section 2.2.4, page 32). A single observation sequence represents many permutations of linking observed properties, see Equation 2.2.4.4

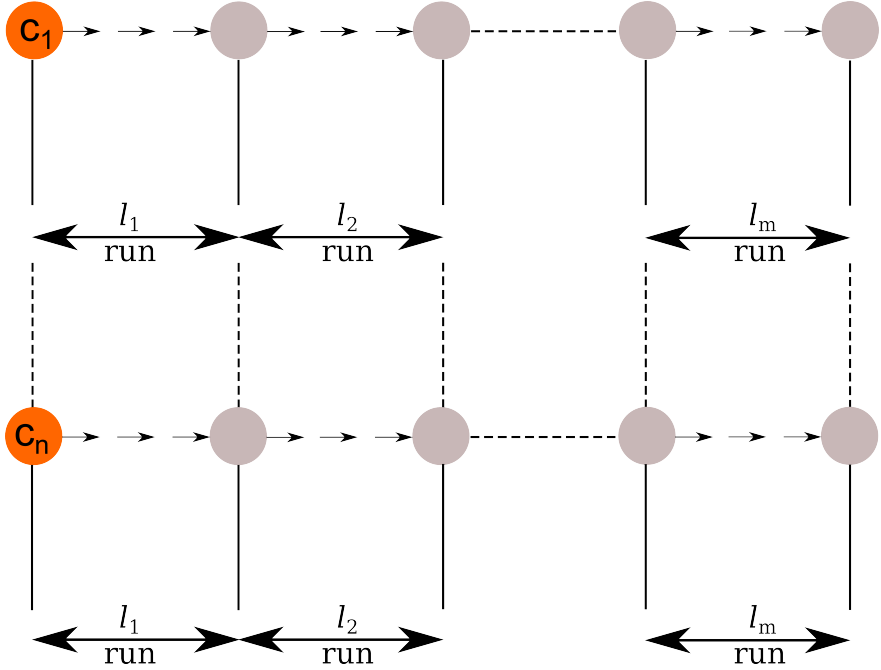
Figure 10: Example of construction of MPR [83]

and Figure 11 for an example [83, 135].

$$os_{AB2} = ((A, 1, 3), (B, 1, 2)) \tag{2.2.4.4}$$

| | | |
|---|---|---|
| $AB$ | $ABB$ | $ABBB$ |
| $AAB$ | $AABB$ | $AABBB$ |
| $AAAB$ | $AAABB$ | $AAABBB$ |
| $AAAAB$ | $AAAABB$ | $AAAABBB$ |

Figure 11: Generic observation sequence permutations example

As can be seen, there are twelve possible variants of linking properties, which is essentially a cross-product of the lengths [83, 135]. Every permutation can be represented by a fixed length observation sequence. The meaning of $os_{AB2}$ is, then, the union of the meanings of each variant [83, 135].

$$(A, 1, 3) = U = \{(A, 1, 0), (A, 2, 0), (A, 3, 0), (A, 4, 0)\}$$
$$(B, 1, 2) = V = \{(B, 1, 0), (B, 2, 0), (B, 3, 0)\}$$
$$\langle (A, 1, 3)(B, 1, 2) \rangle = U \times V$$

For instance, $(A, 2, 0)(B, 3, 0) \in U \times V$ is written as $AABBB$, i.e., an observation sequence that has the $A$-property for two steps followed by three steps that have the $B$-property [83, 135]. Gladyshev has shown how to calculate the meaning of a fixed-length observation sequence (see Section 2.2.4.4.4). In order to calculate the meaning of a variable length observation sequence, one computes the union of the meanings of each fixed-length observation sequence. This can be modeled as a set of MPRs. Each MPR is the meaning of a fixed-length observation sequence [83, 135].

### 2.2.4.6 Meaning of an Evidential Statement

First, the meanings of individual observation sequences are computed as described earlier. Second, Gladyshev combines the meanings of observation sequences into the meaning of the entire evidential statement. To satisfy the evidential statement, a run must satisfy all of its observation sequences [83, 135]. Then Gladyshev states a problem of identifying the subset of runs whose partitionings are present in the meanings of all observation sequences [83, 135].

#### 2.2.4.6.1 Towards the Meaning of an Evidential Statement.

To recapitulate, the meaning of an observation sequence is a set of MPRs. Subsequently, each MPR is the meaning of a fixed-length observation sequence. Further, an evidential statement is a set of observation sequences [83, 135]. Therefore, per Gladyshev, the meaning of an evidential statement is the *combination of the sequence of MPRs* with each MPR being the meaning of a particular fixed-length observation sequence of the evidential statement. Accordingly, the meaning of an evidential statement computation reduces to a problem of how to combine two MPRs [83, 135].

Let $pm_a = (len_a, C_a)$ and $pm_b = (len_b, C_b)$ be two MPRs. A run $r$ can be partitioned by both $pm_a$ and $pm_b$ if and only if two conditions hold [83, 135]:

1. The initial computation of run $r$ belongs to initial computation sets of both MPRs: $r_0 \in C_a$ and $r_0 \in C_b$ [83, 135].

2. Both MPRs have equal total number of computation steps: $\sum len_a = \sum len_b$. If $\sum len_a \neq \sum len_b$, then the two MPRs have no common runs. Otherwise, the common runs are determined by the common set of initial computations $C_a \cap C_b$ [83, 135].

**2.2.4.6.2  Map of Sequence of Partitioned Runs.** Ascending further per Gladyshev [135], a *map of a sequence of partitioned runs* (MSPR), written out as:

$$mspr = ((len_0, len_1, \ldots, len_n), C)$$

is a representation for a set of sequences of partitioned runs [83, 135]. $C$ is the set of initial computations, and $len_0, len_1, \ldots, len_n$ are lists of lengths that describe how to partition runs generated from the elements of $C$. An MSPR is proper if and only if $\sum len_0 = \sum len_1 = \ldots = \sum len_n$. The combination of two MPRs is defined by function $comb()$ that takes two MPRs and returns a proper MSPR [83, 135]:

$$comb(pm_x, pm_y) = \begin{cases} \emptyset & \text{if } \sum len_x \neq \sum len_y \text{ or } C_x \cap C_y = \emptyset \\ ((len_x, len_y), C_x \cap C_y) & \text{otherwise} \end{cases}$$
(2.2.4.5)

**2.2.4.6.3  Use of Combination.** Gladyshev gives a typical usage example of the combination (Equation 2.2.4.5) in the example that follows [135]. Suppose that the meanings of two observation sequences $os_a$ and $os_b$ are represented by two sets of MPRs called $PM_a$ and $PM_b$ respectively. The meaning of the evidential statement $es = (os_a, os_b)$ is expressed by the set of proper MSPRs, which is obtained by combining every MPR from $PM_a$ with every MPR from $PM_b$ [83, 135]:

$$\forall x \in PM_a, \forall y \in PM_b, SPM_{es} = \bigcup comb(x, y) \quad (2.2.4.6)$$

This procedure can be extended to an arbitrary number of observation sequences, thus providing a way to calculate the meaning of an arbitrary evidential statement [83, 135]. As a result, the below is the representation of the previously calculated meaning as an MPR in Section 2.2.4.4.3:

$$MPR(os_{AB}) = (\langle 2, 2 \rangle, \{c_{10}, c_{12}\})$$

**2.2.4.6.4 Computing Combinations of MPRs.** More examples are illustrated further in the consideration of the following MPRs [83, 135]:

$$MPR_1 = (\langle 2, 1, 4\rangle, \{c_1, c_2, c_3\})$$
$$MPR_2 = (\langle 3, 4\rangle, \{c_4, c_5, c_6\})$$
$$MPR_3 = (\langle 4, 4\rangle, \{c_1, c_2, c_3, c_4\})$$
$$MPR_4 = (\langle 5, 2, 1\rangle, \{c_2, c_3, c_5\})$$
$$comb(MPR_4, MPR_3) = ((\langle 4, 4\rangle, \langle 5, 2, 1\rangle), \{c_2, c_3\})$$
$$comb(MPR_1, MPR_2) = \emptyset$$

## 2.2.5 Investigation Cases Examples Using the FSA Approach

Further in this section, what follows is the summary of the FSA approach from [136, 137] applied to two case studies to show its inner workings [305]. Gladyshev's printing case review is in Section 2.2.5.1 and the blackmail case review is in Section 2.2.5.2 accordingly.

We extract the parameters and terms defined in the mentioned works, such as the formalization of the various pieces of evidence and witnesses telling their stories of a particular incident. The goal is to put such statements together to make the description of the incident as precise as possible and focus on the modeling of the core aspect of the case under the investigation. As a result, to establish that a certain claim may be true, the investigator has to demonstrate that there are some meaningful explanations of the evidence that agree with the claim. Conversely, to disprove the claim, the investigator has to show there are no explanations of evidence that agree with that claim [83, 136].

Gladyshev did initial proof-of-concept realization of the algorithms' implementation in CMU COMMON LISP [137] that we improve by re-writing it in FORENSIC LUCID [305]. We subsequently re-model these cases using the new approach in Section 9.3 and Section 9.4, to show for it to be more usable in the actual investigator's work and serve as a basis to further development in the area [305] (such as a GUI front end based on data-flow graphs [89, 309]).

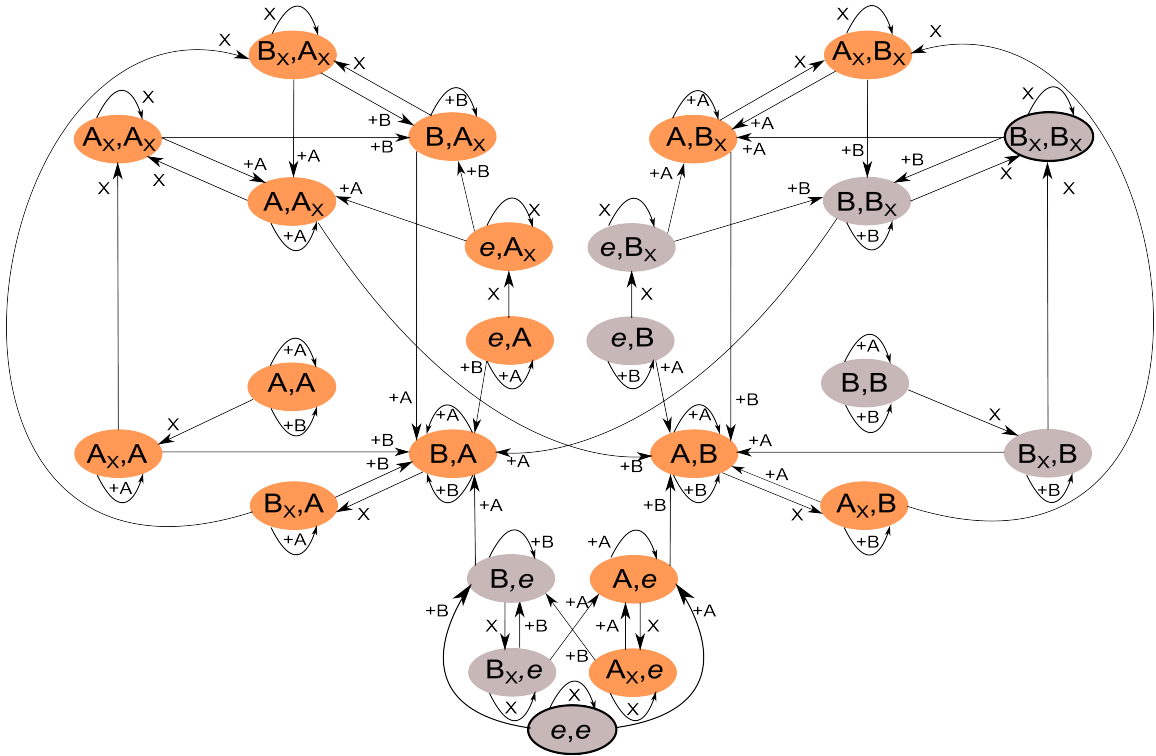

Figure 12: Printer Case state machine [135]

### 2.2.5.1 ACME Printer Case in FSA

This one is the first of the cases we re-examine from Gladyshev's FSA/LISP approach [137].

> *The local area network at some company called ACME Manufacturing consists of two personal computers and a networked printer. The cost of running the network is shared by its two users Alice (A) and Bob (B). Alice, however, claims that she never uses the printer and should not be paying for the printer consumables. Bob disagrees, he says that he saw Alice collecting printouts. According to the manufacturer, the printer works as follows [137]:*
>
> 1. *When a print job is received from the user, it is stored in the first unallocated directory entry of the print job directory.*
> 2. *The printing mechanism scans the print job directory from the beginning and picks the first active job.*
> 3. *After the job is printed, the corresponding directory entry is marked as "deleted", but the name of the job owner is preserved.*

4. The printer can accept only one print job from each user at a time.

5. Initially, all directory entries are empty.

The investigator finds the current state of the printer's buffer as:

1. Job From B Deleted

2. Job From B Deleted

3. Empty

4. Empty

5. ...

**2.2.5.1.1 Investigative Analysis.** If Alice never printed anything, only one directory entry must have been used, because the printer accepts only one print job from each user at a time [137]. However, two directory entries have been used and there are no other users except Alice and Bob. Therefore, it must be the case that both Alice and Bob submitted their print jobs in the same time frame. The trace of Alice's print job was overwritten by Bob's subsequent print jobs. As a result, a finite state machine is constructed to model the situations as in the FSA [137] in Figure 12 to indicate the initial state and other possible states and how to arrive to them when Alice or Bob would have submitted a job and a job would be deleted [83, 137]. (Each state has two print job directory entries, where $e$ is empty, A—print job from Alice, B—print job from Bob, $A_X$—deleted print job from Alice, $B_X$—deleted print job from Bob. The edges denote events of +A or +B corresponding to the addition of the print jobs from Alice or Bob respectively, whereas X corresponds to taking the job for printing by the printer). The FSM presented in [137] covers the entire case with all possible events and transitions resulted due to those events. The FSM is modeled based on the properties of the investigation, in this case the printer queue's state according to the manufacturer specifications and the two potential users. The modeling is assumed to be done by the investigator in the case in order to perform a thorough analysis. It also doesn't really matter how actually it so happened that the Alice's print job was overwritten by Bob's subsequent jobs as is not a concern for this case any further. Assume, this behavior is derived from the manufacturer's specification and the evidence found. The investigator will have to make similar assumptions in the real case [137].

The authors of [137] provided a proof-of-concept implementation of this case in COMMON LISP (the code is not recited in here) which takes about 6–12 pages of printout depending on the printing options set and column format. Using our proposed solution in Section 9.3, we rewrite the example in FORENSIC LUCID and show the advantages of a much finer conciseness and added benefit of the implicit context-driven expression and evaluation, and parallel evaluation that the COMMON LISP implementation lacks entirely [305, 312].

**2.2.5.1.2   Formalization of System Functionality.**   Gladyshev's formalization of the Printing Case has to do with the definition of a number of problem domain elements [83, 137]. This formalization comes from [137] as does most of this section. Here we summarize the formal approach notation and definitions applied to this case from Section 2.2.4.

- $Q$ represents a finite set of all possible states of the Printing Case FSM

- $C_T$ represents a final set of finite computations within Printing Case FSM $T$

- $I$ represents a final set of all possible events that occur in the system (said to be *computations*)

- $o = (P, min, opt)$ is an observation of a property $P$ for the duration of time between $[min, min + opt]$

- $os = (o_1, \ldots, o_n) = ((P_1, min_1, opt_1), \ldots, (P_n, min_n, opt_N))$ is an observation sequence that constitutes an uninterrupted story told by a witness or a piece of evidence

- $es = (os_1, \ldots, os_n)$ is an evidential statement that is composed of all stories that the investigator put together. *es* need not to be ordered, but is used as a part of explanation and event reconstruction algorithm [83, 137]

- $DIR$ is a case-specific set of states of the printer queue slot that the investigator is interested in

Thus, concretely for the Printing Case problem [83, 137] the investigator defines:

- States:
$$DIR = \{A, B, A\_Deleted, B\_Deleted, empty\}, \qquad (2.2.5.1)$$

$$Q = DIR \times DIR \qquad (2.2.5.2)$$

- Events:
$$I = \{add\_A, add\_B, take\} \qquad (2.2.5.3)$$

- Formalization of the evidence [83, 137]:

  - The initial state of the print job directory:

  $$os_{manufacturer} = ((P_{empty}, 1, 0), (C_T, 0, infinitum)) \qquad (2.2.5.4)$$

  $$P_{empty} = \{c \mid c \in C_T, c_0^q = (empty, empty)\} \qquad (2.2.5.5)$$

  - The final state:

  $$P_{B\_Deleted} = \{c \mid c \in C_T, c_0^q = (B\_Deleted, B\_Deleted)\} \qquad (2.2.5.6)$$

  $$os_{final} = ((C_T, 0, infinitum), (P_{B\_Deleted}, 1, 0)) \qquad (2.2.5.7)$$

  - The complete collection of stories [83, 137]:

  $$es_{ACME} = (os_{final}, os_{manufacturer}) \qquad (2.2.5.8)$$

#### 2.2.5.1.3 Testing Investigative Hypothesis.

**Claim of Alice.** Alice's claim is that Alice did not print anything until the investigator examined the printer [83, 137]:

$$P_{Alice} = \{c \mid c \in C_T, c_i^l \neq add\_A\} \qquad (2.2.5.9)$$

$$os_{Alice} = ((P_{Alice}, 0, infinitum), (P_{B\_Deleted}, 1, 0)) \qquad (2.2.5.10)$$

In order to verify Alice's claim, Gladyshev includes it in the final evidential statement in order to try to find an explanation for it.

$$es' = (os_{Alice}, os_{final}, os_{manufacturer}) \qquad (2.2.5.11)$$

There is no explanation for it [83, 137] as shown through the meaning computations.

**Meaning of the Final State.** The meaning of

$$os_{final} = ((C_T, 0, infinitum), (P_{B\_Deleted}, 1, 0)) = (P_{B\_Deleted}, 1, 0) \qquad (2.2.5.12)$$

is the final state itself as observed by the investigator [83, 137]. The meaning also includes all the paths in the graph in Figure 12 that lead to $(B\_Deleted, B\_Deleted)$ (i.e. $(C_T, 0, infinitum)$) [83, 137].

$$M_{final} = \{MPR_{f,1}, MPR_{f,2}, \ldots, MPR_{f,infinitum}\} \qquad (2.2.5.13)$$

$$\begin{aligned}
MPR_{f,1} &= (len_{f,1}, C_{f,1}) \\
len_{f,1} &= \langle 1 \rangle \\
C_{f,1} &= \{\text{all paths arriving at } (empty, empty) \text{ with the length of } 1\} \\
&= \{(*, (B\_Deleted, B\_Deleted))\}
\end{aligned} \qquad (2.2.5.14)$$

$$\begin{aligned}
MPR_{f,2} &= (len_{f,2}, C_{f,2}) \\
len_{f,2} &= \langle 1, 1 \rangle \\
C_{f,2} &= \{\text{all paths arriving at } (empty, empty) \text{ with the length of } 2\} \\
&= \{((take, (B\_Deleted, B)), (*, (B\_Deleted, B\_Deleted))), \\
&\quad ((take, (B, B\_Deleted)), (*, (B\_Deleted, B\_Deleted)))\}
\end{aligned} \qquad (2.2.5.15)$$

$$MPR_{f,3} = (len_{f,3}, C_{f,3})$$

$$len_{f,3} = \langle 2, 1 \rangle$$

$$C_{f,3} = \{\text{all paths leading to } (B\_Deleted, B\_Deleted) \text{ with the length of } 3\}$$

$$= \{((add\_B, (empty, B\_Deleted)),$$

$$(take, (B, B\_Deleted)),$$

$$(*, (B\_Deleted, B\_Deleted))),$$

$$((take, (B, B)), (take, (B\_Deleted, B)), (*, (B\_Deleted, B\_Deleted)))\}$$

$$(2.2.5.16)$$

where:

- $M_{final}$ is the meaning of the final state represented by an infinite collection of associated MPRs for all possible lengths [83, 137].

- $MPR_{f,x}$ represents a map of partitioned runs for the final state $f$ of lengths of $x$. (Please refer to Section 2.2.4.4.4 for the complete definition of an MPR [83, 137].)

- $len_{f,x}$ represents the set of observed path lengths property for the maximum length of $x$ for the final state $f$. The notation $\langle 2, 1 \rangle$ means $\langle min, opt \rangle$, i.e., the lengths of paths between 2 and $2 + 1 = 3$ [83, 137].

In the analysis above and further in the example, the maximum length of 3 is used because it covers paths long enough to reach from the $(empty, empty)$ state to $(B\_Deleted, B\_Deleted)$ or back. Otherwise, the problem would be unfeasible to compute for all infinitely possible MPRs [83, 137].

**Meaning of the Manufacturer's Specifications.** The meaning of the observation sequence corresponding to the manufacturer

$$os_{manufacturer} = ((P_{empty}, 1, 0), (C_T, 0, infinitum)) \quad (2.2.5.17)$$

is comprised of a meaning of the observations of $(P_{empty}, 1, 0)$ and $(C_T, 0, infinitum)$. The former corresponds to just the initial state $(empty, empty)$, and the later to all the paths that originate from the initial state [83, 137].

$$M_{manufacturer} = \{MPR_{m,1}, MPR_{m,2}, \ldots, MPR_{m,infinitum}\} \quad (2.2.5.18)$$

$$\begin{aligned} MPR_{m,1} &= (len_{m,1}, C_{m,1}) \\ len_{m,1} &= \langle 1 \rangle \\ C_{m,1} &= \{\text{all paths initiating from } (empty, empty) \text{ with the length of } 1\} \\ &= \{(*, (empty, empty))\} \end{aligned} \quad (2.2.5.19)$$

$$\begin{aligned} MPR_{m,2} &= (len_{m,2}, C_{m,2}) \\ len_{m,2} &= \langle 1, 1 \rangle \\ C_{m,2} &= \{\text{all paths initiating from } (empty, empty) \text{ with the length of } 2\} \\ &= \{((add\_A, (empty, empty)), (*, (A, empty))), \\ &\quad ((add\_B, (empty, empty)), (*, (B, empty)))\} \end{aligned} \quad (2.2.5.20)$$

$$\begin{aligned} MPR_{m,3} &= (len_{m,3}, C_{m,3}) \\ len_{m,3} &= \langle 1, 2 \rangle \\ C_{m,3} &= \{\text{all paths initiating from } (empty, empty) \text{ with the length of } 3\} \\ &= \{((add\_A, (empty, empty)), (take, (A, empty)), (*, (A\_Deleted, empty))), \\ &\quad ((add\_A, (empty, empty)), (add\_B, (A, empty)), (*, (A, B))), \\ &\quad ((add\_B, (empty, empty)), (take, (B, empty)), (*, (B\_Deleted, empty))), \\ &\quad ((add\_B, (empty, empty)), (add\_A, (B, empty)), (*, (B, A)))\} \end{aligned}$$
$$(2.2.5.21)$$

**Meaning of the Incident.** Gladyshev obtains the meaning of the evidential statement

$$es_{ACME} = (os_{final}, os_{manufacturer}) \quad (2.2.5.22)$$

by combining every MPR from $M_{final}$ with every MPR from $M_{manufacturer}$ [83, 137]:

$$\forall x \in M_{final}, \forall y \in M_{manufacturer} : SPM_{ACME} = \bigcup comb(x, y) \quad (2.2.5.23)$$

The `comb()` operator combines the computations from each MPR with every other MPR and is fully defined in Section 2.2.4.6.2. Since the combination may produce unnecessary duplicates, they are removed using the union of the combination, which is also a minimal set of representing the meaning of the incident as recorded based on the evidence in the form of the final state and a story of the manufacturer as a credible expert witness of how printer operates according to its specification [305].

**Meaning of the Alice's Claim.** The meaning of the observation sequence as stated by the Alice's claim in the $os_{Alice} = ((P_{Alice}, 0, infinitum), (P_{B\_Deleted}, 1, 0))$ is comprised of the meanings of the parts $(P_{Alice}, 0, infinitum)$ and $(P_{B\_Deleted}, 1, 0)$, where the latter simply indicates the printer state as found by the investigator, and the former represents all the paths leading to $(B\_Deleted, B\_Deleted)$ that do no involve any *add_A* event [83, 137].

**Conclusion.** None of the paths in Alice's story originate in $(empty, empty)$ thereby making Alice's story questionable at the very least as shown in Figure 13 [83, 137].

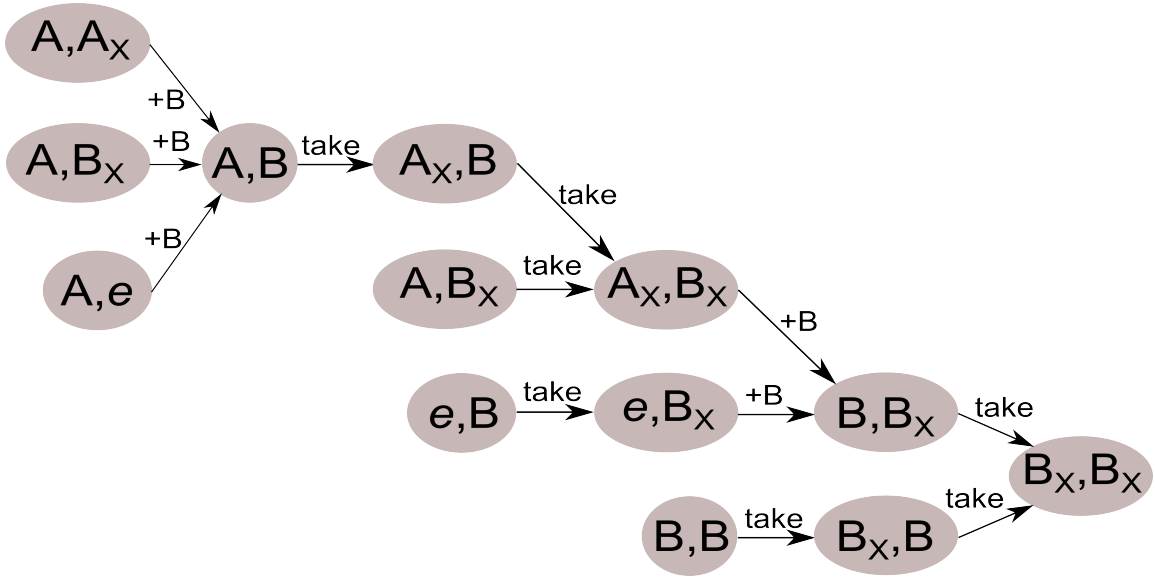

Figure 13: Paths leading to $(B_X, B_X)$

### 2.2.5.2 Initial Blackmail Case Modeling

The following case description in this section is cited from Gladyshev's [136]. It's subsequent realization in FORENSIC LUCID is in Section 9.4.

*A managing director of some company, Mr. C, was blackmailed. He contacted the police and handed them evidence in the form of a floppy disk that contained a letter with a number of allegations, threats, and demands.*

*The message was known to have come from his friend Mr. A. The police officers went to interview Mr. A and found that he was on [holidays] abroad. They seized the computer of Mr. A and interviewed him as soon as he returned into the country. Mr. A admitted that he wrote the letter, but denied making threats and demands. He explained that, while he was on [holidays], Mr. C had access to his computer. Thus, it was possible that Mr. C added the threats and demands into the letter himself to discredit Mr. A. One of the blackmail fragments was found in the slack space of another letter unconnected with the incident. When the police interviewed the person to whom that letter was addressed, he confirmed that he had received the letter on the day that Mr. A had gone abroad on [holidays]. It was concluded that Mr. A must have added the threats and demands into the letter before going on [holidays], and that Mr. C could not have been involved [136].*

In Figure 14 is the initial view of the incident as a diagram illustrating cluster data of the blackmail and unconnected letters [136, 307] since the investigation centers around that particular disk space.

**2.2.5.2.1 Modeling the Investigation.** In the blackmail example, the functionality of the last cluster of a file was used to determine the sequence of events and, hence, to disprove Mr. A's alibi [83, 136]. Thus, the scope of the model is restricted to the functionality of the last cluster in the unrelated file (see Figure 14). The last cluster model can store data objects of only three possible lengths: $L = \{0, 1, 2\}$. Zero length means that the cluster is unallocated. The length of 1 means that the cluster contains the object of the size of the unrelated letter tip. The length of 2 means that the cluster contains the object of the size of the data block with the threats. In Figure 15 is, therefore, the simplified model of the investigation [83, 136].

Figure 14: Cluster data with Blackmail fragments

Simplified cluster model:

possible data lengths:

| 0 | 1 | 2 |
|---|---|---|
| $P_L$—left part | $P_R$—right part | |

possible data values:  
$u$—unrelated  
$t_1$—threats-obscured part $\quad t_2$—threats in slack  
$o_1$—other data left part $\quad\;\, o_2$—other data right part

Observed final state:  
$L = 1$  
$P_L = \{u, t_1, o_1\}$  
$P_R = \{t_2, o_2\}$  
$Q = L \times P_L \times P_R$

| $(u)$ unrelated | $(t_2)$ threats in slack |
|---|---|

Figure 15: Simplified view of the cluster model [136]

**2.2.5.2.2 Modeling Events.** The state of the last cluster can be changed by three types of events [83, 136]:

1. Ordinary writes into the cluster:

$$W = \{(u), (t_1), (o_1), (u, t_2), (u, o_2), (t_1, t_2), (t_1, o_2), (o_1, t_2), (o_1, o_2)\} \quad (2.2.5.24)$$

2. Direct writes into the file to which the cluster is allocated (bypassing the OS):

$$W_d = \{d(u, t_2), d(u, o_2), d(o_1), d(t_1, t_2), d(t_1, o_2), d(o_1, t_2), d(o_1, o_2)\} \quad (2.2.5.25)$$

3. Deletion of the file $D$ sets the length of the file to zero. Therefore, all writes and the deletion comprise $I$:

$$I = W \bigcup W_d \bigcup D \qquad (2.2.5.26)$$

**Formalization of the Evidence.** The final state observed by the investigators is $(1, u, t_2)$ [83, 136]. Gladyshev lets $O_{final}$ denote the observation of this state. The entire final sequence of observations is then [83, 136]:

$$os_{final} = (\$, O_{final}). \qquad (2.2.5.27)$$

The observation sequence $os_{unrelated}$ specifies that the unrelated letter was created at some time in the past, and that it was received by the person to whom it was addressed is:

$$os_{unrelated} = (\$, O_{unrelated}, \$, (C_T, 0, 0), \$) \qquad (2.2.5.28)$$

where $O_{unrelated}$ denotes the observation that the "unrelated" letter tip $(u)$ is being written into the cluster. The evidential statement is then the composition of the two stories [83, 136]:
$es_{blackmail} = \{os_{final}, os_{unrelated}\}$

$$es_{blackmail} = \{os_{final}, os_{unrelated}\} \qquad (2.2.5.29)$$

**Modeling an Explanation of Mr. A's Theory.** Mr. A's theory, encoded using the proposed notation, is:

$$os_{Mr.~A} = (\$, O_{unrelated-clean}, \$, O_{blackmail}, \$) \qquad (2.2.5.30)$$

where $O_{unrelated-clean}$ denotes the observation that the "unrelated" letter $(u)$ is being written into the cluster and, at the same time, the cluster does not contain the blackmail fragment; $O_{blackmail}$ denotes the observation that the right part of the model now contains the blackmail fragment $(t_2)$ [83, 136].

**Modeling Complete Explanations.** There are two most logically possible explanations that can be represented by a state machine [136]. See the corresponding state diagram for the blackmail case in Figure 16 [83, 136].

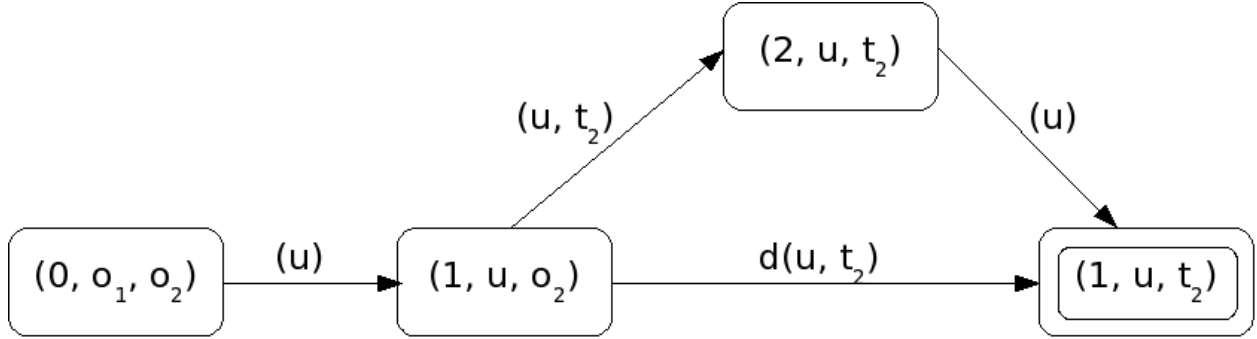

Figure 16: Blackmail Case state machine [307]

1. The first explanation [136]:

$$\ldots \xrightarrow{(u)} (1, u, o_2) \xrightarrow{(u, t_2)} (2, u, t_2) \xrightarrow{(u)} (1, u, t_2) \qquad (2.2.5.31)$$

- Finding the unrelated letter, which was written by Mr. A earlier;
- Adding threats into the last cluster of that letter by editing it "in-place" with a suitable text editor (such as ViM [325]);
- Restoring the unrelated letter to its original content by editing it "in-place" again [83, 136]:

    *"To understand this sequence of events, observe that certain text editors (e.g., ViM [325]) can be configured to edit text "in-place". In this mode of operation, the modified file is written back into the same disk blocks that were allocated to the original file. As a result, the user can forge the file's slack space by (1) appending the desired slack space content to the end of the file, (2) saving it, (3) reverting the file back to the original content, (4) saving it again."* [136]

2. The second explanation [136]:

$$\ldots \xrightarrow{(u)} (1, u, o_2) \xrightarrow{d(u, t_2)} (1, u, t_2) \qquad (2.2.5.32)$$

- The threats are added into the slack space of the unrelated letter by writing directly into the last cluster using, for example, a low-level disk editor [136].

## 2.3  Other Formal Approaches

There are other notable formal approaches to the cyberforensic analysis. For example, Arasteh and Debbabi [21] use control-flow graphs (CFGs), push down systems (PDS), and the ADM process logic [4] (a logic having dynamic, linear, temporal, and modal characteristics originally designed for evaluation of security protocols) to construct code models from the executable in-memory code and call-stack residues (like leftover slack space similar to the Gladyshev's blackmail case study earlier).

Arasteh *et al.* [22] subsequently extend their formal approach to log analysis along with a tree of labeled terms of $\sum$-algebra and a tableau-based proof system for the ADM logic arriving at a case study of a SYN attack detection from various log sources.

Clear parallels and relationships can be made between these approaches, the intensional logic (Section 3.2), and Cartesian programming [378] as formalization tools.

## 2.4  Summary

We reviewed the notion of cyberforensics and more specifically, the background on the formal approach to the digital evidence analysis, which is central to this thesis. We reviewed the Gladyshev's arguably the first comprehensive formal approach to evidence modeling and event reconstruction as well as two sample case studies we further use to test our approach in Chapter 7.

# Chapter 3

# Intensional Logic, Programming, Reasoning, Uncertainty and Credibility

This chapter reviews the background and related work necessary for formal methods, formal logic-based [195, 253, 416] specification and context-orientation, dealing with uncertainty and credibility aspects that are relevant to cyberforensic investigations, reasoning, and event reconstruction in developing soft-computing intelligent systems [204], autonomic self-forensic units (Appendix D), and expert systems [186].

The logic-based formal specifications are used for requirements specification [257, 482, 521] of critical components and systems, knowledge representation and derivation (keeping a knowledge base of a case), validation and verification, and testing hypotheses. As a formal system, logic can be used in the court of law as a part of an expert witness testimony when challenged about the validity and verifiability of the system and system-based argumentation.

Consequently, this chapter summarizes the related background in intensional logic, uncertainty, hierarchical credibility, and so on as these concepts are fundamental in establishing theoretical setting for the intensional programming realized by the GIPSY (Chapter 6) and the formalization of FORENSIC LUCID (Chapter 7). The reader is assumed to have some familiarity with or exposure to the concepts introduced here. Nevertheless, we attempt to summarize the relevant content in a coherent manner in referencing the material actually used subsequently in this thesis. As a result, we survey the related works and then review their fusion.

## 3.1 Overview

The surveyed literature explores a connection between multidimensional logics, such as intensional logics with the theory of mathematical evidence. These are being used (or considered for use) in a variety of application domains for automated reasoning in context-oriented environments where the knowledge may be imprecise and incomplete and credibility is a factor (e.g., formal digital crime investigations). Some of these logical systems (e.g., intensional logics) and probabilistic reasoning systems were not given enough attention, especially when viewed together and parallels are drawn to finite state machines and the set theory [119, 237]. Thus, more specifically, we review various relatively recent literature on intensional logics of multidimensional nature where context can be formalized, combined with with the Dempster–Shafer theory of mathematical evidence along with hierarchical credibility and possible connection to the multidimensional description logics (DLs by themselves are presently more studied and understood in literature and popular in ontological frameworks, but (the author conjectures) can be formalized in intensional logics). We also review the evolution of the intensional logics research into Cartesian programming and intensionality in mathematics (e.g., see a recent workshop of that title held in May 2013 [391]) including a number of references from there of interest discussing Fregean abstract objects; soundness, reflection and intensionality; mathematical intensions and intensionality in mathematics; intensional side of the algebraic topological theorems; knowledge acquisition of numbers by children; formal mathematical sense of mathematical intensions; and others [45, 82, 115, 125, 166, 337, 412].

This chapter is structured as follows: the notion of intensional logics and programming are explored in Section 3.2, uncertainty and mathematical evidence modeling are discussed in Section 3.3, where specifically the Dempster–Shafer theory is further reviewed in Section 3.3.2 with its connection to the research described. We summarize our overall findings in Section 3.4 with the use of material presented here and a conjecture.

## 3.2 Intensional Logics and Programming

Why *Intensional Logic*? It can be traced all the way back to Aristotle [508] to represent sense (meaning) and its change between different possibilities (possible worlds, or contexts) having true statements available there. Concrete realization of the intensional logic ideas resulted in *Intensional Programming*. What follows is the justification and some historical remarks for both.

Many problem domains are intensional in nature and aspects that surround us, e.g., natural language understanding and learning [412], particle in cell simulation [376], World Wide Web [510, 512], computation of differential and tensor equations [361], versioning [246], temporal computation and temporal databases [360, 453], multidimensional signal processing [5, 272], context-driven computing [514], constraint programming [124, 513], negotiation protocols [513], automated reasoning in cyberforensics discussed in this thesis [269, 300, 304], multimedia and pattern recognition [272] among others [364]. The current mainstream programming languages are not well adapted for the natural expression of the intensional aspects of such problems, requiring the expression of the intensional nature of the problem statement into a procedural (and therefore sequential) approach in order to provide a computational solution [302].

What follows is a brief history of *Intensional Logic* synthesized from [112, 508] for the curious reader.

In essence, the intensional concepts in times of Aristotle appeared in response to the need to separate conceptual senses (e.g., the "Prime Minister of Canada") from their concrete context-dependent instantiations (i.e., extensions, e.g., at the time of this writing in July 2013 "Prime Minister of Canada" evaluates to the extension "Stephen Harper") and modal logic. Fast-forward to Gottlob Frege who was making distinctions between "senses" (abstract descriptions of objects) and their "denotations" (concrete "real" instances of those objects) along the same lines. Fast forward again to Carnap [57] who introduced the word "intensional" in 1930s [508] to refer to Frege's "senses". Around the same time (1930s) Lewis *et al.* were reviving the modal logic (that fell out of popularity sometime before that) [508] from the extensional propositional logic by adding $\prec$ connective to mean "strict implication" to avoid some paradoxes (*entailment, explosion*, etc. [40]) associated with material implication → [508]. Then they realized the now standard modal operators ◇ ("possible") and □ ("necessary") are easier to work with and redefined $(P \prec Q)$ as $\neg \Diamond (P \land \neg Q)$ meaning $P$ implies $Q$ if and only if it's impossible for $P$ to be true and $Q$ be false simultaneously [508]. Likewise, Aristotle's notion of "everything necessary is possible" is formalized as $\Box P \to \Diamond P$. Then, everything necessary is true ($\Box P \to P$) and everything that is true is possible ($P \to \Diamond P$) [508].

Fast forward another 20 years or so (per Montague [324] and Melvin [112]), Church had a go at the intensionality with his work on the simple theory of types circa 1951 in *A Formulation of the Logic of Sense and Denotation* subsequently inter-influencing further developments in the area with Carnap [112]. Saul Kripke subsequently in the 60s [218, 219] formalized semantics for modal logic. His core idea is that of the *possible world semantics*; that is statements can be true in certain worlds, but not necessarily in all of them [508]. The world $w$ does not need to be fully defined, but serve as indices into possible interpretations [53] of each statement, e.g., $\Box P$ is true at a particular world $w$ if $P$ is true in *some* worlds $w'$ accessible from $w$ [508]. As William W. Wadge puts it, "Kripke models were a turning point in development of the intensional logic." [508]. That enabled simple temporal logic, e.g., where ◇ would mean *sometimes* and □ would mean *always*, etc. [508]. Temporal logic made use of those operators, but had only a single dimension of time points. Making it multi-dimensional later with intensional operators made it what is its today as *intensional temporal logic* (see page 62 for an informal example).

Many other later formalism followed (70s-80s), such as that of Marcus then on to codecessor works of Montague, Tichý, Bressan, and Gallin [127] exploring various intensional logic aspects [112, 480]. Richard Montague deliberated on the pragmatics and intensional logic in 1970 [91, 324] with a critical discussion and raising problems of interest. Hobbs and Rosenschein subsequently published an approach to make computational sense of Montague's Intensional Logic [171] in 1978. Finally, Dana Scott's 1969 work (as Wadge suggested [508], "perhaps mistitled") *Advice on Modal Logic* laid down more formalisms by extending Kripke's possible worlds with Carnap's distinction between extension and intension (later known as the *Scott-Montague model*) defining a non-empty set $I$ of points of reference (possible worlds) without requirements on accessibility relations [508]. Then $\phi$ in each world has an extension (the truth value) and an intension (the mapping of $w$ to the extension of $\phi$ at $w$, making it an element of the power set $2^I$ [508]. Scott maps (via a unary operator) of intensions to others as $2^I \to 2^I$ either as an accessibility or any other similar relation and Scott's models are not restricted to propositional logic, and can specify worlds at any hierarchical level of types (strings, numbers, people, organizations, etc.) [508]. Scott calls such individuals as virtual individuals (*intensional objects*) grouped in a collection $D$, and these individuals $D^I$ can denote different extensions in different worlds [508]. This approach inspired Wadge and Ashcroft to create LUCID in 1974 and the notion of intensional programming has emerged [25].

William W. Wadge placed a beautifully succinct introduction to Intensional Logic in context in a tutorial in 1999 [508], most of which is still valid and is a highly recommended reading. A lot of background history is also provided in the special issue articles in 2008 by Blanca Mancilla and John Plaice bringing the historical record more up-to-date with possible world versioning and Cartesian programming [246, 378]. Subsequently, Melvin Fitting in his updated 2012 entry [112] provided another broad encyclopedic account on Intensional Logic. All are very recommended readings.

Both Intensional Programming volumes (I and II) from 1995 [350] and 1999 [131] also detail a series of articles on various temporal logic aspects applied to program verification, constraint logic and its semantics, flexible agent grouping, temporal meaning representation and reasoning [18, 214, 244, 358, 394] as well as a proposal of an asynchronous calculus based on absence of actions [220] followed by the more recent 2008 work by Orgun *et al.* [352] on knowledge representation and reasoning with diverse granularity clocks. Fitting in 2005 provided axiomatization of the first-order intensional logic (FOIL) [113]. Intensional logic is used in the present day to describe various natural language phenomena and beyond, e.g., as in the recent (2010) work by Duží *et al.* [98] on procedural semantics of Hyperintensional Logic as a survey of the foundations and applications of Transparent Intensional Logic of Pavel Tichý.

Thus, what is intensional logic, again? Wadge summarized the definition of it as: *"...intensional logic—possible worlds and intensional operators..."* [508].

### 3.2.1 Multidimensional Logics and Programming

Here we mention two major multidimensional branches of modal logic and pick the one of prime use and interest to us. The earlier kind, that began development as such in the 1940-50s is the *Intensional Logic* informally described in the previous section; the later kind that appeared in 1980s [524] is the *Description Logic*, both are being actually families of logics for knowledge representation and reasoning incorporating natural language constructs of possibility, necessity, belief, and others. We further concentrate in some more detail the *multidimensional intensional logic* as the most fundamental and intuitive to use. From the modal to temporal uni-dimensional logic, multidimensional logic, knowledge representation, and programming paradigms emerged. Back to the Scott's 1969 tuple [508] $i \in I$ included:

$$i = (w, p, t, a) \qquad (3.2.1.1)$$

> the notion of the possible world extension $w$ (however big or small, does not need to be fully defined), $p$ is a context point in space, e.g., dimensions of $(x, y, z)$, $a$ being the agent, and $t$ is time, and $i$ is an index into all these coordinates [508].

Fast forward from 1969 to 1991–1993, Faustini and Jagannathan described multidimensional problem solving in INDEXICAL LUCID [109]. Ashcroft *et al.*, subsequently summarized the concepts of *multidimensional programming* in their book with the same title [24]. Baader and Ohlbach present a multi-dimensional terminological knowledge representation language [34] in 1995. Ashcroft discussed multidimensional program verification in terms of reasoning about programs that deal with multidimensional objects [23], also in 1995. Orgun and Du described theoretical foundations of multi-dimensional logic programming in 1997 [351]. Wolter and Zakharyaschev in 1999 [524] discuss multi-dimensional description logics.

**Temporal Intensional Logic Example**

Temporal intensional logic is an extension of temporal logic that allows to specify the time in the future or in the past [361]. What follows is an informal example.

Let's take $E_1$ from Figure 17. The context is a collection of the dimensions (e.g., as in $E_1$'s `place` and `time`) paired with the corresponding tags (**here** and **today** respectively).

1. $E_1 :=$ it is raining **here today**

   Context: {`place:`**`here`**, `time:`**`today`**}

2. $E_2 :=$ it was raining **here** *before*(**today**) = *yesterday*

3. $E_3 :=$ it is going to rain *at* (altitude **here** + 500 m) *after*(**today**) = *tomorrow*

Figure 17: Natural-language contextual expression [361, 380]

Then let us fix **here** to **Montreal** and assume it is a *constant*. In the month of **April 2013**, with a granularity of one day, for every day, we can evaluate $E_1$ to either *true* or *false*, as shown in Figure 18 [300, 304, 312].

```
Tags days in April:   1 2 3 4 5 6 7 8 9 ...
Values (raining?):    F F T T T F F F T ...
```

Figure 18: 1D example of tag-value contextual pairs [361, 380]

If one starts varying the **here** dimension (which could even be broken down into finer $X, Y, Z$), one gets a two-dimensional (or at higher level of detail 4D: $(X, Y, Z, t)$ respectively) evaluation of $E_1$, as shown in Figure 19 [302, 312].

```
Place/Time  1 2 3 4 5 6 7 8 9 ...
Montreal    T F T F T F T F T ...
Quebec      F F T T T F F F T ...
Ottawa      F T T T T T F F F ...
New York    F F F F T T T F F ...
Toronto     F T T T T T F F F ...
Sydney      F F T T T F F F T ...
Moscow      F F F F T T T F F ...
```

Figure 19: 2D example of tag-value contextual pairs [264, 282]

Even with these toy examples we can immediately illustrate the hierarchical notion of the dimensions in the context: so far the place and time we treated as atomic values fixed at days and cities. In some cases, we need finer subdivisions of the context evaluation, where, e.g., time can become fixed at hour, minute, second and finer values, and so is the place broken down into boroughs, regions, streets, etc. and finally the $X, Y, Z$ coordinates in the Euclidean space with the values of millimeters or finer. This notion becomes more apparent and important, e.g., in FORENSIC LUCID, where the temporal components can be, e.g., log entries and other registered events and observations from multiple sources [302, 312].

### 3.2.2 Intensional Programming

*Intensional programming* (IP) is based on multidimensional intensional logics [171], which, in turn, are based on Natural Language Understanding aspects (such as time, situation, direction, etc.) mentioned earlier. IP brings in *dimensions* and *context* to programs (e.g., *space* and *time*). Since *intensional logic* adds dimensions to logical expressions, a non-intensional logic can be seen as a constant or a snapshot in all possible dimensions. To paraphrase, *intensions* are certain statements, whose extensions in possible worlds are true or false (or have some other than Boolean values). *Intensional operators* are operators that allow us to navigate within these contextual dimensions [361]. *Higher-order Intensional Logic* (HOIL) [300, 302, 405] is behind functional programming of LUCID with multidimensional dataflows which intensional programs can query and alter through an explicit notion of contexts as first-class values [300, 302, 304, 312, 365, 513].

From another side, intensional programming [131, 350, 380], in the sense of the latest evolutions of LUCID (Chapter 4), is a programming language paradigm based on the notion of declarative programming *where* the declarations are evaluated in an inherent multidimensional context space. The context space being in the general case infinite, intensional programs are evaluated using a lazy demand-driven model of execution—*eduction* [110, 380], the precept of which is the referential transparency enabling scalable caching at the implementation level. There the program identifiers are evaluated in a restricted context space, in fact, a *point* in space, where each demand is generated, propagated, computed and stored as an *identifier-context* pair [241, 302]. The subsequent demands for the same context, can simply be satisfied by fetching the previously computed value from the store. Plaice and Paquet provided a lucid (no pun intended) explanation of Intensional Programming in a tutorial [380] in 1995, that still holds today. At ISLIP (after several iterations), in 1995 [350] is where per Wadge [508] the intensional logic, the *Advice on Modal Logic* of Scott emerged together to have a broader community actively developing practical languages and systems based on these formalism. Wadge and Ashcroft, based to Scott's models then defined the LUCID language with intensional operators FIRST, NEXT, FBY in 1974, with NEXT, e.g., denoting from $D^I \to D^I$ (page 60), such that `NEXT`$(X) = \lambda n.X(n+1)$ [508].

Intensional programming can be used to solve widely diversified problems mentioned in this chapter as well as in Chapter 4, which can be expressed using diversified languages of intensional nature. There also has been a wide array of flavors of LUCID languages developed over the years. Yet, very few of these languages have made it to the pragmatic implementation level. The GIPSY project (Chapter 6) aims at the creation of a programming environment encompassing compiler generation for all flavors of LUCID, and a generic run-time system enabling the execution of programs written in all flavors of LUCID. Its goal is to provide a flexible platform for the investigation on programming languages of intensional nature, in order to prove the applicability of intensional programming to solve important problems [302]. We continue this trend in this work with FORENSIC LUCID (Chapter 7).

## 3.3 Uncertainty, Evidence, and Credibility

The works reviewed in this section contribute to the enhancement of the cyberforensic analysis with options to automatically reason in the presence of uncertainty and non-binary degrees of belief, credibility (reliability of witnesses) of each witness story and a piece of evidence as an additional artifact of the proposed FORENSIC LUCID system. We subsequently briefly review the related literature in this category involving in particular the *Mathematical Theory of Evidence* by Dempster–Shafer [422] *Probabilistic Argumentation Systems* by Haenni *et al.* [153] and some related probabilistic logic work [87], and the links to intensional logic. (Such probabilistic reasoning provides additional foundations for a future elaborate expert system to train the investigator personnel in presence of uncertainty.)

### 3.3.1 Probabilistic Logics

It is worthwhile to mention some logical foundations to probabilistic reasoning to represent uncertainty, including in modal and intensional aspects [87]. Previously mentioned Carnap himself in 1950 wrote on *Logical Foundations of Probabilities* [87] while working on all aspects of logic including his work on intensional aspects. Roughly, there are two main approaches in probabilistic logic [87] to represent probabilistic uncertainty: *qualitative* (notion of possibility) and *quantitative* (numerical probability) [87]. After Hamblin in 1959, Gärdenfors (along with Segerberg) in 1975 proposed treating qualitative probability as an Intensional Logic [130] as we have seen the intensional logic provided the notion of "possible". Quantitative approaches included treading numerical values of probability spaces. Modal probability logics [87, Section 4] subsequently were introduced bringing back the Kripke possible world semantics that was not possible for some initial probability logic works. That included Fagin and Harpen's work on this in 1988 and 1994 later [87]; indexing and interpretation were introduced for the possible world states with probabilities. Combining qualitative and quantitative [87] then seemed like a natural progression. As an example, quoting Demey *et al.* [87]: $\neg \Box h \wedge (\neg \Box P(h) = 1/2) \wedge (\Diamond P(h) = 1/2)$ *would read as "it is not known that h is true, and it is not known that the probability of h is 1/2, but it is possible that the probability of h is 1/2"*. More recently, in 2006, Halpern and Pucella presented a logic for reasoning about evidence [158] following earlier works about reasoning about uncertainty and knowledge [106, 156, 157]. Most recently (2011), Haenni *et al.* [155], explored the notion probabilistic logics and probabilistic networks.

### 3.3.2 Dempster–Shafer Theory of Evidence and Probabilistic Reasoning Systems

Semantics and interpretation of truth were always of interest to reasoning in any logical system [53]. This includes uncertain knowledge representation and manipulation [156, 425]. This section subsequently reviews the Dempster–Shafer theory that helps us to reason with the presence of uncertainty in beliefs about the system's state. The theory has been extensively discussed and extended; the presentation of which is found in Expert Systems [186] as well as the work by Haenni *et al.* [153, 154] about probabilistic argumentation.

#### 3.3.2.1 Dempster-Shafer Theory of Mathematical Evidence

The Dempster–Shafer theory as a mathematical theory of evidence (DSTME) is there to give machinery to combine evidence from different sources to come up with a *degree of belief* (represented by *a belief function* [423]) that takes into account all the available evidence. The initial theory was a product of the work by Arthur P. Dempster [88] in 1968 on his rule of combination and Glenn Shafer's mathematical evidence theory [422] of 1976. Since then DSTME was applied to different application domains altering sometimes the Dempster's rule of combination (Section 3.3.2.1.3, page 72) to produce better (more intuitive, correct) results in that domain or situation.

Quoting Shafer [424]:

> *The Dempster-Shafer theory, also known as the theory of belief functions, is a generalization of the Bayesian theory of subjective probability. Whereas the Bayesian theory requires probabilities for each question of interest, belief functions allow us to base degrees of belief for one question on probabilities for a related question. These degrees of belief may or may not have the mathematical properties of probabilities; how much they differ from probabilities will depend on how closely the two questions are related.*
>
> *. . .*
>
> *The Dempster-Shafer theory is based on two ideas: the idea of obtaining degrees of belief for one question from subjective probabilities for a related question, and Dempster's rule for combining such degrees of belief when they are based on independent items of evidence.–G. Shaffer, 2002 [424]*

Thus, we review these aspects further.

**3.3.2.1.1 Formal Definition.** Below we quote [88, 422] the formal definition with the symbols adapted to match our needs:

- $Q$ is the universal set representing all possible states $q$ of a system under consideration.

- $2^Q$ is the power set of all subsets of $Q$ (including the empty set $\emptyset$). The elements of $2^Q$ represent propositions concerning the actual state of the system, by containing all and only the states, in which the propositions are true.

- $m$ is a function denoting DSTME's assignment of a belief mass to each element of $2^Q$ (called a *basic belief assignment* (BBA)).

$$m : 2^Q \to [0, 1] \quad (3.3.2.1)$$

It has two properties:

   - the mass of the empty set is zero: $m(\emptyset) = 0$
   - the masses of the remaining members add up to a total of 1: $\sum_{A \in 2^X} m(A) = 1$

The mass $m(A)$ of $A \subset 2^Q$ denotes the fraction of all relevant available evidence supporting the claim that the actual state $q$ belongs to $A$ ($q \in A$). The value of $m(A)$ corresponds *only* to the set $A$ itself.

- bel($A$) and pl($A$) are *belief* and *plausibility* denoting the upper and lower bounds of the probability interval that contains the precise probability of a set of interest, and is bounded by two non-additive continuous bel($A$) and pl($A$):

$$\mathrm{bel}(A) \leq P(A) \leq \mathrm{pl}(A) \quad (3.3.2.2)$$

The belief bel($A$) is the sum of all the masses in $A$ of subsets of the set of interest $A$:

$$\mathrm{bel}(A) = \sum_{B \mid B \subseteq A} m(B) \quad (3.3.2.3)$$

The plausibility pl($A$) is the sum of all the masses of the sets $B$ that intersect $A$:

$$\mathrm{pl}(A) = \sum_{B \mid B \cap A \neq \emptyset} m(B) \quad (3.3.2.4)$$

Belief and plausibility are related as follows:

$$\mathrm{pl}(A) = 1 - \mathrm{bel}(\overline{A}) \quad (3.3.2.5)$$

Conversely, for a finite $A$, given the belief bel($B$) for all subsets $B$ of $A$, one can find the masses $m(A)$ with the following *inverse function*:

$$m(A) = \sum_{B|B \subseteq A} (-1)^{|A-B|} \operatorname{bel}(B) \qquad (3.3.2.6)$$

where $|A - B|$ is the difference of the cardinalities of the two sets [421], From Equation 3.3.2.5 and Equation 3.3.2.6, for a finite set $Q$, one needs to know only one of the mass, belief, or plausibility to deduce the other two. In the case of an infinite $Q$, there can be well-defined belief and plausibility functions but no well-defined mass function [156], but in our cyberforensic investigations $Q$ is always finite.

**3.3.2.1.2  Examples.**   The Wikipedia page on the theory [520] as well as the quoted reference above from Shafer give a good number of examples where the theory would be applicable and how it works.

*To illustrate the idea of obtaining degrees of belief for one question from subjective probabilities for another, suppose I have subjective probabilities for the reliability of my friend Betty. My probability that she is reliable is 0.9, and my probability that she is unreliable is 0.1. Suppose she tells me a limb fell on my car. This statement, which must true if she is reliable, is not necessarily false if she is unreliable. So her testimony alone justifies a 0.9 degree of belief that a limb fell on my car, but only a zero degree of belief (not a 0.1 degree of belief) that no limb fell on my car. This zero does not mean that I am sure that no limb fell on my car, as a zero probability would; it merely means that Betty's testimony gives me no reason to believe that no limb fell on my car. The 0.9 and the zero together constitute a belief function.*

*To illustrate Dempster's rule for combining degrees of belief, suppose I also have a 0.9 subjective probability for the reliability of Sally, and suppose she too testifies, independently of Betty, that a limb fell on my car. The event that Betty is reliable is independent of the event that Sally is reliable, and we may multiply the probabilities of these events; the probability that both are reliable is $0.9 \times 0.9 = 0.81$, the probability that neither is reliable is $0.1 \times 0.1 = 0.01$, and the probability that at least one is reliable*

*is* $1 - 0.01 = 0.99$. *Since they both said that a limb fell on my car, at least [one] of them being reliable implies that a limb did fall on my car, and hence I may assign this event a degree of belief of* $0.99$.

*Suppose, on the other hand, that Betty and Sally contradict each other—Betty says that a limb fell on my car, and Sally says no limb fell on my car. In this case, they cannot both be right and hence cannot both be reliable—only one is reliable, or neither is reliable. The prior probabilities that only Betty is reliable, only Sally is reliable, and that neither is reliable are 0.09, 0.09, and 0.01, respectively, and the posterior probabilities (given that not both are reliable) are 9/19, 9/19, and 1/19, respectively. Hence we have a 9/19 degree of belief that a limb did fall on my car (because Betty is reliable) and a 9/19 degree of belief that no limb fell on my car (because Sally is reliable).*

*In summary, we obtain degrees of belief for one question (Did a limb fall on my car?) from probabilities for another question (Is the witness reliable?). Dempster's rule begins with the assumption that the questions for which we have probabilities are independent with respect to our subjective probability judgments, but this independence is only a priori; it disappears when conflict is discerned between the different items of evidence.*–G. Shaffer, 2002

We adapt one of the examples to re-state one of Gladyshev's examples (Section 2.2.5.1, page 44): "Did Alice print anything?" from the questions "Is Alice a reliable witness?", "Is Bob a reliable witness?" and "Is the printer manufacturer a reliable witness?" following the belief mass assignments and the rule of combination. Suppose the investigator has a subjective probability the manufacturer is 0.9 reliable and 0.1 unreliable in their printer mechanism work specification. Since both Alice and Bob are roughly under the same investigation questions, assume the investigator's subjective probability of their reliability is 0.5 that is no preference is given to either Alice or Bob, or either could be truthful or lying in their testimony, whereas the printer manufacturer is treated as an expert witness. Since Alice and Bob are strictly not independent, their testimony have a degree of conflict, the introduction of the manufacturer's testimony is necessary since both Bob and Alice are equally (un)reliable before the incident, so either one of them can be true (or neither), but not both. These belief assignments of course do not resolve alone the case in question; for that we need the combined approach with the intensional logic presented earlier for event reconstruction from multiple observations in witness accounts. We will review that further in Chapter 7.

We solve a similar operational question in Section 9.5 for the *MAC Spoofer Investigation* case using the question "Is there a real MAC spoofer?" given different witness accounts from logs and probes and their reliability that is when a log entry is observed it is assumed to be 1.0 reliable, when log is empty for some reason or a probe fails (e.g., due to firewall restrictions on the probed host), its witness reliability is 0.0; when observed information is partial that is also reflected in the reliability aspects.

Another quoted example is from signal processing of different color sensors from a distant light source where the light color can be colored in any of the three *red*, *green*, or *blue*. Assigning their masses as given [48]:

| Hypothesis | Mass | Belief | Plausibility |
|---|---|---|---|
| Null (neither) | 0.00 | 0.00 | 0.00 |
| Red | 0.35 | 0.35 | 0.56 |
| Yellow | 0.25 | 0.25 | 0.45 |
| Green | 0.15 | 0.15 | 0.34 |
| Red or Yellow | 0.06 | **0.66** | **0.85** |
| Red or Green | 0.05 | 0.55 | 0.75 |
| Yellow or Green | 0.04 | 0.44 | 0.65 |
| Any | 0.10 | 1.00 | 1.00 |

allows to compute belief and plausibility. *Any* denotes *Red or Yellow or Green* is a catch-all case basically saying there is some evidence "there is a light", but the color is uncertain. Null-hypothesis ("no light", but for completeness) is always given 0 mass. This can be used to model various situations (e.g., the evidence received from a color-blind person or a malfunctioning CCD). Below are two examples of computing belief and plausibility given mass:

$$\text{bel}(Red \text{ or } Yellow) = m(null) + m(Red) + m(Yellow) + m(Red \text{ or } Yellow)$$
$$= 0 + 0.35 + 0.25 + 0.06 = \mathbf{0.66}$$
$$\text{pl}(Red \text{ or } Yellow) = 1 - \text{bel}(\neg(Red \text{ or } Yellow))$$
$$= 1 - \text{bel}(Green)$$
$$= 1 - m(null) - m(Green)$$
$$= 1 - 0 - 0.15 = \mathbf{0.85}$$

**3.3.2.1.3  Dempster's Rule of Combination.**  Once evidential data are in from multiple sources, they need to be combined somehow [421]. Dempster created such a rule as a part of his work on the generalization of the Bayesian inference in 1968 [88]. When viewed under this interpretation, the priors and conditionals are not required to be specified (as opposed to traditional Bayesian, e.g., assigning 0.5 probabilities to values, for which no prior information is available). The rule ignores any such information unless it can be obtained during the overall computation. This can be seen as allowing DSTME to formulate a degree of ignorance vs. the absolute necessity to provide prior probabilities [200, 421].

Thus, we need to combine independent sets of probability mass assignments. Specifically, the *combination* (called as *joint mass*) is calculated from the two mass sets $m_1(B)$ and $m_2(C)$ (where $A = B \cap C$) as follows:

$$m_{1,2}(\emptyset) = 0 \tag{3.3.2.7}$$

$$m_{1,2}(A) = (m_1 \oplus m_2)(A) = \frac{1}{1-K} \sum_{A=B \cap C \neq \emptyset} m_1(B) m_2(C) \tag{3.3.2.8}$$

$$K = \sum_{B \cap C = \emptyset} m_1(B) m_2(C) \tag{3.3.2.9}$$

where $K$ is a measure of the amount of conflict between the two mass sets [422].

This rule was criticized for some scenarios by Zadeh [536] (the recognized father of *Fuzzy Logic* [537]) and others where it produced counter-intuitive results in some cases of high or low conflict, especially when the sources are not independent enough. (See Zadeh's examples of Alice and Bob deciding to pick a move vs. two doctors diagnosing brain tumor in a patient, or a toy murder trial suspect case [200].) Many researchers subsequently proposed various way of combining (fusing) beliefs that suit a particular situation or application (sensor fusion, opinion fusion, etc.) better using the appropriate *fusion operators* by Jøsang and others [198, 199, 200].

#### 3.3.2.2 Probabilistic Argumentation Systems

In the recent work, a lot of attention was devoted to various probabilistic argumentation systems. Notably, Haenni *et al.* consistently discussed the topic [153, 154] in 1999–2003 as well as probabilistic logics and probabilistic networks [155] in 2011. Halpern and Pucella in 2006 provided a logic to reason about evidence as well [158] in 2006. Jøsang in 2012 did a thorough review of different ways to apply the combination rule [200] and provided the relevant background proofs. Shi *et al.* [428] in 2011 provided a hybrid system combining intuitionistic fuzzy description logics with intuitionistic fuzzy logic programs.

Xu *et al.* [174] in 2009 made a strong point for the necessity of attribute reduction in ordered information systems with a possible solution. They exploited the relationship between the Dempster–Shafer theory of mathematical evidence and the Pawlak's *rough set theory* as well as their plausibility and belief functions [286]. This is of relevance to this work as well. Reduction of attributes can become an important step point to reduce state explosion problem in evidence analysis and event reconstruction by applying Occam's razor. Grünwald and Halpern certainly agree with that in their 2004 work *When ignorance is bliss* [147]. This is pertinent to us in the sense of irrelevant context reduction and elimination with a lot of digital evidence to reduce the computational effort and/or reduce confusion, e.g., as filtering is done in FORENSIC LUCID encoders in Section 9.5.7.2, page 266 later on.

## 3.4 Summary

This chapter provided a quick overview of the reasoning foundations and the corresponding related work on intensional logics and the Dempster–Shafer theory of evidence. In this research, we combine these reasoning tools. It serves the purpose to encode forensic knowledge with credibility and reason about forensic knowledge in a context-oriented manner in a formal system for cyberforensic reasoning and event reconstruction in investigator's work where rigor and formality are a must to be used in the court of law, but the reasoning may include incomplete, imprecise, or untrustworthy evidence.

The presented related work around Dempster–Shafer, has a strong correlation with intensional logic [480] (Section 3.2, page 59) and the LUCID programming language (Chapter 4) as well its derivative LUCX [513] where the ordered information system is a context of evaluation, a context set of $\langle dimension : tag \rangle$ pairs, that has seen a number of applications (such as iHTML [510], the contribution of this thesis—FORENSIC LUCID, and others). FORENSIC LUCID [310] (Chapter 7) also blends the Dempster–Shafer theory, and intensional logic and LUCID dialects for ordered evidential contexts [286], such as observation sequences.

**Conjecture.** Not in a mathematical or logic terms, but I would like to conjecture a philosophical point for discussion: *In the metalogical [58, 175][1] sense the (meta?) intensional logic can be used to reason about logical systems governing them all.* (We shall see how that fits with Tichý's work on TIL and his followers [98].)

Since intensional logic dealt with senses and semantics from the beginning, it makes sense to describe other logics with it. Every logic $L$ is a possible world $w$ (however defined, nested, complex, complete, sound, consistent or not). The philosophical aspects aside, the statements of consistency can be true for $w = L$. There may or may not be accessibility transitions between the possible logic worlds, but if there is one, then one logic can be represented (translated to) into another in a whole or in part. Accepting the fact that some statements may not retain their truth values in such a process, that is evaluating a proposition in one logical world may yield one answer, another in another, or be undecidable or undefined in the third.

The same can be applied to more concrete and restricted entities and instantiations the logics may describe, such as the syntax and semantics of natural ($w = L_{NL}$) or programming languages ($w = L_{PL}$). Linguistically, it is known people of different cultures can learn multiple languages, while most NL concepts are universal and fully inter-translatable, but without learning fully the culture, one cannot translate all possible expressions, proverbial or not, so not all statements can be translated between languages, or would have different meanings when attempted. The same (though weaker) is true for the more restricted programming languages (though the situation here is better due to the restricted nature of PLs). While the most universal algebras and commands can be easily inter-translated via a basic assembly form, not all assembly instances support all features of all architectures. Not all high-language constructs can be translated to others (some kind of virtualization and wrappers can be done to make statements from one PL available in another).

---

[1] http://en.wikipedia.org/wiki/Metalogic

Thus, complete translation may not be possible between logics, just like between natural or programming languages since some statements do not have universal meaning across these logics (or languages). However, any two minimal inter-translatable logic systems describing the same model may represent indistinguishable possible logic world intensional individuals [508]. This assertion is in part supported by Assels's refutation of the logic of the global conventionalism (GC) by showing GC's inconsistency [30] in 1985.

# Chapter 4

# The LUCID Programming Language Family

In this chapter we review the background on the notion of the LUCID-family of languages (which we often collectively refer to as "LUCID" instead of implying the specific "ORIGINAL LUCID" of Ashcroft and Wadge from 70s-90s) as instantiations of various HOIL realizations from the overview (Section 4.1), to the historical background and the related work (Section 4.2), to the dialects spectrum (Section 4.3). We then summarize the findings in Section 4.4. LUCID is central to this thesis for its lazy stream processing semantics and ease of expression as well as its solid theoretical and mathematical research base make it ideal for scalable knowledge representation and reasoning of any streamed data.

## 4.1 Lucid Overview

LUCID [24, 25, 26, 27, 509] is a dataflow intensional and functional programming language. In fact, it is now a family of languages that are built upon intensional logic (which in turn can be understood as a multidimensional generalization of temporal logic, cf. Section 3.2, page 59) promoting context-aware demand-driven parallel computation model [304]. A program written in some LUCID dialect is an expression that may have subexpressions that need to be evaluated at a certain *context*. Given a set of dimensions $DIM = \{dimension_i\}$, in which an expression varies, and a corresponding set of indexes, or, *tags*, defined as placeholders over each dimension, the context is represented as a set of $\langle dimension : tag \rangle$ mappings. Each variable in Lucid, often called a *stream*, is evaluated in that defined context that may also evolve using context operators [305, 307, 312, 365, 473, 513, 515]. The first generic version of Lucid, the General Intensional Programming Language (GIPL) [361], defines two basic operators @ and # to navigate (switch and query) in the context space [305, 307, 312]. The GIPL is the first (with the second being Lucx [365, 473, 513, 515], and third is TransLucid [379]) generic programming language of all intensional languages, defined by the means of only two mentioned intensional operators—@ and # [305, 307, 312]. It has been proven that other intensional programming languages of the Lucid family can be translated into the GIPL [305, 307, 312, 361]. A recent similar notion to GIPL called TransLucid was designed by Plaice and colleagues [90, 378, 379, 396]. Plaice and Paquet give a succinct yet comprehensive introduction to Lucid and intensional programming in their tutorial [380] in 1995, which is still a recommended reading today. A more recent (2008) overview is by Plaice and colleagues can be found in [246, 378].

### 4.1.1 Sample Syntax and Semantics

The fundamental syntax and semantics of Lucid are rather simple allowing easier compiler implementation and human comprehension of Lucid programs. This also allows flexible extension in various dialects for application domain specific purposes and use the core baseline as a sound and complete building block. Similarly, as pioneered by GLU, Lucid is relatively easy to marry to imperative programming languages to allow mutual advantage of eductive evaluation and availability of rich libraries. These are the aspects that contributed to the choice of this language to be extended in this thesis. What follows are examples of syntax and semantics of the Lucid dialects relevant to us. For illustratory purposes, in the example excerpts that follow we unify the definitions in a hypothetical language that we call a $G_\geq$, which is the union of necessary features of the Lucid dialects we inherit from. (Further in this thesis, we define Forensic Lucid in much more concrete terms and detail in Chapter 7). To say it simply, $\boxed{G_{>=} \equiv \min \bigcup (\text{GIPL}, \text{Indexical Lucid}, \text{Objective Lucid}, \text{JOOIP}, \text{MARFL})}$. We briefly discuss these dialects in this chapter, and present their concrete features we borrow in Chapter 7.

#### 4.1.1.1 Lucid Syntax

Example syntaxes of $G_\geq$ for expressions, definitions, and operators are presented in Figure 20 and Figure 21 for both GIPL and INDEXICAL LUCID respectively to remind the reader of their structure [305]. The concrete syntax of the GIPL in Figure 20 has been amended to support the `isoed` operator of INDEXICAL LUCID for completeness and is influenced by the productions from LUCX [515] to allow contexts as first-class values while maintaining backward compatibility to the GIPL language proposed by Paquet earlier [305, 361]. In Figure 22 is a simple program illustrating a LUCID expression [305].

```
E   ::=   id
      |   E(E,...,E)
      |   E[E,...,E](E,...,E)
      |   if E then E else E fi
      |   # E
      |   E @ [E:E]
      |   E @ E
      |   E where Q end;
      |   [E:E,...,E:E]
      |   iseod E;
Q   ::=   dimension id,...,id;
      |   id = E;
      |   id(id,....,id) = E;
      |   id[id,...,id](id,....,id) = E;
      |   QQ
```

Figure 20: Sample $G_\geq$ syntax expressions

#### 4.1.1.2 Operational Semantics for LUCID

In the implementing system, GIPSY (see Chapter 6), GIPL is the generic counterpart of all the LUCID programming languages. Like INDEXICAL LUCID, which it is derived from, it has only the two standard intensional operators: `E @ C` for evaluating an expression `E` in context `C`, and `#d` for determining the position in dimension `d` of the current context of evaluation in the context space [361]. SIPLs are LUCID dialects (Specific Intensional Programming Languages) with their own attributes and objectives. Theoretically, all SIPLs can be translated into the GIPL [361]. Here for convenience we provide the semantic rules of GIPL and INDEXICAL LUCID [361] and LUCX [513]. The operational semantics of GIPL is presented in Figure 23. The excerpt of semantic rules of LUCX is then presented as a

$$
\begin{aligned}
op &::= \text{intensional-op} \\
&\mid \text{data-op} \\
\\
\text{intensional-op} &::= \text{i-unary-op} \\
&\mid \text{i-binary-op} \\
\\
\text{i-unary-op} &::= \texttt{first} \mid \texttt{next} \mid \texttt{prev} \\
\text{i-binary-op} &::= \texttt{fby} \mid \texttt{wvr} \mid \texttt{asa} \mid \texttt{upon} \\
\\
\text{data-op} &::= \text{unary-op} \\
&\mid \text{binary-op} \\
\\
\text{unary-op} &::= \texttt{!} \mid \texttt{-} \mid \texttt{iseod} \\
\text{binary-op} &::= \text{arith-op} \\
&\mid \text{rel-op} \\
&\mid \text{log-op} \\
\text{arith-op} &::= \texttt{+} \mid \texttt{-} \mid \texttt{*} \mid \texttt{/} \mid \texttt{\%} \\
\text{rel-op} &::= \texttt{<} \mid \texttt{>} \mid \texttt{<=} \mid \texttt{>=} \mid \texttt{==} \mid \texttt{!=} \\
\text{log-op} &::= \texttt{\&\&} \mid \texttt{||}
\end{aligned}
$$

Figure 21: Sample $G_\geq$ operators

```
N @.d 2
where
    dimension d;
    N = 42 fby.d (N + 1);
end
```

Figure 22: Classical natural-numbers example [361]

conservative extension to GIPL in Figure 24 [300, 304].

Following is the description of the GIPL semantic rules as presented in [361]:

$$\mathcal{D} \vdash E : v \qquad (4.1.1.1)$$

tells that under the *definition environment* $\mathcal{D}$, expression $E$ would evaluate to value $v$.

$$\mathcal{D}, \mathcal{P} \vdash E : v \qquad (4.1.1.2)$$

specifies that in the definition environment $\mathcal{D}$, and in the *evaluation context* $\mathcal{P}$ (sometimes also referred to as a *point* in the context space), expression $E$ evaluates to $v$. The definition environment $\mathcal{D}$ retains the definitions of all of the identifiers that appear in a LUCID program, as created with the semantic rules 4.1.1.16–4.1.1.19 in Figure 23. It is therefore a partial

function

$$\mathcal{D} : \mathbf{Id} \rightarrow \mathbf{IdEntry} \tag{4.1.1.3}$$

where **Id** is the set of all possible identifiers and **IdEntry**, summarized in Table 1, has five possible kinds of values, one for each of the kinds of identifier [300, 304]:

- *Dimensions* define the coordinate pairs, in which one can navigate with the `#` and `@` operators. Their **IdEntry** is simply (`dim`) [361].

- *Constants* are external entities that provide a single value, regardless of the context of evaluation. Examples are integers and Boolean values. Their **IdEntry** is (`const`, $c$), where $c$ is the value of the constant [361].

- *Data operators* are external entities that provide memoryless functions. Examples are the arithmetic and Boolean functions. The constants and data operators are said to define the *basic algebra* of the language. Their **IdEntry** is (`op`, $f$), where $f$ is the function itself [361].

- *Variables* carry the multidimensional streams. Their **IdEntry** is (`var`, $E$), where $E$ is the LUCID expression defining the variable. It should be noted that this semantics makes the assumption that all variable names are unique. This constraint is easy to overcome by performing compile-time renaming or using a nesting level environment scope when needed [361].

- *Functions* are non-recursive user-defined functions. Their **IdEntry** is (`func`, $id_i$, $E$), where the $id_i$ are the formal parameters to the function and $E$ is the body of the function [361].

Table 1: Possible identifier types [361]

| type | form |
|---|---|
| dimension | (`dim`) |
| constant | (`const`, $c$) |
| operator | (`op`, $f$) |
| variable | (`var`, $E$) |
| function | (`func`, $id_i$, $E$) |

The evaluation context $\mathcal{P}$, which is changed when the @ operator is evaluated, or a dimension is declared in a `where` clause, associates a *tag* (i.e., an index) to each relevant dimension. It is, therefore, a partial function

$$\mathcal{P} : \mathbf{Id} \to \mathbf{N} \qquad (4.1.1.4)$$

Each type of identifier can only be used in the appropriate situations. Identifiers of type `op`, `func`, and `dim` evaluate to themselves (Figure 23, rules 4.1.1.6, 4.1.1.7, 4.1.1.8). Constant identifiers (`const`) evaluate to the corresponding constant (Figure 23, rule 4.1.1.5). Function calls, resolved by the $\mathbf{E_{fct}}$ rule (Figure 23, rule 4.1.1.11), require the renaming of the formal parameters into the actual parameters (as represented by $E'[id_i \leftarrow E_i]$). The function $\mathcal{P}' = \mathcal{P} \dagger [id \mapsto v'']$ specifies that $\mathcal{P}'(x)$ is $v''$ if $x = id$, and $\mathcal{P}(x)$ otherwise. The rule for the `where` clause, $\mathbf{E_w}$ (Figure 23, rule 4.1.1.16), which corresponds to the syntactic expression $E$ `where` $Q$, evaluates $E$ using the definitions $Q$ therein. The additions to the definition environment $\mathcal{D}$ and context of evaluation $\mathcal{P}$ made by the $\mathbf{Q}$ rules (Figure 23, rules 4.1.1.17, 4.1.1.18, 4.1.1.19) are local to the current `where` clause. This is represented by the fact that the $\mathbf{E_w}$ rule returns neither $\mathcal{D}$ nor $\mathcal{P}$. The $\mathbf{Q_{dim}}$ rule adds a dimension to the definition environment and, as a convention, adds this dimension to the context of evaluation with the tag 0 (Figure 23, rule 4.1.1.17). The $\mathbf{Q_{id}}$ and $\mathbf{Q_{fid}}$ simply add variable and function identifiers along with their definition to the definition environment (Figure 23, rules 4.1.1.18, 4.1.1.19) [300, 304, 361].

As a conservative extension to GIPL, LUCX's semantics introduced the *context as a first-class value*, as described by the rules in Figure 24. The semantic rule 4.1.1.22 (Figure 24) creates a context as a semantic item and returns it as a context $\mathcal{P}$ that can then be used by the rule 4.1.1.23 to navigate to this context by making it override the current context. The semantic rule 4.1.1.21 expresses that the `#` symbol evaluates to the current context. When used as a parameter to the context calculus operators, this allows for the generation of contexts relative to the current context of evaluation [361, 513, 515]

$$\mathbf{E_{cid}} \;:\; \frac{\mathcal{D}(id) = (\mathtt{const}, c)}{\mathcal{D}, \mathcal{P} \vdash id : c} \tag{4.1.1.5}$$

$$\mathbf{E_{opid}} \;:\; \frac{\mathcal{D}(id) = (\mathtt{op}, f)}{\mathcal{D}, \mathcal{P} \vdash id : id} \tag{4.1.1.6}$$

$$\mathbf{E_{did}} \;:\; \frac{\mathcal{D}(id) = (\mathtt{dim})}{\mathcal{D}, \mathcal{P} \vdash id : id} \tag{4.1.1.7}$$

$$\mathbf{E_{fid}} \;:\; \frac{\mathcal{D}(id) = (\mathtt{func}, id_i, E)}{\mathcal{D}, \mathcal{P} \vdash id : id} \tag{4.1.1.8}$$

$$\mathbf{E_{vid}} \;:\; \frac{\mathcal{D}(id) = (\mathtt{var}, E) \quad \mathcal{D}, \mathcal{P} \vdash E : v}{\mathcal{D}, \mathcal{P} \vdash id : v} \tag{4.1.1.9}$$

$$\mathbf{E_{op}} \;:\; \frac{\mathcal{D}, \mathcal{P} \vdash E : id \quad \mathcal{D}(id) = (\mathtt{op}, f) \quad \mathcal{D}, \mathcal{P} \vdash E_i : v_i}{\mathcal{D}, \mathcal{P} \vdash E(E_1, \ldots, E_n) : f(v_1, \ldots, v_n)} \tag{4.1.1.10}$$

$$\mathbf{E_{fct}} \;:\; \frac{\mathcal{D}, \mathcal{P} \vdash E : id \quad \mathcal{D}(id) = (\mathtt{func}, id_i, E') \quad \mathcal{D}, \mathcal{P} \vdash E'[id_i \leftarrow E_i] : v}{\mathcal{D}, \mathcal{P} \vdash E(E_1, \ldots, E_n) : v} \tag{4.1.1.11}$$

$$\mathbf{E_{c_T}} \;:\; \frac{\mathcal{D}, \mathcal{P} \vdash E : true \quad \mathcal{D}, \mathcal{P} \vdash E' : v'}{\mathcal{D}, \mathcal{P} \vdash \mathtt{if}\ E\ \mathtt{then}\ E'\ \mathtt{else}\ E'' : v'} \tag{4.1.1.12}$$

$$\mathbf{E_{c_F}} \;:\; \frac{\mathcal{D}, \mathcal{P} \vdash E : false \quad \mathcal{D}, \mathcal{P} \vdash E'' : v''}{\mathcal{D}, \mathcal{P} \vdash \mathtt{if}\ E\ \mathtt{then}\ E'\ \mathtt{else}\ E'' : v''} \tag{4.1.1.13}$$

$$\mathbf{E_{tag}} \;:\; \frac{\mathcal{D}, \mathcal{P} \vdash E : id \quad \mathcal{D}(id) = (\mathtt{dim})}{\mathcal{D}, \mathcal{P} \vdash \#E : \mathcal{P}(id)} \tag{4.1.1.14}$$

$$\mathbf{E_{at}} \;:\; \frac{\mathcal{D}, \mathcal{P} \vdash E' : id \quad \mathcal{D}(id) = (\mathtt{dim}) \quad \mathcal{D}, \mathcal{P} \vdash E'' : v'' \quad \mathcal{D}, \mathcal{P}\dagger[id \mapsto v''] \vdash E : v}{\mathcal{D}, \mathcal{P} \vdash E\ @E'\ E'' : v} \tag{4.1.1.15}$$

$$\mathbf{E_w} \;:\; \frac{\mathcal{D}, \mathcal{P} \vdash Q : \mathcal{D}', \mathcal{P}' \quad \mathcal{D}', \mathcal{P}' \vdash E : v}{\mathcal{D}, \mathcal{P} \vdash E\ \mathtt{where}\ Q : v} \tag{4.1.1.16}$$

$$\mathbf{Q_{dim}} \;:\; \frac{}{\mathcal{D}, \mathcal{P} \vdash \mathtt{dimension}\ id\ :\ \mathcal{D}\dagger[id \mapsto (\mathtt{dim})], \mathcal{P}\dagger[id \mapsto 0]} \tag{4.1.1.17}$$

$$\mathbf{Q_{id}} \;:\; \frac{}{\mathcal{D}, \mathcal{P} \vdash id = E\ :\ \mathcal{D}\dagger[id \mapsto (\mathtt{var}, E)], \mathcal{P}} \tag{4.1.1.18}$$

$$\mathbf{Q_{fid}} \;:\; \frac{}{\mathcal{D}, \mathcal{P} \vdash id(id_1, \ldots, id_n) = E\ :\ \mathcal{D}\dagger[id \mapsto (\mathtt{func}, id_i, E)], \mathcal{P}} \tag{4.1.1.19}$$

$$\mathbf{QQ} \;:\; \frac{\mathcal{D}, \mathcal{P} \vdash Q\ :\ \mathcal{D}', \mathcal{P}' \quad \mathcal{D}', \mathcal{P}' \vdash Q'\ :\ \mathcal{D}'', \mathcal{P}''}{\mathcal{D}, \mathcal{P} \vdash Q\ Q'\ :\ \mathcal{D}'', \mathcal{P}''} \tag{4.1.1.20}$$

Figure 23: Extract of operational semantics rules of GIPL [361]

### 4.1.2 Streaming and Basic Operators

The origins of LUCID date back to 1974 [361, 380]. At that time, Ashcroft and Wadge were working on a purely declarative language, in which iterative algorithms could be expressed naturally, which eventually resulted in non-procedural iterative LUCID [25, 26, 27]. Their work further fit into the broad area of research into program semantics and verification. Later it turned out that their work is also relevant to the dataflow networks and coroutines

$$\mathbf{E_{\#(cxt)}} \quad : \quad \overline{\mathcal{D}, \mathcal{P} \vdash \# : \mathcal{P}} \tag{4.1.1.21}$$

$$\mathbf{E_{construction(cxt)}} \quad : \quad \frac{\begin{array}{c}\mathcal{D}, \mathcal{P} \vdash E_{d_j} : id_j \quad \mathcal{D}(id_j) = (\mathtt{dim})\\ \mathcal{D}, \mathcal{P} \vdash E_{i_j} : v_j \quad \mathcal{P}' = \mathcal{P}_0 \dagger [id_1 \mapsto v_1]\dagger \ldots \dagger [id_n \mapsto v_n]\end{array}}{\mathcal{D}, \mathcal{P} \vdash [E_{d_1} : E_{i_1}, E_{d_2} : E_{i_2}, \ldots, E_{d_n} : E_{i_n}] : \mathcal{P}'} \tag{4.1.1.22}$$

$$\mathbf{E_{at(cxt)}} \quad : \quad \frac{\mathcal{D}, \mathcal{P} \vdash E' : \mathcal{P}' \quad \mathcal{D}, \mathcal{P}\dagger\mathcal{P}' \vdash E : v}{\mathcal{D}, \mathcal{P} \vdash E \ @ \ E' : v} \tag{4.1.1.23}$$

$$\mathbf{E_{.}} \quad : \quad \frac{\mathcal{D}, \mathcal{P} \vdash E_2 : id_2 \quad \mathcal{D}(id_2) = (\mathtt{dim})}{\mathcal{D}, \mathcal{P} \vdash E_1.E_2 : tag(E_1 \downarrow \{id_2\})} \tag{4.1.1.24}$$

$$\mathbf{E_{tuple}} \quad : \quad \frac{\mathcal{D}, \mathcal{P} \vdash E : id \quad \mathcal{D}\dagger[id \mapsto (\mathtt{dim})] \quad \mathcal{P}\dagger[id \mapsto 0] \quad \mathcal{D}, \mathcal{P} \vdash E_i : v_i}{\mathcal{D}, \mathcal{P} \vdash \langle E_1, E_2, \ldots, E_n \rangle E : v_1 \ fby.id \ v_2 \ fby.id \ \ldots \ v_n \ fby.id \ \mathtt{eod}} \tag{4.1.1.25}$$

$$\mathbf{E_{select}} \quad : \quad \frac{E = [\mathtt{d} : \mathtt{v'}] \quad E' = \langle \mathtt{E_1}, \ldots, \mathtt{E_n} \rangle \mathtt{d} \quad \mathcal{P}' = \mathcal{P}\dagger[d \mapsto v'] \quad \mathcal{D}, \mathcal{P}' \vdash E' : v}{\mathcal{D}, \mathcal{P} \vdash select(E, E') : v} \tag{4.1.1.26}$$

$$\mathbf{E_{at(s)}} \quad : \quad \frac{\mathcal{D}, \mathcal{P} \vdash \mathcal{C} : \{\mathcal{P}_1, \ldots, \mathcal{P}_2\} \quad \mathcal{D}, \mathcal{P}_{i:1\ldots m} \vdash E : v_i}{\mathcal{D}, \mathcal{P} \vdash E \ @\mathcal{C} : \{v_1, \ldots, v_m\}} \tag{4.1.1.27}$$

$$\mathbf{C_{box}} \quad : \quad \frac{\begin{array}{c}\mathcal{D}, \mathcal{P} \vdash E_{d_i} : id_i \quad \mathcal{D}(id_i) = (\mathtt{dim})\\ \{E_1, \ldots, E_n\} = dim(\mathcal{P}_1) = \ldots = dim(\mathcal{P}_m)\\ E' = \mathtt{f}_p(\mathtt{tag}(\mathcal{P}_1), \ldots, \mathtt{tag}(\mathcal{P}_m)) \quad \mathcal{D}, \mathcal{P} \vdash E' : true\end{array}}{\mathcal{D}, \mathcal{P} \vdash Box[E_1, \ldots, E_n | E'] : \{\mathcal{P}_1, \ldots, \mathcal{P}_m\}} \tag{4.1.1.28}$$

$$\mathbf{C_{set}} \quad : \quad \frac{\mathcal{D}, \mathcal{P} \vdash E_{w:1\ldots m} : \mathcal{P}_m}{\mathcal{D}, \mathcal{P} \vdash \{E_1, \ldots, E_m\} : \{\mathcal{P}_1, \ldots, \mathcal{P}_w\}} \tag{4.1.1.29}$$

$$\mathbf{C_{op}} \quad : \quad \frac{\mathcal{D}, \mathcal{P} \vdash E : id \quad \mathcal{D}(id) = (\mathtt{cop}, f) \quad \mathcal{D}, \mathcal{P} \vdash C_i : v_i}{\mathcal{D}, \mathcal{P} \vdash E(C_1, \ldots, C_n) : f(v_1, \ldots, v_n)} \tag{4.1.1.30}$$

$$\mathbf{C_{sop}} \quad : \quad \frac{\mathcal{D}, \mathcal{P} \vdash E : id \quad \mathcal{D}(id) = (\mathtt{sop}, f) \quad \mathcal{D}, \mathcal{P} \vdash C_i : \{v_{i_1}, \ldots, v_{i_k}\}}{\mathcal{D}, \mathcal{P} \vdash E(C_1, \ldots, C_n) : f(\{v_{1_1}, \ldots, v_{1_s}\}, \ldots, \{v_{n_1}, \ldots, v_{n_m}\})} \tag{4.1.1.31}$$

Figure 24: Extract of operational semantics of LUCX [473, 513]

of Kahn and MacQueen [202, 203]. In the original LUCID (whose operators are in THIS FONT), streams were defined in a pipelined manner, with two separate definitions: one for the initial element, and another one for the subsequent elements [361, 380]. For example, the equations

$$\mathtt{FIRST} \ X \ = \ 0$$
$$\mathtt{NEXT} \ X \ = \ X + 1$$

define the variable $X$ to be a stream, such that

$$x_0 \ = \ 0$$
$$x_{i+1} \ = \ x_i + 1$$

In other words,
$$0 = (0, 0, 0, ..., 0, ...)$$
$$X = (x_0, x_1, \ldots, x_i, \ldots) = (0, 1, \ldots, i, \ldots)$$

Similarly, the equations
$$\text{FIRST } X = X$$
$$\text{NEXT } Y = Y + \text{NEXT } X$$

define variable $Y$ to be the running sum of $X$, i.e.,

$$y_0 = x_0$$
$$y_{i+1} = y_i + x_{i+1}$$

That is,
$$Y = (y_0, y_1, \ldots, y_i, \ldots) = \left(0, 1, \ldots, \tfrac{i(i+1)}{2}, \ldots\right)$$

According to Paquet, then it became clear that a "new" operator at the time, `fby` (*followed by*) can be used to define such typical situations [361, 380] allowing the above two variables be defined as follows:
$$X = 0 \text{ FBY } X + 1$$
$$Y = X \text{ FBY } Y + \text{NEXT } X$$

As a result, Plaice and Paquet summarized the three basic operators of the original LUCID [361, 380]:

> **Definition L1**
> If $X = (x_0, x_1, \ldots, x_i, \ldots)$ and $Y = (y_0, y_1, \ldots, y_i, \ldots)$, then
>
> $$(1) \quad \text{FIRST } X \stackrel{\text{def}}{=} (x_0, x_0, \ldots, x_0, \ldots)$$
> $$(2) \quad \text{NEXT } X \stackrel{\text{def}}{=} (x_1, x_2, \ldots, x_{i+1}, \ldots)$$
> $$(3) \quad X \text{ FBY } Y \stackrel{\text{def}}{=} (x_0, y_0, y_1, \ldots, y_{i-1}, \ldots)$$

Paquet and Plaice further drew parallels to the list operations of LISP dialects, where `first` corresponds to `head`, `next` corresponds to `tail`, and `fby` corresponds to `cons` [361, 380]. This is especially useful, when translating Gladyshev's COMMON LISP implementation (see [137]) into FORENSIC LUCID (Section 9.3, page 250). They further state when these operators are combined with Landin's ISWIM [223] (*If You See What I Mean*), which is essentially typed $\lambda$-calculus with some syntactic sugar, it becomes possible to define complete LUCID programs [361, 380]. The following three derived operators have turned out to be very useful (we will use them later in the text) [361, 380]:

---

**Definition L2**

(1) $X$ WVR $Y$ $\stackrel{\text{def}}{=}$ if FIRST $Y$ then $X$ FBY ( NEXT $X$ WVR NEXT $Y$ )

else ( NEXT $X$ WVR NEXT $Y$ )

(2) $X$ ASA $Y$ $\stackrel{\text{def}}{=}$ FIRST $(X$ WVR $Y)$

(3) $X$ UPON $Y$ $\stackrel{\text{def}}{=}$ $X$ FBY (if FIRST $Y$ then ( NEXT $X$ UPON NEXT $Y$ )

else ( $X$ UPON NEXT $Y$ ))

---

Where `wvr` stands for *whenever*, `asa` stands for *as soon as* and `upon` stands for *advances upon*.

### 4.1.3 Random Access to Streams

In the beginning, with the original LUCID operators, one could only define programs with pipelined dataflows, i.e., in where the $(i+1)$-th element in a stream is only computed once the $i$-th element has been computed. This situation was deemed potentially wasteful, since the $i$-th element might not necessarily be needed. More importantly, it only allows sequential access into streams [361, 380].

In order to have random access into streams, an index `#` corresponding to the current position, the current context of evaluation was created [361, 380]. This step made the infinite extensions (streams), constrained intensions, defining computation according to a context (originally a simple single integer) [361, 380]. Effectively, intensional programming was born [361, 380]. As such, Paquet and Plaice redefined all ORIGINAL Lucid operators in terms of the operators `#` and `@` (and Paquet [361] established their equivalence):

**Definition L3**

$$\begin{align}(1) \quad &\texttt{\#} &\overset{\text{def}}{=}& \quad \texttt{0 FBY (\# + 1)} \\ (2) \quad &X \mathrel{\texttt{@}} Y &\overset{\text{def}}{=}& \quad \texttt{if } Y = 0 \texttt{ then FIRST } X \\ & & & \quad \texttt{else ( NEXT } X) \mathrel{\texttt{@}} (Y-1)\end{align}$$

We re-examine again these and other operators in FORENSIC LUCID in Chapter 7.

### 4.1.4 Eductive Model of Computation

The first operational model for computing Lucid programs was designed independently by Cargill at the University of Waterloo and May at the University of Warwick, based directly on the formal semantics of LUCID, itself based on Kripke models and possible-worlds semantics [218, 219]. This technique was later extended by Ostrum for the implementation of the LUTHID interpreter [354]. While LUTHID was tangential to standard LUCID, its implementation model was later adopted as a basis for the design of the PLUCID interpreter by Faustini and Wadge [110]. This program evaluation model is now called *eduction* and opens doors for distributed execution [302] of such programs [161, 271, 362, 452, 454, 455].

In Figure 25 [264] is the trace represented as an eduction tree during execution of the OBJECTIVE LUCID program in Figure 26. In this figure, the outermost boxes labeled {d:0} etc. represent the current context of evaluation, gray rectangular boxes with expressions represent demands for values to be computed under that context, and the red[1] boxes with the terminal bullets next to them represent the results of the computation of the expressions. In our proposed solution, such eduction tree can be adapted to back-tracing in forensic evaluation, e.g., when the inner-most results are traced back to the final result of the entire computation [305].

The concept of *eduction* can be described as a "tagged-token demand-driven dataflow" [503] computing paradigm (whereupon LUCID influenced a popular media platform and language called PureData [388]). The central concept to this model of execution is the notion of generation, propagation, and consumption of *demands* and their resulting *values*. Lucid programs are declarative programs where every identifier is defined as a HOIL expression using other identifiers and an underlying algebra. An initial demand for the value of a certain identifier is generated, and the eduction engine, using the defining expression of this identifier, generates demands for the constituting identifiers of this expression, on which operators are applied in their embedding expressions. These demands in turn generate other demands, until some demands eventually evaluate to some values, which are then propagated back in the chain of demands, operators are applied to compute expression values, until eventually the value of the initial demand is computed and returned [302].

---

[1] "Red" in the color image; or dark gray in black-and-white image.

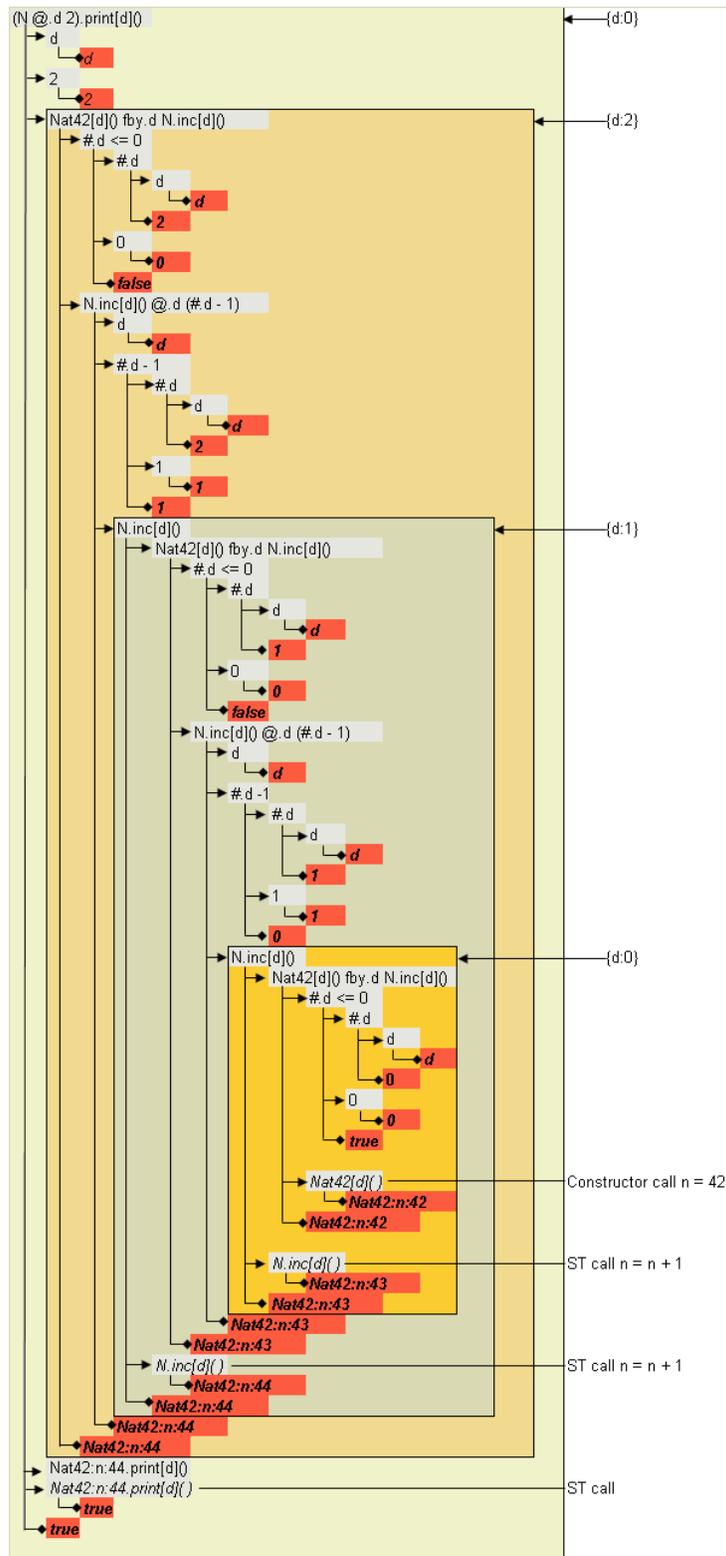

Figure 25: Eduction tree as a trace for the natural-numbers problem in OBJECTIVE LUCID [264]

Lucid identifiers and expressions inherently vary in a *multidimensional context space*, i.e., any identifier or expression can be evaluated in a multidimensional context, thus leading to have identifiers and expressions representing a set of values, one value for each possible context in which the identifier or expression can be evaluated. This is brining the notion of *intensionality*, where identifiers are defined by intensional expressions, i.e., expressions whose evaluation varies in a multidimensional context space, which can then be constrained by a particular multidimensional context specification. Note that Lucid variables and expressions represent "dimensionally abstract" concepts, i.e., they do not explicitly mention their dimensionality. For example, Newton's Law of Universal Gravitation

$$F = (G \cdot m_1 \cdot m_2)/r \cdot r \qquad (4.1.4.1)$$

can be written literally in LUCID as [302]:
F = (G * m1 * m2) / r * r;
and can then be evaluated in different dimensional manifolds (i.e., $n$-dimensional spaces), keeping the same definition, but being evaluated in contexts varying in their dimensionality. For example, F can be evaluated in a one-dimensional space, yielding a single scalar, or in a three-dimensional manifold, yielding a three-dimensional vector. Note that a `time` dimension could also be added where, for example, the masses (`m1` and `m2`) and/or the distance between them (`r`) can be defined as to vary in time. In this case, the expression will inherently vary in the time dimension since some of its constituents vary in this dimension [302].

### 4.1.5 Related Languages

Some other languages can be referred to as intensional even though they may not refer to themselves as such, and were born after LUCID (LUCID began in 1974). Examples include hardware-description languages (HDLs, appeared in 1977) where the notion of time (often the only "dimension", and usually progresses only forward), e.g., Verilog and VHDL [342]. Another branch of newer languages for the becoming popular is aspect-oriented programming (AOP) languages (e.g., ASPECTJ [29] and SHL [228]), Web Service Description Language (WSDL). that can have a notion of context explicitly, but primarily focused on software engineering aspect of software evolution and maintainability [300]. Yet another set of stream-processing languages influenced by LUCID include Miller Puckette's PureData [388] (open-source) and its commercial equivalent Jitter in Max/MSP [74, 75].

## 4.2 Related Work

In this section we briefly review some of the related work on LUCID and its dialects and developments. Some of that work is explained in detail further in Section 4.3 that we draw on the most as well as in [364] where a list of publications, dialects, and related work is being maintained and periodically updated. A lot of the related work was done earlier and summarized in the ISLIP proceedings [131, 350] and some of this work was reviewed in the preceding chapter and is still of relevance today.

LUCID was influenced by Landin's ISWIM [223] in his work *700 programming languages*, producing in the end contexts, the `where` clause, etc. (ISWIM had influence elsewhere as well like SQL, Z, and others). It was designed to model change that happens to various abstract objects and phenomena [380].

A very good overview of LUCID, INDEXICAL LUCID, etc., in the context of WWW as an illustrative example is done in the tutorial by Plaice and Paquet in [380], including multi-dimensional contexts, higher-order functions, and examples. Wadge discusses the Kripke's possible worlds semantics of LUCID in the OO context [507]. Wadge and Yoder further reviewed the possible world semantics in their work *Possible-World Wide Web* [512].

One of the application domains LUCID was proposed for was for verification and reasoning about multidimensional programs by Ashcroft [23]. Ma and Orgun proposed MULTRAN program verification with temporal logic [244]. Yamane proposed real-time OO specification and verification techniques [533].

There were naturally a number of works on the hybrid intensional-imperative programming. The two paradigms have a generally poor interface among each other: on the one hand are conventional imperative programming languages that have no room for multidimensionality or intensional or demand-driven evaluation; on the other hand, existing multidimensional languages that cannot take advantage of imperative features and techniques. Developed over years of research, the combination typically result in much better performance [526]. Liu and Staples proposed introduction of logic constructs into the imperative procedural languages [238] and Rondogiannis followed up with multidimensional additions to the procedural languages [406]. Rondogiannis and Wadge also proposed a way to extend what they call the *intensionalization algorithm* to a broader class of higher-order programs [408]. The GLU# approach embeds a small multidimensional core in a mainstream object-oriented programming language (C++) [359]. By this way, without changing the syntax and semantics of the host language, multidimensionality can be supported. GLU# is a small subset of GLU and it is implemented as a collection of C++ classes and class templates. All operators in LUCID appear as C++ functions. GLU# does not support Lucid functions; however, programmers are able to use lazy method templates in C++ to use C++'s functions in GLU#. GLU# provided a bridge between LUCID and OO [526]. The concept about objects in LUCID first appeared in [122] in the early 1990s. In the later 1990s, Peter Kropf and John Plaice talked about this topic in their paper "intensional objects" [221]. In this paper, intensional objects are considered as openable boxes labeled by LUCID contexts. That work focuses on intensional versioning whose task is to build a system from versioned components, which are already sitting in the *intensional value warehouse* (a cache of the previously computed results). This warehouse is different as the warehouse in intensional programming. The latter is like a cache to improve the performance. The former contains the source of everything, it is like a "catalog" or a "repository", in which the boxes are put into. Each box is of some contents and a tag that is context. Thus, in this approach, these labeled boxes are called intensional objects, which are re-openable and re-packageable [526]. In [93], there is another discussion on issues about object-oriented implementation of intensional languages. In this approach, each variable in a Lucid program is considered as a class and an object of a class is a variable in a context. Each variable definition in a Lucid program is compiled into a C++ class definition that has the same name as the variable. This approach focuses on the implementation-level by creating a class for each Lucid variable, it helps the system to execute in a distributed manner. However, the objects introduced here does not contain information from C++ variables [526]. Lu created similar identifier-context (IC) classes in the multithreaded GEE PoC implementation [241]. Grogono proposed ONYX for multidimensional lazy arrays [144].

On the scientific application side Plaice proposed particle in-cell simulations with LUCID [376] and Paquet with his *Scientific Intensional Programming* work [361] proposed TENSOR LUCID for plasma physics simulations.

Paquet and Plaice proposed the use of LUCID constructs to handle fine amount records in relational databases [360, 367].

Le proposed a notion of a Fuzzy Temporal Prolog. While we are not into Prolog in this work, the fuzzy temporal aspect is of relevance to FORENSIC LUCID. Knowledge representation and temporal knowledge based simulation subsequently proposed by Liu and Orgun in Chronolog is also a related work to FORENSIC LUCID. Panayiotopoulos also proposed temporal reasoning with TRL [358]. At the same time. Androutsopoulos discussed the temporal meaning representation in a natural language front-end [18].

Paquet and Plaice followed by Wan investigated the semantics of dimensions as values [368, 513]. For nested hierarchical contexts the related work includes the work by Schraefel *et al.* [417, 511] on various hypertext representational and modeling issues.

Besnard *et al.* [42] proposed a design of a multi-formalism application and distribution in a dataflow context via an example. Gagné and Plaice have further expanded on the topic of real-time demand-driven computing [126].

Faustini established the equivalence of denotational and operational semantics in pure dataflows [108], which can be useful when relying on the relevant denotational formalism elsewhere (e.g., [532]) and most of the LUCID's semantics was traditionally done in the operational fashion. Uustalu and Vene in a more recent review go over the notion of dataflow programming [478].

On the compiler and toolset side, Swoboda and Wadge produced a set of intensionalizations tools like `vmake`, `libintense` [455], Mark did a basic PoC interpreter in Haskell [250].

## 4.3 Lucid Dialects

Here we briefly review some of the core influential dialects contributing to the construction of FORENSIC LUCID in addition to the information provided in Section 4.1. A great deal is devoted to the dialects and related publications in [364].

### 4.3.1 LUCX

Wan's LUCX [513, 515] (which stands for *Lucid enriched with context* is a fundamental extension of GIPL and the LUCID family as a whole that promotes the contexts to the first-class values thereby creating a "true" generic LUCID language. We recited its semantics in Figure 24 in Section 4.1, page 76. Wan [513, 515] defined a new collection of set operators (e.g., `union`, `intersection`, `box`, etc.) on the multidimensional contexts, which will help with the multiple explanations of the evidential statements in forensic evaluation where the context sets are often defined as cross products (boxes), intersections, and unions. LUCX's further specification, refinement, and implementation details were produced by Tong [365, 473] in 2007–2008 based on Wan's design [302, 305].

### 4.3.2 JLucid, Objective Lucid, and JOOIP

#### 4.3.2.1 JLUCID

JLUCID [145, 264] was the first attempt on intensional arrays and "free Java functions" in the GIPSY environment. The approach used the LUCID language as the driving main computation, where Java methods were peripheral and could be invoked from the Lucid segment, but not the other way around. This was the first instance of hybrid intensional-imperative programming within GIPSY. The semantics of this approach was not completely defined, plus, it was only a single-sided view (LUCID-to-JAVA) of the problem. JLUCID did not support objects of any kind, but introduced the wrapper class idea to contain the freely appearing Java methods (in either GIPL or Indexical Lucid) and served as a precursor to Objective Lucid [304, 312, 526].

#### 4.3.2.2 Objective Lucid

Objective Lucid [262, 264] was a natural extension of the JLucid language mentioned in the previous section that inherited all of the JLucid's features and introduced Java objects to be available for use by Lucid. Objective Lucid expanded the notion of the Java object (a collection of members of different types) to the array (a collection of members of the same type) and first introduced the dot-notation in the syntax and operational semantics in GIPSY (Figure 27). Like in JLucid, Objective Lucid's focus was on the Lucid part being the "main" program and did not allow Java to call intensional functions or use intensional constructs from within a Java class. Objective Lucid was the first in GIPSY to introduce the more complete operational semantics of the hybrid OO intensional language [304, 312, 526]. Having the arrays and objects allows grouping of the related data items (of the same

```
#typedecl
Nat42;

#JAVA
class Nat42
{
    private int n;

    public Nat42()
    {
        n = 42;
    }

    public Nat42 inc()
    {
        n++;
        return this;
    }

    public void print()
    {
        System.out.println("n = " + n);
    }
}

#OBJECTIVELUCID

(N @.d 2).print[d]()
where
    dimension d;
    N = Nat42[d]() fby.d N.inc[d]();
end
```

Figure 26: The natural-numbers problem in Objective Lucid [305]

type or different types) together and evaluate them under the same context. In Figure 26 is the modified example of a demand-driven evaluation of a simple natural numbers problem re-written in OBJECTIVE LUCID [262, 264, 305]. In Figure 25 is the modified example of a demand-driven evaluation of a similar natural numbers problem in Figure 26 re-written in OBJECTIVE LUCID [264]. In this work, such eduction tree can be adapted to back-tracing in forensic evaluation, e.g., when the inner-most results are traced back to the final result of the entire computation.

$$E_{c-vid} : \frac{\begin{array}{c}\mathcal{D},\mathcal{P} \vdash E : id \quad \mathcal{D},\mathcal{P} \vdash E' : id' \\ \mathcal{D}(id) = (\texttt{class, cid, \underline{cdef}}) \quad \mathcal{D}(id') = (\texttt{classv, cid.cvid, \underline{vdef}}) \\ \mathcal{D},\mathcal{P} \vdash \texttt{<cid.cvid>} : v\end{array}}{\mathcal{D},\mathcal{P} \vdash E.E' : v} \quad (4.3.2.1)$$

$$E_{c-fct} : \frac{\begin{array}{c}\mathcal{D},\mathcal{P} \vdash E : id \quad \mathcal{D},\mathcal{P} \vdash E' : id' \quad \mathcal{D},\mathcal{P} \vdash E_1,\ldots,E_n : v_1,\ldots,v_n \\ \mathcal{D}(id) = (\texttt{class, cid, \underline{cdef}}) \quad \mathcal{D}(id') = (\texttt{classf, cid.cfid, \underline{fdef}}) \\ \mathcal{D},\mathcal{P} \vdash \texttt{<cid.cfid}(v_1,\ldots,v_n)\texttt{>} : v\end{array}}{\mathcal{D},\mathcal{P} \vdash E.E'(E_1,\ldots,E_n) : v} \quad (4.3.2.2)$$

$$E_{ffid} : \frac{\begin{array}{c}\mathcal{D},\mathcal{P} \vdash E : id \quad \mathcal{D},\mathcal{P} \vdash E_1,\ldots,E_n : v_1,\ldots,v_n \\ \mathcal{D}(id) = (\texttt{freefun, ffid, \underline{ffdef}}) \\ \mathcal{D},\mathcal{P} \vdash \texttt{<ffid}(v_1,\ldots,v_n)\texttt{>} : v\end{array}}{\mathcal{D},\mathcal{P} \vdash E(E_1,\ldots,E_n) : v} \quad (4.3.2.3)$$

$$\#\text{JAVA}_{\text{objid}} : \frac{\underline{\texttt{cdef}} = \texttt{Class cid} \{\ldots\}}{\mathcal{D},\mathcal{P} \vdash \underline{\texttt{cdef}} : \mathcal{D}\dagger[\texttt{cid} \mapsto (\texttt{class, cid, \underline{cdef}})], \mathcal{P}} \quad (4.3.2.4)$$

$$\#\text{JAVA}_{\text{objvid}} : \frac{\underline{\texttt{cdef}} = \texttt{Class cid} \{\ldots \underline{\texttt{vdef}} \ldots\} \quad \underline{\texttt{vdef}} = \texttt{public } \textit{type } \texttt{vid};}{\mathcal{D},\mathcal{P} \vdash \underline{\texttt{cdef}} : \mathcal{D}\dagger[\texttt{cid.vid} \mapsto (\texttt{classv, cid.vid, \underline{vdef}})], \mathcal{P}} \quad (4.3.2.5)$$

$$\#\text{JAVA}_{\text{objfid}} : \frac{\begin{array}{c}\underline{\texttt{cdef}} = \texttt{Class cid} \{\ldots \underline{\texttt{fdef}} \ldots\} \\ \underline{\texttt{fdef}} = \texttt{public } \textit{frttype } \texttt{fid}(\textit{fargtype}_1 \textit{ farg}_{id_1},\ldots,\textit{fargtype}_n \textit{ farg}_{id_n})\end{array}}{\mathcal{D},\mathcal{P} \vdash \underline{\texttt{cdef}} : \mathcal{D}\dagger[\texttt{cid.fid} \mapsto (\texttt{classf, cid.fid, \underline{fdef}})], \mathcal{P}} \quad (4.3.2.6)$$

$$\#\text{JAVA}_{\text{ffid}} : \frac{\underline{\texttt{ffdef}} = \textit{frttype } \texttt{ffid}(\textit{fargtype}_1 \textit{ farg}_{id_1},\ldots,\textit{fargtype}_n \textit{ farg}_{id_n})}{\mathcal{D},\mathcal{P} \vdash \underline{\texttt{ffdef}} : \mathcal{D}\dagger[\texttt{ffid} \mapsto (\texttt{freefun, ffid, \underline{ffdef}})], \mathcal{P}} \quad (4.3.2.7)$$

Figure 27: Extract of operational semantics of OBJECTIVE LUCID [304]

#### 4.3.2.3 JOOIP

Wu's JOOIP [526, 528] greatly complements OBJECTIVE LUCID by allowing JAVA to call the intensional language constructs closing the gap and making JOOIP a complete hybrid OO intensional programming language within the GIPSY environment. JOOIP's semantics further refines in a greater detail the operational semantics rules of LUCID and OBJECTIVE LUCID in the attempt to make them complete [304, 312, 526]. JOOIP's approach following GLU# is natural since object-oriented languages are known by literally all computer scientists and software engineers. Especially, JAVA is a very popular and widely used language in today's application domains. JOOIP increases the visibility of Intensional Programming [131, 350] (see Section 3.2) is to make it more mainstream via a marriage between Object-Oriented Programming and Intensional Programming paradigms, allowing a broader audience to be exposed to the benefits of Intensional Programming [507, 526] within their favorite OO language. To show the similarities and differences between JOOIP and GLU#, Wu [528] provided the translation of some of the examples given in [359] into JOOIP for the comparison reasons and to show its advantages. A similar embedding of multidimensional characteristics in a conventional programming language has been proposed by Rondogiannis [407]. In his approach, JAVA is used as the host language and intensional languages are embedded into JAVA as a form of definitional lazy multidimensional arrays. Integration of FORENSIC LUCID with JOOIP becomes more relevant when considered together for hybrid-intensional-imperative programming in particular for self-forensics presented in Appendix D.

### 4.3.3 MARFL

*MARF Language* (MARFL) was conceived and designed to manage the MARF system' runtime (see Section 5.2) configuration as collections of nested name-value configuration option pairs [272]. While not strictly of the Lucid family or GIPSY, MARFL [272] was nearly entirely influenced by LUCID. It is based on the overloaded `@` and `#` operators and allows to navigate into the depth of the higher-order contextual space using the dot operator (see Figure 28) [304]. The latter was indirectly (re-invented in part) influenced by iHTML and `libintense` [452, 455]. For detailed discussion on MARFL and its semantics please refer to Appendix C. FORENSIC LUCID adapts this idea of the hierarchical context navigation and querying with the overloaded `@` and `#` for evidential statements, observation sequences, and individual observations.

$$\mathbf{E_{E.did}} \quad : \quad \frac{\mathcal{D}(E.id) = (\mathtt{dim})}{\mathcal{D}, \mathcal{P} \vdash E.id : id.id} \quad (4.3.3.1)$$

Figure 28: Higher-order context Dot operator of MARFL [304]

## 4.4 Summary

Since the LUCID family of languages thrived around intensional logic that makes the notion of context explicit and central, and relatively recently in LUCX, a first class value [365, 473, 513, 515] (that can be passed around as function parameters or as return values and have a set of operators defined upon), we greatly draw on this notion by formalizing our evidence and the witness stories as a contextual specification of the incident to be tested for consistency against the incident model specification. In our specification model we require more than just atomic context values—we need a higher-order context hierarchy to specify different level of detail of the incident and being able to navigate into the "depth" of such a context. A similar provision has been made in [272] and earlier works of Swoboda and colleagues in [452, 453, 454, 455] that needed some modifications to the expressions of the cyberforensic context [300, 304].

To summarize, expressions written in virtually all LUCID dialects correspond to higher-order intensional logic (HOIL) expressions with some dialect-specific instantiations. They all can alter the context of their evaluation given a set of operators and in some cases types of contexts, their rank, range, and so on. HOIL combines functional programming and intensional logics, e.g., temporal intensional logic (Section 3.2, page 59). The contextual expression can be passed as parameters and returned as results of a function and constitute the multi-dimensional constraint on the LUCID expression being evaluated. The corresponding context calculus [365, 473, 513] defines a comprehensive set of context operators, most of which are set operators and the baseline operators are `@` and `#` that allow to switch the current context or query it, respectively. Other operators allow to define a context space and a point in that context corresponding to the current context. The context can be arbitrary large in its rank. The identified variables of the dimension type within the context can take on any data type, e.g., an integer, or a string, during lazy binding of the resulting context to a dimension identifier [302].

# Chapter 5

# Data Mining and Pattern Recognition

This chapter discusses the relevant background on the data mining and pattern recognition facet of this research to present some of the supporting related work in the area as well as a detailed discussion of the devised tools and their results. This discussion is important and relevant because the data mining techniques are used substantially in digital investigation with imprecise or encrypted data in files or network packets (netflows) and the PoC tools presented here are some of the sources of encoded fuzzy classification evidential data that can be used in reasoning in digital investigation cases. The other motivation is the augmentation of the tools themselves to support self-forensics (Appendix D) in autonomic environments.

In the detailed discussion we primarily focus on MARF, `fileType`, MARFCAT, and MARFPCAT produced and maintained primarily by the author Mokhov. All these are designed to export any of their classification findings and data structures in the declarative FORENSIC LUCID format via their corresponding encoders to be a part of the witness testimonies aiding investigations of incidents such as involving vulnerable software (browsers, etc.) or malicious software (Chapter 9), and self-forensic investigations (Appendix D). MARFCAT and MARFPCAT also have problem-specific DGT and DWT in GIPSY (see Chapter 6) for scalable demand-driven evaluation using GIPSY's distributed middleware.

This chapter is organized as follows. The related work is referenced in Section 5.1. Brief description of MARF is in Section 5.2, of MARFCAT in Section 5.4, and of MARFPCAT in Section 5.5 respectively. Some example classification results for all approaches are presented as well. Then follows a brief summaryconcluding remarks in Section 5.6.

## 5.1 Related Work

There is a great number of related works in the area published over the last two decades or so. We briefly focus on the subgenre of data mining and pattern recognition works related to signal, audio, and NL processing, switching to the various aspects of the data mining aspects applied to computer and network security and digital forensics. The classical works of data mining, patter recognition, signal processing, NLP and the related disciplines are sub-disciplines of AI [81, 182, 186, 201, 248, 353, 411]. In reference to the previous background chapter on LUCID (Chapter 4), multidimensional intensional signal processing was proposed by Agi using GLU in 1995 [5] and MARFL for MARF by the author Mokhov (Appendix C).

In typical pipelines when data are loaded and preprocessed (filtered, etc.), the next major step is to do feature extraction of features of interest. This can be done automatically and/or manually to improve the classification precision and recall metrics. For this to work, the best features types have to be properly selected and a few proposals for feature selection and extraction algorithms were put forward [140, 243] including continual feature selection [426]. The classification stage comes into play in learning and classification using a variety of proposed classifiers and their combinations [63, 129, 208].

Chen in 2006 [64] in his book summarized the issues about information sharing and data mining as intelligence and security informatics disciplines.

### 5.1.1 Open-Source Tools

There are a number of open-source tools and frameworks primarily done in JAVA to do data mining, pattern recognition, signal processing, etc. and that can also integrate with commercial tools and toolboxes, such as that of MATLAB [418] and others. MARF that began in 2002 is detailed in the further sections. It started off as an audio recognition framework by the author and colleagues, but went beyond that and was applied for image, text, code, data analysis, and different classification applications, some of which are detailed further in Section 5.3, Section 5.4, and Section 5.5. At around the same time frame, another open-source system emerged, also done in JAVA, at CMU called *CMU Sphinx* [468], which focused on the reliable and robust tasks for speech recognition (speech-to-text). On around 2006 *Weka* [469] was created and emerged; its source code base originally followed very much the MARF's architecture. This system has had a lot of developer support and maintenance and significantly grew in popularity since. The author Mokhov (while working on the MARFPCAT branch of the project (detailed further in this chapter) as a part of a malware classification based on packet-capture (*pcap*) data project [49]) developed initial MARF plug-in support to allow Weka's rich classifier set to be used in the MARF's pipeline. GATE [462] is a tool with rich customizable pipeline of processing resources (PRs) support written in JAVA originally developed as *General Architecture for Text Engineering*, more specifically to enable NLP developers to create easier NLP applications. It has outgrown the text aspect of it allowing various other technologies being added for other classification tasks and knowledge representation such as Weka, ontologies, and others. MARF's plug-in support for it was also planned [281].

### 5.1.2 Network Forensics

There was a number of data mining, machine learning, and resulting classification approaches used for forensic investigation and analysis of network traffic, especially the encrypted kind over SSL/TLS, `ssh`, etc. to identify network-active applications and malware. For example, *Density Based Spacial Clustering of Application with Noise* (DBSCAN) was proposed [531] in 2008 to use clustering algorithms to identify various FTP clients, VLC media player, and UltraVNC traffic over encrypted channels. That was followed by Alshammari *et al.* [15, 16] to identify `ssh` and Skype encrypted traffic (without looking at payload, port numbers, and IP addresses). Additionally, comparison of algorithms and approaches for network traffic classification were proposed separately by Alshammari *et al.* [14] in 2008 and Okada *et al.* [346] in 2011 surveying and comparing various machine learning algorithms for encrypted traffic analysis. Prior to that, in 2000, Lee *et al.* [232] introduced the notion of adaptive intrusion detection with data mining techniques, followed by Bloedorn [44] in 2001 of MITRE with their technical report on data mining for network intrusion detection with data mining as well. Livadas *et al.* [239] in 2006 used machine learning techniques to classify botnet traffic followed by Binsalleeh *et al.* [43] in 2010 with the very detailed analysis of the Zeus botnet crimeware toolkit. Simon *et al.* [430], likewise in 2006, proposed to detect non-obvious brute-force network scanning using a data mining approach as well. Finally, Boukhtouta *et al.* (including the author) [49] proposed network malware identification by machine-learning captured network malware traffic pcaps as well as benign traffic (regardless the fact if the payload encrypted or not) via a select set of features and comparing it to the captured traffic currently going through the network.

### 5.1.3 Malware Analysis and Classification for Investigations

Some of the malware and malicious code techniques described here are also used in the network anomaly analysis described in the previous section especially for the network-enabled malware that propagates, so some techniques described there are also applicable here. In part our methodology has some similarities in common with the related work on automatic classification of new, unknown malware and malware in general such as viruses, web malware, worms, spyware, and others where AI pattern recognition and expert system techniques are successfully used for automatic classification [290].

Schultz *et al.* [419] in 2001 proposed the data mining techniques to detect new malicious executables as opposed to traditional signature-based approaches. Sung *et al.* [451] proposed the static analysis of vicious executables (SAVE) framework in 2004. In 2007, Bailey, Nazario, *et al.* [36] came up with automated analysis and classification of various Internet malware. Provos *et al.* [387] in the same year did the web-based malware "ghost in the browser" analysis. Suenaga proposed malware analysis through linguistics techniques by searching for ethnic words [445] and Hnatiw *et al.* [170] proposed techniques for parsing malicious and malformed executables as well, both in 2007. Rafique *et al.* [395] followed suit in 2008 and proposed another approach for automatic adjudication of new malware as well.

In 2007, Hwang *et al.* [176] proposed an anti-malware expert system. Spectral techniques are used for pattern scanning in malware detection by Eto *et al.* in [105] in 2009 where they propose a malware distinction method based on scan patterns. Subsequently, Inoue *et al.* [104, 183] proposed a general data mining system for incident analysis with data mining engines called NICTER based on data mining techniques.

Classification results may be encoded as FORENSIC LUCID constructs as an evidence in investigations to support or disprove reasoning about investigative claims in incidents involving malware. Additionally, FORENSIC LUCID-based reasoners are planned to be integrated into some of these tool as a part of the future work (Section 10.4).

### 5.1.4 Code Analysis for Investigations

Arguably, to the author's knowledge MARFCAT (Section 5.4) in 2010 was the first time a machine learning approach was attempted to static code analysis for vulnerable/weak code classification with the first results demonstrated during the SATE2010 workshop [284, 285, 347]. In the same year, a somewhat similar approach independently was presented [52] for vulnerability classification and prediction using machine learning and SVMs, but working with a different set of data [314].

Additional related work (to various degree of relevance or use) can be found below (this list is not exhaustive) [287, 314]: A taxonomy of Linux kernel vulnerability solutions in terms of patches and source code as well as categories for both are found in [297]. The core ideas and principles behind the MARF's pipeline and testing methodology for various algorithms in the pipeline adapted to this case are found in [270, 281] as well as in Section 5.2 as it was the easiest implementation available to accomplish the task. There also one can find the majority of the core options used to set the configuration for the pipeline in terms of algorithms used. A binary analysis using machine a learning approach for quick scans for files of known types in a large collection of files is described in [290] as well as the NLP and machine learning for NLP tasks in DEFT2010 [279, 283] with the corresponding `DEFT2010App` and its predecessor for hand-written image processing `WriterIdentApp` [318]. Tlili's 2009 PhD thesis covers topics on automatic detection of safety and security vulnerabilities in open source software [472]. Statistical analysis, ranking, approximation, dealing with uncertainty, and specification inference in static code analysis are found in the works of Engler's team [215, 216, 217]. Kong *et al.* further advance static analysis (using parsing, etc.) and specifications to eliminate human specification from the static code analysis in [213]. Hanna *et al.* describe a synergy between static and dynamic analysis for the detection of software security vulnerabilities in [163] paving the way to unify the two analysis methods. Other researchers propose the MEDUSA system for metamorphic malware dynamic analysis using API signatures in [330].

Some of the statistical NLP techniques we used are described at length in [248]. BitBlaze (and its web counterpart, WebBlaze) are other recent tools that do fast static and dynamic binary code analysis for vulnerabilities, developed at Berkeley [435, 436]. For wavelets, for example, Li *et al.* [234] have shown wavelet transforms and $k$-means classification can be used to identify communicating applications fast on a network and is relevant to the study of the code in text or binary form [314].

## 5.2 MARF

MARF (*Modular A\* Recognition Framework*) as a framework (and its instance as a library) has been covered in a number of work since its inception in 2002 [260, 268, 270, 283, 290, 465] for various classification tasks originally targeting audio applications, but eventually outgrowing that domain (hence the change from *Audio* to *A\** in the name). The research into MARF has led to connections with other disciplines such as presented here intensional programming with the design of MARFL, a MARF configuration specification and manipulation language (Appendix C), design of the code analysis and network packet analysis applications MARFCAT and MARFPCAT presented further, forensic analysis of file types [290], and NLP aspects [281]. MARF was integrated in various capacities with GIPSY (Chapter 6) since circa 2005, had a PoC design of the distributed [271, 296] and autonomic versions [320, 494, 495, 496]. MARF's data structure declarations (as well as that of other middleware [322]) as well as its applications are a subject of encoding and export in FORENSIC LUCID (Section 8.5.1.1) to serve as additional witness accounts in investigations.

### 5.2.1 MARF Overview

MARF is an open-source project that provides pattern recognition APIs with sample implementation for (un)supervised machine learning and classification, including biometric forensic identification, written in JAVA [260, 268, 270, 465] by the author and collaborators in 2002. As one of its roles, it serves as a testbed to verify common and novel algorithms for sample loading, preprocessing, feature extraction, training and classification stages. In this role MARF provides researchers with a tool for the practical comparison of the algorithms in a uniform environment and allows for dynamic module selection via reflection [142] based on a wide array of configuration options supplied by MARF applications. Within few years MARF accumulated a fair number of implementations for each of the pipeline stages (cf. Figure 29, page 103) allowing comparative studies of algorithm combinations, studying their behavior and other properties when used for various pattern recognition tasks. MARF, its derivatives, and applications were also used beyond audio processing tasks, as in this work, due to the generality of the design and implementation in [265, 271, 292] and several other works [290, 319].

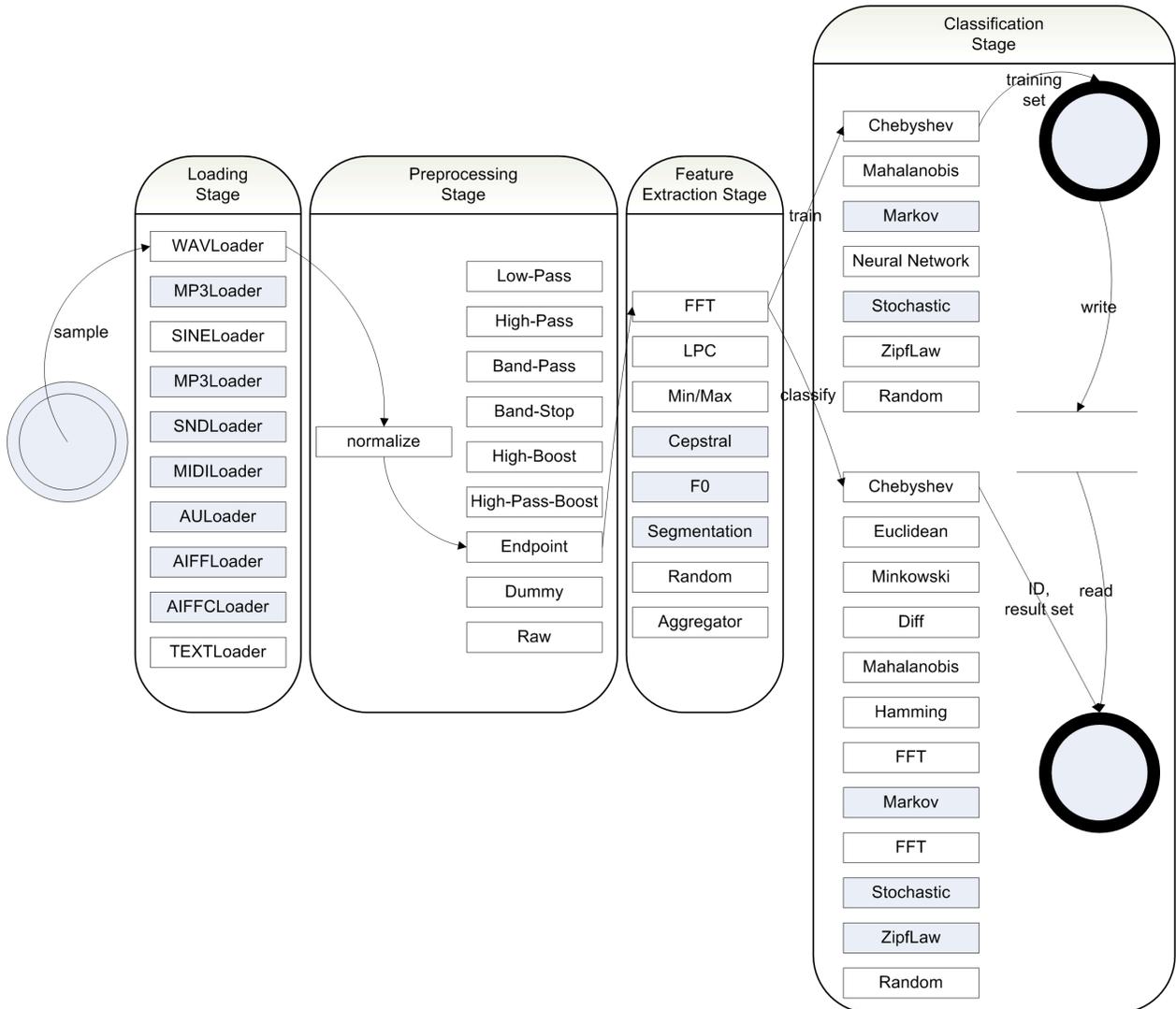

Figure 29: MARF's pattern-recognition pipeline [272]

Some of the MARF's architectural design influenced GIPSY ([264], Chapter 6) and MARF's utility modules are likewise in use by GIPSY. Both distributed version of MARF (DMARF) and GIPSY were together proposed case studies from the security [271] and self-forensics [321] standpoints.

### 5.2.2 MARF Architecture

The *Modular A\* Recognition Framework* (MARF) [260, 268, 270, 465] is a JAVA framework, and an open-source research platform and a collection of pattern recognition, signal processing, and natural language processing (NLP) algorithms written in JAVA and put into a modular and extensible framework facilitating addition of new algorithms for use and experimentation by scientists. A MARF instance can run distributively [265] over a network, run stand-alone, or may just act as a simple library in applications. MARF has a number of algorithms implemented for various pattern recognition and some signal processing tasks [271, 322].

The backbone of MARF consists of pipeline stages that communicate with each other to get the data they need in a chained manner. MARF's pipeline of algorithm implementations is illustrated in Figure 29, where the implemented algorithms are in white boxes, and the stubs or in-progress algorithms are in gray. The pipeline consists of four basic stages: sample loading, preprocessing, feature extraction, and training/classification [271, 322].

There are a number of applications that test MARF's functionality and serve as examples of how to use or to test MARF's modules. One of the most prominent applications is `SpeakerIdentApp`—Text-Independent Speaker Identification (who, gender, accent, spoken language, etc.) [317]. Its derivative, `FileTypeIdentApp`, was used to employ MARF's capabilities for forensic analysis of file types [290] as opposed to [271, 322] the Unix `file` utility [77, 78].

### 5.2.3 Pattern Recognition Pipeline

The conceptual pattern recognition pipeline design presented in Figure 29 depicts the core of the data flow and transformation between the stages in MARF [260, 465]. Generally, the classical pattern recognition process starts by loading a sample (e.g., an audio recording, text, image file, pcap, or virtually any regular file), preprocessing it somehow (e.g., normalization and filtering out noisy and "silent" data), then extracting the most prominent features, and, finally either training the system such that it learns a new set of features of a given subject or actually classifies what/who the subject is [290].

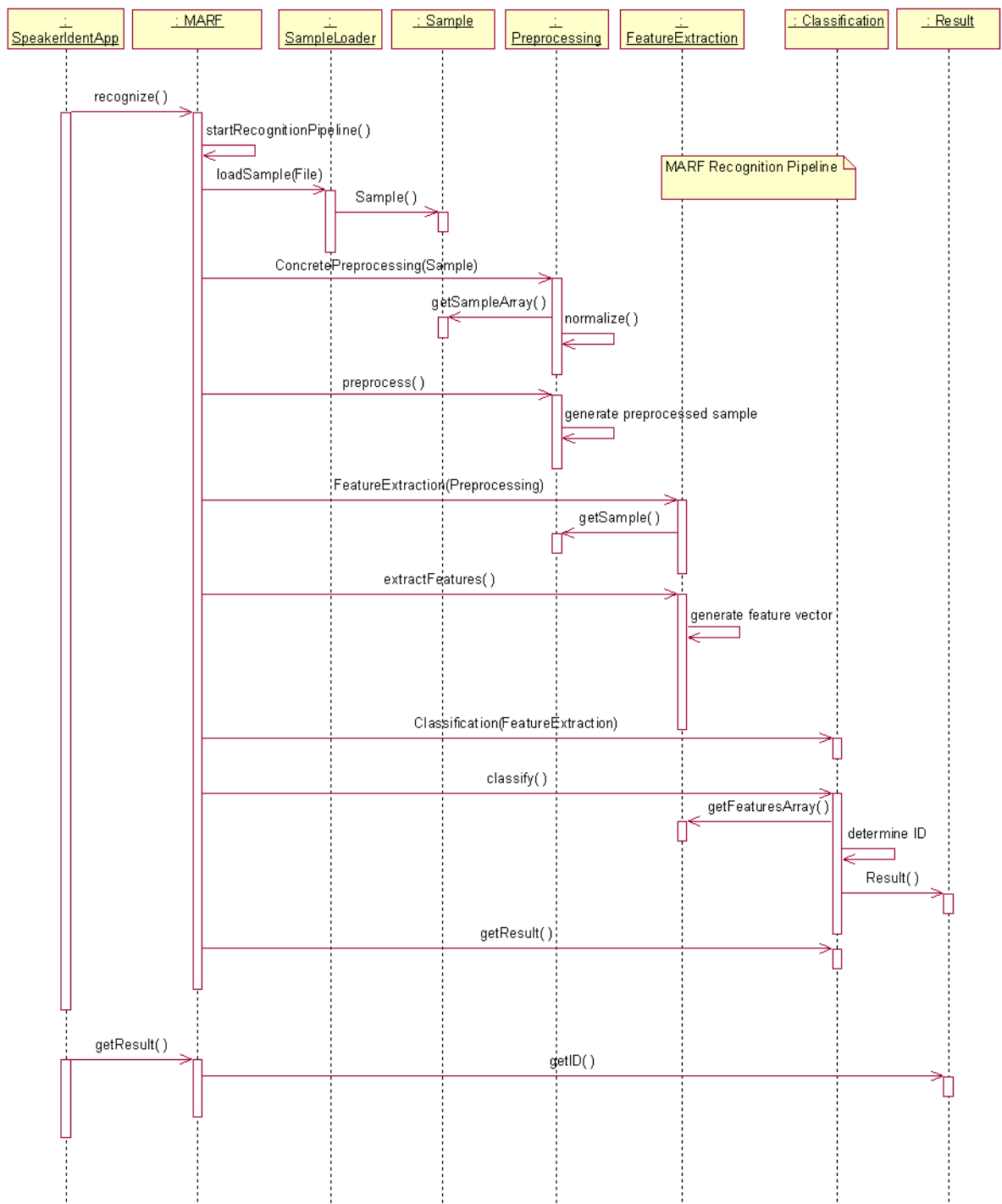

Figure 30: MARF's pattern-recognition pipeline sequence diagram [290]

The outcome of training is either a collection of some form of feature vectors or their mean or median clusters [270], called *training sets*, which are stored per every learned subject. The outcome of classification is an instance of the `ResultSet` data structure, which is a sorted collection of IDs (`int`) and their corresponding outcome values (`double`); the sorting is done from most likely outcome to least likely. The most likely one is the ultimate outcome to be interpreted by the application. Some of the details of such processing of classification are illustrated on the actual sequence of events and method calls within the main `MARF` module is shown in Figure 30 [290].

### 5.2.4 Distributed MARF (DMARF)

DMARF [265] is based on the classical MARF whose pipeline stages were made into distributed nodes [322]. Specifically, the classical MARF presented earlier was extended [265] to allow the stages of the pipeline to run as distributed nodes as well as their front-ends, as shown in Figure 31 in a high-level overview. The basic stages and the front-ends were designed to support, but implemented without backup recovery or hot-swappable capabilities. They only support communication over Java RMI [523], CORBA [446], and XML-RPC WebServices [296, 447]. Later, DMARF was further extended to allow management of its nodes with SNMP [293] by implementing the proxy SNMPv2 [165] agents and translating some of the management information to DMARF's "native" operations. There is also an undergoing project on the intensional configuration scripting language, MARFL [272] (see Appendix C) to script MARF tasks and applications and allows them to be run distributively either using MARF's own infrastructure or over a GIPSY network instance (Chapter 6). Being distributed, DMARF has new data structures and data flow paths that are not covered by the FORENSIC LUCID specification of the classical MARF, so we contribute an extension in this work [322] in Section 8.5.1.1 and Appendix D.4.6.1.

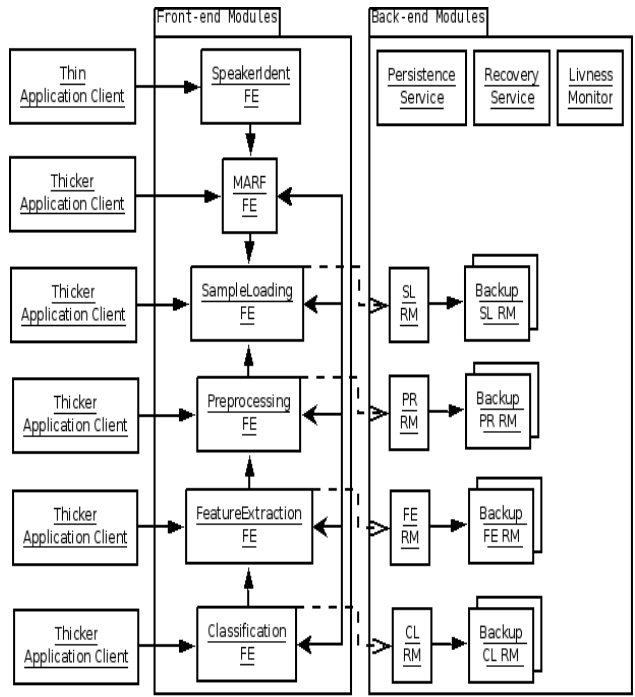

Figure 31: The distributed MARF pipeline

## 5.3 `fileType`

The Unix `file` utility determines file types of regular files by examining usually the first 512 bytes of the file that often contain some magic header information or typical header information for binary files or common text file fragments; otherwise, it defers to the OS-dependent `stat()` system call. It combines heuristics with the common file extensions to give the final result of classification. While `file` is standard, fast and small, and its magic database is "serviceable" by expert users, for it to recognize new file types, perhaps with much finer granularity it requires code and/or magic database updates and a patch release from the core developers to recognize new file types correctly. MARF-based `fileType` was proposed in 2008 an alternative `file`-like utility in determining file types with much greater flexibility that can learn new types on the user's side and be integrated into forensic toolkits as a plug-in that relies on the `file`-like utility and uses signal processing techniques to compute the "spectral signatures" of file types. What follows is an overview of the design of such a tool based on MARF's collection of algorithms and the selection of the best combination and the integration of the tool into a forensic toolkit to enhance the tool, called `fileType` with the automatic machine learning capabilities of the new file types. Some of the advantages and disadvantages of this tool are compared with the `file` utility in terms of various metrics [290].

It also served as a predecessor to the MARF Forensic Data Sniffer case study. A similar file type analysis has recently (2012) been included with Sourcefire's SIM and FireSIGHT into their toolset to report or block certain file types.

### 5.3.1 Overview

`fileType` follows an approach using the MARF's collection of algorithms to determine file types in various ways and compare them using signal processing and NLP techniques, both supervised and unsupervised machine learning, and various file format loaders. MARF and its application `SpeakerIdentApp` [317] were shown to be used as a proof-of-concept for biometric forensic analysis of the phone-quality audio recordings to classify the identities of speakers irrespective of what speakers say on voice recordings, their gender, and spoken accent [268, 270]. `fileType` adapts MARF's pattern recognition pipeline, the `SpeakerIdentApp` application, and the magic database of `file` to be used together, in the resulting application called in JAVA `FileTypesIdentApp`, with a shorthand invocation of `fileType` [290].

MARF conveniently has a class, `ByteArrayFileReader` that reads a file from a file system or an URI (or any `Reader` or `InputStream` for that matter). `fileType` employs this class to read the file data (either the first 512 bytes or the entire file as options) and the values of the byte array become features for classification (spectral or otherwise). It then may optionally do the regular signal pattern recognition techniques [270] of preprocessing and feature extraction to remove all the unwanted noise and silence and extract more discriminating features [290].

The `FileTypesIdentApp` application, a.k.a `fileType`, is capable of understanding some of the `file`'s options [77], and the work is under way to be able to experiment with `file`'s magic database. `fileType` has its own database that it can augment throughout its lifetime automatically using machine learning techniques. The statistics of the algorithm combinations tried and their recognition accuracy performance along with the run-time are stored in a comma-separated values (CSV) file, per each major technique [290].

Table 2: File-type identification top 10 results, bigrams ([290])

| Guess | Rank | Configuration | GOOD | BAD | Precision, % |
|---|---|---|---|---|---|
| 1st | 1 | -wav -raw -lpc -cheb | 147 | 54 | 73.13 |
| 1st | 1 | -wav -silence -noise -raw -lpc -cheb | 147 | 54 | 73.13 |
| 1st | 1 | -wav -noise -raw -lpc -cheb | 147 | 54 | 73.13 |
| 1st | 1 | -wav -norm -lpc -cheb | 147 | 54 | 73.13 |
| 1st | 1 | -wav -silence -raw -lpc -cheb | 147 | 54 | 73.13 |
| 1st | 2 | -wav -silence -norm -fft -cheb | 129 | 72 | 64.18 |
| 1st | 3 | -wav -bandstop -fft -cheb | 125 | 76 | 62.19 |
| 1st | 3 | -wav -silence -noise -norm -fft -cheb | 125 | 76 | 62.19 |
| 1st | 3 | -wav -silence -low -fft -cheb | 125 | 76 | 62.19 |
| 1st | 4 | -wav -silence -norm -lpc -cheb | 124 | 77 | 61.69 |

Table 3: File-type identification top 10 results, 2nd best, bigrams ([290])

| Guess | Rank | Configuration | GOOD | BAD | Precision, % |
|---|---|---|---|---|---|
| 2nd | 1 | -wav -raw -lpc -cheb | 166 | 35 | 82.59 |
| 2nd | 1 | -wav -silence -noise -raw -lpc -cheb | 166 | 35 | 82.59 |
| 2nd | 1 | -wav -noise -raw -lpc -cheb | 166 | 35 | 82.59 |
| 2nd | 1 | -wav -norm -lpc -cheb | 166 | 35 | 82.59 |
| 2nd | 1 | -wav -silence -raw -lpc -cheb | 166 | 35 | 82.59 |
| 2nd | 2 | -wav -silence -norm -fft -cheb | 137 | 64 | 68.16 |
| 2nd | 3 | -wav -bandstop -fft -cheb | 130 | 71 | 64.68 |
| 2nd | 3 | -wav -silence -noise -norm -fft -cheb | 140 | 61 | 69.65 |
| 2nd | 3 | -wav -silence -low -fft -cheb | 140 | 61 | 69.65 |
| 2nd | 4 | -wav -silence -norm -lpc -cheb | 176 | 25 | 87.56 |

Table 4: File-type identification top 10 results, bigrams, per file type ([290])

| Guess | Rank | File type | GOOD | BAD | Precision, % |
|---|---|---|---|---|---|
| 1st | 1 | Mach-O filetype=10 i386 | 64 | 0 | 100.00 |
| 1st | 2 | HTML document text | 64 | 0 | 100.00 |
| 1st | 3 | TIFF image data; big-endian | 64 | 0 | 100.00 |
| 1st | 4 | data | 64 | 0 | 100.00 |
| 1st | 5 | ASCII c program text; with very long lines | 64 | 0 | 100.00 |
| 1st | 6 | Rich Text Format data; version 1; Apple Macintosh | 128 | 0 | 100.00 |
| 1st | 7 | ASCII English text | 64 | 0 | 100.00 |
| 1st | 8 | a /sw/bin/ocamlrun script text executable | 516 | 60 | 89.58 |
| 1st | 9 | perl script text executable | 832 | 192 | 81.25 |
| 1st | 10 | NeXT/Apple typedstream data; big endian; version 4; system 1000 | 255 | 65 | 79.69 |

### 5.3.2 Sample Results

In [290], an experiment was conducted to use a MARF-based `FileTypeIdentApp` for bulk forensic analysis of file types using signal processing techniques. Certain results were quite encouraging precision/recall-wise for the first and second best top 10 statistics extracts in Table 2 and Table 3, as well as statistics per file type in Table 4 [278].

### 5.3.3 Limitations and Drawbacks

In the current implementation of the `fileType` there are several drawbacks and limitations (that were planned to be eliminated or reduced as a part of the future work) [290]. These are listed in part as follows:

- The presented here technique of machine learning and the tool are more effective to use with the regular files only, and are not suited for special files and devices that are file-system specific. For those, the same approach as `file` does using the `stat()` system call [483] can be used (but this is no longer machine learning, effectively deferring such tasks to `file` that does it better) [290].

- Training the system with noisy data samples can deteriorate the recognition accuracy of the tool, which may lead to the problem of over-fitting [162]. This can be either accidental (local) or malicious (system-wide, CAF) by supplying a file of one type for training, but telling it is another. This is a general problem with any machine-learning tools and applications. A way of dealing with this partly is to validate each of the incoming training samples by classifying them first and comparing with the class specified for training and in the case of mismatch, report to the user of a potential problem; in the safe mode refuse to train in the case of mismatch. This will prevent the accidental training on the wrong data for misclassification. The latter only partly solves the problem, as the system can be cheated at the beginning when the new file type being inserted for the first time and is mistrained on [290].

- To be seriously considered in a real investigation toolset and environment, legally, the tool has to be proved to be correct in its design and implementation as well as components it relies on, such as MARF, e.g., by using JML [55, 229, 230] and Isabelle [372] for this task later in the project [290].

## 5.4  MARFCAT

We elaborate on the details of the methodology and the corresponding results of application of the machine learning techniques along with signal processing and NLP alike to static source and binary code analysis in search for and investigation on program weaknesses and vulnerabilities [287]. Here we review the tool, named *MARFCAT*, a *MARF-based Code Analysis Tool* [285], first exhibited at the Static Analysis Tool Exposition (SATE) workshop in 2010 [347] to machine-learn from the (Common Vulnerabilities and Exposures) CVE-based vulnerable as well as synthetic CWE-based cases to verify the fixed versions as well as non-CVE based cases from the projects written in various programming languages. The second iteration of this work was prepared for SATE IV [348] and used its updated data set [314].

### 5.4.1 Overview

We review our machine learning approach to static code analysis and fingerprinting for weaknesses related to security, software engineering, and others using the open-source MARF framework and the MARFCAT application based on it for the NIST's SATE 2010 and SATE IV [348] static analysis tool exposition workshop's data sets that include additional test cases, including large synthetic cases [287, 314]. To aid detection of weak or vulnerable code, including source or binary on different platforms the machine learning approach proved to be fast and accurate for such tasks. We use signal and NLP processing techniques in our approach to accomplish the identification and classification tasks. MARFCAT's design from the beginning in 2010 was made independent of the language being analyzed, be it source code, bytecode, or binary. We evaluated also additional algorithms that were used to process the data [314]. This work is imperative in digital investigations and that's why MARF itself and MARFCAT were designed to export evidence in the FORENSIC LUCID format (Section 8.5.1.1, page 234, [314]).

In 2010, at the core of the workshop there were C/C++-language and JAVA language tracks comprising CVE-selected cases as well as stand-alone cases. The CVE-selected cases had a vulnerable version of a software in question with a list of CVEs attached to it, as well as the most known fixed version within the minor revision number. One of the goals for the CVE-based cases is to detect the known weaknesses outlined in CVEs using static code analysis and also to verify if they were really fixed in the "fixed version" [287, 347].

The test cases at the time included CVE-selected: C: Wireshark 1.2.0 (vulnerable) and Wireshark 1.2.9 (fixed); C++: Chrome 5.0.375.54 (vulnerable) and Chrome 5.0.375.70 (fixed); JAVA: Tomcat 5.5.13 (vulnerable) and Tomcat 5.5.29 (fixed), and non-CVE selected: C: Dovecot; JAVA: Pebble 2.5-M2. They were later expanded to other cases and newer versions and a PHP test case was added (Wordpress). For more information on the data sets see Section 9.1.1.1.1, page 245. The open-source MARFCAT tool itself [285] was developed to machine-learn from the CVE-based vulnerable cases and verify the fixed versions as well as non-CVE based cases from similar programming languages [287].

At the time, the presented machine learning approach was novel and highly beneficial in static analysis and routine testing of any kind of code, including source code and binary deployments for its efficiency in terms of speed, relatively high precision, robustness, and being a complementary tool to other approaches that do in-depth semantic analysis, etc., by prioritizing those tools' targets. All these techniques can be used in an automated manner in diverse distributed and scalable demand-driven environments (e.g., GIPSY, Chapter 6) in order to ensure the code safety, especially the mission critical software code in all kinds of systems. It uses spectral, acoustic and language models to learn and classify such a code [314].

### 5.4.2 Core Principles

The core methodology principles include:

- Machine learning and dynamic programming

- Spectral and signal processing techniques

- NLP $n$-gram and smoothing techniques (add-$\delta$, Witten-Bell, MLE, etc.)

MARFCAT uses signal processing techniques (i.e., no syntactic parsing or otherwise work at the syntax and semantics levels). MARFCAT treats the source code as a "signal", equivalent to binary, where each $n$-gram ($n = 2$ presently, i.e., two consecutive characters or, more generally, bytes) are used to construct a sample amplitude value in the signal. In the NLP pipeline, it similarly treats the source code as "characters", where each $n$-gram ($n = 1..3$) is used to construct the language model [314].

The MARFCAT system is shown the examples of files with weaknesses and it learns them by computing spectral signatures using signal processing techniques or various language models (based on options) from CVE-selected test cases. When some of the mentioned techniques are applied (e.g., filters, silence/noise removal, other preprocessing and feature extraction techniques), the line number information is lost as a part of this process [314].

When testing, MARFCAT computes either how similar or distant each file is from the known trained-on weakness-laden files or compares the trained language models with the unseen language fragments in the NLP pipeline. In part, the methodology can approximately be seen as some fuzzy signature-based "antivirus" or IDS software systems detect bad signature, except that with a large number of machine learning and signal processing algorithms and fuzzy matching, we test to find out which combination gives the highest precision and best run-time [314].

At the present, however, MARFCAT processes the whole files instead of parsing the finer-grain details of patches and weak code fragments. This aspect lowers the precision, but is relatively fast to scan all the code files [314].

### 5.4.3 CVEs and CWEs – the Knowledge Base

The CVE-selected test cases serve as a source of the knowledge base to gather information of how known weak code "looks like" in the signal form [347], which are stored as spectral signatures clustered per CVE or CWE (Common Weakness Enumeration). The introduction by the SAMATE team of a large synthetic code base with CWEs, serves as a part of knowledge base learning as well [314]. Thus, we:

- Teach the system from the CVE-based cases
- Test on the CVE-based cases
- Test on the non-CVE-based cases

For synthetic cases, similarly:

- Teach the system from the CWE-based synthetic cases
- Test on the CWE-based synthetic cases
- Test on the CVE and non-CVE-based cases for CWEs from synthetic cases

We created index files in XML in the format similar to that of SATE to index all the file of the test case under study. The CVE-based cases after the initial index generation are manually annotated from the NVD database before being fed to the system [314].

### 5.4.4 Categories for Machine Learning

The two primary groups of classes MARFCAT is trained and tested on include naturally the CVEs [340, 341] and CWEs [484]. The advantages of CVEs is the precision and the associated meta knowledge from [340, 341] can be all aggregated and used to scan successive versions of the same software or derived products (e.g., WebKit in multiple browsers). CVEs are also generally uniquely mapped to CWEs. The CWEs as a primary class, however, offer broader categories, of kinds of weaknesses there may be, but are not yet well assigned and associated with CVEs, so we observe the loss of precision. Since there is no syntactic parsing, MARFCAT generally cannot deduce weakness types or even simple-looking aspects like line numbers where the weak code may be, it resorts to the secondary categories, that are usually tied into the first two, which we also machine-learn along, such as issue types (*sink*, *path*, *fix*) and line numbers [314].

### 5.4.5 Algorithms

In the methodology systematic tests and selection of the best (a tradeoff between speed and accuracy) combination(s) of the algorithm implementations available is conducted. The subsequent runs then use only the selected algorithms for subsequent testing. This methodology includes the cases when the knowledge base for the same code type is learned from multiple sources (e.g., several independent C test cases) [314].

#### 5.4.5.1 Signal Pipeline

Algorithmically-speaking, the steps that are performed in the machine-learning signal based analysis are in Figure 1. The specific algorithms come from the classical literature and other sources and are detailed in [270], related works therein, and in Section 5.2. In the context of MARFCAT the loading typically refers to the interpretation of the files being scanned in terms of bytes forming amplitude values in a signal (as an example, 8kHz or 16kHz frequency) using either uni-gram, bi-gram, or tri-gram approach. Then, the preprocessing allows to be none at all ("raw", or the fastest), normalization, traditional frequency domain filters, wavelet-based filters, etc. Feature extraction involves reducing an arbitrary length signal to a fixed length feature vector of what thought to be the most relevant features are in the signal (e.g., spectral features in FFT, LPC), min-max amplitudes, etc. The classification stage is then separated either to train by learning the incoming feature vectors (usually as $k$-means clusters, median clusters, or plain feature vector collections, combined with, for example, neural network training) or testing them against the previously learned models [314] (cf., Section 5.2).

---
```
   // Construct an index mapping CVEs to files and locations within files
 1 Compile meta-XML index files from the CVE reports (line numbers, CVE, CWE, fragment size, etc.). Partly done by a
   Perl script and partly annotated manually;
 2 foreach source code base, binary code base do
       // Presently in these experiments we use simple mean clusters of feature vectors or unigram language
           models per default MARF specification [270, 465]
 3     Train the system based on the meta index files to build the knowledge base (learn);
 4     begin
 5         Load (interpret as a wave signal or n − gram);
 6         Preprocess (none, FFT-filters, wavelets, normalization, etc.);
 7         Extract features (FFT, LPC, min-max, etc.);
 8         Train (Similarity, Distance, Neural Network, etc.);
 9     end
10     Test on the training data for the same case (e.g., Tomcat 5.5.13 on Tomcat 5.5.13) with the same annotations to
       make sure the results make sense by being high and deduce the best algorithm combinations for the task;
11     begin
12         Load (same);
13         Preprocess (same);
14         Extract features (same);
15         Classify (compare to the trained k-means, or medians, or language models);
16         Report;
17     end
18     Similarly test on the testing data for the same case (e.g., Tomcat 5.5.13 on Tomcat 5.5.13) without the
       annotations as a sanity check;
19     Test on the testing data for the fixed case of the same software (e.g., Tomcat 5.5.13 on Tomcat 5.5.33);
20     Test on the testing data for the general non-CVE case (e.g., Tomcat 5.5.13 on Pebble or synthetic);
21 end
```
---

**Algorithm 1:** Machine-learning-based static code analysis testing algorithm using the signal pipeline [314]

#### 5.4.5.2 NLP Pipeline

The steps that are performed in NLP and the machine-learning based analyses are presented in Figure 2. The specific algorithms again come from the classical literature (e.g., [248]) and are detailed in [281] and the related works. To be more specific for this background overview, the loading typically refers to the interpretation of the files being scanned in terms of $n$-grams: uni-gram, bi-gram, or tri-gram approach and the associated statistical smoothing algorithms, the results of which (a vector, 2D or 3D matrix) are stored [314].

### 5.4.6 Binary and Bytecode Analysis

MARFCAT also does preliminary JAVA bytecode and compiled C code static analysis and produces results using the same signal processing, NLP, combined with machine learning and data mining techniques. At this writing, the NIST SAMATE synthetic reference data set

```
 1  Compile meta-XML index files from the CVE reports (line numbers, CVE, CWE, fragment size, etc.). Partly done by a
    Perl script and partly annotated manually;
 2  foreach source code base, binary code base do
        // Presently these experiments use simple unigram language models per default MARF
           specification [281]
 3      Train the system based on the meta index files to build the knowledge base (learn);
 4      begin
 5          Load (n-gram);
 6          Train (statistical smoothing estimators);
 7      end
 8      Test on the training data for the same case (e.g., Tomcat 5.5.13 on Tomcat 5.5.13) with the same annotations to
        make sure the results make sense by being high and deduce the best algorithm combinations for the task;
 9      begin
10          Load (same);
11          Classify (compare to the trained language models);
12          Report;
13      end
14      Similarly test on the testing data for the same case (e.g., Tomcat 5.5.13 on Tomcat 5.5.13) without the
        annotations as a sanity check.
15      Test on the testing data for the fixed case of the same software (e.g., Tomcat 5.5.13 on Tomcat 5.5.33);
16      Test on the testing data for the general non-CVE case (e.g., Tomcat 5.5.13 on Pebble or synthetic);
17  end
```

**Algorithm 2:** Machine-learning-based static code analysis testing algorithm using the NLP pipeline [314]

for JAVA and C was used. The algorithms presented in Section 5.4.5 are used as-is in this scenario with the modifications to the index files. The modifications include removal of the line numbers, source code fragments, and lines-of-text counts (which are largely meaningless and ignored. The byte counts may be recomputed and capturing a byte offset instead of a line number was projected. The filenames of the index files were updated to include `-bin` in them to differentiate from the original index files describing the source code [314].

### 5.4.7 Wavelets

During MARFCAT design and development as a part of a collaboration project wavelet-based signal processing for the purposes of noise filtering is being introduced into MARF to compare it to no-filtering or FFT-based classical filtering. It's been also shown in [234] that wavelet-aided filtering could be used as a fast preprocessing method for a network application identification and traffic analysis [236] as well [314]. That implementation relies in part on the algorithm and methodology found in [1, 211, 212, 420], and at this point only a separating 1D discrete wavelet transform (SDWT) has been tested [314].

## 5.4.8 Demand-Driven Distributed Evaluation with GIPSY

To enhance the scalability of the approach, we convert the MARFCAT stand-alone application to a distributed one using an eductive model of computation (demand-driven) implemented in the General Intensional Programming System (GIPSY)'s multi-tier run-time system [160, 191, 362, 501], which can be executed distributively using Jini (Apache River), or JMS [193] (see Section 6) [314].

To adapt the application to the GIPSY's multi-tier architecture, we create a problem-specific generator and worker tiers (PS-DGT and PS-DWT respectively) for the MARFCAT application. The generator(s) produce demands of what needs to be computed in the form of a file (source code file or a compiled binary) to be evaluated and deposit such demands into a store managed by the demand store tier (DST) as pending. Workers pickup pending demands from the store, and them process then (all tiers run on multiple nodes) using a traditional MARFCAT instance. Once the result (a `Warning` instance) is computed, the PS-DWT deposit it back into the store with the status set to *computed*. The generator "harvests" all computed results (warnings) and produces the final report for a test cases. Multiple test cases can be evaluated simultaneously or a single case can be evaluated distributively. This approach helps to cope with large amounts of data and avoid recomputing warnings that have already been computed and cached in the DST [314]. Rabah also contribute a graphical GMT [393], and MARFCAT configuration is rendered in Figure 38 as an example.

In this setup a demand represents a file (a path) to scan (actually an instance of the `FileItem` object), which is deposited into the DST. The PS-DWT picks up that and checks the file per training set that's already there and returns a `ResultSet` object back into the DST under the same demand signature that was used to deposit the path to scan. The result set is sorted from the most likely to the list likely with a value corresponding to the distance or similarity. The PS-DGT picks up the result sets and does the final output aggregation and saves report in one of the desired report formats (see Section 5.4.9 picking up the top two results from the result set and testing against a threshold to accept or reject the file (path) as vulnerable or not. This effectively splits the monolithic MARFCAT application in two halves in distributing the work to do where the classification half is arbitrary parallel [314].

Simplifying assumptions:

- Test case data and training sets are present on each node (physical or virtual) in advance (via a copy or a CIFS or NFS volume), so no demand driven training occurs, only classification

- The demand assumes to contain only the file information to be examined (`FileItem`)

- PS-DWT assumes a single pre-defined configuration, i.e., the configuration for MARF-CAT's option is not a part of the demand

- PS-DWT assume CVE or CWE testing based on its local settings and not via the configuration in a demand

### 5.4.9 Export and Encoding

#### 5.4.9.1 SATE

By default MARFCAT produces the report data in the SATE XML format, according to the SATE IV requirements. In this iteration other formats are being considered and realized. To enable multiple format output, the MARFCAT report generation data structures were adapted case-based output [314].

#### 5.4.9.2 FORENSIC LUCID

MARFCAT began FORENSIC LUCID export support, the core topic of this thesis (Chapter 7). Following the data export in FORENSIC LUCID in the preceding work [269, 304, 310] we use it as a format for evidential processing of the results produced by MARFCAT. The Chapter 7 provides details of the language; it will suffice to mention here that the report generated by MARFCAT in FORENSIC LUCID is a collection of warnings as observations with the hierarchical notion of nested context of warning and location information. These will form an evidential statement in FORENSIC LUCID (see, e.g., [314, Appendix]). The example scenario where such evidence compiled via a MARFCAT FORENSIC LUCID report would be in web-based applications and web browser-based incident investigations of fraud, XSS, buffer overflows, etc. linking CVE/CWE-based evidence analysis of the code (binary or source) security bugs with the associated web-based malware propagation or attacks to provide possible events where specific attacks can be traced back to the specific security vulnerabilities [314].

### 5.4.10 Sample Results

This section illustrates some sample classification results on a few sample test cases from the SATE workshop.

#### 5.4.10.1 Chrome 5.0.375.54

This version's CVE testing result of Chrome 5.0.375.54 is in Table 5. The results are as good as the training data given; if there are mistakes in the data selection then the results will also have mistakes accordingly [287].

Table 5: Sample CVE classification stats for Chrome 5.0.375.54 [287]

| guess | run | algorithms | good | bad | % |
|---|---|---|---|---|---|
| 1st | 1 | `-nopreprep -raw -fft -eucl` | 10 | 1 | 90.91 |
| 1st | 2 | `-nopreprep -raw -fft -cos` | 10 | 1 | 90.91 |
| 1st | 3 | `-nopreprep -raw -fft -diff` | 10 | 1 | 90.91 |
| 1st | 4 | `-nopreprep -raw -fft -cheb` | 10 | 1 | 90.91 |
| 1st | 5 | `-nopreprep -raw -fft -mink` | 9 | 2 | 81.82 |
| 1st | 6 | `-nopreprep -raw -fft -hamming` | 9 | 2 | 81.82 |
| 2nd | 1 | `-nopreprep -raw -fft -eucl` | 11 | 0 | 100.00 |
| 2nd | 2 | `-nopreprep -raw -fft -cos` | 11 | 0 | 100.00 |
| 2nd | 3 | `-nopreprep -raw -fft -diff` | 11 | 0 | 100.00 |
| 2nd | 4 | `-nopreprep -raw -fft -cheb` | 11 | 0 | 100.00 |
| 2nd | 5 | `-nopreprep -raw -fft -mink` | 10 | 1 | 90.91 |
| 2nd | 6 | `-nopreprep -raw -fft -hamming` | 10 | 1 | 90.91 |
| guess | run | class | good | bad | % |
| 1st | 1 | CVE-2010-2301 | 6 | 0 | 100.00 |
| 1st | 2 | CVE-2010-2300 | 6 | 0 | 100.00 |
| 1st | 3 | CVE-2010-2299 | 6 | 0 | 100.00 |
| 1st | 4 | CVE-2010-2298 | 6 | 0 | 100.00 |
| 1st | 5 | CVE-2010-2297 | 6 | 0 | 100.00 |
| 1st | 6 | CVE-2010-2304 | 6 | 0 | 100.00 |
| 1st | 7 | CVE-2010-2303 | 6 | 0 | 100.00 |
| 1st | 8 | CVE-2010-2295 | 10 | 2 | 83.33 |
| 1st | 9 | CVE-2010-2302 | 6 | 6 | 50.00 |
| 2nd | 1 | CVE-2010-2301 | 6 | 0 | 100.00 |
| 2nd | 2 | CVE-2010-2300 | 6 | 0 | 100.00 |
| 2nd | 3 | CVE-2010-2299 | 6 | 0 | 100.00 |
| 2nd | 4 | CVE-2010-2298 | 6 | 0 | 100.00 |
| 2nd | 5 | CVE-2010-2297 | 6 | 0 | 100.00 |
| 2nd | 6 | CVE-2010-2304 | 6 | 0 | 100.00 |
| 2nd | 7 | CVE-2010-2303 | 6 | 0 | 100.00 |
| 2nd | 8 | CVE-2010-2295 | 10 | 2 | 83.33 |
| 2nd | 9 | CVE-2010-2302 | 12 | 0 | 100.00 |

### 5.4.10.2 Tomcat 5.5.13

This example of a MARFCAT classification run represents CVE-based testing on training for Tomcat 5.5.13. Classifiers corresponding to `-cheb` (Chebyshev distance) and `-diff` (Diff distance) continue to dominate as in the other test cases [287]. These CVE-based results are summarized in Table 6 [287].

Table 6: Sample CVE stats for Tomcat 5.5.13 [287]

| | | | | | |
|---|---|---|---|---|---|
| 1st | 1 | -nopreprep -raw -fft -diff | 36 | 7 | 83.72 |
| 1st | 2 | -nopreprep -raw -fft -cheb | 36 | 7 | 83.72 |
| 1st | 3 | -nopreprep -raw -fft -cos | 37 | 9 | 80.43 |
| 1st | 4 | -nopreprep -raw -fft -eucl | 34 | 9 | 79.07 |
| 1st | 5 | -nopreprep -raw -fft -mink | 28 | 15 | 65.12 |
| 1st | 6 | -nopreprep -raw -fft -hamming | 26 | 17 | 60.47 |
| 2nd | 1 | -nopreprep -raw -fft -diff | 40 | 3 | 93.02 |
| 2nd | 2 | -nopreprep -raw -fft -cheb | 40 | 3 | 93.02 |
| 2nd | 3 | -nopreprep -raw -fft -cos | 40 | 6 | 86.96 |
| 2nd | 4 | -nopreprep -raw -fft -eucl | 36 | 7 | 83.72 |
| 2nd | 5 | -nopreprep -raw -fft -mink | 31 | 12 | 72.09 |
| 2nd | 6 | -nopreprep -raw -fft -hamming | 29 | 14 | 67.44 |
| guess | run | algorithms | good | bad | % |
| 1st | 1 | CVE-2006-7197 | 6 | 0 | 100.00 |
| 1st | 2 | CVE-2006-7196 | 6 | 0 | 100.00 |
| 1st | 3 | CVE-2006-7195 | 6 | 0 | 100.00 |
| 1st | 4 | CVE-2009-0033 | 6 | 0 | 100.00 |
| 1st | 5 | CVE-2007-3386 | 6 | 0 | 100.00 |
| 1st | 6 | CVE-2009-2901 | 3 | 0 | 100.00 |
| 1st | 7 | CVE-2007-3385 | 6 | 0 | 100.00 |
| 1st | 8 | CVE-2008-2938 | 6 | 0 | 100.00 |
| 1st | 9 | CVE-2007-3382 | 6 | 0 | 100.00 |
| 1st | 10 | CVE-2007-5461 | 6 | 0 | 100.00 |
| 1st | 11 | CVE-2007-6286 | 6 | 0 | 100.00 |
| 1st | 12 | CVE-2007-1858 | 6 | 0 | 100.00 |
| 1st | 13 | CVE-2008-0128 | 6 | 0 | 100.00 |
| 1st | 14 | CVE-2007-2450 | 6 | 0 | 100.00 |
| 1st | 15 | CVE-2009-3548 | 6 | 0 | 100.00 |
| 1st | 16 | CVE-2009-0580 | 6 | 0 | 100.00 |
| 1st | 17 | CVE-2007-1355 | 6 | 0 | 100.00 |
| 1st | 18 | CVE-2008-2370 | 6 | 0 | 100.00 |
| 1st | 19 | CVE-2008-4308 | 6 | 0 | 100.00 |
| 1st | 20 | CVE-2007-5342 | 6 | 0 | 100.00 |
| 1st | 21 | CVE-2008-5515 | 19 | 5 | 79.17 |
| 1st | 22 | CVE-2009-0783 | 11 | 4 | 73.33 |
| 1st | 23 | CVE-2008-1232 | 13 | 5 | 72.22 |
| 1st | 24 | CVE-2008-5519 | 6 | 6 | 50.00 |
| 1st | 25 | CVE-2007-5333 | 6 | 6 | 50.00 |
| 1st | 26 | CVE-2008-1947 | 6 | 6 | 50.00 |
| 1st | 27 | CVE-2009-0781 | 6 | 6 | 50.00 |
| 1st | 28 | CVE-2007-0450 | 5 | 7 | 41.67 |
| 1st | 29 | CVE-2007-2449 | 6 | 12 | 33.33 |
| 1st | 30 | CVE-2009-2693 | 2 | 6 | 25.00 |
| 1st | 31 | CVE-2009-2902 | 0 | 1 | 0.00 |

## 5.5 MARFPCAT

As a part of the fingerprinting network malware project [49] by analyzing the pcap traces as a branch the MARFPCAT (*MARF-based PCap Analysis Tool*) [289] was created as an extension of the latest revision of MARFCAT [314]. Specifically, to improve detection and classification of the malware in the network traffic or otherwise we employ fast MARF-based machine learning approach to static pcap analysis, fingerprinting, and subsequent investigation. Similarly to the other tools, MARFPCAT is first trained on the known malware pcap data and then measures the detection precision. Then we test it on the unseen data during training, but known to the investigator, and we select the best available machine learning algorithm combination to do so in subsequent investigations. MARFPCAT, like MARFCAT has PS-DWT and -DGT backends to run over a GIPSY network and FORENSIC LUCID export capability. In Section 10.4, page 284 it considered to be used as a evidence feed tool for network forensics related investigations about malware and scanning.

### 5.5.1 Overview

The MARFPCAT work elaborates on the details of the earlier methodology and the corresponding results of application of the machine learning techniques along with signal processing and NLP alike to network packet analysis in search for malicious code in the packet capture (pcap) data. Most of the ideas in Section 5.4 [287] are still applicable here where the same approach was used to machine-learn, detect, and classify vulnerable or weak code fast and with relatively high precision.

We show the system the examples of pcap files with malware and MARFPCAT learns them by computing spectral signatures using signal processing techniques. When we test, we compute how similar or distant each file is from the known trained-on malware-laden files. At the present, however, we are looking at the whole pcap files. This aspect lowers the precision, but is fast to scan all the files.

MARFPCAT was first designed to use JNetPcap, a Java wrapper of `libpcap` as one of the loaders to extract headers and other packet data structure items to refine the classification precision when not using the whole-file training and classification. This work was further refined in [49] by Boukhtouta *et al*. For comparative studies with this work MARF added wrapper plug-ins to allow Weka's classifiers to be available to the MARF pipeline and MARF applications.

### 5.5.2 The Knowledge Base

The GFI malware database with known malware, the reports, etc. serves as a knowledge base to machine-learn from in this experiment. Thus, we primarily:

- Teach the system from the known cases of malware from their pcap data

- Test on the known cases

- Test on the unseen cases

### 5.5.3 Categories for Machine Learning

The primary category is the malware class, e.g., "Virus1"', "Trojan2", etc., which are internally enumerated. The known data is indexed via a PERL script creating an XML file and MARFPCAT uses such files for training and testing.

### 5.5.4 Sample Results

Some classification results follow (Table 7, Table 9, and Table 11) using various algorithm combinations including wavelets. The precision results are designed to be assigned to the credibility (confidence) weights $w$ encoded in FORENSIC LUCID observations.

## 5.6 Summary

Of particular interest to this thesis are the results that are supplied as an evidence encoded in observations and observation sequences in FORENSIC LUCID with the precision/confidence value assigned to the credibility value of each observation. We review the current results of this experimental work, its current shortcomings, advantages, and practical implications.

Table 7: Top 6 distance malware classifiers and 32 results, FFT feature vectors

| guess | run | algorithms | good | bad | % |
|---|---|---|---|---|---|
| 1st | 1 | `-dynaclass -binary -nopreprep -raw -fft -cos -flucid` | 67 | 154 | 30.32 |
| 1st | 2 | `-dynaclass -binary -nopreprep -raw -fft -diff -flucid` | 55 | 166 | 24.89 |
| 1st | 3 | `-dynaclass -binary -nopreprep -raw -fft -cheb -flucid` | 55 | 166 | 24.89 |
| 1st | 4 | `-dynaclass -binary -nopreprep -raw -fft -eucl -flucid` | 50 | 171 | 22.62 |
| 1st | 5 | `-dynaclass -binary -nopreprep -raw -fft -hamming -flucid` | 37 | 184 | 16.74 |
| 1st | 6 | `-dynaclass -binary -nopreprep -raw -fft -mink -flucid` | 34 | 187 | 15.38 |
| 2nd | 1 | `-dynaclass -binary -nopreprep -raw -fft -cos -flucid` | 92 | 129 | 41.63 |
| 2nd | 2 | `-dynaclass -binary -nopreprep -raw -fft -diff -flucid` | 77 | 144 | 34.84 |
| 2nd | 3 | `-dynaclass -binary -nopreprep -raw -fft -cheb -flucid` | 77 | 144 | 34.84 |
| 2nd | 4 | `-dynaclass -binary -nopreprep -raw -fft -eucl -flucid` | 73 | 148 | 33.03 |
| 2nd | 5 | `-dynaclass -binary -nopreprep -raw -fft -hamming -flucid` | 46 | 175 | 20.81 |
| 2nd | 6 | `-dynaclass -binary -nopreprep -raw -fft -mink -flucid` | 47 | 174 | 21.27 |
| guess | run | class | good | bad | % |
| 1st | 1 | VirTool.Win32.VBInject.gen.bp (v) | 6 | 0 | 100.00 |
| 1st | 2 | Trojan.Win32.Agent.roei | 6 | 0 | 100.00 |
| 1st | 3 | BehavesLike.Win32.Malware.dls (mx-v) | 6 | 0 | 100.00 |
| 1st | 4 | Worm.Win32.AutoRun.dkch | 6 | 0 | 100.00 |
| 1st | 5 | Trojan-FakeAV.Win32.Agent.det | 6 | 0 | 100.00 |
| 1st | 6 | FraudTool.Win32.FakeRean | 6 | 0 | 100.00 |
| 1st | 7 | VirTool:Win32/Obfuscator.WJ (suspicious) | 6 | 0 | 100.00 |
| 1st | 8 | Trojan.Win32.Vilsel.ayyw | 6 | 0 | 100.00 |
| 1st | 9 | Worm:Win32/Yeltminky.A!dll | 6 | 0 | 100.00 |
| 1st | 10 | Trojan.Win32.Meredrop | 6 | 0 | 100.00 |
| 1st | 11 | TrojanDownloader:Win32/Allsum | 12 | 0 | 100.00 |
| 1st | 12 | Virtumonde | 6 | 0 | 100.00 |
| 1st | 13 | Backdoor.Win32.Hupigon.nndu | 6 | 0 | 100.00 |
| 1st | 14 | VirTool:WinNT/Protmin.gen!C [generic] | 6 | 0 | 100.00 |
| 1st | 15 | PWS:Win32/Fareit.gen!C [generic] | 6 | 0 | 100.00 |
| 1st | 16 | Trojan-Dropper.Win32.Injector.cxqb | 6 | 0 | 100.00 |
| 1st | 17 | Trojan.Win32.Menti.mlgp | 6 | 0 | 100.00 |
| 1st | 18 | Trojan.Win32.Buzus (v) | 6 | 0 | 100.00 |
| 1st | 19 | Trojan.Win32.FakeAV.lcpt | 12 | 0 | 100.00 |
| 1st | 20 | Trojan.Win32.Agent.rlot | 6 | 0 | 100.00 |
| 1st | 21 | Trojan-Spy.Win32.SpyEyes.aecv | 6 | 0 | 100.00 |
| 1st | 22 | Trojan:Win32/Swrort.A | 11 | 1 | 91.67 |
| 1st | 23 | TrojanDownloader:Win32/Carberp.C | 11 | 1 | 91.67 |
| 1st | 24 | PWS:Win32/Lolyda.BF | 15 | 3 | 83.33 |
| 1st | 25 | Trojan.Win32.Yakes.qjn | 8 | 4 | 66.67 |
| 1st | 26 | Trojan.Win32.Agent.rlnz | 5 | 7 | 41.67 |
| 1st | 27 | Trojan.Win32.VBKrypt.fkvx | 6 | 12 | 33.33 |
| 1st | 28 | VirTool:Win32/VBInject.OT | 6 | 12 | 33.33 |
| 1st | 29 | HomeMalwareCleaner.FakeVimes | 36 | 264 | 12.00 |
| 1st | 30 | Trojan.Win32.Generic!BT | 56 | 598 | 8.56 |
| 1st | 31 | Trojan.FakeAlert | 6 | 108 | 5.26 |
| 1st | 32 | Trojan.Win32.Generic.pak!cobra | 0 | 18 | 0.00 |

### 5.6.1 Shortcomings

Following are the most prominent issues with the presented tools and approaches. Some of them are more "permanent", while others are solvable and intended to be addressed in the future work [314].

Looking at a signal is less intuitive visually for code analysis by humans. (However, the approach can produce an easily identifiable problematic spectrogram in some cases). For source code analysis, line numbers are a problem (easily "filtered out" as high-frequency "noise", etc.). As a result, a whole "relativistic" and machine learning methodology was developed for the line numbers in [284] to compensate for that. Generally, when CVEs are the primary class, by accurately identifying the CVE number one can get all the other pertinent details from the CVE database, including patches and line numbers making this a lesser issue. Accuracy depends on the quality of the knowledge base collected. Some of this collection and annotation is manually done to get the indexes right, and, hence, error prone. Should there be mistakes and errors, the output quality will suffer. To detect more of the useful CVE or CWE signatures in non-CVE and non-CWE cases requires large knowledge bases (human-intensive to collect), which can perhaps be shared by different vendors via a common format, such as FORENSIC LUCID. For MARFCAT, no path tracing (since no parsing is present); no slicing, semantic annotations, context, locality of reference, etc. are presently possible. Therefore, the corresponding attribute "sink", "path", and "fix" results found in the reports also have to be machine-learned. There is a significant number of algorithms and their combinations to try (currently ≈ 1800 permutations) to get the best top $N$ precise result. This is, however, also an advantage of the approach as the underlying framework can quickly allow for such testing. In most of the cases, only file-level training vs. fragment-level training is done—presently the classes are trained based on the entire files instead of the known file fragments of interest. The latter would be more fine-grained and precise than whole-file classification, but slower. However, overall the file-level processing is a man-hour limitation than a technological one.

Table 9: Top 6 distance malware classifiers and 32 results, wavelet filter preprocessing

| guess | run | algorithms | good | bad | % |
|---|---|---|---|---|---|
| 1st | 1 | `-dynaclass -binary -noprepep -sdwt -fft -cos -flucid` | 55 | 146 | 27.36 |
| 1st | 2 | `-dynaclass -binary -noprepep -sdwt -fft -diff -flucid` | 41 | 180 | 18.55 |
| 1st | 3 | `-dynaclass -binary -noprepep -sdwt -fft -mink -flucid` | 41 | 180 | 18.55 |
| 1st | 4 | `-dynaclass -binary -noprepep -sdwt -fft -cheb -flucid` | 41 | 180 | 18.55 |
| 1st | 5 | `-dynaclass -binary -noprepep -sdwt -fft -eucl -flucid` | 41 | 180 | 18.55 |
| 1st | 6 | `-dynaclass -binary -noprepep -sdwt -fft -hamming -flucid` | 30 | 191 | 13.57 |
| 2nd | 1 | `-dynaclass -binary -noprepep -sdwt -fft -cos -flucid` | 75 | 126 | 37.31 |
| 2nd | 2 | `-dynaclass -binary -noprepep -sdwt -fft -diff -flucid` | 56 | 165 | 25.34 |
| 2nd | 3 | `-dynaclass -binary -noprepep -sdwt -fft -mink -flucid` | 67 | 154 | 30.32 |
| 2nd | 4 | `-dynaclass -binary -noprepep -sdwt -fft -cheb -flucid` | 55 | 166 | 24.89 |
| 2nd | 5 | `-dynaclass -binary -noprepep -sdwt -fft -eucl -flucid` | 58 | 163 | 26.24 |
| 2nd | 6 | `-dynaclass -binary -noprepep -sdwt -fft -hamming -flucid` | 44 | 177 | 19.91 |
| guess | run | class | good | bad | % |
| 1st | 1 | VirTool.Win32.VBInject.gen.bp (v) | 6 | 0 | 100.00 |
| 1st | 2 | Trojan.Win32.Agent.roei | 6 | 0 | 100.00 |
| 1st | 3 | BehavesLike.Win32.Malware.dls (mx-v) | 6 | 0 | 100.00 |
| 1st | 4 | Worm.Win32.AutoRun.dkch | 6 | 0 | 100.00 |
| 1st | 5 | Trojan-FakeAV.Win32.Agent.det | 6 | 0 | 100.00 |
| 1st | 6 | FraudTool.Win32.FakeRean | 6 | 0 | 100.00 |
| 1st | 7 | VirTool:Win32/Obfuscator.WJ (suspicious) | 6 | 0 | 100.00 |
| 1st | 8 | Trojan.Win32.Vilsel.ayyw | 6 | 0 | 100.00 |
| 1st | 9 | Worm:Win32/Yeltminky.A!dll | 6 | 0 | 100.00 |
| 1st | 10 | Trojan.Win32.Meredrop | 6 | 0 | 100.00 |
| 1st | 11 | Virtumonde | 6 | 0 | 100.00 |
| 1st | 12 | Backdoor.Win32.Hupigon.nndu | 6 | 0 | 100.00 |
| 1st | 13 | VirTool:WinNT/Protmin.gen!C [generic] | 6 | 0 | 100.00 |
| 1st | 14 | PWS:Win32/Fareit.gen!C [generic] | 6 | 0 | 100.00 |
| 1st | 15 | Trojan-Dropper.Win32.Injector.cxqb | 6 | 0 | 100.00 |
| 1st | 16 | Trojan.Win32.Menti.mlgp | 6 | 0 | 100.00 |
| 1st | 17 | Trojan.Win32.Buzus (v) | 6 | 0 | 100.00 |
| 1st | 18 | Trojan.Win32.Agent.rlot | 6 | 0 | 100.00 |
| 1st | 19 | Trojan-Spy.Win32.SpyEyes.aecv | 6 | 0 | 100.00 |
| 1st | 20 | Trojan.Win32.FakeAV.lcpt | 11 | 1 | 91.67 |
| 1st | 21 | TrojanDownloader:Win32/Allsum | 10 | 2 | 83.33 |
| 1st | 22 | Trojan.Win32.Yakes.qjn | 10 | 2 | 83.33 |
| 1st | 23 | Trojan.Win32.Agent.rlnz | 9 | 3 | 75.00 |
| 1st | 24 | Trojan:Win32/Swrort.A | 6 | 6 | 50.00 |
| 1st | 25 | TrojanDownloader:Win32/Carberp.C | 6 | 6 | 50.00 |
| 1st | 26 | Trojan.Win32.VBKrypt.fkvx | 5 | 11 | 31.25 |
| 1st | 27 | VirTool:Win32/VBInject.OT | 5 | 11 | 31.25 |
| 1st | 28 | HomeMalwareCleaner.FakeVimes | 46 | 250 | 15.54 |
| 1st | 29 | Trojan.FakeAlert | 8 | 104 | 7.14 |
| 1st | 30 | Trojan.Win32.Generic.pak!cobra | 1 | 17 | 5.56 |
| 1st | 31 | Trojan.Win32.Generic!BT | 18 | 626 | 2.80 |
| 1st | 32 | PWS:Win32/Lolyda.BF | 0 | 18 | 0.00 |

Table 11: Top 6 distance malware classifiers and 32 results, low-pass filter preprocessing

| guess | run | algorithms | good | bad | % |
|---|---|---|---|---|---|
| 1st | 1 | `-dynaclass -binary -nopreprep -low -fft -cos -flucid` | 60 | 161 | 27.15 |
| 1st | 2 | `-dynaclass -binary -nopreprep -low -fft -cheb -flucid` | 54 | 167 | 24.43 |
| 1st | 3 | `-dynaclass -binary -nopreprep -low -fft -diff -flucid` | 54 | 167 | 24.43 |
| 1st | 4 | `-dynaclass -binary -nopreprep -low -fft -eucl -flucid` | 46 | 175 | 20.81 |
| 1st | 5 | `-dynaclass -binary -nopreprep -low -fft -hamming -flucid` | 35 | 186 | 15.84 |
| 1st | 6 | `-dynaclass -binary -nopreprep -low -fft -mink -flucid` | 33 | 188 | 14.93 |
| 2nd | 1 | `-dynaclass -binary -nopreprep -low -fft -cos -flucid` | 88 | 133 | 39.82 |
| 2nd | 2 | `-dynaclass -binary -nopreprep -low -fft -cheb -flucid` | 74 | 147 | 33.48 |
| 2nd | 3 | `-dynaclass -binary -nopreprep -low -fft -diff -flucid` | 74 | 147 | 33.48 |
| 2nd | 4 | `-dynaclass -binary -nopreprep -low -fft -eucl -flucid` | 69 | 152 | 31.22 |
| 2nd | 5 | `-dynaclass -binary -nopreprep -low -fft -hamming -flucid` | 49 | 172 | 22.17 |
| 2nd | 6 | `-dynaclass -binary -nopreprep -low -fft -mink -flucid` | 48 | 173 | 21.72 |
| guess | run | class | good | bad | % |
| 1st | 1 | Trojan:Win32/Swrort.A | 12 | 0 | 100.00 |
| 1st | 2 | VirTool.Win32.VBInject.gen.bp (v) | 6 | 0 | 100.00 |
| 1st | 3 | Trojan.Win32.Agent.roei | 6 | 0 | 100.00 |
| 1st | 4 | BehavesLike.Win32.Malware.dls (mx-v) | 6 | 0 | 100.00 |
| 1st | 5 | Worm.Win32.AutoRun.dkch | 6 | 0 | 100.00 |
| 1st | 6 | Trojan-FakeAV.Win32.Agent.det | 6 | 0 | 100.00 |
| 1st | 7 | FraudTool.Win32.FakeRean | 6 | 0 | 100.00 |
| 1st | 8 | VirTool:Win32/Obfuscator.WJ (suspicious) | 6 | 0 | 100.00 |
| 1st | 9 | Trojan.Win32.Vilsel.ayyw | 6 | 0 | 100.00 |
| 1st | 10 | Worm:Win32/Yeltminky.A!dll | 6 | 0 | 100.00 |
| 1st | 11 | Trojan.Win32.Meredrop | 6 | 0 | 100.00 |
| 1st | 12 | Virtumonde | 6 | 0 | 100.00 |
| 1st | 13 | Backdoor.Win32.Hupigon.nndu | 6 | 0 | 100.00 |
| 1st | 14 | VirTool:WinNT/Protmin.gen!C [generic] | 6 | 0 | 100.00 |
| 1st | 15 | PWS:Win32/Fareit.gen!C [generic] | 6 | 0 | 100.00 |
| 1st | 16 | Trojan-Dropper.Win32.Injector.cxqb | 6 | 0 | 100.00 |
| 1st | 17 | Trojan.Win32.Menti.mlgp | 6 | 0 | 100.00 |
| 1st | 18 | Trojan.Win32.Buzus (v) | 6 | 0 | 100.00 |
| 1st | 19 | Trojan.Win32.FakeAV.lcpt | 12 | 0 | 100.00 |
| 1st | 20 | Trojan.Win32.Agent.rlot | 6 | 0 | 100.00 |
| 1st | 21 | Trojan-Spy.Win32.SpyEyes.aecv | 6 | 0 | 100.00 |
| 1st | 22 | TrojanDownloader:Win32/Allsum | 11 | 1 | 91.67 |
| 1st | 23 | TrojanDownloader:Win32/Carberp.C | 10 | 2 | 83.33 |
| 1st | 24 | PWS:Win32/Lolyda.BF | 15 | 3 | 83.33 |
| 1st | 25 | Trojan.Win32.Yakes.qjn | 8 | 4 | 66.67 |
| 1st | 26 | Trojan.Win32.Agent.rlnz | 6 | 6 | 50.00 |
| 1st | 27 | Trojan.Win32.VBKrypt.fkvx | 6 | 12 | 33.33 |
| 1st | 28 | VirTool:Win32/VBInject.OT | 6 | 12 | 33.33 |
| 1st | 29 | HomeMalwareCleaner.FakeVimes | 37 | 263 | 12.33 |
| 1st | 30 | Trojan.Win32.Generic.pak!cobra | 2 | 16 | 11.11 |
| 1st | 31 | Trojan.FakeAlert | 8 | 106 | 7.02 |
| 1st | 32 | Trojan.Win32.Generic!BT | 35 | 619 | 5.35 |

These shortcomings may affect the credibility/confidence score in the data mining analysis encoded into the observations in FORENSIC LUCID. The lower the score, the less likely the evidence from this analysis is to be used to support or refute claims in the case at hand. Thus, addressing these shortcomings is an important aspect to improve.

### 5.6.2 Advantages

There are some key advantages of the approaches and tools presented, which follow. The approach is relatively fast (e.g., MARFCAT's processing of Wireshark's ≈ 2400 files to train and test completes in about three minutes) on a now-commodity desktop or a laptop. As signal processing tools, they are language- and protocol-independent (no syntactic parsing)— given enough examples one can apply them to any data type (language, pcap, etc.), i.e., the methodology is the same no matter C, C++, JAVA or any other source or binary languages (PHP, C#, VB, PERL, bytecode, assembly, etc.) are used, or using other data types such as images, pcaps, etc. As a machine-learning based approach, the tools can automatically learn a large knowledge base to test on known and unknown cases as well as can learn, for example, from previous SATE 2008–2013 reports that are publicly available [348]. We can also use the tools to quickly pre-scan project data for further analysis by humans or other tools that do in-depth parsing and semantic analyses as a means to prioritize large data sets for such tools. We often get high (or good enough) precision and recall in CVE and CWE detection, even at the file level and good enough precision at the whole-pcap processing. There are many algorithms and their combinations to select the best for a particular classification task after initial test runs. The approach can cope with altered code or code clones used in other projects (e.g., a lot of problems in Chrome were found it WebKit, used by several browsers).

The fast and relatively high precision results can guide any forensic investigation when big data are involved faster to enable investigators to focus. The higher precision results can prioritize the FORENSIC LUCID observations related to them in the process.

### 5.6.3 Practical Implications

We outline some practical implications of the presented approach [314]. MARFCAT can be used on any target language without modifications to the methodology or knowing the syntax of the language. Thus, it scales to any popular and new language analysis with a very small amount of effort [314]. MARFPCAT can likewise scale for various data/code clones, malware, or application identification. MARFCAT was easily adapted to the compiled binaries and bytecode to be able detect vulnerable deployments and installations—akin to virus scanning of binaries, but instead scanning for infected binaries, one would scan for security-weak binaries on site deployments to alert system administrators to upgrade their packages [314]. It can likewise learn from binary signatures from other tools like Snort [438]. As a result both tools are extendable to the embedded code and mission-critical code found in aircraft, spacecraft, and various autonomous systems [314] for incident prevention or investigations. Spectral signatures and machine learning techniques are very beneficial for file type analysis as well, especially when there is a need to bulk-preprocess a large collection of files for preliminary classification of "files of interest" on suspect's hard drive [290] (e.g., with the evidence in the form of audio and text files recovered from a suspect's computer).

All the tools are designed in JAVA with the easier plug-in-like integration of the tools into JAVA-based plug-in frameworks, such as JPF, Eclipse, and others [290] in mind. MARF, MARFCAT, and MARFPCAT were already updated to export their results and data structures in the FORENSIC LUCID format in order to allow their inclusion into the existing investigative cases.

# Chapter 6

# The General Intensional Programming System

This background chapter covers the General Intensional Programming System (GIPSY) [161, 241, 264, 302, 361, 362, 363, 366, 369, 370, 499, 529], which is is an open-source platform implemented primarily in JAVA to investigate properties of the LUCID [25, 26, 27] (see Chapter 4) family of intensional programming languages and beyond. GIPSY is being developed and maintained by the *GIPSY Research and Development Group* at Concordia University, Montreal, Canada. As a multi-tier distributed system, it evaluates LUCID programs following a demand-driven distributed generator-worker architecture, and is designed as a modular collection of frameworks where components related to the development (RIPE[1]), compilation (GIPC[2]), and execution/evaluation (GEE[3]) of LUCID [27] programs are decoupled allowing easy extension, addition, and replacement of the components and subcomponents. The high-level general architecture of GIPSY is presented in Figure 32 as well as the high-level structure of the GIPC framework is in Figure 33 [271, 304, 305, 307, 322].

This background chapter is compiled from a list of the related cited works by the *GIPSY Research and Development Group* and the author Mokhov. In this chapter we present the general GIPSY overview (Section 6.1, page 129) including the necessary historical notes of its conception, evolution, and the subsequent extensions for the use in this thesis as well as the unaware reader. We follow through with its key architectural design points (Section 6.2, page 136) to provide an in-depth background overview to subsequent GIPSY contributions in Part II.

---

[1] Run-time Integrated Programming Environment, implemented in `gipsy.RIPE`
[2] General Intensional Programming Compiler, implemented in `gipsy.GIPC`
[3] General Eduction Engine, implemented in `gipsy.GEE`

## 6.1 Overview

GIPSY [241, 264, 282, 302, 363, 366, 369, 370, 499, 529] is a continued effort for the design and development of a flexible and adaptable multi-lingual programming language development framework aimed at the investigation on the LUCID family of intensional programming (Section 3.2) languages [24, 25, 26, 354, 361, 364, 379, 396, 509]. Using this platform, programs written in various flavors of LUCID can be compiled and executed in a variety of ways [161, 301, 302, 315]. The framework approach adopted is aimed at providing the possibility of easily developing compiler components for other languages of intensional nature, and to execute them on a generally language-independent run-time system. With LUCID being a functional "data-flow" language (cf. Chapter 4), its programs can be executed in a distributed processing environment [315]. As a result, by being multi-lingual, GIPSY's design incorporates the mentioned flexible compilers framework and a run-time system to allow processing of programs written in multiple dialects of LUCID as well as mixing them with common imperative languages, such as JAVA, potentially all in the same source code "program" or a source code file comprising a semantic unit of interrelated program fragments written in multiple languages and interacting with each other. This is what makes GIPSY different from being "just a Lucid dialect" into a complete programming system for multiple languages though glued together by the type system (described in depth in Appendix B) and the "meta" preprocessor language of various declarations to aid compilation [264, 282, 301].

### 6.1.1 Related Work

The ideas relevant to GIPSY are covered in a number of works from various research teams about intensional and multidimensional programming, distributed, parallel as well as with hybrid intensional/imperative paradigms. The most prominent is perhaps GLU [5, 188, 189, 359] that prompted GIPSY's existence. The GLU (Granular Lucid) system, developed at the Stanford Research Institute in the 1990s, was arguably the first large hybrid intensional-procedural system to allow INDEXICAL LUCID programs to use C or FORTRAN data structures and procedures [188, 189]. GLU then relied on a distributed Generator-Worker execution architecture to evaluate such programs. Due to the lack of flexibility in its architectural design, GLU was not able to adapt to further evolutions of LUCID, or to interface with object-oriented languages [361, 526] (at least not until later appearance of GLU# in 2004 to interface with C++ [359]). In that context, GIPSY's design started off with a similar model as GLU's, but with flexibility and adaptability in mind [363, 366, 370]. Using a framework approach, the GIPSY has been used to develop compilers for different variants of LUCID, which are allowed to coexist in the same program [264, 399, 527]. Moreover, its design additionally permits Lucid programs to use procedures or methods defined in virtually any procedural language [526] as long as a corresponding compiler plug-in is added to the framework. A similar model is also successfully adopted by the prominent *GNU Compiler Collection* (GCC) [485].

The distributed and parallel indexical and intensional program evaluation was studied in several works, such as Du's work in 1999 [92, 94] on indexical parallel programming and Swoboda's and Plaice's work on distributed context computing [454] followed by formalization of distributed intensional programming paradigm by Swoboda in his PhD thesis [452] in 2004. A slightly more recent work on scheduling and evaluation of multidimensional programs in parallel/distributed environments was performed by Ben Hamed [159] in 2008. The data-flow aspect of intensional programming was also explored in graph-based distributed systems [56] back in 1995 by Cao *et al*. Subsequently, a number of formalisms appeared including Bensnard *et al*. [42] presenting an example of a multi-formalism application design and distribution in a data-flow context, Swoboda's work [452, 454], and Fourtounis [118] in 2011 about formal specification and analysis of a parallel virtual machine for lazy functional languages (that refers to some of the GIPSY and the mentioned related work). Fisher and Kakoudakis did flexible agent grouping in executable temporal logic [111] and Gagné and Plaice looked at the demand-driven real-time computing [126]. Ranganathan and Campbell in 2003 also proposed a middleware for context-aware agents in ubiquitous computing environments [397]. Peralta *et al.* proposed an approach similar to GIPSY for automatic synthesis and deployment of intensional Kahn process networks [375].

The hybridification (mixing intensional/logic programming with imperative languages) aspects were discussed in several works: Liu and Stables [238] proposed inclusion of logic expressions into the procedural programming languages in 1995; in 1999 Rondogiannis subsequently proposed adding multidimensionality to imperative programming languages [406]; while Swoboda and Wadge proposed tools to intensionalize software systems in general [455]. As time passed, Ditu and Plaice proposed the general TRANSLUCID [90] language with elaborate typing system and intent to support multicore processors, Cartesian programming model [377] and some of its PoC eager [378, 379] and multithreaded [396] implementation by Rahilly in 2007–2008.

In relationship to the aspect-oriented programming (AOP), Du pondered of the links between AOP and the intensional programming models [95] by drawing parallels between the notion of context in two paradigms.

All of this work and related contributions are accumulated and summarized in [364], which is being regularly updated as a reference resource. The GIPSY's software architecture is able to constantly change and accept new ideas from the GIPSY members as well as the cited works of others above for comparative studies and evaluation.

### 6.1.2 Historical Notes

Historically, the concept of GIPSY was conceived as a very modular collection of frameworks and artifacts geared towards sustainable support for the intensional programming languages and embracing continuous iterative revision and development overcoming issues of the earlier GLU system [5, 188, 189] that did not survive for very long due to its inflexibility to extend to the newer dialects, and its unmaintainability defects [161, 302, 361, 362, 363, 370].

The initial GIPSY and GIPL ideas were featured in Paquet's PhD thesis [361] in 1999. Then those ideas were architecturally expanded by Paquet and Kropf in 2000 [363]. Subsequently, the project moved to Concordia where it resides at the present day. In the meantime, Grogono defined GIPC increments [143] in 2002. Ren, also in 2002 [399], produced the first INDEXICAL LUCID and GIPL compilers as a part of the first PoC GIPC written in JAVA and JavaCC [506]. Wu in the same year provided the first version of the semantic analyzer and translation rules to the INDEXICAL LUCID and GIPL compilers [527]. Wu, Paquet, and Grogono expanded further the detailed design of the GIPC framework in 2003 [530]. Alagar, Paquet, and Wan described the notion of Intensional Programming for agent communication in 2004 [7]. Lu, Grogono, and Paquet subsequently proposed and defined the initial GEE LUCID interpreter in C++ subsequently rewritten in JAVA by Lu in multithreaded and distributed Java RMI (Remote Method Invocation) prototypes [241, 242]. In 2004, Grogono proposed ONYX for lazy multidimensional arrays [144]. Then, in the same year, Paquet, Wu, and Grogono expanded on the GIPC architecture further [369]. After that, also in 2004, Tao provided the first realization of the data warehousing and garbage collection in the GEE [459]. Following that, Ding [89] provided a PoC implementation of the automated translation between graphical and textual representations of intensional programs for INDEXICAL LUCID and GIPL within GIPC. In the meantime, in 2005–2006 Wan developed a full context theory for intensional programming and the corresponding artifact of the LUCX language [513, 514, 515]. Vassev looked into the distributed aspects in his follow-up work on the GEE by designing the Demand Migration Framework (DMF) and its concrete instance of Jini DMS (Demand Migration System) to incorporate the demand store and the distribution architecture middleware to work together [498, 499, 501]. Wu and Paquet reviewed the GIPSY architecture [370, 529] and began pondering about including intensional expressions into JAVA. In 2003–2005, the author Mokhov (along with Paquet and Grogono), re-integrated and retrofitted all the components with the original design by Paquet and defined the `Preprocessor` and introduced the first instance of hybrid intensional-imperative compilation frameworks of GIPC with JLUCID and OBJECTIVE LUCID (with sequential threads and communication procedures as well as operational semantics) prototype compilers, initial type system, and other aspects integrating all the components under a common CVS [146] repository [145, 261, 262, 264] that led to a more general framework for greater interoperability between intensional and imperative programming languages—GICF and a web-based editor RIPE servlet front-end. In 2007–2008, Pourteymour *et al.* provided another PoC instance of the DMS, implemented using JMS [383, 384, 385]. Around the same time frame, Tong *et al.* implemented a LUCX compiler [365, 473, 474]. Also in 2007, the author

Mokhov proposed the use of GIPSY as a platform for the *Intensional Cyberforensics* and FORENSIC LUCID [266] eventually culminating in this thesis (see Section 6.1.4.3 and further). The language MARFL is proposed by the author to manage the MARF's (*Modular Audio Recognition Framework*) configuration [272] with the hierarchical contexts in it further in 2008. He also performed the first preliminary security evaluation of the earlier GICF's design and the distributed aspects of GIPSY [271]. Vassev in the same year 2008, moved on to propose self-management properties of GIPSY with the design of the Autonomic GIPSY (AGIPSY) [500]. Subsequently, in around 2007–2010 Paquet proposed a new multi-tier architecture for GEE [362] where Han *et al.* provided its initial implementation [160, 161], and Ji and the author Mokhov completely unified the Jini and JMS DMS'es [193] and Ji did a scalability study of the two [191]. Concurrently, in 2008–2009, Wu *et al.* completed design and implementation of a compiler for JOOIP to embed LUCID expressions into JAVA [526, 528] such that JAVA classes can instantiate intensional variables, and have the LUCID expressions access the JAVA class properties (methods, variables (local, instance, class)). Meanwhile, the author Mokhov investigated the GIPSY's use for the HOIL support [302] and provided a complete *GIPSY Type System* (Appendix B) specification [301, 315] alongside with Tong and Paquet in the context of the multi-tier work and LUCX compiler implementation. The author Mokhov *et al.* then moved on to the proposal of *self-forensics* aspects within GIPSY and other systems [321], which is also an ongoing work (see Appendix D). In 2011–2013, the author Mokhov developed problem-specific generator and worker tiers for MARFCAT (Section 5.4) and MARFPCAT [287, 289, 314] and for genome sequencing as a case study with Rabah [393]. At the same time, in 2011–2012, Rabah designed the first prototype of a graph-based RIPE component to manage distributed GIPSY networks [393]. The author Mokhov (along with Rabah's help) built the GIPSY cluster lab environment (detailed in Section 8.6, page 238) for higher-performance and scalability evaluations. Subsequently, the author Mokhov expanded the compiler and run-time support onto the FORENSIC LUCID language as a part of this thesis.

As previously mentioned, the GIPSY's design is centered around the compiler framework (GIPC) following good principles of compiler design [240], the eduction execution engine (GEE), i.e., the run-time execution environment (akin to a virtual machine for execution of the intensional logic expressions), and the programming environment (RIPE). The former of the three is responsible to support multiple compilers in a similar compiler framework that all produce a consistent, well-agreed on binary format, essentially a compiled GIPSY program, as a binary output. The second performs lazy demand-driven (potentially parallel/distributed) evaluation of the compiled LUCID programs, or as we call them, GIPSY programs [302]. This idea of the GIPSY's framework approach is to provide an infrastructure to develop compiler and run-time components for other languages of intensional nature easier as well as to execute them on a relatively language-independent run-time system [315]. As discussed earlier, LUCID programs in general can be naturally executed in a distributed processing environment because its constructs and expressions do not impose sequentiality. However, the standard LUCID algebra (i.e., types and operators) is extremely fine-grained and can hardly benefit from distributed evaluation of the operands. Adding granularity to the data elements manipulated by LUCID comes through the addition of coarser-grained data types and their corresponding operators [315] (Appendix B.4). With LUCID semantics being defined as typeless, a solution to the granularity problem consists in adding a hybrid counterpart to LUCID to allow an external language to define an algebra of coarser-grained types and operators [301, 315].

### 6.1.3 From Sequentiality to Concurrency with DMS and Multi-Tier Architecture

The eventual availability of the Demand Migration Framework and System (DMF and DMS) [384, 385, 498, 499] turned the GIPSY's PoC RMI implementation [241] into a true distributed system [271]. These developments prompted a number of related research subdirections some of which are still active today. Specifically, in [264] when a notion of prototype JLUCID and OBJECTIVE LUCID languages was conceived, it led to a more general framework for greater interoperability between intensional and imperative programming languages— GICF [271]. However, with that greater flexibility that the languages, GICF, and DMS brought, there came to be the issues of security of the embedded code, and demand monitoring, etc. [271].

The subsequent evolution presented in [161, 191, 193] furthered the original architecture for the run-time system of the GIPSY (as hinted in [361], and elaborated in [302, 362, 363, 370]). The architecture proposed in these works was itself developed following the generator-worker architecture adopted successfully by GLU [188, 189], as mentioned earlier, where its run-time system was not as scalable and flexible [161] as the solutions presented in the cited GIPSY works. In addition, the communication procedures of the distributed run-time system implemented in GLU were implemented using RPC only [161]. The GIPSY solution proposed a much more flexible approach by integrating demand migration and storage by using the DMF, which can be concretely instantiated using various middleware technologies, such as Jini [498, 499] and JMS [384, 385] and others as they become available and integrated allowing for even heterogeneous mixed use of such technologies simultaneously fulfilling different communication requirements and availability [161, 191, 193]. Further design and development has GIPSY planned to try other/new architectures (e.g., Hadoop, PlanetLab [466], and others) and prompted an establishment of a dedicated cluster setup (see Section 8.6, page 238). The distributed and scalable evaluation becomes very important in large scale reasoning and evaluation of large amounts of digital evidence in FORENSIC LUCID case evaluations.

### 6.1.4 Context-Oriented Reasoning with LUCID in GIPSY

The reasoning aspect of GIPSY is a particularity of a LUCID dialect (Chapter 4) rather than its entire architecture. The architecture is general enough to go beyond pure an evaluation of intensional logic expressions. If those expressions form a language dialect that helps us with reasoning (such as FORENSIC LUCID to reason about cybercrime incidents and claims).

#### 6.1.4.1 Reasoning in Hybrid OO Environment

Object-orientation came to GIPSY with the concept of OBJECTIVE LUCID [264, 282] that allowed for rudimentary support of LUCID programs to manipulate JAVA objects as first class values. Its generalization resulted in JOOIP [526, 528] offering LUCID fragments within JAVA code and allowing the Lucid code to reference to JAVA methods and variables, along with the corresponding type system extensions and providing context-aware JAVA objects [302] (see

Section 4.3.2, page 92).

#### 6.1.4.2 Reasoning in Autonomic Environment

This aspect is important for the project on self-forensics described in Appendix D. Vassev and Paquet designed an Autonomic GIPSY [500] (AGIPSY) version of the platform with the corresponding ASSL toolset [486, 493, 502] as a research case study where GIPSY is turned into an autonomic system via an ASSL specification to allow it running unattended with self-healing, self-protecting, self-configuring, and self-optimizing autonomic properties. Unattended reasoning environment is essential for long-running problem solving programs with minimal intervention and maintenance [302].

#### 6.1.4.3 FORENSIC LUCID Compilation and Evaluation Platform

GIPSY is the proposed testing and investigation platform for the compilation and distributed cyberforensics evaluation of the FORENSIC LUCID programs [300, 304] (introduced later in Chapter 7) [271, 304, 305, 307, 322]. GEE is the component where the distributed demand-driven evaluation takes place, subtasked to different evaluation engines implemeting the GEE framework (see Figure 58 and Chapter 8). We rely on the GIPSY's compilers for the intensional languages like GIPL [361], LUCX [513], OBJECTIVE LUCID [264], and JOOIP [526]. We reach out to the syntax and operational semantics of those languages implemented in GIPSY and draw from them some ideas for the simple context specification of dimensions and tags, the navigational operators @ and #, and the "dot-notation" for object properties and apply it to context spaces. The dialects referred to cover by themselves a large scope of issues FORENSIC LUCID capitalizes on. This is what this thesis is about, in particular Chapter 7, Chapter 8, and Chapter 9 which we defer to the reader about.

## 6.2 GIPSY's Architecture

Intensional programming (see Section 3.2, [131, 350]), in the context of the LUCID programming language, implies a declarative programming paradigm. The declarations are evaluated in an inherent multi-dimensional context space [161, 473]. GIPSY evolved from a modular collection of frameworks for local execution into a multi-tier architecture [161, 362]. Back in the early days, with the bright but short-lived story of GLU in mind, efforts were made to design a system with similar capacities, but significantly more flexible in order to cope with the fast evolution and diversity of the LUCID family of languages, thus necessitating a flexible compiler architecture, and a language-independent run-time system for the execution of LUCID programs. As a result, the GIPSY project's architecture [241, 264, 282, 302, 370] aims at providing such a flexible platform for the investigation on intensional and hybrid intensional-imperative programming [161]. The architecture of the General Intensional Programming Compiler (GIPC) is framework-based, allowing the modular development of compiler components (e.g., parser, semantic analyzer, and translator). It is based on the notion of the Generic Intensional Programming Language (GIPL) [361, 365], which is the core run-time language into which all other flavors of the LUCID (a family of intensional programming languages) language can be translated to [302]. The notion of a generic language also solved the problem of language-independence of the run-time system by allowing a common representation for all compiled programs, the Generic Eduction Engine Resources (GEER), which is a dictionary of run-time resources compiled from a GIPL program, that had been previously generated from the original program using semantic translation rules defining how the original LUCID program can be translated into the GIPL [161, 302]. The design of the GIPSY compiler framework went through several iterations [264, 282, 369, 399, 527]. The architecture necessitates the presence of the intensional-imperative type system and support links to imperative languages [301, 302, 315]. A generic distributed run-time system has been proposed in [161, 362].

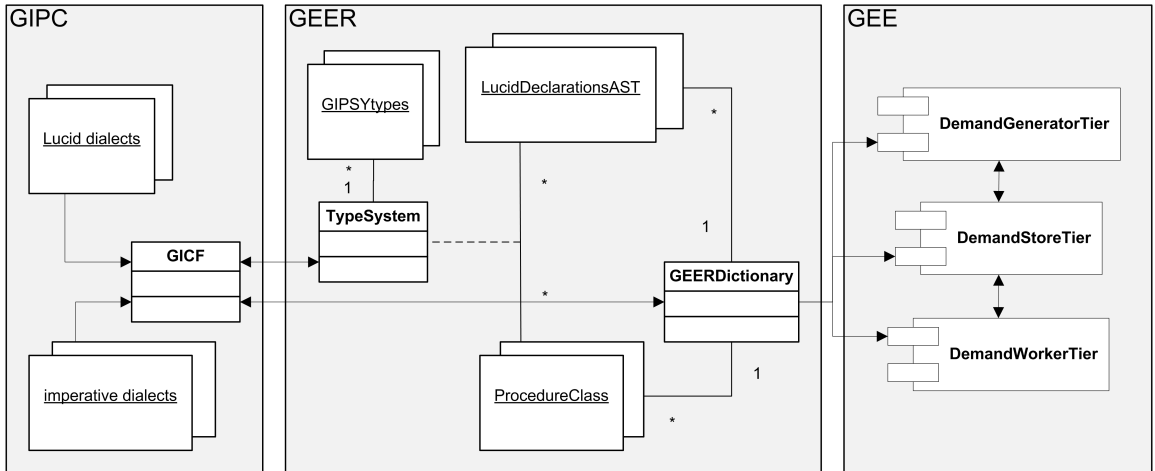

Figure 32: High-level structure of GIPSY's GEER flow overview [315]

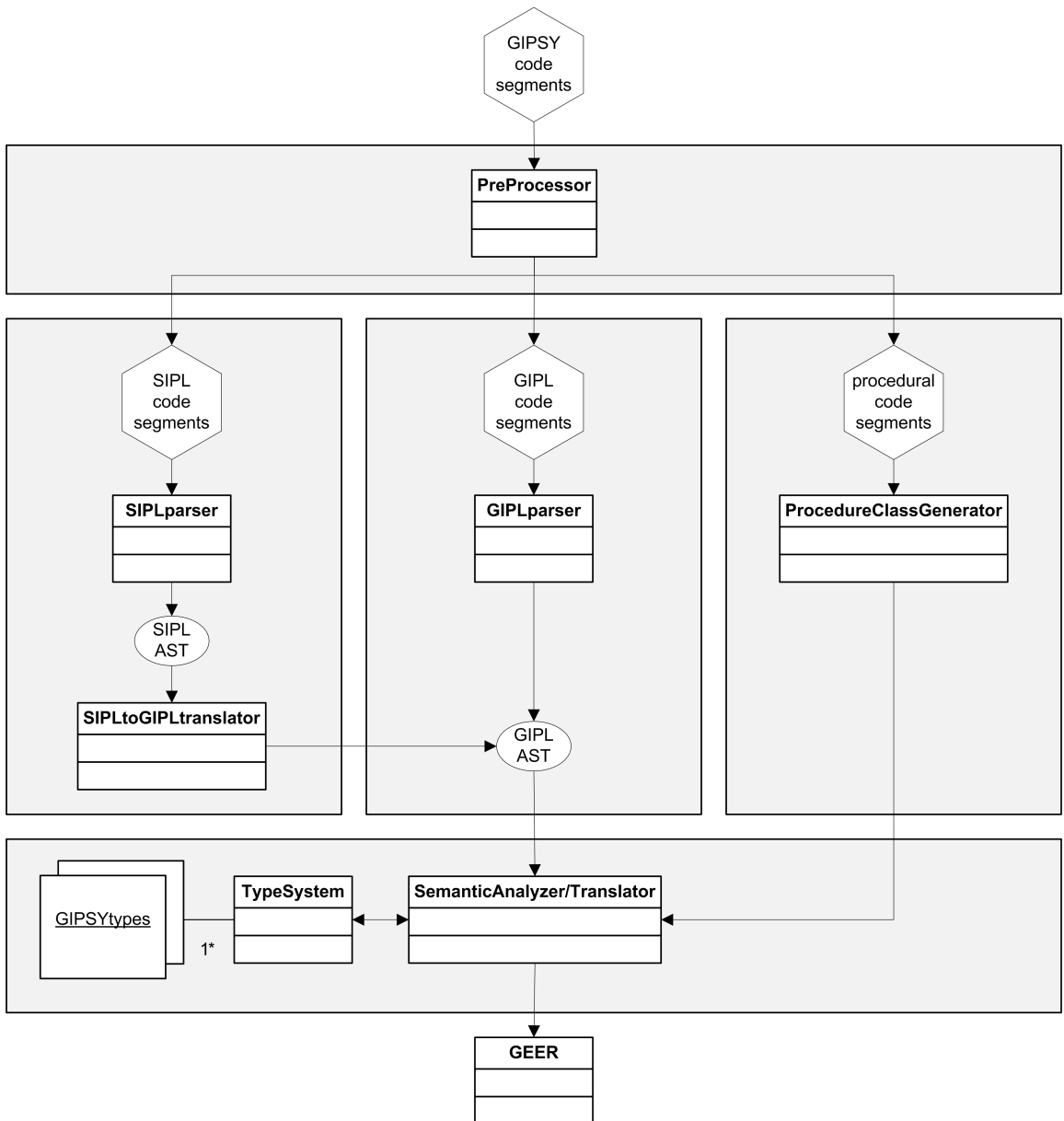

Figure 33: High-level structure of the GIPC framework [315]

GIPSY has a collection of compilers under the (GIPC, see Figure 33) framework and the corresponding run-time environment under the eduction execution engine (GEE) among other things that communicate through the GEE Resources (GEER) (see the high-level architecture in Figure 32). These two modules are the major primary components for compilation and execution of intensional programs, which require amendments for the changes proposed in this work of intensional forensics [271, 305, 322].

### 6.2.1 General Intensional Program Compiler (GIPC)

The more detailed architecture of GIPC is conceptually represented at the higher level in Figure 33. It hosts the type abstractions and implementations that are located in the `gipsy.lang` package and serve as a glue between the compiler (the GIPC—a General Intensional Program Compiler) and the run-time system (known as the GEE—a General Eduction Engine) to do the static and dynamic semantic analyses and evaluation respectively. Since GIPSY is a modular system, the majority of its components can be replaced as long as they comply with some general architectural interface/API. One of such API interfaces is the `GIPSYProgram` (conceptually represented as a GEER—the GEE Resource—a dictionary of run-time resources) that contains among other things, the type annotations that can be statically inferred during compilation. At run-time, the engine does its own type checking and evaluation when traversing the abstract syntax tree (AST) stored in the GEER and evaluating expressions represented in the tree. Since both the GIPC and the GEE use the same type system to do their analysis, they consistently apply the semantics and rules of the type system with the only difference that the GEE, in addition to the type checks, does the actual evaluation [302, 315].

#### 6.2.1.1 GIPC Preprocessor

The `Preprocessor` [264, 282] is a component that is invoked first by the GIPC (see Figure 33) on incoming GIPSY program's source code stream. The `Preprocessor`'s role is to do preliminary program analysis, processing, and splitting the source GIPSY program into "chunks", each potentially written in a different language and identified by a *language tag*. In a very general view, a GIPSY program is a hybrid program written in different language variants in one or more source file. Consequently, there has to be an interface to glue all these code segments together to ensure proper evaluation. Thus, the `Preprocessor` after some initial parsing (using its own preprocessor syntax) and producing the initial parse tree, constructs a preliminary dictionary of symbols used throughout all parts of the program. This is the basis for type matching and semantic analysis applied later on. This is also where the first step of type assignment occurs, especially on the boundary between typed and typeless parts of the program, e.g., JAVA (typed) and a specific LUCID dialect (typeless). The `Preprocessor` then splits the code segments of the GIPSY program into chunks preparing them to be fed to the respective concrete compilers for those chunks. The chunks are represented through the `CodeSegment` class, instances of which the `GIPC` collects [302, 315].

**GIPSY Program Segments.** There are four baseline types of segments defined in a GIPSY program [315]. These are:

- `#funcdecl` (in a way similar to C's `extern`) declares function prototypes written as imperative language functions defined later or externally from this program to be used by the intensional language part. The syntactical form of these prototypes is particular to GIPSY programs and need not resemble the actual function prototype declaration they describe in their particular programming language. They serve as a basis for static and dynamic type assignment and checking within the GIPSY type system with regards to procedural functions called by other parts of the GIPSY program, e.g., the LUCID code segments [315].

- `#typedecl` lists all user-defined data types that can potentially be used by the intensional part, e.g., classes. These are the types that do not explicitly appear in the matching table (in Table 17, Appendix B) describing the basic data types allowed in GIPSY programs [315].

- `#<IMPERATIVELANG>` declares that this is a code segment written in whatever IMPERATIVELANG may be, e.g., `#JAVA` for JAVA, `#CPP` for C++, `#FORTRAN` for FORTRAN, `#PERL` for PERL, and `#PYTHON` for PYTHON, etc. [315].

- `#<INTENSIONALLANG>` declares that what follows is a code segment written in whatever INTENSIONALLANG may be, for example `#GIPL`, `#LUCX`, `#JOOIP`, `#INDEXICALLUCID`, `#JLUCID`, `#OBJECTIVELUCID`, `#TENSORLUCID`, `#TRANSLUCID`, `#FORENSICLUCID` [300], and `#ONYX` [144], etc., as specified by the available GIPSY implementations and stubs. An example of a hybrid program is presented in Listing 6.1. The preamble of the program with the type and function declaration segments are the main source of type information that is used at compile time to annotate the nodes in the tree to help both static and semantic analyses [315].

```
#typedecl
myclass;

#funcdecl
myclass foo(int,double);
float bar(int,int):"ftp://localhost/cool.class":baz;
int f1();

#JAVA
myclass foo(int a, double b) {
   return new myclass(new Integer((int)(b + a)));
}
class myclass {
   public myclass(Integer a) {
      System.out.println(a);
   }
}

#CPP
#include <iostream>
int f1(void) {
   cout << "hello";
   return 0;
}

#OBJECTIVELUCID
A + bar(B, C)
where
   A = foo(B, C).intValue();
   B = f1();
   C = 2.0;
end;
```

Listing 6.1: Example of a hybrid GIPSY program

### 6.2.1.2 GICF Overview

The *General Imperative Compiler Framework* (GICF) [261] is the particular GIPSY's compiler framework that allows for a generalized way of inclusion of any imperative languages into intensional variants within the GIPSY environment and allowing the syntactical co-existence of the intensional and imperative languages in one source file by providing a `Preprocessor` that splits the intensional and imperative code chunks that are fed to their respective compilers, and then the results are gathered and linked together to form a compiled hybrid program as an instance of GEER [526].

GLU [188, 189], JLUCID, and OBJECTIVE LUCID [264] prompted the development of GICF. The framework targets the integration of different imperative languages into GIPSY programs for I/O, portability, extensibility, and flexibility reasons [271]. GLU promoted C and FORTRAN functions within; JLUCID/OBJECTIVE LUCID/JOOIP promoted embedded JAVA. Since GIPSY targets to unite most intensional paradigms in one research system, it makes an effort to be as general as possible and as compatible as possible and pragmatic at the same time [271].

The GICF is there if we need to be able to run, for example, GLU programs with minimum modifications to the code base. GIPSY's GIPC would be extended with a compiler module in this case to support C and FORTRAN functions as it does for JAVA. GICF is made extensible such that later on the language support for C++, PERL, PYTHON, shell scripts, and so on can be relatively easily added. With GICF it is also possible to have a multi-segment multi-language GIPSY program with embedded code [271].

### 6.2.2 General Eduction Engine (GEE)

The primary purpose of the GEE is to evaluate compiled LUCID programs following their operational semantics (see Section 4.1.1.2) either locally or distributively using the lazy demand-driven model (i.e., eduction). The research in this area covers various extensions and applications as well as comparative studies of various middleware technologies.

To address run-time scalability concerns, GEE is the component where the distributed demand-driven evaluation takes place by relying on the Demand Migration System (DMS) [383, 501] and on the multi-tier architecture overall [161, 271, 322, 362]. The distributed system [73, 133] design architecture adopted for the run-time system is a distributed multi-tier architecture, where each tier can have any number of instances [302]. The architecture bears resemblance with a peer-to-peer architecture [161, 362], where [302]:

- Demands are propagated without knowing where they will be processed or stored.

- Any tier or node can fail without the system to be fatally affected.

- Nodes and tiers can seamlessly be added or removed on the fly as computation is happening.

- Nodes and tiers can be affected at run-time to the execution of any GIPSY program, i.e., a specific node or tier could be computing demands for different programs.

The founding works cited earlier [161, 264, 362, 366, 383, 384, 385, 498, 499, 501] cover the initial design, and proof-of-concept implementations of the DMF—Jini- and JMS—based as well as the surrounding integration effort of them in to the GEE [192].

#### 6.2.2.1 Multi-Tier Demand Migration System

The *Demand Migration System* (DMS) is an implementation of the *Demand Migration Framework* (DMF) introduced first by Vassev and then extended by Pourteymour in [383, 384, 385, 499, 501]. The initial version of the DMS relied solely on Jini [194] (now known as *Apache River* [20]) for transport and storage of the demands and the results with a JavaSpaces [245] repository acting as a data warehouse cache for the most frequently demanded computations and their results (i.e., the demand store). The DMF is an architecture that is centered around the demand store with transport agents (TAs) that implement a particular protocol (as a proof-of-concept Jini and JMS [449] TAs are used [384, 385]) to deliver demands between the demand store, workers (that do the actual computation primarily for procedural demands), and generators (that request the computation to be done and collect results) [271, 322]. Thus, GIPSY has some implementation of RMI [523], Jini [194], and JMS [449] and the one that relies on the DMS for its transport needs [271, 322].

##### 6.2.2.1.1 Multi-Tier Unification of Jini and JMS DMS.

Initially, when Vassev and Pourteymour provided the Jini and JMS implementations respectively, they were rather disparate and did not integrate fully well with the updated multi-tier architecture put forward by Paquet at the design and implementation level. However, it was important to bring them under the same frameworked rooftop for comparative studies as well as concurrent use in order to be able to gain insights and recommend a particular architecture for a given GIPSY network setup. Pourteymour [383] initially researched the JMS and its properties in comparison with Jini in the published literature and tutorials [11, 20, 65, 101, 114, 164, 197]. Ji and the author Mokhov subsequently did a design update and refactoring to enable smooth scripted selection of either Jini- or JMS-based tiers (or both middleware technologies used simultaneously) to participate in a single GIPSY computation network [192, 193]. Ji afterward did an in-depth scalability study of the two [191] in the integrated environment.

**6.2.2.1.2 Generic Eduction Engine Resources.** One of the central concepts of the GIPSY's' solution is *language independence* of the run-time system. In order to achieve that, the design relies on an intermediate representation that is generated by the compiler: the Generic Eduction Engine Resources (GEER). The GIPC compiles a program into an instance of the GEER(s), including a dictionary of identifiers extracted from the program [264, 282, 369]. Since the compiler framework provides with the potential to allow additions of any flavor of the LUCID language to be added through automated compiler generation taking semantic translation rules in input [527], the compiler authors need to provide a parser and a set of rules to compile and link a GEER, often by translating a specific LUCID dialect to GIPL first [302].

As the name suggests, the GEER structure is generic, in the sense that the data structure and semantics of the GEER are independent of the source language. This is necessitated by the fact that the engine was designed to be "source-language independent", an important feature made possible by the presence of the Generic Intensional Programming Language (GIPL) as a generic language in the Lucid family of languages [161, 302, 362]. Thus, the compiler first translates the source program (written in any flavor of LUCID) into the "generic LUCID" [364], then generates the GEER run-time resources for this program, which are then made available at run-time to the various tiers upon demand. The GEER contains, for all Lucid identifiers in a given program, typing information, rank (i.e., dimensionality information), as well as an abstract syntax tree (AST) representation of the declarative definition of each identifier [161, 302, 362]. It is this latter tree that is traversed later on by the demand generator tier in order to proceed with demand generation. In the case of hybrid LUCID programs, the GEER also contains a dictionary of procedures called by the LUCID program, known as *Sequential Procedure Classes*, as they in fact are *wrapper classes* wrapping procedures inside a JAVA class in cases where the functions being called are not written in JAVA [264, 282, 302] using JNI [442].

**6.2.2.1.3 GIPSY Tier.** The architecture adopted for the most recent evolution of the GIPSY is a multi-tier architecture where the execution of GIPSY programs is divided in three different tasks assigned to separate tiers [161, 362]. Each GIPSY tier is a separate process that communicates with other tiers using *demands*, i.e., the *GIPSY Multi-Tier Architecture* operational mode is fully *demand-driven*. The demands are generated by the tiers and migrated to other tiers using the *Demand Store Tier*. We refer to a *tier* as an abstract and generic entity that represents a computational unit independent of other tiers and that collaborates with other tiers to achieve program execution as a group (GIPSY network) [161, 302, 362]. The author Mokhov made the tier architecture extensible to the application-specific domains as well allowing problem-specific tier instances to use the architecture (e.g., see Section 8.3 and Section 5.4). In Figure 34 is the context use case diagram describing user interaction with the nodes and tiers to get them started to form a GIPSY software network. The user interaction is done either via command line or GUI support in the `RIPE` package interfacing the `GMT` [393].

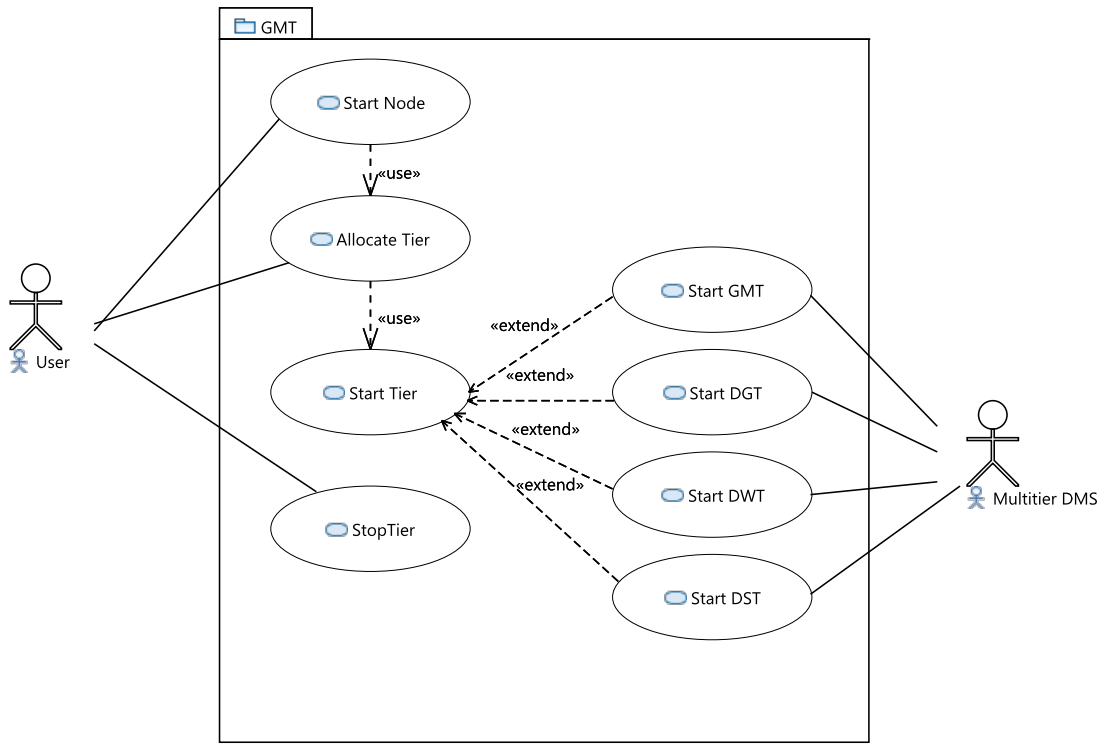

Figure 34: GMT context use case diagram

**6.2.2.1.4 GIPSY Node.** Abstractly, a *GIPSY Node* is a computer (physical or virtual) that has registered for the hosting of one or more *GIPSY Tiers*. GIPSY Nodes are registered through a *GIPSY Manager Tier* (GMT) instance. Technically, a GIPSY Node is a controller that wraps GIPSY Tier instances, and that is remotely reporting and being controlled by a GIPSY Manager Tier [161, 362]. Operationally, a GIPSY Node hosts one tier controller for each kind of Tier (see Figure 35). The *Tier Controller* acts as a factory that will, upon necessity, create *instances* of this Tier, which provide the concrete operational features of the Tier in question. This model permits scalability of computation by allowing the creation of new Tier instances as existing tier instances get overloaded or lost [161, 302, 362].

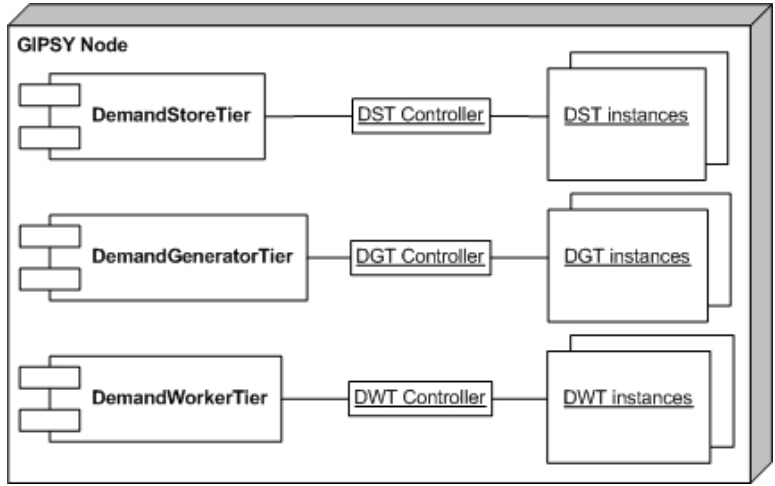

Figure 35: Design of the GIPSY node [161, 362]

**6.2.2.1.5 GIPSY Instance.** A *GIPSY Instance* is a set of interconnected GIPSY Tiers deployed on GIPSY Nodes executing GIPSY programs by sharing their respective GEER instances. A GIPSY Instance can be executing across different GIPSY Nodes, and the same GIPSY Node may host GIPSY Tiers that are members of separate GIPSY Instances [161, 302, 362]. In Figure 35 is Paquet's rending of the described design [362]. GIPSY Instances form so called *GIPSY software networks*, similar in a way to software-defined networks [117].

**6.2.2.1.6 Demand Generator Tier.** The *Demand Generator Tier* (DGT) generates demands according to the program declarations and definitions stored in one of the instances of GEER that it hosts. The demands generated by the Demand Generator Tier instance can be further processed by other Demand Generator Tier instances (in the case of *intensional demands*) or *Demand Worker Tier* instances (in the case of *procedural demands*), the demands being migrated across tier instances through a *Demand Store Tier* instance. Each DGT instance hosts a set of GEER instances that corresponds to the LUCID programs it can process demands for. A demand-driven mechanism allows the Demand Generator Tier to issue *system demands* requesting for additional GEER instances to be added to its *GEER Pool* (a local collection of cached GEERs it has learned), thus enabling DST instances to process demands for additional programs in execution on the GIPSY networks they belong to [161, 302, 362]. The author Mokhov additionally introduced the notion of problem-specific DGTs (e.g., `MARFCATDGT` discussed later) to show the wide array of applications that are possible using the multi-tier architecture as a middleware platform.

**6.2.2.1.7  Demand Store Tier.**  The *Demand Store Tier* (DST) acts as a tier middleware in order to migrate demands between tiers. In addition to the migration of the demands and values across different tiers, the Demand Store Tiers provide persistent storage of demands and their resulting values, thus achieving better processing performances by not having to re-compute the value of every demand every time it is re-generated after having been processed. From this latter perspective, it is equivalent to the historical notion of an *intensional value warehouse* [363, 459] in the eductive model of computation (Section 4.1.4, page 86). A centralized communication point or warehouse is likely to become an execution bottleneck for large long-running computations. In order to avoid that, the Demand Store Tier is designed to incorporate a peer-to-peer architecture as needed and a mechanism to connect all Demand Store Tier instances in a given GIPSY network instance. This allows any demand or its resulting value to be stored on any available DST instance, but yet allows abstract querying for a specific demand value on any of the DST instances. If the demanded value is not found in the DST instance receiving the demand, it will contact its DST peers using a peer-to-peer mechanism. This mechanism allows to see the Demand Store abstractly as a single store that is, behind the scenes, a distributed one [161, 302, 362].

**6.2.2.1.8  Demand Worker Tier.**  The *Demand Worker Tier* (DWT) processes primarily *procedural demands*, i.e., demands for the execution of functions or methods defined in a procedural language, which are only present in the case where hybrid intensional programs are being executed. The DGT and DWT duo is an evolution of the generator-worker architecture adopted in GLU [5, 188, 189]. It is through the operation of the DWT that the increased granularity of computation is achieved. Similarly to the DGT, each DWT instance hosts a set of compiled resident procedures (sequential thread procedure classes) that corresponds to the procedural demands it can process pooled locally. A demand-driven mechanism allows the Demand Worker Tier to issue *system demands* requesting for additional GEERs to be added to its GEER pool, thus achieving increased processing knowledge capacity over time, eductively [161, 302, 362].

**6.2.2.1.9  General Manager Tier.**  A *General Manager Tier* (GMT) is a component that enables the registration of GIPSY Nodes and Tiers, and to allocate them to the GIPSY network instances that it manages. The General Manager Tier interacts with the allocated tiers in order to determine if new tiers and/or nodes are necessary to be created, and issue system demands to GIPSY Nodes to spawn new tier instances as needed. In order to ease the node registration, the General Manager Tier can be implemented using a web interface and/or a stand-alone graph-based UI [393], so that users can register nodes using a standard web browser, rather than requiring a client. As with DSTs, multiple GMTs are designed to be peer-to-peer components, i.e., users can register a node through any GMT, which will then inform all the others of the presence of the new node, which will then be available for hosting new GIPSY Tiers at the request of any of the GMT currently running. The GMT uses *system demands* to communicate with Nodes and Tiers [161, 302, 362], in a way similar to SNMP get and set requests [165, 251, 440]. In Figure 36 is a more detailed GMT-oriented use case diagram of Figure 34 (where other tiers are implied with the focus on the GMT). Once started, the GMT acts as a service using the DMS, which is designed to play an active role in managing the GIPSY software network instance along with the interacting user.

### 6.2.2.2  Scalability

The GIPSY multi-tier architecture after extensive design revision, implementation and refactoring [160, 161, 191, 193] was put to scalability tests by Ji [191]. The need for such is also emphasized in in the recent work by Fourtounis *et al.* [118] where eduction and massive parallelism of intensional evaluation are discussed in a formal model and PoC implementation for analysis.

Scalability is an important attribute of any computing system as it represents the ability to achieve long-term success when the system is facing growing demands load. The multi-tier architecture was adopted for the GIPSY runtime system for research goals such as scalability [161, 362]; therefore, upon implementation of the PoC Jini and JMS DMS, the scalability of the GIPSY runtime system needed to be assessed for the validation of the implementation, which Ji has done in his mater's thesis in 2011 [191, 192].

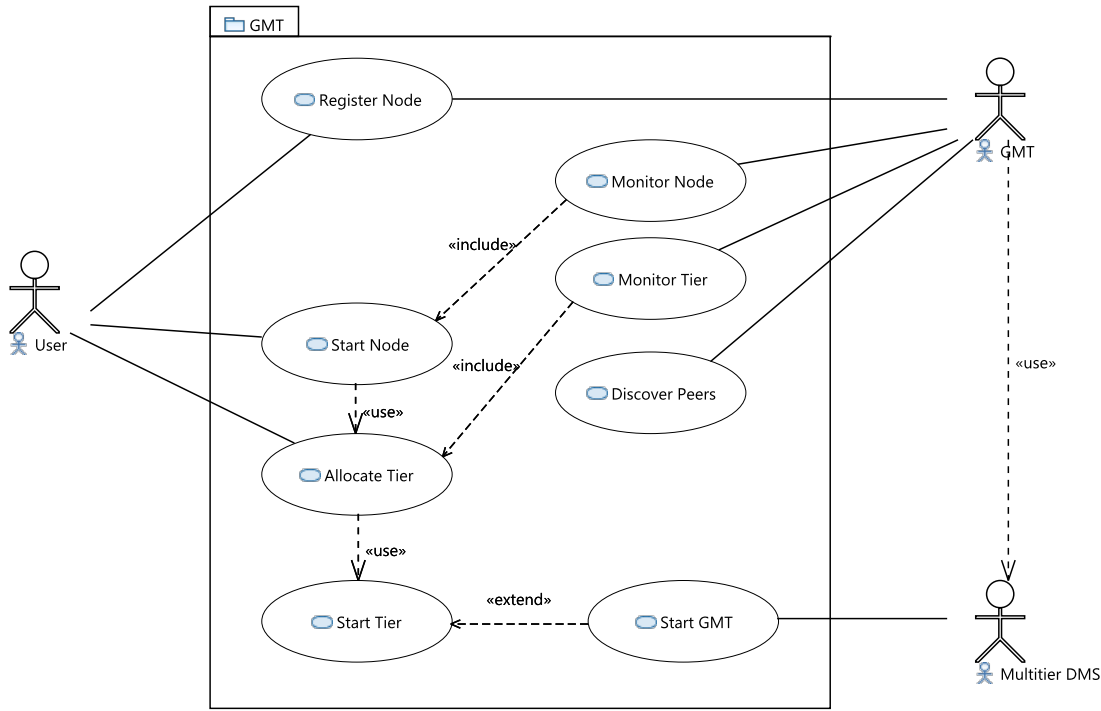

Figure 36: Detailed GMT context use case diagram

While scalability is an important term, Ji discovered during his research that although the term *scalability* is widely used and treated as an important attribute in the computing community, there was no commonly accepted definition of scalability available [96]. Without a standardized definition, researchers used *scalability* to denote the capability for the long-term success of a system in different aspects, such as the ability of a system to hold increasing amount of data, to handle increasing workload gracefully, and/or to be enlarged easily [46, 191, 192]. Ji did test under which circumstances Jini DMS was better or worse over the JMS circumstances in terms of scaling out of GIPSY Nodes onto more physical computers and their available CPUs for the tiers, then the amount of memory used on the nodes before DSTs run out of memory, amount of demands they could handle, network delays and turnaround time, and other characteristics [191].

The scalability aspect is very important to this work as there is always a possibility of a need to do heavy-weight data mining and process vast amounts of digital evidence in an investigation efficiently, robustly, and effectively, a problem akin to a human genome sequence alignment and reconstruction [100].

### 6.2.3 Run-time Interactive Programming Environment—(RIPE)

RIPE's place in the GIPSY's architecture [264, 363] has to do with the user interaction aspect of the system. While presently it is the most underdeveloped aspect, it has had some valuable contributions made to it. The first of them is the PoC realization of the bidirectional translation of INDEXICAL LUCID and GIPL into visual Graphviz-based data-flow graphs [89] based on the corresponding earlier assertion by Paquet [361] that such multidimensional translation is possible (see Chapter E). In Figure 37 is the context use case diagram for RIPE illustrating user goal-level activities the user needs. The latest one of the contributions is

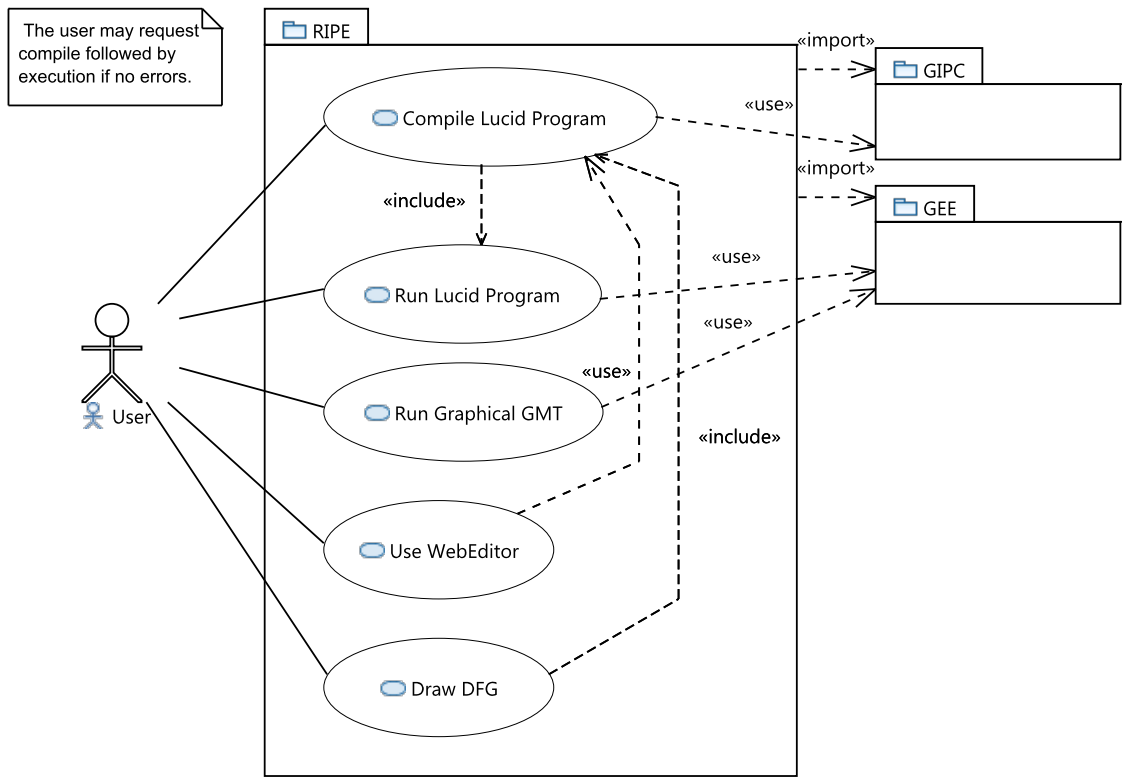

Figure 37: RIPE context use case diagram

the graph-based visual GMT (GGMT) integration by Rabah [393]. Its purpose is to allow start up and configuration management of the GIPSY network instances initially developed

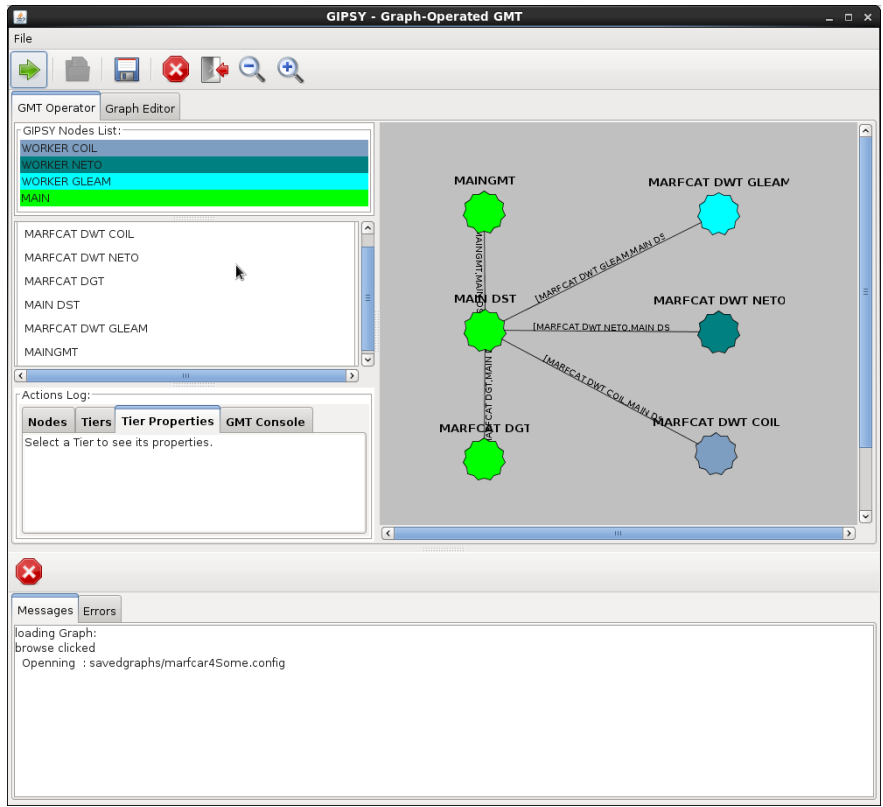
Figure 38: MARFCAT GIPSY network graph

mostly with the command-line interface (CLI) by Ji and a simple Simulator GUI by Vassev. This more flexible approach made it significantly more usable to operate GIPSY networks. See for examples its connectivity for the MARFCAT test case with the corresponding graph in Figure 38.

The author Mokhov additionally contributed to the general refactoring and standardization of the RIPE package hierarchy as well as contributing a servlet-based web front-end to launch the compilation and execution process via a browser on a server side [264] (see Figure 39). CLI is also supported for RIPE primarily for invocation by various scripts. The *MAC Spoofer Analyzer* case relies on this in particular (see Section 9.5) to start the analysis process.

## 6.3 Summary

To summarize we have motivated the discussion and necessity of the General Intensional Programming System (GIPSY) and its core GIPC, GEE, and RIPE components and reviewed its existence in the historical context of GLU as well as follow-up design and development iterations. Particular level of detail was given to the GIPC and GEE frameworks with the goal of flexibility (data types, languages, replaceable components) and scalability (multi-tier DMS) in mind, important for the Forensic Lucid project presented in this thesis, especially its compiler and extended run-time system design and development described in Chapter 8.

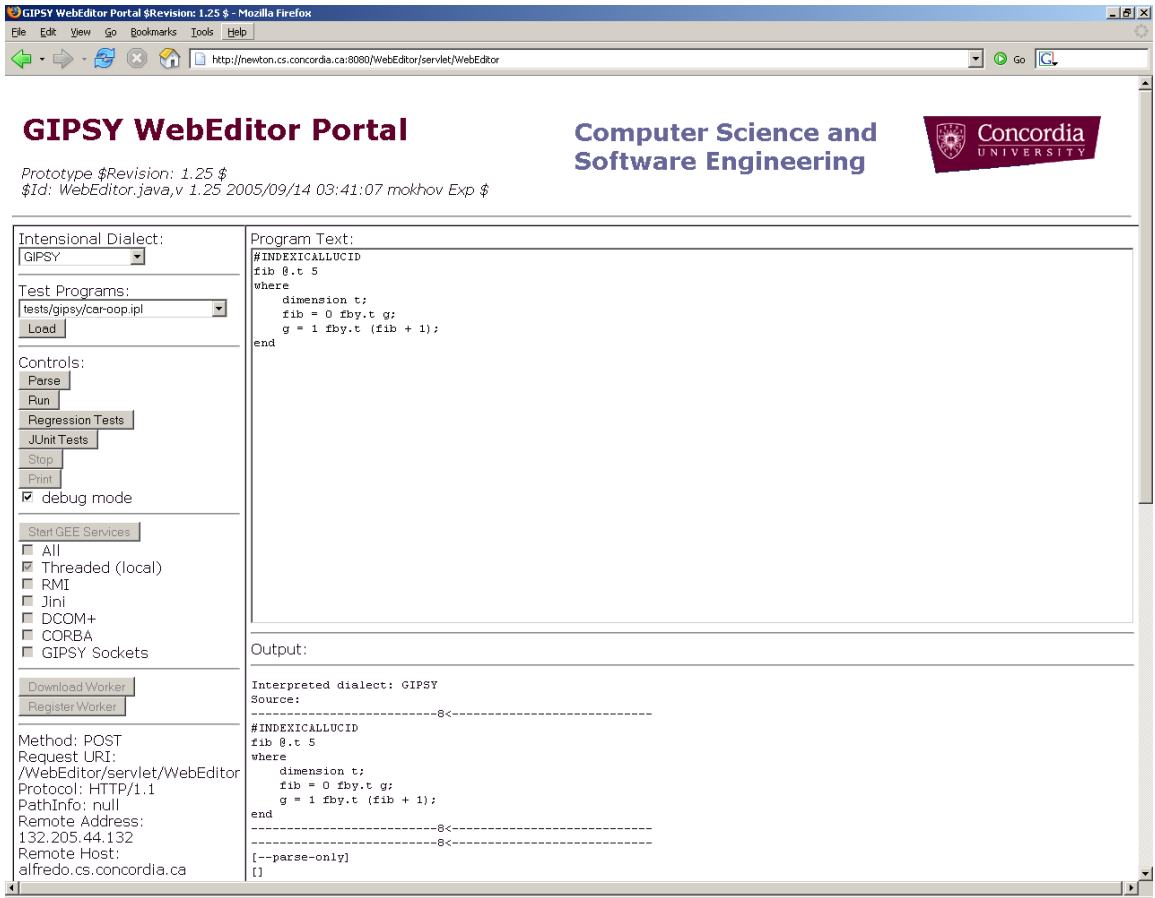

Figure 39: GIPSY `WebEditor` interface [264]

# Part II

# Methodology

## Chapter 7

# Forensic Lucid Design and Specification

This chapter discusses the core contribution of the thesis—the FORENSIC LUCID language. This includes the methodology on the formalization of the necessary forensic constructs, the syntax, semantics, and augmentation in dereference to the background work in Part I. This chapter is an extended compilation of a number of the published and unpublished works [269, 290, 300, 304, 307, 312] detailing various aspects of FORENSIC LUCID construction and case studies including a number of supporting works.

FORENSIC LUCID [269, 300, 304, 309, 310] is a forensic case specification language for automatic deduction and event reconstruction in digital crime incidents. The language itself is general enough to specify any events (in terms of their properties and description), duration, as well as the context-oriented system model [277, 308, 321]. FORENSIC LUCID is based on LUCID [24, 25, 26, 379, 509] and its various dialects that allow natural expression of various phenomena, inherently parallel, and most importantly, context-aware, i.e., the notion of context is specified as a first-class value [365, 473, 513]. FORENSIC LUCID is also significantly influenced by and is meant to be a more usable improvement of the work of Gladyshev *et al.* on formal forensic analysis and event reconstruction using finite state automate (FSA) to model incidents and reason about them [136, 137] by also including trustworthiness and credibility factors into the equation using the Dempster–Shafer theory [277, 308, 321].

As previously mentioned (Chapter 1), the first formal approach to cybercrime investigation was the finite-state automata (FSA) approach by Gladyshev *et al.* [136, 137] (Section 2.2, page 29). Their approach, however, is unduly complex to use and to understand for non-theoretical-computer science or equivalently minded investigators [300, 311]. The aim of FORENSIC LUCID is to alleviate those difficulties, expressive and usable, and to be able to specify credibility.

Thus, this chapter presents the summary of the requirements, design, and formalization of the syntax and semantics of our proposed language. That includes an overview (Section 7.1), design and requirements considerations of the FORENSIC LUCID language (Section 7.2), including higher-order contexts (Section 7.2.2), syntax (Section 7.3) and semantics (Section 7.4), and discussion on related topics in Section 7.5, such as mapping of the FORENSIC LUCID concepts to the background work on intensional logic (cf. Section 3.2, page 59), Gladyshev's formalisms (cf. Section 2.4, page 56), and the Dempster–Shafer evidence theory (cf. Section 3.3.2, page 66) in Section 7.5.1.

## 7.1 Approach Overview

As the reader may recall, Gladyshev created a finite-state-automata (FSA) model [136, 137] to encode the evidence and witness accounts (Section 2.2, page 29) in order to combine them into an *evidential statement*, then model the FSA of a particular case, and given the FSA verify if a certain claim agrees with the evidential statement or not (via backtracing and the event reconstruction algorithm) and if it does what were possible event sequences that explain that claim [274, 313]. Based on the formal parameters and terms defined in that work [135, 136, 137], we likewise model various pieces of evidence and witnesses telling their own stories of an incident. The goal is to put them together to make the description of the incident as precise as possible [307, 312]. To demonstrate that a certain claim may be true, an investigator has to show that there are some explanations of the evidence that agree with the claim. To disprove the claim, the investigator has to show there are no explanations of evidence that agree with the claim [136, 137]. On the other hand, the work by Dempster–Shafer and others [153, 422] defined a mathematical theory of evidence (Section 3.3.2, page 66), where factors like credibility and trustworthiness play a role in the evaluation of mathematical expressions [274, 313]. Thirdly, a body of work on intensional logics and programming (Section 3.2, page 59) provided a formal model that throughout years of development placed the context as a first-class value in logical and programming expressions in the LUCID family (Chapter 4) of languages [274, 313].

Thus, in this work we augment Gladyshev's formalization with the credibility weight and other properties derived from the mathematical theory of evidence and we encode it as an evidential context in the FORENSIC LUCID language for forensic case management, evaluation, and event reconstruction.

## 7.2 The FORENSIC LUCID Language Requirements and Design Considerations

This section presents concepts and considerations in the design of the FORENSIC LUCID language. The end goal is to define our FORENSIC LUCID language where its syntactic constructs and expressions concisely model cyberforensic evidence and stories told by witnesses as a context of evaluations, which normally correspond to the initial state of the case (e.g., initial printer state when purchased from the manufacturer, as in [137], see further Section 9.3), towards what we have actually observed (as corresponding to the final state in the Gladyshev's FSM, e.g., when an investigator finds the printer with two queue entries ($B_{deleted}, B_{deleted}$) in Section 2.2.5.1, page 44). The implementing system (i.e., GIPSY [366], Chapter 6, page 128) is designed to backtrace intermediate results in order to provide the corresponding event reconstruction path, if it exists. The fundamental result of a FORENSIC LUCID expression in its basic form is either *true* or *false*, i.e., "guilty" or "not guilty" given the evidential evaluation context per explanation with the backtrace(s). There can be multiple backtraces, that correspond to the explanation of the evidence (or lack thereof) [300, 304, 305, 312].

### 7.2.1 Core Language Properties, Features, and Requirements

We define and use FORENSIC LUCID to model the evidential statements and other expressions representing the evidence and observations as a higher-order hierarchical context of evaluation [304, 307, 312]. One of the goals is to be able to "script" the evidence and the stories as expressions and run that "script" through an evaluation system that provides the results of such an evaluation [305]. An execution trace of a running FORENSIC LUCID program is designed to expose the possibility of the proposed claim to be true with the reconstructed event sequence backtrace between the final observed event to the beginning of the events. Intensionally, this means the initial possible world $q_0$ is accessible from the final possible world $q_{final}$ via one or more accessibility relations and possibly other worlds (states) (cf. Section 3.2, page 59). FORENSIC LUCID capitalizes in its design by aggregating the features and semantics of multiple LUCID dialects mentioned in Chapter 4 needed for these tasks along with its own extensions [300, 304, 305, 312].

LUCX's context calculus with contexts as first-class values [513] and operators on simple contexts and context sets (`union`, `intersection`, etc., defined further) are augmented to manipulate hierarchical contexts in FORENSIC LUCID (Section 7.3.3, page 183). Additionally, FORENSIC LUCID inherits the properties (described subsequently in detail in this chapter) of MARFL (see Appendix C), OBJECTIVE LUCID and JOOIP (see Section 4.3.2, page 92) and their comprising dialects for the arrays and structural representation of data for modeling the case data structures such as events, observations, and groupings and correlation of the related data [300, 304, 305]. Hierarchical contexts in FORENSIC LUCID follow the example of MARFL [272] using a dot operator and by overloading both `@` and `#` (defined further in Section 7.3, page 166) to accept different types as their left and right arguments [300, 304, 307, 312] (cf. Table 15, page 203).

One of the basic requirements in FORENSIC LUCID design is that the final language is a conservative extension of the previous LUCID dialects (valid programs in those languages are valid programs in FORENSIC LUCID). This is helpful when complying with the compiler (GIPC) and the run-time subsystem (GEE) frameworks within the implementing system, the General Intensional Programming System (GIPSY) (cf., Section 6.1) [304, 307, 312, 366, 370]. The partial translation rules (Section 7.3.2.2) provided when implementing the language compiler within GIPSY, such that the run-time environment (General Eduction Engine, or GEE) can execute it with minimal changes to GEE's implementation [305, 312].

### 7.2.2 Higher Order Context

*Higher-order context*s (HOCs) represent essentially nested contexts, e.g., as conceptually shown in Figure 40 modeling evidential statement for forensic specification evaluation. Using the $\langle dimension : tag \rangle$ representation, a HOC looks like:

{es : {os1 : {o1 : (P1, min, max), o2 : (P2, min, max), o2 : (P3, min, max)}, os2 : {...}, os3 : {...}}

The early notion and specification of nested contexts first appeared Swoboda's works [452, 454, 455], but there the evaluation has taken place only at the leaf context nodes. Another, more recent work on the configuration specification as a context in the intensional manner was the MARFL language (Appendix C) of the author [269, 272], allowing evaluation at arbitrary nesting of the configuration context with some defaults in the intermediate and leaf context nodes [302].

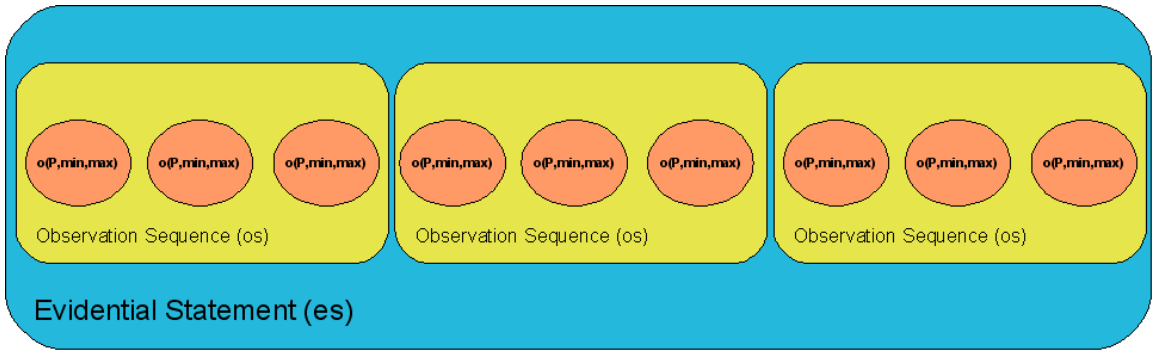

Figure 40: Nested context hierarchy example for cyberforensic investigation [300, 304]

FORENSIC LUCID is context-oriented where a crime scene model comprises a state machine of evaluation and the forensic evidence and witness accounts comprise the context for its possible worlds. Formally, the basic context entities comprise an observation $o$ (in Equation 7.2.2.1), observation sequence $os$ (in Equation 7.2.2.2), and the evidential statement (in Equation 7.2.2.3). These terms are inherited from Gladyshev's work [136, 137] (Section 2.1.4, page 29) and represent the forensic context of evaluation in FORENSIC LUCID. An observation of a property $P$ has a duration between $[\min, \min + \max]$ per Section 2.1.4, page 29.

This original definition of $o$ is extended with $w$ to amend each observation with weight factor (to indicate probability, credibility, or belief mass) to model trustworthiness of observations, evidence, and witness accounts in accordance with the Dempster–Shafer mathematical theory of evidence [422]. $t$ is an optional timestamp as in a forensic log for that property used to refine event co-relation in time for temporal data [277, 308, 321].

An observation sequence $os$ represents a partial description of an incident told by evidence or a witness account (electronic or human). It is formally a chronologically ordered collection of observations representing a story witnessed by someone or something (e.g., a human witness, a sensor, or a logger). It may also encode a description of any digital or physical evidence found. All these "stories" (observation sequences) all together represent an evidential statement about an incident (its knowledge base). The evidential statement $es$ is an unordered collection of observation sequences. The property $P$ itself can encode anything of interest—an element of any data type or even another FORENSIC LUCID expression, a nested object instance hierarchy, or an event [277, 308, 321].

$$o = (P, \min, \max, w, t) \qquad (7.2.2.1)$$

$$os = \{o_1, \ldots, o_n\} \qquad (7.2.2.2)$$

$$es = \{os_1, \ldots, os_m\} \qquad (7.2.2.3)$$

Having constructed the context in the form of the evidential statement, one needs to build a *transition function* $\psi$ and its inverse $\Psi^{-1}$ to describe the "crime scene" at the center of the incident (or the *incident scene* to generalize the term a little bit to include accidents, malfunctions, etc. that are not necessarily criminal or malicious in nature).

### 7.2.3 Formal Syntax and Semantics

The FORENSIC LUCID system incorporates the necessary syntax and semantics constructs [313] as detailed in the following sections.

#### 7.2.3.1 Definitions and Terms

This section defines common terminology used subsequently in the chapter throughout the formal definition of FORENSIC LUCID syntax and semantics and their description.

**7.2.3.1.1 Structural Operational Semantics.** We use structural operational semantics (see examples in Chapter 4) to describe the semantics of FORENSIC LUCID. We justify this choice for a number of reasons.

- Plotkin proposed the notion in 1981 [381]. While in 1979–1982 Ashcroft and Wadge argued for prescriptive denotational semantics for their language [28], Faustini in his thesis the same year established the equivalence of a denotational and an operational semantics of pure dataflows [108] (refer to Section 4.1.2, page 82 for information on dataflows).

- Mosses in 2001 [326] discussed the differences between denotational and operational semantics (among others). In the same year, Degano and Priami argued operational semantics is *close to intuition, mathematically simple, and allows easy and early prototyping* [86].

- Paquet [361] in 1999 has given a structural operational semantics of GIPL in 1999 (recited earlier in Figure 23, page 82). The subsequently produced work on LUCX [473, 513], JOOIP [528], and previous material by the author Mokhov used this semantics style consistently ever since.

- A lot of JAVA-related publications use structured operational semantics. GIPSY (Chapter 6) is implemented in JAVA and it is easier to relate to JAVA semantics. GEE's main interpreter follows this style for GIPL programs. Not to mention the hybrid LUCID-JAVA approaches such that of OBJECTIVE LUCID and JOOIP mentioned earlier that need to specify joint semantics.

- PRISM's language semantics is also described in the structural operational semantics style ([190], Section 8.2.2.1, page 215).

The basic operational semantics description has rules in the form of $\frac{Premises}{Conclusions}$ with *Premises* defined as possible computations, which, if take place, *Conclusions* will also take place [86, 513].

Following the steps presented in Section 4.1.1.2, page 78, [361, 513] we thus define the following:

Table 13: FORENSIC LUCID identifier types in $\mathcal{D}$

| type | form |
|---|---|
| dimension | (`dim`) |
| constant | (`const`, $c$) |
| operator | (`op`, $f$) |
| variable | (`var`, $E$) |
| function | (`func`, $id_i$, $E$) |
| class | (`class, cid, cdef`) |
| member variable | (`classV, cid.cvid, vdef`) |
| member method | (`classF, cid.cfid, fdef`) |
| free procedure | (`freefun, ffid, ffdef`) |
| context operator | (`cop`, $f$) |
| context set operator | (`sop`, $f$) |
| **observation** | (`odim`, $E$, $\min$, $\max$, $w$, $t$) |
| **observation sequence** | (`osdim`, `ordered` $odim_i$) |
| **evidential statement** | (`esdim`, `unordered` $osdim_i$) |
| **forensic operator** | (`fop`, $f$) |

**Definition environment:** The *definition environment* $\mathcal{D}$ stores the definitions of all of the identifiers that appear in a FORENSIC LUCID program. As in Equation 4.1.1.3, it is a partial function

$$\mathcal{D} : \mathbf{Id} \to \mathbf{IdEntry} \qquad (7.2.3.1)$$

where **Id** is the set of all possible identifiers and **IdEntry**, summarized in Table 13, has possible kinds of values for each identified kind. (This table is an extended version of Table 1 with additional entries we define further.) Per Equation 4.1.1.1, $\mathcal{D} \vdash E : v$ means the expression $E$ evaluates to the value $v$ under $\mathcal{D}$ [361].

**Evaluation context:** The current *evaluation context* $\mathcal{P}$ (sometimes also referred to as a *point* in the context space) is an additional constraint put on evaluation, so per Equation 4.1.1.2, $\mathcal{D}, \mathcal{P} \vdash E : v$ specifies that given $\mathcal{D}$, and in $\mathcal{P}$, expression $E$ evaluates to $v$ [361].

**Identifier kinds:** The following identifier kinds can be found in $\mathcal{D}$:

1. *Dimensions* define simple coordinate pairs, which one can query and navigate with the `#` and `@` operators. Their **IdEntry** is simply (`dim`) [361].

2. *Constants* (**IdEntry** = (`const`, $c$), where $c$ is the value of the constant) are external entities that provide a single value independent of the context of evaluation. Examples are integers, strings, and Boolean values, etc. [361].

3. *Data operators* are external entities that provide memoryless functions. Examples are the arithmetic and Boolean functions. The constants and data operators are said to define the *basic algebra* of the language. Their **IdEntry** is (`op`, $f$), where $f$ is the function itself [361].

4. *Variables* carry the multidimensional streams (described in Section 4.1.2, page 82). Their **IdEntry** is (`var`, $E$), where $E$ is the FORENSIC LUCID expression defining the variable. Variable names are unique [361]. Finite variable streams can be bound by the special beginning-of-data (`bod`) and end-of-data (`eod`) markers (`eod` was introduced in [360]). Scoping and nesting is resolved by renaming and via the dot "." operator.

5. *Functions* (`func`, $id_i$, $E$) are FORENSIC LUCID user-defined functions, where the $id_i$ are the formal parameters to the function and $E$ is the body of the function [361].

6. *Classes* (`class`, `cid`, <u>`cdef`</u>) in our work are user-defined JAVA classes, members of which can be accessed from FORENSIC LUCID (come from OBJECTIVE LUCID and JOOIP) to support JAVA-LUCID hybrid intensional-imperative programming [264, 528]. This integration is needed, e.g., for the self-forensics project (Appendix D). The same applies to member variables and methods described further. In general, the *Classes* entry does not need to be restricted to JAVA classes, but can be expanded to be defined in any language that supports the notion and the run-time supports hybridification. `cid` is the class identifier, and <u>`cdef`</u> is a class definition in the imperative language (JAVA in our case, so <u>`cdef`</u> = <u>`JavaCDef`</u>).

7. *Member variables* (`classV`, `cid.cvid`, <u>`vdef`</u>) are user-defined JAVA data members which can be accessed from FORENSIC LUCID (are originally from OBJECTIVE

LUCID and JOOIP) to support JAVA-LUCID hybrid intensional-imperative programming [264, 528]. `cid.cvid` is the variable identifier in the class `cid`, and <u>vdef</u> is a member variable definition in the imperative language (JAVA in our case, so <u>vdef</u> = <u>JavaVDef</u>).

8. *Member methods* (`classF, cid.cfid,` <u>fdef</u>) are user-defined JAVA methods (are originally from OBJECTIVE LUCID and JOOIP) to support JAVA-LUCID hybrid intensional-imperative programming [264, 528]. `cid.cfid` is the method identifier in the class `cid`, and <u>fdef</u> is a member method definition in the imperative language (JAVA in our case, so <u>fdef</u> = <u>JavaFDef</u>).

9. *Free procedures* (also known as *free functions* in the previous work [264, 528]) (`freefun, ffid,` <u>ffdef</u>) are user-defined JAVA methods written "freely" without an explicit container class directly in a FORENSIC LUCID program (the notion comes from JLUCID) [264]. (It is a responsibility of the compiler to generate a wrapper class containing any such procedures [264].)

10. *Context operators* (`cop`, $f$) is a LUCX-style simple context operator type [473, 513]. These operators help us to manipulate context components in set-like operations, such as union, intersection, difference, and the like. Both simple context and context set operators are defined further.

11. *Context set operators* (`sop`, $f$) is a LUCX-style context set operator type [473, 513]. Both simple context and context set operators are defined further.

12. *Observations* (`odim`, $E$, min, max, $w, t$) are FORENSIC LUCID dimensions each encoding an observation of a property defined by a FORENSIC LUCID expression $E$, duration [min, min + max], basic belief mass assignment $w$, and a wall-clock timestamp $t$. This is a basic unit of a forensic context. One can manipulate these with the context operators.

13. *Observation sequences* (`osdim`, $odim_i$) are FORENSIC LUCID dimensions encoding each observation sequence, as an ordered collection of observations $odim_i$. This is a part of a forensic context. One can manipulate these with the context operators.

14. *Evidential statements* (`esdim`, $osdim_i$) are FORENSIC LUCID dimensions encoding

an unordered collection of observation sequences $osdim_i$. This is a part of a forensic context. One can manipulate these with the context operators.

15. *Forensic operators* (`fop`, $f$) are FORENSIC LUCID context operators [473, 513] defined further in Section 7.3.4, page 191.

### 7.2.3.1.2 Context Types.
A general definition of context is [473, 513]:

**Definition 1.** Context: *A context c is a finite subset of the relation: $c \subset \{(d,x)|d \in DIM \wedge x \in T\}$, where DIM is the set of all possible dimensions, and T is the set of all possible tags [473, 513].*

**Definition 2.** Tag: *Tags are the indices to mark positions in dimensions [473, 513].*

**Definition 3.** Simple context: *A simple context is a collection of $\langle dimension : tag \rangle$ pairs, where there are no two such pairs having the same dimension component. Conceptually, a simple context represents a point in the context space, i.e., this is the traditional* GIPL *context $\mathcal{P}$. A simple context having only one $\langle dimension : tag \rangle$ pair is called a* micro context. *It is the building block for all the context types [473, 513].*

Syntactically, simple context is $[E : E, ..., E : E]$ [473, 513], i.e., $[E_{d_1} : E_{t_1}, ..., E_{d_n} : E_{t_n}]$ where $E_{d_i}$ evaluate to dimensions and $E_{t_i}$ evaluate to tags.

**Definition 4.** Context set: *A context set (also known as "non-simple context") is a set of simple contexts. Context sets represent regions of the context space, which can be seen as a set of points, considering that the context space is discrete. Formally speaking, context set is a set of $\langle dimension : tag \rangle$ mappings that are not defined by a function [473, 513].*

Syntactically, context set is $\{E, ..., E\}$, where $E \rightarrow [E : E, ..., E : E]$.

### 7.2.3.1.3 Forensic Context.

**Definition 5.** Forensic context: *Forensic context represents an evidential statement of a case under consideration. It's a hierarchical encoding of evidence in terms of the* `evidential statement`, `observation sequence`, *and* `observation` *constructs.*

Forensic contexts form the basis of cyberforensic reasoning in FORENSIC LUCID and are elaborated further in detail in terms of syntax and semantics.

#### 7.2.3.1.4 Context Calculus.

**Definition 6.** Context calculus: *Context calculus is a calculus defining representation and types of contexts and operators on them [473, 513].*

In LUCX terms, context calculus defines a set of context operators on simple contexts and context sets [513] presented earlier. In FORENSIC LUCID terms, the context calculus is extened to support forensic contexts and additional operators and lifting of LUCX contexts into forensic contexts. All the new constructs are defined in the sections that follow.

#### 7.2.3.1.5 Tag Set Types.

**Definition 7.** Tag set: *A tag set $T$ is a collection of all possible tags attached to a dimension [473].*

Traditionally, in many earlier LUCID dialects $T$ was always an infinite ordered set of natural numbers $\mathbb{N}$. However, it is not always convenient in some application domains [473], such as forensic computing. In LUCX's compiler implementation by Tong [473] flexible tag set types were designed to allow more application domains. Tags, therefore, can be of any given desired type. It is up to the particular semantic analyzer to determine validity of tag types [473]. By default each dimension's tag set is both ordered and infinite, to maintain backward compatibility [473].

**Definition 8.** Ordered Tag Set: *A tag set on which a relation $R$ satisfies the following three properties [237, 237, 473]:*

1. *Reflexive: For any $a \in S$, we have $aRa$*

2. *Antisymmetric: If $aRb$ and $bRa$, then $a = b$*

3. *Transitive: If $aRb$ and $bRc$, then $aRc$*

*which is essentially a partial order set (*post*) [431, Chapter 4]. Ordered tag set can be either finite or infinite [473].*

**Definition 9.** Unordered Tag Set: *A tag set which is not ordered is called an unordered set. Unordered tag set can be either finite or infinite [473].*

**Definition 10.** Finite Tag Set: *A tag set I is called finite and more strictly, inductive, if there exists a positive integer n such that I contains just n members. Finite tag set can be either ordered or unordered [473].*

**Definition 11.** Infinite Tag Set: *A tag set, which is not finite is called an infinite set. Infinite tag set can be either ordered or unordered [473].*

**Definition 12.** Periodic Tag Set: *A tag set where its ordered subset represents a period repeated in a pattern two or more times. Can be combined with the (un)ordered, finite, and infinite types [473].*

**Definition 13.** Non-periodic Tag Set: *A tag set that is not periodic. Can be combined with the (un)ordered, finite, and infinite types [473].*

Syntactically, `ordered`, `unordered`, `finite`, `infinite`, `periodic`, and `nonperiodic` are defined in the language to declare a desired permutation of a tag set type declared alongside `dimension` [473]. FORENSIC LUCID inherits these definitions from LUCX. Certain combinations are of not much practical use (e.g., `unordered infinite` tag set types), but they are provided for completeness by Tong [473]. In FORENSIC LUCID, `ordered` and `unordered finite` tag sets are the most common in usage.

Following the above fundamental definitions, Section 7.3 is devoted to the complete syntax specification and Section 7.4 is devoted to the operational semantics specification of FORENSIC LUCID.

## 7.3 Concrete FORENSIC LUCID Syntax

The concrete syntax of the FORENSIC LUCID language is summarized in several figures. It is influenced by the productions from LUCX [513, 515] (see Section 4.3.1, page 92), JLUCID and OBJECTIVE LUCID [264], GIPL and INDEXICAL LUCID [361] (see Section 4.1, page 76), and the hierarchical contexts from MARFL [272] (see Appendix C for details).

In Figure 41 are common top-level syntax expressions $E$ of the language from identifiers, to function calls, operators, arrays, the `where` clause, dot-notation, and so on. In Figure 42 are the $Q$ `where` productions containing various types of declarations, including the common

LUCX dimensions and their tag sets as well as the new constructs for forensic dimension types. In Figure 43 we elaborate on the syntactical details on the hierarchical forensic context specification of the observations, observation sequences, and evidential statements including different observation types. In Figure 44 is a syntax specification of different types of operators, such as arithmetic, logic, and intensional operators, in their unary and binary forms.

The `embed` operator, the notion of which is inherited from JLUCID (Section 4.3.2.1, page 92), is adapted in FORENSIC LUCID primarily to include large evidential specifications from external to the main program sources, such as encoded log files.

### 7.3.1 Syntactical Extension for Observations

The syntactical notation of the unit of observation in FORENSIC LUCID extends the observation context definition with the optional $w$ and $t$ components. $w$ is always defined and by default is 1, and $t$ can be left undefined (set to `eod`) as it is not always needed. When defined, $t$ is a wall-clock timestamp of the beginning of the observation, which can be used to be mapped to from log files or other real-time sensitive data and to present in the event reconstruction reports. $t$ can also be used in reactive systems with clocks and self-forensics, but this aspect is not discussed here. Thus, the following would be an actual expression in the language given all the symbols are defined [313] and expressions declared.

```
observation o = (P, min, max, w, t);
```

or more concretely:

```
observation o = ("A printed", 1, 0, 0.85);
```

where $t$ is omitted, $w = 85\%$ confidence/belief, and the rest as in the original definition by Gladyshev (cf., Section 2.2). Next,

```
observation o = P;
P = "A printed";
```

is equivalent to

$$
\begin{array}{rll}
E & ::= & id \hfill (7.3.0.2) \\
& | & E(E,...,E) \hfill (7.3.0.3) \\
& | & E[E,...,E](E,...,E) \hfill (7.3.0.4) \\
& | & \texttt{if } E \texttt{ then } E \texttt{ else } E \texttt{ fi} \hfill (7.3.0.5) \\
& | & \texttt{\#}\, E \hfill (7.3.0.6) \\
& | & E \texttt{ @ } E \hfill (7.3.0.7) \\
& | & E\langle E,\ldots,E\rangle \hfill (7.3.0.8) \\
& | & \texttt{select}(E,E) \hfill (7.3.0.9) \\
& | & \texttt{Box}[E,\ldots,E|E] \hfill (7.3.0.10) \\
& | & E \texttt{ where } Q \texttt{ end;} \text{ (see Figure 42)} \hfill (7.3.0.11) \\
& | & [E:E,...,E:E] \hfill (7.3.0.12) \\
& | & E \textit{ bin-op } E \hfill (7.3.0.13) \\
& | & \textit{un-op } E \hfill (7.3.0.14) \\
& | & E \textit{ i-bin-op } E \mid E \textit{ i-bin-op}_w E \hfill (7.3.0.15) \\
& | & \textit{i-un-op } E \mid \textit{i-un-op}_w E \hfill (7.3.0.16) \\
& | & E \textit{ cxtop } E \hfill (7.3.0.17) \\
& | & \textit{bounds} \hfill (7.3.0.18) \\
& | & \texttt{embed}(URI, METHOD, E, E, ...) \hfill (7.3.0.19) \\
& | & E[E,...,E] \hfill (7.3.0.20) \\
& | & [E,...,E] \hfill (7.3.0.21) \\
& | & E.E \hfill (7.3.0.22) \\
& | & E.E(E,...,E) \hfill (7.3.0.23) \\
\textit{bounds} & ::= & \texttt{eod} \mid \texttt{bod} \mid \texttt{+INF} \mid \texttt{-INF} \hfill (7.3.0.24)
\end{array}
$$

Figure 41: Concrete FORENSIC LUCID syntax ($E$) [304, 305]

```
observation o = (P, 1, 0, 1.0);
```

P in the example can be any FORENSIC LUCID expression. If the wall-clock timestamp is desired, it can be produced as a string or an integer in one of the standard date/time formats, e.g., "Mon Jul 29 11:58:16 EDT 2013" or 16581129611312091 (as a number of seconds from the epoch). Internally, they are uniformly represented. These can be supplied by human

$$
\begin{aligned}
Q \quad ::= \quad & \texttt{dimension } id, ..., id; & (7.3.0.25) \\
| \quad & \texttt{dimension } id : \texttt{ordered finite [periodic | nonperiodic] } \{E, ..., E\} & (7.3.0.26) \\
| \quad & \texttt{dimension } id : \texttt{ordered finite [periodic | nonperiodic] } \{E \texttt{ to } E \texttt{ [step } E\texttt{] }\} \\
& & (7.3.0.27) \\
| \quad & \texttt{dimension } id : \texttt{ordered infinite [periodic | nonperiodic] } \{E \texttt{ to } E \texttt{ [step } E\texttt{] }\} \\
& & (7.3.0.28) \\
| \quad & \texttt{dimension } id : \texttt{ordered infinite } = E & (7.3.0.29) \\
| \quad & \texttt{dimension } id : \texttt{unordered finite [periodic | nonperiodic] } \{E, ..., E\} \\
& & (7.3.0.30) \\
| \quad & \texttt{dimension } id : \texttt{unordered infinite } = E & (7.3.0.31) \\
| \quad & \texttt{evidential statement [unordered [finite]] } id \texttt{ [}= ES\texttt{]}; & (7.3.0.32) \\
| \quad & \texttt{observation sequence [ordered [finite]] } id \texttt{ [}= OS\texttt{]}; & (7.3.0.33) \\
| \quad & \texttt{observation } id \texttt{ [}= O\texttt{]}; & (7.3.0.34) \\
| \quad & id = E; & (7.3.0.35) \\
| \quad & id(id, ...., id) = E; & (7.3.0.36) \\
| \quad & E.id = E; & (7.3.0.37) \\
| \quad & QQ & (7.3.0.38)
\end{aligned}
$$

Figure 42: Concrete FORENSIC LUCID syntax ($Q$) [304, 305]

investigators, but more often by FORENSIC LUCID-generating logging facilities, such as that of *MAC Spoofer Investigation* (Section 9.5, page 257). In the cases, when the wall-clock timestamp is not needed, it is `null`.

```
observation o = ("A printed", 1, 0, 0.85, "Mon Jul 29 11:58:16 EDT 2013");
```

See Section 7.4 for the discussion on the semantic aspects of these declarations.

### 7.3.2 Core Operators

The basic set of the classic intensional operators [361] is extended with the similar operators, but inverted in one of their aspects: either negation of truth or reversal of direction of navigation (or both). While some of such operators can be defined using the existing operators, just as we define them further, their introduction as a syntactical sugar allows for

$$
\begin{array}{lllr}
ES & ::= & \{OS, ..., OS\} \quad\quad\quad\quad \text{// evidential statement} & (7.3.0.39)\\
& | & OS & (7.3.0.40)\\
& | & E & (7.3.0.41)\\
\\
OS & ::= & \{O, ..., O\} \quad\quad\quad\quad\quad \text{// observation sequence} & (7.3.0.42)\\
& | & O & (7.3.0.43)\\
& | & E & (7.3.0.44)\\
\\
O & ::= & (E, E, E, E, E) \quad \text{// observation (P, min, max, w, t)} & (7.3.0.45)\\
& | & (E, E, E, E) \quad\quad\quad\quad\quad \text{// (P, min, max, w)} & (7.3.0.46)\\
& | & (E, E, E) \quad\quad\quad\quad\quad\quad\quad \text{// (P, min, max)} & (7.3.0.47)\\
& | & (E, E) \quad\quad\quad\quad\quad\quad\quad\quad\quad\quad \text{// (P, min)} & (7.3.0.48)\\
& | & E \quad\quad\quad\quad\quad\quad\quad\quad\quad\quad\quad\quad\quad\quad \text{// P} & (7.3.0.49)\\
& | & \$ \quad\quad\quad\quad \text{// no-observation (Ct, 0, INF+)} & (7.3.0.50)\\
& | & \backslash 0(E) \quad\quad\quad\quad\quad \text{// zero-observation (P, 0, 0)} & (7.3.0.51)\\
\end{array}
$$

Figure 43: Concrete FORENSIC LUCID syntax ($ES, OS, O$) [304, 305]

better expressibility. First, we provide a definition of these new operators alongside with the classical ones to familiarize the reader what they do.

### 7.3.2.1 Definitions of Core Operators

The operators are defined below to give a more complete picture. The classical operators `first`, `next`, `fby`, `wvr`, `upon`, and `asa` were previously defined in [361] and earlier works (see [131, 350] and papers and references therein). The other complimentary, inverse, and negation operators are added in this work (while most of the these seem obvious to have in any LUCID dialect), so they were defined and revised from [303, 304]. In this list of operators, especially the reverse ones, we make an assumption that the streams we are working with are finite, which is sufficient for our tasks. Thus, our streams of tags (context values) can be bounded between `bod` and `eod` constructs (similarly as in intensional databases [360, 367]) and be of a finite tag set type. For the summary of the application of the defined further

| | | | |
|---|---|---|---|
| *bin-op* | ::= | *arith-op* \| *logical-op* \| *bitwise-op* | (7.3.0.52) |
| *un-op* | ::= | + \| - | (7.3.0.53) |
| | | | |
| *arith-op* | ::= | + \| - \| * \| / \| % \| ^ | (7.3.0.54) |
| *logical-op* | ::= | < \| > \| >= \| <= \| == \| in \| && \| \|\| \| ! | (7.3.0.55) |
| *bitwise-op* | ::= | \| \| & \| ~ \| !\| \| !& | (7.3.0.56) |
| | | | |
| *i-bin-op* \| *i-bin-op$_w$* | ::= | @ \| *i-bin-op-forw* \| *i-bin-op-back* | (7.3.0.57) |
| | \| | *i-logic-bitwise-op* \| *i-forensic-op* | (7.3.0.58) |
| | | | |
| *i-bin-op-forw* | ::= | fby \| upon \| asa \| wvr | (7.3.0.59) |
| | \| | nfby \| nupon \| nasa \| nwvr | (7.3.0.60) |
| | | | |
| *i-bin-op-back* | ::= | pby \| rupon \| ala \| rwvr | (7.3.0.61) |
| | \| | npby \| nrupon \| nala \| nrwvr | (7.3.0.62) |
| | | | |
| *i-logic-bitwise-op* | ::= | and \| or \| xor | (7.3.0.63) |
| | \| | nand \| nor \| nxor | (7.3.0.64) |
| | \| | band \| bor \| bxor | (7.3.0.65) |
| | | | |
| *i-un-op* \| *i-un-op$_w$* | ::= | # \| *i-bin-un-forw* \| *i-bin-un-back* | (7.3.0.66) |
| | | | |
| *i-bin-un-forw* | ::= | first \| next \| iseod | (7.3.0.67) |
| | \| | second \| nnext \| neg \| not | (7.3.0.68) |
| | | | |
| *i-bin-un-back* | ::= | last \| prev \| isbod | (7.3.0.69) |
| | \| | prelast \| nprev | (7.3.0.70) |
| | | | |
| *cxtop* | ::= | \isSubContext \| \difference | (7.3.0.71) |
| | \| | \intersection \| \projection | (7.3.0.72) |
| | \| | \hiding \| \override \| \union \| \in | (7.3.0.73) |
| *i-forensic-op* | ::= | combine \| product \| bel \| pl | (7.3.0.74) |

Figure 44: Concrete FORENSIC LUCID syntax (operators) [304, 305]

operators' examples [300, 305], please refer to Section 7.3.2.3, page 183.

A less mundane version of the core intensional operators is added to account for the belief mass assignments $w$ in observations (denoted as $\texttt{op}_w$ in the examples, but syntactically they are expressed identically to the classical versions in a FORENSIC LUCID program). These augmented operator versions act just as normal operators would, but with the additional requirement $w \geq 1/2$. Such behavior is only defined over the forensic contexts when a run-time determination is made over the context type used. When $w < 1/2$, $\texttt{eod}$ is returned. When $w = 1$, such operators behave using their classical sense. It is possible to mix forensic and non-forensic contexts, in such a case the non-forensic contexts are lifted to forensics contexts (see the semantics in Section 7.4) before the application of the operator.

**Definition 14.** *Let $X$ be a* FORENSIC LUCID *stream of natural numbers (Section 4.1.2, page 82) and $O_X$ is a stream of observations of $X$ (effectively,* lifting *every element of $X$ to an observation; additionally a stream of observations is in effect an observation sequence as we shall see), where $\min = 1, \max = 0, w = 1.0$. Let $Y$ be another stream of Boolean values;* true *is cast to 1 and* false *to 0 when used together with $X$ in one stream. $O_Y$ is a stream of observations of $Y$.*

$$X = (x_0, x_1, \ldots, x_i, \ldots)$$
$$= (0, 1, \ldots, i, \ldots)$$
$$Y = (y_0, y_1, \ldots, y_i, \ldots)$$
$$= (true, false, \ldots, true, \ldots)$$
$$O_X = (o_0(x_0, 1, 0, 1.0), o_1(x_1, 1, 0, 1.0), \ldots, o_i(x_i, 1, 0, 1.0), \ldots)$$
$$= (o_0(0, 1, 0, 1.0), o_1(1, 1, 0, 1.0), \ldots, o_i(i, 1, 0, 1.0), \ldots)$$
$$O_Y = (o_0(y_0, 1, 0, 1.0), o_1(y_1, 1, 0, 1.0), \ldots, o_i(y_i, 1, 0, 1.0), \ldots)$$

**Definition 15.** *$first$: a stream of the first element of the argument stream.*

$$\texttt{first } X = (x_0, x_0, \ldots, x_0, \ldots)$$
$$\texttt{first}_w \ O_X = \texttt{first } O_X \ \&\&\ o_0.w \geq 1/2 = (o_0, o_0, \ldots, o_0, \ldots)$$

**Definition 16.** `second`: *a stream of the second element of the argument stream.*

$$\texttt{second } X = (x_1, x_1, \ldots, x_1, \ldots)$$
$$\texttt{second}_w\, O_X = \texttt{second } O_X \,\&\&\, o_1.w \geq 1/2 = (o_1, o_1, \ldots, o_1, \ldots)$$

**Definition 17.** `last`: *a stream of the last element of the argument stream.*

$$\texttt{last } X = (x_n, x_n, \ldots, x_n, \ldots)$$
$$\texttt{last}_w\, O_X = (o_n, o_n, \ldots, o_n, \ldots)$$

This informal definition of the `last` operator relies on the earlier stated assumption that a lot of our streams can be explicitly finite for the language we designed. This affects some of the follow-up operators that rely in that fact just as well. `last` works with `finite` tag sets.

**Definition 18.** `prelast`: *a stream of elements one before the last one of the argument stream.*

$$\texttt{prelast } X = (x_{n-1}, x_{n-1}, \ldots, x_{n-1}, \ldots)$$
$$\texttt{prelast}_w\, O_X = (o_{n-1}, o_{n-1}, \ldots, o_{n-1}, \ldots)$$

**Definition 19.** `next`: *a stream of elements of the argument stream after the first.*

$$\texttt{next } X = (x_1, x_2, \ldots, x_{i+1}, \ldots)$$
$$\texttt{second } X = \texttt{first next } X$$
$$\texttt{next}_w\, O_X = \texttt{second } O_X \,\&\&\, o_i.w \geq 1/2 = (o_1, o_2, \ldots, o_n, \ldots)$$
$$\texttt{second}_w\, O_X = \texttt{first}_w\, \texttt{next}_w\, O_X$$

**Definition 20.** *prev*: *a stream of elements of the argument stream before the last.*

$$\text{prev } X = (x_{n-1}, \ldots, x_{i+1}, x_i, x_{i-1}, \ldots)$$

$$\text{prelast } X = \text{first prev } X$$

$$\text{prev}_w \; O_X = (o_{n-1}, \ldots, o_{i+1}, o_i, o_{i-1}, \ldots)$$

$$\text{prelast}_w \; O_X = \text{first}_w \; \text{prev}_w \; O_X$$

**Definition 21.** *fby*: *the first element of the first argument stream followed by all of the second argument stream.*

$$X \text{ fby } Y = (x_0, y_0, y_1, \ldots, y_{i-1}, \ldots)$$

$$O_X \text{ fby}_w \; Y = O_X \text{ fby } (o_i(Y, 1, 0, 1.0))$$

$$O_X \text{ fby}_w \; O_Y = O_X \text{ fby } (O_Y \text{ wvr } O_Y.w \geq 1/2)$$

$$X \text{ fby}_w \; O_Y = (o_0(X, 1, 0, 1.0)) \text{ fby } (O_Y.P \text{ wvr } O_Y.w \geq 1/2)$$

In the definitions above mixing $O_X$ and $Y$ shows lifting of $Y$ to a default observation dimension per semantics described in Section 7.4, page 192. A similar technique applies to all FORENSIC LUCID intensional operators, so we omit the repetition in subsequent definitions.

**Definition 22.** *pby*: *the first element of the first argument preceded by all of the second.*

$$X \text{ pby } Y = (y_0, y_1, \ldots, y_{i-1}, \ldots, y_n, x_0)$$

$$O_X \text{ pby}_w \; O_Y = O_X \text{ pby } (O_Y.P \text{ wvr } O_Y.w \geq 1/2)$$

$$\ldots = \ldots$$

**Definition 23.** *neg*: *a stream of negated arithmetic values of the argument.*

$$\text{neg } X = (-x_0, -x_1, -x_2, \ldots, -x_{i+1}, \ldots)$$

**Definition 24.** *not*: *a stream of inverted truth values of the argument.*

$$\text{not } X = (!x_0, !x_1, !x_2, ..., !x_{i+1}, ...)$$

**Definition 25.** *and*: *a logical AND stream of truth values of its arguments.*

$$X \text{ and } Y = (x_0 \text{ \&\& } y_0, x_1 \text{ \&\& } y_1, x_2 \text{ \&\& } y_2, \ldots, x_{i+1} \text{ \&\& } y_{i+1}, \ldots)$$

**Definition 26.** *or*: *a logical OR stream of truth values of its arguments.*

$$X \text{ or } Y = (x_0 \text{ || } y_0, x_1 \text{ || } y_1, x_2 \text{ || } y_2, \ldots, x_{i+1} \text{ || } y_{i+1}, \ldots)$$

**Definition 27.** *xor*: *a logical XOR stream of truth values of its arguments.*

$$X \text{ xor } Y = (x_0 \oplus y_0, x_1 \oplus y_1, x_2 \oplus y_2, \ldots, x_{i+1} \oplus y_{i+1}, \ldots)$$

**Definition 28.** *wvr* (*stands for* whenever): *wvr chooses from its left-hand-side operand only values in the current dimension where the right-hand-side evaluates to* true.

$$X \text{ wvr } Y =$$
$$\text{if first } Y \neq 0$$
$$\text{then } X \text{ fby } (\text{next } X \text{ wvr next } Y)$$
$$\text{else } (\text{next } X \text{ wvr next } Y)$$
$$O_X \text{ wvr}_w O_Y = O_X \text{ wvr } (O_Y.P \text{ \&\& } O_Y.w \geq 1/2)$$
$$\ldots = \ldots$$

**Definition 29.** *rwvr* (*stands for* retreat whenever): *rwvr chooses from its left-hand-side operand backwards only values in the current dimension where the right-hand-side evaluates to* true.

$$X \text{ rwvr } Y =$$
$$\textbf{if } \texttt{last } Y \neq 0$$
$$\textbf{then } X \text{ pby } (\text{prev } X \text{ rwvr prev } Y)$$
$$\textbf{else } (\text{prev } X \text{ rwvr prev } Y)$$
$$O_X \text{ rwvr}_w O_Y = O_X \text{ rwvr } (O_Y.P \text{ \&\& } O_Y.w \geq 1/2)$$
$$\ldots = \ldots$$

**Definition 30.** *nwvr (stands for* not whenever*): nwvr chooses from its left-hand-side operand only values in the current dimension where the right-hand-side evaluates to* false.

$$X \text{ nwvr } Y = X \text{ wvr not } Y =$$
$$\textbf{if } \texttt{first } Y == 0$$
$$\textbf{then } X \text{ fby } (\text{next } X \text{ nwvr next } Y)$$
$$\textbf{else } (\text{next } X \text{ nwvr next } Y)$$
$$O_X \text{ nwvr}_w O_Y = O_X \text{ nwvr } (O_Y.P \text{ \&\& } O_Y.w \geq 1/2)$$
$$\ldots = \ldots$$

**Definition 31.** *nrwvr (stands for* not to retreat whenever*): nrwvr chooses from its left-hand-side operand backwards only values in the current dimension where the right-hand-side evaluates to* false.

$$X \text{ nrwvr } Y = X \text{ rwvr not } Y =$$
$$\textbf{if } \texttt{last } Y == 0$$
$$\textbf{then } X \text{ pby } (\text{prev } X \text{ nrwvr prev } Y)$$
$$\textbf{else } (\text{prev } X \text{ nrwvr prev } Y)$$
$$O_X \text{ nrwvr}_w O_Y = O_X \text{ nrwvr } (O_Y.P \text{ \&\& } O_Y.w \geq 1/2)$$
$$\ldots = \ldots$$

**Definition 32.** *asa (stands for* as soon as*): asa returns the value of its left-hand-side as a first point in that stream as soon as the right-hand-side evaluates to* true.

$$X \text{ asa } Y = \text{first } (X \text{ wvr } Y)$$

$$O_X \text{ asa}_w O_Y = \text{first}_w (O_X \text{ wvr}_w O_Y)$$

$$\ldots = \ldots$$

**Definition 33.** *ala (stands for* as late as *(or reverse of* as soon as*)): ala returns the value of its left-hand-side as the last point in that stream when the right-hand-side evaluates to* true *for the last time.*

$$X \text{ ala } Y = \text{last } (X \text{ wvr } Y)$$

$$O_X \text{ ala}_w O_Y = \text{last}_w (O_X \text{ wvr}_w O_Y)$$

$$\ldots = \ldots$$

**Definition 34.** *nasa (stands for* not as soon as*): nasa returns the value of its left-hand-side as a first point in that stream as soon as the right-hand-side evaluates to* false.

$$X \text{ nasa } Y = \text{first } (X \text{ nwvr } Y)$$

$$O_X \text{ nasa}_w O_Y = \text{first}_w (O_X \text{ nwvr}_w O_Y)$$

$$\ldots = \ldots$$

**Definition 35.** *nala (stands for* not as late as *(or reverse of* not a soon as*)): nala returns the value of its left-hand-side as the last point in that stream when the right-hand-side evaluates to* false *for the last time.*

$$X \text{ nala } Y = \text{last } (X \text{ nwvr } Y)$$

$$O_X \text{ nala}_w Y = \text{last}_w (O_X \text{ nwvr}_w O_Y)$$

$$\ldots = \ldots$$

**Definition 36.** *upon* (stands for advances upon): unlike `asa`, *upon* switches context of its left-hand-side operand if the right-hand side is true.

$$X \text{ upon } Y = X \text{ fby } ($$
$$\textbf{if } \texttt{first } Y \neq 0$$
$$\textbf{then } (\texttt{next } X \text{ upon next } Y)$$
$$\textbf{else } (X \text{ upon next } Y))$$
$$O_X \text{ upon}_w O_Y = O_X \text{ upon } (O_Y.P \text{ \&\& } O_Y.w \geq 1/2)$$
$$\ldots = \ldots$$

**Definition 37.** *rupon* (stands for retreats upon): *rupon* switches context backwards of its left-hand-side operand if the right-hand side is true.

$$X \text{ rupon } Y = X \text{ pby } ($$
$$\textbf{if } \texttt{last } Y \neq 0$$
$$\textbf{then } (\texttt{prev } X \text{ rupon prev } Y)$$
$$\textbf{else } (X \text{ rupon prev } Y))$$
$$O_X \text{ rupon}_w O_Y = O_X \text{ rupon } (O_Y.P \text{ \&\& } O_Y.w \geq 1/2)$$
$$\ldots = \ldots$$

**Definition 38.** *nupon* (stands for not advances upon, or, rather advances otherwise): *nupon* switches context of its left-hand-side operand if the right-hand side is false.

$$X \text{ nupon } Y = X \text{ upon not } Y = X \text{ fby } ($$
$$\textbf{if } \texttt{first } Y == 0$$
$$\textbf{then } (\texttt{next } X \text{ nupon next } Y)$$
$$\textbf{else } (X \text{ nupon next } Y))$$
$$O_X \text{ nupon}_w O_Y = O_X \text{ nupon } (O_Y.P \text{ \&\& } O_Y.w \geq 1/2)$$
$$\ldots = \ldots$$

**Definition 39.** *nrupon (stands for not retreats upon): **nrupon** switches context backwards of its left-hand-side operand if the right-hand side is false.*

$$X \text{ nrupon } Y = X \text{ rupon not } Y = X \text{ pby } ($$
$$\text{if last } Y == 0$$
$$\text{then } (\text{prev } X \text{ nrupon prev } Y)$$
$$\text{else } (X \text{ nrupon prev } Y))$$
$$O_X \text{ nrupon}_w O_Y = O_X \text{ nrupon } (O_Y.P \text{ \&\& } O_Y.w \geq 1/2)$$
$$\ldots = \ldots$$

**Definition 40.** *"." (dot): is a scope membership navigation operator.*

The "." operator is employed in multiple uses, so we include it into our syntax and semantics.

- From indexical expressions [351, 361] with the operators defined in the preceding and following sections it facilitates operations on parts of the evaluation context to denote the specific dimensions of interest to work with.

- From JOOIP/OBJECTIVE LUCID expressions, . allows nested class/object membership access.

- Additional use in FORENSIC LUCID (inspired from MARFL) is the contextual depth navigation, similar in a way to the OO membership navigation, such as $ES.OS.O.P$ (see summary in Table 15).

**Definition 41.** *#: queries the current context of evaluation.*

Traditionally ([361], Section 4.1.3, page 85) # is defined as:

$$\# = 0 \text{ FBY } (\# + 1)$$

for ℕ tag set, that were infinite and ordered, essentially being akin to array indices into the current dimension $X$, and #$X$ returned the current implicit index within $X$. Subsequently,

$$\text{\#}.E = 0 \text{ fby}.E(\text{\#}+1)$$
$$\text{\#}C = C$$
$$\text{\#}S = S$$
$$\text{\#}F = F$$
$$\text{\#}.F = 0 \text{ fby}.F.(\text{\#}+1)$$
$$\text{\#}_w \equiv \text{\#}$$

#.$E$ was introduced to allow querying any part of the current multidimensional context where $E$ evaluates to a dimension, a part of the context. $C$ and $S$ are Lucx simple contexts and context sets.

Since the tag set types were first augmented in Lucx to contain arbitrary enumerable tag set values (cf. Section 7.2.3.1.5, page 165, [473]), such as strings, or non-numerically ordered sequences, # is augmented to support such tag sets by returning the current tag of any type in the current dimension indexing them in the order of their declaration in a Forensic Lucid program.

In Forensic Lucid, # is augmented to support querying hierarchical forensic contexts $F$. The indexing mechanism is same as in Lucx Thus, for an evidential statement $ES$, # $ES$ returns the current observation sequence $OS$, which is a tag in $ES$; likewise # $OS$ returns the current observation $O$, which is a tag in $OS$; # $O$ returns the current tuple $\langle P, \min, \max, w, t \rangle$. When combined with the Forensic Lucid dot . operator: $ES.\text{\#}$ returns the current $OS$, $ES.\text{\#}.\text{\#}$ returns the current $O$, $ES.\text{\#}.\text{\#}.P$ returns the current $O.P$, and so on. $\text{\#}_w$ is equivalent to # as it is a query and not a navigation operator, so it does not take $w$ into account when querying the current context.

**Definition 42.** `@` : *switches the current context of evaluation.*

Traditionally ([361], Section 4.1.3, page 85) `@` is defined as:

$$X \ @ \ Y = \text{if } Y = 0 \text{ then } \textsc{first } X$$
$$\text{else } (\textsc{next } X) \ @ \ (Y - 1)$$

for $\mathbb{N}$ tag set in a single dimension. Subsequently,

$$X \ @ \ .dE$$
$$X \ @ \ C$$
$$X \ @ \ S$$
$$X \ @_w \ F$$
$$X \ @_w \ F.F$$

are also provided. $X \ @ \ .dE$ is the indexical way to work with parts of the context [361]; $C$ and $S$ are the simple contexts and context sets [513], and $F$ are forensic contexts.

Augmentation to Lucx's implementation extension of the tag sets [473] has the same underlying principles as `#` in the preceding section: the order of declaration of the tags in a tag set becomes and implicit index in that tag collection, and a corresponding mapping of custom tags produces the desired effect with `@`. For example, if an unordered finite tag set of a dimension `color` has three strings it ranges over, {"red","green","blue"}, it is legal to say, e.g., `c@[color : "green"]`, where internally, "green" is indexed 1 in the order of declaration in the tag set, and a mapping exists "green" $\leftrightarrow$ 1. "Unordered" in this sense simply means the order in which the tags are declared is not important, but such a mapping is necessary for operators defined earlier (`first`, `next`, `fby`, etc.) to work intuitively and consistently with the classical $\mathbb{N}$ ones. (As an analogy, this is similar to the SQL standard and the way results are returned in a `SELECT` query without the `ORDER BY` clause for subsequent processing by the application.)

@ is likewise adapted to work with forensic contexts yielding different types of forensic contexts depending on its types of arguments, as, e.g., illustrated in Table 15. Unlike $\#_w$, $@_w$ is does take $w$ into the account when switching forensic contexts:

$$X \ @_w \ O = X \ @ \ (O \ \&\& \ O.w \geq 1/2),$$

similarly to the previously presented operators. For observation sequences and evidential statements, $@_w$ is more elaborate:

$$X \ @_w \ OS = X \ @ \ (OS \ \&\& \ \texttt{bel} \ OS \geq 1/2)$$
$$X \ @_w \ ES = X \ @ \ (ES \ \&\& \ \texttt{bel} \ ES \geq 1/2)$$

where the belief `bel` is defined in Section 7.3.4, page 191.

### 7.3.2.2 Definition of Core Operators by Means of @ and #

This step follows the same tradition as the most of the SIPLs in GIPSY (Chapter 6), which are translated into GIPL (Section 4.1.1.2, page 78). The existing rewrite rules as well as the new rules are applied here for the majority of our operators to benefit from the existing interpretation available in GEE (Section 6.2.2, page 142) for evaluation purposes. This translation may also be helpful to the developers of similar systems other than GIPSY. Following the steps similar to Paquet's [361], we further represent the definition of the operators via @ and # (this includes the $@_w$ and $\#_w$ interpretations of the operators). Again, there is a mix of classical operators that were previously defined in [350, 361], such as `first`, `next`, `fby`, `wvr`, `upon`, and `asa` as well as the new operators from this work [304]. The collection of the translated operators is summarized in Figure 45.

For the same reasons of (im)practicality as in the PoC implementation of LUCX's parser and semantic analyzer [473], we don't translate some FORENSIC LUCID constructs however. Additionally, not all translations are currently possible, such as those dealing with the credibility factors. Therefore, for these we define a new set of operational semantic rules presented further in Section 7.4, page 192. Instead, dedicated interpreter plug-ins are designed to work directly with the untranslated constructs, such as the forensic contexts and their operators. This is useful, e.g., for various interpretation backends, such as eductive vs. PRISM, and so on (Chapter 8).

The primitive operators are founding blocks to construct more complex case-specific functions that represent a particular investigation case as well as more elaborate *forensic operators* [300].

#### 7.3.2.3 Summary of the Core Operators' Examples

Here we illustrate a few basic examples of application of the FORENSIC LUCID operators (both, classical LUCID and the newly introduced operators). Assume we have two bounded (between `bod` and `eod`) streams $X$ and $Y$ of ten elements. The $X$ stream is just an ordered sequence of natural numbers between 1 and 10. If queried for values below 1 an beginning-of-data (`bod`) marker would be returned; similarly if queried beyond 10, the end-of-data marker (`eod`) is returned. The $Y$ stream is a sequence of ten truths values (can be replaced with 0 for "false" and 1 for "true"). The operators applied to these streams may return bounded or unbounded streams of the same or different length than the original depending on the definition of a particular operator. Also assume the current tag is 0 ($\#\texttt{X} = 0, \#\texttt{Y} = 0$). The resulting table showing the application of the classical and the new operators is in Table 14 [300, 305].

### 7.3.3 Context Calculus Operators

What follows are the definitions of the core context calculus operators: `\isSubContext`, `\difference`, `\intersection`, `\projection`, `\hiding`, `\override`, and `\union` based on LUCX and its compiler implementation [473, 513] and expanded to include forensic contexts and tag sets through lifting. The context and context set operations defined below are the LUCX original definitions [473, 513] provided for completeness. Their forensic context and tag set versions are newly defined. While the concept of tag sets (Section 7.2.3.1.5, page 165) is explored in detail at the implementation level by Tong [473], the operators on this section were never defined on them. We define these operators to compose a complete picture. This is achieved through the concept of *lifting* introduced by Guha [148] applied to contextual knowledge to allow the use of known formulas from one context type or its portion to another. We also add the convenient \in operator. Through lifting simple contexts are converted to observations and context sets to observation sequences when required.

Table 14: Example of application of FORENSIC LUCID operators to bounded streams [305]

| stream/index | -1 | 0 | 1 | 2 | 3 | 4 | 5 | 6 | 7 | 8 | 9 | 10 | 11 |
|---|---|---|---|---|---|---|---|---|---|---|---|---|---|
| X | bod | 1 | 2 | 3 | 4 | 5 | 6 | 7 | 8 | 9 | 10 | eod | eod |
| Y | bod | T | F | F | T | F | F | T | T | F | T | eod | eod |
| first X | | 1 | 1 | 1 | 1 | 1 | 1 | 1 | 1 | 1 | 1 | | |
| last X | | 10 | 10 | 10 | 10 | 10 | 10 | 10 | 10 | 10 | 10 | | |
| next X | | | 2 | 3 | 4 | 5 | 6 | 7 | 8 | 9 | 10 | eod | eod |
| prev X | | bod | | | | | | | | | | | |
| X fby Y | | 1 | T | F | F | T | F | F | T | T | F | T | eod |
| X pby Y | | T | F | F | T | F | F | T | T | F | T | 1 | eod |
| X wvr Y | | 1 | | | 4 | | | 7 | 8 | | 10 | | |
| X rwvr Y | | 10 | | | 8 | | | 7 | 4 | | 1 | | |
| X nwvr Y | | | 2 | 3 | | 5 | 6 | | | 9 | | | |
| X nrwvr Y | | | 9 | 6 | | 5 | 3 | | | 2 | | | |
| X asa Y | | 1 | 1 | 1 | 1 | 1 | 1 | 1 | 1 | 1 | 1 | | |
| X nasa Y | | 2 | 2 | 2 | 2 | 2 | 2 | 2 | 2 | 2 | 2 | | |
| X ala Y | | 10 | 10 | 10 | 10 | 10 | 10 | 10 | 10 | 10 | 10 | | |
| X nala Y | | 9 | 9 | 9 | 9 | 9 | 9 | 9 | 9 | 9 | 9 | | |
| X upon Y | | 1 | 2 | 2 | 2 | 3 | 3 | 3 | 4 | 5 | 5 | eod | |
| X rupon Y | | 10 | 9 | 9 | 8 | 7 | 7 | 7 | 6 | 6 | 6 | bod | |
| X nupon Y | | 1 | 1 | 2 | 3 | 3 | 4 | 5 | 5 | 5 | 6 | 6 | eod |
| X nrupon Y | | 10 | 10 | 9 | 9 | 9 | 8 | 7 | 7 | 6 | 5 | 5 | bod |
| neg X | | -1 | -2 | -3 | -4 | -5 | -6 | -7 | -8 | -9 | -10 | eod | eod |
| not Y | | F | T | T | F | T | T | F | F | T | F | eod | eod |
| X and Y | | 1 | 0 | 0 | 1 | 0 | 0 | 1 | 1 | 0 | 1 | eod | eod |
| X or Y | | 1 | 2 | 3 | 5 | 5 | 6 | 7 | 9 | 9 | 11 | eod | eod |
| X xor Y | | 0 | 2 | 3 | 5 | 5 | 6 | 6 | 9 | 9 | 11 | eod | eod |

**Definition 43.** \isSubContext

- If $C_1$ and $C_2$ are simple contexts and every micro context of $C_1$ is also a micro context of $C_2$, then $C_1$ \isSubContext $C_2$ returns **true**. An empty simple context is the sub-context of any simple context [473]. As in the concept of subset in set theory, $C_1 \subseteq C_2$ [473].

- If $S_1$ and $S_2$ are context sets, then $S_1$ \isSubContext $S_2$ returns **true** if every simple

context of $S_1$ is also a simple context of $S_2$. An empty context set is the sub-context of any context set [473]. Likewise, $S_1 \subseteq S_2$.

- If $F_1$ and $F_2$ are forensic contexts, then $F_1$ `\isSubContext` $F_2$ returns `true` if every nested context of $F_1$ is also a nested context of $F_2$.

**Definition 44.** `\difference`

- If $C_1$ and $C_2$ are simple contexts, then $C_1$ `\difference` $C_2$ returns a simple context that is a collection of all micro contexts that are members of $C_1$, but not members of $C_2$: $C_1$ `\difference` $C_2 = \{m_i \mid m_i \in C_1 \wedge m_i \notin C_2\}$ [473]. If $C_1$ `\isSubContext` $C_2$ is `true`, then the returned simple context is `empty`. `\difference` is also valid if two simple contexts have no common micro context; the returned simple context is simply $C_1$ [473].

- If $S_1$ and $S_2$ are context sets, `\difference` returns a context set $S$, where every simple context $C \in S$ is computed as $C_1$ `\difference` $C_2$ : $S = S_1$ `\difference` $S_2 = \{C \mid C = C_1$ `\difference` $C_2 \wedge C \neq \emptyset \wedge C_1 \in S_1 \wedge C_2 \in S_2\} \vee S = \emptyset$ [473]. If for every $C_1$ and $C_2$, $C_1$ `\difference` $C_2 = \emptyset$, then $S_1$ `\difference` $S_2 = \emptyset$ [473]. However, if there's at least one pair of $C_1$ and $C_2$ where $C_1$ `\difference` $C_2 \neq \emptyset$, then the result is not empty [473].

- If $T_1$ and $T_2$ are finite tag sets, then $T_1$ `\difference` $T_2$ returns a tag set that is a collection of all tags that are members of $T_1$, but not members of $T_2$: $T_1$ `\difference` $T_2 = \{t_i \mid t_i \in T_1 \wedge t_i \notin T_2\}$. If $T_1$ `\in` $T_2$ is `true` (`\in` is defined further on page 190), then the returned tag set is `empty`. `\difference` is also valid if two tag sets have no common tag; the returned tag set is then $T_1$.

- If $F_1$ and $F_2$ are forensic contexts, `\difference` returns $F$, where every nested context $C \in F$ is computed as $C_1$ `\difference` $C_2$ : $F = F_1$ `\difference` $F_2 = \{C \mid C = C_1$ `\difference` $C_2 \wedge C \neq \emptyset \wedge C_1 \in F_1 \wedge C_2 \in F_2\} \vee F = \emptyset$.

$$
\begin{align}
\text{first } X &= X@0 & (7.3.2.1)\\
\text{last } X &= X@(\#@(\#\text{iseod}(\#)-1)) & (7.3.2.2)\\
\text{next } X &= X@(\#+1) & (7.3.2.3)\\
\text{prev } X &= X@(\#-1) & (7.3.2.4)\\
X \text{ fby } Y &= \text{if } \#=0 \text{ then } X \text{ else } Y@(\#-1) & (7.3.2.5)\\
X \text{ pby } Y &= \text{if iseod } \# \text{ then } X \text{ else } Y@(\#+1) & (7.3.2.6)\\
X \text{ wvr } Y &= X@T \text{ where} & (7.3.2.7)\\
&\quad T = U \text{ fby } U@(T+1)\\
&\quad U = \text{if } Y \text{ then } \# \text{ else next } U\\
&\quad \text{end}\\
X \text{ rwvr } Y &= X@T \text{ where} & (7.3.2.8)\\
&\quad T = U \text{ pby } U@(T-1)\\
&\quad U = \text{if } Y \text{ then } \# \text{ else prev } U\\
&\quad \text{end}\\
X \text{ nwvr } Y &= X@T \text{ where} & (7.3.2.9)\\
&\quad T = U \text{ fby } U@(T+1)\\
&\quad U = \text{if } Y == 0 \text{ then } \# \text{ else next } U\\
&\quad \text{end}\\
X \text{ rnwvr } Y &= X@T \text{ where} & (7.3.2.10)\\
&\quad T = U \text{ pby } U@(T-1)\\
&\quad U = \text{if } Y == 0 \text{ then } \# \text{ else prev } U\\
&\quad \text{end}\\
X \text{ asa } Y &= \text{first } (X \text{ wvr } Y) & (7.3.2.11)\\
X \text{ nasa } Y &= \text{first } (X \text{ nwvr } Y) & (7.3.2.12)\\
X \text{ ala } Y &= \text{last } (X \text{ rwvr } Y) & (7.3.2.13)\\
X \text{ nala } Y &= \text{last } (X \text{ nrwvr } Y) & (7.3.2.14)\\
X \text{ upon } Y &= X@W \text{ where} & (7.3.2.15)\\
&\quad W = 0 \text{ fby } (\text{if } Y \text{ then } (W+1) \text{ else } W)\\
&\quad \text{end}\\
X \text{ rupon } Y &= X@W \text{ where} & (7.3.2.16)\\
&\quad W = 0 \text{ pby } (\text{if } Y \text{ then } (W-1) \text{ else } W)\\
&\quad \text{end}\\
X \text{ nupon } Y &= X@W \text{ where} & (7.3.2.17)\\
&\quad W = 0 \text{ fby } (\text{if } Y == 0 \text{ then } (W+1) \text{ else } W)\\
&\quad \text{end}\\
X \text{ nrupon } Y &= X@W \text{ where} & (7.3.2.18)\\
&\quad W = 0 \text{ pby } (\text{if } Y == 0 \text{ then } (W-1) \text{ else } W)\\
&\quad \text{end}\\
\text{neg } X &= -X & (7.3.2.19)\\
\text{not } X &= \text{if } X \text{ then } !X \text{ else } X & (7.3.2.20)\\
X \text{ and } Y &= X \&\& Y & (7.3.2.21)\\
X \text{ or } Y &= X || Y & (7.3.2.22)\\
X \text{ xor } Y &= \text{not}((X \text{ and } Y) \text{ or not } (X \text{ or } Y)) & (7.3.2.23)
\end{align}
$$

Figure 45: Operators translated to GIPL-compatible definitions [304, 305]

**Definition 45.** $\backslash intersection$

- If $C_1$ and $C_2$ are simple contexts, then $C_1 \backslash intersection\ C_2$ returns a new simple context, which is the collection of those micro contexts that belong to both $C_1$ and $C_2$: $C_1 \backslash intersection\ C_2 = \{m_i \mid m_i \in C_1 \land m_i \in C_2\}$. If $C_1$ and $C_2$ have no common micro context, the result is an *empty* simple context [473].

- If $S_1$ and $S_2$ are context sets, then the resulting intersection context set is defined as $S = S_1 \backslash intersection\ S_2 = \{C \mid C = C_1 \backslash intersection\ C_2 \land C \neq \emptyset \land C_1 \in S_1 \land C_2 \in S_2\} \lor S = \emptyset$. If for every $C_1$ and $C_2$, $C_1 \backslash intersection\ C_2 = \emptyset$, then $S_1 \backslash intersection\ S_2 = \emptyset$. However, if there's at least one pair of $C_1$ and $C_2$ where $C_1 \backslash intersection\ C_2 \neq \emptyset$, the result is not empty [473].

- If $T_1$ and $T_2$ are finite tag sets, then $T_1 \backslash intersection\ T_2$ returns a new unordered tag set, which is the collection of tags that belong to both $T_1$ and $T_2$: $T_1 \backslash intersection\ T_2 = \{t_i \mid t_i \in T_1 \land t_i \in T_2\}$. If $T_1$ and $T_2$ have no common tag, the result is an *empty* tag set.

- If $F_1$ and $F_2$ are forensic contexts, then the resulting intersection of the nested contexts $F = F_1 \backslash intersection\ F_2 = \{C \mid C = C_1 \backslash intersection\ C_2 \land F \neq \emptyset \land C_1 \in F_1 \land C_2 \in F_2\} \lor F = \emptyset$.

**Definition 46.** $\backslash projection$

- If $C$ is a simple context and $D$ is a set of dimensions, this operator filters only those micro contexts in $C$ that have their dimensions in set $D$ : $C \backslash projection\ D = \{m \mid m \in C \land dim(m) \in D\}$. The result is *empty*, if there's no micro context having the same dimension as in the dimension set. *dim(m)* returns the dimension of micro context *m* [473].

- The projection of a dimension set onto a context set is a context set, which is a collection of all simple contexts having a $\backslash projection$ on the dimension set [473]. If $S$ is a context set, $D$ is a dimension set; $S' = S \backslash projection\ D = \{n \mid n = C \backslash projection\ D \land n \neq \emptyset \land C \in S\} \lor S' = \emptyset$.

- If $C$ is a simple context and $T$ is a finite set of tags, this operator filters only those micro contexts in $C$ that have their tags in set $T$: $C \text{ \textbackslash projection } T = \{m \mid m \in C \land tag(m) \in T\}$. The result is **empty**, if there's no micro context having the same tag as in the tag set. **tag(m)** returns the tag of micro context **m**. The projection of a tag set onto a context set is a context set, which is a collection of all simple contexts having a \projection on the tag set. If $S$ is a context set, $S' = S \text{ \textbackslash projection } T = \{n \mid n = C \text{ \textbackslash projection } T \land n \neq \emptyset \land C \in S\} \lor S' = \emptyset$.

- Projection of a dimension set onto a forensic context is a forensic context containing all nested contexts filtered through the \projection on the dimension set. If $F$ is a forensic context, $D$ is a dimension set; $F' = F \text{ \textbackslash projection } D = \{n \mid n = C \text{ \textbackslash projection } D \land n \neq \emptyset \land C \in F\} \lor F' = \emptyset$.

**Definition 47.** \hiding

- If $C$ is a simple context and $D$ is a dimension set, this operator removes all the micro contexts in $C$ whose dimensions are in $D$: $C \text{ \textbackslash hiding } D = \{m \mid m \in C \land dim(m) \notin D\}$. If $D$ contains all the dimensions appeared in $C$, the result is an empty simple context. Additionally, $C \text{ \textbackslash projection } D \bigcup C \text{ \textbackslash hiding } D = C$ [473].

- For context set $S$, and dimension set $D$, the \hiding operator constructs a context set $S'$ where $S'$ is obtained by \hiding each simple context in $S$ on the dimension set $D$: $S' = S \text{ \textbackslash hiding } D = \{n \mid n = C \text{ \textbackslash hiding } D \land n \neq \emptyset \land C \in S\} \lor S' = \emptyset$.

- If $C$ is a simple context and $T$ is a finite tag set, this operator removes all the micro contexts in $C$ whose tags are in $T$: $C \text{ \textbackslash hiding } T = \{m \mid m \in C \land tag(m) \notin T\}$. If $T$ contains all the tags appearing in $C$, the result is an empty simple context. For context set $S$, the \hiding operator constructs a context set $S'$ where $S'$ is obtained by \hiding each simple context in $S$ on the tag set $T$: $S' = S \text{ \textbackslash hiding } T = \{n \mid n = C \text{ \textbackslash hiding } T \land n \neq \emptyset \land C \in S\} \lor S' = \emptyset$.

- For forensic context $F$ and dimension set $D$, \hiding constructs a forensic context $F'$ by hiding each nested context in $F$ on dimension set $D$: $F' = F \text{ \textbackslash hiding } D = \{n \mid n = C \text{ \textbackslash hiding } D \land n \neq \emptyset \land C \in F\} \lor F' = \emptyset$.

**Definition 48.** `\override`

- If $C_1$ and $C_2$ are simple contexts, then $C_1$ `\override` $C_2$ returns a new simple context $C$, which is the result of the conflict-free union of $C_1$ and $C_2$, as defined as follows: $C = C_1$ `\override` $C_2 = \{m \mid (m \in C_1 \land dim(m) \notin dim(C_2)) \lor m \in C_2\}$ [473].

- For every pair of context sets $S_1, S_2$, `\override` returns a set of contexts $S$, such that every context $C \in S$ is computed as $C_1$ `\override` $C_2$; $C_1 \in S_1, C_2 \in S_2 : S = S_1$ `\override` $S_2 = \{C \mid C = C_1$ `\override` $C_2 \mid C_1 \in S_1 \land C_2 \in S_2 \land C \neq \emptyset\} \lor S = \emptyset$ [473].

- For every pair of forensic contexts $F_1, F_2$, `\override` returns a nested context $F$: $C_1 \in F_1, C_2 \in F_2 : F = F_1$ `\override` $F_2 = \{C \mid C = C_1$ `\override` $C_2 \mid C_1 \in F_1 \land C_2 \in F_2 \land C \neq \emptyset\} \lor F = \emptyset$.

**Definition 49.** `\union`

- If $C_1$ and $C_2$ are simple contexts, then $C_1$ `\union` $C_2$ returns a new simple context $C$, for every micro context $m$ in $C$: $m$ is an element of $C_1$ or $m$ is an element of $C_2$: $C_1$ `\union` $C_2 = \{m \mid m \in C_1 \lor (m \in C_2 \land m \notin C_1)\}$ [473]. If there is at least one pair of micro contexts in $C_1$ and $C_2$ sharing the same dimension and these two micro contexts are not equal then the result is a non-simple context translated into a context set: for a non-simple context $C$, construct the set $Y = \{y_d = C$ `\projection` $\{d\} \mid d \in dim(C)\}$. Denoting the elements of set $Y$ as $y_1, \ldots, y_p$, construct the set $S(C)$ of simple contexts: $S(C) = \{m_1$ `\override` $m_2$ `\override` $\ldots$ `\override` $m_p \mid m_1 \in y_1 \land m_2 \in y_2 \land \ldots m_p \in y_p\}$, The non-simple context is viewed as the set $S(C)$, such that $S(C) = \{s \in S \mid dim(s) = dim(C) \land s \subset C\}$ [473].

- If $C_1$ and $C_2$ are context sets, then $C = C_1$ `\union` $C_2$ is computed as follows [473]:

    1. $D_1 = \{dim(m) \land m \in C_1\}, D_2 = \{dim(m) \land m \in C_2\}, D_3 = D_1 \bigcap D_2$

    2. Compute $X_1 : X_1 = \{m_i \bigcup (m_j$ `\hiding` $D_3) \land m_i \in C_1 \land m_j \in C_2\}$

    3. Compute $X_2 : X_2 = \{m_j \bigcup (m_i$ `\hiding` $D_3) \land m_i \in C_1 \land m_j \in C_2\}$

4. Finally: $C = X_1 \bigcup X_2$

   This procedure ensures contexts sets always have the same well-defined context structure [473, 513].

- If $T_1$ and $T_2$ are finite tag sets of tags of the same type, $T$ is their non-order-preserving union: $T = T_1 \text{\textbackslash union } T_2 = \{t \mid t \in T_1 \vee (t \in T_2 \wedge t \notin T_1)\}$.

- If forensic contexts are observations $O_1$ and $O_2$ their union is an observation sequence $OS$ if the observations are ordered in wall-clock time: $OS = O_1 \text{\textbackslash union } O_2 = \{O_1.t, O_2.t \mid \{O_1, O_2\} \wedge O_1.t < O_2.t\}$. If the observations are conflicting in time (equal, or undefined), the are lifted to an evidential statement containing two constructed observation sequences that have one observation each: $ES = O_1 \text{\textbackslash union } O_2 = \{O_1 \in OS_1, O_2 \in OS_2\}$. If forensic contexts are observation sequences $OS_1$ and $OS_2$, their union is $OS = OS_1 \text{\textbackslash union } OS_2$, i.e., a fusion merge of the two observation sequences, if $OS_1$ and $OS_2$ contain non-conflicting observations $O_i$, ordered chronologically by $O_i.t$ not sharing a common $t$. If observations are conflicting or $O_i.t$ are undefined in any of $OS_1$ or $OS_2$, then the union of such observation sequences is an evidential statement $ES = OS_1 \text{\textbackslash union } OS_2 = \{OS_1, OS_2\}$. If forensic contexts are evidential statements $ES_1$ and $ES_2$, is a simple union of all observation sequences from both statements $ES = ES_1 \text{\textbackslash union } ES_2 = \{os \mid os \in ES_1 \vee (os \in ES_2 \wedge os \notin ES_1)\}$. We admit conflicting observation sequences as they do happen in real-life investigations.

**Definition 50.** $\text{\textbackslash in}$

- If $C_1$ and $C_2$ are simple contexts, then $C_1 \text{\textbackslash in } C_2 = C_1 \text{\textbackslash isSubContext } C_2$.

- If $S_1$ and $S_2$ are context sets, then $S_1 \text{\textbackslash in } S_2 = S_1 \text{\textbackslash isSubContext } S_2$.

- If $F_1$ and $F_2$ are forensic contexts, then $F_1 \text{\textbackslash in } F_2 = F_1 \text{\textbackslash isSubContext } F_2$.

- If $D_1 \subset DIM$ and $D_2 \subset DIM$ are dimension sets, then the operator $D_1 \text{\textbackslash in } D_2$ returns **true** if $D_1 \subset D_2$.

- If $T_1$ and $T_2$ are finite tag sets, then the operator $T_1$ `\in` $T_2$ returns **true** if $T_1 \subseteq T_2$. If $T_1$ is infinite and $T_2$ is finite, `\in` always returns **false**. If $T_2$ is infinite and $T_1$ is either finite or infinite, the set membership can be determined by a function. This last case is of little interest to us any further.

While this operator hasn't appeared at the syntax level for any context types or their parts, at the implementation level Tong provided `isInTagSet()` while developing LUCX parser and semantic analyzer [473].

### 7.3.4 Forensic Context Operators

While the operators presented so far in the preceding sections are designed to support forensics context in addition to their traditional behavior, the operators presented here were designed to work originally with the forensic contexts only specific to FORENSIC LUCID and lift any non-forensic contexts if used.

The operators presented here are based on the discussion of the combination [137] function and others to support the required case implementation [304]. The discussed earlier (see Section 2.2.4.6.2, page 42) Gladyshev's `comb()` operator [136, 137] needs to be realized in the general manner in FORENSIC LUCID for combining our analogies of multiple partitioned runs (MPRs) [136, 137], which in our case are higher-level contexts and context sets, in the new dimension types [300, 304, 305, 312] (see Table 13).

**Definition 51.** `combine` *corresponds to the comb function as originally described by Gladyshev ([137], Section 2.2.4.6.2, page 42).*

**Definition 52.** `product` *corresponds to the cross-product [137] of contexts.*

Belief and plausibility operators `bel` and `pl` are introduced to deal with the evidence evaluation in the Dempster–Shafer terms (Section 3.3.2, page 66). The basic argument to these operators is an observation $o$; in this case they simply evaluate into the $o.w$ component of $o$. `bel` and `pl` and be computed of any two observations $o_1$ and $o_2$ of the same property ($o_1.P = o_2.P$) can be made according $o_1.w$ and $o_2.w$. The latter is a founding block to compute `bel` and `pl` of any number of observations relating to that property.

**Definition 53.** `bel` *corresponds to belief computation based on DSTME (Section 3.3.2, page 66) adapted to forensic contexts as follows: given an observation $O$, `bel` $O = O.w$; if $O = \$$, `bel` $\$ = 1$; if $O = \backslash 0$, `bel` $\backslash 0 = 0$; if $C$ is a simple context or $S$ is a context set, `bel` $C =$ `bel` $S = 1$; given an observation sequence $OS$ and its observations $O_i$, `bel` $OS = \{O_i \in OS \mid \sum_i O_i.w/i\}$; given an evidential statement $ES$ and its subset of interest of observation sequences $ES_B$, `bel` $ES = \sum_{ES_B|ES_B \subseteq ES} w(ES_B)$.*

Thus, the credibility values of $OS$ and $ES$ are derived from the beliefs assigned to the contained observations.

**Definition 54.** `pl` *corresponds to plausibility computation based on DSTME adapted to forensic contexts as follows: given an observation $O$, `pl` $O =$ `bel` $O$; given an observation sequence $OS$ and its observations $O_i$, `pl` $OS =$ `bel` $OS$; given an evidential statement $ES$ and its subset of interest of observation sequences $ES_B$, `bel` $ES = \sum_{ES_B|ES_B \cap ES \neq \emptyset} w(ES_B) = 1 -$ `bel`$(\overline{ES})$.*

## 7.4 Operational Semantics

Like the syntax, the operational semantics of FORENSIC LUCID capitalizes on the semantic rules of GIPL [361] (Section 4.1.1.2, page 78), INDEXICAL LUCID [109], OBJECTIVE LUCID [264], and LUCX [513], JOOIP [528], and an inspiration from MARFL (Figure 84, page 357) augmented with with the new operators, probabilities, and definitions [300, 304, 312]. We specify resulting semantic definitions in FORENSIC LUCID along with the explanation of the rules and the notation. The new rules of the operational semantics of FORENSIC LUCID cover the operators primarily, including the reverse and logical stream operators as well as forensic-specific operators [300, 304, 305]. We use the same notation as the referenced languages to maintain consistency in defining our rules [300, 304].

### 7.4.1 General Semantic Rules

The rules are grouped in several figures: the basic core rules are in Figure 46, the core context-related rules are in Figure 47, the hybrid FORENSIC LUCID-JAVA interaction rules



are in Figure 48, the rules related to observations, observation sequences, and evidential statements are in Figure 49, Figure 51, Figure 52, Figure 53, and in Table 15 respectively. What follows are notes on the additional details of rules of interest.

1. The evaluation context environment $\mathcal{P}$ defines a point in the multidimensional context space at which an expression $E$ is to be evaluated [361, 526]. $\mathcal{P}$ changes when the @ operator is evaluated, or a dimension is declared in a `where` clause. It is a set of $\langle dimension : tag \rangle$ mappings, associating each dimension with a tag index over this dimension. $\mathcal{P}$ is thus a partial function:

$$\mathcal{P} : \mathbf{E_I} \to \mathbf{T} \qquad (7.4.1.1)$$

where $\mathbf{E_I}$ is a set of possible dimension identifiers, declared statically or computed dynamically. $\mathbf{T}$ is a tag set.

In traditional LUCID semantics $\mathbf{E_I} = \mathbf{Id}$ ([361, 513], Section 4.1.1.2, page 78). In our case, the extended $\mathbf{E_I}$ includes $\mathbf{Id}$ and $\mathbf{E.Id}$, where the latter form allows nested OO-like identifiers, both declared explicitly or be a result of evaluation of E into a dynamic identifier allowing for intuitive representation of context hierarchies.

In traditional LUCID semantics, the tag set $\mathbf{T} = \mathbb{N}$, i.e., a set of natural numbers [361]. In LUCX's extension, $\mathbf{T} = \mathbf{U}$ is any enumerable tag set [513], that may include various orderings of numerical or string tag sets, both finite and infinite ([473], cf. Section 7.2.3.1.5, page 165). In FORENSIC LUCID, $\mathbf{T} = \mathbf{U_f}$ that includes $\mathbb{N}, \mathbf{U}$, as well as tag values associated with observations, their sequences, and evidential statements.

2. The initial definition environment $\mathcal{D}_0$ includes the predefined operators, the constants, and $\mathcal{P}_0$ defines the initial context of evaluation [361, 513]. Thus, $\mathcal{D}_0, \mathcal{P}_0 \vdash E : v$ represents the computation of any FORENSIC LUCID expression $E$ resulting in value $v$.

3. The semantic operator † represents the addition of a mapping in the definition environment $\mathcal{D}$, associating an identifier with its corresponding semantic record, here represented as a tuple [361, 528].



4. Each type of identifier (cf. Table 13, page 161) can only be used in the appropriate situations. Identifiers of type `op`, `func`, and `dim` evaluate to themselves (Figure 46, rules 7.4.1.3, 7.4.1.19, 7.4.1.4).

5. Constant identifiers (`const`) evaluate to the corresponding constant (Figure 46, rule 7.4.1.2).

6. Function calls, resolved by the $\mathbf{E_{fct}}$ rule (Figure 46, rule 7.4.1.14), require the renaming of the formal parameters into the actual parameters (as represented by $E'[id_i \leftarrow E_i]$).

7. Arrays are resolved by the $\mathbf{E_{array}}$ rule (Figure 46, rule 7.4.1.6), which is a collection of expressions $E_i$ evaluating into their values $v_i$ under the current context. Evaluating an array @ certain context evaluates each element of the array at that context. When needed, an array can be lifted to a tuple (the described further $\mathbf{E_{tuple}}$ rule).

8. The function $\mathcal{P}' = \mathcal{P}\dagger[id \mapsto v'']$ specifies that $\mathcal{P}'(x)$ is $v''$ if $x = id$, and $\mathcal{P}(x)$ otherwise.

9. The rule for the `where` clause, $\mathbf{E_w}$ (Figure 46, rule 7.4.1.15), which corresponds to the syntactic expression $E$ `where` $Q$, evaluates $E$ using the definitions $Q$ therein.

10. The additions to the definition environment $\mathcal{D}$ and context of evaluation $\mathcal{P}$ made by the $\mathbf{Q}$ rules (Figure 47, rule 7.4.1.29; Figure 46, rules 7.4.1.16, 7.4.1.17) are local to the current `where` clause. This is represented by the fact that the $\mathbf{E_w}$ rule returns neither $\mathcal{D}$ nor $\mathcal{P}$.

11. The $\mathbf{Q_{dim}}$ rule adds a dimension to the definition environment and (as a default convention for $\mathbb{N}$ tags maintained for GIPL compatibility), adds this dimension to the context of evaluation with the tag 0 (Figure 47, rule 7.4.1.29). This allows us to maintain compatibility with the predecessor dialects. For the tag set types other than $\mathbb{N}$ where the first tag is either non-zero or not a number, that first tag is added to the context of evaluation with its underlying collection index of 0 (cf. Definition 42, page 181) (where the collections at the implementation level is `marf.util.FreeVector` in our case).

12. The $\mathbf{Q_{id}}$ and $\mathbf{Q_{fid}}$ simply add variable and function identifiers along with their definition to the definition environment (Figure 46, rules 7.4.1.16, 7.4.1.17) [300, 304, 361].



13. The semantic rule $\mathbf{E_{tuple}}$ (7.4.1.26) evaluates a tuple as a finite stream whose dimension is explicitly indicated as $E$ in the corresponding syntax rule $\langle E_1, \ldots, E_n \rangle id$. Accordingly, the semantic rule $\mathbf{E_{select}}$ (7.4.1.27) picks up one element indexed by $E$ from the tuple $E'$ [513].

14. The evaluation rule for the navigation operator @, $\mathbf{E_{at(cxt)}}$ (7.4.1.24), which corresponds to the syntactic expression $E$ @ $E'$, evaluates $E$ in context $E'$ [513].

15. The evaluation rule for the set navigation operator @, $\mathbf{E_{at(set)}}$ (7.4.1.28), which corresponds to the syntactic expression $E$ @ $E'$, evaluates $E$ in a set of contexts $C$. Therefore, the evaluation result is a collection of results of evaluating $E$ at each element of $C$ [513].

16. The semantic rule $\mathbf{E_{construction(cxt)}}$ (7.4.1.23, Figure 47) evaluates $[E_{d_1} : E_{i_1}, \ldots, E_{d_n} : E_{i_n}]$ to a simple context [513]. It specifically creates a context as a semantic item and returns it as a context $\mathcal{P}$ that can then be used by the rule 7.4.1.24 to navigate to this context by making it override the current context.

17. The semantic rule $\mathbf{E_{construction(set)}}$ (7.4.1.31) correspondingly constructs $\{E_1, \ldots, E_m\}$ as a context set [513].

18. The semantic rule 7.4.1.13 is valid for the definition of the context operators, where the actual parameters evaluate to values $v_i$ that are contexts $\mathcal{P}_i$.

19. The semantic rule 7.4.1.22 expresses that the # symbol evaluates to the current context. When used as a parameter to the context calculus operators, this allows for the generation of contexts relative to the current context of evaluation [300, 304, 361, 513, 515].

20. When intensional operators are used and one of their arguments happens to be of types (odim), (osdim), or (esdim), their $\text{op}_w$ version is assumed (taking $w$ into the account as exemplified in Section 7.3.2, page 169. When binary intensional operators are used and one of the arguments is of a forensic context type and another is a LUCX context type, the latter is *lifted* (similar to type casting in classical imperative languages) to the forensic context as described in Section 7.4.2, page 197 and in subsequent sections.



21. $\mathcal{T}_G(E)$ corresponds to mapping of FORENSIC LUCID and JAVA data types by the *GIPSY Type System* (Appendix B).

22. For hybrid intensional-imperative glue for FORENSIC LUCID and JAVA we define the rules in Figure 48. This set of rules enables FORENSIC LUCID to interact with JAVA (and vice versa), for example, to enable the use of FORENSIC LUCID in JOOIP for purposes of self-forensics (Appendix D).

    **J<sub>CDef</sub>** semantically identifies a JAVA class associates this class declaration to the identifier `cid`, and stores it in the definition environment $\mathcal{D}$. A class can contain member variables (<u>JavaVDef</u>) and member functions (<u>JavaFDef</u>). These are processed in a similar manner by the two following semantic rules [526, 528]. And the rules that follow define the semantics of the evaluation of those.

23. **J<sub>VDef</sub>** specifies a JAVA member variable in a JAVA class <u>JavaCDef</u> by the syntactical specification: `public type vid` ... inside the class' declaration. The specification (`classV`, `cid.vid`, <u>JavaVDef</u>) is used to represent a JAVA class data member `vid` declared inside a class declaration <u>JavaVDef</u> for the class `cid` [526, 528].

24. **J<sub>FDef</sub>** specifies a JAVA member method in a JAVA class <u>JavaCDef</u> from the syntactic specification: `public ft fid(fpt₁ fp₁, ..., fptₙ fpₙ){...}`. Likewise, the specification (`classF`, `cid.fid`, <u>JavaFDef</u>) is used to represent a JAVA member method `fid` declared inside a class declaration <u>JavaCDef</u> for the class `cid` [526, 528].

25. **J<sub>FFDef</sub>** specifies the JLUCID notion of "free JAVA function", which is a method not explicitly defined in a given class by a FORENSIC LUCID programmer. Its syntactical specification is: `ft ffid(fpt₁ fp₁, ..., fptₙ fpₙ){...}`. The semantic specification (`freefun`, `ffid`, <u>JavaFreeFDef</u>) represents the "free JAVA function" `ffid`, i.e., a function that is directly available in the FORENSIC LUCID program, and that is not a member of any class [526]. Since free functions are not allowed in standard JAVA, in terms of implementation, these "free functions" are all put inside a generated wrapper class by the compiler to be part of the GEER of the execution engine as originally defined in [264].



26. **L$_{\text{objV}}$** specifies the semantics of the evaluation of a reference to a class data member by a FORENSIC LUCID expression using the object-oriented dot operator (Section 7.3.2, page 169, page 179). The rule's premises insure that, in $E.vid$: (1) the FORENSIC LUCID expression $E$ evaluates to a value $v$ that is an object of type cid, as being associated in the definition environment $\mathcal{D}$ to the tuple (class, cid, <u>JavaCDef</u>); (2) the variable vid is a public member of the class cid. Once this is established as holding, the Java Virtual Machine can be called upon to evaluate $v.vid$ (noted as **JVM**$[[v.vid]]$), to yield a value $v_r$ [526, 528].

27. **L$_{\text{objF}}$** specifies of the evaluation of a reference to a class member method by a FORENSIC LUCID expression using the dot operator. The rule's premises insure that in $E.fid(E_1, \ldots, E_n)$: (1) the FORENSIC LUCID expression $E$ evaluates to a value $v$ that is an object of type cid, as being associated in the definition environment $\mathcal{D}$ to the tuple (class, cid, <u>JavaCDef</u>); (2) the method fid is a public member of the class cid. Once this is established as holding, all actual parameters are evaluated to values $v_1, \ldots, v_n$, the JVM can be called upon to evaluate $v.fid(v_1, \ldots, v_n)$ (denoted as **JVM**$[[v.fid(v_1, \ldots, v_n)]]$), to yield a value $v_r$ [526, 528].

28. **L$_{\text{FF}}$** speficies the evaluation of free JAVA functions. The rule is a simpler version of **L$_{\text{objF}}$** with no class type identifiers present, and no object to compute upon. As mentioned earlier, JAVA does not have free functions, so all free functions are wrapped in a "free function wrapper" class at compilation, with all free functions inserted in it as static functions [145, 264], which are then processed by the JVM [526, 528]. The **J$_{\text{FFDef}}$** rule is inserting all the free functions in this wrapper class, which we called ffw. Then, upon calling such "free functions", this rule is called and assumes that the "free functions" have been wrapped as static functions into the ffw class, then call the appropriate function [526, 528].

### 7.4.2 Extended Observation

We augment the notion of observation based on Equation 7.2.2.1 to formalize it as illustrated in Figure 49. In general, all components of an observation are stored with the assumption



$$\mathbf{E_{cid}} \quad : \quad \frac{\mathcal{D}(id) = (\texttt{const}, c)}{\mathcal{D}, \mathcal{P} \vdash id : c} \qquad (7.4.1.2)$$

$$\mathbf{E_{opid}} \quad : \quad \frac{\mathcal{D}(id) = (\texttt{op}, f)}{\mathcal{D}, \mathcal{P} \vdash id : id} \qquad (7.4.1.3)$$

$$\mathbf{E_{fid}} \quad : \quad \frac{\mathcal{D}(id) = (\texttt{func}, id_i, E)}{\mathcal{D}, \mathcal{P} \vdash id : id} \qquad (7.4.1.4)$$

$$\mathbf{E_{vid}} \quad : \quad \frac{\mathcal{D}(id) = (\texttt{var}, E) \quad \mathcal{D}, \mathcal{P} \vdash E : v}{\mathcal{D}, \mathcal{P} \vdash id : v} \qquad (7.4.1.5)$$

$$\mathbf{E_{array}} \quad : \quad \frac{\mathcal{D}, \mathcal{P} \vdash E : [E_1, \ldots, E_m] \quad \mathcal{D}, \mathcal{P} \vdash E_i : v_i}{\mathcal{D}, \mathcal{P} \vdash [E_1, \ldots, E_m] : [v_1, \ldots, v_m]} \qquad (7.4.1.6)$$

$$\mathbf{E_{c_T}} \quad : \quad \frac{\mathcal{D}, \mathcal{P} \vdash E : \textit{true} \quad \mathcal{D}, \mathcal{P} \vdash E' : v'}{\mathcal{D}, \mathcal{P} \vdash \texttt{if } E \texttt{ then } E' \texttt{ else } E'' : v'} \qquad (7.4.1.7)$$

$$\mathbf{E_{c_F}} \quad : \quad \frac{\mathcal{D}, \mathcal{P} \vdash E : \textit{false} \quad \mathcal{D}, \mathcal{P} \vdash E'' : v''}{\mathcal{D}, \mathcal{P} \vdash \texttt{if } E \texttt{ then } E' \texttt{ else } E'' : v''} \qquad (7.4.1.8)$$

$$\mathbf{E_{\infty+}} \quad : \quad \frac{\mathcal{D}, \mathcal{P} \vdash E : \texttt{INF+}}{\mathcal{D}, \mathcal{P} \vdash \texttt{INF+} : \mathcal{T}_G(\texttt{Long.MAX\_VALUE})} \qquad (7.4.1.9)$$

$$\mathbf{E_{\infty-}} \quad : \quad \frac{\mathcal{D}, \mathcal{P} \vdash E : \texttt{INF-}}{\mathcal{D}, \mathcal{P} \vdash \texttt{INF-} : \mathcal{T}_G(\texttt{Long.MIN\_VALUE})} \qquad (7.4.1.10)$$

$$\mathbf{E_{eod}} \quad : \quad \frac{\mathcal{D}, \mathcal{P} \vdash E : \texttt{eod}}{\mathcal{D}, \mathcal{P} \vdash \texttt{eod} : \textit{null}} \qquad (7.4.1.11)$$

$$\mathbf{E_{bod}} \quad : \quad \frac{\mathcal{D}, \mathcal{P} \vdash E : \texttt{bod}}{\mathcal{D}, \mathcal{P} \vdash \texttt{bod} : \textit{null}} \qquad (7.4.1.12)$$

$$\mathbf{E_{op}} \quad : \quad \frac{\mathcal{D}, \mathcal{P} \vdash E : id \quad \mathcal{D}(id) = (\texttt{op}, f) \quad \mathcal{D}, \mathcal{P} \vdash E_i : v_i}{\mathcal{D}, \mathcal{P} \vdash E(E_1, \ldots, E_n) : f(v_1, \ldots, v_n)} \qquad (7.4.1.13)$$

$$\mathbf{E_{fct}} \quad : \quad \frac{\mathcal{D}, \mathcal{P} \vdash E : id \quad \mathcal{D}(id) = (\texttt{func}, id_i, E') \quad \mathcal{D}, \mathcal{P} \vdash E'[id_i \leftarrow E_i] : v}{\mathcal{D}, \mathcal{P} \vdash E(E_1, \ldots, E_n) : v} \qquad (7.4.1.14)$$

$$\mathbf{E_w} \quad : \quad \frac{\mathcal{D}, \mathcal{P} \vdash Q : \mathcal{D}', \mathcal{P}' \quad \mathcal{D}', \mathcal{P}' \vdash E : v}{\mathcal{D}, \mathcal{P} \vdash E \texttt{ where } Q : v} \qquad (7.4.1.15)$$

$$\mathbf{Q_{id}} \quad : \quad \frac{}{\mathcal{D}, \mathcal{P} \vdash id = E \ : \ \mathcal{D}\dagger[id \mapsto (\texttt{var}, E)], \mathcal{P}} \qquad (7.4.1.16)$$

$$\mathbf{Q_{fid}} \quad : \quad \frac{}{\mathcal{D}, \mathcal{P} \vdash id(id_1, \ldots, id_n) = E \ : \ \mathcal{D}\dagger[id \mapsto (\texttt{func}, id_i, E)], \mathcal{P}} \qquad (7.4.1.17)$$

$$\mathbf{QQ} \quad : \quad \frac{\mathcal{D}, \mathcal{P} \vdash Q : \mathcal{D}', \mathcal{P}' \quad \mathcal{D}', \mathcal{P}' \vdash Q' : \mathcal{D}'', \mathcal{P}''}{\mathcal{D}, \mathcal{P} \vdash Q \ Q' \ : \ \mathcal{D}'', \mathcal{P}''} \qquad (7.4.1.18)$$

Figure 46: Operational semantics rules of FORENSIC LUCID: $E$ and $Q$ Core

of default values when not all of the comprising components were specified, similarly to the allowed syntactical constructs in Figure 43 where $P = E$ with the $w$ being the credibility or trustworthiness weight of that observation, and the $t$ being an optional wall-clock timestamp. With $w = 1$ the $o$ is equivalent to the original model proposed by Gladyshev [313].

The simple contexts and context sets are lifted (cf. Section 7.3.3, page 183) to observations via $P$ as illustrated by the rule $\mathbf{C_{op(flift)}}$ (7.4.2.13, in Figure 50), (with the defaults assigned



$$\mathbf{E_{did}} \quad : \quad \frac{\mathcal{D}(id) = (\texttt{dim})}{\mathcal{D}, \mathcal{P} \vdash id : id} \tag{7.4.1.19}$$

$$\mathbf{E_{E.did}} \quad : \quad \frac{\mathcal{D}(E.id) = (\texttt{dim})}{\mathcal{D}, \mathcal{P} \vdash E.id : id.id} \tag{7.4.1.20}$$

$$\mathbf{E_{tag}} \quad : \quad \frac{\mathcal{D}, \mathcal{P} \vdash E : id \qquad \mathcal{D}(id) = (\texttt{dim})}{\mathcal{D}, \mathcal{P} \vdash \#E : \mathcal{P}(id)} \tag{7.4.1.21}$$

$$\mathbf{E_{\#(cxt)}} \quad : \quad \overline{\mathcal{D}, \mathcal{P} \vdash \# : \mathcal{P}} \tag{7.4.1.22}$$

$$\mathbf{E_{construction(cxt)}} \quad : \quad \frac{\begin{array}{c}\mathcal{D}, \mathcal{P} \vdash E_{d_j} : id_j \qquad \mathcal{D}(id_j) = (\texttt{dim}) \\ \mathcal{D}, \mathcal{P} \vdash E_{i_j} : v_j \qquad \mathcal{P}' = \mathcal{P}_0\dagger[id_1 \mapsto v_1]\dagger\ldots\dagger[id_n \mapsto v_n]\end{array}}{\mathcal{D}, \mathcal{P} \vdash [E_{d_1} : E_{i_1}, E_{d_2} : E_{i_2}, \ldots, E_{d_n} : E_{i_n}] : \mathcal{P}'} \tag{7.4.1.23}$$

$$\mathbf{E_{at(cxt)}} \quad : \quad \frac{\mathcal{D}, \mathcal{P} \vdash E' : \mathcal{P}' \qquad \mathcal{D}, \mathcal{P}\dagger\mathcal{P}' \vdash E : v}{\mathcal{D}, \mathcal{P} \vdash E @ E' : v} \tag{7.4.1.24}$$

$$\mathbf{E_{dot}} \quad : \quad \frac{\mathcal{D}, \mathcal{P} \vdash E_2 : id_2 \qquad \mathcal{D}(id_2) = (\texttt{dim})}{\mathcal{D}, \mathcal{P} \vdash E_1.E_2 : tag(E_1 \downarrow \{id_2\})} \tag{7.4.1.25}$$

$$\mathbf{E_{tuple}} \quad : \quad \frac{\mathcal{D}, \mathcal{P} \vdash E : id \qquad \mathcal{D}\dagger[id \mapsto (\texttt{dim})] \qquad \mathcal{P}\dagger[id \mapsto 0] \qquad \mathcal{D}, \mathcal{P} \vdash E_i : v_i}{\mathcal{D}, \mathcal{P} \vdash \langle E_1, E_2, \ldots, E_n \rangle E : v_1 \; \texttt{fby.}id \; v_2 \; \texttt{fby.}id \; \ldots \; v_n \; \texttt{fby.}id \; \texttt{eod}} \tag{7.4.1.26}$$

$$\mathbf{E_{select}} \quad : \quad \frac{E = [\texttt{d}:\texttt{v'}] \qquad E' = \langle \texttt{E}_1, \ldots, \texttt{E}_n \rangle \texttt{d} \qquad \mathcal{P}' = \mathcal{P}\dagger[d \mapsto v'] \qquad \mathcal{D}, \mathcal{P}' \vdash E' : v}{\mathcal{D}, \mathcal{P} \vdash select(E, E') : v} \tag{7.4.1.27}$$

$$\mathbf{E_{at(set)}} \quad : \quad \frac{\mathcal{D}, \mathcal{P} \vdash \mathcal{C} : \{\mathcal{P}_1, \ldots, \mathcal{P}_2\} \qquad \mathcal{D}, \mathcal{P}_{i:1\ldots m} \vdash E : v_i}{\mathcal{D}, \mathcal{P} \vdash E @C : \{v_1, \ldots, v_m\}} \tag{7.4.1.28}$$

$$\mathbf{Q_{dim}} \quad : \quad \overline{\mathcal{D}, \mathcal{P} \vdash \texttt{dimension} \; id \; : \; \mathcal{D}\dagger[id \mapsto (\texttt{dim})], \mathcal{P}\dagger[id \mapsto 0]} \tag{7.4.1.29}$$

$$\mathbf{C_{construction(box)}} \quad : \quad \frac{\begin{array}{c}\mathcal{D}, \mathcal{P} \vdash E_{d_i} : id_i \qquad \mathcal{D}(id_i) = (\texttt{dim}) \\ \{E_1, \ldots, E_n\} = dim(\mathcal{P}_1) = \ldots = dim(\mathcal{P}_m) \\ E' = \texttt{f}_p(\texttt{tag}(\mathcal{P}_1), \ldots, \texttt{tag}(\mathcal{P}_m)) \qquad \mathcal{D}, \mathcal{P} \vdash E' : true\end{array}}{\mathcal{D}, \mathcal{P} \vdash Box[E_1, \ldots, E_n | E'] : \{\mathcal{P}_1, \ldots, \mathcal{P}_m\}} \tag{7.4.1.30}$$

$$\mathbf{C_{construction(set)}} \quad : \quad \frac{\mathcal{D}, \mathcal{P} \vdash E_{w:1\ldots m} : \mathcal{P}_m}{\mathcal{D}, \mathcal{P} \vdash \{E_1, \ldots, E_m\} : \{\mathcal{P}_1, \ldots, \mathcal{P}_w\}} \tag{7.4.1.31}$$

$$\mathbf{C_{op(cxt)}} \quad : \quad \frac{\mathcal{D}, \mathcal{P} \vdash E : id \qquad \mathcal{D}(id) = (\texttt{cop}, f) \qquad \mathcal{D}, \mathcal{P} \vdash C_i : v_i}{\mathcal{D}, \mathcal{P} \vdash E(C_1, \ldots, C_n) : f(v_1, \ldots, v_n)} \tag{7.4.1.32}$$

$$\mathbf{C_{sop(set)}} \quad : \quad \frac{\mathcal{D}, \mathcal{P} \vdash E : id \qquad \mathcal{D}(id) = (\texttt{sop}, f) \qquad \mathcal{D}, \mathcal{P} \vdash C_i : \{v_{i_1}, \ldots, v_{i_k}\}}{\mathcal{D}, \mathcal{P} \vdash E(C_1, \ldots, C_n) : f(\{v_{1_1}, \ldots, v_{1_s}\}, \ldots, \{v_{n_1}, \ldots, v_{n_m}\})} \tag{7.4.1.33}$$

Figure 47: Operational semantics rules of FORENSIC LUCID: $E$ and $Q$ Core Context

to $\min, \max, w, t$ per the rules in Figure 49).

### 7.4.3 Observation Sequence

An observation sequence is an ordered collection of observations; the related semantic rules are in Figure 51. The rule 7.4.3.3 corresponds to the most common case of declaration of a witness account. The rule 7.4.3.4 serves to initialize the empty observation sequence dimension; such a sequence will need to be populated at the program execution time via



$$
\mathbf{L_{objV}} \;:\; \frac{\begin{array}{c} \mathcal{D},\mathcal{P} \vdash E : v \quad \mathcal{T}_G(v) = \mathcal{D}(cid) = (\texttt{class},\ \texttt{cid},\ \underline{\texttt{JavaCDef}}) \\ \mathcal{D},\mathcal{P} \vdash vid : vid \quad \mathcal{D}(cid.vid) = (\texttt{classV},\ \texttt{cid.vid},\ \underline{\texttt{JavaVDef}}) \\ \mathcal{D},\mathcal{P} \vdash \mathbf{JVM}[\![v.vid]\!] : v_r \end{array}}{\mathcal{D},\mathcal{P} \vdash E.vid : v_r} \quad (7.4.1.34)
$$

$$
\mathbf{L_{objF}} \;:\; \frac{\begin{array}{c} \mathcal{D},\mathcal{P} \vdash E : v \quad \mathcal{T}_G(v) = \mathcal{D}(cid) = (\texttt{class},\ \texttt{cid},\ \underline{\texttt{JavaCDef}}) \\ \mathcal{D},\mathcal{P} \vdash fid : fid \quad \mathcal{D}(cid.fid) = (\texttt{classF},\ \texttt{cid.fid},\ \underline{\texttt{JavaFDef}}) \\ \mathcal{D},\mathcal{P} \vdash E_1,\ldots,E_n : v_1,\ldots,v_n \\ \mathcal{D},\mathcal{P} \vdash \mathbf{JVM}[\![v.fid(v_1,\ldots,v_n)]\!] : v_r \end{array}}{\mathcal{D},\mathcal{P} \vdash E.fid(E_1,\ldots,E_n) : v_r} \quad (7.4.1.35)
$$

$$
\mathbf{L_{FF}} \;:\; \frac{\begin{array}{c} \mathcal{D}(\mathit{ffid}) = (\texttt{freefun},\ \texttt{ffid},\ \underline{\texttt{JavaFFDef}}) \\ \mathcal{D},\mathcal{P} \vdash E_1,\ldots,E_n : v_1,\ldots,v_n \\ \mathcal{D},\mathcal{P} \vdash \mathbf{JVM}[\![\mathit{ffw}.\mathit{ffid}(v_1,\ldots,v_n)]\!] : v_r \end{array}}{\mathcal{D},\mathcal{P} \vdash \mathit{ffid}(E_1,\ldots,E_n) : v_r} \quad (7.4.1.36)
$$

$$
\mathbf{J_{CDef}} \;:\; \frac{\underline{\texttt{JavaCDef}} = \texttt{class cid }\{\ldots\}}{\mathcal{D},\mathcal{P} \vdash \underline{\texttt{JavaCDef}} \;:\; \mathcal{D}\dagger[cid \mapsto (\texttt{class},\ \texttt{cid},\ \underline{\texttt{JavaCDef}})],\ \mathcal{P}} \quad (7.4.1.37)
$$

$$
\mathbf{J_{VDef}} \;:\; \frac{\begin{array}{c} \underline{\texttt{JavaCDef}} = \texttt{class cid }\{\ldots \underline{\texttt{JavaVDef}} \ldots\} \\ \underline{\texttt{JavaVDef}} = \texttt{public type vid }\ldots; \end{array}}{\mathcal{D},\mathcal{P} \vdash \underline{\texttt{JavaVDef}} \;:\; \mathcal{D}\dagger[cid.vid \mapsto (\texttt{classV},\ \texttt{cid.vid},\ \underline{\texttt{JavaVDef}})],\ \mathcal{P}} \quad (7.4.1.38)
$$

$$
\mathbf{J_{FDef}} \;:\; \frac{\begin{array}{c} \underline{\texttt{JavaCDef}} = \texttt{class cid }\{\ldots \underline{\texttt{JavaFDef}} \ldots\} \\ \underline{\texttt{JavaFDef}} = \texttt{public ft fid}(\texttt{fpt}_1\ \texttt{fp}_1,\ \ldots,\ \texttt{fpt}_n\ \texttt{fp}_n)\{\ldots\} \end{array}}{\mathcal{D},\mathcal{P} \vdash \underline{\texttt{JavaFDef}} \;:\; \mathcal{D}\dagger[cid.fid \mapsto (\texttt{classF},\ \texttt{cid.fid},\ \underline{\texttt{JavaFDef}})],\ \mathcal{P}} \quad (7.4.1.39)
$$

$$
\mathbf{J_{FFDef}} \;:\; \frac{\begin{array}{c} \underline{\texttt{JavaFFWCDef}} = \texttt{class ffw }\{\ldots \underline{\texttt{JavaFFDef}} \ldots\} \\ \underline{\texttt{JavaFFDef}} = \texttt{ft ffid}(\texttt{fpt}_1\ \texttt{fp}_1,\ \ldots,\ \texttt{fpt}_n\ \texttt{fp}_n)\{\ldots\} \end{array}}{\mathcal{D},\mathcal{P} \vdash \underline{\texttt{JavaFFDef}} \;:\; \mathcal{D}\dagger[\mathit{ffid} \mapsto (\texttt{freefun},\ \texttt{ffid},\ \underline{\texttt{JavaFFDef}})],\ \mathcal{P}} \quad (7.4.1.40)
$$

Figure 48: Operational semantics of FORENSIC LUCID (OO)

declaration of observations or observation expressions using context operators. The other rules are defined as per usual. As described in Section 7.3.3, page 183 a context set is lifted to an observation sequence (rule 7.4.2.14 in Figure 50), where each set element is lifted to an observation as per preceding section.

### 7.4.4 Evidential Statement

Evidential statement is a collection of observation sequences where, unlike with observation sequences, ordering is not important. Its semantics is in Figure 52. The expression and declarations rules defined as per usual including common declaration and empty evidential statements. The rule $\mathbf{Q_{esdim}}(\times)$ in particular is designed to handle the generic observation sequences promoting them to an evidential statement forensic context type (Section 7.2.3.1.3, page 164) when any of their contained observations have their max $> 0$.

Furthermore, in Table 15 are results of application of different arguments to @, #, and "." (dot) starting from the evidential statement all the way down to the observation components



$$\mathbf{E_{odid}} \;:\; \frac{\mathcal{D}(id) = (\mathtt{odim})}{\mathcal{D}, \mathcal{P} \vdash id : id} \tag{7.4.2.1}$$

$$\mathbf{E_{ovid}} \;:\; \frac{\mathcal{D}(id) = (\mathtt{odim}, E, \min, \max, w, t) \quad \mathcal{D}, \mathcal{P} \vdash E : o_v}{\mathcal{D}, \mathcal{P} \vdash id : o_v} \tag{7.4.2.2}$$

$$\mathbf{Q_{odim1}} \;:\; \frac{(\mathtt{odim}, E, \min, \max, w, t) \quad \mathcal{D}, \mathcal{P} \vdash E : o_v}{\mathcal{D}, \mathcal{P} \vdash \mathtt{observation}\ id = (E, \min, \max, w, t) \;:\; \begin{bmatrix} \mathcal{D}\dagger[id \mapsto (\mathtt{odim})], \\ \mathcal{P}\dagger[id.P \mapsto o_v, \\ id.min \mapsto \min, \\ id.max \mapsto \max, \\ id.w \mapsto w, \\ id.t \mapsto t] \end{bmatrix}} \tag{7.4.2.3}$$

$$\mathbf{Q_{odim2}} \;:\; \frac{(\mathtt{odim}, E, \min, \max, w) \quad \mathcal{D}, \mathcal{P} \vdash E : o_v}{\mathcal{D}, \mathcal{P} \vdash \mathtt{observation}\ id = (E, \min, \max, w) \;:\; \begin{bmatrix} \mathcal{D}\dagger[id \mapsto (\mathtt{odim})], \\ \mathcal{P}\dagger[id.P \mapsto o_v, \\ id.min \mapsto \min, \\ id.max \mapsto \max, \\ id.w \mapsto w, \\ id.t \mapsto \mathtt{eod}] \end{bmatrix}} \tag{7.4.2.4}$$

$$\mathbf{Q_{odim3}} \;:\; \frac{(\mathtt{odim}, E, \min, \max) \quad \mathcal{D}, \mathcal{P} \vdash E : o_v}{\mathcal{D}, \mathcal{P} \vdash \mathtt{observation}\ id = (E, \min, \max) \;:\; \begin{bmatrix} \mathcal{D}\dagger[id \mapsto (\mathtt{odim})], \\ \mathcal{P}\dagger[id.P \mapsto o_v, \\ id.min \mapsto \min, \\ id.max \mapsto \max, \\ id.w \mapsto 1.0, \\ id.t \mapsto \mathtt{eod}] \end{bmatrix}} \tag{7.4.2.5}$$

$$\mathbf{Q_{odim4}} \;:\; \frac{(\mathtt{odim}, E, \min) \quad \mathcal{D}, \mathcal{P} \vdash E : o_v}{\mathcal{D}, \mathcal{P} \vdash \mathtt{observation}\ id = (E, \min) \;:\; \begin{bmatrix} \mathcal{D}\dagger[id \mapsto (\mathtt{odim})], \\ \mathcal{P}\dagger[id.P \mapsto o_v, \\ id.min \mapsto \min, \\ id.max \mapsto 0, \\ id.w \mapsto 1.0, \\ id.t \mapsto \mathtt{eod}] \end{bmatrix}} \tag{7.4.2.6}$$

$$\mathbf{Q_{odim5}} \;:\; \frac{(\mathtt{odim}, E) \quad \mathcal{D}, \mathcal{P} \vdash E : o_v}{\mathcal{D}, \mathcal{P} \vdash \mathtt{observation}\ id = E \;:\; \begin{bmatrix} \mathcal{D}\dagger[id \mapsto (\mathtt{odim})], \\ \mathcal{P}\dagger[id.P \mapsto o_v, \\ id.min \mapsto 1, \\ id.max \mapsto 0, \\ id.w \mapsto 1.0, \\ id.t \mapsto \mathtt{eod}] \end{bmatrix}} \tag{7.4.2.7}$$

Figure 49: Operational semantics of FORENSIC LUCID: an observation



$$\mathbf{E_{at(o)}} \quad : \quad \frac{\mathcal{D}, \mathcal{P} \vdash E' : \mathcal{O} \qquad \mathcal{D}, \mathcal{P}\dagger\mathcal{O} \vdash E : o_v}{\mathcal{D}, \mathcal{P} \vdash E \mathbin{@} E' : o_v.\langle P, \min, \max, w, t\rangle} \qquad (7.4.2.8)$$

$$\mathbf{E_{at(os)}} \quad : \quad \frac{\mathcal{D}, \mathcal{P} \vdash \mathcal{OS} : \{\mathcal{O}_1, \ldots, \mathcal{O}_m\} \qquad \mathcal{D}, \mathcal{O}_{i:1\ldots m} \vdash E : o_{v_i}}{\mathcal{D}, \mathcal{P} \vdash E \mathbin{@} OS : \mathtt{ordered}\ \{o_{v_1}, \ldots, o_{v_m}\}} \qquad (7.4.2.9)$$

$$\mathbf{E_{at(es)}} \quad : \quad \frac{\mathcal{D}, \mathcal{P} \vdash \mathcal{ES} : \{\mathcal{OS}_m, \ldots, \mathcal{OS}_m\} \qquad \mathcal{D}, \mathcal{OS}_{i:1\ldots m} \vdash E : os_{v_i}}{\mathcal{D}, \mathcal{P} \vdash E \mathbin{@} ES : \{os_{v_1}, \ldots, os_{v_m}\}} \qquad (7.4.2.10)$$

$$\mathbf{C_{fop(os)}} \quad : \quad \frac{\mathcal{D}, \mathcal{P} \vdash E : id \qquad \mathcal{D}(id) = (\mathtt{fop}, f) \qquad \mathcal{D}, \mathcal{P} \vdash O_i : o_{v_i}}{\mathcal{D}, \mathcal{P} \vdash E(O_1, \ldots, O_n) : f(o_{v_1}, \ldots, o_{v_n})} \qquad (7.4.2.11)$$

$$\mathbf{C_{fop(es)}} \quad : \quad \frac{\mathcal{D}, \mathcal{P} \vdash E : id \qquad \mathcal{D}(id) = (\mathtt{fop}, f) \qquad \mathcal{D}, \mathcal{P} \vdash OS_i : \{o_{v_{i_1}}, \ldots, o_{v_{i_k}}\}}{\mathcal{D}, \mathcal{P} \vdash E(OS_1, \ldots, OS_n) : f(\{o_{v_{1_1}}, \ldots, o_{v_{1_s}}\}, \ldots, \{o_{v_{n_1}}, \ldots, o_{v_{n_m}}\})} \qquad (7.4.2.12)$$

$$\mathbf{C_{op(flift)}} \quad : \quad \frac{\mathcal{D}, \mathcal{P} \vdash E : id \quad \mathcal{D}(id) = (\mathtt{cop}, f) \quad \mathcal{D}, \mathcal{P} \vdash C : v \quad \mathcal{D}, \mathcal{P} \vdash O : o_v}{\mathcal{D}, \mathcal{P} \vdash E(O) : f(o_v.P = v)} \qquad (7.4.2.13)$$

$$\mathbf{C_{sop(flift)}} \quad : \quad \frac{\mathcal{D}, \mathcal{P} \vdash E : id \quad \mathcal{D}(id) = (\mathtt{sop}, f) \quad \mathcal{D}, \mathcal{P} \vdash C_i : \{v_{i_1}, \ldots, v_{i_k}\} \quad \mathcal{D}, \mathcal{P} \vdash OS : \{o_{v_i}\}}{\mathcal{D}, \mathcal{P} \vdash E(O_1, \ldots, O_n) : f(o_{v_1}.P = \{v_{1_1}, \ldots, v_{1_s}\}, \ldots, o_{v_n}.P = \{v_{n_1}, \ldots, v_{n_m}\})} \qquad (7.4.2.14)$$

Figure 50: Operational semantics of FORENSIC LUCID: forensic operators and lifting

$$\mathbf{E_{osdid}} \quad : \quad \frac{\mathcal{D}(id) = (\mathtt{osdim})}{\mathcal{D}, \mathcal{P} \vdash id : id} \qquad (7.4.3.1)$$

$$\mathbf{E_{osvid}} \quad : \quad \frac{\mathcal{D}(id) = (\mathtt{osdim}) \qquad \mathcal{D}, \mathcal{P} \vdash E : os_v}{\mathcal{D}, \mathcal{P} \vdash id : os_v} \qquad (7.4.3.2)$$

$$\mathbf{Q_{osdim}} \quad : \quad \frac{(\mathtt{osdim}, E_1, \ldots, E_n) \qquad \mathcal{D}, \mathcal{P} \vdash E_i : o_i \qquad \mathcal{D}, \mathcal{P} \vdash E : os_v}{\mathcal{D}, \mathcal{P} \vdash \mathtt{observation\ sequence}\ id = \{E_1, \ldots, E_n\} : \begin{bmatrix} \mathcal{D}\dagger[id \mapsto (\mathtt{osdim})], \\ \mathcal{P}\dagger[id \mapsto \mathtt{ordered}\langle id.o_{v_1}, \ldots, id.o_{v_n}\rangle] \end{bmatrix}} \qquad (7.4.3.3)$$

$$\mathbf{Q_{osdim(\emptyset)}} \quad : \quad \frac{(\mathtt{osdim}) \qquad \mathcal{D}, \mathcal{P} \vdash E : os_v}{\mathcal{D}, \mathcal{P} \vdash \mathtt{observation\ sequence}\ id : \begin{bmatrix} \mathcal{D}\dagger[id \mapsto (\mathtt{osdim})], \\ \mathcal{P}\dagger[id \mapsto \emptyset] \end{bmatrix}} \qquad (7.4.3.4)$$

Figure 51: Operational semantics of FORENSIC LUCID: an observation sequence

following the definitions in Section 7.3.2, page 169. In case #$ES$ and #$OS$ angle brackets $\langle \ldots \rangle$ simply denote the tuple of either observation sequences or observations of which the current one is actually returned, of type $OS$ or $O$ respectively. While observation sequences in $ES$ are not strictly speaking ordered as mentioned, the order of declaration and updates by context operators is used in the underlying implementation.

### 7.4.5 Belief and Plausibility

The semantics of the `bel` and `pl` forensic operators is in Figure 53. The semantics follows the details explained in the operator definitions in Section 7.3.4, page 191. Belief and plausibility



$$\mathbf{E_{esdid}} \quad : \quad \frac{\mathcal{D}(id) = (\texttt{esdim})}{\mathcal{D}, \mathcal{P} \vdash id : id} \tag{7.4.4.1}$$

$$\mathbf{E_{esvid}} \quad : \quad \frac{\mathcal{D}(id) = (\texttt{esdim}) \quad \mathcal{D}, \mathcal{P} \vdash E : es_v}{\mathcal{D}, \mathcal{P} \vdash id : es_v} \tag{7.4.4.2}$$

$$\mathbf{Q_{esdim}} \quad : \quad \frac{(\texttt{esdim}, E_1, \ldots, E_n) \quad \mathcal{D}, \mathcal{P} \vdash E_i : os_i \quad \mathcal{D}, \mathcal{P} \vdash E : es_v}{\mathcal{D}, \mathcal{P} \vdash \texttt{evidential statement } id = \{E_1, \ldots, E_n\} : \begin{bmatrix} \mathcal{D}\dagger[id \mapsto (\texttt{esdim})], \\ \mathcal{P}\dagger[id \mapsto \langle id.os_{v_1}, \ldots, id.os_{v_n}\rangle] \end{bmatrix}} \tag{7.4.4.3}$$

$$\mathbf{Q_{esdim}(\emptyset)} \quad : \quad \frac{(\texttt{esdim}) \quad \mathcal{D}, \mathcal{P} \vdash E : es_v}{\mathcal{D}, \mathcal{P} \vdash \texttt{evidential statement } id : \begin{bmatrix} \mathcal{D}\dagger[id \mapsto (\texttt{esdim})], \\ \mathcal{P}\dagger[id \mapsto \emptyset] \end{bmatrix}} \tag{7.4.4.4}$$

$$\mathbf{Q_{esdim}(\times)} \quad : \quad \frac{(\texttt{esdim}, E_1, \ldots, E_n) \quad \mathcal{D}, \mathcal{P} \vdash E_i : os_i \quad \exists \mathcal{D}, \mathcal{P} \vdash E_i : os_i.o_i.max > 0 \quad \mathcal{D}, \mathcal{P} \vdash E : es_v}{\mathcal{D}, \mathcal{P} \vdash \texttt{evidential statement } id = \{E_1, \ldots, E_n\} : \begin{bmatrix} \mathcal{D}\dagger[id \mapsto (\texttt{esdim})], \\ \mathcal{P}\dagger[id \mapsto \langle id.os_{v_1} \times id.os_{v_i}\rangle] \end{bmatrix}} \tag{7.4.4.5}$$

Figure 52: Operational semantics of FORENSIC LUCID: an evidential statement

Table 15: Types of context operators' arguments and resulting types

| Left Type | operator | Right Type | $\rightarrow$ | Resulting Type |
|---|---|---|---|---|
| $O$ | @ | $OS$ | $\rightarrow$ | $O$ |
| $O$ | @ | $ES$ | $\rightarrow$ | $O$ |
| $OS$ | combine | $OS$ | $\rightarrow$ | $OS$ |
| $OS$ | combine | $ES$ | $\rightarrow$ | $ES$ |
| $ES$ | combine | $ES$ | $\rightarrow$ | $ES$ |
| $OS$ | product | $OS$ | $\rightarrow$ | $ES$ |
| | # | $ES$ | $\rightarrow$ | $\langle os_1, \ldots, os_n\rangle$ |
| | # | $OS$ | $\rightarrow$ | $\langle o_1, \ldots, o_n\rangle$ |
| | # | $O$ | $\rightarrow$ | $(P, \min, \max, w, t)$ |
| | # | $O.P$ | $\rightarrow$ | $E$ |
| | # | $O.w$ | $\rightarrow$ | $FLOAT$ |
| | # | $O.min$ | $\rightarrow$ | $INTEGER$ |
| $ES$ | . | $OS$ | $\rightarrow$ | $OS$ |
| $OS$ | . | $O$ | $\rightarrow$ | $O$ |
| $O$ | fby | $O$ | $\rightarrow$ | $OS$ |
| $O$ | stream-op | $O$ | $\rightarrow$ | $OS$ |

of no-observations $ basically say everything is believable and plausible (like the *Any* catch-all case in the DSTME examples). Belief and plausibility of zero-observations \0, therefore, correspond to the null-hypothesis, so they are set to 0.0.

## 7.5 Discussion

This section provides an additional insight into the just defined syntax and semantics and how it all fits together with the material presented in the background (Part I).



$$\mathbf{E_{bel}} \quad : \quad \frac{\begin{array}{c}\mathcal{D},\mathcal{P} \vdash E : id \quad \mathcal{D}(id) = (\mathtt{fop}, \mathtt{bel}) \\ \mathcal{D},\mathcal{P} \vdash E' : id' \quad \mathcal{D},\mathcal{P} \vdash E' : o_v \\ \mathcal{D}\dagger[id' \mapsto (\mathtt{odim})] \quad \mathcal{P}\dagger[id' \mapsto id'.w]\end{array}}{\mathcal{D},\mathcal{P} \vdash \mathtt{bel}(E') : o_v.w} \quad (7.4.5.1)$$

$$\mathbf{E_{bel(\$)}} \quad : \quad \frac{\begin{array}{c}\mathcal{D},\mathcal{P} \vdash E : id \quad \mathcal{D}(id) = (\mathtt{fop}, \mathtt{bel}) \\ \mathcal{D},\mathcal{P} \vdash E' : id' \quad \mathcal{D},\mathcal{P} \vdash E' : \$ \\ \mathcal{D}\dagger[id' \mapsto (\mathtt{odim})]\end{array}}{\mathcal{D},\mathcal{P} \vdash \mathtt{bel}(E') : 1.0} \quad (7.4.5.2)$$

$$\mathbf{E_{bel(\backslash 0)}} \quad : \quad \frac{\begin{array}{c}\mathcal{D},\mathcal{P} \vdash E : id \quad \mathcal{D}(id) = (\mathtt{fop}, \mathtt{bel}) \\ \mathcal{D},\mathcal{P} \vdash E' : id' \quad \mathcal{D},\mathcal{P} \vdash E' : \backslash 0 \\ \mathcal{D}\dagger[id' \mapsto (\mathtt{odim})]\end{array}}{\mathcal{D},\mathcal{P} \vdash \mathtt{bel}(E') : 0.0} \quad (7.4.5.3)$$

$$\mathbf{E_{bel(EE)}} \quad : \quad \frac{\begin{array}{c}\mathcal{D},\mathcal{P} \vdash E : id \quad \mathcal{D}(id) = (\mathtt{fop}, \mathtt{bel}) \\ \mathcal{D},\mathcal{P} \vdash E' : id' \quad \mathcal{D},\mathcal{P} \vdash E' : o'_v \\ \mathcal{D},\mathcal{P} \vdash E'' : id' \quad \mathcal{D},\mathcal{P} \vdash E'' : o''_v \\ \mathcal{D}\dagger[id' \mapsto (\mathtt{odim})] \quad \mathcal{P}\dagger[id' \mapsto id'.w] \\ \mathcal{D}\dagger[id'' \mapsto (\mathtt{odim})] \quad \mathcal{P}\dagger[id'' \mapsto id''.w] \\ \mathcal{D},\mathcal{P} \vdash E' : o'_v.P = \mathcal{D},\mathcal{P} \vdash E'' : o''_v.P\end{array}}{\mathcal{D},\mathcal{P} \vdash \mathtt{bel}(E', E'') : o'_v.w + o''_v.w} \quad (7.4.5.4)$$

$$\mathbf{E_{pl(\$)}} \quad : \quad \frac{\begin{array}{c}\mathcal{D},\mathcal{P} \vdash E : id \quad \mathcal{D}(id) = (\mathtt{fop}, \mathtt{pl}) \\ \mathcal{D},\mathcal{P} \vdash E' : id' \quad \mathcal{D},\mathcal{P} \vdash E' : \$ \\ \mathcal{D}\dagger[id' \mapsto (\mathtt{odim})]\end{array}}{\mathcal{D},\mathcal{P} \vdash \mathtt{pl}(E') : 1.0} \quad (7.4.5.5)$$

$$\mathbf{E_{pl(\backslash 0)}} \quad : \quad \frac{\begin{array}{c}\mathcal{D},\mathcal{P} \vdash E : id \quad \mathcal{D}(id) = (\mathtt{fop}, \mathtt{pl}) \\ \mathcal{D},\mathcal{P} \vdash E' : id' \quad \mathcal{D},\mathcal{P} \vdash E' : \backslash 0 \\ \mathcal{D}\dagger[id' \mapsto (\mathtt{odim})]\end{array}}{\mathcal{D},\mathcal{P} \vdash \mathtt{pl}(E') : 0.0} \quad (7.4.5.6)$$

Figure 53: Operational semantics of FORENSIC LUCID: belief and plausibility

### 7.5.1 Mapping FORENSIC LUCID, Gladyshev and Dempster–Shafer Theories

We illustrate how the FORENSIC LUCID computations fit in the context of formalization introduced by Gladyshev (Section 2.2.2, page 30) as well as the use of the Dempster–Shafer theory (Section 3.3.2, page 66).

1. Gladyshev's finite collection of all possible events $I$ (see Section 2.2.2) forms a tag set to denote all possible computations $c$.

2. Every $q$ in Gladyshev's state machine $Q$ is a possible world in our view. $q$ is the current contextual point in space $\mathcal{P}$.

    In the case of the *Printer Case* (Section 2.2.5.1.3, page 48) the notion of the printer is the intension, its various possible world extensions include the states $q$ of the printer



queue, where following the events $I$ make transitions from world to world. Likewise, in the *Blackmail Case* (Section 2.2.5.2, page 51) the intension of the disk cluster is instantiated in different extensions of the letter fragment states.

3. Gladyshev's runs and their partitionings have a length. Since LUCID streams are generally infinite we do not have a length, we denote finite streams end with the `eod` (end-of-data) and `bod` (beginning-of-data) stream markers as well as the query operators such as `iseod` and `isbod` (see the corresponding syntax productions: 7.3.0.24, 7.3.0.67, and 7.3.0.69).

4. For convenience and consistency of expression we define

$$\max \equiv \mathtt{opt}$$

and

$$infinitum \equiv \mathtt{INF+}$$

in Gladyshev's formalization of observation (Section 2.2.4, page 32).

5. $os = o$ simply means an observation sequence $os$ containing simply a single observation $o$ as in $os = \{o\}$.

6. The no-observation in Section 2.2.4, $\$ = (C_T, 0, infinitum)$ is like a catch-all case in the Dempster–Shafer in Section 3.3.2 to complement belief mass assignment. $C_T$ is a finite tag set, and $infinitum$ (a constant longer than the longest run in Gladyshev's formalism) is `INF+` in our case, which maps to JAVA `Long.MAX_VALUE` within the `GIPSYInteger` type in the GIPSY Type System.

7. In general, $o = (E_1, E_2, E_3, E_4, E_5)$, in practice, $\min = E_2$ and $\max = E_3$ evaluate to integers, $w = E_4$ to a floating point number $[0\ldots 1]$, and $t = E_5$ is an optional date/time expression when an activity began (consistently align to some epoch, e.g., a Unix timestamp, which is a number of seconds since the epoch, so it is technically an integer as well, but standard string timestamps are also syntactically accepted to facilitate encoding of log data).

8. We use $\text{fby}_w$ for probabilistic `fby` where the syntax is the same as the regular `fby`, but $o$'s $w$ is taken into account, when $w = 1$, $\text{fby}_w$ behaves like regular `fby`. The same for `wvr`, $X$ is whenever $Y$'s $w$ is sufficiently high ($Y.w > 1/2$) and similarly for other operators defined in Section 7.3.2, page 169. $w$ puts an additional constraint on the transitions that use it to avoid taking unreliable evidence into the account. Having a low $w$ effectively may force an `eod` in a stream of observations earlier than the actual end of stream discarding the evidence after a low-credibility observation as non-reliable and therefore non-admissible. It is certainly possible for a given observation sequence (testimony) to contain alternating sub-sequences of high- and low-credibility observations; therefore, a full-stop at the first encounter of $o.w < 1/2$ may not be desirable. In this case the investigator has to review the observation sequence and break it up into individual sequences containing series of credible and non-credible observations. The manual aspect can be automated if the observation sequences come from software tools, logs, services, or monitoring agents, or as a preprocessing tool or script before running the FORENSIC LUCID program.

### 7.5.2 Forward Tracing vs. Back-tracing

Naturally, the GEE (Section 6.2.2, page 142) makes demands in the demand-driven evaluation in the order the tree (AST) of an intentional program is traversed. Tracing of the demand requests in this case will be "forward tracing" (e.g., see the first top half of the Figure 25, page 87). Such tracing is useful in the debugging and visual verification but is less useful than the mentioned back-tracing when demands are resolved, when dealing with the back-tracing in forensic investigation in an attempt to reconstruct events from the final state observations back to the initial state. Back-tracing is also naturally present when demands are computed and return results, which would be the second half of Figure 25, page 87. The latter may not be sufficient in the forensic evaluation, so a set of reverse operators to `next`, `fby`, `asa`, etc. were needed.

### 7.5.3 Constructing Forensic Context of Evaluation

We need to provide an ability to encode the "stories" told by the evidence and witnesses. These constitute the primary context of evaluation that gives meaning to the case under investigation. The more complete "return value" of the forensic expression evaluation is a collection of backtraces (may be empty), which contain the "paths of fuzzy truth". If a given path trace contains values considered as true, it's an explanation of a story. If there is no such path (i.e., the trace is empty), there is not enough supporting evidence of the entire claim to be true [300, 304, 305, 307, 312]. The backtraces are augmented with the plausibility values using the Dempster–Shafer approach (Section 3.3.2, page 66) computed from the belief mass assignments to the contained observations (in an equivalent Gladyshev's backtraces, the plausibility and belief are always 1.0).

The context spaces (Section 3.2.2, page 64) are finite and can be navigated through in all directions of along the dimension indexes. The finiteness is not a stringent requirement (as normal LUCID's tag sets are infinite in nature), but in the cyberforensic investigation there is always a finite number of elements in the stories told by witnesses and the evidence, such that the investigator can process them in humanly reasonably time and arrive at some conclusion [305].

We, therefore, defined streams of observations (i.e., observation sequences $os_i$). In fact, in FORENSIC LUCID we defined higher-order dimensions and lower-order dimensions [300, 304]. The highest-level one is the *evidential statement es*, which is a `finite unordered` set of observation sequences *os*. The *observation sequence os* is a `finite ordered` set of observations *o*. The *observation o* is an "eyewitness" of a particular property *P* along with the *duration* of the said observation [300, 304]. We mentioned before (Section 7.3.2, page 181), we admit navigating `unordered` tag sets with operators like `next`, `prev`, `fby`, `@` and dependent operators by using the underlying collection index in the order they were stored meaning the order of declaration is not important, but it is sequential. The order may change when using context calculus operators (Section 7.2.3.1.4, page 165), but this is done prior navigation. This is a FORENSIC LUCID extension to the LUCX's context types [473]. Thus, it is not important in which order the investigators lists the observation sequences in the evidential statement. It is important, however, within the observation sequences, the observations are strictly ordered.

### 7.5.4 Transition Function

A *transition function* (described in [136, 137], Section 2.2.2, page 30) derived from Gladyshev *et al.* [135, 136, 137] determines how the context of evaluation changes during forensic computation. It represents in part the case's crime scene modeling as a possible world state. A transition function $\psi$ is investigation-case-specific (i.e., depends on a particular forensic case being modeled, and, therefore, is not readily available). In general, it is to be provided by the investigator, just like in Gladyshev's COMMON LISP implementation. In the FSA approach, the transition function is the labeled graph itself ([137], Figure 12, page 44).

In general, we already have basic intensional operators to query and navigate from one possible world state to another (see Chapter 4, page 76). These operators represent the basic "built-in" transition functions in themselves (the intensional operators such as `@`, `#`, `iseod`, `first`, `next`, `fby`, `wvr`, `upon`, and `asa` as well as their inverse operators [304] defined earlier). However, a specific problem being modeled requires more specific transition function than just plain intensional operators. In this case the transition function is a FORENSIC LUCID function where the matching state transition is modeled through a sequence of invocation of intensional operators [300, 304, 305, 307, 312]. That is each state is a possible world and intensional operators enable transitions between the states.

See the *ACME Printing Case* and *Blackmail Case* remodeled in FORENSIC LUCID in Section 9.3 and Section 9.4 respectively in Chapter 9. There we provide the first FORENSIC LUCID implementation of $\psi$, $\Psi^{-1}$, and the "main()" (program entry point) in Listing 9.5, Listing 9.6, and Listing 9.4 respectively [303, 304] from Gladyshev's cases.

### 7.5.5 Generic Observation Sequences

The generic observation sequence context contains observations whose properties' duration is not fixed to the min value alone (as in $(P, \min, 0)$, omitting $w, t$). The third position in the observation, $\max \neq 0$ in the generic observation, and, as a result, in the containing observation sequence (e.g., $os = (P_1, 1, 2)(P_2, 1, 1)$). (Please refer to Section 2.2.4.5 and [135, 136, 137] for a more detailed example of a generic observation sequence [304, 305, 312].) We adopt a simple way of modeling generic observation sequences [137] by lifting the observation sequence type to an evidential statement type enumerating all variants in the min = min + max arguments for all possible max in the original and setting in the generated set max = 0 via the semantic rule $\mathbf{Q_{esdim}}(\times)$ in Figure 52, page 203.

### 7.5.6 Conservative Extension

We claim FORENSIC LUCID is a conservative extension [253] of GIPL and other comprising LUCID dialects despite its comparatively large feature base. That is programs written in those dialects are also valid programs (but not necessarily forensically interesting) in FORENSIC LUCID, i.e., any virtual machine interpreting FORENSIC LUCID programs during evaluation should be capable of interpreting GIPL, INDEXICAL LUCID, JLUCID, OBJECTIVE LUCID, LUCX, and JOOIP programs.

The extension covers three main aspects: the additional **IdEntry**'s in Table 13, syntax, and semantics. However, the extensions do not alter the meaning of the programs in predecessor LUCID systems.

1. Additional **IdEntry**'s:

   Wan already established LUCX [513] is a conservative extension of GIPL with the *simple context operators* and *context set operators*. The JAVA member extensions were added by OBJECTIVE LUCID *et al.*, again conservatively to GIPL and INDEXICAL LUCID (never explicitly told so, but we may as well do that here). The new forensic entries covering the forensic contexts and operators are likewise additional entries.

2. Syntax extensions:

   The reverse, forensic, probabilistic, and dot operators, as well as the observation, observation sequence, and evidential statement constructs are additional entities that did not exist in the comprising dialects, as such they do not affect consistency of the original programs.

3. Semantics extensions:

We augment the semantics of intensional operators with the credibility weight factor $w$ as well as ability to navigate forensic hierarchical contexts. If $w = 1$, the operators behave according to their nominal definitions. Any operations on the forensic contexts is new to FORENSIC LUCID and does not affect the predecessor dialects.

Thus, we can conclude FORENSIC LUCID does not introduce additional inconsistencies to the predecessor dialects.

## 7.6 Summary

FORENSIC LUCID is a contribution that fuses intensional programming and the DSTME. It offers new formalization of observations and its impact on observation sequences, and evidential statements. This formalization addresses the shortcoming in Gladyshev's theory by making a more realistic evidence representation with the credibility assessment [313] as well as more scalable and accessible to a wider audience due to the simpler nature of LUCID-based languages.

From the logic perspective, it was shown one can model computations as logic [222]. When armed with context and a demand-driven model adopted in the implementation of the LUCID family of languages that limits the scope of evaluation in a given set of dimensions and their tags, we come to the intensional programming artifact. In essence, we formalize our forensic computation units in an intensional manner. We see a lot of potential for this work to be successful and beneficial for cyberforensics as well as intensional programming communities.

The authors of the FSA approach did a proof-of-concept implementation of the proposed algorithms in CMU COMMON LISP (cf. [137, Appendix]) that we improve the usability of by re-writing in FORENSIC LUCID in Section 9.3, page 250 and Section 9.4, page 254. FORENSIC LUCID's software architecture design aspects within GIPSY are discussed further in Chapter 8, page 211 and various use-cases and scenarios are discussed in Chapter 9, page 244.

## Chapter 8

## Software Architecture Design

This chapter discusses the proposed software design and implementation aspects behind FORENSIC LUCID (presented in the preceding chapter). This includes specific contributions to GIPSY in terms of its GIPC and GEE frameworks redesign to support the FORENSIC LUCID compilation and run-time. The architectural design centers around the FORENSIC LUCID parser and semantic analyzer, various re-design details of GEE to support ASPECTJ and PRISM backends and the multi-evaluation-backend framework in general, as well as production of various data-to-FORENSIC LUCID encoders. We also discuss the related background work where applicable. We present the necessary architectural design concepts, frameworks, and some of their PoC implementation. While our main target evaluation platform is GIPSY (Chapter 6, page 128), the design is meant to be general enough for any FORENSIC LUCID-implementing system. We review related work where appropriate that was not mentioned in earlier chapters that impacts the proposed design decisions in this chapter as well as in Chapter 9.

### 8.1 FORENSIC LUCID Compiler

The general design approach (Section 6.2.1, page 139) for adding a new SIPL compiler calls for implementing the `IIntensionalCompiler` interface augmented with a specific IPL parser, semantic analyzer, and optionally a translator. Accordingly, we add the corresponding new FORENSIC LUCID compiler framework to GIPC of GIPSY. One needs to create a JavaCC [506] grammar and FORENSIC LUCID-to-GIPL translation rules where applicable (e.g., see Section 7.3.2.2, page 182) and a possible JAVA-based implementation of some of the new operators [305] and constructs that were not translated, such that GEE can evaluate them at run time. We likewise introduce the `FORENSICLUCID FormatTag` to handle FORENSIC LUCID-specific constructs. These constitute annotations of the AST nodes (similarly to previously introduced `ImperativeNode`s in [264]) that allow GEE's evaluation engines to deal appropriately with them when an interpreter encounters such nodes. The `FORENSICLUCID` annotation allows to invoke the appropriate operators from the GIPSY type system at runtime.

### 8.1.1 FORENSIC LUCID Parser

Following the tradition of many GIPC's parsers, FORENSIC LUCID's grammar (in accordance with its syntax presented in Section 7.3, page 166) is specified in a particular grammar format. We use *Java Compiler Compiler* (JavaCC) [506] to generate the parser for FORENSIC LUCID. The resulting current grammar specification is in GIPSY's CVS repository [366].

### 8.1.2 FORENSIC LUCID Semantic Analyzer

FORENSIC LUCID's semantic analyzer's design calls for it to be primarily an extension of the LUCX's semantic analyzer [473], primarily because LUCX is not fully translated into GIPL and because FORENSIC LUCID adds new constructs, such as forensic contexts and the DSTME that don't have yet any known translation algorithm into GIPL. Thus, the programming artifact `ForensicLucidSemanticAnalyzer` is created to account for the new node types in AST corresponding to the extensions (primarily the forensic and LUCX context types and context calculus operators presented in Section 7.2.3.1.3, page 164 and Section 7.3.3, page 183 respectively). `ForensicLucidSemanticAnalyzer` capitalizes on the earlier semantic analyzers implemented by Tong [473] and Wu [527].

`ForensicLucidSemanticAnalyzer`'s responsibility is not ensure the static compiled AST and the `Dictionary` of identifiers adhere to the declarative aspects of the FORENSIC LUCID language's operational semantics described in Section 7.4, page 192. The semantic analyzer traverses the AST that came out of the JJTree tool of JavaCC top-down/depth-first to do the static type-checking, identifier scope and definition checks, initial rank analysis, and other typical semantic analysis tasks [527]. The differences from the traditional GIPL- and INDEXICAL LUCID-based dialects, additional type checks are done for LUCX-derived simple contexts, context sets, and specifically tag sets [473].

Since not all of the semantic checks can be done at the compile time, the run-time evaluation engines do the run-time checks during program execution. However, the majority of the **QQ** rule's declarations in FORENSIC LUCID (rule 7.4.1.18, Figure 46, page 198) that correspond to the evidential context specification are usually all statically specified from either log files or manually by the investigator.

## 8.2 FORENSIC LUCID **Run-time System Design**

In FORENSIC LUCID run-time `GEE` is instantiated to spawn concurrent threads of evaluation of the same FORENSIC LUCID GEER by any or all designed backends by default as detailed further. ASPECTJ (`--aspectj`, implies `--flucid`) is for forward tracing, potentially as an optimization for just tracing the normal flow of execution $\psi$ instead of backtracing. The PRISM backend is invoked with `--prism` in addition to or to skip the normal backend. A regular eductive `ForensicLucidInterpreter` for traditional GIPL-like evaluation is made available via `ForensicGEE` (Figure 61, page 225). What follows is the detailed design description surrounding these components.

### 8.2.1 ASPECTJ **Backend Design**

Aspect-oriented programming (AOP) has ties with the intensional programming paradigm when it comes to the notion of context [95]. Specifically, Du previously [95] proposed a relationship between the two programming paradigms in 2005. The ASPECTJ language [29] is the extension of the JAVA language to add AOP capabilities to JAVA programs. While AOP's implementation in ASPECTJ is mostly software engineering practices-oriented tool, it can help resolving implementation issues with tracing the forensic evaluation process.

The AOP paradigm has also been applied to systematic security hardening aspects in software systems design patterns by Laverdi'e by reproposing a related *Security Hardening Language* (SHL) in 2007 [228].

#### 8.2.1.1 Tracing Program Execution

The GEE (Section 6.2.2, page 142) of GIPSY is the intensional demand-driven evaluation engine. When it executes a LUCID-dialect program, it is natural to trace it. The trace can be of that of the intensional program itself as well as its hybrid components, e.g., written in JAVA. The proposed design covers the implementation of an eductive system in JAVA and ASPECTJ. The designed example of an execution trace of an OBJECTIVE LUCID [264] program is in Figure 25, page 87. Such a trace is, for example, very useful for the proposed cyberforensic evaluation [267, 291].

#### 8.2.1.2 ASPECTJ Run-time Environment

The GEE's executor's (`gipsy.GEE.Executor`) design is adjusted to produce the required traces similarly to the debug mode. Additionally, the ASPECTJ's joint points [29], like *before* and *after* triggers are specifically of use, of which the GEE itself is unaware, and such an execution trace can be used as the explanation of a story told by a the evidence and witnesses. The exact place of putting such tracing is in the implementation of intensional operators. Due to the GIPL engine's fine granularity of the execution (only `@` and `#` intensional operators along with context set operators and classical operators are executed), another instance (extension) of the engine has to be made, implemented in ASPECTJ, that is capable of doing tracing on the higher-level forensic operators to provide a more meaningful trace. Its design also accommodates for the new forensic context types [291, 305].

An ASPECTJ wrapper is applied to the GEE, specifically around the engine's implementation of the run-time system's core that does the actual evaluation. A problem was identified with the outlined designed approach is that the GEE's interpreter implementation either too coarse-grained when it comes to procedural demands too fine grained when it comes to all the intensional operators translated to `@` and `#`. Therefore, the design of the new intensional backend is undertaken for better tracing of forensic evaluation and its presentation. However, the advantage of using ASPECTJ for this task is that an ASPECTJ wrapper of the engine can be written without much, if any, alteration of any current engine running.

## 8.2.2 Probabilistic Model Checking Backend Design

Model checking has to do with defining a model of a system or component or concept usually as a state machine, and then stating a hypothesis or a claim to see if it agrees with the model [35]. Model checking was applied in various areas, including specifically autonomic computing [502], security metrics for network security [123, 457], and UML/SysML models [190]. Probabilistic model checking is an extension that takes into account probabilities associated with the model and its transitions.

Ritchley was one of the first to use a model checker to analyze network vulnerabilities [402]. The network is modeled as a graph; along with goal states (representing the desired attack target), initial conditions, and other necessary details defined. Then, an assumption is made that there is no way an attacker can reach a goal state. If indeed there is no such a path, the model checker would respond with *true*. If there is a path, the model checker in Ritchley's case would respond by giving a *single path* that an attacker can follow to get to the goal. Subsequently, Sheyner *et al.*'s [427] methods improved Ritchley's work by altering the model checker such that instead of giving only a single counterexample, the model checker will give all the paths leading to the goal. The main critique of these model-checking methods was the eventual state explosion problem for large and complex problems (networks) and the methods were "abandoned". Forensic computing approach presented in Gladyshev's work may suffer from the same issues. The attack graph community in the above examples switched to use DAGs and graph search techniques to avoid such problems. We believe the eductive context-oriented manner of computation, however, can address some of these problems by computing the paths that are only needed by the context specification in a demand-driven manner with the resulting values cached in the scalable DST for later use even if the underlying problem is complex. (This may bring the model-checking aspect back to the attack graph security researchers as well).

#### 8.2.2.1 PRISM

PRISM [467] is a probabilistic model checker with its simple input language syntax and semantics. Internally it implements various modules for decision processes, such as, e.g., Markov Decision Process (MDP), and others. The basic core syntax of PRISM is in Figure 54, which is in essence a series of declarations of variables and the associated actions and guards (similar to the UML state diagrams) and the associated probabilities $p_i$, and a reward/cost specification [190]. In Figure 55, Figure 56, and Figure 57 are formal syntax and operational semantics of PRISM recited from Jarraya's specification [190].

```
[action] guard → p₁: update₁ + ⋯ + pₙ: updateₙ;
rewards ''rewardname''
guard : reward;
  [action] guard : reward;
endrewards
```

Figure 54: PRISM input language basic syntax [190]

```
prism_model    ::=  model_type
                    global_declaration     (Global Declarations)
                    modules                (Modules Specification)

modules        ::=  module  module_name
                    localvar_dec           (Local Variables Declarations)
                    c                      (Commands)
                    endmodule
                 |  modules || modules     (Modules Composition)

global_declaration ::= const_dec            (Constants Declarations)
                    formula_dec             (Formulas Declarations)
                    globalvar_dec           (Global Variables Declarations)

model_type     ::=  mdp
                 |  ctmc
                 |  dtmc
```

Figure 55: PRISM input language syntax (1) [190]

#### 8.2.2.2 Model-Checking with PRISM

Once the probabilistic model in FORENSIC LUCID is built (evidential statement and transition functions), to verify it (concurrently with the general evaluation) is to check the model with the probabilistic model-checking tool PRISM [467]. As a result, the design of one of the GIPSY evaluation engine backends of a FORENSIC LUCID program generates a translated PRISM code, which is then passed on to the PRISM tool itself [313].

We considered and decided against another possibility of translating directly from the FORENSIC LUCID specification to PRISM at compile time (i.e., in GIPC). That translation does not necessarily retain the operational semantics and prohibits the traditional eductive and parallel evaluation of FORENSIC LUCID programs that would normally be available in

$$
\begin{array}{rlll}
\mathcal{C} \ni c & ::= & [\alpha]\, w \rightharpoonup u & \text{(Command)} \\
& | & c \cup c & \\
\mathcal{W} \ni w & ::= & e & \text{(Guard)} \\
\mathcal{U} \ni u & ::= & \lambda : d & \text{(Update)} \\
& | & u_1 + u_2 & \\
\mathcal{D} \ni d & ::= & skip & \text{(Update unit)} \\
& | & x' = e & \\
& | & d_1 \wedge d_2 & \\
\mathcal{E} \ni e & ::= & x & \text{(Expression)} \\
& | & v & \\
& | & e_1\, op\, e_2 & \\
& | & \neg\, e_1 & \\
\mathcal{V} \ni v & ::= & i & \text{(Value)} \\
& | & d & \\
& | & b & \\
op & \in & \{*, /, +, -, <, \leq, >, \geq, =, \neq, \wedge, \vee\} & \text{(Operators)} \\
\lambda & \in & [0,1] & \text{(Probability Value)} \\
x & \in & variables & \text{(Variable)} \\
\alpha & \in & actions & \text{(Action)} \\
i & \in & integer & \text{(Integer)} \\
d & \in & double & \text{(Double)} \\
b & \in & boolean & \text{(Boolean)} \\
\end{array}
$$

Figure 56: PRISM input language syntax (2) [190]

(SKIP) $\quad \langle skip, s \rangle \to s$

(UPD-EVAL) $\quad \langle x' = e, s \rangle \to s[x \mapsto [\![e]\!](s)]$

(UPD-PROCESSING) $\quad \dfrac{\langle d_1, s \rangle \to s_1}{\langle d_1 \wedge d_2, s \rangle \to \langle d_2, s_1 \rangle}$

(PROB-UPD) $\quad \dfrac{\langle d, s \rangle \to s_1}{\langle \lambda : d, s \rangle \to \mu_{s_1}^{\lambda}}$

(PROBCHOICE-UPD) $\quad \dfrac{\langle u_1, s \rangle \to \mu_1 \quad \langle u_2, s \rangle \to \mu_2}{\langle u_1 + u_2, s \rangle \to f(\mu_1)(\mu_2)}$

(ENABLED-CMD) $\quad \dfrac{[\![w]\!](s) = true}{\langle [\alpha]\, w \rightharpoonup u, s \rangle \xrightarrow{\alpha} \langle u, s \rangle}$

(CMD-PROCESSING) $\quad \dfrac{\langle c, s \rangle \xrightarrow{\alpha} \langle u, s \rangle \quad \langle u, s \rangle \to \mu}{\langle \{c\} \cup C, s \rangle \xrightarrow{\alpha} \langle \{c\} \cup C, \mu \rangle}$

Figure 57: PRISM input language operational semantics [190]

GEE [313]. It is also a factor that GEE deals with the core intensional operators (Section 7.3.2, page 169) translated into @$_w$ and #$_w$ (Section 7.3.2.3, page 183), which are uniformly translatable to much simpler PRISM constructs.

The design of the `PRISMWrapper` backend includes interpretation of the Forensic Lucid semantics in the compiled GEER while matching it to the PRISM input language operational semantics (recited in Figure 57 as fully introduced and specified by Jarraya in [190]). As the Forensic Lucid interpreter descends down the Forensic Lucid AST (that follows the Forensic Lucid operational semantics), generate appropriate PRISM commands taking into the account $w$ of observations involved. This translation wouldn't be much of use to non-Forensic Lucid programs since it concerns only the forensic contexts that have the $w$ assigned to be used in the PRISM models.

## 8.3 Updates to GIPSY's Frameworks' Design

There were a number of design changes planned and carried out in various GIPSY frameworks, including GIPC and GEE to support Forensic Lucid (among other systems, such as MARFCAT). This includes the type system and algebras support (Appendix B), intensional demands (Chapter 6), and substantial refactoring. GEE's Lucid interpretation module (`Interpreter`) had to be upgraded to be `GIPSYContext`-aware and use other GIPSY Type System's types consistently (as its previous iteration ([241]) was using arrays of integers to represent the context of evaluation and did not support any of the new Lucx constructs). Along with other GIPSY R&D team members, standardizing multi-tier behavioral aspects to enable scalable Forensic Lucid evaluation, the type system, and developing the physical infrastructure and configuration (GIPSY cluster lab setup, Section 8.6) have been undertaken. The author have been responsible for the major physical software architecture re-design and core API specification and unification along with partial implementation of the mentioned frameworks with Tong [473], Han [160], Ji [191] providing complete implementations of their related parts on Lucx compiler and multi-tier DMS in their master's theses and Wu integrating the JOOIP work on syntax and semantic analysis in her PhD thesis [528].

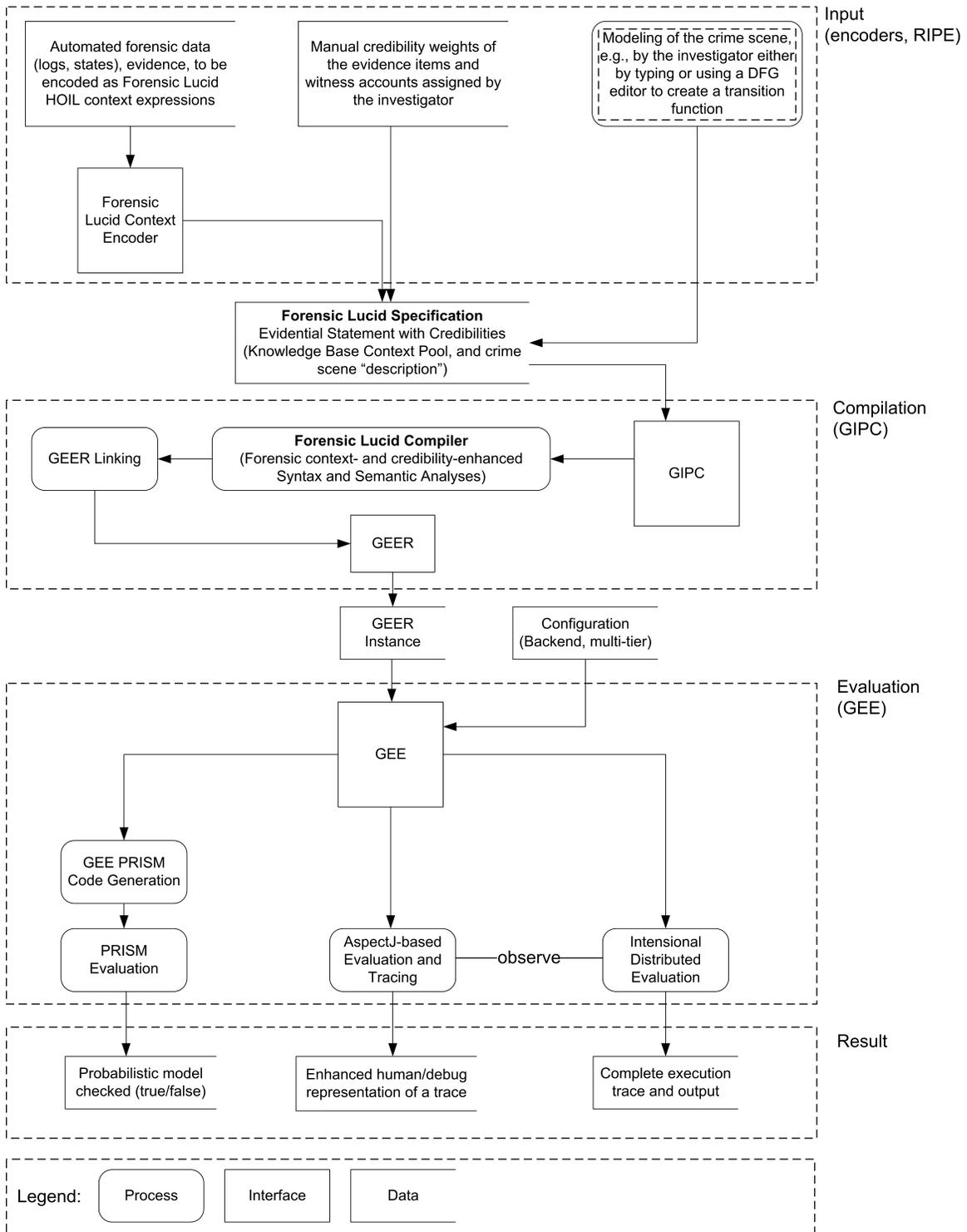

Figure 58: FORENSIC LUCID compilation and evaluation flow in GIPSY [322]

In Figure 58 [322] is a general conceptual design overview of the FORENSIC LUCID compilation and evaluation process involving various components and systems. Of main interest to this work are the inputs to the compiler—the FORENSIC LUCID fragments (hierarchical observation sequence context representing the encoded evidence and witness accounts) and programs (descriptions of the crime scenes as transition functions) can come from different sources as in input. The evidential knowledge of the case and the crime scene model are combined to form a FORENSIC LUCID program. The complete specification is then processed by the compiler depicted as GIPC on the image (the General Intensional Program Compiler) through the invocation of the FORENSIC LUCID SIPL compiler that is aware of the FORENSIC LUCID constructs, such as the forensic contexts, their properties along with credibility weights, etc. and operators detailed in Section 7.3, page 166. The compiler produces an intermediate version of the compiled program as an AST and a contextual dictionary of all identifiers (e.g., observations among other things), encapsulated in a GEER that evaluation engines (under the GEE component) understand. GEER linking is optionally needed if either hybrid compilation of, e.g., JOOIP and FORENSIC LUCID takes place, or a multi-language program is compiled where each compiled fragment has its own AST and they need to be liked together as a single compiled program (see Section 6.2.1, page 139). GEER linking for a pure FORENSIC LUCID program is a no-op. The compiled GEER and engine configuration (either a default, or a custom file) are fed to GEE for processing by one or more evaluation engines. The configuration tells GEE which engines and their run-time setting to use. The said FORENSIC LUCID evaluation engines are designed to use the traditional eduction, ASPECTJ-based tracing, and probabilistic model checking with PRISM [467] to cover the comprehensive subset of computations and provide a backtrace of event reconstruction [311] and can run concurrently processing the same GEER.

In Figure 59 is the updated high-level structure of GIPC (cf. Figure 33, page 138 in Chapter 6). This design incorporates a new framework for semantic analyzers that may behave differently for different SIPLs (Section 4.1.1.2, page 78). The modifications (detailed further) comprise the fact that SIPLs may produce an AST that is not necessarily translated (fully, or partially) into GIPL (as is the case of LUCX and FORENSIC LUCID) bypassing the `SIPLtoGIPLtranslator`. For such cases, the specific semantic analyzers are created to work specifically with such ASTs. The specific semantic analyzers are aware of any additional types in the type system particular to the language in question and rely on its validating methods, such as, e.g., statically declared tag set types, observation's components are proper types (credibility $w$ is a float, $\min, \max$ are integers, etc.). In figure, the semantic analyzers, translators, and GEER linkers are conceptually represented as one class simply because these modules work together closely. At the implementation level, they are separate JAVA classes.

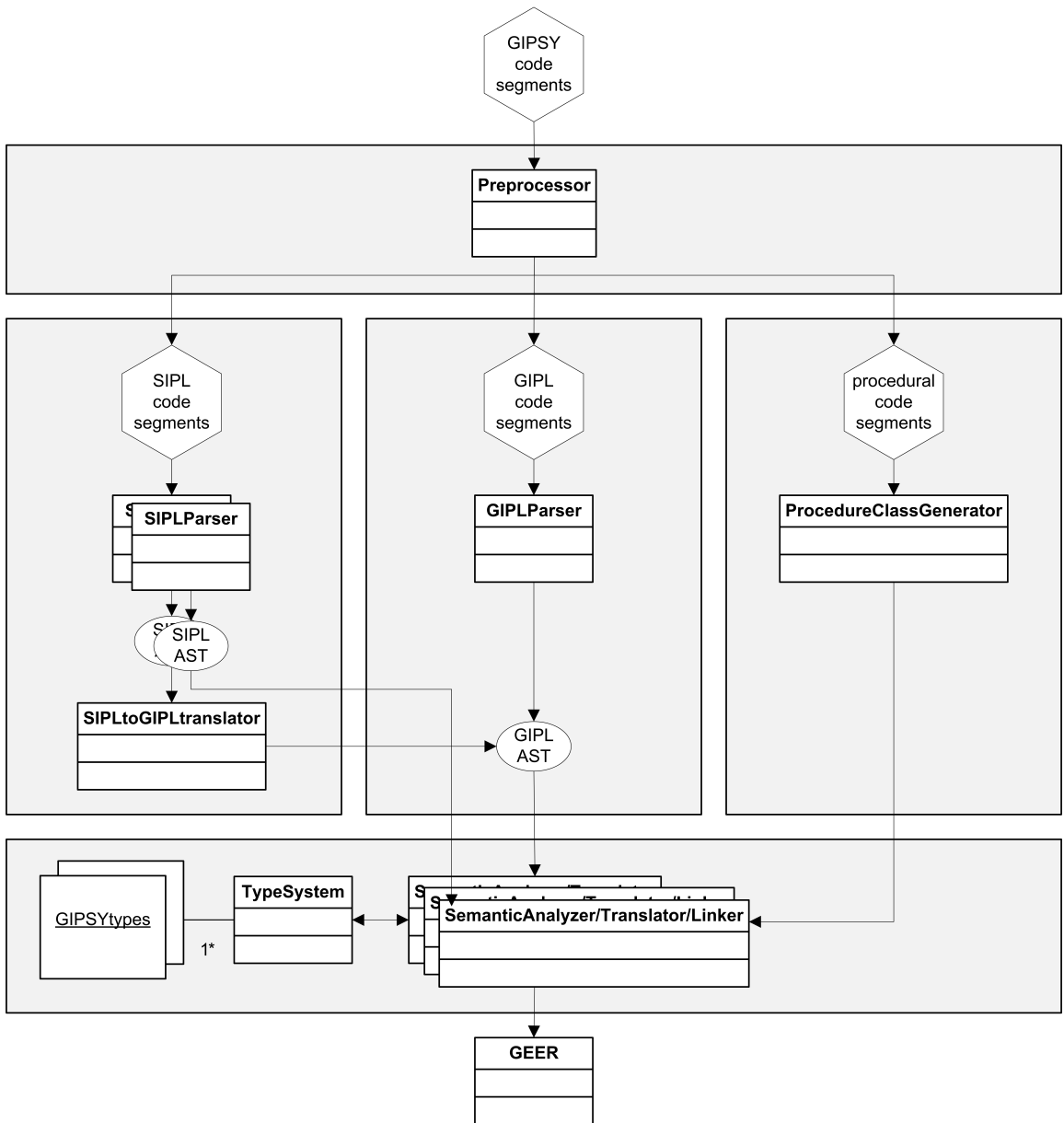

Figure 59: Updated high-level structure of the GIPC framework

At present, the Semantic Analyzers Framework is represented by the `ISemanticAnalyzer`

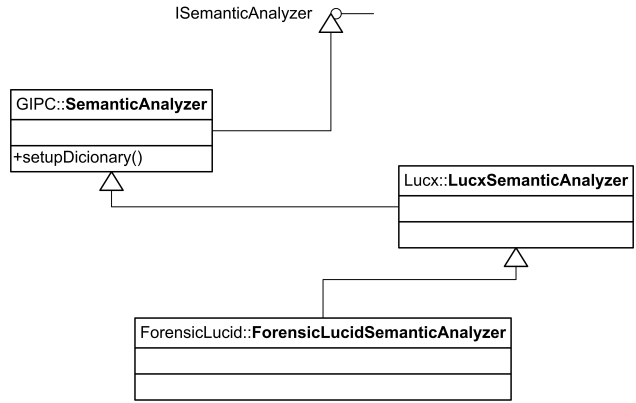

Figure 60: Semantic analyzers framework

interface and has three concrete instances implementing it: `SemanticAnalyzer`, which is the general analyzer that is originally capable of handling of classical GIPL, INDEXICAL LUCID, JLUCID, OBJECTIVE LUCID, and JOOIP programs (combined with the procedural code linking for the hybrid dialects in `ProcedureClassGenerator`) contributed to by Wu [527, 528] and Mokhov [264]. Then follow `LucxSemanticAnalyzer`, produced by Tong [473], and the `ForensicLucidSemanticAnalyzer` for the work presented here on FORENSIC LUCID following the inheritance hierarchy for convenience and code re-use for many of the semantic rules. All semantic analyzers adhere to the same API and produce a GEER, corresponding to their syntactical and semantic annotations. In Figure 60 is the class diagram corresponding to the just described framework. `ForensicLucidSemanticAnalyzer` primarily inherits from `LucxSemanticAnalyzer` to work with tag set, simple context, and context set type checks as well as context calculus operators. Additional checks are added to work with forensic contexts and operators. Any class implementing `ISemanticAnalyzer` can now be added to the GIPC framework as a plug-in should there be need in doing so for future extensions, dialects, and experimentation.

### 8.3.1 Engine Redesign

Semantically speaking, the LUCID program interpretation happens at the generator components, DGTs (Section 6.2.2.1.6, page 146), because this is where the eductive evaluation engines reside. Then it can take various routes: be local, multi-threaded, or distributed via the multi-tier DMS (Section 6.2.2.1, page 143). The aforementioned compiled `GIPSYProgram` (a GEER instance) containing the annotated AST as well as the `Dictionary` of identifiers (corresponding to the initial $\mathcal{D}_0$), and definitions, is passed to GEE (Figure 58, page 219). GEE then hands them over to an interpreter. Traditionally, there was only a single interpreter, with compiled GIPL as its only input language. Then this design morphed into a more general architecture to allow more than one interpreter and demand generator types. Thus, to add interpretation non-translated-to-GIPL dialects (e.g., LUCX and FORENSIC LUCID), one first and foremost has to replace the `Interpreter` component in the DGT based on the either a command-line option or the annotated node tag in the GEER's AST (e.g., a `FormatTag` [264] of `FORENSICLUCID` in our case). In this process the DST, the primary demand propagation machinery, does not change. However, the worker side of DWT may occasionally change, depending on the type of procedural demands (Section 6.2.2.1.8, page 147) to be processed (such as in the case of MARFCAT DWT).

A framework support for this was needed and was provided to allow multiple evaluation engines interpreting new input language material. In the case of FORENSIC LUCID, we override DGT's `Executor` to be able to spawn one of the three evaluation engines, and the eductive `ForensicLucidInterpreter` overrides `Interpreter`. As a side benefit, this also includes support for so called "null"-languages of, e.g., applications (found in `gipsy.apps`) such as MARFCAT (`gipsy.apps.MARFCAT`) and MARFPCAT (`gipsy.apps.MARFPCAT`) from Chapter 5, `OCTMARF` [319] (`gipsy.apps.OCTMARF`), or genome sequence aligning application (`gipsy.apps.memocode.genome`) that do not currently have a compilable LUCID-family language front-end, but could still use the multi-tier DMS middleware of GEE for distributed computation. For mentioned applications, problem-specific (PS) DGT and DWT were created wrapping the functionality of the applications into the demand-driven model, e.g., `MARFCATDGT`, `MARFCATDWT`, and the like.

Thus, non-GIPL engine software layer differentiation became necessary for problem-specific-tiers as well as non-GIPL languages to be able to take advantage of the GEE's distributed middleware infrastructure by improving the GEE framework making it more flexible.

That allows various programming language paradigms to be included into GIPSY that are not necessarily compatible with GIPL or the translation effort to undertake is too great to be effective, error-free, or efficient [473]. This can allow, for example, pLucid, TransLucid, and MARFL run-times. However, the specific interpreter run-times need to be inserted into the GEE framework by implementing the documented interfaces and the operational semantics of these languages.

#### 8.3.1.1 Evaluation Engines' Components

The following is the design related to various evaluation engine components. The invocation of the desired evaluation can be configured or directly invoked by the user via the GMT or its RIPE front-end allowing selection of the evaluation engines from the GUI or via command-line options, of which currently the latter is implemented. In Figure 61 is the overall design connecting various related GEE components.

- `IEvaluationEngine`—all run-time evaluation engines within the GEE framework are to implement this API. It also allows for external plug-ins.

- `IDemandGenerator` is the interface annotating any component that can participate in demand generation activity and `DemandGenerator` its generic implementation. All Lucid interpreters are by default meant to generate demands for eductive execution (local non-eductive implementations simply implement no-op demand generation functionality). `DemandGenerator` also connects interpreters with the multi-tier DMS (discussed further).

- `Executor` + `LegacyInterpreter`—compose the classical GIPL AST interpreter refitted back into the new framework. `LegacyEductiveInterpreter` is an update to `LegacyInterpreter` to actually benefit from the demand generation as the classical `LegacyInterpreter` did everything locally.

- `Interpreter` is a classical interpreter rewritten to use directly the GIPSY Type System.

- `ForensicLucidInterpreter` is forensic-context and context-operators aware extension of the `Interpreter`.

- `ForensicGEE` is simply an extension of the `Executor` designed to work by default with the `ForensicLucidInterpreter`.

- `AspectGEE`—a designed AOP extension stub of the `Executor`s.

- `PRISMWrapper`—a designed non-eductive stub for evaluation of the `GIPSYProgram` such that its "execution" yields a PRISM program that eventually is fed to the PRISM itself [467].

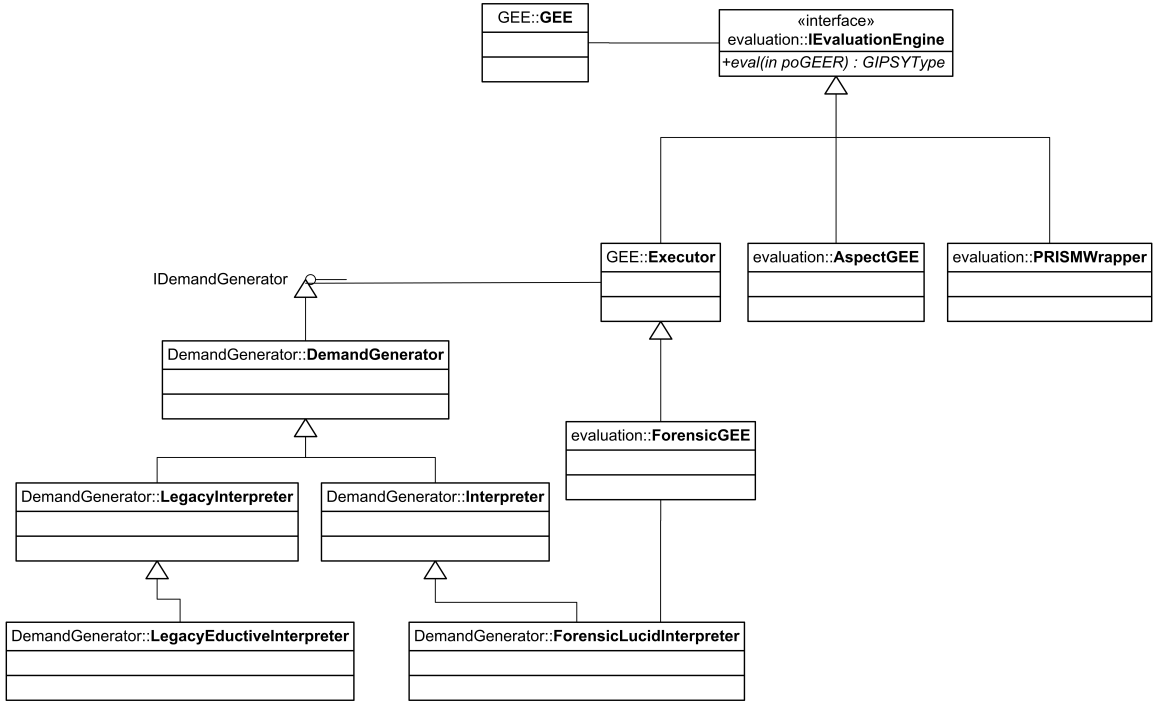

Figure 61: Evaluation engines class diagram

Thus, the FORENSIC LUCID `GIPSYProgram` GEER, depending on the configuration, is simultaneously passed to `ForensicGEE` and `PRISMWrapper` independent evaluation backends; with `AspectGEE` optionally observing `ForensicGEE` for tracing purposes in our design.

#### 8.3.1.2 Backtracing

The design of the backtracing in the `ForensicGEE` interpreting backend includes the execution of the reverse FORENSIC LUCID operators, which is an extension of the classical `Executor` backend (Figure 61).

#### 8.3.1.3 Refactoring Demand Generators vs. Dispatchers within DGT

To enable the different evaluation engines and interpreters as presented in the previous sections in Figure 61 a refactoring re-design effort had to be undertaken [193] to allow FORENSIC

LUCID computation (and as a by-product problem-specific application DGTs mentioned earlier). The illustration of some of the concepts described in this section is in Figure 62. As a part of the refactoring work, we discerned the relationship between and clearly defined the roles of the demand generators and the middle-ware specific demand dispatchers within the DGT. Since generators do not necessarily use the distributed DMS's middleware (Jini and JMS) in their interpreters, the dispatching logic was redefined to be invoked by the generators when needed to make a demand by the interpreter. The notion of the dispatcher was originally designed by Vassev [501] and had the roles of the dispatcher and generator intertwined and reversed requiring decoupling of the generators from the knowledge of the specific middleware technology used.

In accordance with this refined design, a Demand Generator in DGT maintains a local list of demands, "My pending demands" for FORENSIC LUCID program identifiers. The demands on that list are said to be generated ("issued"), but have not come back yet with results, i.e., they preserve their `PENDING` state, in case this list ever to be re-issued to another DST. Issuing of the demands is generally non-blocking, but is subject to the program semantics incorporated into the GEER's AST. An AYDY (*are-you-done-yet*) thread in the Generator periodically checks if any results have come back for the demands on the local pending list with the Dispatcher. They get removed from the list when the result comes back and returned to the user.

The local demand list in the Generator is expected to be most frequently a size of one (at a time) for a single FORENSIC LUCID program, i.e., the root node of the AST. This may change, however, if the Generator picks up intensional demands (effectively subtrees) of own or others' programs that are in the associated store. It is technically also possible issue demands for identifiers without data dependencies, e.g., a demand for $B + C$ could have a demand for $B$ and a demand for $C$ be in the pending demands list of the generator.

The Generator (`ForensicLucidInterpreter` in our case) is aware of the FORENSIC LUCID program semantics and the AST structure, but is unaware of the transport agents (TAs), DSTs, communication mechanisms, and other lower-level middleware technological aspects (Section 6.2.2.1, page 143). The Dispatcher is unaware of the program's overall structured and semantics (except for any annotations pertinent to scheduling not described here), but knows about TAs and communication protocols and scheduling strategies. The Generator runs at the CPU speed; the Dispatcher runs at the communication speed including all the latencies, marshaling/demarshaling costs. Thus, the Generator-Dispatcher pair of the evaluation engine are analogues to the device driver-halves—the general upper half is device-independent running at the CPU speed and known to all and the lower half is device-dependent, communication driven by the events from the device. The two halves are synchronized via a buffer. In this analogy replacing the device with the communication technology and strategy selection and absorption of the inherent latency is handled by the Dispatcher, and upper half that concerns more with the semantics of the GIPSY program (e.g., a FORENSIC LUCID application) is the Generator.

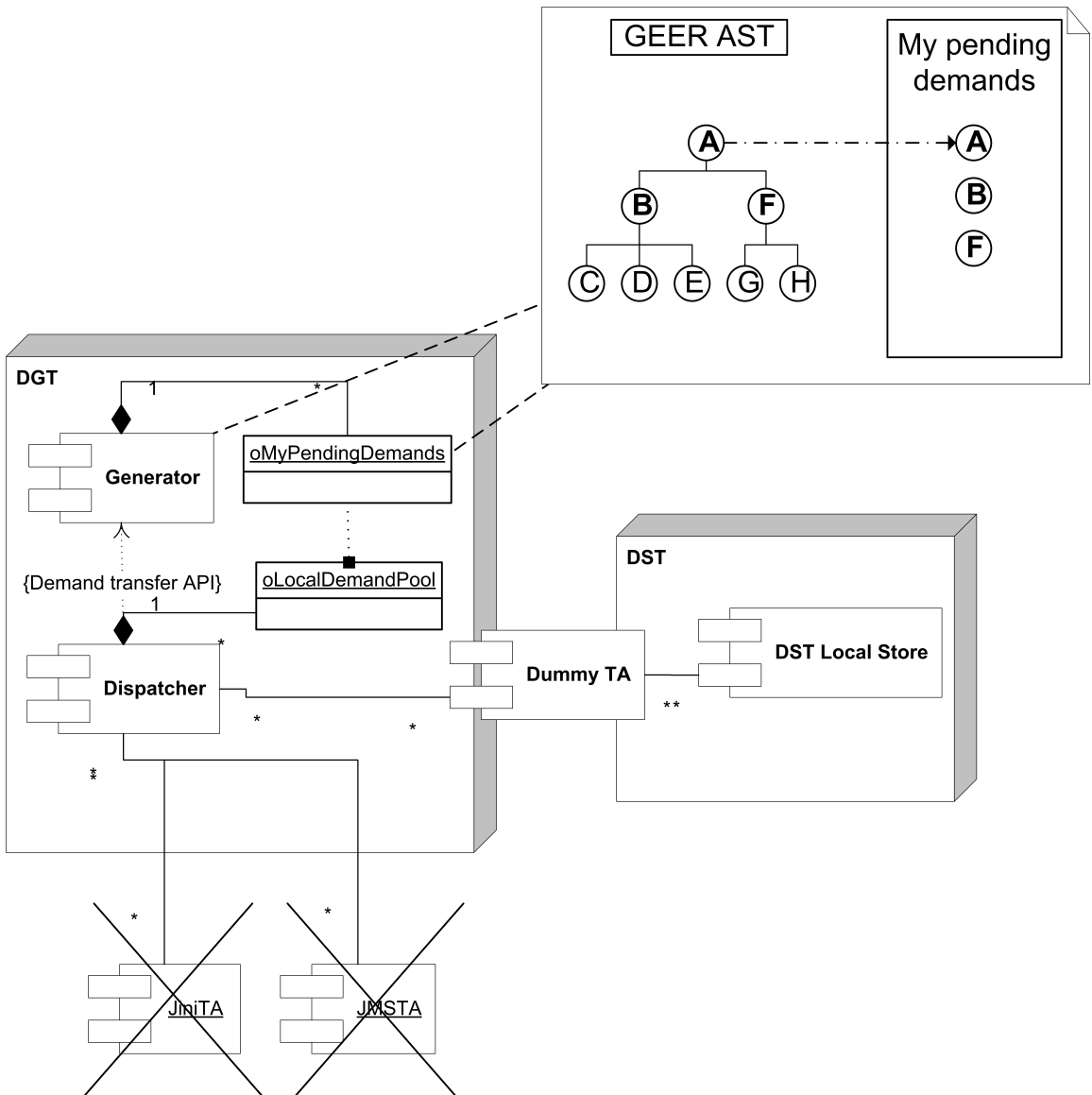

Figure 62: Demand Generator and Dispatcher relationship

The Generator issues the demands to the Dispatcher via the Dispatcher's API. The Dispatcher may be aware of some preliminary scheduling strategies and the associated communication TAs for the demand transfer to the DST. The Dispatcher always maintains a link to the local DST via a *Dummy* TA [191, 501]. Additionally, it keeps its own "Local demand pool" [160, 362], as a buffer, to which the `PENDING` demands from the Generator are deposited via the Dispatcher's API. The TAs are then, in their separate threads of execution, work with that local pool buffer to deliver the demands lazily to a DST with an appropriate TA. The pool is an input/output buffer, that also harvests the evaluation results from the DST before the Generator's AYDY thread picks it up for delivery to the main application.

If the Generator dies, its local pending demands list goes with the Generator. However, assuming the demands were successfully dispatched, stored, computed and results deposited in the DST, a subsequent restart of the Generator by the user for the same program would just return the results cached in the DST. As a result of the Generator's crash and the loss of the Generator's state is tolerable and makes the forensic computation more reliable. If the Dispatcher thread dies and its local demand pool buffer goes with it, assuming everything else is healthy, the AYDY thread in the Generator with a timeout will re-issue the Generator's pending demands again refilling the Dispatcher's local demand pool, which will be filled in with the results later from the DST the Dispatcher communicates via TAs. The local DST store also harbors system demands for the Generator-Dispatcher unit within their DGT container.

This design was a necessary refinement of Paquet's conceptual multi-tier design and Mokhov's integration of that design into the actual code base with some implementation followed by Han [160] and Ji [191] detailed implementation efforts and further refinement of the multi-tier aspects of the design. Without it, the FORENSIC LUCID backends would have to be unjustifiably aware of Jini, JMS, RMI, or any new communication middleware technology added to GEE later. It also benefits non-FORENSIC LUCID backends as well.

The adoption of such an architecture for FORENSIC LUCID programs is particularly important for scalable processing of large evidential statements, including avoiding computational processing any duplicate forensic context that may happen frequently in large logs.

### 8.3.2 Configuration Management

Following the ideas of `marf.Configuration` and MARFL, `gipsy.Configuration` was created to contain name-value pairs of the currently running updatable nested configuration of GEE. In FORENSIC LUCID run-time instance, the configuration tells the main GEE which backends to start and any of their specific configuration details. This includes configuring the multi-tier DMS (e.g., picking Jini or JMS middleware and their settings) to evaluate large evidential data collections, configuration of the DST, an the overall GIPSY network the investigator anticipates is needed to evaluate the problem at hand efficiently. In Figure 58, page 219, configuration is provided to GEE by the investigators (along with the compiled GEER) according the their requirements and the size of the case problem to evaluate. There are default settings that are otherwise used for local non-distributed evaluation.

- `oConfigurationSettings` is a `Properties` type of object encapsulated by the class. It is properly `synchronized` and `Serializable`.

- Some of the `Properties`' API is mirrored onto `Configuration`, so it can also be serialized as a text name=value file or XML, conveniently provided by JAVA.

- Upon initialization it sets initial defaults for the root path of the classes for dynamic discovery as well as policy configuration files for Jini and JMS.

- For FORENSIC LUCID, sets the `ca.concordia.cse.gipsy.GEE.backends.flucid` to `all` by default causing the startup of the three evaluation backends presented. Options `aspectj`, `prism`, and `eductive` are also individually available and can be specified via comma-separated values. Thus, `all` is a shorthand for `aspectj,prism,eductive` in our design.

Ji made later use of this design and elaborated its concrete implementation further as required for his scalability study [191].

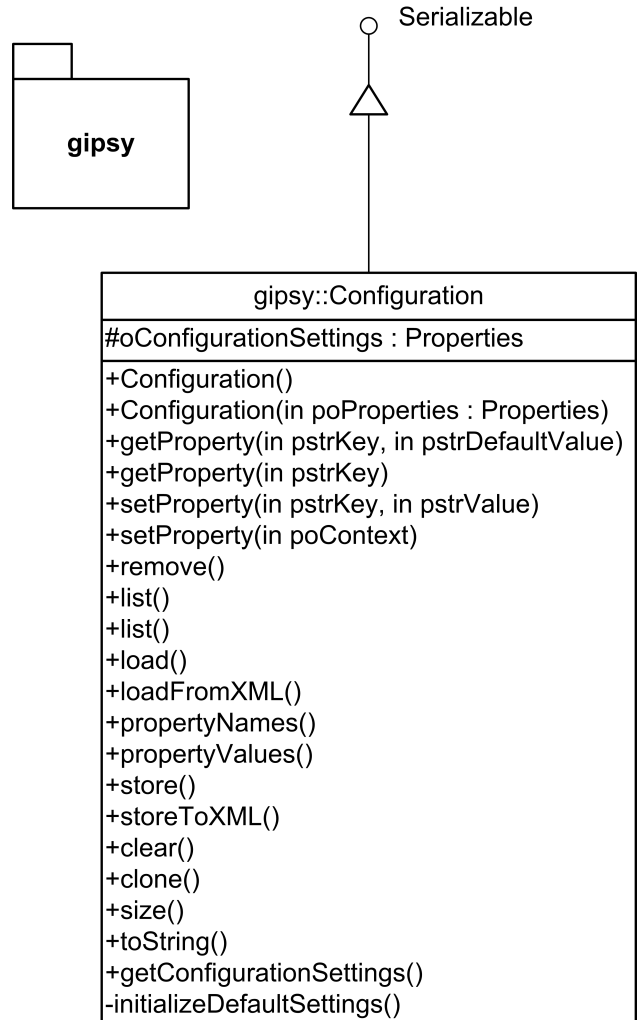
Figure 63: Configuration UML class diagram

### 8.3.3 Compile-Time Annotations

Compile-time annotations were mentioned to annotate imperative nodes, LUCX nodes, or FORENSIC LUCID nodes in the AST. Imperative nodes were originally introduced to denote coarse-grained units of computation, such as JAVA methods (that become procedural demands in GIPSY's multi-tier DMS presented in Section 6.2.2.1, page 143). The author the introduced the annotations framework's design and the corresponding interface originally to allow different scheduling strategies and metrics within the GEE run-time to implement the ideas in Ben-Hamed's thesis [159] for follow-up scalability investigations that are a part of the future work (Section 10.4.3, page 289). The annotations became useful in FORENSIC LUCID (and LUCX) run-time design to invoke the appropriate evaluation engine and type system components that are capable of processing untranslated (to GIPL) FORENSIC LUCID nodes.

- `IAnnotation`, in the package `gipsy.interfaces`

- `NodeAnnotation` (of the AST nodes), as per Figure 64

- An example annotation from [264] is the `ImperativeNode`, instances of which are procedural demand-related.

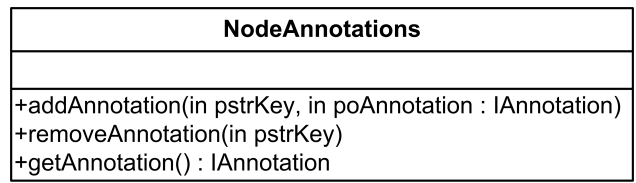

Figure 64: Annotations class diagram

## 8.3.4 Type System

Implementation-wise, FORENSIC LUCID relies on the *GIPSY Type System* (Appendix B) first introduced in [264] and refined in [301, 315, 365, 473]. The new context types are encoded correspondingly to the subclasses of `GIPSYContext`, `Dimension`, and `TagSet` types (defined in [473]). These include the dimension types `observation` (`Observation`), `observation sequence` (`ObservationSequence`), and `evidential statement` (`EvidentialStatement`) that all extend `Dimension`. Likewise, `GIPSYForensicContext` is provided. The instances of the type `GIPSYOperator` are designed to hold the definitions of the new operators [305], including, e.g., `bel` and `pl`. While Appendix B discusses the GIPSY type system in detail, the forensic extensions are illustrated in Figure 65. The package `gipsy.lang.context.forensic` is added to contain the physical classes corresponding to the forensic contexts.

LUCX's `TagSet` and its derivatives are updated to ise `marf.util.FreeVector`, which is an extension of `java.util.Vector` that allows theoretically vectors of infinite length (bounded by the memory available to the JVM), so it is possible to set or get an element of the vector beyond its current physical bounds [264]. Getting an element beyond the boundaries returns `null`, as if the object at that index was never set [264]. Setting an element beyond bounds automatically grows the vector to that element. In the GIPSY, `marf.util.FreeVector` is used as a base for `Dictionary` [264] and as a collection of `Dimension`s with their tag sets in `GIPSYContext`. It's the index within an instance of this class is what determines the behavior of getting next, previous, etc. elements of finite tag sets, regardless if declared ordered or unordered in FORENSIC LUCID.

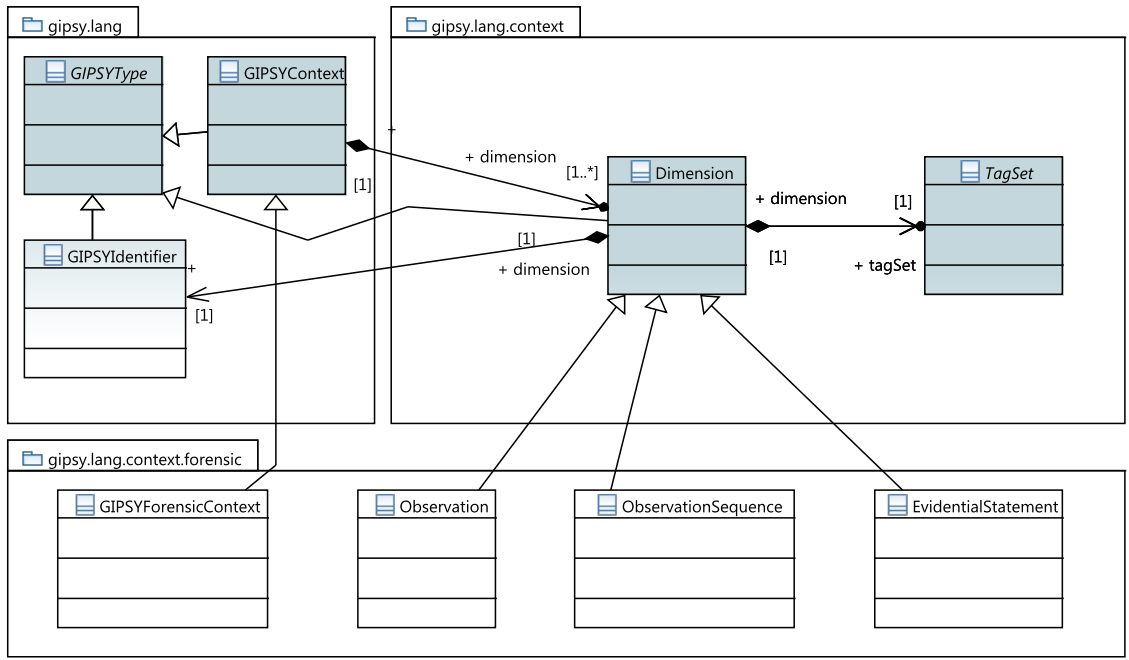

Figure 65: Forensic extensions to the GIPSY Type System

## 8.4 FORENSIC LUCID Component Testing

The testing framework design for FORENSIC LUCID largely follows the testing principles used by the previous GIPSY-related projects of using modules to test-drive main code the `tests` package, including plain tests, source input files, and JUnit-based tests. We also update the `Regression` application to automate some of the non-shell-scripting aspects of testing. The following tests are designed for the FORENSIC LUCID compiler:

- Test all GIPL examples still parse

- Test all GIPL examples still pass semantic check

- Test all LUCX examples still parse

- Test all LUCX examples still pass semantic check

- Test all FORENSIC LUCID examples still parse

- Test all FORENSIC LUCID examples still pass semantic check

Other tests are designed according to the examples presented for the operators that work on bound streams, e.g., Table 14 is as an input for testing of any implementation.

## 8.5 FORENSIC LUCID Encoders

The purpose of such encoders is to extract evidential information from various sources and encode it in FORENSIC LUCID to be ready for use in investigations.

Specifically, exporting and encoding is necessary to present the evidence in the common format (FORENSIC LUCID) for the case when the systems are at least partially automated. Some software systems can be taught to log their data structure states directly in FORENSIC LUCID or after-the-fact transcribing log entries into FORENSIC LUCID format. This is more evident in the Chapter 9's *MAC Spoofer Investigation* when we generate and export a collection of observation sequences from various sources. This particular section illustrates some of the encoded evidence for the software systems mentioned to be exported during their operation, should a fault, incident, bug, or a problem occur, the forensic log (encoded with FORENSIC LUCID) can be used in a subsequent investigation. This is particularly useful in Self-Forensics applications and application domains detailed in Appendix D.

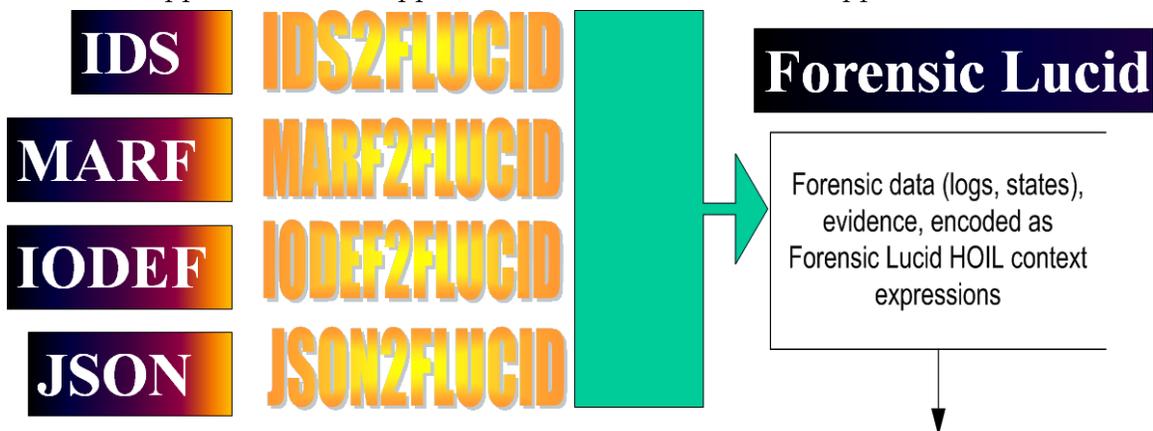

Figure 66: Conceptual illustration of FORENSIC LUCID encoders

There were a number of exporters proposed to encode the evidence data from various data sources (some conceptualized in Figure 66) (Section 9.1.1, page 244) [269, 310, 321] as well as in the mentioned MAC spoofer investigations in Section 9.5.

### 8.5.1 MARF

MARF is one of potent evidential data sources mentioned in Figure 66. Since MARF, as a pattern recognition system, along with its applications described in Chapter 5, able to provide classification data (vulnerable, malicious code, or biometric classification tasks) and assign the confidence in reliability of its classification results. These data can complement other evidence in investigator's possession and automate some of the process.

#### 8.5.1.1 Classical MARF's Evidence

The evidence extracted from the classification analysis results of MARF comes from the several internal data structures [269], namely `Result`, `ResultSet`, `TrainingSet` (stored trained-on data), and MARF instance's `Configuration` (Chapter 5). The `Result` consists of tuples containing $\langle ID, outcome \rangle$, which are the properties of a single classification result. The result set, `ResultSet`, is a collection of such tuples. Processed samples (e.g., utterances, text, imagery, code; stored feature vectors or clusters of `double`s, see Chapter 5 for details), alongside with the training file names and IDs comprise the training set data, and configuration is a collection of processing settings that led to the current results given the training set [269]. We used this way of modeling of the forensic context to extend it to our other case studies [322] (Appendix D).

The property $P$ is specified in the three main categories: configuration, training set, and the result set in MARF. Its observed duration is set as a default of $(1,0)$, as the notion of real duration varies per configuration details, but we are interested in how we arrive from the given configuration and training set to the results [269] in this case. The finer-grained details, including the actual duration may be specified inside $P$. `observation o = P` as syntactically written, is therefore equivalent to $o = (P, 1, 0)$ as mentioned in Chapter 7. The observation sequence $os$ is defined as a sequence of three observations, each observation per category. The observations are ordered as (1) configuration `configo`, (2) training set `tseto`, and (3) the classification result `resulto`. The meaning of this observation sequence is that given some MARF configuration settings and the existing training set, the system produces the classification result. During the training, the observation sequence is slightly different, but also has three observations: configuration, incoming sample, and the resulting training set, which are encoded accordingly as all the necessary primitives for that are already defined [269]. The complete exportable FORENSIC LUCID expression is a 3-observation sequence as presented in Listing 67. With the simplifying assumption, the $(1, 0)$ syntactical constructs can be dropped, only keeping the $P$, which in this case is a higher-order context specification, as shown in Figure 68 [269]. Such a contextual specification of MARF internals FORENSIC LUCID inherited from the MARFL language [272, 322].

```
MARFos = { confo, tseto, resulto } =
{
  ([
    sample loader      : WAV [ channels: 2, bitrate: 16, encoding: PCM, f : 8000 ],
    preprocessing      : LOW_PASS_FFT_FILTER [ cutoff: 2024, windowsize: 2048 ],
    feature extraction : LPC [ poles: 40, windowsize: 2048 ],
    classification     : MINKOWSKI_DISTANCE [ r : 6 ]
  ], 1, 0),

  ([data:{[5.2,3.5,7.5],[3.6,2.5,5.5,6.5]}, files:[''/foo/bar.wav'',''/bar/foo.wav'']], 1, 0),

  ([ID:5, outcome:1.5], 1, 0)
}
```

Figure 67: Example of a three-observation sequence context exported from MARF to FORENSIC LUCID [322]

```
MARFos = { confo, tseto, resulto } =
{
  [
    sample loader      : WAV [ channels: 2, bitrate: 16, encoding: PCM, f : 8000 ],
    preprocessing      : LOW_PASS_FFT_FILTER [ cutoff: 2024, windowsize: 2048 ],
    feature extraction : LPC [ poles: 40, windowsize: 2048 ],
    classification     : MINKOWSKI_DISTANCE [ r : 6 ]
  ],

  [data:{[5.2,3.5,7.5],[3.6,2.5,5.5,6.5]}, files:[''/foo/bar.wav'',''/bar/foo.wav'']],

  [ID:5, outcome:1.5]
}
```

Figure 68: Example of a simplified three-observation sequence context exported from MARF to FORENSIC LUCID [322]

### 8.5.2 MARFCAT and MSA Encoders

Other encoder examples include MARFCAT (Section 5.4.1, page 111) shown in Listing 8.1 (see [314, Appendix] for an extensive example); MARFPCAT; and MAC Spoofer Analyzer (MSA) modules (Section 9.5, page 257). In Listing 8.2, Listing 8.3, Listing 8.4 are examples of the switch management, Argus netflows, and `arp` related evidence encoded in FORENSIC LUCID as observation sequences.

```
// ...
weakness_25 @ [id:25, tool_specific_id:25, cweid:20, cwename:"Input Validation (CWE20)"]
where
  dimension id, tool_specific_id, cweid, cwename;
  observation sequence weakness_25 = (locations_wk_25, 1, 0, 0.001495003320843141);
  locations_wk_25 = locations @ [tool_specific_id:25, cweid:20, cwename:"Input Validation
      (CWE20)"];
    observation location_id_2012( [line => 89, path => "wireshark-1.2.0/plugins/docsis/
        packet-bpkmreq.c")
    textoutput="";
    observation grade = ([ severity => 1, tool_specific_rank => 668.8948352542972], 1, 0,
        0.001495003320843141);
end;
// ...
```

Listing 8.1: MARFCAT encoded evidence fragment example

While *modi operandi* of MSA are detailed in Section 9.5, page 257, we briefly describe the encoding process used by its modules here. Effectively, there are two types of encoders in MSA: the ones that work with the typical log data from services of interest ("dead forensics") and real-time probes ("live forensics") (Section 2.1, page 24).

```
// ...
// 'swm' evidence, encoded: Tue Jul 13 15:37:22 2013
observation sequence swm_os = o_swm_switch fby os_swm_entries;
observation o_swm_switch = ([switch:"switch1"], 1, 0);
observation sequence os_swm_entries =
{
  ([port:"Fa0/1", port-state:"up/up", mac:"00:1b:63:b5:f8:0f", hostname:"flucid-44.encs.
      concordia.ca"] => "STATIC 666 secured,restrict,pfast # [Auto]")
};
// end of 'swm' evidence
// ...
```

Listing 8.2: `swm` encoded evidence example

The log-based encoders usually work with a relatively structured log files, which begin with a timestamp, service, event logged, and details pertinent to each particular event, e.g., an IP address, a MAC address, a username, etc. These data are first filtered for the data of potential interest (e.g., using `grep` and PERL regexp [403]). Each log line corresponds to an observation $o$. The duration in such cases is commonly $(\min, \max) = (1, 0)$ indicating presence of the observed property $P$ with reliability $w = 1.0$, and $t$ being set to the timestamp extracted from the log file. $P$ is then a collection of $\langle dimension : tag \rangle$ pairs encoding the details found on the log line. An observation sequence is an ordered finite collection of such observations of log lines of interest from the same log file. Listing 8.3 is an example of the log-based encoded evidence. In MSA there are several log files, so the encoding is made uniform via the `EncodingUtils` module that formats and normalizes the data that have more than one way of legitimate representation in use (most common examples are timestamps, MAC addresses, and hostnames).

The live-probe encoders are more dynamic and challenging as they work with often less structured data than that of log entries. The data are also time-sensitive due to the liveness aspect and are subject to what the possible spoofer is doing *right now* on the network. As a result the live probes can come up with different degree of output collected with each run. A set of the most pertinent details was identified from the most complete experimental runs by a human expert. Thus, each live-probe encoder type tries to extract as much evidence as possible based on the set of criteria set by the expert. Depending on how complete the extracted data are, the encoder places different values of the reliability weight $w$ encoded in its observations. Typically, each live-probe's encoders collector's line also corresponds to an observation in FORENSIC LUCID terms, but that data can sometimes span more than one line of text pertaining to a single observation. $P$ is then encoded similarly to the log-based encoders with the $\langle dimension : tag \rangle$ pairs with the data of potential interest. Listing 8.2 and Listing 8.4 are examples of the live-probe encoded evidence.

No-observations ($) are also encoded when the log search or probe come up empty for whatever reason. Partial observations also occur when only some of the data are available, then often $w$ is set to values less than 1.0, based on the expert human estimates.

The quality of the encoders follows good common design and development practices and reviewed by more than one person.

```
// ...
// Argus evidence, encoded: Tue Jul 13 15:37:21 2013
observation sequence argus_os =
{
  netflow_o_1,
  netflow_o_2,
  netflow_o_3,
  // ...
  netflow_o_1019
};

observation netflow_o_1 = ([flow-start:"Tue Jul 13 11:24:53 2013", flow-end:"Tue Jul 13
    11:24:53 2013", protocol:"tcp", src-mac:"02:5f:f8:93:80:00", dst-mac:"02:29:bb:29:d9:1a"
    , src-ipaddr:"aaa.aa.aa.aaa", src-port:136, direction:"->", dst-ipaddr:"132.205.44.252",
     dst-port:252, packets:2, src-bytes:120, dst-bytes:0, state:"REQ"], 1, 0, 1.0, "Tue Jul
    13 11:24:53 2013");

observation netflow_o_2 = ([flow-start:"Tue Jul 13 11:49:05 2013", flow-end:"Tue Jul 13
    11:49:05 2013", protocol:"tcp", src-mac:"02:5f:f8:93:80:00", dst-mac:"02:29:bb:29:d9:1a"
    , src-ipaddr:"bb.bb.bb.bbb", src-port:212, direction:"->", dst-ipaddr:"132.205.44.252",
    dst-port:252, packets:2, src-bytes:124, dst-bytes:0, state:"REQ"], 1, 0, 1.0, "Tue Jul
    13 11:49:05 2013");

observation netflow_o_3 = ([flow-start:"Tue Jul 13 12:27:34 2013", flow-end:"Tue Jul 13
    12:27:34 2013", protocol:"tcp", src-mac:"02:5f:f8:93:80:00", dst-mac:"02:29:bb:29:d9:1a"
    , src-ipaddr:"cc.ccc.ccc.ccc", src-port:152, direction:"->", dst-ipaddr:"132.205.44.252"
    , dst-port:252, packets:2, src-bytes:120, dst-bytes:0, state:"REQ"], 1, 0, 1.0, "Tue Jul
     13 12:27:34 2013");
// ...
```

Listing 8.3: Argus encoded evidence example

```
// ...
// arp log evidence, encoded: Tue Aug 13 15:37:13 2013
observation sequence arp_os =
{
  arp_o_1
};

observation arp_o_1 = ([ipaddr:"132.205.44.252", mac:"00:1b:63:b5:f8:0f"], 1, 0, 1.0);
// end of arp log evidence
// ...
```

Listing 8.4: `arp` encoded evidence example

## 8.6 GIPSY Cluster Lab Setup

### 8.6.1 Experimental Setup and Script Design

Here we design our initial experimental setup, modeling and simulation, and the testbed environments.

We describe the GIPSY cluster environment design, hardware and software, to manage and to run the experiments as well script our testing environment around GIPSY, MARF-CAT, and related platforms.

#### 8.6.1.1 Hardware

1. 16 hosts (Dell Precision 390), running a mix of Linux and Windows

2. 2 Cisco Catalyst 2950 switches

3. Local DNS server (following the principles of [31])

4. Local router/firewall

5. Local DHCP server

6. GIPSY run-time deployed across the hosts

#### 8.6.1.2 Software

The software tools included in the GIPSY cluster design are listed below. `iptables` and BIND are the most essential installed on the routers to manage the local DNS names and network address translation between the physical GIPSY network (192.168.88.0/24, VLAN88) and the ENCS network (132.205.8.224/28, VLAN629) to the outside world.

1. MRTG [345]

2. Argus [390]

3. `iptables` [389, 398]

4. BIND [8]

### 8.6.2 Topological Setup

The physical network design consists of 2 48-port Cisco Catalyst 2950 switches a machine running DNS and DHCP servers, a machine running `iptables` for firewall and routing, a monitoring machine running MRTG, Argus, Snort, and other tools of our own the uplink to the Internet where all the clients would want to connect to. The purpose is to run scalable computations running GIPSY networks for forensic and data mining computations. The network schematic is in Figure 69, and the corresponding IP configuration is in Table 16.

Table 16: GIPSY cluster IP assignment

| Node | DNS name | IP |
|---|---|---|
| $S_p$ | storm.gipsy.private | 192.168.88.1 |
| $S2_p$ | storm2.gipsy.private | 192.168.88.2 |
| $D_p$ | detroit.gipsy.private | 192.168.88.250 |
| $E_p$ | elsinore.gipsy.private | 192.168.88.251 |
| $N$ | newton.encs.concordia.ca | 132.205.8.230 |
| $S$ | storm.encs.concordia.ca | 132.205.8.238 |
| $\alpha$ | alpha.gipsy.private | 192.168.88.65 |
| $\beta$ | beta.gipsy.private | 192.168.88.66 |
| $\gamma$ | gamma.gipsy.private | 192.168.88.67 |
| $\delta$ | delta.gipsy.private | 192.168.88.68 |
| $\epsilon$ | epsilon.gipsy.private | 192.168.88.69 |
| $\zeta$ | zeta.gipsy.private | 192.168.88.70 |
| $\eta$ | eta.gipsy.private | 192.168.88.71 |
| $\theta$ | theta.gipsy.private | 192.168.88.72 |
| $\iota$ | iota.gipsy.private | 192.168.88.73 |
| $\kappa$ | kappa.gipsy.private | 192.168.88.74 |
| $\lambda$ | lambda.gipsy.private | 192.168.88.75 |
| $\mu$ | mu.gipsy.private | 192.168.88.76 |
| $\nu$ | nu.gipsy.private | 192.168.88.77 |
| $\xi$ | xi.gipsy.private | 192.168.88.78 |
| $o$ | omicron.gipsy.private | 192.168.88.79 |

Additionally, we include the forensic and log-analysis tools with machine learning, data mining, and statistical analysis capabilities. The GIPSY's DSTs are designed to be typically located at Storm and Storm2 as two powerful servers; however, when Storms are unavailable, any node can have DSTs on them. Linux machines running Scientific Linux 5.x have an additional VMWare virtual to support more hosts as well as the SATE experiment test bed for MARFCAT code analysis described in Section 5.4, page 110. Each of the 390 node or its virtual is typically a worker node DWT. A portion of the cluster was allocated to Rabah to run XEN virtuals for virtual network experiments [392].

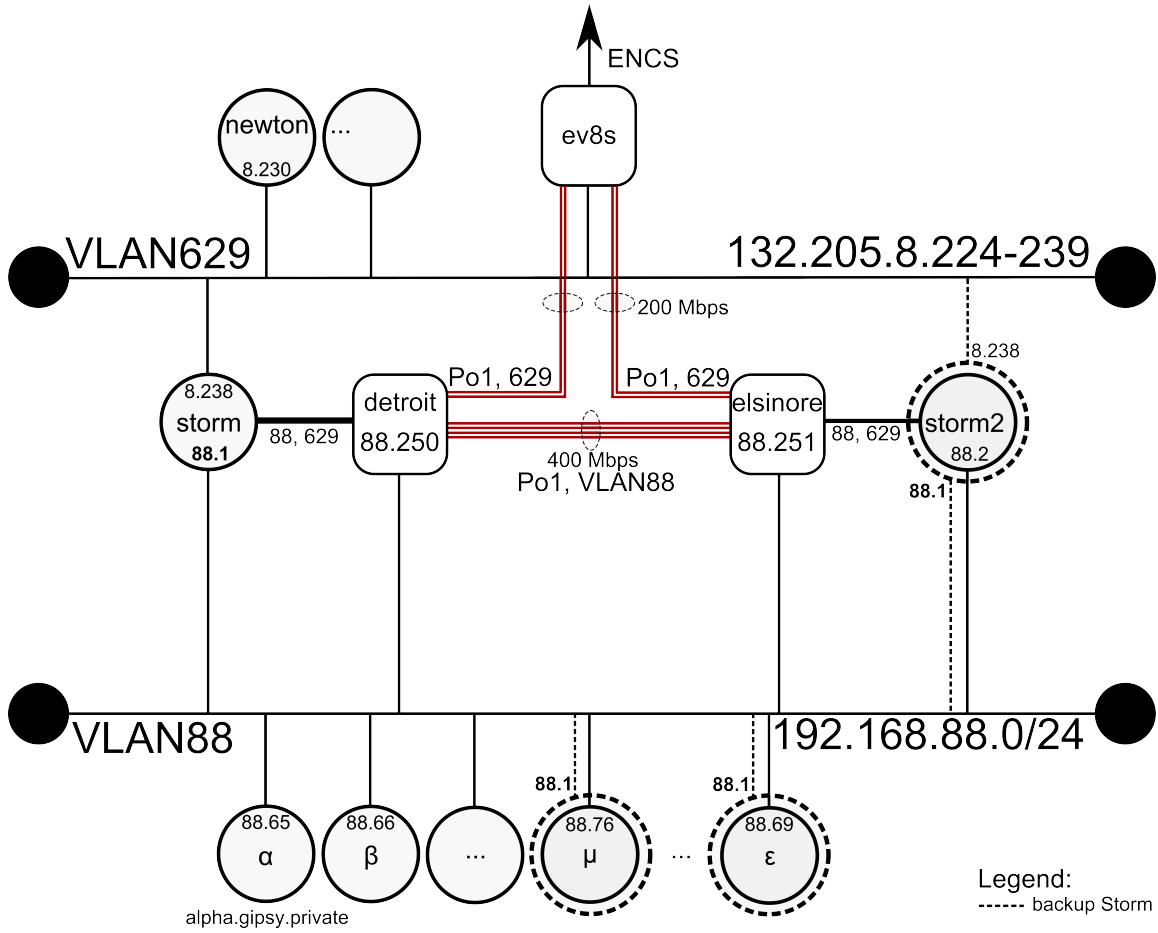
Figure 69: GIPSY cluster network topology

## 8.7 Summary

We described architectural design details for the evaluating system of GIPSY along with the related work and PoC implementation aspects. This includes the necessary updates, modifications, and contributions to the design of the *GIPSY Type System*; compiler (GIPC) and the semantic analyzers framework; the run-time system (GEE) and the additional engines for FORENSIC LUCID: intensional/eductive backend (traditional evaluation), ASPECTJ (whose primary role here is of tracing and human-readable trace presentation), and PRISM model-checking backend to represent a probabilistic state machine similar to Gladyshev's; and the testing framework (to automate compilation and semantic analysis tests of the new FORENSIC LUCID and old LUCX components and JUnit tests to make sure they remain valid within the FORENSIC LUCID system). We provided examples of encoders that provide translation of data structures (or log entries or probes in the case of MSA in Section 9.5, page 257) into the FORENSIC LUCID format in order to do follow-up investigations either manually or automatically (e.g., in self-forensics first briefly introduced in Section 1.2, page 4 and Section 2.1, page 24 and detailed in Appendix D). The JavaCC grammar specification of the FORENSIC LUCID parser `ForensicLucid.jjt` is in GIPSY's source code repository. Its generated compiler sources are integrated in `gipsy.GIPC.intensional.SIPL.ForensicLucid`. Finally, we described the GIPSY compute cluster physical design and network architecture to support larger scale experiments.

# Part III

# Conclusion

## Chapter 9

# Evaluation Applications

This chapter presents example applications of FORENSIC LUCID based on the pertinent case studies discussed prior (Section 2.2.5, page 43) and further. We first present an overview (Section 9.1), including the background description of the data sources used (Section 9.1.1) and the concept of misuse cases (use cases with hostile intent) in Section 9.1.2 followed by various examples demonstrating the use of FORENSIC LUCID constructs. We then summarize our findings in Section 9.6.

## 9.1 Overview

After the data and misuse case description we present the DSTME illustratory example encoded in FORENSIC LUCID in Section 9.2. We then rewrite Gladyshev's COMMON LISP implementation of the *ACME Printing Case* and the *Blackmail Investigation* with the FORENSIC LUCID programs presented in this chapter in Section 9.3, page 250 and Section 9.4, page 254. Additional cases other than these classical two are also presented, including the design of the elaborate live MAC spoofer analysis and investigation in Section 9.5, page 257.

### 9.1.1 Data Sources and Data Description

In this section we provide some background information on the data and data sources used for some of the case-related experiments in this thesis. These data complement the research carried out for the data mining-related aspects presented in Chapter 5 including aspects that have to do with the data gathering (such as the classification data from the classifiers, or network and file data, or log data, or data taken with the live probes) and encoding it (such as, e.g., from MARF and MARFCAT in Section 8.5.1.1, Section 9.5, Appendix D.4.6) into FORENSIC LUCID for the purposes of subsequent analysis and automated reasoning, and evaluation. Data are also something we make inferences from when representing knowledge and what we visualize for comprehension and management properties (Appendix E).

#### 9.1.1.1 Data Sets

##### 9.1.1.1.1 MARFCAT.

For the MARFCAT-based (Section 5.4, page 110) investigations, we use the SAMATE data set; this includes the source code samples and a number of associated CVEs or CWEs [284, 287]. The SAMATE reference data set contains C/C++, JAVA, and PHP language tracks comprising CVE-selected cases as well as stand-alone cases and the large generated synthetic C and JAVA test cases (CWE-based, with a lot of variants of different known weaknesses). SATE IV expanded some cases from SATE2010 by increasing the version number, and dropped some other cases (e.g., Chrome) [314].

The C/C++ and JAVA test cases of various client and server OSS software are compilable into the binary and object code, while the synthetic C and JAVA cases generated for various CWE entries provided for greater scalability testing (also compilable). The CVE-selected cases had a vulnerable version of a software in question with a list of CVEs attached to it, as well as the most known fixed version within the minor revision number. One of the goals for the CVE-based cases is to detect the known weaknesses outlined in CVEs using static code analysis and also to verify if they were really fixed in the "fixed version" [348]. The cases with known CVEs and CWEs were used as the training models described in the methodology. The summary below is a union of the data sets from SATE2010 and SATE IV [314].

The initial list of the CVEs to locate in the test cases were collected from the NVD [340, 348] for Wireshark 1.2.0, Dovecot, Tomcat 5.5.13, Jetty 6.1.16, and Wordpress 2.0 [314].

The specific test cases with versions and language included CVE-selected [314]:

- C: Wireshark 1.2.0 (vulnerable) and Wireshark 1.2.18 (fixed, up from Wireshark 1.2.9 in SATE2010)

- C: Dovecot (vulnerable) and Dovecot (fixed)

- C++: Chrome 5.0.375.54 (vulnerable) and Chrome 5.0.375.70 (fixed)

- JAVA: Tomcat 5.5.13 (vulnerable) and Tomcat 5.5.33 (fixed, up from Tomcat 5.5.29 in SATE2010)

- JAVA: Jetty 6.1.16 (vulnerable) and Jetty 6.1.26 (fixed)

- PHP: Wordpress 2.0 (vulnerable) and Wordpress 2.2.3 (fixed)

originally non-CVE selected in SATE2010 [314]:

- C: Dovecot

- JAVA: Pebble 2.5-M2

Synthetic CWE cases produced by the SAMATE team:

- C: Synthetic C covering 118 CWEs and $\approx$ 60K files

- JAVA: Synthetic JAVA covering $\approx$ 50 CWEs and $\approx$ 20K files

**9.1.1.1.2 Other Data Sources.** Some experimental, empirical, real and meta data are used from the Faculty of ENCS network, some of which is described in detail in [31]. Various other data sources for encoding and experiments include:

1. Netflows and Argus [390] logs for various network-based investigations.

2. For the *MAC Spoofer Investigation* (Section 9.5, page 257) case in particular the data sources used in probes of the live hosts as well as central logs of host activity, switch activity, services likes DHCP, as well that of monitoring tools [31] and *Nessus* [460] scans.

3. Data sources include faculty-wide installations of Linux `/encs/bin` binary software. The use of these data works in conjunction with the MARFCAT (Section 5.4, page 110) data described earlier in pseudo-hypothetical scenario in vulnerable software investigations.

4. The network-enabled malware investigations used in MARFPCAT (see Section 5.5, page 121) including a very closely recent work [49] used the *GFI SandBox* [132] (since March 2013 known as *ThreatAnalyzer* [470]. It supplies an extensive feed of malware organized into a database and some preliminary analysis reports combined with pcap data traces captured in the sandbox. The database at the time comprised more than 1.5 million malware samples that serve as a good source of machine learning in search for insight and flag suspicious traffic. The malware was run in controlled environment yielding $\approx 100000$ pcap files labeled with hashes of malware [49].

5. *ForensicDataSniffer*-based investigations (the tool is an offshoot of the `fileType` (Section 5.3, page 107) and MARFCAT applications) relies on the common data sets provided by NIST and others such as collections of publicly released government documents to detect such documents and types on hard disk and encode the resulting findings as the results to compose an evidential statement in FORENSIC LUCID. The rest of the data sources in file type analysis come from typical and atypical OS installations to make an addition input of data in an investigation.

### 9.1.2 Misuse Cases

Ian Alexander [9, 10, 521] in 2002 introduced the notion of *Misuse Cases* in software requirements specification [233, 482] extending the terms of Use Case (UC) diagrams to visualize adversarial cases that threaten the normal functionality of the innocent use cases. The misuse cases make the non-functional requirements more apparent telling why and where they occur and how and where they can be mitigated. We make use of the Misuse Case in our UC context diagrams in the *MAC Spoofer Investigation* case study further in Section 9.5.

## 9.2 Toy DSTME Examples in FORENSIC LUCID

```
[bel(es), pl(es)]
where
  evidential statement es = { betty, sally };

  observation sequence betty = { oBetty };
  observation sequence sally = { oSally };

  observation oBetty = ("limb on my car", 1, 0, 0.99);
  observation oSally = ("limb on my car", 1, 0, 0.99);
end
```

Listing 9.1: Limb Example

Here for illustrative purposes we translate the previously presented DSTME example of Shafer in Section 3.3.2.1.2, page 69 in Listing 9.1. The result of this program is an array of two values containing the belief and plausibility of the Betty-Sally evidential statement.

In Listing 9.2 is a quick raining example written in FORENSIC LUCID roughly corresponding to the raining example in Figure 19, page 63. It illustrates several FORENSIC LUCID concepts, such as the use of LUCX simple contexts and tag sets combined with the forensic contexts and lifting in the use of the \intersection operator. Forensic context illustrate either full explicit specification of a raining observation or "plain" raining observation. The observed property itself here is a simple context as well. The result of this program is an entire observation.

In Listing 9.3 the same raining example is rewritten to make the use of the observation $w$ component to represent the raining or not fact. In that listing we dropped the optional timestamp and keywords for the tag sets that are otherwise redundant to specify when it is unambiguous what tag set type to use. This way of representing the "raining" condition can be used to allow expressing rain uncertainty chances other than 0.0 or 1.0 in future weather predictions (e.g., 50% rain chance would be 0.5) or certainty of the sensors that captured the data in the past, or represent that it rained only for a part of the day that day when it did. This representation can be used with belief and plausibility computations accordingly.

```
raining @ [city:"Montreal", month:"Sep", day:4]
where
        dimension city: unordered finite nonperiodic {"Montreal", "Ottawa", "Quebec"};
        dimension day: ordered finite nonperiodic {1 to 31};

        evidential statement es_raining = {os_montreal_raining, os_ottawa_raining,
            os_quebec_raining};

        observation sequence os_montreal_raining =
        {
                ([city:"Montreal", month:"Sep", day:1, raining:true], 1, 0, 1.0, "September
                    1, 2013"),
                [city:"Montreal", month:"Sep", day:2, raining:false],
                [city:"Montreal", month:"Sep", day:3, raining:true],
                [city:"Montreal", month:"Sep", day:4, raining:false],
                [city:"Montreal", month:"Sep", day:5, raining:true],
                [city:"Montreal", month:"Sep", day:6, raining:false],
                [city:"Montreal", month:"Sep", day:7, raining:true],
                [city:"Montreal", month:"Sep", day:8, raining:false],
                [city:"Montreal", month:"Sep", day:9, raining:true]
        };

        observation sequence os_quebec_raining =
        {
                ([city:"Quebec", month:"Sep", day:1, raining:false], 1, 0, 1.0, "September
                    1, 2013"),
                [city:"Quebec", month:"Sep", day:2, raining:false],
                [city:"Quebec", month:"Sep", day:3, raining:true],
                [city:"Quebec", month:"Sep", day:4, raining:true],
                [city:"Quebec", month:"Sep", day:5, raining:true],
                [city:"Quebec", month:"Sep", day:6, raining:false],
                [city:"Quebec", month:"Sep", day:7, raining:false],
                [city:"Quebec", month:"Sep", day:8, raining:false],
                [city:"Quebec", month:"Sep", day:9, raining:true]
        };

        observation sequence os_ottawa_raining =
        {
                ([city:"Ottawa", month:"Sep", day:1, raining:false], 1, 0, 1.0, "September
                    1, 2013"),
                [city:"Ottawa", month:"Sep", day:2, raining:true],
                [city:"Ottawa", month:"Sep", day:3, raining:true],
                [city:"Ottawa", month:"Sep", day:4, raining:true],
                [city:"Ottawa", month:"Sep", day:5, raining:true],
                [city:"Ottawa", month:"Sep", day:6, raining:true],
                [city:"Ottawa", month:"Sep", day:7, raining:false],
                [city:"Ottawa", month:"Sep", day:8, raining:false],
                [city:"Ottawa", month:"Sep", day:9, raining:false]
        };

        // The result is an observation "([city:"Montreal", month:"Sep", day:4, raining:true
            ], 1, 0, 1.0)"
        raining = ([city:#city,month:#month,day:#day] \intersection es_raining);
end
```

Listing 9.2: Raining Example

```
raining @ [city:"Montreal", month:"Sep", day:4]
where
        dimension city: {"Montreal", "Ottawa", "Quebec"};
        dimension day: {1 to 31};

        evidential statement es_raining = {os_montreal_raining, os_ottawa_raining,
            os_quebec_raining};

        observation sequence os_montreal_raining =
        {
                ([city:"Montreal", month:"Sep", day:1], 1, 0, 1.0),
                ([city:"Montreal", month:"Sep", day:2], 1, 0, 0.0),
                ([city:"Montreal", month:"Sep", day:3], 1, 0, 1.0),
                ([city:"Montreal", month:"Sep", day:4], 1, 0, 0.0),
                ([city:"Montreal", month:"Sep", day:5], 1, 0, 1.0),
                ([city:"Montreal", month:"Sep", day:6], 1, 0, 0.0),
                ([city:"Montreal", month:"Sep", day:7], 1, 0, 1.0),
                ([city:"Montreal", month:"Sep", day:8], 1, 0, 0.0),
                ([city:"Montreal", month:"Sep", day:9], 1, 0, 1.0)
        };

        observation sequence os_quebec_raining =
        {
                ([city:"Quebec", month:"Sep", day:1], 1, 0, 0.0),
                ([city:"Quebec", month:"Sep", day:2], 1, 0, 0.0),
                ([city:"Quebec", month:"Sep", day:3], 1, 0, 1.0),
                ([city:"Quebec", month:"Sep", day:4], 1, 0, 1.0),
                ([city:"Quebec", month:"Sep", day:5], 1, 0, 1.0),
                ([city:"Quebec", month:"Sep", day:6], 1, 0, 0.0),
                ([city:"Quebec", month:"Sep", day:7], 1, 0, 0.0),
                ([city:"Quebec", month:"Sep", day:8], 1, 0, 0.0),
                ([city:"Quebec", month:"Sep", day:9], 1, 0, 1.0)
        };

        observation sequence os_ottawa_raining =
        {
                ([city:"Ottawa", month:"Sep", day:1], 1, 0, 0.0),
                ([city:"Ottawa", month:"Sep", day:2], 1, 0, 1.0),
                ([city:"Ottawa", month:"Sep", day:3], 1, 0, 1.0),
                ([city:"Ottawa", month:"Sep", day:4], 1, 0, 1.0),
                ([city:"Ottawa", month:"Sep", day:5], 1, 0, 1.0),
                ([city:"Ottawa", month:"Sep", day:6], 1, 0, 1.0),
                ([city:"Ottawa", month:"Sep", day:7], 1, 0, 0.0),
                ([city:"Ottawa", month:"Sep", day:8], 1, 0, 0.0),
                ([city:"Ottawa", month:"Sep", day:9], 1, 0, 0.0)
        };

        // The result is an observation "([city:"Montreal", month:"Sep", day:3], 1, 0, 1.0)"
        raining = ([city:#city,month:#month,day:#day] \intersection es_raining);
end
```

Listing 9.3: DSTME Raining Example

## 9.3 ACME Printing Case in FORENSIC LUCID

### 9.3.1 Investigative Analysis

The investigative analysis follows the same principles used in Gladyshev's FSA/LISP approach described in Section 2.2.5.1, page 44 (cf., the COMMON LISP implementation in [137]).

We remodel the case in FORENSIC LUCID as further illustrated in this section. Resulting FORENSIC LUCID program fragments corresponding to this case are in Listing 9.5 modeling $\psi$, Listing 9.6 modeling $\Psi^{-1}$, and Listing 9.4 modeling the "main" program with all initial evidential declarations.

### 9.3.2 Sample Forensic Lucid Specification

The simulated printer case is specified in FORENSIC LUCID as follows. $\psi$ is implemented in Listing 9.5. We then provide the implementation of $\Psi^{-1}$ [304] in Listing 9.6. Finally, the "main program" is modeled in Listing 9.4 that sets up the context hierarchy and the invokes $\Psi^{-1}$. This specification is the translation of the COMMON LISP implementation by Gladyshev described earlier [137] and described in Section 2.2.5.1 in semi-structured English [312].

```
alice_claim @ es
where
  // Consistent evidential statement
  evidential statement es = { printer, manuf };

  observation sequence printer = F;
  observation sequence manuf = {Oempty, $};
  observation sequence alice = {Oalice, F};

  observation F = ("B_deleted", 1, 0);
  observation Oalice = (P_alice, 0, INF+);
  observation Oempty = ("empty", 1, 0);

  // No "add_A" event per claim from Alice
  dimension P_alice : unordered finite nonperiodic {"add_B", "take"};

  dimension S : unordered finite nonperiodic {"empty", "A_deleted", "B_deleted"};

  alice_claim = invpsiacme[S](es \union alice);

  // ... function declarations
end
```

Listing 9.4: Developing the Printing Case: "main" [304, 305, 312]

**The "Main Program"**

In Listing 9.4 is where the computation begins in our FORENSIC LUCID example. This is an equivalent of `main()`, or the program entry point, in other (mainstream) languages like JAVA, C, or C++. The goal of this fragment is to setup (declare) the context of evaluation, which is core to the case—the evidential statement `es`. The `es` is the highest level dimension in LUCID parlance, and it is hierarchical. This is an unordered list (set) of stories and witness accounts of the incident (themselves known as observation sequences); their ordering is arbitrary. The es is a context where in ⟨*dimension* : *tag*⟩ pairs dimensions are denoted by their identifiers `alice`, `printer`, and `manuf`, and their tags are declared nested values, i.e., expanding the declarations, we arrive at `alice:{Oalice, F}`, `printer:{F}`, and `manuf:{Oempty, $}`. The relevant stories to the incident are that of Alice (`alice`), the evidence of the printer's final state as found by the investigator (`printer`), and the "expert testimony" by the manufacturer of how the printer works. These observation sequences are in turn defined as ordered collections of observations nesting one level deeper into the context. The printer's final state dimension `F` is the only observation for the printer found by the investigator, which is an observation of the property of the printer's queue "Bob's job deleted last" syntactically written as "B_deleted" as inherited from Gladyshev's notation. Its duration `((..., 1, 0))` merely indicates that it was simply present. The `manuf` observation sequence is dictated by the manufacturer's specification that the printer's queue state was empty initially for an undetermined period of time (`$`) when the printer was delivered. These are two observations, following sequentially in time. Alice's timeline (also two observations) is that from the beginning Alice did not perform any actions signified by the computation properties $P$ such as "add_B" or "take" (implying the computation "add_A" has never happened (0 duration for the "infinity", i.e., until the investigator examined the printer). This is effectively Alice's claim. Thus, `alice_claim` is a collection of Boolean results for possible explanations or lack thereof for Alice's claim in this case at the context of all the declared evidence and as subsequently evaluated by `invpsiacme` ($\Psi^{-1}$). If Alice's claim were to validate, their results would be "true" and "false" otherwise [312]. As we know, her claim was not consistent with the evidential statement (Figure 13, page 51).

**Modeling the Forward Transition Function $\psi$**

In Listing 9.5 is $\psi$ illustrating the normal forward flow of operations to model the scene. It is also a translation from the COMMON LISP implementation by Gladyshev [137] using the FORENSIC LUCID syntax and operators described earlier ([304], Section 7.3, page 166). The function is modeled per manufacturer specifications of the printer queue. "A" corresponds to "Alice" and "B" to "Bob" along with their prompted queue actions to add or delete print jobs. The code is a rather straightforward translation of the FSA/LISP code from [137]. `S` is a collection of computational all possible properties observed. `c` is a "computation" action to add or take print jobs by the printer's spooler [312].

```
  acmepsi[S](c, s)
  where
    d1 = first s;
    d2 = second s;

    // Add a print job from Alice
    if c == "add_A"
    then
      if d1 == "A" || d2 == "A" then s
      else
        // d1 in S
        if (d1 \in S) then "A" fby d2
        else
          if (d2 \in S) then d1 fby "A"
          else s fi;
        fi;
      fi;

    // Add a print job from Bob
    else if c == "add_B" then
      if (d1 == "B") || (d2 == "B") then s
      else
        if (d1 \in S) then "B" fby d2
        else
          if (d2 \in S) then d1 fby "B"
          else s fi;

    // Printer takes the job per manufacturer specification
    else if c == "take"
      if d1 == "A" then "A_deleted" fby d2
      else
        if d1 == "B" then "B" fby d2
        else
          if d2 == "A" then d1 fby "A_deleted";
          else
            if d2 == "B" then d1 fby "B_deleted";
            else s fi

    // Done
    else s fby eod fi;
  end; // psi
```

Listing 9.5: Transition function $\psi$ in FORENSIC LUCID for the ACME Printing Case

**Modeling the Inverse Transition Function $\Psi^{-1}$**

In Listing 9.6 is the inverse $\Psi^{-1}$ backtracking implementation with the purpose of event reconstruction, also translated from COMMON LISP to FORENSIC LUCID like the preceding fragments using the FORENSIC LUCID operators. It is naturally more complex than $\psi$ due to a possibility of choices (non-determinism) when going back in time, so these possibilities have to be explored. This backtracking, if successful, for any claim, would provide the Gladyshev's "explanation" of that claim, i.e., the claim attains its meaning and is validated within the provided evidential statement. $\Psi^{-1}$ is based on the traversal from F to the initial observation of the printer's queue as defined in "main". If such a path were to exist, then Alice's claim would have had an explanation. `pby` (*preceded by*) is the FORENSIC LUCID inverse operator of classical LUCID's `fby` (*followed by*). `backtraces` is an array of event backtracing computations identified with variables; their number and definitions depend on the crime scene and are derived from the state machine of Gladyshev [312].

The transition functions presented have to be written by the investigator. While they are significantly shorter and more manageable than the COMMON LISP implementation by Gladyshev [137], the usability of the language can still be improved by allowing visual DFG composition of these resorting to the normal $\psi$ flow as a part of the future work (Appendix E). $\Psi^{-1}$ is defined by the investigator exploring known possibilities from the final state to the possible preceding states; if the graph is available, following the transitions back from the final state.

## 9.4 Blackmail Case in FORENSIC LUCID

The blackmail case example of the implementation steps (Section 2.2.5.2, page 51) modeled in FORENSIC LUCID is in Listing 9.7. At the top of the example we construct the hierarchical context representing the evidential statement and comprising observations. The syntax is made to relate to the mathematical description of Gladyshev's FSA, but with the semantics of FORENSIC LUCID. Any event property can also be mapped to a human-readable description via `=>` that can be printed out in a trace. `invtans` corresponds to $\Psi^{-1}$; given all states, the evidential statement, and Mr. A's claim as an argument it attempts to find possible backtrace explanations within the disk cluster model. The function `trans` corresponds to $\psi$ [307].

```
invpsiacme[S](s)
where
  backtraces = [A, B, C, D, E, F, G, H, I, J, K, L, M ];

  d1 = first s;
  d2 = second s;

  A = if d1 == "A_deleted"
      then d2 pby "A" pby "take" else bod fi;
  B = if d1 == "B_deleted"
      then d2 pby "B" pby "take" else bod fi;
  C = if (d2 == "A_deleted") && (d1 != "A") && (d2 != "B")
      then d1 pby "A" pby "take" else bod fi;
  D = if (d2 == "B_deleted") && (d1 != "A") && (d2 != "B")
      then d1 pby "B" pby "take" else bod fi;
  // d1 in S and d2 in S
  E = if (d1 \in S) && (d2 \in S)
      then s pby "take" else bod fi;
  F = if (d1 == "A") && (d2 != "A")
      then
        [ d2 pby "empty" pby "add_A",
          d2 pby "A_deleted" pby "add_A",
          d2 pby "B_deleted" pby "add_A" ]
      else bod fi;
  G = if (d1 == "B") && (d2 != "B")
      then
        [ d2 pby "empty" pby "add_B",
          d2 pby "A_deleted" pby "add_B",
          d2 pby "B_deleted" pby "add_B" ]
      else bod fi;
  H = if (d1 == "B") && (d2 == "A")
      then
        [ d1 pby "empty" pby "add_A",
          d1 pby "A_deleted" pby "add_A",
          d1 pby "B_deleted" pby "add_A" ]
      else bod fi;
  I = if (d1 == "A") && (d2 == "B")
      then
        [ d1 pby "empty" pby "add_B",
          d1 pby "A_deleted" pby "add_B",
          d1 pby "B_deleted" pby "add_B" ]
      else bod fi;
  J = if (d1 == "A") || (d2 == "A")
      then s pby "add_A" else bod fi;
  K = if (d1 == "A") && (d2 == "A")
    then s pby.d "add_B" else bod fi;
  L = if (d1 == "B") && (d2 == "A")
    then s pby "add_A" else bod fi;
  M = if (d1 == "B") || (d2 == "B")
    then s pby "add_B" else bod fi;
end; // inv
```

Listing 9.6: Inverse transition function $\Psi^{-1}$ in FORENSIC LUCID for the ACME Printing Case

```
os_mra @ es_blackmail
where
  // Core context of evaluation
  evidential statement es_blackmail = { os_final, os_unrelated };

  // List of all possible dimension tags
  dimension W : unordered finite nonperiodic
  {
    "(u)", "(t1)", "(o1)",
    "(u,t2)", "(u,o2)",
    "(t1,t2)", "(t1,o2)",
    "(o1,t2)", "(o1,o2)"
  };

  dimension Wd : unordered finite nonperiodic
  {
    "d(u,t2)", "d(u,o2)", "d(o1)", "d(t1,t2)",
    "d(t1,o2)", "d(o1,t2)", "d(o1,o2)"
  };

  I = W \union Wd;

  // Mr. A's story
  observation sequence os_mra = { $, o_unrelated_clean, $, o_blackmail, $ };

  // Crime scene description
  observation sequence os_final = { $, o_final };
  observation sequence os_unrelated = { $, o_unrelated, $, \0(I), $ };

  observation o_final = ("(u,t2)" => "threats in slack of unrelated letter", 1);
  observation o_unrelated_clean = ("(u,o1)", 1);

  backtraces = invtrans[I](es_blackmail \union os_mra);

  trans[I](c, s)
  where
    if( c == "(u)" && s.#.# == ("(o1,o2)", 0))
    then ("(u,o2)", 1) fby trans[next I](next s.#);

    else if( c == "(u,t2)" && s.#.# == ("(u,o2)", 1))
    then ("(u,t2)", 2) fby trans[next I](next s.#);

    else if( c == "d(u,t2)" && s.#.# == ("(u,o2)", 1))
    then ("(u,t2)", 1) fby eod;

    else if( c == "(u)" && s.#.# == ("(u,t2)", 2))
    then ("(u,t2)", 1) fby eod;
  end;

  invtrans[I](es)
  where
    backtraces = [ A, B ];
    A = o_final       pby [es.#, I:"(u)"]
        ("(u,t2)", 2) pby [es.#, I:"(u,t2)"]
        ("(u,o2)", 1) pby [es.#, I:"(u)"]
        ("(o1,o2)", 0);

    B = o_final pby [es.#, I:"d(u,t2)"] ("(u,o2)", 1) pby [es.#, I:"(u)"] ("(o1,o2)", 0);

    invtrans[next I](next es.#]);
  end;
end
```

Listing 9.7: Blackmail Case modeling in FORENSIC LUCID

## 9.5 MAC Spoofer Investigation

> *There are programs that can be used to spoof an address but to physically change it is hard. The address is coded into the [network] device on the computer. If you changed it to another one, there is a chance of conflict with a valid address on some other device. Should you get a major hub device, you will be getting a service disconnect as soon as they trace it. Depending on what you are doing when detected, you may get visited by people driving plain looking cars, wearing nice suits with bulges under the coats. —Yahoo! Answers user "Laid back"[1]*

The case presented here is designed to work in an operational environment and covers a significant number of the FORENSIC LUCID features in encoding the related evidence from the realistic data. The author Mokhov designed this case and its resulting programming artifact of gathering and encoding of the evidence as a part of his work in the Faculty network administration group (NAG).

A *Media Access Control* (MAC) address [180][2] spoofer[3] analysis and investigation has to do with the automated processing and reasoning about the *possible MAC spoofer* alerts on the Faculty of ENCS network [31]. It's fairly easy to spoof a MAC address by changing the network card settings on a PC or a laptop or via a virtual machine in most operating systems, or most house-grade routers that allow setting an arbitrary MAC address for a variety of legitimate needs [329]. However, on the analyst-managed subnets a MAC spoofer presents a great security concern if a MAC address of a MAC-based port security setting of a tightly secured desktop in a managed lab or office gets spoofed by a visiting laptop (which the user is in full control of) as it may potentially gain unauthorized access to restricted services/resources and/or increase the Faculty's liability if, for example, the laptop is infested with malware.

The existing *MAC Spoofer Watcher* tool, `msw`, presently does a lot of work in detecting such possibilities and alerting the network group's security RT (*RT: Request Tracker* [504][4]) queue (along with cellphone notifications) of a possible ongoing incident after a series of probes and checks. It goes out of its way to filter out the likely false positives (identified commonly through several iterations of the deployed tool in daily network operations). However, despite all the filtering, more often than not, a false positive report is generated. Analysts have to manually (and sequentially) double-check if it is a real spoofer or a false positive right after the alert is generated to the RT ticket as well as to the analysts' cellphones at any time of the day or night. A possible slew of such events can overwhelm the limited human resources dealing with such events while attending to many other duties.

---

[1] http://answers.yahoo.com/question/index?qid=20080322213119AAY8yUf
[2] http://en.wikipedia.org/wiki/MAC_address
[3] http://en.wikipedia.org/wiki/MAC_spoofing
[4] http://www.bestpractical.com/?rt=4.0.8

The `msw`'s processing is triggered by switch port events registered in the `switchlog` that `msw` observes, when LINK-UP events occur at Layer 2 (of the OSI model [79]). If all of its preliminary false-positiveness tests incrementally fail, `msw` generates the RT email alert at the time. This email is designed to trigger a follow-up investigation by the MSA tool presented in this section to automate the process and establish the likelihood of a genuine spoofer or a false positive with high confidence to assist the network security crew in their daily operations by supplying the results of the analysis and event reconstruction back to the same RT ticket.

This follow-up investigation includes two types of digital forensics: live and "dead" (passive) [59, 252, 373], simultaneously (see also Section 2.1, page 24). The process re-examines the evidence used by `msw` as well as performing additional live probes as explained further. The live forensics includes re-examining the switch port status, ARP [382][5] cache, `ping`, and `finger` replies, etc. In parallel, the past logs on the switch port link activity, user activity syslog, DHCP logs, netflows surrounding the time of the incident are also examined. These tools, daemons, and their logs are the witnesses and their witness accounts of the relevant events are a part of the evidential statement in our case. All evidence is gathered in the common format of FORENSIC LUCID, and is fed to the "analysis engine" (i.e., GIPSY presented earlier in Chapter 6).

It is possible that more than one RT ticket/alert is generated for the same host in a short period of time—currently that means all "offences" will be processed possibly concurrently with the gradually growing evidence with each offence. That means the current and the recent past evidence will have a disadvantage of multiple handlings of logically the same case;

---

[5]http://en.wikipedia.org/wiki/Address_Resolution_Protocol

the subsequent runs, however, would gather all the previous and the new evidence, thereby potentially increasing the confidence in the subsequent analyses. Currently, there is no synchronization between the multiple concurrently running investigations working independently on the same logical case; however, the "phase 2" design calls to leverage GIPSY's implementation of the eductive computation model (Chapter 6), where the already-computed event reconstruction results at the same forensic contexts (Section 7.2.3.1.3, page 164) are cached in the DST (Section 6.2.2.1.7, page 147) speeding up later runs. This implies, a GIPSY instance should be running on a standby waiting to process such things, whereas currently we spawn a GIPSY instance.

### 9.5.1 MAC Spoofer RT Alert

In essence, when a MAC spoofer report/RT ticket (see, e.g., a shortened version in Figure 70) arrives, it constitutes a primary claim "there is a MAC spoofer" with some initial facts and evidence from the ticket (encoded in FORENSIC LUCID as exemplified in Listing 9.9). We want to automatically verify its true- or false-positiveness to some degree of certainty by gathering all the supporting evidence from various log files (e.g., `switchlog`, activity log, argus/netflow data, etc.) in one common format of FORENSIC LUCID and reason about it. Subsequently, we send a follow-up report message with the analysis and detail event reconstruction results. In case of a likely true positive, other things are to be possibly automatically verified/established, such as who is the likely perpetrator and their malignity, whether NFS access was attempted, etc.

The proposed analyzer of such reports is designed to work complimentary to the mentioned MAC Spoofer Watcher (`msw`) tool [31].

### 9.5.2 Speculative Claim Evaluation of *Possible*

Here we present what we call a *two-claim solution* to the possible MAC spoofer analysis problem. The automation assumes two main parallel routes of evaluation of a *possible* MAC spoofer by evaluating two derived claims speculatively and simultaneously: (1) "there is a MAC spoofer" and (2) "the spoofer claim is a false positive". The expected outcomes to be

```
From: NOBODY@encs.concordia.ca
Subject: [encs.concordia.ca #259835] Possible MAC spoofer: flucid.encs.concordia.ca (switch1 Fa0/16)
Date: Wed, 1 Aug 2012 07:07:07 -0400

Wed Aug 01 07:07:07 2012: Request 259835 was acted upon.
Transaction: Ticket created by NOBODY@encs.concordia.ca
       Queue: EX-Security
     Subject: Possible MAC spoofer: flucid.encs.concordia.ca (switch1 Fa0/16)
       Owner: Nobody
  Requestors: NOBODY@encs.concordia.ca
      Status: new

123.45.67.89    flucid.encs.concordia.ca
Room: H123
Jack: 22
Switch Port: switch1 Fa0/16
PID of the testing macspoofwatch flucid.encs.concordia.ca [123.45.67.89] (switch1 Fa0/16): 22814
Test data:
    result   refused
```

Figure 70: Example RT ticket alerting of a possible MAC spoofing attempt

proved with the event reconstruction trace are either (a) "(1) is true and (2) is false", or (b) "(1) is false and (2) is true". (a) or (b), if successful, are then said to be consistent, in essence, cross-validating and supplementing each other. Such a parallel evaluation can also give preliminary partial results faster and is novel in comparison to traditional approaches [135, 136, 137].

There are of course pathological cases of high conflict of (c) and (d) where both (1) and (2) are either true or false respectively, leading to contradiction (the claim is true and it is a false positive or the claim is false and it is not a false positive). This means either the evidence is insufficient or the "crime scene" modeled is not correct. Both (1) and (2) work with the same evidence (analogous to the cases built by the prosecution and defence in a court of law scenario), but their starting claims are of opposing views.

### 9.5.3 Report Analysis by Human Experts

A human expert manually examining the evidence and doing an investigation traditionally goes sequentially through some of the steps occasionally taking shortcuts by omitting some of the steps if something is obvious or performing them to increase confidence in the analysis results:

1. Check the switch port is still up

2. Check the host responds to `telnet` with `memory` and `hardware` commands (the custom commands are designed to return expected values)

3. Check `switchlog` for LINK-UP events regarding the switch port in question

4. Delete ARP entry for the host from the ARP cache with `arp` (for further checks)

5. Check the host responds to `ping`, `finger` (both should respond as expected, else the firewall configuration is non-standard)

    (a) If `ping` answers, check the TTL values for likely Linux (64) or Windows (128) hosts [6], or someone plugged in a router (63 or 127, i.e., decremented by 1 at each hop) [39, 456]

    (b) No answer to `ping` nor ARP cache entry re-appears, likely machine no longer up due to a quick reboot or ghosting; often leading to a false-positive

    (c) No answer to `ping`, but ARP query returns successfully. A good indicator of a likelihood of a real MAC spoofer (the machine acquired an IP, is up, but does not reply as it should)

6. Check the host responds to `nbtscan` for Windows hosts (should return one of the known strings)

7. Attempt to `ssh` to the host (should accept certain logins and respond appropriately)

8. Check activity log for boot and login events for the host

9. Optionally check for swpvios for this port in the recent past

10. Check argus/netflow logs for connections made from the host after the spoofer report for potential illicit network activity

---

[6]`http://www.kellyodonnell.com/content/determining-os-type-ping`

### 9.5.4 Augmented Parallel Algorithm

We propose to automate and enhance the manual process presented in the preceding section with a parallel algorithm for MSA that does extra verification, evidence gathering, and subsequent reporting, all in near-realtime.

The augmented checking/investigation algorithm for the automated solution is in Algorithm 3. In the general case, it consists of both live- (probing and checking the active MAC-spoofer-accused on the network) and dead-forensics (examining some of the logs after the fact) techniques simultaneously. It should be noted while the algorithm is depicted in the traditional sequential manner, many of steps related to both live and dead forensics data gathering are designed to be done concurrently, which will commonly yield a performance improvement compared to the human expert doing the same.

Additional evidence is gathered with an `nmap` scan to fingerprint the OS and ports. Then, gather the DHCP requests made to see if the offending laptop by default (suddenly) queries DHCP for the IP while not too long ago the legitimate host did its own DHCP requests (with known intervals or, if rebooted frequently, with the additional log entries "booted successfully in OS"). This will not work if the laptop has been intentionally configured with static IP, but could provide additional confidence, if the DHCP was used (which is the default in many cases). An additional query to `nbtscan` for additional evidence, *à la* `msw` is also made.

### 9.5.5 Use and Misuse Cases

The UML Use Case diagram in Figure 71 depicts most common use and misuse cases of the ENCS network in the context of MAC spoofer investigations. Misuse cases are in gray. Following Alexander's notion of Misuse Cases for requirements specification ([9], Section 9.1.2, page 247) to visualize the threat and mitigation scenarios as well as the primary and supporting actors involved. A lot of investigative work is shared between the network administrator and the *MAC Spoofer Analyzer* (MSA), where the latter automatically gathers all the evidence in one place, encodes, invokes the analysis computation via GIPSY and then generates two reports: the summarized evidence and designed upcoming automated reasoning report. The MAC spoofer misuse cases illustrate the typical actions a person with a UM laptop would

```
 1  begin
 2  │   Bootstrap by an RT alert notification of a possible MAC spoofer via a procmail handler;
    │   // In the below live and dead forensics processings in the begin-end blocks is done parallel
    │       processes, including the individual checks.
    │   // Each check/evidence collector as a product, encodes its output in FORENSIC LUCID as an
    │       observation sequence.  All such sequences are later combined into an evidential statement for
    │       further analysis.
    │   // Live network MAC spoofer forensics
 3  │   begin
 4  │   │   Check the switch port is still up;
    │   │   // SL5 and Windows 7 only should be there
 5  │   │   Check host OS and ports open with nmap;
 6  │   │   Check how the host responds to telnet with memory and hardware commands;
    │   │   // Should respond as expected, else the firewall configuration is non-standard
 7  │   │   Delete ARP entry for the host from the ARP cache with arp;
 8  │   │   Check how the host responds to ping, finger;
 9  │   │   begin
10  │   │   │   if ping answers then
11  │   │   │   │   Check the TTL values for likely Linux (64) or Windows (128) hosts, or someone plugged in a router
    │   │   │   │   (63 or 127);
12  │   │   │   end
13  │   │   │   else
14  │   │   │   │   No answer to ping nor ARP cache entry re-appears, likely machine no longer up due to a quick
    │   │   │   │   reboot or ghosting; often leading to a false-positive;
15  │   │   │   end
    │   │   │   // May increase confidence in false-positiveness
16  │   │   │   begin
17  │   │   │   │   Reaffirm possible boot and patching in the activity log;
18  │   │   │   end
19  │   │   │   No answer to ping, but ARP query returns successfully. A good indicator of a likelihood of a real MAC
    │   │   │   spoofer (the machine acquired an IP, is up, but does not reply as it should);
20  │   │   end
21  │   │   Check how the host responds to nbtscan;
    │   │   // Should be allowed internally
22  │   │   Attempt to ssh to the host;
23  │   end
    │   // ``Dead'' network MAC spoofer forensics
24  │   begin
25  │   │   Check switchlog for LINK-UP events regarding the switch port in question;
26  │   │   Check activity log for boot and login events for the host;
27  │   │   Optionally check for swpvios [31] for this port in the recent past;
28  │   │   Check argus/netflow logs for connections made from the host after the spoofer report for potential illicit
    │   │   network activity;
29  │   │   Check the DHCP requests for the host;
30  │   end
    │   // Analysis
31  │   begin
32  │   │   Gather evidence from all the checks and encode it in FORENSIC LUCID;
33  │   │   Invoke analysis engine;
34  │   end
35  │   Generate a report;
36  end
```

**Algorithm 3:** Augmented MAC spoofer checking/investigation algorithm

do to spoof a MAC address. While presently MSA is not downing the switch port, the design includes this use case to be complete; this action depends on the result of the analysis and confidence in it. In Section 9.5.8 is the description of the corresponding sequence diagram (Figure 73).

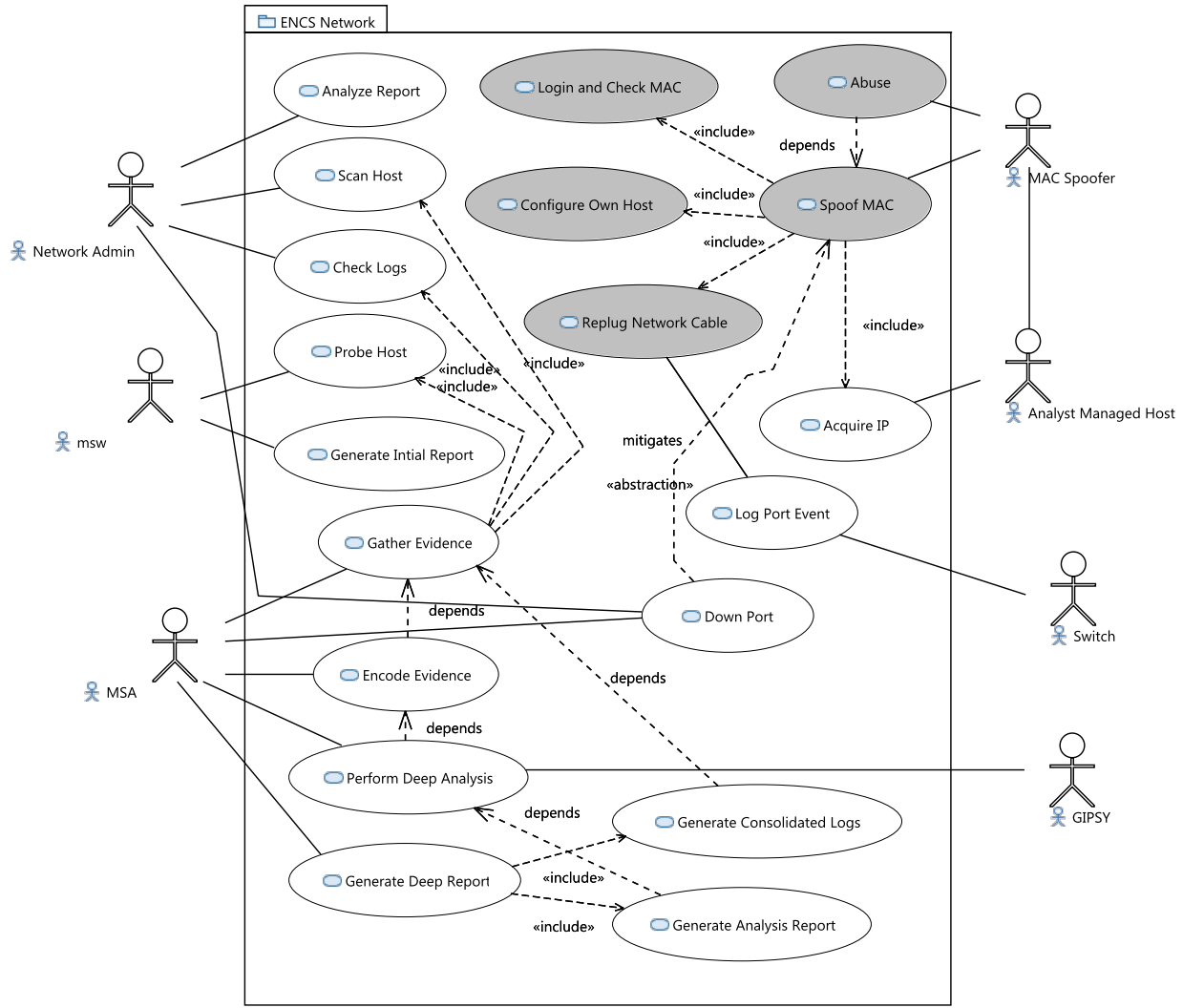

Figure 71: Use and Misuse Cases in MAC spoofer investigations

### 9.5.6 False Positives

The typical false positive spoofer detections include:

1. Unusually slow booting (and often patching right afterward and rebooting of) Windows hosts.

2. Older outdated images of OS's that were rarely booted to and unpatched/not re-imaged for a long time.

3. Accidental user-managed (UM) host (with the UM image) on a trusted virtual LAN (VLAN) [31, 70] (either mistaken OS image/IP address or switch port configuration settings).

4. Host in the middle of ghosting or in the ghosting console without a recognized NetBIOS string on a trusted VLAN.

5. Exceptional printers on the trusted VLAN.

6. Malfunctioning host (hard drive or memory going bad).

While not all can be handled in an automated manner easily, the proposed MSA tool is designed to be able to identify cases 1, 2, and 6 in its follow-up investigation.

### 9.5.7 Components

The MSA system's design consists of the following components/modules acting together in unison (sequentially or in parallel):

1. Procmail [481] handler

2. RT ticket email to FORENSIC LUCID encoder

3. Live and log data collector/FORENSIC LUCID encoder (activity, switchlog, argus [390], DHCP, and others)

4. (under design consideration) Network database query component for previously known information on the hosts, locations, switch ports, and MAC addresses

5. FORENSIC LUCID processor (GIPSY) invocator

6. Output/conclusion report generator

#### 9.5.7.1 Procmail Handler

This component consists of two parts: the .procmailrc recipe for procmail [481][7] (see Figure 72) looking for `Possible MAC spoofer` and `From:` matching a regular expression `.*(nobody|root(\+[a-z\d-]+)?|nag(db)?)@` [403] and handing over a copy of that RT

---
[7]http://en.wikipedia.org/wiki/Procmail

ticket email to the handler (the variable `$HOME` is appropriately set to the installation directory); and the handler is a script that (1) parses the `Subject:` header for RT ticket number, host, switch and port number, (2) checks in its internal storage if this ticket has already been handled to avoid duplicate or useless handling of replies or comments to the ticket by people in case the `From:` didn't filter enough in (1), (3) if this is a new spoofer report claim, save the RT email into a `.rt` file, which is a text file with the name of the RT ticket number containing the RT email, e.g., `123456.rt` (see Figure 70 for example), (4) `ssh` to primary compute and control host and start the Collector/Encoder component there.

```
:0c:mac-spoofer-analyzer.lock
* ^Subject: .*Possible MAC spoofer
* ^From: .*(nobody|root(\+[a-z\d-]+)?|nag(db)?)@
| $HOME/bin/mac-spoofer-alert-handler | ...
```

Figure 72: Procmail handler trigger recipe (rule)

#### 9.5.7.2 Collector/Encoder

The collector/encoder script consists of multiple modules and subcomponents that collect information from various data sources (mostly logs or probes on different hosts) and convert them into an evidential statement *es* context format encoded in FORENSIC LUCID. The script, `mac-spoofer-evidence-collector`, is a multiprocess PERL application that calls upon various modules (listed below) to gather the corresponding evidence available.

There are three primary API methods that the main collector/encoder dynamically discovers and calls for each of the following modules are `collect()`, `encode()`, and `getType()`. The `collect()` method implements a specific data gathering mechanism (from a log file on some host or a direct live probe around the *initial contextual point* of interest in meta-LUCID terms[8] as initial filtering criteria), and stores the gathered data (RT-number.collector-type). The `encode()` methods selectively encode the collected data into FORENSIC LUCID with some preprocessing and filtering following the Occam's razor principles and normalization (for easier co-relation of timestamps and other data items of the same type, such as MAC

---

[8]i.e., as a $\mathcal{P}$ (Section 7.2.3, page 159) analogy in `mac-spoofer-evidence-collector` in PERL

addresses). Something that is not encoded or not recognized by the encoder for a particular data log, is still preserved either as a simple observation in an observation sequence or in a comment section of the corresponding FORENSIC LUCID fragment, such that a human investigator can review and fine-tune it later. The `getType()` is simply a way for the main collector/encoder to tell apart which module was called to use as an extension. All encoded files have an additional extension of `.ctx`, that is context files (that have observation sequences of interest). The main script then check-sums all the collected and encoded data with `sha1sum`.

1. `RTForensicLucidEncoder` is the module that is the very first to be invoked to parse and encode the initial RT ticket into a claim (see Listing 9.9 as an example). In particular, it extracts the date, hostname, IP, switch, and port context point data that are used for subsequent lookups in the related logs, live data, and database entries that follow to gather all relevant contextual evidence surrounding the initial RT report. The other concrete collector/encoder modules search for data temporally near this context point.

2. `SwitchLogForensicLucidEncoder` uses the date, switch, and port information to look for related events in the `switchlog` and encode the events of interest such as link state changes, *swpvio*s, etc.

3. `ActivityLogForensicLucidEncoder` uses the date and host information to look for host bootups, shutdowns, patching, and user logins/logouts.

4. `MSWForensicLucidEncoder` encodes log entries produced by `msw`. The older `msw` did not produce non-STDERR log file entries, only errors and email notifications. To be able to replay some of the data (that are normally real-time) from the past, we need an event trace/decision log. This is especially relevant for development and testing on past cases while no actual spoofing is going on. Thus as a part of this work, `msw` was itself augmented to add extra logging functionality, and, while at it, making it in a more FORENSIC LUCID-friendly manner to simplify extraction by logging sub-expressions directly in the FORENSIC LUCID context format. This approach compensates somewhat for the fact that real-time data for `nbtscan` and other probes and the like are not available in the offline playback. Furthermore, those data serve as additional evidential observations to improve the confidence in the subsequent analysis.

5. `DHCPLogForensicLucidEncoder` uses the date and host information to look for the host's DHCP requests and the corresponding DHCP protocol exchange messages.

6. `ARPPingForensicLucidEncoder` uses `arp` and `ping` to gather the live host presence evidence, which is especially useful in the presence of a firewall on the spoofing host.

7. `NmapForensicLucidEncoder` uses `nmap` to gather the live OS and open ports evidence.

8. `FingerForensicLucidEncoder` uses `finger` to gather the live OS and login evidence.

9. `SSHForensicLucidEncoder` uses `ssh` to gather the live "genuineness" test evidence.

10. `ArgusForensicLucidEncoder` uses Argus commands to gather netflow summaries, etc., as evidence of network activity prior the report arrival time $t_{ar}$, such as at least $t_{ar} - t_{linkup}$ primarily because a number of probes are done between the LINK-UP event [68, 70, 343, 344] and the ticket arrival time to weed out the majority of false positives.

11. `NbtscanForensicLucidEncoder` uses `nbtscan` to gather the live NetBIOS evidence (for Windows hosts) for work groups and the MAC address.

12. `SWMForensicLucidEncoder` uses the switch management `swm` [31] to check the live current port status at the time of the check if it is up.

13. `EncodingUtils` module was developed as a helper module for many of the above to uniformly encode data such as timestamps, MAC addresses, and hostnames, which sometimes vary in their lexical representation in different logs or probes. It, therefore, has `timestampformat()`, `macformat()`, and `hostnameformat()` methods. For example, timestams formatted similar to `"Jul 7 15:10:18"`, `"2013-07-07 16:24 EDT"`, `"2013-07-07 13:10:55497751"`, or `"201307020450"`, become formatted like `"Tue Jul`

9 08:56:56 2013" in the human-readable form for reporting and as a long epoch integer internally. Likewise, MAC addresses have different legal lexical representations used by different vendors, like `00bb335588ff` in raw, `0:bb:33:55:88:ff` in Argus (stripping leading zeroes), `00-BB-33-55-88-FF` by Microsoft, `00bb.3355.88ff` by Cisco are folded into the DCHP ethers format `00:bb:33:55:88:ff`. Hosts are simply formatted into fully-qualified domain names (FQDNs).

**9.5.7.2.1  Encoding the RT Ticket Claim.**  The RT ticket claim's evidence is primarily the following:

- Ticket arrival timestamp $t_{ar}$, e.g., `Wed, 1 Aug 2012 07:07:07 -0400`, serves as a contextual point in time from which to move forward and backward to extract relevant events in other evidential sources.

- *ipaddr* – possible spoofer's IP address; filter for logs that use IP addresses only. This works well if the spoofer uses the same IP address as the legitimate host, via DHCP or statically. This partially breaks if the spoofer changes an IP address to be of another host on the same subnet after checking, which IPs are unused. (Such behavior is a sign of extreme deliberation and possibly malice of someone who knows what they are doing as opposed to "scriptkiddie" tools for mere "Internet access" with a laptop.) This case can still be caught via the other evidential sources and logs that do not use IPs, and primarily host/switch/port based checks. The extra `arp` checks for the investigation if the MAC address in DHCP matches the expected IP address talking on the port or not, will confidently tell if the spoofer is genuine.

- *hostname* – possible spoofer's DNS name; filter for logs that use hostnames.

- *switch* – uplink switch DNS name where the host is/was connected to.

- *port* – uplink port on the switch where the host is/was connected to.

- *mac* – the MAC address being spoofed; for lookups in DHCP, switch, database, and other sources that have it.

```
// ...
observation msw_o = (Pspoofer, 1, 0, 1.0, t_ar);
observation sequence msw_claim = { $, msw_o };
// ...
observation sequence msw_counter_claim = { $, not(msw_o), $ };

Pspoofer =
{
    [host:
    {
      [hostname:"flucid.encs.concordia.ca"],
      [IP:"123.45.67.89"],
      [mac:"00:aa:22:bb:44:cc"],
      [room:"H123"],
      [jack:"22"]
    }
    ],
    [switch:"switch1"],
    [port:"Fa0/16"],
    [t_ar:"Wed, 1 Aug 2012 07:07:07 -0400"]
};
// ...
```

Listing 9.9: MAC Spoofer Analyzer's RT evidential statement context encoding example

The secondary context, encoded hierarchically (primarily for reporting to humans and possibly other uses) includes a room number, jack number on the wall, and (possibly in the future) hardware/memory/OS information extracted.

The primary context point is used to construct the two-claim solution (see Section 9.5.2) and gather the evidence from the log files surrounding that context filtered by $t_{ar} \pm 24hrs$, *ipaddr*, *hostname*, *switch*, *port*, and *mac* further. The RT two-claim is made indirectly by `msw`, as a "prosecution witness".

**9.5.7.2.2  Encoding Log and Probe Evidence.**   Collection of other pertinent evidence depends on the context of the initial report described in the preceding section. The modules described earlier collect both live and dead evidence from probes and logs. Not everything possible is collected to avoid unnecessary complexity in encoding and evaluation, so only information is kept that is likely to be helpful in the decision making. The data selection is made by the typical criteria a human expert selects the data for the same task. Following the examples presented earlier in Section 8.5.2, page 236, we illustrate some of the examples of the encoded evidence in FORENSIC LUCID for the modules mentioned. Context calculus operators (Section 7.3.3, page 183) help with additional filtering of deemed unnecessary data during computation.

In Listing 9.10 is the `msw` evidence encoding example. In Listing 9.11 is the empty (no-observation) activity log example. In the regular cases, activity log features operating system booted and users logged on. `perp_o` in the presence of users can be set to the most likely perpetrator in a follow-up investigation. `finger` and `ssh` live evidence supplements the activity log evidence with similar information on expected banners and possible users, if available. No-observations are recorded similarly.

```
// ...
// msw evidence, encoded: Jul 22 09:17:31 2013
observation sequence msw_os =
{
  msw_encsldpd_o_1,
  msw_ghost_o_2,
  msw_arp_o_3
};

observation msw_encsldpd_o_1 = ([switch:"switch1",port:"FastEthernet0/1",ipaddr:"
    132.205.44.252",hostname:"flucid-44.encs.concordia.ca",encsldpd:false], 1, 0, 1.0, "Jul
    13 14:33:37 2013");
observation msw_ghost_o_2 = ([switch:"switch1",port:"FastEthernet0/1",ipaddr:"132.205.44.252
    ",hostname:"flucid-44.encs.concordia.ca",ghost:false], 1, 0, 1.0, "Jul 13 14:33:38 2013"
    );
observation msw_arp_o_3 = ([switch:"switch1",port:"FastEthernet0/1",ipaddr:"132.205.44.252",
    hostname:"flucid-44.encs.concordia.ca",arp:true], 1, 0, 1.0, "Jul 13 14:33:39 2013");
// end of msw evidence
// ...
```

Listing 9.10: `msw` encoded evidence example

```
// ...
// activity log evidence, encoded: Jul 22 09:17:25 2013
observation perp_o = $;
observation sequence activity_os =
{
  activity_o_1
};

observation activity_o_1 = $;
// end of activity log evidence
// ...
```

Listing 9.11: `activity` encoded no-observation evidence example

Live probes by `nmap` (and similarly complementary by `nbtscan`) give a list of open ports and other aspects that are compared to the minimum expected ports and other values. In Listing 9.12 is an example of captured and encoded `nmap` evidence. The samples of ignored lines are in the comment section; they play no role in evaluation but recorded anyway in case the investigator wants to include some of that data later on.

In Listing 9.13 is an example of the evidence encoded from the DHCP logs to supplement the investigation and provide visibility into the situation.

```
// ...
// 'nmap' evidence, encoded: Jul 14 21:09:23 2013
observation sequence nmap_os = os_nmap_entries;

observation sequence os_nmap_entries =
{
  ([protocol_port:135, protocol:"tcp"], 1, 0),
  ([protocol_port:139, protocol:"tcp"], 1, 0),
  ([protocol_port:445, protocol:"tcp"], 1, 0),
  ([protocol_port:49152, protocol:"tcp"], 1, 0),
  ([protocol_port:49157, protocol:"tcp"], 1, 0),
  ([protocol_port:6002, protocol:"tcp"], 1, 0),
  ([protocol_port:49153, protocol:"tcp"], 1, 0),
  ([protocol_port:49154, protocol:"tcp"], 1, 0),
  ([protocol_port:49156, protocol:"tcp"], 1, 0),
  ([protocol_port:7001, protocol:"tcp"], 1, 0),
  ([protocol_port:7002, protocol:"tcp"], 1, 0),
  ([protocol_port:49155, protocol:"tcp"], 1, 0),
  ([protocol_port:135, protocol:"tcp"] => "open   msrpc         Microsoft Windows RPC", 1, 0),
  ([protocol_port:139, protocol:"tcp"] => "open   netbios-ssn", 1, 0),
  ([protocol_port:445, protocol:"tcp"] => "open   netbios-ssn", 1, 0),
  ([protocol_port:6002, protocol:"tcp"] => "open   http          SafeNet Sentinel License
      Monitor httpd 7.3", 1, 0),
  ([protocol_port:7001, protocol:"tcp"] => "open   tcpwrapped", 1, 0),
  ([protocol_port:7002, protocol:"tcp"] => "open   hbase-region Apache Hadoop Hbase 1.3.0 (
      Java Console)", 1, 0),
  ([protocol_port:49152, protocol:"tcp"] => "open   msrpc         Microsoft Windows RPC", 1,
      0),
  ([protocol_port:49153, protocol:"tcp"] => "open   msrpc         Microsoft Windows RPC", 1,
      0),
  ([protocol_port:49154, protocol:"tcp"] => "open   msrpc         Microsoft Windows RPC", 1,
      0),
  ([protocol_port:49155, protocol:"tcp"] => "open   msrpc         Microsoft Windows RPC", 1,
      0),
  ([protocol_port:49156, protocol:"tcp"] => "open   msrpc         Microsoft Windows RPC", 1,
      0),
  ([protocol_port:49157, protocol:"tcp"] => "open   msrpc         Microsoft Windows RPC", 1,
      0),
  ([mac:"00:13:72:xx:xx:xx"] => "(Dell)", 1, 0),
  ([os:"Microsoft Windows 7|2008"], 1, 0),
  ([hops:1], 1, 0)
};

// Unencoded data
/*
  Starting Nmap 6.25 ( http://nmap.org ) at 2013-07-14 21:08 EDT
  NSE: Loaded 106 scripts for scanning.
  NSE: Script Pre-scanning.
  Initiating ARP Ping Scan at 21:08
  Scanning xxx.xxx.xx.xx [1 port]
  Completed ARP Ping Scan at 21:08, 0.00s elapsed (1 total hosts)
  Initiating SYN Stealth Scan at 21:08
  Scanning xxx.xxx.xx.xx [1000 ports]
*/
// end of 'nmap' evidence
// ...
```

Listing 9.12: `nmap` encoded evidence example

#### 9.5.7.2.3 Collection/Encoding Summary.

All the encoded evidence, e.g., for the ticket RT12345 is saved into the appropriate files: `12345.rt.ctx` (e.g., as in Listing 9.9,

```
// ...
// DHCP evidence, encoded: Jul 14 21:08:13 2013
observation sequence dhcpd_os =
{
  dhcp_log_o_1,
  dhcp_log_o_2,
  dhcp_log_o_3,
  dhcp_log_o_4,
  dhcp_log_o_5
};

observation dhcp_log_o_1 = ([dhcpmsg:"DHCPOFFER", direction1:"on", ipaddr:"xxx.xxx.xx.xx",
    direction2:"to", mac:"xx:xx:xx:xx:xx:xx", via:"xxx.xxx.xx.x"], 1, 0, 1.0, "Jul 14
    11:58:03 2013");
observation dhcp_log_o_2 = ([dhcpmsg:"DHCPREQUEST", direction1:"for", ipaddr:"xxx.xxx.xx.xx"
    , dhcpd:"xxx.xxx.xx.xxx", direction2:"from", mac:"xx:xx:xx:xx:xx:xx", via:"xxx.xxx.xx.x"
    ], 1, 0, 1.0, "Jul 14 11:58:03 2013");
observation dhcp_log_o_3 = ([dhcpmsg:"DHCPACK", direction1:"on", ipaddr:"xxx.xxx.xx.xx",
    direction2:"to", mac:"xx:xx:xx:xx:xx:xx", via:"xxx.xxx.xx.x"], 1, 0, 1.0, "Jul 14
    11:58:03 2013");
observation dhcp_log_o_4 = ([dhcpmsg:"DHCPINFORM", direction1:"from", ipaddr:"xxx.xxx.xx.xx"
    , via:"xxx.xxx.xx.x"], 1, 0, 1.0, "Jul 14 11:58:07 2013");
observation dhcp_log_o_5 = ([dhcpmsg:"DHCPACK", direction1:"to", ipaddr:"xxx.xxx.xx.xx", mac
    :"xx:xx:xx:xx:xx:xx", via:"bond0"], 1, 0, 1.0, "Jul 14 11:58:07 2013");
// end of DHCP evidence
// ...
```

Listing 9.13: `dhcp` log encoded evidence example

`12345.switchlog.ctx`, `12345.activity.ctx`, `12345.nmap.ctx`, and others plus the incident modeling transition functions $\psi$ (forward tracing) and $\Psi^{-1}$ (optimized backtracing, see Section 7.5.4 for their description and definition) in the file `mac-spoofer-transition.ipl` for the case into the case file `12345.spoofer.ipl` for the primary claim "there is a spoofer" and `12345.notspoofer.ipl` as "defence" claim that "this is a false positive" for parallel evaluation.

In case of reasonable true-positiveness, the design calls for subclaims to be created and evaluated as well: `12345.perp.ipl`, `12345.nfs.ipl` to determine who (*attribution*) and how malicious they are based on the previously extracted evidence (e.g., via activity and Argus logs).

At the end of its operation, Collector/Encoder (after checksumming everything) passes all the collected and encoded data to the FORENSIC LUCID processor (see the following section) to do the actual reasoning and event reconstruction computations. A GIPSY instance is spawned per claim to be evaluated.

#### 9.5.7.3 FORENSIC LUCID **Processor**

The FORENSIC LUCID Processor presently includes the `mac-spoofer-flucid-processor` script feeding the encoded evidence to GIPSY (see Section 6.1, page 129 and Chapter 8) that has a FORENSIC LUCID parser and a distributed run-time system, implemented in JAVA. This component is designed to do the heavy weight computation. The MSA design includes a provision to act on the results of analysis if the confidence is high by, e.g., shutting down the switch port or quarantining the IP address.

#### 9.5.7.4 **Output/Conclusion Generator**

In MSA, this corresponds to the `mac-spoofer-reporter` to report the findings to system administrators. Presently, 1 is already in production; 2 reports in the development environment.

- Decision tree, findings, conclusions; mail with the proper RT subject (under active development).

- Multi-stage reporting mailings as data become available:

    1. Gathered evidence first; grouped together in one place. (This is already in operation and is of help earlier on to any active network security people watching in.)

    2. Analysis, that may computationally take a longer time, will be be delivered in a follow-up analysis update.

### 9.5.8 **Sequence Diagram**

The UML sequence diagram shown in Figure 73 depicts the design of a series of synchronous and asynchronous calls by the modules involved. It is based on the components mentioned earlier in Section 9.5.7.2 and further, their API, and on Algorithm 3. All the encoders work asynchronously as child processes (`fork()`) started by `mac-spoofer-evidence-collector`. This is because there is no particular ordering for their execution required (except for the RT ticket encoding as there is a data dependency on the context point from it used by the other modules as well as preferential start up of live forensics modules first to do their probes sooner). All modules produce intermediate raw files and encoded timestamped files; the latter are subsequently collected into one FORENSIC LUCID script when all the modules return, which is then fed to GIPSY to compile and evaluate for reasoning purposes. When GIPSY returns, the final analysis report is generated.

### 9.5.9 Challenges and Limitations

- Short-lived spoofer sessions, so not much live forensic data are available.

- Need to measure true negatives, false negatives, false positives.

- Sysadmin mistakes (assigning a managed VLAN or replacing the OS image from managed to UM without updates to the VLAN), which is nearly impossible to tell apart from genuine spoofing.

### 9.5.10 Related Work

There is hardly any related work of that nature. However, naturally sysadmins and security community alike tend to produce tools for various degrees of automation.

On of such open-source tools is `mac-san` [3] from Andrew Adams at Pittsburgh Supercomputing Center designed to scan hosts on subnets and VLANs for vulnerabilities. It uses both SNMP (to poll switches) and ICMP (via `nmap`) to determine active hosts on the network and then rush at them with Nessus [460]. It uses a database to for historical security scans as well as previous associations between MACs and IPs. It also has a notification machinery if the scan fails for some reason in the last two scans or so. It also checks for MACs/IPs tuple mismatches between scans. It is roughly equivalent to a combinations of the tools we use from our scripset [31] as well, including Nessus with daily scans of last active hosts that are recorded regularly into the database. Like us, they allow for scan exceptions for a given MAC/IP address.

A commercial equivalent of what we are doing includes the recently emerged FireSIGHT with a FireAMP agent from SourceFire that is normally installed on all client computers.

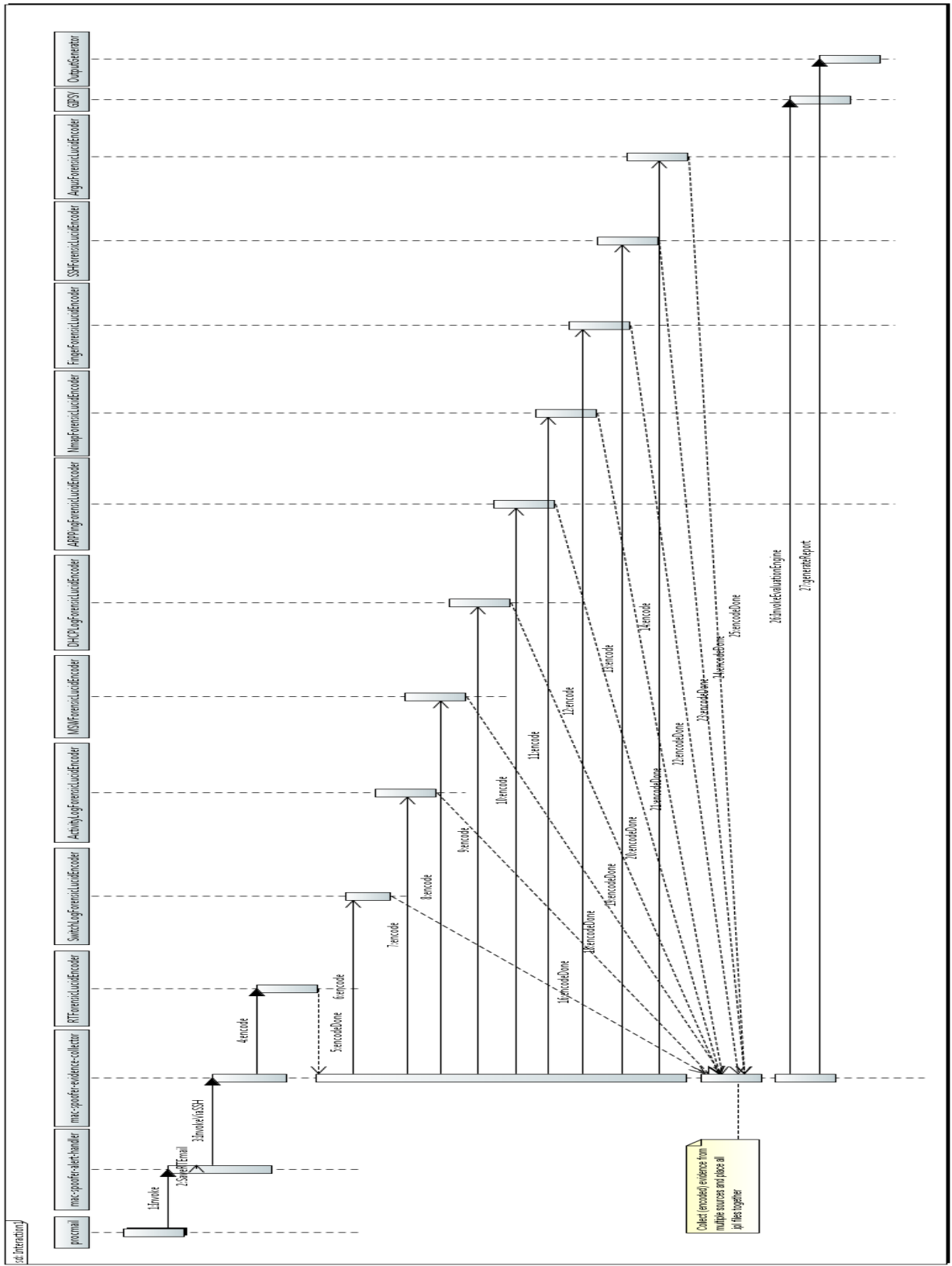

Figure 73: MAC spoofer analyzer UML sequence diagram

We are considering possible integration of FireAMP with our toolset to define an appropriate security policy to mandate FireAMP's presence on all machines, to cover the user-managed land when connecting to our network, and deny connection if after a certain grace time period FireAMP is not responsive from a given UM IP.

### 9.5.11 Ongoing and Future Work

MSA is already a part of the nagtools set [31] in production that is being actively designed, developed, and maintained in iterative builds. This ongoing process includes the FORENSIC LUCID reasoning aspects as well as immediate future plans that follow:

- Plan to provide a direct hook for `msw` to invoke the reasoner not via RT. In this case one can group relevant `.ctx`, etc. files by the hostname, IP, or MAC addresses instead of RT ticket numbers. Some adjustments will need to be made about duplicate handling for the same host for possible prior investigations.

- Plan to provide a hook to `swm` to shutdown the switch port in question in case of a high confidence in the detection of the spoofing activity.

- Plan to provide a hook to the account management in case of high severity and high confidence, autoblock the account of the offender (long term future goal).

## 9.6 Summary

We presented a few application examples of use of Forensic Lucid to show the use of some of its intensional and probabilistic constructs.

The toy examples in the beginning encoding simple cases or statements follow by a more concrete detailed case of the *MAC Spoofer Investigation* MSA tool design where numerous evidence sources are encoded into Forensic Lucid format exhibiting both intensional and probabilistic DSTME features.

Code size reduction from Common Lisp ([137]) to Forensic Lucid (Section 2.2.5.1.3, page 48) is approximately 4-fold (545 lines of Common Lisp (not including DOT Graphviz visualization output of event reconstruction) vs. 113 of Forensic Lucid). Forensic Lucid code is arguably more readable and easier to maintain.

We briefly summarized the data sources used of various types data used in investigation experiments and tests.

As of March 2013, the implementation in automatic gathering and alerting of the related evidence in real-time in one place received positive feedback from NAG avoiding a lot of manual lookups and data gathering outlined in Section 9.5.3.

## Chapter 10

# Concluding Remarks and Future Work

This thesis addresses a notion of digital forensics investigation of swaths of digital and non-digital evidence, case and crime scene modeling, and event reconstruction of a crime scene involving computing equipment incorporating credibility and trustworthiness factors of the evidence and the witnesses into the equation. In crime investigation the investigators focus on reconstructing what has possibly happened after arriving at the scene, gathering evidence, and interviewing witnesses. Digital investigation bears some similarities with the traditional process including proper handling of the evidence (storage, logs) reports, etc. However, the problem with digital evidence is that there are too many devices and too much data to humanly analyze to attempt to reconstruct what has happened, especially in a crime involving computer and network equipment. FORENSIC LUCID is provided, along with a proposed design of the evaluation GIPSY platform, to alleviate those difficulties based on science and formal methods (to be credible in court of law), to represent the knowledge of the case, including evidential description, witness accounts, assign credibility values to them, all in a common shape and form and then validate hypotheses claims against that evidential data.

We likewise presented in-depth background setting for this work in Part I to provide the necessary historical context on the tools and techniques used, data sources, and where the approach comes from as well as to acknowledge the work of predecessors and to provide a reasonable self-contained reference resource to the readers. We also included complete definition of the syntax and semantics of FORENSIC LUCID and further explored relevant case studies, design and implementation aspects.

## 10.1 Objectives Achieved

In dereference to the primary thesis objectives stated in Section 1.5.1 we have designed the FORENSIC LUCID incident specification/scripting language, its formal syntax and semantics, and related case studies. Specifically:

- FORENSIC LUCID (Chapter 7)

    - FORENSIC LUCID syntax (Section 7.3, page 166)
    - FORENSIC LUCID operational semantics (Section 7.4, page 192)
    - Hierarchical higher-order context specification (Section 7.2.2, page 158)
    - Operators and transition functions (Section 7.3.4, page 191)
    - Observation specification with credibility (Section 7.4.2, page 197)

- FORENSIC LUCID parser and semantic analyzer (Section 8.1, page 211)

- Design of the run-time evaluation framework and FORENSIC LUCID encoders (Section 8.2, page 213)

- Example cases (see Chapter 9, page 244)

## 10.2 Summary

Through the series of discussions, definitions of the syntax, operators, semantics, and some examples of the FORENSIC LUCID applications we reached a milestone to show the benefits of the intensional approach over the FSA. As far as the implementing system concerned, it has advantages of parallelizing the computation and introduces the notion of context and eduction that are absent in the FSA/LISP approach of Gladyshev *et al.* [135, 136, 137].

Context and eduction allow a better expression of or constraints on the evidence as well as its more optimized evaluation respectively. We took advantage of the existing concepts, syntax and semantic rules and constructs from the intensional programming and the Lucid family of languages [161, 304, 305, 362].

We presented the fundamentals of Forensic Lucid, its concepts, ideas, and dedicated purpose—to model, specify, and evaluate digital forensics cases. The process of doing so is simpler and more manageable than the previously proposed FSA model and its Common Lisp realization (Section 9.6, page 278). We re-wrote in Forensic Lucid (Section 9.3, page 250) one of the sample cases initially modeled by Gladyshev in the FSA and Common Lisp to show the specification is indeed more manageable and comprehensible than the original [137, 312]. We likewise modeled a specification of the example blackmail case Forensic Lucid (Section 9.4, page 254). The notions and notations used in Forensic Lucid re-use some of those used by Gladyshev ([136], Section 2.2.5.2, page 51) achieving uniformity and comparability of the approaches [307]. We have likewise written a few simple DSTME-oriented examples illustrating a concrete use of various language constructs together matching with the background-set examples in Section 9.2, page 248.

We implemented a number of encoders and encoding techniques to automatically represent the evidence from various data sources in the Forensic Lucid format as a case knowledge base for the designed case studies. The most complete of these is the *MAC Spoofer Analyzer* case detailed in Section 9.5, page 257 and MARFCAT (Section 5.4, page 110).

Our language is based on more than 35 years of research on correctness and soundness of programs [25] and the corresponding mathematical foundations of the Lucid language (Chapter 4), which is a significant factor should a Forensic Lucid-based analysis be presented in court in the future. The logical formalisms behind Forensic Lucid date even further back in time (see Chapter 3).

We likewise presented a modular intensional programming research platform, GIPSY, for reasoning tasks of HOIL expressions. The concept of context as a first-class value is central in the programming paradigms GIPSY is built to explore, such as the family of the Lucid programming languages. At the time of this writing, GIPSY has various degree of support for compilation of GIPL, Indexical Lucid, Lucx, JLucid, Objective Lucid, JOOIP, and now FORENSIC LUCID [302]. We augmented the GIPSY Type System to support forensic contexts. We also refactored substantially its GIPC and GEE frameworks to support multiple semantic analyzers and evaluation backends respectively and designed the API for the new backends based on FORENSIC LUCID, PRISM, and ASPECTJ in Chapter 8.

## 10.3 Limitations

We realize by looking at the presented examples the usability aspect is still desired to be improved upon further for easier adoption by investigators, especially when modeling $\psi$ and $\Psi^{-1}$, prompting one of the future work items to address it further [312] using visualization techniques (Appendix E).

In Section 3.3.2.1.3, page 72, we mentioned points of criticism of the Dempster's rule of combination as a belief fusion operator in the DSTME. That rule favors a shared belief between observations of the same property and ignores the others in the evidential statement from different observation sequences. We partially address this with the averaging fusion operator within an observation sequence and cumulative fusion between any two observations. This appears to be appropriate enough for our purposes, but more formal investigation is needed into selection of the most appropriate fusion operator. Thus, we defined the total credibility of an observation sequence as an average of all the weights in this observation sequence ([313], Definition 53, page 192):

$$W_{avg} = \frac{\sum(w_i)}{n} \qquad (10.3.0.1)$$

A less naive way of calculating weights is using some pre-existing functions. What comes to mind is the activation functions used in artificial neural networks (ANNs), e.g. [313]:

$$W_{ANN} = \sum \frac{1}{(1 + e^{-nw_i})} \qquad (10.3.0.2)$$

The witness stories or evidence with higher scores of $W$ have higher credibility. The lower scores, therefore, indicate less credibility or more tainted evidence [313]. Such a representation would coincide well with machine learning and data mining techniques when the credibility weights can be learned and updated throughout the investigative run automatically, as done for ANNs.

Another potential limitation that has emerged is the hardcoded notion of $\geq 1/2$ from [87] meaning to us as "credible enough", such as in use in the $\mathtt{op}_w$ operator types. There may be forensic applications and investigations that require higher thresholds, making $1/2$ an investigator settable parameter. A possible solution to this is to introduce the parameter, e.g., reserved dimension type $w_{DSTME}$ by default evaluating $\mathcal{D}_0, \mathcal{P}_0 \vdash w_{DSTME} : 1/2$ that can be altered at run-time to the desired value.

An issue of the credibility/belief weight assignment has inherent limitations in the reliability of the investigators assigning them or the encoders of the tools assigning them based on expert estimates. As a corollary, the FORENSIC LUCID language with any of its implementing toolsets at this stage form a forensic assistant to an investigator or an expert, and not an absolute proof system.

The credibility weight assignment also impacts the usability issues for an investigator deciding between assigning, e.g., 93% or 97%. A possible solution to this is to add syntactically pre-defined keywords available to investigators to be used consistently, e.g., as presented in research by Kent at CIA in 1964 [206] by quantifying uncertainty with ranges, e.g., `almost_certain` ($93\% \pm 6\%$), `probable` ($75\% \pm 12\%$), `even_chances` ($50\% \pm 10\%$), `probably_not` ($30\%\pm10\%$), `almost_certainly_not` ($30\%\pm10\%$). The ranges can be adapted to be handled with the observation's $\min, \max$ parameters in this case.

It is important to mention, per Section 10.1, page 280, we provide static syntax and semantic analysis and the type system within the GIPSY system at this stage with the run-time completion being a part of the near-future work. Likewise, complete formal mechanization of the language in a proof assistant (started in [300]) is also beyond the scope of this thesis and is also a part of the near-future work to establish a high degree of confidence in the correctness of the semantics of FORENSIC LUCID.

Finally, we have not completed the complexity analysis and the resulting metrics and measurements for the presented work's run-time and scalability. Such analysis and evaluation are multifold. In the LUCID-family, one of such analyses is *rank analysis* (dimensionality of expressions), which helps in run-time optimization and was performed by several research publications in the past (see references in Chapter 4). The GIPSY's scalability of the distributed run-time of GEE has been extensively studied by Ji and colleagues (see references in Chapter 6). The event reconstruction (ERA, including $\psi$ and $\Psi^{-1}$) implementation within the FORENSIC LUCID backend is the other major aspect. We began working on the metrics framework within the GIPSY run-time (where `Metric` itself is designed as an annotation type part of the GIPSY type system) to study such aspects as well as optimize run-time scheduling of GIPSY networks. Thus, rank analysis, run-time scalability analysis, and the ERA complexity analysis (including the individual context operators) form the various facets of complexity analysis to be completed in the future work.

## 10.4  Future Work

Future work will focus on resolving the limitations presented in the preceding section as well as will include ongoing related projects.

In general, the proposed practical approach in the cyberforensics field can also be used to model and evaluate normal investigation process involving crimes not necessarily associated with information technology or any other types of incident investigations [304, 307] as long as there is a way to encode the evidence, hypotheses, and the incident scene. Combined with an expert system (e.g., implemented in CLIPS [400]), it can also be used in training new staff in investigation techniques [305, 307, 312].

There are, therefore, a number of topics of interest to investigate in subsequent research endeavors. The R&D in digital forensics and forensic computing are more relevant and evident in the digital age and digital investigation. The need for standards and automation in investigation processes to aid human investigators is paramount. There is a lot of research to be done still before the formal concepts mature and become a standard practice in the field. A particular aspect under consideration is the idea of *self-forensics* with the proposed FORENSIC LUCID language standard for autonomic software and hardware systems in all kinds of vehicles, craft, and on the Internet (details of that project propositions are in Appendix D).

### 10.4.1 Tool Integration

There are a number of tools that the FORENSIC LUCID approach can be integrated with for better incident investigation. While MSA (Section 9.5, page 257) is already integrated, there are a number of others where it makes sense to integrate or develop plug-ins for:

- Augment common software packages and custom scripts to log centrally (e.g., via syslog) directly in the FORENSIC LUCID format as an option. This includes apache, Tomcat, identd/authd, sshd, iptables, Event Log, etc. Additionally, a plug-in for Splunk [439] to be FORENSIC LUCID-aware in its searches.

- A natural integration with the HIVE (an open infrastructure for malware collection and analysis) [62] as well as with an incident analysis system NICTER and its engines based on data mining techniques [104, 183].

- FORENSIC LUCID is also a knowledge representation language. Translation from other knowledge representation languages, such as Vassev's *KnowLang* is also envisioned for the purposes of a common format for investigations.

- Forensic context specification with FORENSIC LUCID for case analysis, evaluation, and event reconstruction in ubiquitous computing platforms.

- Integration with Ftklipse [225], GATE [462] and JPF ([21, 22, 85], Section 2.1.3, page 28).

- Integration with JDSF ([271, 294], Section D.3.3, page 363).

- Design and implement the Forensic Data Flow Graph tool as an extension to RIPE in GIPSY to improve the usability of the system. The activity of programming a transition function, presented in Section 7.5.4, for non-computer programmer experts may be tedious, but a LUCID program can be modeled as a data-flow graph. In GIPSY, there was an implementation of a DFG tool [89] that allowed basic programming in INDEXICAL LUCID using graphical DFG representation, which provided bidirectional translation from a DFG to INDEXICAL LUCID code and back. We propose to extend the tool to include FORENSIC LUCID constructs and make it more convenient to use for non-programmers. This would enable an easier construction of the transition functions graphically by an investigator [305].

Appendix E is dedicated to further discussion and progress on this aspect.

### 10.4.2 More on Formalization

- Refine the semantics of LUCX's context sets and their operators to be more sound, including *Box*, so some of our formalisms can be based on it.

- `not` applied to a transition event $e \in I$ (where $I$ is a set of all events) would mean $\text{not } e \Leftrightarrow I \setminus e$, i.e., all the other events currently in $I$, except $e$ for easier specification. For example, if $I = \{$"add_A", "add_B", "take"$\}$, $e =$ "add_A", then `not` $e$ would mean $\{$"add_B", "take"$\}$ using `\difference`. $I$ itself is fully specified in the program by the investigator, but could be made to be learned dynamically as the FORENSIC LUCID program executes.

- Complete an RFC[1] on FORENSIC LUCID and release a full standard FORENSIC LUCID specification (similar to that of JAVA [141]).

- **Transition Function as an Operator Stream**.

  In order to model composite transition functions, we will define a stream of operators of application in a scenario. Thus, each path in the graph is a finite stream of application of FORENSIC LUCID operators. The operator streams can be derived from the corresponding DFGs modeled by the investigator. Since TRANSLUCID [378] (mentioned earlier in Section 4.1, page 76) does have notions of function streams and ranges, this opens a possible area of collaboration.

- `cluster` **operator**.

  Discrete statistical contextual points is of interest to define and navigate to represent median and mean clusters (e.g., as in signal processing in MARF).

---

[1]http://www.ietf.org/rfc.html, http://www.rfc-editor.org/

- **HOIFL**.

  With the resulting combination of logics and theories presented here we may arrive at something we call *Higher-Order Intensional Fuzzy Logic* (HOIFL). The founding formalism we borrow are instantiated in FORENSIC LUCID (see Chapter 7). More specifically, a HOIFL variant in this thesis combines intensionality an its concrete instantiation in the GIPSY type system (Appendix B), Dempster–Shafer belief functions, the joined and modified rules of weighted combination of computations in Gladyshev's approach, and intensional programming in LUCID.

  While giving a complete rigorous treatment of HOIFL is beyond the scope of this thesis, we can roughly sketch that HOIFL includes the necessary intensional logic constructs presented earlier, belief mass assignment and probability operators, hierarchical contexts, first-order intensional logic (FOIL), and a type system, in addition to the standard higher-order extensional logic constructs. To visualize HOIFL as a conservative minimal extension:

  $$\boxed{\min(\;\boxed{\text{HOL}}\;\boxed{\text{FOIL}}\;\boxed{\text{Theory of Types}}\;\text{HOIL+DSTME}\;)\;\text{HOIFL}}$$

  A sketch (including Figure 74) of the formalism thus would begin as:

  - Model $M = \langle Q, I, R, D, D^I, m \rangle$.
  - $Q$ is the set of possible worlds (states).
  - $I$ are statements or events.
  - $D$ is a set of intensional objects ([508], Section 3.2, page 59); $D^I$ is the set of their extensions.
  - $R$ is a set of accessibility relations $D^I \to D^I$ (such as transitions, or transition functions, or operators, such as `NEXT`, `FBY`, $\psi, \Psi, \ldots \in R$) between states $q \in Q$.
  - $m$ is belief mass interpretation that is assigned to each object in the domain $w = m(I, \Gamma), \Gamma \in Q$.
  - $V$ is a *valuation function* assigning propositions from $m$ to each world $q$.

  A complete formulation of axiomatization is to be done in the future work beyond this thesis. We can capitalize on a number of earlier and recent referenced work [112]. For

$$M, \Gamma \models_v^w P(x_1, \ldots) \Leftrightarrow V(<(v(x_1), \ldots)>, \Gamma) = w \quad (10.4.2.1)$$

$$M, \Gamma \models_v^w X \wedge Y \Leftrightarrow M, \Gamma \models_v^w X \text{ and } M, \Gamma \models_v^w Y \quad (10.4.2.2)$$

$$\ldots \Leftrightarrow \ldots$$

$$M, \Gamma \models_v^w \Box X \Leftrightarrow M, \Gamma \models_v^w X \text{ for every } \Delta \in Q \text{ with } \Gamma R \Delta \quad (10.4.2.3)$$

$$M, \Gamma \models_v^w \Diamond X \Leftrightarrow M, \Gamma \models_v^w X \text{ for some } \Delta \in G \text{ with } \Gamma R \Delta \quad (10.4.2.4)$$

$$\ldots \Leftrightarrow \ldots$$

$$M, \Gamma \models_v^w \texttt{NEXT}(X) \Leftrightarrow M, \Gamma \models_v^w \lambda n.X(X+1) \text{ and } w \geq 1/2 \quad (10.4.2.5)$$

$$\ldots \Leftrightarrow \ldots$$

$$M, \Gamma \models_v^w P(X) \geq P(Y) \Leftrightarrow M, \Gamma \models_v^w X \text{ at least as probable as } Y \quad (10.4.2.6)$$

$$M, \Gamma \models_v^w P(X) \geq \top \Leftrightarrow X \text{ has a probability of } 1.0 \quad (10.4.2.7)$$

$$M, \Gamma \models_v^w P(X) \geq \neg P(X) \Leftrightarrow X \text{ has a probability of at least } 1/2 \quad (10.4.2.8)$$

$$\ldots \Leftrightarrow \ldots$$

$$M, \Gamma \models_v^w \text{bel}(X) \Leftrightarrow M, \Gamma \models_v^w X \text{ has a belief of } \sum_{B | B \subseteq X} m(B) \leq P(X) \quad (10.4.2.9)$$

$$M, \Gamma \models_v^w \text{pl}(X) \Leftrightarrow M, \Gamma \models_v^w X \text{ has a plausibility of } \sum_{B | B \subseteq X} m(B) \leq P(X) \quad (10.4.2.10)$$

$$\ldots \Leftrightarrow \ldots$$

Figure 74: HOIFL initial sketch formulation

example, Muskens proposed intensional models for the Theory of Types [17, 328] in 2007. Our logic axiomatization will draw from the recent (2003–2011) existing axiomatizations and semantic formalizations of FOIL [113], probabilistic logics [87, 155], ADM process logic [4], Halpern's and Pucella's logic for reasoning about evidence [158].

- Isabelle/HOL [339, 372, 518] is a proof assistant that works with Higher-Order Logics (HOLs). The author is using it to assist with formalizing and proving FORENSIC LUCID constructs, derived from the semantic rules, operators, and core language constructs. The end goal of this exercise is to establish a solid formal provable base and avoid mistakes of an equivalent manual exercise [300]. Thus, we will explore the theoretical results, e.g., soundness, completeness of FORENSIC LUCID, by formally proving of the correctness of the semantic rules of FORENSIC LUCID using Isabelle/HOL [372] and their equivalence to the FSA approach through push-down systems. (As a side benefit, if the proofs are done for core FORENSIC LUCID, they will automatically extend to GIPL, INDEXICAL LUCID, LUCX, and others [305].) Additionally, the culmination of this aspect is to publish the FORENSIC LUCID proof theory (`ForensicLucid.thy`-in-progress along with other LUCID dialects) in the *Archive of Formal Proofs* [209].

### 10.4.3 Distributed Systems, Network and Security Research, and Forensics

- Cf. Section 5.1.2, page 99, it is important to handle digital forensics of IPv6 traffic, including evidence encoding and case analysis with FORENSIC LUCID. As IPv6 is becoming more prominent, IPv6-based security investigations with a lot of unknowns and recently identified security problems [120, 255] are necessary to cope with.

- Similarly to IPv6, the digital investigation, evidence modeling, and case specification for wireless, mobile and ad-hoc networks is another major research area to follow up on.

- Extend the work to digital evidence modeling and investigation in botnets.

- Similarly to the MSA tool, automate Faculty network scanning investigations, where the network is undergoing port-scanning attacks.

- Ben Hamed [159] and Du proposed a scheduler and the corresponding heuristics, graph annotation and analysis as well as their implementation under the scheduler for a distributed intensional evaluation engine. Paquet *et al.* proposed GEE on the other hand within the GIPSY with the baseline language and a set of algebras in the engines. This idea generalizes the notion of the Ben Hamed's scheduler and proposes a compile-time and run-time design within GIPSY to include the scheduler component for various optimization analyses, e.g., rank and graph analyses with the generic annotation support that can be used for various dialects and their intermediate representations.

## 10.5 Acknowledgments

This research work took about 6-7 years to reach this point. Throughout the years the author had the pleasure to work with a lot of people on the various facets of the research, get inspiration from, and who provided diverse support and encouragement. While the author is an inclusionist, please forgive if he forgot to mention someone as there are many people to acknowledge.

First of all, an enormous *thank you* goes to my supervisors, Drs. Joey Paquet and Mourad Debbabi for their extraordinary support and extreme patience and tolerance allowing me to explore various research aspects in this thesis and beyond. Their contribution to the success of this thesis is immense with their combined expertise in intensional programming, distributed systems and frameworks, security and cyberforensics. Thanks to Joey Paquet for extensive in-depth reviewing and comments with vast amounts of patience in doing so. Thanks to Mourad Debbabi for introducing me to digital investigation, PRISM, realistic data and test cases and being an excellent teacher during my M.Eng in Information Systems Security where I discovered my topic in around 2006.

A lot of thanks go to Dr. Peter Grogono, who has been very supportive of my work all the way back from my master's studies and for the recommendation to use the Dempster–Shafer theory. I would like to appreciate the additional examining committee members who agreed to devote their valuable time to judge this thesis: Drs. Weichang Du, Terry Fancott, and Amr Youssef. Thanks to Dr. Patrice Chalin for being there in the past with his critical to-the-point reviews in my master's thesis and the doctoral thesis proposal and the follow-up seminar. I would like to also thank Dr. John Plaice for very thorough and helpful reviews of Chapter 1, Chapter 4, and Chapter 6 in this thesis.

Thanks to Dr. Aiman Hanna for long-time support and advice back from my undergraduate days up to the presentation of this very thesis. Thanks to Dr. Ching Y. Suen whose course back in 2002 inspired the work on MARF and its applications that are still around today (Chapter 5). Another token of acknowledgment goes to Dr. Benjamin Fung for comments on Chapter 5. Thanks to Drs. John McKay, Lingyu Wang, A. Ben Hamza, Chadi Assi, and René Witte for various support of my work through courses, publications, advice, morally, or funding. Thanks to Drs. Sabine Bergler and Leila Kosseim and the CLaC lab for the journey through the internals of natural language processing related to the intensional side of this work. Some of the practical implementation resulting from those courses went into MARF and its applications as well (e.g., [283]). Thanks to Frank Rudzicz [410] as well. Thanks to Dr. Brigitte Jaumard for contributing the node hardware and a rack for the GIPSY cluster design (Section 8.6, page 238).

Big thanks goes to my professional Academic IT Services (AITS) unit in the Faculty of ENCS. A special thanks goes to Michael J. Assels, the manager of Networks and Security who approved and thoroughly reviewed the *MAC Spoofer Analyzer* material (Section 9.5, page 257) in several iterations as a part of my work on this project and my thesis. It's an immense pleasure to work with him as a colleague for my professional AITS duties side of things. Thanks to the rest of the AITS crew! It's been a pleasure to work with all the members of NAG (Michael Spanner and Manny Taveroff), Faculty Information Systems (FIS) group, Desktop Operations Group (DOG), System Administration Group (SAG), User Services and Service Desk (USG and SD). Thanks to Joel Krajden, Stan Swiercz, Francois Carrière, Sigmund Lam, Paul Gill and Frank Maselli among others.

Thanks to many of the past and current GIPSY R&D team members for their valuable contributions, collaboration, suggestions, and reviews, especially Sleiman Rabah, and including Touraj Laleh, Arash Khodadadi, Yi Ji, Bin Han, Dr. Aihua Wu, Dr. Emil Vassev, Xin Tong, Amir Pourteymour, Dr. Kaiyu Wan, Chun Lei Ren, Dr. Bo Lu, Yimin Ding, Lei Tao. The author would like to acknowledge the people previously involved with the MARF and JDSF projects, namely: Stephen Sinclair, Ian Clement, Dimitrios Nicolacopoulos, Lee Wei Huynh, Jian Li, Farid Rassai, and others.

Thanks go out to the Computer Security Lab team members: Amine Boukhtouta, Claude Fachkha, Hamad Binsalleeh, Sujoy Ray, Nour-Eddine Lakhdari, William Nzoukou, and many others for their strong team work and support. Thanks to Andrei Soeanu for presenting the work on the initial *ACME Printing Case* in FORENSIC LUCID [312] at ICDF2C'2011 on behalf of the author.

This research and development work was funded in part by NSERC, the Faculty of Engineering and Computer Science (ENCS), Concordia University (Montreal, Canada), the NCFTA Canada, and NIST SAMATE group (a MARFCAT lecture). AITS donated two switches and provided gigabit connectivity to the GIPSY cluster (Section 8.6, page 238) to ENCS. As a visiting scholar at the Visualization and Graphics Lab, Department of Computer Science and Technology, Tsinghua University, the author was a recipient of the Canada-China Scholars' Exchange Program (CCSEP) scholarship.

Thanks to the open-source communities that produce quality open-source software and inspired the author to do the same. Thanks to the communities of Eclipse, Apache, Open/Libre Office, Linux (and Scientific Linux specifically), various LaTeX distributions and tools, compilers, and many other OSS contributors. Likewise, thanks to various open-access resources, such as Wikipedia, TeX.SE, and Scholarpedia among others.

Of course, to conclude this section it is important to mention my family again who were there for me and suffered my long-duration absences from the family life while supporting me and my wife with home chores as well as our wonderful kids Deschy and Timmy (sorry daddy was very busy absent for long hours)—thank you my parents-in-law and brother-in-law: Fang Liu, Lin Song, and Liu Song— without your support I could not have completed this work.

# Bibliography

[1] A. F. Abdelnour and I. W. Selesnick. Nearly symmetric orthogonal wavelet bases. In *Proc. IEEE Int. Conf. Acoust., Speech, Signal Processing (ICASSP)*, May 2001.

[2] AccessData. FTK – Forensic Toolkit. [online], 2008–2013. `http://www.accessdata.com/products/digital-forensics/ftk`.

[3] A. K. Adams. `mac-scan` – scan hosts on a VLAN or network for vulnerabilities. [online], Pittsburgh Supercomputing Center, 2009–2013. `http://www.psc.edu/index.php/networking/647-mac-scan`.

[4] K. Adia, M. Debbabi, and M. Mejri. A new logic for electronic commerce protocols. *Int J Theor Comput Sci (TCS)*, 291(3):223–283, Jan. 2003.

[5] I. Agi. GLU for multidimensional signal processing. In Orgun and Ashcroft [350]. ISBN: 981-02-2400-1.

[6] D. Agrawal et al. Autonomic computing expressing language. Technical report, IBM Corporation, 2005.

[7] V. S. Alagar, J. Paquet, and K. Wan. Intensional programming for agent communication. In J. Leite, A. Omicini, P. Torroni, and P. Yolum, editors, *Declarative Agent Languages and Technologies II*, volume 3476 of *Lecture Notes in Computer Science*, pages 239–255. Springer Berlin Heidelberg, 2005.

[8] P. Albitz and C. Liu. *DNS and BIND*. O'Reilly, 3 edition, 1998. ISBN: 1-56592-512-2.

[9] I. Alexander. Misuse Cases: Use Cases with hostile intent. [online], Nov. 2002. `http://www-dse.doc.ic.ac.uk/Events/BCS-RESG/Aleksander.pdf`.

[10] I. Alexander and L. Beus-Dukic. *Discovering Requirements*. Wiley, 2009.

[11] J. Allard, V. Chinta, S. Gundala, and G. G. R. III. JINI meets UPnP: An architecture for JINI/UPnP interoperability. In *Proceedings of the 2003 International Symposium on Applications and the Internet 2003*. SAINT, 2003.

[12] B. Allen. Monitoring hard disks with SMART. *Linux Journal*, 117, Jan. 2004. `http://www.linuxjournal.com/article/6983`, last viewed May 2012.

[13] G. Allwein and J. Barwise, editors. *Logical reasoning with diagrams*. Oxford University Press, Inc., New York, NY, USA, 1996.

[14] R. Alshammari and A. N. Zincir-Heywood. Investigating two different approaches for encrypted traffic classification. In *Proceedings of the Sixth Annual Conference on Privacy, Security and Trust (PST'08)*, pages 156–166. IEEE Computer Society, Oct. 2008.

[15] R. Alshammari and A. N. Zincir-Heywood. Machine learning based encrypted traffic classification: Identifying SSH and Skype. In *Proceedings of the IEEE Symposium on Computational Intelligence for Security and Defense Applications (CISDA 2009)*, pages 1—8. IEEE, July 2009.

[16] R. A. Alshammari. *Automatically Generating Robust Signatures Using a Machine Learning Approach To Unveil Encrypted VOIP Traffic Without Using Port Numbers, IP Addresses and Payload Inspection*. PhD thesis, Dalhousie University, Halifax, Nova Scotia, Canada, May

2012.

[17] P. Andrews. Church's type theory. In E. N. Zalta, editor, *The Stanford Encyclopedia of Philosophy*. Stanford, spring 2009 edition, 2009. http://plato.stanford.edu/archives/spr2009/entries/type-theory-church/.

[18] I. Androutsopoulos. Temporal meaning representation in a natural language front-end. In Gergatsoulis and Rondogiannis [131], pages 197–213.

[19] S. Anson, S. Bunting, R. Johnson, and S. Pearson. *Mastering Windows Network Forensics and Investigation*. Sybex, 2 edition, June 2012.

[20] Apache River Community. Apache River. [online], 2010. http://river.apache.org/index.html.

[21] A. R. Arasteh and M. Debbabi. Forensic memory analysis: From stack and code to execution history. *Digital Investigation Journal*, 4(1):114–125, Sept. 2007.

[22] A. R. Arasteh, M. Debbabi, A. Sakha, and M. Saleh. Analyzing multiple logs for forensic evidence. *Digital Investigation Journal*, 4(1):82–91, Sept. 2007.

[23] E. A. Ashcroft. Multidimensional program verification: Reasoning about programs that deal with multidimensional objects. In Orgun and Ashcroft [350], pages 30–41. Invited Contribution.

[24] E. A. Ashcroft, A. A. Faustini, R. Jagannathan, and W. W. Wadge. *Multidimensional Programming*. Oxford University Press, London, Feb. 1995. ISBN: 978-0195075977.

[25] E. A. Ashcroft and W. W. Wadge. Lucid – a formal system for writing and proving programs. *SIAM J. Comput.*, 5(3), 1976.

[26] E. A. Ashcroft and W. W. Wadge. Erratum: Lucid – a formal system for writing and proving programs. *SIAM J. Comput.*, 6(1):200, 1977.

[27] E. A. Ashcroft and W. W. Wadge. Lucid, a nonprocedural language with iteration. *Communications of the ACM*, 20(7):519–526, July 1977.

[28] E. A. Ashcroft and W. W. Wadge. $R$ for semantics. *ACM Transactions on Programming Languages and Systems*, 4(2):283–294, Apr. 1982.

[29] AspectJ Contributors. *AspectJ: Crosscutting Objects for Better Modularity*. eclipse.org, 2007. http://www.eclipse.org/aspectj/.

[30] M. J. Assels. The logic of global conventionalism. Master's thesis, Department of Philosophy, Concordia University, Montreal, Canada, Mar. 1985. Online at http://spectrum.library.concordia.ca/5204/.

[31] M. J. Assels, D. Echtner, M. Spanner, S. A. Mokhov, F. Carrière, and M. Taveroff. Multi-faceted faculty network design and management: Practice and experience. In B. C. Desai, A. Abran, and S. Mudur, editors, *Proceedings of $C^3S^2E'11$*, pages 151–155, New York, USA, May 2010–2011. ACM. Short paper; full version online at http://www.arxiv.org/abs/1103.5433.

[32] AT&T Labs Research and Various Contributors. The DOT language. [online], 1996–2012. http://www.graphviz.org/pub/scm/graphviz2/doc/info/lang.html.

[33] AT&T Labs Research and Various Contributors. Graphviz – graph visualization software. [online], 1996–2012. http://www.graphviz.org/.

[34] F. Baader and H. J. Ohlbach. A multi-dimensional terminological knowledge representation language. *Journal of Applied Non-Classical Logics*, 5(2), 1995.

[35] C. Baier and J.-P. Katoen. *Principles of Model Checking*. Massachusetts Institute of Technology, 2008. ISBN: 978-0-262-02649-9.

[36] M. Bailey, J. Oberheide, J. Andersen, Z. M. Mao, F. Jahanian, and J. Nazario. Automated classification and analysis of Internet malware. Technical report, University of Michigan, Apr. 2007. http://www.eecs.umich.edu/techreports/cse/2007/CSE-TR-530-07.pdf.

[37] C. D. Ball. Helping lawyers master technology. [online], blog, column, publications, 2006–2013. http://www.craigball.com/Ball_Technology.

[38] R. Bardohl, M. Minas, G. Taentzer, and A. Schürr. Application of graph transformation to visual languages. In *Handbook of Graph Grammars and Computing by Graph Transformation: Applications, Languages, and Tools*, volume 2, pages 105–180. World Scientific Publishing Co., Inc., River Edge, NJ, USA, 1999.

[39] R. Bejtlich. *The Tao of Network Security: Beyond Intrusion Detection*. Addison-Wesley, 2005. ISBN: 0-321-24677-2.

[40] J. Bennett. *A Philosophical Guide to Conditionals*. Oxford: Clarendon Press, 2003.

[41] T. Berners-Lee, R. Fielding, U. C. Irvine, and L. Masinter. RFC 2396: Uniform Resource Identifiers (URI): Generic Syntax. [online], Aug. 1998. http://www.ietf.org/rfc/rfc2396.txt, viewed in November 2007.

[42] L. Besnard, P. Bourani, T. Gautier, N. Halbwachs, S. Nadjm-Tehrani, and A. Ressouche. Design of a multi-formalism application and distribution in a data-flow context: An example. In Gergatsoulis and Rondogiannis [131], pages 149–167.

[43] H. Binsalleeh, T. Ormerod, A. Boukhtouta, P. Sinha, A. M. Youssef, M. Debbabi, and L. Wang. On the analysis of the Zeus botnet crimeware toolkit. In *Eighth Annual Conference on Privacy, Security and Trust, PST 2010, August 17-19, 2010, Ottawa, Ontario, Canada*, pages 31–38. IEEE, 2010.

[44] E. Bloedorn, A. D. Christiansen, W. Hill, C. Skorupka, L. M. Talbot, and J. Tivel. Data mining for network intrusion detection: How to get started. Technical report, The MITRE Corporation, 2001.

[45] F. Boccuni. A theory of Fregean abstract objects. In Quinon and Antonutti [391]. Online at http://www.fil.lu.se/index.php?id=18879.

[46] A. B. Bondi. Characteristics of scalability and their impact on performance. In *Proceedings of the 2nd international workshop on Software and performance*, pages 195–203, 2000.

[47] G. Booch, J. Rumbaugh, and I. Jacobson. *The Unified Modeling Language User Guide*. Addison-Wesley, 1999. ISBN: 0201571684.

[48] C. Borgelt, M. Steinbrecher, and R. R. Kruse. *Graphical Models: Representations for Learning, Reasoning and Data Mining*. Wiley, second edition, Sept. 2009.

[49] A. Boukhtouta, N.-E. Lakhdari, S. A. Mokhov, and M. Debbabi. Towards fingerprinting malicious traffic. In *Proceedings of ANT'13*, volume 19, pages 548–555. Elsevier, June 2013.

[50] J. R. Boyd. Destruction and creation. [online], U.S. Army Command and General Staff College, Sept. 1976. http://www.goalsys.com/books/documents/DESTRUCTION_AND_CREATION.pdf.

[51] J. R. Boyd. The essence of winning and losing. [online], June 1995. A five slide set by Boyd; http://www.danford.net/boyd/essence.htm.

[52] M. Bozorgi, L. K. Saul, S. Savage, and G. M. Voelker. Beyond heuristics: Learning to classify vulnerabilities and predict exploits. In *Proceedings of the 16th ACM SIGKDD international conference on Knowledge Discovery and Data Mining*, KDD'10, pages 105–114, New York, NY, USA, 2010. ACM.

[53] M. Bunge. Interpretation. In *Semantics II: Interpretation and Truth*, volume 2 of *Treatise on Basic Philosophy*, pages 1–41. Springer Netherlands, 1974.

[54] S. Bunting. *EnCase Computer Forensics – The Official EnCE: EnCase Certified Examiner Study Guide*. Sybex, 3 edition, Sept. 2012.

[55] L. Burdy et al. An overview of JML tools and applications. *International Journal on Software Tools for Technology Transfer*, 7(3):212–232, 2005.

[56] J. Cao, L. Fernando, and K. Zhang. Programming distributed systems based on graphs. In

Orgun and Ashcroft [350], pages 83–95.

[57] R. Carnap. *Meaning and Necessity: a Study in Semantics and Modal Logic*. University of Chicago Press, Chicago, USA, 1947.

[58] R. Carnap. *Introduction to Symbolic Logic and its Applications*. Dover Publications, June 1958.

[59] B. D. Carrier. Risks of live digital forensic analysis. *Communications of the ACM*, 49(2):57–61, Feb. 2006. http://www.d.umn.edu/~schw0748/DigitalForensics/p56-carrier.pdf.

[60] B. D. Carrier. Autopsy forensic browser. [online], 2006–2013. http://www.sleuthkit.org/autopsy/.

[61] B. D. Carrier. The Sleuth Kit. [online], 2006–2013. http://www.sleuthkit.org/sleuthkit/.

[62] D. Cavalca and E. Goldoni. HIVE: an open infrastructure for malware collection and analysis. In *Proceedings of the 1st Workshop on Open Source Software for Computer and Network Forensics*, pages 23–34, 2008.

[63] P. R. Cavalin, R. Sabourin, and C. Y. Suen. Dynamic selection of ensembles of classifiers using contextual information. In *Multiple Classifier Systems*, LNCS 5997, pages 145–154, Mar. 2010.

[64] H. Chen. *Intelligence and Security Informatics for International Security: Information Sharing and Data Mining*. Integrated Series in Information Systems. Springer-Verlag New York, Inc., Secaucus, NJ, USA, 2006.

[65] S. Chen and P. Greenfield. QoS evaluation of JMS: An empirical approach. In *Proceedings of the 37th Hawaii International Conference on System Sciences*, 2004.

[66] Y. Chen and W.-T. Tsai. *Service-Oriented Computing and Web Data Management: from Principle to Development*. Kendall Hunt Publishing Company, 2 edition, 2008.

[67] Z. Chlondowski. S.M.A.R.T. Linux: Attributes reference table. [online], S.M.A.R.T. Linux project, 2007–2012. http://smartlinux.sourceforge.net/smart/attributes.php, last viewed May 2012.

[68] Cisco Systems, Inc. *Catalyst 2950 Switch Hardware Installation Guide*, Oct. 2003.

[69] A. Clark and E. Dawson. Optimisation heuristics for the automated cryptanalysis of classical ciphers. *Journal of Combinatorial Mathematics and Combinatorial Computing*, 1998.

[70] K. Clark and K. Hamilton. *Cisco LAN Switching*. Cisco Press, 1999. ISBN: 1-57870-094-9.

[71] CNN. California bus crash. [online], Apr. 2009. http://www.cnn.com/2009/US/04/29/california.crash/index.html.

[72] CNN. 'Catastrophic failure' caused North Sea copter crash. [online], Apr. 2009. http://www.cnn.com/2009/WORLD/europe/04/11/scotland.helicopter.crash.failure/index.html.

[73] G. Coulouris, J. Dollimore, and T. Kindberg. *Distributed Systems: Concepts and Design*. Addison-Wesley, 4 edition, 2005. ISBN: 0-321-26354-5.

[74] Cycling '74. Jitter 1.5. [online], 2005. http://www.cycling74.com/products/jitter.html.

[75] Cycling '74. Max/MSP. [online], 2005. http://www.cycling74.com/products/maxmsp.html.

[76] K. Dahbur and B. Mohammad. The anti-forensics challenge. In *Proceedings of the 2011 International Conference on Intelligent Semantic Web-Services and Applications (ISWSA'11)*, pages 14:1–14:7, New York, NY, USA, Apr. 2011. ACM.

[77] I. F. Darwin, J. Gilmore, G. Collyer, R. McMahon, G. Harris, C. Zoulas, C. Lowth, E. Fischer, and Various Contributors. `file` – determine file type, BSD General Commands Manual, file(1). BSD, Jan. 1973–2007. man file(1).

[78] I. F. Darwin, J. Gilmore, G. Collyer, R. McMahon, G. Harris, C. Zoulas, C. Lowth, E. Fischer, and Various Contributors. `file` – determine file type. [online], Mar. 1973–2008. ftp://ftp.astron.com/pub/file/, last viewed April 2008.

[79] J. D. Day. The (un)revised OSI reference model. *SIGCOMM Comput. Commun. Rev.*, 25(5):39–55, 1995.

[80] J. D. Day and H. Zimmermann. The OSI reference model. In *Proceedings of the IEEE*, volume 71, pages 1334—1340, Washington, DC, USA, Dec. 1983. IEEE Computer Society.

[81] T. Dean, J. Allen, and Y. Aloimonos, editors. *Artificial Intelligence: Theory and Practice*. Benjamin/Cummings, 1995. ISBN 0-8053-2547-6.

[82] W. Dean. Soundness, reflection, and intensionality. In Quinon and Antonutti [391]. Online at http://www.fil.lu.se/index.php?id=18879.

[83] M. Debbabi. INSE 6150: Lecture 6: Formal analysis (II). Concordia Institute for Information Systems Engineering, Concordia University, Montreal, Canada, 2006. http://www.ciise.concordia.ca/~debbabi.

[84] M. Debbabi. INSE 6150: Lecture notes. Concordia Institute for Information Systems Engineering, Concordia University, Montreal, Canada, 2006.

[85] M. Debbabi, A. R. Arasteh, A. Sakha, M. Saleh, and A. Fry. A collection of JPF forensic plugins. Computer Security Laboratory, Concordia Institute for Information Systems Engineering, 2007–2008.

[86] P. Degano and C. Priami. Enhanced operational semantics: A tool for describing and analyzing concurrent systems. *ACM Computing Surveys*, 33(2):135–176, 2001.

[87] L. Demey, B. Kooi, and J. Sack. Logic and probability. In E. N. Zalta, editor, *The Stanford Encyclopedia of Philosophy*. Stanford, spring 2013 edition, 2013. http://plato.stanford.edu/archives/spr2013/entries/logic-probability/.

[88] A. P. Dempster. A generalization of Bayesian inference. *Journal of the Royal Statistical Society*, 30:205–247, 1968.

[89] Y. Ding. Automated translation between graphical and textual representations of intensional programs in the GIPSY. Master's thesis, Department of Computer Science and Software Engineering, Concordia University, Montreal, Canada, June 2004. http://newton.cs.concordia.ca/~paquet/filetransfer/publications/theses/DingYiminMSc2004.pdf.

[90] G. Ditu. *The Programming Language TransLucid*. PhD thesis, University of New South Wales, Australia, 2007.

[91] D. Dowty, R. Wall, and S. Peters. *Introduction to Montague Semantics*. D. Reidel, Dordrecht, The Netherlands, 1981.

[92] W. Du. *Indexical Parallel Programming*. PhD thesis, Department of Computer Science, Victoria University, Canada, 1991.

[93] W. Du. Object-oriented implementation of intensional language. In *Proceedings of the 7th International Symposium on Lucid and Intensional Programming*, pages 37–45, Menlo Park, California, USA, Sept. 1994. SRI International.

[94] W. Du. Toward an intensional model for programming large scale distributed systems. In Gergatsoulis and Rondogiannis [131], pages 244–258.

[95] W. Du. On the relationship between AOP and intensional programming through context, July 2005. Keynote talk at the Intensional Programming Session of PLC'05.

[96] L. Duboc, D. S. Rosenblum, and T. Wicks. A framework for characterization and analysis of software system scalability. In I. Crnkovic and A. Bertolino, editors, *ESEC/SIGSOFT FSE*, pages 375–384. ACM, Sept. 2007.

[97] E. Dulaney. *CompTIA Security+ Study Guide: Exam SY0-301*. Sybex, 5 edition, June 2011.

[98] M. Duží, B. Jespersen, and P. Materna. *Procedural Semantics for Hyperintensional Logic: Foundations and Applications of Transparent Intensional Logic*, volume 17 of *Logic, Epistemology, and the Unity of Science*. Springer Science + Business Media, Inc., 2010.

[99] Eclipse contributors et al. Eclipse Platform. eclipse.org, 2000–2013. http://www.eclipse.

org, last viewed January 2013.

[100] S. A. Edwards. MEMOCODE 2012 hardware/software codesign contest: DNA sequence aligner. [online], 2012. Online at http://memocode.irisa.fr/2012/2012-memocode-contest.pdf;; Reference implementation at http://memocode.irisa.fr/2012/2012-memocode-contest.tar.gz.

[101] R. Eggen and M. Eggen. Efficiency of distributed parallel processing using Java RMI, sockets, and CORBA. In *Proceedings of the 2001 International Conference on Parallel and Distributed Processing Techniques and Applications (PDPTA'01)*. PDPTA, June 2001.

[102] M. Eissele, M. Kreiser, and T. Erd. Context-controlled flow visualization in augmented reality. In *Proceedings of Graphics Interface 2008 (GI'08)*, pages 89–96, Windsor, ON, Canada, May 2008. Canadian Human-Computer Communications Society.

[103] M. Endrei, J. Ang, A. Arsanjani, et al. *Patterns: Service-Oriented Architecture and Web Services*. IBM, 2004. IBM Red Book; online at http://www.redbooks.ibm.com/abstracts/sg246303.html.

[104] M. Eto, D. Inoue, J. Song, J. Nakazato, K. Ohtaka, and K. Nakao. NICTER: a large-scale network incident analysis system: case studies for understanding threat landscape. In *Proceedings of the First Workshop on Building Analysis Datasets and Gathering Experience Returns for Security*, BADGERS'11, pages 37–45, New York, NY, USA, 2011. ACM.

[105] M. Eto, K. Sonoda, D. Inoue, K. Yoshioka, and K. Nakao. A proposal of malware distinction method based on scan patterns using spectrum analysis. In *Proceedings of the 16th International Conference on Neural Information Processing: Part II*, ICONIP'09, pages 565–572, Berlin, Heidelberg, 2009. Springer-Verlag.

[106] R. Fagin, J. Y. Halpern, Y. Moses, and M. Y. Vardi. *Reasoning About Knowledge*. MIT Press, Cambridge, MA, USA, 2003.

[107] W. Farmer. The seven virtues of simple type theory. *Journal of Applied Logic*, 6(3):267–286, Sept. 2008.

[108] A. A. Faustini. *The Equivalence of a Denotational and an Operational Semantics of Pure Dataflow*. PhD thesis, University of Warwick, Computer Science Department, Coventry, United Kingdom, 1982.

[109] A. A. Faustini and R. Jagannathan. Multidimensional problem solving in Lucid. Technical Report SRI-CSL-93-03, SRI International, 1993.

[110] A. A. Faustini and W. W. Wadge. An eductive interpreter for the language Lucid. *SIGPLAN Not.*, 22(7):86–91, 1987.

[111] M. Fisher and T. Kakoudakis. Flexible agent grouping in executable temporal logic. In Gergatsoulis and Rondogiannis [131], pages 93–105.

[112] M. Fitting. Intensional logic. In E. N. Zalta, editor, *The Stanford Encyclopedia of Philosophy*. Stanford, winter 2012 edition, 2012. http://plato.stanford.edu/archives/win2012/entries/logic-intensional/.

[113] M. C. Fitting. FOIL axiomatized. [online], Aug. 2005. http://comet.lehman.cuny.edu/fitting/bookspapers/pdf/papers/FOILAxioms.pdf.

[114] R. Flenner. *Jini and JavaSpaces Application Development*. Sams, 2001.

[115] J. Folina. Mathematical intensions and intensionality in mathematics. In Quinon and Antonutti [391]. Online at http://www.fil.lu.se/index.php?id=18879.

[116] B. Fonseca. Shuttle Columbia's hard drive data recovered from crash site. [online], May 2008. http://www.computerworld.com/action/article.do?command=viewArticleBasic&articleId=9083718.

[117] N. Foster, M. J. Freedmany, A. Guha, R. Harrisonz, N. P. Kattay, C. Monsantoy, J. Reichy,

M. Reitblatt, J. Rexfordy, C. Schlesingery, A. Story, and D. Walkery. Languages for software-defined networks. In *Proceedings of IEEE COMS'13*. IEEE, 2013.

[118] G. Fourtounis, P. C. Ölveczky, and N. Papaspyrou. Formally specifying and analyzing a parallel virtual machine for lazy functional languages using Maude. In *Proceedings of the 5th International Workshop on High-Level Parallel Programming and Applications*, HLPP'11, pages 19–26, New York, NY, USA, 2011. ACM.

[119] A. A. Fraenkel. *Abstract set theory*. North-Holland Pub. Co., New York, Amsterdam, 4th revised edition, 1976.

[120] S. Frankel, R. Graveman, J. Pearce, and M. Rooks. Guidelines for the secure deployment of IPv6. Technical Report Special Publication 800-119, NIST, Dec. 2010. http://csrc.nist.gov/publications/nistpubs/800-119/sp800-119.pdf.

[121] E. Freeman, E. Freeman, K. Sierra, and B. Bates. *Head First Design Patterns*. O'Reilly & Associates, Inc., first edition, Oct. 2004. http://www.oreilly.com/catalog/hfdesignpat/toc.pdf, http://www.oreilly.com/catalog/hfdesignpat/chapter/index.html.

[122] B. Freeman-Benson. Lobjcid: Objects in Lucid. In *Proceedings of the 1991 Symposium on Lucid and Intensional Programming*, pages 80–87, Menlo Park, California, USA, Apr. 1991. SRI International.

[123] M. Frigault and L. Wang. Measuring network security using bayesian network-based attack graphs. In *COMPSAC*, pages 698–703, 2008.

[124] T. Frühwirth. Constraint solving with constraint handling rules. In Gergatsoulis and Rondogiannis [131], pages 14–30. Tutorial.

[125] M. Gabbay. Making sense of maths: a formalist theory of mathematical intension. In Quinon and Antonutti [391]. Online at http://www.fil.lu.se/index.php?id=18879.

[126] J.-R. Gagné and J. Plaice. Demand-driven real-time computing. In Gergatsoulis and Rondogiannis [131], pages 168–181. ISBN: 981-02-4095-3.

[127] D. Gallin. *Intensional and Higher-Order Modal Logic: With Applications to Montague Semantics*. North-Holland, Amsterdam, The Netherlands, 1975.

[128] E. Gamma, R. Helm, R. Johnson, and J. Vlissides. *Design Patterns: Elements of Reusable Object-Oriented Software*. Addison-Wesley, 1995. ISBN: 0201633612.

[129] E. Garcia. Cosine similarity and term weight tutorial. [online], 2006. http://www.miislita.com/information-retrieval-tutorial/cosine-similarity-tutorial.html.

[130] P. Gärdenfors. Qualitative probability as an intensional logic. *Journal of Philosophical Logic*, 4:171–185, 1975.

[131] M. Gergatsoulis and P. Rondogiannis, editors. *Proceedings of ISLIP'99*, volume Intensional Programming II. World Scientific, June 1999. ISBN: 981-02-4095-3.

[132] GFI, Inc. Malware analysis with GFI SandBox (formerly CWSandbox). [online], 2010–2012. http://www.threattracksecurity.com/enterprise-security/sandbox-software.aspx.

[133] S. Ghosh. *Distributed Systems – An Algorithmic Approach*. CRC Press, 2007. ISBN: 978-1-58488-564-1.

[134] J.-Y. Girard. Linear logic. *Theoretical Computer Science*, 50(1):1–101, 1987.

[135] P. Gladyshev. *Formalising Event Reconstruction in Digital Investigations*. PhD thesis, Department of Computer Science, University College Dublin, Aug. 2004. Online at http://www.formalforensics.org/publications/thesis/index.html.

[136] P. Gladyshev. Finite state machine analysis of a blackmail investigation. *International Journal of Digital Evidence*, 4(1), 2005.

[137] P. Gladyshev and A. Patel. Finite state machine approach to digital event reconstruction. *Digital Investigation Journal*, 2(1), 2004.

[138] O. Göbel, S. Frings, D. Günther, J. Nedon, and D. Schadt, editors. *IT-Incidents Management*

*& IT-Forensics - IMF 2008, Conference Proceedings, September 23-25, 2008, Mannheim, Germany*, volume 140 of *LNI*. GI, 2008.

[139] O. Göbel, S. Frings, D. Günther, J. Nedon, and D. Schadt, editors. *IT-Incidents Management & IT-Forensics - IMF 2009, Conference Proceedings, September 15-17, 2009, Stuttgart, Germany*. IEEE Computer Society, 2009.

[140] S. Goodman and A. Hunter. Feature extraction algorithms for pattern classification. In *Proceedings of Ninth International Conference on Artificial Neural Networks*, volume 2, pages 738–742, 1999.

[141] J. Gosling, B. Joy, G. Steele, and G. Bracha. *Java Language Specification*. Addison-Wesley Professional, 3 edition, 2005. ISBN 0321246780.

[142] D. Green. Trail: Java reflection API. [online], 2001–2012. http://docs.oracle.com/javase/tutorial/reflect/index.html.

[143] P. Grogono. GIPC increments. Technical report, Department of Computer Science and Software Engineering, Concordia University, Montreal, Canada, Apr. 2002.

[144] P. Grogono. Intensional programming in Onyx. Technical report, Department of Computer Science and Software Engineering, Concordia University, Montreal, Canada, Apr. 2004.

[145] P. Grogono, S. Mokhov, and J. Paquet. Towards JLucid, Lucid with embedded Java functions in the GIPSY. In *Proceedings of the 2005 International Conference on Programming Languages and Compilers (PLC 2005)*, pages 15–21. CSREA Press, June 2005.

[146] D. Grune, B. Berliner, D. D. Z. Zuhn, J. Polk, L. Jones, D. R. Price, M. D. Baushke, B. Murphy, C. T. Pino, F. U. M. ao, J. Hyslop, and J. Meyering. Concurrent Versions System (CVS). [online], 1989–2012. http://savannah.nongnu.org/projects/cvs/.

[147] P. D. Grünwald and J. Y. Halpern. When ignorance is bliss. In *Proceedings of the 20th Conference on Uncertainty in Artificial Intelligence*, UAI'04, pages 226–234, Arlington, Virginia, United States, 2004. AUAI Press.

[148] R. V. Guha. *Contexts: A Formalization and Some Applications*. PhD thesis, Stanford University, Feb. 1995.

[149] Guidance Software. EnCase. [online], 2006–2013. http://www.encase.com/.

[150] C. Gundabattula and V. G. Vaidya. Building a state tracing Linux kernel. In O. Göbel, S. Frings, D. Günther, J. Nedon, and D. Schadt, editors, *Proceedings of the IT Incident Management and IT Forensics (IMF'08)*, LNI140, pages 173–196, Sept. 2008.

[151] V. Haarslev, Y. Lu, and N. Shiri. ONTOXPL – intelligent exploration of OWL ontologies. In *2004 IEEE/WIC/ACM International Conference on Web Intelligence (WI 2004)*, pages 624–627. IEEE Computer Society, Sept. 2004.

[152] R. Hadjidj, M. Debbabi, H. Lounis, F. Iqbal, A. Szporer, and D. Benredjem. Towards an integrated e-mail forensic analysis framework. *Digital Investigation*, 5(3-4):124–137, 2009.

[153] R. Haenni, J. Kohlas, and N. Lehmann. Probabilistic argumentation systems. Technical report, Institute of Informatics, University of Fribourg, Fribourg, Switzerland, Oct. 1999.

[154] R. Haenni and N. Lehmann. Probabilistic argumentation systems: a new perspective on the dempster-shafer theory. *International Journal of Intelligent Systems*, 18(1):93–106, 2003.

[155] R. Haenni, J.-W. Romeijn, G. Wheeler, and J. Williamson. *Probabilistic Logics and Probabilistic Networks*, volume 350 of *Synthese Library: Studies in Epistemology, Logic, Methodology, and Philosophy of Science*. Springer Science+Business Media B.V., 2011.

[156] J. Y. Halpern. *Reasoning about Uncertainty*. MIT Press, Cambridge, MA, USA, 2003.

[157] J. Y. Halpern, R. Fagin, Y. Moses, and M. Y. Vardi. *Reasoning About Knowledge*. MIT Press, 1995. http://www.cs.rice.edu/~vardi/papers/book.pdf.

[158] J. Y. Halpern and R. Pucella. A logic for reasoning about evidence. *J. Artif. Int. Res.*, 26(1):1–34, May 2006.

[159] K. M. B. Hamed. *Multidimensional Programs on Distributed Parallel Computers: Analysis and Implementation.* PhD thesis, Computer Science, the University of New Brunswick, Feb. 2008.

[160] B. Han. Towards a multi-tier runtime system for GIPSY. Master's thesis, Department of Computer Science and Software Engineering, Concordia University, Montreal, Canada, 2010.

[161] B. Han, S. A. Mokhov, and J. Paquet. Advances in the design and implementation of a multi-tier architecture in the GIPSY environment with Java. In *Proceedings of SERA 2010*, pages 259–266. IEEE Computer Society, 2010. Online at http://arxiv.org/abs/0906.4837.

[162] J. Han, M. Kamber, and J. Pei. *Data Mining: Concepts and Techniques.* Morgan Kaufmann Publishers Inc., San Francisco, CA, USA, 3rd edition, 2011.

[163] A. Hanna, H. Z. Ling, X. Yang, and M. Debbabi. A synergy between static and dynamic analysis for the detection of software security vulnerabilities. In R. Meersman, T. S. Dillon, and P. Herrero, editors, *OTM Conferences (2)*, volume 5871 of *Lecture Notes in Computer Science*, pages 815–832. Springer, 2009.

[164] M. Hapner, R. Burridge, R. Sharma, J. Fialli, and K. Stout. *Java(TM) Message Service API Tutorial and Reference.* Prentice Hall PTR, 2002. ISBN 0201784726.

[165] D. Harrington, R. Presuhn, and B. Wijnen. RFC 2571: An Architecture for Describing SNMP Management Frameworks. [online], Apr. 1999. http://www.ietf.org/rfc/rfc2571.txt, viewed in January 2008.

[166] J. Heylen. Peano numerals as buck-stoppers. In Quinon and Antonutti [391]. Online at http://www.fil.lu.se/index.php?id=18879.

[167] M. G. Hinchey, J. L. Rash, W. Truszkowski, C. Rouff, and R. Sterritt. Autonomous and autonomic swarms. In *Software Engineering Research and Practice*, pages 36–44. CSREA Press, 2005.

[168] R. Hinden, B. Carpenter, and L. Masinter. RFC 2732: Format for Literal IPv6 Addresses in URL's. [online], Dec. 1999. http://www.ietf.org/rfc/rfc2732.txt, viewed in November 2007.

[169] H. Hiss. Checking the satisfiability of XML-specifications. Technical Report 2008-1-Ait Mohamed, Department of Electrical and Computer Engineering, Concordia University, Montreal, Canada, Aug. 2008. In Theorem Proving in Higher Order Logics (TPHOLs2008): Emerging Trends Proceedings.

[170] N. Hnatiw, T. Robinson, C. Sheehan, and N. Suan. Pimp my PE: Parsing malicious and malformed executables. In H. Martin, editor, *Proceedings of the 17th Virus Bulletin International Conference*, pages 9–17, Vienna, Austria: The Pentagon, Abingdon, OX143YP, England, Sept. 2007.

[171] J. R. Hobbs and S. J. Rosenschein. Making computational sense of Montague's intensional logic. *Artificial Intelligence*, 9:287–306, 1978. http://www.stanford.edu/class/linguist289/hobbs78.pdf.

[172] Honeynet Project. Honeynet forensics project scans. [online], 2002–2013. http://honeynet.org/scans.

[173] P. Horn. Autonomic computing: IBM's perspective on the state of information technology. Technical report, IBM T. J. Watson Laboratory, Oct. 2001.

[174] W. hua Xu, X. yan Zhang, J. min Zhong, and W. xiu Zhang. Attribute reduction in ordered information systems based on evidence theory. *Knowl Inf Syst*, 25:169–184, Sept. 2009.

[175] G. Hunter. *Metalogic: An Introduction to the Metatheory of Standard First-Order Logic.* University of California Press, 1971.

[176] K. Hwang and D. Jung. Anti-malware expert system. In H. Martin, editor, *Proceedings of the 17th Virus Bulletin International Conference*, pages 9–17, Vienna, Austria: The Pentagon,

Abingdon, OX143YP, England, Sept. 2007.

[177] IBM, BEA Systems, Microsoft, SAP AG, and Siebel Systems. Business Process Execution Language for Web Services version 1.1. [online], IBM, Feb. 2007. http://www.ibm.com/developerworks/library/specification/ws-bpel/.

[178] IBM Corporation. An architectural blueprint for autonomic computing. Technical report, IBM Corporation, 2006.

[179] IBM Tivoli. Autonomic computing policy language. Technical report, IBM Corporation, 2005.

[180] IEEE. 802-1990: IEEE standards for local and metropolitan networks: Overview and architecture. [online], Sept. 2004. http://grouper.ieee.org/groups/802/802overview.pdf.

[181] IEEE. 754-2008: IEEE standard for floating-point arithmetic. [online], Aug. 2008. http://ieeexplore.ieee.org/servlet/opac?punumber=4610933.

[182] E. C. Ifeachor and B. W. Jervis. *Speech Communications*. Prentice Hall, New Jersey, USA, 2002.

[183] D. Inoue, K. Yoshioka, M. Eto, M. Yamagata, E. Nishino, J. Takeuchi, K. Ohkouchi, and K. Nakao. An incident analysis system NICTER and its analysis engines based on data mining techniques. In *Proceedings of the 15th International Conference on Advances in Neuro-Information Processing – Volume Part I*, ICONIP'08, pages 579–586, Berlin, Heidelberg, 2009. Springer-Verlag.

[184] Internet Assigned Numbers Authority (IANA). Private enterprise numbers: Smi network management private enterprise codes. [online], IANA, June 2013. http://www.iana.org/assignments/enterprise-numbers.

[185] G. M. Jackson. *Predicting Malicious Behavior: Tools and Techniques for Ensuring Global Security*. Wiley, 1 edition, June 2012.

[186] P. Jackson, editor. *Introduction to Expert Systems*. Addison-Wesley, third edition, 1995. ISBN 0-201-87686-8.

[187] R. Jagannathan. Intensional and extensional graphical models for GLU programming. In Orgun and Ashcroft [350], pages 63–75.

[188] R. Jagannathan and C. Dodd. GLU programmer's guide. Technical report, SRI International, Menlo Park, California, 1996.

[189] R. Jagannathan, C. Dodd, and I. Agi. GLU: A high-level system for granular data-parallel programming. In *Concurrency: Practice and Experience*, volume 1, pages 63–83, 1997.

[190] Y. Jarraya. *Verification and Validation of UML and SysML Based Systems Engineering Design Models*. PhD thesis, Department of Electrical and Computer Engineering, Concordia University, Montreal, Canada, Feb. 2010.

[191] Y. Ji. Scalability evaluation of the GIPSY runtime system. Master's thesis, Department of Computer Science and Software Engineering, Concordia University, Montreal, Canada, Mar. 2011.

[192] Y. Ji, S. A. Mokhov, and J. Paquet. Design for scalability evaluation and configuration management of distributed components in GIPSY. Unpublished, 2010–2013.

[193] Y. Ji, S. A. Mokhov, and J. Paquet. Unifying and refactoring DMF to support concurrent Jini and JMS DMS in GIPSY. In B. C. Desai, S. P. Mudur, and E. I. Vassev, editors, *Proceedings of the Fifth International C\* Conference on Computer Science and Software Engineering (C3S2E'12)*, pages 36–44, New York, NY, USA, June 2010–2013. ACM. Online e-print http://arxiv.org/abs/1012.2860.

[194] Jini Community. Jini network technology. [online], Sept. 2007. http://java.sun.com/developer/products/jini/index.jsp.

[195] R. Johnsonbaugh. *Advanced Engineering Mathematics*. Pearson Prentice Hall, 7th edition,

2009. ISBN: 978-0-13-159318-3.

[196] H. F. Jordan and G. Alaghband. *Fundamentals of Parallel Processing*. Pearson Education, Inc., 2003. ISBN 0-13-901158-7.

[197] R. Joshi. A comparison and mapping of Data Distribution Service (DDS) and Java Message Service (JMS). Real-Time Innovations, Inc., 2006.

[198] A. Jøsang, J. Diaz, and M. Rifqi. Cumulative and averaging fusion of beliefs. *Information Fusion*, 11(2):192–200, 2010.

[199] A. Jøsang and R. Hankin. Interpretation and fusion of hyper opinions in subjective logic. In *Proceedings of the 15th International Conference on Information Fusion (FUSION)*, pages 1225–1232, 2012.

[200] A. Jøsang and S. Pope. Dempster's rule as seen by little colored balls. *Computational Intelligence*, 28(4), May 2012.

[201] D. S. Jurafsky and J. H. Martin. *Speech and Language Processing*. Prentice-Hall, Inc., Pearson Higher Education, Upper Saddle River, New Jersey 07458, 2000. ISBN 0-13-095069-6.

[202] G. Kahn. The semantics of a simple language for parallel processing. In *Proceedings of the IFIP Congress '74*, pages 471–475, Amsterdam, 1974. Elsevier North-Holland.

[203] G. Kahn and D. B. MacQueen. Coroutines and networks of parallel processes. In *Proceedings of the IFIP Congress '77*, pages 993–998, Amsterdam, 1977. Elsevier North-Holland.

[204] F. O. Karray and C. de Silva. *Soft Computing and Intelligent Systems Design: Theory, Tools, and Applications*. Person Education Ltd. / Addison Wesley, 2004. ISBN: 0-321-11617-8.

[205] M. Kaufmann and J. S. Moore. An ACL2 tutorial. In Mohamed et al. [259], pages 17–21.

[206] S. Kent. Words of estimative probability. [online], CIA, 1964. https://www.cia.gov/library/center-for-the-study-of-intelligence/csi-publications/books-and-monographs/sherman-kent-and-the-board-of-national-estimates-collected-essays/6words.html.

[207] J. O. Kephart and D. M. Chess. The vision of autonomic computing. *IEEE Computer*, 36(1):41–50, 2003.

[208] M. Khalifé. Examining orthogonal concepts-based micro-classifiers and their correlations with noun-phrase coreference chains. Master's thesis, Department of Computer Science and Software Engineering, Concordia University, Montreal, Canada, 2004.

[209] G. Klein, T. Nipkow, and L. C. Paulson. The archive of formal proofs. SourceForge.net, 2008. http://afp.sourceforge.net/, last viewed: April 2008.

[210] D. Koenig. Web services business process execution language (WS-BPEL 2.0): The standards landscape. Presentation, IBM Software Group, 2007.

[211] M. Kokare, P. K. Biswas, and B. N. Chatterji. Texture image retrieval using new rotated complex wavelet filters. *IEEE Transaction on Systems, Man, and Cybernetics-Part B: Cybernetics*, 6(35):1168–1178, 2005.

[212] M. Kokare, P. K. Biswas, and B. N. Chatterji. Rotation-invariant texture image retrieval using rotated complex wavelet filters. *IEEE Transaction on Systems, Man, and Cybernetics-Part B: Cybernetics*, 6(36):1273–1282, 2006.

[213] Y. Kong, Y. Zhang, and Q. Liu. Eliminating human specification in static analysis. In *Proceedings of the 13th International Conference on Recent Advances in Intrusion Detection*, RAID'10, pages 494–495, Berlin, Heidelberg, 2010. Springer-Verlag.

[214] C. D. Koutras and C. Nomikos. On the computational complexity of stratified negation in linear-time temporal logic programming. In Gergatsoulis and Rondogiannis [131], pages 106–117.

[215] T. Kremenek, K. Ashcraft, J. Yang, and D. Engler. Correlation exploitation in error ranking. In *Foundations of Software Engineering (FSE)*, 2004.

[216] T. Kremenek and D. Engler. Z-ranking: Using statistical analysis to counter the impact of static analysis approximations. In *SAS 2003*, 2003.

[217] T. Kremenek, P. Twohey, G. Back, A. Ng, and D. Engler. From uncertainty to belief: Inferring the specification within. In *Proceedings of the 7th Symposium on Operating System Design and Implementation*, 2006.

[218] S. A. Kripke. A completeness theorem in modal logic. *Journal of Symbolic Logic*, 31(2):276–277, 1966.

[219] S. A. Kripke. Semantical considerations on modal logic. *Journal of Symbolic Logic*, 34(3):501, 1969.

[220] P. Krishnan. An asynchronous calculus based on absence of actions. In Orgun and Ashcroft [350], pages 234–248.

[221] P. Kropf and J. Plaice. Intensional objects. In *International symposium on Languages for Intensional Programming*, pages 37–45, Athens, Greece, June 1999. Demokrits Institute.

[222] R. Lalement. *Computation as Logic*. Prentice Hall, 1993. C.A.R. Hoare Series Editor. English translation from French by John Plaice.

[223] P. J. Landin. The next 700 programming languages. *Communications of the ACM*, 9(3):157–166, 1966.

[224] C. Larman. *Applying UML and Patterns: An Introduction to Object-Oriented Analysis and Design and Iterative Development*. Pearson Education, third edition, Apr. 2006. ISBN: 0131489062.

[225] M.-A. Laverdière, S. A. Mokhov, D. Bendredjem, and S. Tsapa. Ftkplipse – Forensic Toolkits Eclipse Plug-ins. SourceForge.net, 2005–2008. http://ciisesec.svn.sourceforge.net/viewvc/ciisesec/forensics, last viewed April 2008.

[226] M.-A. Laverdière, S. A. Mokhov, S. Tsapa, and D. Benredjem. Ftklipse–design and implementation of an extendable computer forensics environment: Software requirements specification document, 2005–2009. http://arxiv.org/abs/0906.2446.

[227] M.-A. Laverdière, S. A. Mokhov, S. Tsapa, and D. Benredjem. Ftklipse–design and implementation of an extendable computer forensics environment: Specification design document, 2005–2009.

[228] M.-A. Laverdière-Papineau. Towards systematic software security hardening. Master's thesis, Concordia University, 2007. ISBN: 9780494344446; http://spectrum.library.concordia.ca/975561/.

[229] G. T. Leavens. The Java modeling language (JML). [online], 2007. http://www.jmlspecs.org/.

[230] G. T. Leavens and Y. Cheon. Design by contract with JML. Technical report, Formal Systems Laboratory (FSL) at UIUC, 2006.

[231] H. C. Lee. The utilization of forensic evidence in IED incidents. In *Proceedings of the European Intelligence and Security Informatics Conference (EISIC) 2012*, pages 1–2, Aug. 2012.

[232] W. Lee, S. J. Stolfo, and K. W. Mok. Adaptive intrusion detection: A data mining approach. *Artificial Intelligence Review*, 14:533–567, 2000.

[233] D. Leffingwell and D. Widrig. *Managing Software Requirements: A Use Case Approach*. Addison-Wesley, 2 edition, 2003. ISBN: 0-321-12247-X.

[234] R. Li, O.-J. Xi, B. Pang, J. Shen, and C.-L. Ren. Network application identification based on wavelet transform and k-means algorithm. In *Proceedings of the IEEE International Conference on Intelligent Computing and Intelligent Systems (ICIS2009)*, volume 1, pages 38–41, Nov. 2009.

[235] M. Ligh, S. Adair, B. Hartstein, and M. Richard. *Malware Analyst's Cookbook and DVD: Tools and Techniques for Fighting Malicious Code*. Wiley, 1 edition, Nov. 2010.

[236] K. Limthong, F. Kensuke, and P. Watanapongse. Wavelet-based unwanted traffic time series analysis. In *2008 International Conference on Computer and Electrical Engineering*, pages 445–449. IEEE Computer Society, 2008.

[237] S. Lipschutz. *Schaum's Outline of Theory and Problems of Set Theory and Related Topics*. Schaum's Outlines. McGraw-Hill, New York, 2nd edition, 1998. ISBN: 978-0070381599.

[238] Y. Liu and J. Staples. Building logic constructs into procedural programming languages. In Orgun and Ashcroft [350], pages 96–109.

[239] C. Livadas, R. Walsh, D. E. Lapsley, and W. T. Strayer. Using machine learning techniques to identify botnet traffic. In *LCN*, pages 967–974, Washington, DC, USA, 2006. IEEE Computer Society.

[240] K. C. Louden. *Compiler Construction: Principles and Practice*. PWS Publishing Company, 1997. ISBN 0-564-93972-4.

[241] B. Lu. *Developing the Distributed Component of a Framework for Processing Intensional Programming Languages*. PhD thesis, Department of Computer Science and Software Engineering, Concordia University, Montreal, Canada, Mar. 2004.

[242] B. Lu, P. Grogono, and J. Paquet. Distributed execution of multidimensional programming languages. In *Proceedings of the 15th IASTED International Conference on Parallel and Distributed Computing and Systems (PDCS 2003)*, volume 1, pages 284–289. International Association of Science and Technology for Development, Nov. 2003.

[243] Y. Lu, I. Cohen, X. S. Zhou, and Q. Tian. Feature selection using principal feature analysis. In *Proceedings of the 15th International Conference on Multimedia*, pages 301–304, Augsburg, Germany, 2007. ACM.

[244] W. Ma and M. A. Orgun. Verifying MULTRAN programs with temporal logic. In Orgun and Ashcroft [350], pages 186–206.

[245] Q. H. Mamoud. Getting started with JavaSpaces technology: Beyond conventional distributed programming paradigms. [online], July 2005. http://java.sun.com/developer/technicalArticles/tools/JavaSpaces/.

[246] B. Mancilla and J. Plaice. Possible worlds versioning. *Mathematics in Computer Science*, 2(1):63–83, 2008.

[247] K. Mandia, C. Prosise, and M. Pepe. *Incident Response and Computer Forensics*. McGraw-Hill, 2nd edition, 2003.

[248] C. D. Manning and H. Schutze. *Foundations of Statistical Natural Language Processing*. MIT Press, 2002.

[249] D. Mares. Software links for forensics investigative tasks. [online], 2006. http://www.dmares.com/maresware/SITES/tasks.htm.

[250] E. Mark et al. Lucid (PoC in Haskell). [online], HaskellWiki, 2006–2011. http://www.haskell.org/haskellwiki/Lucid.

[251] D. R. Mauro and K. J. Schmidt. *Essential SNMP*. O'Reilly, 2001. ISBN: 0-596-00020-00.

[252] M. McDougal. Live forensics on a Windows system: Using Windows Forensic Toolchest (WFT). [online], 2003–2006. http://www.foolmoon.net/downloads/Live_Forensics_Using_WFT.pdf.

[253] E. Mendelson. *Introduction to Mathematical Logic*. Chapman & Hall, 4 edition, 1997.

[254] W. J. Meng, J. Rilling, Y. Zhang, R. Witte, S. Mudur, and P. Charland. A context-driven software comprehension process model. In *Proceedings of the IEEE Software Evolvability Workshop (SE'06)*, Sept. 2006.

[255] M. Messner. Pen testing on IPv6 networks: In through the back door. *Linux Magazine*, 143:16–20, Oct. 2012. http://www.linux-magazine.com/Online/Features/IPv6-Pen-Testing.

[256] W. Metcalf. Snort in-line. [online], 2011. http://snort-inline.sourceforge.net/.

[257] B. Meyer. On formalism in specifications. *IEEE Software*, 2(1):6–26, 1985.

[258] N. G. Miller. *A Diagrammatic Formal System for Euclidean Geometry*. PhD thesis, Cornell University, U.S.A, 2001.

[259] O. A. Mohamed, C. A. M. noz, and S. Tahar, editors. *Theorem Proving in Higher Order Logics, 21st International Conference, TPHOLs 2008, Montreal, Canada, August 18-21, 2008*, volume 5170 of *LNCS*. Springer, 2008.

[260] S. Mokhov, I. Clement, S. Sinclair, and D. Nicolacopoulos. Modular Audio Recognition Framework. Department of Computer Science and Software Engineering, Concordia University, Montreal, Canada, 2002–2003. Project report, http://marf.sf.net, last viewed April 2012.

[261] S. Mokhov and J. Paquet. General imperative compiler framework within the GIPSY. In *Proceedings of the 2005 International Conference on Programming Languages and Compilers (PLC 2005)*, pages 36–42. CSREA Press, June 2005.

[262] S. Mokhov and J. Paquet. Objective Lucid – first step in object-oriented intensional programming in the GIPSY. In *Proceedings of the 2005 International Conference on Programming Languages and Compilers (PLC 2005)*, pages 22–28. CSREA Press, June 2005.

[263] S. A. Mokhov. Lucid, the intensional programming language and its semantics in PVS. Department of Computer Science and Software Engineering, Concordia University, Montreal, Canada, Apr. 2004. Semantics of Programming Languages Course Project Report.

[264] S. A. Mokhov. Towards hybrid intensional programming with JLucid, Objective Lucid, and General Imperative Compiler Framework in the GIPSY. Master's thesis, Department of Computer Science and Software Engineering, Concordia University, Montreal, Canada, Oct. 2005. ISBN 0494102934; online at http://arxiv.org/abs/0907.2640.

[265] S. A. Mokhov. On design and implementation of distributed modular audio recognition framework: Requirements and specification design document. [online], Aug. 2006. Project report, http://arxiv.org/abs/0905.2459, last viewed April 2012.

[266] S. A. Mokhov. *Intensional Cyberforensics – a PhD Proposal*. Department of Computer Science and Software Engineering, Concordia University, Montreal, Canada, Dec. 2007.

[267] S. A. Mokhov. Intensional forensics – the use of intensional logic in cyberforensics. Technical report, Concordia Institute for Information Systems Engineering, Concordia University, Montreal, Canada, Jan. 2007. ENGR6991 Technical Report.

[268] S. A. Mokhov. Choosing best algorithm combinations for speech processing tasks in machine learning using MARF. In S. Bergler, editor, *Proceedings of the 21st Canadian AI'08*, LNAI 5032, pages 216–221, Berlin Heidelberg, May 2008. Springer-Verlag.

[269] S. A. Mokhov. Encoding forensic multimedia evidence from MARF applications as Forensic Lucid expressions. In T. Sobh, K. Elleithy, and A. Mahmood, editors, *Novel Algorithms and Techniques in Telecommunications and Networking, proceedings of CISSE'08*, pages 413–416, University of Bridgeport, CT, USA, Dec. 2008. Springer. Printed in January 2010.

[270] S. A. Mokhov. Study of best algorithm combinations for speech processing tasks in machine learning using median vs. mean clusters in MARF. In B. C. Desai, editor, *Proceedings of C3S2E'08*, pages 29–43, Montreal, Quebec, Canada, May 2008. ACM.

[271] S. A. Mokhov. Towards security hardening of scientific distributed demand-driven and pipelined computing systems. In *Proceedings of the 7th International Symposium on Parallel and Distributed Computing (ISPDC'08)*, pages 375–382. IEEE Computer Society, July 2008.

[272] S. A. Mokhov. Towards syntax and semantics of hierarchical contexts in multimedia processing applications using MARFL. In *Proceedings of the 32nd Annual IEEE International Computer Software and Applications Conference (COMPSAC)*, pages 1288–1294, Turku, Finland, July

2008. IEEE Computer Society.

[273] S. A. Mokhov. WriterIdentApp – Writer Identification Application. Unpublished, 2008–2013.

[274] S. A. Mokhov. Enhancing the formal cyberforensic approach with observation modeling with credibility factors and mathematical theory of evidence. [online], also in *;login: vol. 34, no. 6, p. 101*, Dec. 2009. Presented at WIPS at USENIX Security'09, `http://www.usenix.org/events/sec09/wips.html`.

[275] S. A. Mokhov. Java Data Security Framework (JDSF) and its applications: API design refinement. [online], also in *;login: vol. 34, no. 6, p. 93*, Dec. 2009. Poster at USENIX Security'09, `http://www.usenix.org/events/sec09/poster.html`.

[276] S. A. Mokhov. The role of self-forensics modeling for vehicle crash investigations and event reconstruction simulation. In J. S. Gauthier, editor, *Proceedings of the Huntsville Simulation Conference (HSC'09)*, pages 342–349. SCS, Oct. 2009. Online at `http://arxiv.org/abs/0905.2449`.

[277] S. A. Mokhov. Towards improving validation, verification, crash investigations, and event reconstruction of flight-critical systems with self-forensics. [online], June 2009. A white paper submitted in response to NASA's RFI NNH09ZEA001L, `http://arxiv.org/abs/0906.1845`, mentioned in `http://ntrs.nasa.gov/archive/nasa/casi.ntrs.nasa.gov/20100025593_2010028056.pdf`.

[278] S. A. Mokhov. Combining and comparing multiple algorithms for better learning and classification: A case study of MARF. In S. Jabin, editor, *Robot Learning*, chapter 2, pages 17–42. InTech, Aug. 2010. ISBN: 978-953-307-104-6, online at `http://www.intechopen.com/download/pdf/pdfs_id/12131`.

[279] S. A. Mokhov. Complete complimentary results report of the MARF's NLP approach to the DEFT 2010 competition. [online], June 2010. `http://arxiv.org/abs/1006.3787`.

[280] S. A. Mokhov. Cryptolysis: A Framework for Verification of Optimization Heuristics for the Automated Cryptanalysis of Classical Ciphers and Natural Language Word Segmentation. In *Proceedings of SERA 2010*, pages 295–302. IEEE Computer Society, May 2010.

[281] S. A. Mokhov. Evolution of MARF and its NLP framework. In *Proceedings of C3S2E'10*, pages 118–122. ACM, May 2010.

[282] S. A. Mokhov. *Hybrid Intensional Computing in GIPSY: JLucid, Objective Lucid and GICF*. LAP - Lambert Academic Publishing, Mar. 2010. ISBN 978-3-8383-1198-2.

[283] S. A. Mokhov. L'approche MARF à DEFT 2010: A MARF approach to DEFT 2010. In *Proceedings of the 6th DEFT Workshop (DEFT'10)*, pages 35–49. LIMSI / ATALA, July 2010. DEFT 2010 Workshop at TALN 2010; online at `http://deft.limsi.fr/actes/2010/pdf/2_clac.pdf`.

[284] S. A. Mokhov. The use of machine learning with signal- and NLP processing of source code to fingerprint, detect, and classify vulnerabilities and weaknesses with MARFCAT. [online], Oct. 2010. Online at `http://arxiv.org/abs/1010.2511`.

[285] S. A. Mokhov. MARFCAT – MARF-based Code Analysis Tool. Published electronically within the MARF project, `http://sourceforge.net/projects/marf/files/Applications/MARFCAT/`, 2010–2013. Last viewed April 2012.

[286] S. A. Mokhov. Review of *"Attribute reduction in ordered information systems based on evidence theory. Xu W., Zhang X., Zhong J., Zhang W. Knowledge and Information Systems 25(1): 169-184, 2010"*. In Computing Reviews [174]. CR138828 (1109-0972); online at `http://computingreviews.com/review/review_review.cfm?review_id=138828`.

[287] S. A. Mokhov. The use of machine learning with signal- and NLP processing of source code to fingerprint, detect, and classify vulnerabilities and weaknesses with MARFCAT. Technical Report NIST SP 500-283, NIST, Oct. 2011. Report: `http://www.nist.gov/manuscript-`

publication-search.cfm?pub_id=909407, online e-print at http://arxiv.org/abs/1010.2511.

[288] S. A. Mokhov. Review of *"The Anti-Forensics Challenge. Dahbur K., Mohammad B., ISWSA'11, ACM. April 18–20, 2011, Amman, Jordan"*. In *Computing Reviews* [76]. CR139793 (1207-0756); online at http://computingreviews.com/review/review_review.cfm?review_id=139793.

[289] S. A. Mokhov. MARFPCAT – MARF-based PCap Analysis Tool. Published electronically within the MARF project, 2012–2013. http://sourceforge.net/projects/marf/files/Applications/MARFCAT/.

[290] S. A. Mokhov and M. Debbabi. File type analysis using signal processing techniques and machine learning vs. `file` unix utility for forensic analysis. In O. Goebel, S. Frings, D. Guenther, J. Nedon, and D. Schadt, editors, *Proceedings of the IT Incident Management and IT Forensics (IMF'08)*, LNI140, pages 73–85. GI, Sept. 2008.

[291] S. A. Mokhov et al. Intensional Programming for AOP Tasks for Intensional Programming. Unpublished, 2008.

[292] S. A. Mokhov, L. W. Huynh, and J. Li. Managing distributed MARF with SNMP. Concordia Institute for Information Systems Engineering, Concordia University, Montreal, Canada, Apr. 2007. Project report. Hosted at http://marf.sf.net and http://arxiv.org/abs/0906.0065, last viewed February 2011.

[293] S. A. Mokhov, L. W. Huynh, and J. Li. Managing distributed MARF's nodes with SNMP. In *Proceedings of PDPTA'2008*, volume II, pages 948–954, Las Vegas, USA, July 2008. CSREA Press.

[294] S. A. Mokhov, L. W. Huynh, J. Li, and F. Rassai. A Java Data Security Framework (JDSF) for MARF and HSQLDB. Concordia Institute for Information Systems Engineering, Concordia University, Montreal, Canada, Apr. 2007. Project report. Hosted at http://marf.sf.net, last viewed April 2008.

[295] S. A. Mokhov, L. W. Huynh, and L. Wang. The integrity framework within the Java Data Security Framework (JDSF): Design refinement and implementation. In T. Sobh, K. Elleithy, and A. Mahmood, editors, *Novel Algorithms and Techniques in Telecommunications and Networking, Proceedings of CISSE'08*, pages 449–455. Springer, Dec. 2008. Printed in January 2010.

[296] S. A. Mokhov and R. Jayakumar. Distributed Modular Audio Recognition Framework (DMARF) and its applications over web services. In T. Sobh, K. Elleithy, and A. Mahmood, editors, *Proceedings of TeNe'08*, pages 417–422, University of Bridgeport, CT, USA, Dec. 2008. Springer. Printed in January 2010.

[297] S. A. Mokhov, M.-A. Laverdière, and D. Benredjem. Taxonomy of Linux kernel vulnerability solutions. In *Innovative Techniques in Instruction Technology, E-learning, E-assessment, and Education*, pages 485–493, 2007. Proceedings of CISSE/SCSS'07.

[298] S. A. Mokhov, M.-A. Laverdière, N. Hatami, and A. Benssam. Cryptolysis v.0.0.1 – a framework for automated cryptanalysis of classical ciphers. [online], 2005–2013. Project report; http://arxiv.org/abs/1101.1075.

[299] S. A. Mokhov, J. Li, and L. Wang. Simple dynamic key management in SQL randomization. In *Proceedings of NTMS'09*, pages 458–462. IEEE, Dec. 2009. ISBN: 978-1-4244-4765-7.

[300] S. A. Mokhov and J. Paquet. Formally specifying and proving operational aspects of Forensic Lucid in Isabelle. Technical Report 2008-1-Ait Mohamed, Department of Electrical and Computer Engineering, Concordia University, Montreal, Canada, Aug. 2008. In Theorem Proving in Higher Order Logics (TPHOLs2008): Emerging Trends Proceedings. Online at: http://users.encs.concordia.ca/~tphols08/TPHOLs2008/ET/76-98.pdf and

http://arxiv.org/abs/0904.3789.

[301] S. A. Mokhov and J. Paquet. A type system for higher-order intensional logic support for variable bindings in hybrid intensional-imperative programs in GIPSY. In T. Matsuo, N. Ishii, and R. Lee, editors, *9th IEEE/ACIS International Conference on Computer and Information Science, IEEE/ACIS ICIS 2010*, pages 921–928. IEEE Computer Society, May 2010. Presented at SERA 2010; online at http://arxiv.org/abs/0906.3919.

[302] S. A. Mokhov and J. Paquet. Using the General Intensional Programming System (GIPSY) for evaluation of higher-order intensional logic (HOIL) expressions. In *Proceedings of SERA 2010*, pages 101–109. IEEE Computer Society, May 2010. Online at http://arxiv.org/abs/0906.3911.

[303] S. A. Mokhov, J. Paquet, and M. Debbabi. Designing a language for intensional cyberforensic analysis. Unpublished, 2007.

[304] S. A. Mokhov, J. Paquet, and M. Debbabi. Formally specifying operational semantics and language constructs of Forensic Lucid. In O. Göbel, S. Frings, D. Günther, J. Nedon, and D. Schadt, editors, *Proceedings of the IT Incident Management and IT Forensics (IMF'08)*, LNI140, pages 197–216. GI, Sept. 2008. Online at http://subs.emis.de/LNI/Proceedings/Proceedings140/gi-proc-140-014.pdf.

[305] S. A. Mokhov, J. Paquet, and M. Debbabi. Designing Forensic Lucid – an intensional specification language for automated cyberforensic reasoning. Submitted for publication to J.MCS, 2008–2013.

[306] S. A. Mokhov, J. Paquet, and M. Debbabi. Reasoning about a simulated printer case investigation with Forensic Lucid. In J. S. Gauthier, editor, *Proceedings of the Huntsville Simulation Conference (HSC'09)*, page 45. SCS, Oct. 2009. Abstract, fully online at http://arxiv.org/abs/0906.5181.

[307] S. A. Mokhov, J. Paquet, and M. Debbabi. Towards automated deduction in blackmail case analysis with Forensic Lucid. In J. S. Gauthier, editor, *Proceedings of the Huntsville Simulation Conference (HSC'09)*, pages 326–333. SCS, Oct. 2009. Online at http://arxiv.org/abs/0906.0049.

[308] S. A. Mokhov, J. Paquet, and M. Debbabi. Towards formal requirements specification of self-forensics for autonomous systems. Submitted for review to J. Req. Eng., 2009–2013.

[309] S. A. Mokhov, J. Paquet, and M. Debbabi. The need to support of data flow graph visualization of Forensic Lucid programs, forensic evidence, and their evaluation by GIPSY. [online], Sept. 2010. Poster at VizSec'10; online at http://arxiv.org/abs/1009.5423.

[310] S. A. Mokhov, J. Paquet, and M. Debbabi. Towards automatic deduction and event reconstruction using Forensic Lucid and probabilities to encode the IDS evidence. In S. Jha, R. Sommer, and C. Kreibich, editors, *Proceedings of RAID'10*, LNCS 6307, pages 508–509. Springer, Sept. 2010.

[311] S. A. Mokhov, J. Paquet, and M. Debbabi. On the need for data flow graph visualization of Forensic Lucid programs and forensic evidence, and their evaluation by GIPSY. In *Proceedings of the Ninth Annual International Conference on Privacy, Security and Trust (PST), 2011*, pages 120–123. IEEE Computer Society, July 2011. Short paper; full version online at http://arxiv.org/abs/1009.5423.

[312] S. A. Mokhov, J. Paquet, and M. Debbabi. Reasoning about a simulated printer case investigation with Forensic Lucid. In P. Gladyshev and M. K. Rogers, editors, *Proceedings of ICDF2C'11*, number 0088 in LNICST, pages 282–296. Springer, Oct. 2011. Submitted in 2011, appeared in 2012; online at http://arxiv.org/abs/0906.5181.

[313] S. A. Mokhov, J. Paquet, M. Debbabi, and P. Grogono. Enhancing the formal cyberforensic approach with observation modeling with credibility factors and mathematical theory of

evidence. Unpublished, 2008–2013.

[314] S. A. Mokhov, J. Paquet, M. Debbabi, and Y. Sun. MARFCAT: Transitioning to binary and larger data sets of SATE IV. [online], May 2012. Submitted for publication to JSS; online at http://arxiv.org/abs/1207.3718.

[315] S. A. Mokhov, J. Paquet, and X. Tong. A type system for hybrid intensional-imperative programming support in GIPSY. In *Proceedings of C3S2E'09*, pages 101–107, New York, NY, USA, May 2009. ACM.

[316] S. A. Mokhov, F. Rassai, L. W. Huynh, and L. Wang. The authentication framework within the Java data security framework (JDSF): Design refinement and implementation. In T. Sobh, K. Elleithy, and A. Mahmood, editors, *Novel Algorithms and Techniques in Telecommunications and Networking, Proceedings of CISSE'08*, pages 423–429. Springer, Dec. 2008. Printed in January 2010.

[317] S. A. Mokhov, S. Sinclair, I. Clement, D. Nicolacopoulos, and the MARF Research & Development Group. SpeakerIdentApp – Text-Independent Speaker Identification Application. Published electronically within the MARF project, http://marf.sf.net, 2002–2013. Last viewed February 2010.

[318] S. A. Mokhov, M. Song, and C. Y. Suen. Writer identification using inexpensive signal processing techniques. In T. Sobh and K. Elleithy, editors, *Innovations in Computing Sciences and Software Engineering; Proceedings of CISSE'09*, pages 437–441. Springer, Dec. 2009. ISBN: 978-90-481-9111-6, online at: http://arxiv.org/abs/0912.5502.

[319] S. A. Mokhov and Y. Sun. OCT segmentation survey and summary reviews and a novel 3D segmentation algorithm and a proof of concept implementation. [online], 2011–2013. Online at http://arxiv.org/abs/1204.6725.

[320] S. A. Mokhov and E. Vassev. Autonomic specification of self-protection for Distributed MARF with ASSL. In *Proceedings of C3S2E'09*, pages 175–183, New York, NY, USA, May 2009. ACM.

[321] S. A. Mokhov and E. Vassev. Self-forensics through case studies of small to medium software systems. In *Proceedings of IMF'09*, pages 128–141. IEEE Computer Society, Sept. 2009.

[322] S. A. Mokhov, E. Vassev, J. Paquet, and M. Debbabi. Towards a self-forensics property in the ASSL toolset. In *Proceedings of C3S2E'10*, pages 108–113. ACM, May 2010.

[323] M. Monroe, R. Lan, J. M. del Olmo, B. Shneiderman, C. Plaisant, and J. Millstein. The challenges of specifying intervals and absences in temporal queries: a graphical language approach. In *Proceedings of the SIGCHI Conference on Human Factors in Computing Systems*, CHI'13, pages 2349–2358, New York, NY, USA, 2013. ACM.

[324] R. Montague. Pragmatics and intensional logic. *Synthese*, 22(1-2):68–94, 1970.

[325] B. Moolenaar and Contributors. Vim the editor – Vi Improved. [online], 2009. http://www.vim.org/.

[326] P. D. Mosses. The varieties of programming language semantics and their uses. In *4th International Andrei Ershov Memorial Conference on Perspective of System Informatics (PSI'01)*, volume 2244 of *LNCS*, pages 165–190, Berlin, 2001. Springer-Verlag.

[327] R. Murch. *Autonomic Computing: On Demand Series*. IBM Press, Prentice Hall, 2004.

[328] R. Muskens. Intensional models for the theory of types. *Journal of Symbolic Logic*, 72:98–118, 2007.

[329] My Digital Life Editorial Team. How to change or spoof MAC address in Windows XP, Vista, Server 2003/2008, Mac OS X, Unix and Linux. [online], June 2008.

[330] V. P. Nair, H. Jain, Y. K. Golecha, M. S. Gaur, and V. Laxmi. MEDUSA: MEtamorphic malware dynamic analysis using signature from API. In *Proceedings of the 3rd International Conference on Security of Information and Networks*, SIN'10, pages 263–269, New York, NY,
310

USA, 2010. ACM.

[331] NASA. Hubble status report. [online], Dec. 2008. http://www.nasa.gov/mission_pages/hubble/servicing/SM4/news/status_rpt_20081230.html.

[332] NASA. Hubble status report. [online], Dec. 2008. http://www.nasa.gov/mission_pages/hubble/servicing/SM4/news/status_rpt_20081219.html.

[333] NASA. Hubble status report #3: Hubble science operations deferred while engineers examine new issues. [online], Oct. 2008. http://www.nasa.gov/mission_pages/hubble/servicing/SM4/news/status_update_20081017.html.

[334] NASA. Hubble status report #4. [online], Oct. 2008. http://www.nasa.gov/mission_pages/hubble/servicing/SM4/news/status_rpt_4_20081017.html.

[335] NASA. Mars rover team diagnosing unexpected behavior: Mars exploration rover mission status report. [online], Jan. 2009. http://www.nasa.gov/mission_pages/mer/news/mer-20090128.html.

[336] NASA. Kepler mission manager update: December 22 2010 safe mode event investigation. [online], Dec. 2010. http://www.nasa.gov/mission_pages/kepler/news/keplerm-20101230.html.

[337] S. Negri. The intensional side of algebraic-topological representation theorems. In Quinon and Antonutti [391]. Online at http://www.fil.lu.se/index.php?id=18879.

[338] NetBeans Community. NetBeans Integrated Development Environment. [online], 2004–2013. http://www.netbeans.org.

[339] T. Nipkow, L. C. Paulson, and M. Wenzel. *Isabelle/HOL: A Proof Assistant for Higher-Order Logic*, volume 2283. Springer-Verlag, Nov. 2007. http://www.in.tum.de/~nipkow/LNCS2283/, last viewed: December 2007.

[340] NIST. National Vulnerability Database. [online], 2005–2013. http://nvd.nist.gov/.

[341] NIST. National Vulnerability Database statistics. [online], 2005–2013. http://web.nvd.nist.gov/view/vuln/statistics.

[342] T. S. Obuchowicz. *It's Only VHDL (But I Like It)*. Pearson Custom Publishing, 2005. ISBN 0-536-10092-6.

[343] W. Odom. *CCENT/CCNA ICND1: 640-822 Official Cert Guide*. Cisco Press, 3 edition, 2012. ISBN: 978-1-58720-425-8.

[344] W. Odom. *CCNA ICND2: 640-816 Official Cert Guide*. Cisco Press, 3 edition, 2012. ISBN: 978-1-58720-435-7.

[345] T. Oetiker, D. Rand, and the MRTG Community. Tobi oetiker's MRTG – the Multi Router Traffic Grapher. [online], 2008–2011. http://oss.oetiker.ch/mrtg/.

[346] Y. Okada, S. Ata, N. Nakamura, Y. Nakahira, and I. Oka. Comparisons of machine learning algorithms for application identification of encrypted traffic. In *Proceedings of the 10th International Conference on Machine Learning and Applications and Workshops (ICMLA)*, volume 2, pages 358–361, Dec. 2011.

[347] V. Okun, A. Delaitre, P. E. Black, and NIST SAMATE. Static Analysis Tool Exposition (SATE) 2010. [online], 2010. See http://samate.nist.gov/SATE2010Workshop.html.

[348] V. Okun, A. Delaitre, P. E. Black, and NIST SAMATE. Static Analysis Tool Exposition (SATE) IV. [online], Mar. 2012. See http://samate.nist.gov/SATE.html.

[349] OpenESB Contributors. BPEL service engine. [online], 2009. https://open-esb.dev.java.net/BPELSE.html.

[350] M. A. Orgun and E. A. Ashcroft, editors. *Proceedings of ISLIP'95*, volume Intensional Programming I. World Scientific, May 1995. ISBN: 981-02-2400-1.

[351] M. A. Orgun and W. Du. Multi-dimensional logic programming: Theoretical foundations. *Theoretical Computer Science*, 185(2):319–345, 1997.

[352] M. A. Orgun, C. Liu, and A. C. Nayak. Knowledge representation, reasoning and integration using temporal logic with clocks. *Mathematics in Computer Science*, 2(1):143–163, 2008.

[353] D. O'Shaughnessy. *Speech Communications*. IEEE, New Jersey, USA, 2000.

[354] C. B. Ostrum. *The Luthid 1.0 Manual*. Department of Computer Science, University of Waterloo, Ontario, Canada, 1981.

[355] H. Otrok, J. Paquet, M. Debbabi, and P. Bhattacharya. Testing intrusion detection systems in MANET: A comprehensive study. In *CNSR'07: Proceedings of the Fifth Annual Conference on Communication Networks and Services Research*, pages 364–371, Washington, DC, USA, 2007. IEEE Computer Society.

[356] D. Ottenheimer and M. Wallace. *Securing the Virtual Environment: How to Defend the Enterprise Against Attack*. Wiley, 1 edition, May 2012. Included DVD.

[357] G. Palmer (Editor). A road map for digital forensic research, report from first digital forensic research workshop (DFRWS). Technical report, DFRWS, 2001.

[358] T. Panayiotopoulos. Temporal reasoning with TRL. In Gergatsoulis and Rondogiannis [131], pages 133–148.

[359] N. S. Papaspyrou and I. T. Kassios. GLU# embedded in C++: a marriage between multidimensional and object-oriented programming. *Softw., Pract. Exper.*, 34(7):609–630, 2004.

[360] J. Paquet. Relational databases as multidimensional dataflows. Master's thesis, Departement d'Informatique, Université Laval, Québec, Canada, 1995.

[361] J. Paquet. *Scientific Intensional Programming*. PhD thesis, Department of Computer Science, Laval University, Sainte-Foy, Canada, 1999.

[362] J. Paquet. Distributed eductive execution of hybrid intensional programs. In *Proceedings of the 33rd Annual IEEE International Computer Software and Applications Conference (COMPSAC'09)*, pages 218–224, Seattle, Washington, USA, July 2009. IEEE Computer Society.

[363] J. Paquet and P. Kropf. The GIPSY architecture. In *Proceedings of Distributed Computing on the Web*, Quebec City, Canada, 2000.

[364] J. Paquet and S. A. Mokhov. Furthering baseline core Lucid. [online], 2011–2013. http://arxiv.org/abs/1107.0940.

[365] J. Paquet, S. A. Mokhov, and X. Tong. Design and implementation of context calculus in the GIPSY environment. In *Proceedings of the 32nd Annual IEEE International Computer Software and Applications Conference (COMPSAC)*, pages 1278–1283, Turku, Finland, July 2008. IEEE Computer Society.

[366] J. Paquet, S. A. Mokhov, E. I. Vassev, X. Tong, Y. Ji, A. H. Pourteymour, K. Wan, A. Wu, S. Rabah, B. Han, B. Lu, L. Tao, Y. Ding, C. L. Ren, and The GIPSY Research and Development Group. The General Intensional Programming System (GIPSY) project. Department of Computer Science and Software Engineering, Concordia University, Montreal, Canada, 2002–2014. http://newton.cs.concordia.ca/~gipsy/, last viewed January 2014.

[367] J. Paquet and J. Plaice. The intensional relation. In Orgun and Ashcroft [350], pages 214–227.

[368] J. Paquet and J. Plaice. The semantics of dimensions as values. In Gergatsoulis and Rondogiannis [131], pages 259–273.

[369] J. Paquet, A. Wu, and P. Grogono. Towards a framework for the General Intensional Programming Compiler in the GIPSY. In *Proceedings of the 19th Annual ACM Conference on Object-Oriented Programming, Systems, Languages, and Applications (OOPSLA 2004)*, pages 164–165, New York, NY, USA, Oct. 2004. ACM.

[370] J. Paquet and A. H. Wu. GIPSY – a platform for the investigation on intensional programming languages. In *Proceedings of the 2005 International Conference on Programming Languages and Compilers (PLC 2005)*, pages 8–14. CSREA Press, June 2005.

[371] M. Parashar and S. Hariri, editors. *Autonomic Computing: Concepts, Infrastructure and*

*Applications*. CRC Press, Dec. 2006.

[372] L. C. Paulson, T. Nipkow, and M. Wenzel. Isabelle: A generic proof assistant. [online], University of Cambridge and Technical University of Munich, 2007–2013. http://isabelle.in.tum.de/, last viewed March 2013.

[373] C. Pearce. Computing forensics: a live analysis. [online], Apr. 2005. http://www.linux.org.au/conf/2005/security_miniconf/presentations/crpearce-lca2005.pdf.

[374] C. Pearce. Helix: Open-source forensic toolkit. [online], Apr. 2005. http://www.e-fense.com/helix.

[375] M. Peralta, S. Mukhopadhyay, and R. Bharadwaj. Automatic synthesis and deployment of intensional kahn process networks. In D. Ślęzak, T. hoon Kim, S. S. Yau, O. Gervasi, and B.-H. Kang, editors, *Grid and Distributed Computing*, volume 63 of *Communications in Computer and Information Science*, pages 73–87. Springer Berlin Heidelberg, 2009.

[376] J. Plaice. Particle in-cell simulation with Lucid. In Orgun and Ashcroft [350], pages 149–161.

[377] J. Plaice. Cartesian programming. Technical Report UNSW-CSE-TR-1101, University of Grenoble, France, Jan. 2011. Habilitation Thesis, online at ftp://ftp.cse.unsw.edu.au/pub/doc/papers/UNSW/1101.pdf.

[378] J. Plaice, B. Mancilla, and G. Ditu. From Lucid to TransLucid: Iteration, dataflow, intensional and Cartesian programming. *Mathematics in Computer Science*, 2(1):37–61, 2008.

[379] J. Plaice, B. Mancilla, G. Ditu, and W. W. Wadge. Sequential demand-driven evaluation of eager TransLucid. In *Proceedings of the 32nd Annual IEEE International Computer Software and Applications Conference (COMPSAC)*, pages 1266–1271, Turku, Finland, July 2008. IEEE Computer Society.

[380] J. Plaice and J. Paquet. Introduction to intensional programming. In Orgun and Ashcroft [350], pages 1–14. Tutorial.

[381] G. D. Plotkin. A structural approach to operational semantics. Lecture Notes DAIMI FN19, Department of Computer Science, University of Aarhus, 1981.

[382] D. C. Plummer. RFC 826: An Ethernet Address Resolution Protocol. [online], Nov. 1982. http://tools.ietf.org/html/rfc826, viewed in December 2012.

[383] A. H. Pourteymour. Comparative study of Demand Migration Framework implementation using JMS and Jini. Master's thesis, Department of Computer Science and Software Engineering, Concordia University, Montreal, Canada, Sept. 2008.

[384] A. H. Pourteymour, E. Vassev, and J. Paquet. Towards a new demand-driven message-oriented middleware in GIPSY. In *Proceedings of PDPTA 2007*, pages 91–97. PDPTA, CSREA Press, June 2007.

[385] A. H. Pourteymour, E. Vassev, and J. Paquet. Design and implementation of demand migration systems in GIPSY. In *Proceedings of PDPTA 2009*. CSREA Press, June 2008.

[386] N. Provos. Steganography detection with stegdetect. [online], 2004. http://www.outguess.org/detection.php.

[387] N. Provos, D. McNamee, P. Mavrommatis, K. Wang, and N. Modadugu. The ghost in the browser analysis of web-based malware. http://www.usenix.org/events/hotbots07/tech/fullpapers/provos/provos.pdf, 2007.

[388] M. Puckette and PD Community. Pure Data. [online], 2007–2013. http://puredata.org.

[389] G. N. Purdy. *Linux iptables: Pocket Reference*. O'Reilly, 2004. ISBN: 978-0-596-00569-6.

[390] QoSient, LLC. Argus: Auditing network activity. [online], 2000–2013. http://www.qosient.com/argus/.

[391] P. Quinon and M. Antonutti, editors. *Workshop on Intensionality in Mathematics*, May 2013. Online at http://www.fil.lu.se/index.php?id=18879.

[392] S. Rabah. A broker-based resource publication and discovery framework for network virtualization environment. Master's thesis, Department of Computer Science and Software Engineering, Concordia University, Montreal, Canada, Jan. 2014.

[393] S. Rabah, S. A. Mokhov, and J. Paquet. An interactive graph-based automation assistant: A case study to manage the GIPSY's distributed multi-tier run-time system. In C. Y. Suen, A. Aghdam, M. Guo, J. Hong, and E. Nadimi, editors, *Proceedings of the ACM Research in Adaptive and Convergent Systems (RACS 2013)*, pages 387–394, New York, NY, USA, Oct. 2011–2013. ACM. Pre-print: http://arxiv.org/abs/1212.4123.

[394] A. Raffaetà and T. Frühwirth. Two semantics for temporal annotated constraint logic. In Gergatsoulis and Rondogiannis [131], pages 78–92.

[395] A. N. M. E. Rafiq and Y. Mao. A novel approach for automatic adjudification of new malware. In N. Callaos, W. Lesso, C. D. Zinn, J. Baralt, J. Boukachour, C. White, T. Marwala, and F. V. Nelwamondo, editors, *Proceedings of the 12th World Multi-Conference on Systemics, Cybernetics and Informatics (WM-SCI'08)*, volume V, pages 137–142, Orlando, Florida, USA, June 2008. IIIS.

[396] T. Rahilly and J. Plaice. A multithreaded implementation for TransLucid. In *Proceedings of the 32nd Annual IEEE International Computer Software and Applications Conference (COMPSAC)*, pages 1272–1277, Turku, Finland, July 2008. IEEE Computer Society.

[397] A. Ranganathan and R. H. Campbell. A middleware for context-aware agents in ubiquitous computing environments. In M. Endler and D. Schmidt, editors, *Proceedings of Middleware 2003*, volume 2672 of *Lecture Notes in Computer Science*, pages 143–161. Springer Berlin Heidelberg, 2003.

[398] M. Rash. *Linux Firwalls: Attack Detection and Response with iptables, psad, and fwsnort*. No Starch Press, Inc., San Francisco, 3 edition, 2007. ISBN: 978-1-59327-141-1.

[399] C. L. Ren. General intensional programming compiler (GIPC) in the GIPSY. Master's thesis, Department of Computer Science and Software Engineering, Concordia University, Montreal, Canada, 2002.

[400] G. Riley. CLIPS: A tool for building expert systems. [online], 2007–2011. http://clipsrules.sourceforge.net/, last viewed May 2012.

[401] J. Rilling, W. J. Meng, and O. Ormandjieva. Context driven slicing based coupling measure. In *ICSM*, page 532, 2004.

[402] R. W. Ritchey and P. Ammann. Using model checking to analyze network vulnerabilities. In *Proceedings of the 2000 IEEE Symposium on Security and Privacy (SP'00)*, pages 156–, Washington, DC, USA, 2000. IEEE Computer Society.

[403] RJK. Regexp syntax summary. http://www.greenend.org.uk/rjk/2002/06/regexp.html, last viewed May 2008, June 2002.

[404] Y. Rogers, H. Sharp, and J. Preece. *Interaction Design: Beyond Human - Computer Interaction*. Wiley Publishing, 3rd edition, 2011. Online resources: id-book.com.

[405] P. Rondogiannis. *Higher-Order Functional Languages and Intensional Logic*. PhD thesis, Department of Computer Science, University of Victoria, Victoria, Canada, 1994.

[406] P. Rondogiannis. Adding multidimensionality to procedural programming languages. In Gergatsoulis and Rondogiannis [131], pages 274–291.

[407] P. Rondogiannis. Adding multidimensionality to procedural programming languages. *Software: Practice and Experience*, 29(13):1201–1221, 1999.

[408] P. Rondogiannis and W. W. Wadge. Extending the intensionalization algorithm to a broader class of higher-order programs. In Orgun and Ashcroft [350], pages 228–233.

[409] P. Rubin, A. Robbins, J. Kingdon, D. MacKenzie, and R. Smith. `touch` – change file timestamps, touch(1). GNU coreutils 8.10, Feb. 2011. man file(1).

[410] F. Rudzicz and S. A. Mokhov. Towards a heuristic categorization of prepositional phrases in English with WordNet. [online], 2003–2013. http://arxiv.org/abs/1002.1095.

[411] S. J. Russell and P. Norvig, editors. *Artificial Intelligence: A Modern Approach*. Prentice Hall, New Jersey, USA, 1995. ISBN 0-13-103805-2.

[412] B. Sarnecka. The first few numbers: How children learn them and why it matters. In Quinon and Antonutti [391]. Online at http://www.fil.lu.se/index.php?id=18879.

[413] B. Sateli, E. Angius, S. S. Rajivelu, and R. Witte. Can text mining assistants help to improve requirements specifications? In *Proceedings of the Mining Unstructured Data (MUD'12)*, Oct. 2012. http://sailhome.cs.queensu.ca/mud/res/sateli-mud2012.pdf.

[414] B. Sateli and R. Witte. Natural language processing for mediawiki: The semantic assistants approach. In *Proceedings of the 8th International Symposium on Wikis and Open Collaboration (WikiSym'12)*. ACM, Aug. 2012.

[415] D. Savage, J. Harrington, J. Yembrick, M. Curie, and NASA. NASA to discuss Hubble anomaly and servicing mission launch delay. [online], Sept. 2008. http://www.nasa.gov/home/hqnews/2008/sep/HQ_M08-187_HST_Telecon.html.

[416] U. Schöning. *Logic for Computer Scientists*. Birkhäuser Boston, 2008.

[417] M. C. Schraefel, B. Mancilla, and J. Plaice. Intensional hypertext. In Gergatsoulis and Rondogiannis [131], pages 40–54.

[418] R. Schreiber. MATLAB. *Scholarpedia*, 2(6):2929, 2007. http://www.scholarpedia.org/article/MATLAB.

[419] M. G. Schultz, E. Eskin, E. Zadok, and S. J. Stolfo. Data mining methods for detection of new malicious executables. In *Proceedings of IEEE Symposium on Security and Privacy*, pages 38–49, Oakland, 2001.

[420] I. Selesnick, S. Cai, K. Li, L. Sendur, and A. F. Abdelnour. MATLAB implementation of wavelet transforms. Technical report, Electrical Engineering, Polytechnic University, Brooklyn, NY, 2003. Online at http://taco.poly.edu/WaveletSoftware/.

[421] K. Sentz and S. Ferson. Combination of evidence in dempster-shafer theory. Technical Report SAND 2002-0835, Sandia, Apr. 2002. http://www.sandia.gov/epistemic/Reports/SAND2002-0835.pdf.

[422] G. Shafer. *The Mathematical Theory of Evidence*. Princeton University Press, 1976.

[423] G. Shafer. Perspectives on the theory and practice of belief functions. *International Journal of Approximate Reasoning*, 3:1–40, 1990.

[424] G. Shafer. Dempster-Shafer theory. [online], 2002. http://www.glennshafer.com/assets/downloads/articles/article48.pdf.

[425] G. Shafer and J. Pearl, editors. *Readings in Uncertain Reasoning*. Morgan Kaufmann, 1990.

[426] V. Sharma, J. G. Muers, and S. Lewis. Continual feature selection: a cost effective method to enhancing the capabilities. In H. Martin, editor, *Proceedings of the 17th Virus Bulletin International Conference*, pages 9–17, Vienna, Austria: The Pentagon, Abingdon, OX143YP, England, Sept. 2007.

[427] O. Sheyner, J. Haines, S. Jha, R. Lippmann, and J. M. Wing. Automated generation and analysis of attack graphs. In *Proceedings of the 2002 IEEE Symposium on Security and Privacy*, pages 273–, Washington, DC, USA, 2002. IEEE Computer Society.

[428] L. Shi, S. Lu, T. Sun, and D. Ouyang. A hybrid system combining intuitionistic fuzzy description logics with intuitionistic fuzzy logic programs. In *Proceedings of the Eighth International Conference on Fuzzy Systems and Knowledge Discovery (FSKD'11)*, volume 1, pages 60–64, 2011.

[429] B. Shishkov, J. Cordeiro, and A. Ranchordas, editors. *ICSOFT 2009 - Proceedings of the 4th International Conference on Software and Data Technologies*, volume 1. INSTICC Press, July

[430] G. J. Simon, H. Xiong, E. Eilertson, and V. Kumar. Scan detection: A data mining approach. In *Proceedings of SDM 2006*, pages 118–129, 2006. http://www.siam.org/meetings/sdm06/proceedings/011simong.pdf.

[431] D. A. Simovici and C. Djeraba. *Mathematical Tools for Data Mining: Set Theory, Partial Orders, Combinatorics*. Springer Publishing Company, Incorporated, 1st edition, 2008. ISBN: 9781848002005.

[432] M. P. Singh and M. N. Huhns. *Service-Oriented Computing: Semantics, Processes, Agents*. John Wiley & Sons, Ltd, West Sussex, England, 2005.

[433] E. Skoudis and L. Zelster. *Malware: Fighting Malicious Code*. Computer Networking and Distributed Systems. Prentice Hall, 2004.

[434] M. G. Solomon, K. Rudolph, E. Tittel, N. Broom, and D. Barrett. *Computer Forensics JumpStart*. Sybex, 2 edition, Mar. 2011.

[435] D. Song. BitBlaze: Security via binary analysis. [online], 2010. Online at http://bitblaze.cs.berkeley.edu.

[436] D. Song. WebBlaze: New techniques and tools for web security. [online], 2010. Online at http://webblaze.cs.berkeley.edu.

[437] M. Song. *Computer-Assisted Interactive Documentary and Performance Arts in Illimitable Space*. PhD thesis, Special Individualized Program/Computer Science and Software Engineering, Concordia University, Montreal, Canada, Dec. 2012. Online at http://spectrum.library.concordia.ca/975072 and http://arxiv.org/abs/1212.6058.

[438] Sourcefire. Snort: Open-source network intrusion prevention and detection system (IDS/IPS). [online], 1999–2013. http://www.snort.org/.

[439] Splunk Inc. Splunk: Search and analysis engine for IT data. [online], 2005–2012. http://www.splunk.com/.

[440] W. Stallings. *SNMP, SNMPv2, SNMPv3, and RMON 1 and 2*. Addison-Wesley, 3 edition, 1999. ISBN: 0-201-48534-6.

[441] N. Stankovic, M. A. Orgun, W. Cai, and K. Zhang. *Visual Parallel Programming*, chapter 6, pages 103—129. World Scientific Publishing Co., Inc., 2002.

[442] B. Sterns. The Java Native Interface (JNI). [online], 2001–2005. http://java.sun.com/developer/onlineTraining/Programming/JDCBook/jni.html.

[443] W. R. Stevens and S. A. Rago. *Advanced Programming in the UNIX Environment*. Pearson Education, Inc., second edition, June 2005. ISBN 0-201-43307-9.

[444] J. M. Stewart, M. Chapple, and D. Gibson. *CISSP: Certified Information Systems Security Professional Study Guide*. Sybex, 6 edition, July 2012.

[445] M. Suenaga. Virus linguistics – searching for ethnic words. In H. Martin, editor, *Proceedings of the 17th Virus Bulletin International Conference*, pages 9–17, Vienna, Austria: The Pentagon, Abingdon, OX143YP, England, Sept. 2007.

[446] Sun Microsystems, Inc. Java IDL. Sun Microsystems, Inc., 2004. http://java.sun.com/j2se/1.5.0/docs/guide/idl/index.html.

[447] Sun Microsystems, Inc. The Java web services tutorial (for Java Web Services Developer's Pack, v2.0). [online], Feb. 2006. http://download.oracle.com/docs/cd/E17802_01/webservices/webservices/docs/2.0/tutorial/doc/.

[448] Sun Microsystems, Inc. *Class URI*. Sun Microsystems, Inc., 2007. http://java.sun.com/j2se/1.5.0/docs/api/java/net/URI.html, Viewed in November, 2007.

[449] Sun Microsystems, Inc. Java Message Service (JMS). [online], Sept. 2007. http://java.sun.com/products/jms/.

[450] Sun Microsystems, Inc. NetBeans 6.7.1. [online], 2009–2010. http://netbeans.org/

```
downloads/6.7.1/index.html.
```

[451] A. H. Sung, J. Xu, P. Chavez, and S. Mukkamala. Static analyzer of vicious executables (SAVE). In *Proceedings of 20th Annual of Computer Security Applications Conference*, pages 326–334, Dec. 2004.

[452] P. Swoboda. *A Formalisation and Implementation of Distributed Intensional Programming*. PhD thesis, The University of New South Wales, Sydney, Australia, 2004.

[453] P. Swoboda and J. Plaice. An active functional intensional database. In F. Galindo, editor, *Advances in Pervasive Computing*, pages 56–65. Springer, 2004. LNCS 3180.

[454] P. Swoboda and J. Plaice. A new approach to distributed context-aware computing. In A. Ferscha, H. Hoertner, and G. Kotsis, editors, *Advances in Pervasive Computing*. Austrian Computer Society, 2004. ISBN 3-85403-176-9.

[455] P. Swoboda and W. W. Wadge. Vmake and ISE general tools for the intensionalization of software systems. In Gergatsoulis and Rondogiannis [131], pages 310–320. ISBN: 981-02-4095-3.

[456] A. S. Tanenbaum and D. J. Wetherall. *Computer Networks*. Prentice Hall, fifth edition, 2011. ISBN: 978-0-13-212695-3.

[457] W. N. Tankou. A unified framework for measuring a network's mean time-to-compromise. Master's thesis, Concordia Institute for Information Systems Engineering, Concordia University, Montreal, Canada, June 2013.

[458] C. Tao, K. Wongsuphasawat, K. Clark, C. Plaisant, B. Shneiderman, and C. G. Chute. Towards event sequence representation, reasoning and visualization for EHR data. In *Proceedings of the 2nd ACM SIGHIT International Health Informatics Symposium*, IHI'12, pages 801–806, New York, NY, USA, 2012. ACM.

[459] L. Tao. Warehouse and garbage collection in the GIPSY environment. Master's thesis, Department of Computer Science and Software Engineering, Concordia University, Montreal, Canada, 2004.

[460] Tenable Network Security. Nessus: the network vulnerability scanner. [online], 2002–2013. http://www.nessus.org/nessus/.

[461] The Coroner's Toolkit Project. The coroner's toolkit (TCT). [online], 1999. http://www.porcupine.org/forensics/tct.html.

[462] The GATE Team. General Architecture for Text Engineering (GATE). [online], 1995–2013. http://gate.ac.uk/, last viewed April 2012.

[463] The GIPSY Research and Development Group. The General Intensional Programming System (GIPSY) project. Department of Computer Science and Software Engineering, Concordia University, Montreal, Canada, 2002–2013. http://newton.cs.concordia.ca/~gipsy/, last viewed March 2013.

[464] The Honeynet Project. *Know Your Enemy*. Honeynet, 2nd edition, 2004.

[465] The MARF Research and Development Group. The Modular Audio Recognition Framework and its Applications. [online], 2002–2013. http://marf.sf.net and http://arxiv.org/abs/0905.1235, last viewed April 2012.

[466] The PlanetLab Project. PlanetLab – an open platform for developing, deploying, and accessing planetary-scale services. [online], 2007–2013. http://www.planet-lab.org/.

[467] The PRISM Team. PRISM: a probabilistic model checker. [online], 2004–2013. http://www.prismmodelchecker.org/, last viewed January 2013.

[468] The Sphinx Group at Carnegie Mellon. The CMU Sphinx group open source speech recognition engines. [online], 2007–2012. http://cmusphinx.sourceforge.net.

[469] The Weka Project. Weka 3: Data mining with open source machine learning software in Java. [online], 2006–2013. http://www.cs.waikato.ac.nz/ml/weka/.

[470] ThreatTrack Security. ThreadAnalyzer: Dynamic sandboxing and malware analysis (formerly GFI SandBox). [online], 2013. http://www.threattracksecurity.com/enterprise-security/sandbox-software.aspx.

[471] C. Thuen. Understanding counter-forensics to ensure a successful investigation. [online], University of Idaho, 2007. http://citeseerx.ist.psu.edu/viewdoc/summary?doi=10.1.1.138.2196.

[472] S. Tlili. *Automatic detection of safety and security vulnerabilities in open source software*. PhD thesis, Concordia Institute for Information Systems Engineering, Concordia University, Montreal, Canada, 2009. ISBN: 9780494634165.

[473] X. Tong. Design and implementation of context calculus in the GIPSY. Master's thesis, Department of Computer Science and Software Engineering, Concordia University, Montreal, Canada, Apr. 2008.

[474] X. Tong, J. Paquet, and S. A. Mokhov. Complete context calculus design and implementation in GIPSY. [online], 2007–2008. http://arxiv.org/abs/1002.4392.

[475] K. Trinidad, K. Herring, S. Hendrix, and E. C. NASA. NASA sets target shuttle launch date for Hubble servicing mission. [online], Dec. 2008. http://www.nasa.gov/home/hqnews/2008/dec/HQ_08-320_Hubble_May2009.html.

[476] W. Truszkowski, M. Hinchey, J. Rash, and C. Rouff. NASA's swarm missions: The challenge of building autonomous software. *IT Professional*, 6(5):47–52, 2004.

[477] D. G. Ullman. *Making Robust Decisions: Decision Management For Technical, Business, and Service Teams*. Victoria: Trafford, 2007. ISBN: 1-4251-0956-X, http://www.stsc.hill.af.mil/CrossTalk/2007/04/0704Ullman.html.

[478] T. Uustalu and V. Vene. The essence of dataflow programming. In K. Yi, editor, *Proceedings of APLAS'05*, volume 3780 of *Lecture Notes in Computer Science*, pages 2–18. Springer Berlin Heidelberg, 2005.

[479] U. Vahalia. *UNIX Internals: The New Frontiers*. Prentice Hall, Inc., second edition, 1996. ISBN 0-13-101908-2.

[480] J. van Benthem. *A Manual of Intensional Logic*. CSLI Publications, Stanford and The University of Chicago Press, 1988.

[481] S. R. van den Berg and P. A. Guenther. procmail v3.22. [online], Sept. 2001. http://www.procmail.org/.

[482] A. van Lamsweerde. *Requirements Engineering: From System Goals to UML Models to Software Specifications*. Wiley, 2009. ISBN: 978-0-470-01270-3.

[483] Various Contributors. `fstat`, `fstat64`, `lstat`, `lstat64`, `stat`, `stat64` – get file status, BSD System Calls Manual, stat(2). BSD, Apr. 1994. man stat(2).

[484] Various contributors and MITRE. Common Weakness Enumeration (CWE) – a community-developed dictionary of software weakness types. [online], 2006–2013. See http://cwe.mitre.org.

[485] Various Contributors and the GNU Project. GNU Compiler Collection (GCC). [online], 1988–2009. http://gcc.gnu.org/onlinedocs/gcc/.

[486] E. Vassev. *ASSL: Autonomic System Specification Language – A Framework for Specification and Code Generation of Autonomic Systems*. LAP Lambert Academic Publishing, Nov. 2009. ISBN: 3-838-31383-6.

[487] E. Vassev and M. Hinchey. ASSL: A software engineering approach to autonomic computing. *IEEE Computer*, 42(6):90–93, 2009.

[488] E. Vassev and M. Hinchey. ASSL specification model for the image-processing behavior in the NASA Voyager mission. Technical report, Lero - The Irish Software Engineering Research Center, 2009.

[489] E. Vassev, M. Hinchey, and A. J. Quigley. A self-adaptive architecture for autonomic systems developed with ASSL. In Shishkov et al. [429], pages 163–168.

[490] E. Vassev, M. Hinchey, and A. J. Quigley. Towards model checking with Java PathFinder for autonomic systems specified and generated with ASSL. In Shishkov et al. [429], pages 251–256.

[491] E. Vassev, M. G. Hinchey, and J. Paquet. Towards an ASSL specification model for NASA swarm-based exploration missions. In *Proceedings of the 23rd Annual ACM Symposium on Applied Computing (SAC 2008) - AC Track*, pages 1652–1657. ACM, 2008.

[492] E. Vassev, H. Kuang, O. Ormandjieva, and J. Paquet. Reactive, distributed and autonomic computing aspects of AS-TRM. In J. Filipe, B. Shishkov, and M. Helfert, editors, *ICSOFT (1)*, pages 196–202. INSTICC Press, Sept. 2006.

[493] E. Vassev and S. A. Mokhov. An ASSL-generated architecture for autonomic systems. In *Proceedings of C3S2E'09*, pages 121–126, New York, NY, USA, May 2009. ACM.

[494] E. Vassev and S. A. Mokhov. Self-optimization property in autonomic specification of Distributed MARF with ASSL. In B. Shishkov, J. Cordeiro, and A. Ranchordas, editors, *Proceedings of ICSOFT'09*, volume 1, pages 331–335, Sofia, Bulgaria, July 2009. INSTICC Press.

[495] E. Vassev and S. A. Mokhov. Towards autonomic specification of Distributed MARF with ASSL: Self-healing. In *Proceedings of SERA 2010 (selected papers)*, volume 296 of *SCI*, pages 1–15. Springer, 2010.

[496] E. Vassev and S. A. Mokhov. Developing autonomic properties for distributed pattern-recognition systems with ASSL: A Distributed MARF case study. *LNCS Transactions on Computational Science, Special Issue on Advances in Autonomic Computing: Formal Engineering Methods for Nature-Inspired Computing Systems*, XV(7050):130–157, 2012. Accepted in 2010; appeared February 2012.

[497] E. Vassev, O. Ormandjieva, and J. Paquet. ASSL specification of reliability self-assessment in the AS-TRM. In J. Filipe, B. Shishkov, and M. Helfert, editors, *Proceedings of the 2nd International Conference on Software and Data Technologies (ICSOFT 2007)*, volume SE, pages 198–206. INSTICC Press, July 2007.

[498] E. Vassev and J. Paquet. A general architecture for demand migration in a demand-driven execution engine in a heterogeneous and distributed environment. In *Proceedings of the 3rd Annual Communication Networks and Services Research Conference (CNSR 2005)*, pages 176–182. IEEE Computer Society, May 2005.

[499] E. Vassev and J. Paquet. A generic framework for migrating demands in the GIPSY's demand-driven execution engine. In *Proceedings of the 2005 International Conference on Programming Languages and Compilers (PLC 2005)*, pages 29–35. CSREA Press, June 2005.

[500] E. Vassev and J. Paquet. Towards autonomic GIPSY. In *Proceedings of the Fifth IEEE Workshop on Engineering of Autonomic and Autonomous Systems (EASE 2008)*, pages 25–34. IEEE Computer Society, 2008.

[501] E. I. Vassev. General architecture for demand migration in the GIPSY demand-driven execution engine. Master's thesis, Department of Computer Science and Software Engineering, Concordia University, Montreal, Canada, June 2005. ISBN 0494102969.

[502] E. I. Vassev. *Towards a Framework for Specification and Code Generation of Autonomic Systems*. PhD thesis, Department of Computer Science and Software Engineering, Concordia University, Montreal, Canada, 2008.

[503] L. Verdoscia. ALFA fine grain dataflow machine. In Orgun and Ashcroft [350], pages 110–134.

[504] J. Vincent, D. Rolsky, D. Chamberlain, R. Foley, and R. Spier. *RT Essentials*. O'Reilly Media, Inc., Aug. 2005.

[505] P. C. Vinh and J. P. Bowen. On the visual representation of configuration in reconfigurable

computing. *Electron. Notes Theor. Comput. Sci.*, 109:3–15, 2004.

[506] S. Viswanadha and Contributors. Java compiler compiler (JavaCC) - the Java parser generator. [online], 2001–2008. https://javacc.dev.java.net/.

[507] W. W. Wadge. Possible WOOrlds. In Orgun and Ashcroft [350], pages 56–62. Invited Contribution.

[508] W. W. Wadge. Intensional logic in context. In Gergatsoulis and Rondogiannis [131], pages 1–13. Tutorial.

[509] W. W. Wadge and E. A. Ashcroft. *Lucid, the Dataflow Programming Language*. Academic Press, London, 1985.

[510] W. W. Wadge, G. Brown, M. C. Schraefel, and T. Yildirim. Intensional HTML. In *4th International Workshop PODDP'98*, Mar. 1998.

[511] W. W. Wadge and M. C. Schraefel. Putting the hyper back in hypertext. In Gergatsoulis and Rondogiannis [131], pages 31–39.

[512] W. W. Wadge and A. Yoder. The Possible-World Wide Web. In Orgun and Ashcroft [350], pages 207–213.

[513] K. Wan. *Lucx: Lucid Enriched with Context*. PhD thesis, Department of Computer Science and Software Engineering, Concordia University, Montreal, Canada, 2006.

[514] K. Wan, V. Alagar, and J. Paquet. A context theory for intensional programming. In *Workshop on Context Representation and Reasoning (CRR05)*, July 2005.

[515] K. Wan, V. Alagar, and J. Paquet. Lucx: Lucid enriched with context. In *Proceedings of the 2005 International Conference on Programming Languages and Compilers (PLC 2005)*, pages 48–14. CSREA Press, June 2005.

[516] L. Wang, S. Jajodia, and D. Wijesekera. *Preserving Privacy in On-line Analytical Processing (OLAP)*. Springer, Berlin, 2007. ISBN: 0-387-46273-2.

[517] T. D. Wang, A. Deshpande, and B. Shneiderman. A temporal pattern search algorithm for personal history event visualization. *IEEE Trans. on Knowl. and Data Eng.*, 24(5):799–812, May 2012.

[518] M. Wenzel, L. C. Paulson, and T. Nipkow. The Isabelle framework. In Mohamed et al. [259], pages 33–38.

[519] Wikipedia. S.M.A.R.T. — Wikipedia, the free encyclopedia. [Online; accessed 2-May-2012], 2012. http://en.wikipedia.org/w/index.php?title=S.M.A.R.T.&oldid=489447839.

[520] Wikipedia. Dempster–Shafer theory — Wikipedia, The Free Encyclopedia. [Online; accessed 15-July-2013], 2013. http://en.wikipedia.org/w/index.php?title=Dempster-Shafer_theory&oldid=558643087.

[521] R. Witte. SOEN 6481: Systems requirements specification, lecture notes. Department of Computer Science and Software Engineering, Concordia University, Montreal, Canada, 2012. Fall 2012.

[522] R. Witte, Q. Li, Y. Zhang, and J. Rilling. Text mining and software engineering: An integrated source code and document analysis approach. *IET Software Journal, Special Issue on Natural Language in Software Development*, 2(1):3–16, Feb. 2008. http://www.semanticsoftware.info/system/files/witte_etal_iet2008.pdf.

[523] A. Wollrath and J. Waldo. Java RMI tutorial. Sun Microsystems, Inc., 1995–2005. http://java.sun.com/docs/books/tutorial/rmi/index.html.

[524] F. Wolter and M. Zakharyaschev. Multi-dimensional description logics. In T. Dean, editor, *Proceedings of the Sixteenth International Joint Conference on Artificial Intelligence, (IJCAI 1999)*, volume 2, pages 104–109. Morgan Kaufmann, 1999. http://ijcai.org/PastProceedings/IJCAI-99-VOL-1/PDF/016.pdf.

[525] S. W. Wood. A forensic computing framework to fit any legal system. In O. Göbel, S. Frings,

D. Günther, J. Nedon, and D. Schadt, editors, *Proceedings of the IT Incident Management and IT Forensics (IMF'08)*, LNI140, pages 41–54, Sept. 2008.

[526] A. Wu, J. Paquet, and S. A. Mokhov. Object-oriented intensional programming: Intensional Java/Lucid classes. In *Proceedings of SERA 2010*, pages 158–167. IEEE Computer Society, 2010. Online at: http://arxiv.org/abs/0909.0764.

[527] A. H. Wu. Semantic checking and translation in the GIPSY. Master's thesis, Department of Computer Science and Software Engineering, Concordia University, Montreal, Canada, 2002.

[528] A. H. Wu. *OO-IP Hybrid Language Design and a Framework Approach to the GIPC*. PhD thesis, Department of Computer Science and Software Engineering, Concordia University, Montreal, Canada, 2009.

[529] A. H. Wu and J. Paquet. Object-oriented intensional programming in the GIPSY: Preliminary investigations. In *Proceedings of the 2005 International Conference on Programming Languages and Compilers (PLC 2005)*, pages 43–47. CSREA Press, June 2005.

[530] A. H. Wu, J. Paquet, and P. Grogono. Design of a compiler framework in the GIPSY system. In *Proceedings of the 15th IASTED International Conference on Parallel and Distributed Computing and Systems (PDCS 2003)*, volume 1, pages 320–328. International Association of Science and Technology for Development, Nov. 2003.

[531] M.-D. Wu and S. D. Wolfthusen. Network forensics of partial SSL/TLS encrypted traffic classification using clustering algorithms. In O. Göbel, S. Frings, D. Günther, J. Nedon, and D. Schadt, editors, *Proceedings of the IT Incident Management and IT Forensics (IMF'08)*, LNI140, pages 157–172, Sept. 2008.

[532] H. Yahyaoui, M. Debbabi, and N. Tawbi. A denotational semantic model for validating JVML/CLDC optimizations under Isabelle/HOL. In *Proceedings of the 7th International Conference on Quality Software (QSIC'07)*, pages 348–355, 2007.

[533] S. Yamane. Real-time object-oriented specification and verification. In Orgun and Ashcroft [350], pages 162–185.

[534] Y. Yan. Description language and formal methods for web service process modeling. In *Business Process Management: Concepts, Technologies and Applications*, volume Advances in Management Information Systems. M.E. Sharpe Inc., 2008.

[535] Y. Yan and X. Zheng. A planning graph based algorithm for semantic web service composition. In *CEC/EEE 2008*, pages 339–342, 2008.

[536] L. A. Zadeh. On the validity of Dempster's rule of combination of evidence. Technical Report 79/24, University of California, Berkely, CA, 1979.

[537] L. A. Zadeh. Fuzzy logic. *Scholarpedia*, 3(3):1766, 2008. http://www.scholarpedia.org/article/Fuzzy_logic.

[538] Y. Zhang, J. Rilling, and V. Haarslev. An ontology based approach to software comprehension–reasoning about security concerns in source code. In *30th IEEE COMPSAC 2006*, Chicago, Sept. 2006.

[539] Q. Zhao. Implementation of an object-oriented intensional programming system. Master's thesis, Department of Computer Science, University of New Brunswick, Canada, 1997.

[540] C. Zheng and J. R. Heath. Simulation and visualization of resource allocation, control, and load balancing procedures for a multiprocessor architecture. In *MS'06: Proceedings of the 17th IASTED international conference on Modelling and simulation*, pages 382–387, Anaheim, CA, USA, 2006. ACTA Press.

# Part IV

# Appendices

# Appendix A

# Glossary

**ACL** — access control list

**ACL2** — A Computational Logic for Applicative Common Lisp [205]

**ADMARF** — Autonomic DMARF

**ADM** — ADM (Adia–Debbabi–Mejri) process logic [4]

**ACME** — fictitious company name[1]

**AGIPSY** — Autonomic GIPSY [500]

**AOP** — aspect-oriented programming

**API** — application programming interface[2]

**ARP** — Address Resolution Protocol [382][3]

**AE** — autonomic element

**AS** — autonomic system

**AS-TRM** — autonomic system-time reactive model [492, 497]

**ASSL** — Autonomic System Specification Language [502]

**AST** — abstract syntax tree

**AYDY** — *are-you-done-yet* polling

**BIOS** — Basic Input/Output System

**BPEL** — Business Process Execution Language [177, 210, 349]

**CAD** — computer-aided design

**CAF** — computer anti-forensics

**CCD** — charge-coupled device

**CF** — computer forensics

**CFG** — control-flow graph

---

[1] http://en.wikipedia.org/wiki/Acme_Corporation
[2] http://en.wikipedia.org/wiki/API
[3] http://en.wikipedia.org/wiki/Address_Resolution_Protocol

**Cisco IOS** — originally Internetwork Operating System

**CIFS** — Common Internet File System

**CLI** — command-line interface

**CLIPS** — C Language Integrated Production System [400]

**codecessor** — A (technically) co-descendant notion or concept, but not necessarily a close sibling to concepts alike emerged around the same timeframe from the same or similar predecessor concept(s) [300, 304, 312]

**CORBA** — Common Object Requester Broker Architecture

**CPU** — central processing unit

**CSV** — comma-separated values

**CVE** — Common Vulnerabilities and Exposures

**CVS** — Concurrent Versions System [146]

**CWE** — Common Weakness Enumeration [484]

**DAG** — directed acyclic graph

**DBMS** — database management system

**DBSCAN** — Density Based Spacial Clustering of Application with Noise [531]

**DFG** — data-flow graph

**DFRWS** — Digital Forensics Research Workshop [357]

**DGT** — Demand Generator Tier

**DIR** — directory

**DHCP** — Dynamic Host Control Protocol

**DL** — description logic

**DMARF** — Distributed MARF

**DMF** — Demand Migration Framework

**DMS** — Demand Migration System

**DNS** — Domain Name Service

**DSTME** — Dempster–Shafer theory of mathematical evidence

**DST** — Demand Store Tier

**DWT** — Demand Worker Tier

**EHR** — Electronic Health Records [458]

**ENCS** — Faculty of Engineering and Computer Science, Concordia University

**ERA** — event reconstruction algorithm (Section 2.2.4.4)

**FOIL** — first-order intensional logic [113]

**FFT** — Fast Fourier Transform

**FSA** — finite-state automata

**FTK** — Forensic Toolkit [2]

**FTP** — File Transfer Protocol

**GATE** — General Architecture for Text Engineering [462]

**GC** — Global Conventionalism [30]

**GCC** — GNU Compiler Collection [485]

**GEE** — General Eduction Engine

**GEER** — GEE Resources

**GICF** — General Imperative Compiler Framework [261]

**GIPC** — General Intensional Program Compiler, Figure 32, [370, 463]

**GIPL** — General Intensional Programming Language [361, 370, 463]

**GIPSY** — General Intensional Programming System [370, 463]

**GLU** — Granular Lucid [188, 189, 361]

**GMT** — General Manager Tier

**GGMT** — graphical (graph-based) GMT [393]

**GUI** — graphical user interface

**HCI** — human-computer interaction

**HOIFL** — higher-order intensional fuzzy logic

**HOIL** — higher-order intensional logic

**HOL** — higher-order logic

**I/O** — input/output

**IC** — identified-context (class) [241]

**ICMP** — Internet Control Message Protocol

**ICT** — information and communications technology

**IDE** — interactive development environment

**IDS** — intrusion detection system

ıHTML — Intensional HTML [510, 511]

**IME** — interactive modeling environment

**IRSNET** — intelligent RSNET

**IODEF** — the Incident Object Description Exchange Format[4]

**IP** — depending on context of a particular chapter, may refer to

- Intensional Programming [131, 350] (Chapter 1, Section 3.2)
- Internet Protocol Layer-3 address [79, 80] (Section 9.5)

**IPL** — Intensional Programming Language (e.g., FORENSIC LUCID, GLU, LUCID, INDEXICAL LUCID, JLUCID, TENSOR LUCID, OBJECTIVE LUCID, ONYX [144], GIPL, TRANSLUCID)

---

[4] www.ietf.org/rfc/rfc5070.txt

**ISS** — Illimitable Space System [437]

**JavaCC** — Java Compiler Compiler [506]

**JDSF** — JAVA Data Security Framework

**JMS** — JAVA Messaging Service

**JOOIP** — JAVA-based object-oriented intensional programming language

**JPF** — JAVA Plug-in Framework

**JSON** — JavaScript Object Notation

**LAN** — local area network

**LOD** — level of detail

**LPC** — linear predictive coding

**MAC** — Media Access Control address [180][5]

**MARF** — Modular Audio Recognition Framework [465]

**MARFCAT** — MARF-based Code Analysis Tool [285, 287]

MARFL — MARF Language [272]

**MARFPCAT** — MARF-based PCap Analysis Tool

**MRTG** — the Multi Router Traffic Grapher [345]

**NAG** — Network Administration Group (of the Faculty of ENCS, Concordia)

**NICTER** — Network Incident analysis Center for Tactical Emergency Response [104]

**MDP** — Markov Decision Process

**ME** — managed element

**MPR** — map of partitioned runs (see Section 2.2.4.4.4)

**MSPR** — map of sequence of partitioned runs (see Section 2.2.4.6.2)

**MSA** — MAC Spoofer Analyzer (Section 9.5)

**msw** — MAC Spoof Watcher

**NFS** — Network File System [443, 479]

**NLP** — natural language processing

**NVD** — National Vulnerability Database [340]

**OO** — object-oriented

**OODA** — Observe-Orient-Decide-Act loop[6]

**OpenGL** — Open Graphics Library

**OSI** — Open Systems Interconnection model [79, 80][7]

**OSS** — open source software

---

[5]http://en.wikipedia.org/wiki/MAC_address
[6]http://en.wikipedia.org/wiki/OODA_loop
[7]http://en.wikipedia.org/wiki/OSI_reference_model

**OWL** — Web Ontology Language

**PC** — personal computer

**PCM** — pulse-code modulation

**pcap** — packet capture data

**PDS** — push down system

**PA** — Peano arithmetic

**PID** — process identifier [443, 479]

**PL** — programming language

**PoC** — proof-of-concept

**PR** — depending on context of a particular chapter, may refer to

- partitioned run (Section 2.2.2)
- processing resource (as in GATE, Chapter 5)

**PRISM** — probabilistic model checker [467]

**PS** — problem-specific

**QoS** — quality of service

**R&D** — research and development

**RIPE** — Run-time Intensional Programming Environment ("IDE")

**RMI** — Remote Method Invocation

**RPC** — Remote Procedure Call

**RSNET** — road-side network

**RT** — *RT: Request Tracker* [504][8]

**SAMATE** — Software Assurance Metrics And Tool Evaluation NIST project[9]

**SATE** — Static Analysis Tool Exposition [10]

**SAVE** — Static Analysis of Vicious Executables framework [451]

**SDWT** — separating discrete wavelet transform

**self-CHOP** — self-(configuring, healing, optimizing, and protecting) properties

**SFAP** — self-forensics autonomic property (Appendix D)

**SIPL** — Specific IPL (e.g., INDEXICAL LUCID, JLUCID, TENSOR LUCID, OBJECTIVE LUCID, ONYX)

**SHL** — Security Hardening Language [228]

**S.M.A.R.T.** — Self-Monitoring, Analysis, and Reporting Technology [12, 67, 519]

**SNMP** — Simple Network Management Protocol [440]

---

[8] http://www.bestpractical.com/?rt=4.0.8
[9] http://samate.nist.gov
[10] http://samate.nist.gov/SATE.html

**SMI** — Structure of Management Information [184][11]

**SLO** — service-level objective

**SOA** — service-oriented architecture

**SSD** — solid state drive

**SSH** — secure shell

**SSL** — secure socket layer

**STDERR** — standard error stream [443, 479]

**STDIN** — standard input stream [443, 479]

**STDOUT** — standard output stream [443, 479]

**STGT** — Simple Theory of GIPSY Types, an STT extension (Appendix B)

**STT** — Mendelson's Simple Theory of Types [253]

**swpvio** — switch port violation (see Section 9.5)

**SYN** — "synchronize" packet

**SysML** — Systems Modeling Language

**TA** — transport agent

**TIL** — Transparent Intensional Logic [98]

**TTL** — time-to-live

**TCT** — The Coroner's Toolkit [461]

**TLS** — transport layer security

**UEFI** — Unified Extensible Firmware Interface[12]

**UC** — Use Case

**UM** — user-managed

**UML** — Unified Modeling Language [47, 224, 482]

**URI** — Unified Resource Identifier

**VANET** — vehicular ad hoc network

**VCN** — vehicular network

**ViM** — Vi Improved [325]

**VLAN** — virtual LAN [31, 70]

**WAV** — wave audio file format

**WFT** — Windows Forensic Toolkit [252]

**XML** — Extensible Markup Language

---

[11]http://en.wikipedia.org/wiki/Structure_of_Management_Information
[12]http://en.wikipedia.org/wiki/Unified_Extensible_Firmware_Interface

# Appendix B

# A Type System Theory for Higher-Order Intensional Logic Support in Hybrid Intensional-Imperative Programs in GIPSY

We describe a type system and its theory the General Intensional Programming System (GIPSY), designed to support intensional programming languages (built upon higher-order intensional logic) and their imperative counter-parts for the eductive execution model. We extend the Mendelson's simple theory of types (STT) [17, 107, 253] by adding the intensionality axiom to it. The *intensionality* principle covers language expressions that explicitly take into the account a multidimensional context space of evaluation with the context being a first-class value that serves a number of applications that need the notion of context to proceed. The theory is complemented with the software engineering design and implementation study of the GIPSY type system [301]. In GIPSY, the type system glues the static and dynamic typing between intensional and imperative languages in its compiler and run-time environments to support the intensional expressions and their evaluation written in various dialects of the intensional programming language LUCID. We, therefore, describe and discuss the properties of such a type system and the related type theory as well as particularities of the semantics, design and implementation of the GIPSY type system [315].

## B.1 Overview

This work primarily a combination of the previous results describing the concrete GIPSY types specification and implementation in [315] as well as theoretical foundations behind the GIPSY Type System [301].

The General Intensional Programming System (GIPSY) (see Chapter 6 for an in-depth background discussion) has been built around the LUCID family of intensional programming languages (see Chapter 4) that rely on the higher-order intensional logic (HOIL, see

Section 4.4) to provide context-oriented multidimensional reasoning about intensional expressions. HOIL combines functional programming with various intensional logics to allow explicit context expressions to be evaluated as first-class values that can be passed as parameters to functions and return as results with an appropriate set of operators defined on contexts [302]. GIPSY's frameworks are implemented in JAVA as a collection of replaceable components for the compilers of various LUCID dialects and the demand-driven eductive evaluation engine that can run distributively. GIPSY provides support for hybrid programming models that couple intensional and imperative languages for a variety of needs. Explicit context expressions limit the scope of evaluation of math expressions (effectively a LUCID program is a mathematics or physics expression constrained by the context) in tensor physics, regular math in multiple dimensions, etc., and for cyberforensic reasoning as one of the use-cases of interest (Chapter 7). Thus, GIPSY is a support testbed for HOIL-based languages some of which enable such reasoning, as in formal cyberforensic case analysis with event reconstruction. Its type system is designed to support all these hybrid paradigms because the standard LUCID algebra (i.e., types and operators) is extremely fine-grained and can hardly benefit from parallel evaluation of their operands (cf., Chapter 4). Adding granularity to the data elements manipulated by LUCID inevitably comes through the addition of coarser-grained data types and their corresponding operators. LUCID semantics being defined as typeless, a solution to this problem consists in adding a hybrid counterpart to LUCID to allow an external language to define an algebra of coarser-grained types and operators. In turn, this solution raises the need for an elaborated type system to bridge the implicit types of LUCID semantics with the explicit types and operators (i.e., functions) defined in the hybrid counterpart language [315]. This chapter, therefore, presents such a type system used in GIPSY at compile time for static type checking, as well as at run-time for dynamic type checking [301, 315].

### B.1.1 Organization

After our problem statement and short presentation of our proposed solution, in Section B.1.2 follows the concise description of the GIPSY Type System subproject of the GIPSY research and development effort in Section B.2.1. Having briefly covered this introductory material we move onto the actual definition of the Simple Theory of GIPSY Types (STGT) as an extension of the classical Simple Theory of Types (STT) in Section B.3. Further, we describe the properties of our STGT via various categories of types and their applicability to our system in Section B.5 to illustrate where STGT stands [301]. Then, we present the complete GIPSY type system as used by the compiler (the General Intensional Programming Compiler—GIPC, see Section 6.2.1) and the run-time system (the General Eduction Engine—GEE, see Section 6.2.2) when producing and executing a binary *GIPSY program* [264, 282] (called General Eduction Engine Resources—GEER) respectively [161, 301, 315, 362]. Finally, we conclude in Section B.6 describing limitations and future work to address them [301]. (We don't overview the GIPSY project here as it and the related work are detailed in Chapter 6.)

### B.1.2 Summary of the Problem and the Proposed Solution

The beginnings of the GIPSY Type System were bootstrapped by a number of the related works [264, 282, 301, 315] in GIPSY to support hybrid and object-oriented intensional programming [526, 528], and LUCX's context type extension known as context calculus implementation [365, 473] for contexts to act as first-class values [301].

**Problem.** Data types are implicit in LUCID (as well as in its dialects or many functional languages). As such, the type declarations normally never appear in the LUCID programs at the syntactical level. The data type of a value is inferred from the result of evaluation of an expression. In most imperative languages, like JAVA, C++ and the like, the types are explicit and the programmers must *declare* the types of the variables, function parameters and return values before they are used in evaluation [315]. In GIPSY, we needed to allow any LUCID dialect to be able to uniformly invoke functions/methods written in imperative languages and the other way around and perform semantic analysis in the form of type assignment and checking statically at compile time or dynamically at run time, perform any type conversion if needed, and evaluate such hybrid HOIL expressions. At the same time, we need to allow a programmer to specify, or declare, the types of variables, parameters, and return values for both intensional and imperative functions as a binding contract between inter-language invocations despite the fact that LUCID is not explicitly typed [315]. Thus, we need a general type system, well specified and founded in a sound theory, designed and implemented to support such scenarios [301, 315].

**Proposed Solution.** The unique particularity of the type system presented here is that it is above a specific programming language model of either the LUCID family of languages or imperative languages. It is designed to bridge programming language paradigms, the two most general of them would be the intensional and imperative paradigms. GIPSY has a collection of frameworks designed to support a common run-time environment and co-existence of the intensional and imperative languages [315]. Thus, the type system is that of a generic GIPSY program that can include code segments written in a theoretically arbitrary number of intensional and imperative dialects supported by the system vs. being a type system for a specific language [315]. What follows is the details of the proposed solution and the specification of the simple GIPSY type system and its theory [301, 315].

## B.2 The GIPSY Type System

### B.2.1 Introduction to the GIPSY Type System

The introduction of JLUCID, OBJECTIVE LUCID, and the General Imperative Compiler Framework (GICF) [264, 282] prompted the development of the GIPSY Type System as implicitly understood by the LUCID language and its incarnation within GIPSY to handle types in a more general manner as a glue between the imperative and intensional languages within the system. Further evolution of different LUCID variants such as LUCX introducing contexts as first-class values and JOOIP [526, 528] (Java Object-Oriented Intensional Programming language) highlighted the need of the further development of the type system to accommodate the more general properties of the intensional and hybrid languages [301, 315].

The type system is also required to extend the higher-order intensional logic (HOIL) [302] support onto the imperative dialects in the hybrid intensional-imperative programs. This in particular includes the notion of context added to the imperative programs as well as dynamic variables binding and assignment upon intensional type discovery when done evaluating intensional expressions and, for example, assigning their result to a variable declared in a JAVA class. The same applies to the function and method parameters as well as their return results. This further necessitates the type matching rules between LUCID and other languages, similar to the ones defined for example for JAVA in Table 17 per earlier works [264, 282, 301, 315, 365, 526, 528].

### B.2.2 Matching Lucid and Java Data Types

Here we present a case of interaction between LUCID and JAVA [141]. Allowing LUCID to call JAVA methods brings a set of issues related to the data types, especially when it comes to type checks between LUCID and JAVA parts of a hybrid program. This is pertinent when LUCID variables or expressions are used as parameters to JAVA methods and when a JAVA method returns a result to be assigned to a LUCID variable or used in an intensional expression. The sets of types in both cases are not exactly the same. The basic set of LUCID data types as defined by Grogono [143] is `int`, `bool`, `double`, `string`, and `dimension`. LUCID's `int` is of the same size as JAVA's `long`. GIPSY and JAVA `double`, `boolean`, and `String` are equivalent. LUCID `string` and JAVA `String` are simply mapped internally through `StringBuffer`; thus, one can think of the LUCID `string` as a reference when evaluated in the intensional program. Based on this fact, the lengths of a LUCID `string` and JAVA `String` are the same. JAVA `String` is also an object in JAVA; however, at this point, a LUCID program has no direct access to any `String`'s properties (though internally we do and we may expose it later to the programmers). We also distinguish the `float` data type for single-precision floating point operations. The `dimension` index type is said to be an integer or string (as far as its dimension tag values are concerned), but might be of other types eventually, as discussed in [365, 473]. Therefore, we perform data type matching as presented in Table 17. Additionally, we allow the `void` JAVA return type, which will always be matched to a Boolean expression `true` in LUCID as an expression has to always evaluate to something. As for now our types mapping and restrictions are as per Table 17. This is the mapping table for the JAVA-to-IPL-to-JAVA type adapter. Such a table would exist for mapping between any imperative-to-intensional language and back, e.g., the C++-to-IPL-to-C++ type adapter [315].

### B.2.3 Design and Implementation of the Type System

While the main language of GIPSY, LUCID, is polymorphic and does not have explicit types, co-existing with other languages necessitates definition of GIPSY types and their mapping to a particular language being embedded. In Figure 75 is the detailed design of the GIPSY Type System [315].

Each class is prefixed with `GIPSY` to avoid possible confusion with similar definitions in the `java.lang` package. The `GIPSYVoid` type always evaluates to the Boolean `true`, as described earlier in Section B.2.2. The other types wrap around the corresponding JAVA object wrapper classes for the primitive types, such as `Long`, `Float`, etc. Every class keeps a lexeme (a lexical representation) of the corresponding type in a GIPSY program and overrides `toString()` to show the lexeme and the contained value. These types are extensively used by the `Preprocessor`, imperative and intensional compilers, and `SemanticAnalyzer` for the general type of GIPSY program processing, and by the GEE's `Executor` [315].

Figure 75: The GIPSY type system [315]

Table 17: Matching data types between LUCID and JAVA [301, 315]

| Parameter Types Used in LUCID | Corresponding JAVA Types | Internal GIPSY Types |
|---|---|---|
| string | String | GIPSYString |
| float | float | GIPSYFloat |
| double | double | GIPSYDouble |
| int | int | GIPSYInteger |
| dimension | int, String | Dimension |
| bool | boolean | GIPSYBoolean |
| *class* | Object | GIPSYObject |
| *URI* | Object | GIPSYEmbed |
| [] | [] | GIPSYArray |
| *operator* | Method | GIPSYOperator |
| *function* | Method | GIPSYFunction |
| Return Types of JAVA Methods | Types of LUCID Expressions | Internal GIPSY Types |
| int, byte, long | int, dimension | GIPSYInteger |
| float | float | GIPSYFloat |
| double | double | GIPSYDouble |
| boolean | bool | GIPSYBoolean |
| char | char | GIPSYCharacter |
| String | string, dimension | GIPSYString |
| Method | *function* | GIPSYFunction |
| Object | *class* | GIPSYObject |
| Object | *URI* | GIPSYEmbed |
| void | bool::true | GIPSYVoid |

The other special types that have been created are either experimental or do not correspond to a wrapper of a primitive type. `GIPSYIdentifier` type case corresponds to a declaration of some sort of an identifier in a GIPSY program to be put into the dictionary, be it a variable or a function name with the reference to their definition. Constants and conditionals may be anonymous and thereby not have a corresponding explicit identifier. `GIPSYEmbed` is another special type that encapsulates embedded code via a URI parameter and later is exploded into multiple types corresponding to procedural demands (JAVA or any other language methods or functions) [264]. `GIPSYFunction` and its descendant `GIPSYOperator` correspond to the function types for regular operators and user-defined functions. A `GIPSYFunction` can either encapsulate an ordinary LUCID function (which is immutable as in functional programming) or a procedure (e.g., a JAVA method), which may often be mutable (i.e., with side effects). These four types (*identifier, embed, function,* and *operator*) are not directly exposed to a GIPSY programmer and at this point are managed internally. An operation is usually mapped by the actual operators like addition, subtraction, and others via the corresponding algebras. `GIPSYContext` and `Dimension` are a new addition to the type system implementation since its first inception [264]. They represent context-as-first-class-values in the context calculus defined by Wan in [513] and refined and implemented by Tong [473]. The rest of the type system is exposed to the GIPSY programmers in the preamble of a GIPSY program, i.e., the `#funcdecl` and `#typedecl` segments, which result in the embryo of the dictionary for linking, semantic analysis, and execution. Once imperative compilers of procedural demands return, the type data structures (return and parameter types) declared in the preamble are matched against what was discovered by the compilers and if the match is successful, the link is made. By capturing the types such as *identifier, embed, function, operator* and *context, dimension*, the GIPSY type system lays down fundamentals the higher-order intensional logic (HOIL) support that combines functional programming, intensional logic, context calculus, and in some instances hybrid paradigm support, and the corresponding types [315].

We further describe various properties of the concrete GIPSY types and their more detailed specification in Section B.4 and Section B.5. There we detail the inner workings of each type in more detail [315] as well describe some of the properties through the notions of existential, union, intersection, and linear types.

### B.2.4 Use of the GIPSY Type System

Some of the types are in use only in the hybrid programs, that's where they mostly appear visible to a programmer. Another subset of types is internally used by the GIPSY in its compiler (GIPC) and run-time environment (GEE) frameworks primarily for dynamic discovery and semantic checking of the types [315].

**Relationship to GIPC.** The GIPC uses the type system and classes defined for it in the compilation process to do static type checking and type annotation of literal values in source code programs as well as storing the type annotations with the parameter or return values in hybrid dialects to indicate expected data types to be passed or returned when calling a JAVA method from a LUCID dialect and back. Some of the static type declarations come from the `Preprocessor` after parsing the `#funcdecl` and `#typedecl` sections [315].

**Relationship to GEE.** The GEE uses the type system at run-time to do run-time dynamic type checking as well as to perform the actual evaluation of arithmetic, context set, and object operators. When the execution of a given GIPSY program reaches an `ImperativeNode` [264] indicating a call to an imperative procedure, the type annotations and expressions are used to validate the parameter and return types to check whether they match or not and indicate an error if they don't [264, 315].

## B.3 Simple Theory of GIPSY Types

### B.3.1 Simple Theory of Types

Our simple theory of the GIPSY types (STGT) is based on the "Simple theory of types" (STT) by Mendelson [253]. The theoretical and practical considerations are described in the sections that follow. The STT partitions the qualification domain into an ascending hierarchy of types with every individual value assigned a type. The type assignment is dynamic for the intensional dialects as the resulting type of a value in an intensional expression may not be known at compile time. The assignment of the types of constant literals is done at compile-time, however. In the hybrid system, which is mostly statically typed at the code-segment boundaries, the type assignment also occurs at compile-time. On the boundary between the intensional programming languages (IPLs) and imperative languages, prior to an imperative procedure being invoked, the type assignment to the procedure's parameters from IPL's expression is computed dynamically *and* matched against a type mapping table similar to that of Table 17. Subsequently, when the procedure call returns back to the IPL, the type of the imperative expression is matched back through the table and assigned to the intensional expression that expects it. The same table is used when the call is made by the procedure to the IPL and back, but in the reverse order [301].

Further in STT, all quantified variables range over only one type making the first-order logic applicable as the underlying logic for the theory. This also means the all elements in the domain and all co-domains are of the same type. The STT states there is an atomic type that has no member elements in it, and the members of the second-high from the basic atomic type. Each type has a next higher type similarly to `succ` ($\succ$) in Peano arithmetic (PA) and the `next` operator in LUCID. This is also consistent to describe the composite types, such as arrays and objects as they can be recursively decomposed (or "flattened", see [264, 282]) into the primitive types to which the STT applies [301].

Symbolically, the STT uses primed and unprimed variables and the infix set notation of $\in$. The formulas $\Phi(x)$ rely on the fact that the unprimed variables are all of the same type. This is similar to the notion of a LUCID stream with the point-wise elements of the stream having the same type. The primed variables ($x'$) in STT range over the next higher type. There are two atomic formulas in STT of the form of identity, $x = y$, and set membership, $y \in x'$ [301].

### B.3.2 GIPSY Types Axioms

The STT [253] defines the four basic axioms for the variables and the types they can range over: *Identity*, *Extensionality*, *Comprehension*, and *Infinity*. In STGT, we add the *Intensionality* on as the fifth axiom [301]. The variables in the definition of the *Identity* relationship and in the *Extensionality* and *Comprehension* axioms typically range over the elements of one of the two nearby types. In the set membership [119, 237], only the unprimed variables that range over the lower type in the hierarchy can appear on the left of $\in$; conversely, the primed ones that range over higher types can only appear on the right of $\in$ [253]. The axioms are defined as:

1. **Identity**: $x = y \leftrightarrow \forall z'[x \in z' \leftrightarrow y \in z']$

2. **Extensionality**: $\forall x[x \in y' \leftrightarrow x \in z'] \rightarrow y' = z'$

3. **Comprehension**: $\exists z' \forall x[x \in z' \leftrightarrow \Phi(x)]$. This covers objects and arrays as any collection of elements here may form an object of the next, higher type. The STT states the comprehension axiom is schematic with respect to $\Phi(x)$ (which is a first-order formula with $x$ as a free variable) and the types [253]. $\Phi(x)$ works with type hierarchies and is not arbitrary like Church's [17], Russel's, and other type theories avoiding *Russel's paradox*.

4. **Infinity**: $\forall x, y[x \neq y \rightarrow [xRy \bigvee yRx]]$. There exists a non-empty binary relation $R$ over the elements of the atomic type that is transitive, irreflexive, and strongly connected.

5. **Intensionality**: the intensional types and operators (cf. Table 13, page 161) are based on the intensional logic (Section 3.2, page 59) and context calculus (Section 7.2.3.1.4, page 165). These are extensively described in Chapter 3, Chapter 4, Chapter 6, Chapter 7 and the previously cited works [113, 131, 350, 361, 365, 473, 513]. This present type system accommodates the two in a common hybrid execution environment of the GIPSY. A context $c$ is a finite subset of the relation (Section 7.2.3.1.4, page 165, [473, 516]): $c \subset \{(d, x) \mid d \in DIM \land x \in T\}$, where $DIM$ is the set of all possible dimensions, and $T$ is the set of all possible tags [301] (indices). Therefore, $\forall x \text{ @ } c[(x \text{ @ } c) \in y' \leftrightarrow (x \text{ @ } c) \in z'] \rightarrow y' = z'$.

## B.4 Concrete Specification of the GIPSY Types

The following sections describe the behavior, equivalence, implementation, valid range of values, and supported operators of every type in the GIPSY system. We define the lexical specification if applicable, its visibility to (explicit ability to use by) the programmers, and the type implementation details in the GIPSY environment [315].

**Integer.**

- Lexeme: `int`

- Visibility to programmers: explicitly visible only in the hybrid GIPSY programs.

- The valid range of the type is the same as the corresponding value of the type `long` in JAVA. That covers all the JAVA `byte`, `int`, and `long` types.

- Internal implementation: the `GIPSYInteger` class corresponds to this type. Internally, the class encapsulates `java.lang.Long` thereby making all operations available to that type in JAVA to GIPSY.

- Operators supported: arithmetic, logic, equality

**Single-Precision Floating Point Number.**

- Lexeme: `float`

- Visibility to programmers: explicitly visible only in the hybrid GIPSY programs.

- The valid range of the type is the same as the corresponding value of the type `float` of JAVA.

- Internal implementation: the `GIPSYFloat` class corresponds to this type. Internally, the class encapsulates `java.lang.Float` thereby making all operations available to that type in JAVA to GIPSY.

- Operators supported: arithmetic, logic, equality

**Double-Precision Floating Point Number.**

- Lexeme: `double`

- Visibility to programmers: explicitly visible only in the hybrid GIPSY programs.

- The valid range of the type is the same as the corresponding value of the type `double` of JAVA.

- Internal implementation: the `GIPSYDouble` class corresponds to this type. Internally, the class encapsulates `java.lang.Double` thereby making all operations available to that type in JAVA to GIPSY.

- Operators supported: arithmetic, logic, equality

**Character String.**

- Lexeme: `string`

- Visibility to programmers: explicitly visible only in the hybrid GIPSY programs. Implicitly defined in LUCID programs as enclosed into either single (') or double (") quotation marks. In the case of single (') quote, it has to be more than one character to be considered a string, otherwise it is treated as a type Character.

- The valid range of the type is the same as the corresponding value of the type `String` of JAVA.

- Internal implementation: the `GIPSYString` class corresponds to this type. Internally, the class encapsulates an instance of `java.lang.StringBuffer` thereby making all operations available to that type in JAVA to GIPSY.

- Operators supported: concatenation, substrings, subscript, dot, equality

**Character.**

- Lexeme: `char`

- Visibility to programmers: explicitly visible only in the hybrid GIPSY programs. Implicitly defined in LUCID programs as a single character enclosed into single (') quotation marks. It is assignable to the `String` type (in a way similar to integer being assignable to a double) and can participate in the concatenation to form character strings.

- The valid range of the type is the same as the corresponding value of the type `char` of JAVA that is 2-byte UNICODE character set.

- Internal implementation: the `GIPSYCharacter` class type corresponds to this type. Internally, the class encapsulates an instance of `java.lang.Character` thereby making all operations available to that type in JAVA to GIPSY.

- Operators supported: concatenation, logic, equality

**Void.**

- Lexeme: `void`

- Visibility to programmers: explicitly visible only in the hybrid GIPSY programs when declaring the procedure prototypes in the preamble of a GIPSY program's source code. Can only be a "return type".

- The valid range of the type is a Boolean constant `true` as seen as a return value from an intensional language.

- Internal implementation: the `GIPSYVoid` class that extends the `GIPSYBoolean` class with the truth value always set to `true`.

- Operators supported: equality

**Dimension.**

- Lexeme: `dimension`

- Visibility to programmers: explicitly visible in intensional-only and hybrid GIPSY programs when declaring the dimensions in a LUCID dialect program or listing the dimension-type variables, function parameters or return types in the preamble of the hybrid GIPSY programs. The dimension point-wise or set-wise $\langle dimension : tag \rangle$ mappings also appear in the intensional expressions setting up the context of evaluation and are, as such, a part of the `GIPSYContext` type.

- The valid range of the type was traditionally the natural numbers [508]. In Lucx and its context calculus [365, 473, 513] the dimension's tag values internally can be represented as integers or strings; thus, the dimension type implementation includes a general opaque `GIPSYType` to contain the actual type of tags that this dimension represents. This design keeps the door open for addition of other tag value types to dimensions (e.g., floating-point or objects, but these dimension tag types have their set of issues and are not discussed here).

- Internal implementation: the `Dimension` class.

- Operators supported: equality, switch, query

**Context and Context Set.**

- Lexeme: in the general case a set of $\langle dimension : tag \rangle$ mappings in the syntactical form of `{[d:v,...], ...}`. For more details, see [365, 473, 513].

- Visibility to programmers: explicitly visible in intensional-only GIPSY programs in the intensional expressions setting up the context of evaluation.

- The valid range of the type is determined by the constraints set forth by the underlying tag set types specified in [473].

- Internal implementation: the `GIPSYContext` class (depends on `Dimension` and `TagSet` types). According to the definitions of *simple context* and *context set*, we apply the Composite design pattern [121, 128, 224] to organize the context class' implementation. The context calculus operators have been specified via the `GIPSYContext`'s public interface. Due to the recursive nature of the composition, the definition of these operators is also recursive in the case of context sets.

- Operators supported: simple context and context set operators (@, #, and others), equality

**Boolean.**

- Lexeme: `bool`

- Visibility to programmers: explicitly visible only in the hybrid GIPSY programs.

- The valid range of the type is determined by the constants `true` and `false`.

- Internal implementation: the `GIPSYBoolean` class corresponds to this type. Internally, the class encapsulates an instance of `java.lang.Boolean` thereby making all operations available to that type in JAVA to GIPSY.

- Operators supported: logical, equality

**Array.**

- Lexeme: `[]`

- Visibility to programmers: explicitly visible in hybrid and pure intensional GIPSY programs with array support via the `[]` notation.

- The valid range of the type is determined by the types of the elements of the array and array's length for the array's index. Array is a collection of elements of the same type and has a length.

- Internal implementation: the `GIPSYArray` class corresponds to this type. Internally, the class extends `GIPSYObject` as stated in [264] the objects and arrays are treated similarly with the exception that arrays are a collection of elements of the same type vs. objects being a "collection" of elements potentially of different types.

- Operators supported: equality, dot, subscript, set

**Object.**

- Lexeme: `class`

- Visibility to programmers: explicitly visible in the hybrid GIPSY programs that support classes and class-like structures and types (e.g., `struct` in C).

- The valid range of the type is determined by the public operations defined on the objects in their particular language.

- Internal implementation: the `GIPSYObject` class corresponds to this type. Internally, the class encapsulates an instance of `java.lang.Object` to hold the value of the class. The value does not need to be a JAVA object, but, for example, a serialized binary version of a C++ class or its source code (to be lazily compiled on demand potentially on another host and platform).

- Operators supported: equality, dot

**Identifier.**

- Lexeme: a non-reserved word user-defined identifier

- Visibility to programmers: explicitly visible in all GIPSY programs.

- The valid range of the type is the same as specified by a legal subset of lexically valid character strings in GIPSY program identifiers. These are currently UNICODE enabled strings that begin with a letter.

- Internal implementation: the `GIPSYIndentifier` class corresponds to this type. Internally, the class encapsulates the identifier lexeme as `java.lang.String`.

- Operators supported: assignment, equality

**Function.**

- Lexeme: a non-reserved word user-defined identifier that syntactically corresponds to the function prototype declaration or definition, plus the keyword `immutable` that refers to a function definition that is free of side-effects.

- Visibility to programmers: explicitly visible in all GIPSY programs.

- The valid range of the type is the same as specified by a legal subset of lexically and syntactically valid function definitions. Functions have types (range over "functional" (as in functional programming) or "procedural" (sequential threads written in imperative languages) and states (ranging over "immutable" and "volatile"). The states are primarily for the "procedural" functions as the intensional functions are automatically "immutable".

- Internal implementation: the `GIPSYFunction` class corresponds to this type. Internally, the class encapsulates the state and the type mentioned earlier along with the `FunctionItem` (transformed from [527]) class instance that represents the function entry in the dictionary and the AST and the dictionary of the compiled `GIPSYProgram`. That, in turn, encapsulates function parameters, return types, and function identifier thereby making all the available operations on those components.

- Operators supported: equality, evaluation

**Operator.**

- Lexeme: multiple; primarily the classical arithmetic, logic, bitwise, set, and intensional depending on the types applied

- Visibility to programmers: explicitly visible in all GIPSY programs.

- The valid range of the type is the same as specified by a legal subset of lexically and syntactically valid operator kinds definitions.

- Internal implementation: the `GIPSYOperator` class corresponds to this type. Internally, the class extends `GIPSYFunction`. Operators internally are said to be immutable and functional by default.

- Operators supported: equality, evaluation

**Embedded Payload.**

- Lexeme: `embed()` in JLUCID and OBJECTIVE LUCID [264] and a specific syntactically declared function prototype with an URI, function identifier, and arguments in the preamble of a hybrid GIPSY program [264], see Listing 6.1 for the example.

- Visibility to programmers: explicitly visible in JLUCID and OBJECTIVE LUCID ("stand-alone") and general hybrid GIPSY programs.

- The valid range of the type is the same as specified by a legal subset of lexically and syntactically valid URI and argument identifier definitions [41, 168, 448].

- Internal implementation: the `GIPSYEmbed` class corresponds to this type. Internally, the class encapsulates `java.lang.Object` to hold any value of the embedded code and data and its URI using `java.net.URI`.

- Operators supported: equality, evaluation

## B.5 Describing Some GIPSY Types' Properties

To demonstrate most of the pertinent properties of GIPSY types and to put it in perspective for the readers to get a better and more complete understanding of the spectrum of their behavior, we cover them in light of describing types of types and comparing to existential, union, intersection, and linear types [301].

### B.5.1 Types of Types

Types of types are generally referred to as *kinds*. Kinds provide categorization to the types of similar nature. While some type systems provide kinds as first class entities available to programmers, in GIPSY we do not expose this functionality in our type system at this point. However, at the implementation level there are provisions to do so that we may later decide to expose for the use of programmers. Internally, we define several broad kinds of types, presented the sections that follow [301].

#### B.5.1.1 Numeric Kind

The primitive types under this category are numerical values, which are represented by `GIPSYInteger`, `GIPSYFloat`, and `GIPSYDouble`. They provide implementation of the common arithmetic operators, such as addition, multiplication and so on, as well as logical comparison operators of ordering and equality. Thus, for a numerical type $T$, the following common operators are provided. The resulting type of any arithmetic operator is the largest of the two operands in terms of length (the range of `double` of length say $k$ covers the range of `int` with the length say $m$ and if both appear as arguments to the operator, then the resulting value's type is that of without loss of information, i.e., largest in length `double`). The result of the logical comparison operators is always Boolean $B$ regardless the length of the left-hand-side and right-hand-side numerical types [301].

1. $T_{max} : T_1 \geq T_2 \to T_1 \mid T_1 < T_2 \to T_2$

2. $T_{multiply} : T_k \times T_m \to T_{\max(k,m)}$

3. $T_{divide} : T_k / T_m \to T_{\max(k,m)}$

4. $T_{add} : T_k + T_m \to T_{\max(k,m)}$

5. $T_{subtract} : T_k - T_m \to T_{\max(k,m)}$

6. $T_{mod} : T_k \% T_m \to T_{\max(k,m)}$

7. $T_{pow} : T_k \char`\^ T_m \to T_{\max(k,m)}$

8. $T_> : T > T \to B$

9. $T_< : T < T \to B$

10. $T_\geq : T \geq T \to B$

11. $T_\leq : T \leq T \to B$

12. $T_= : T = T \to B$

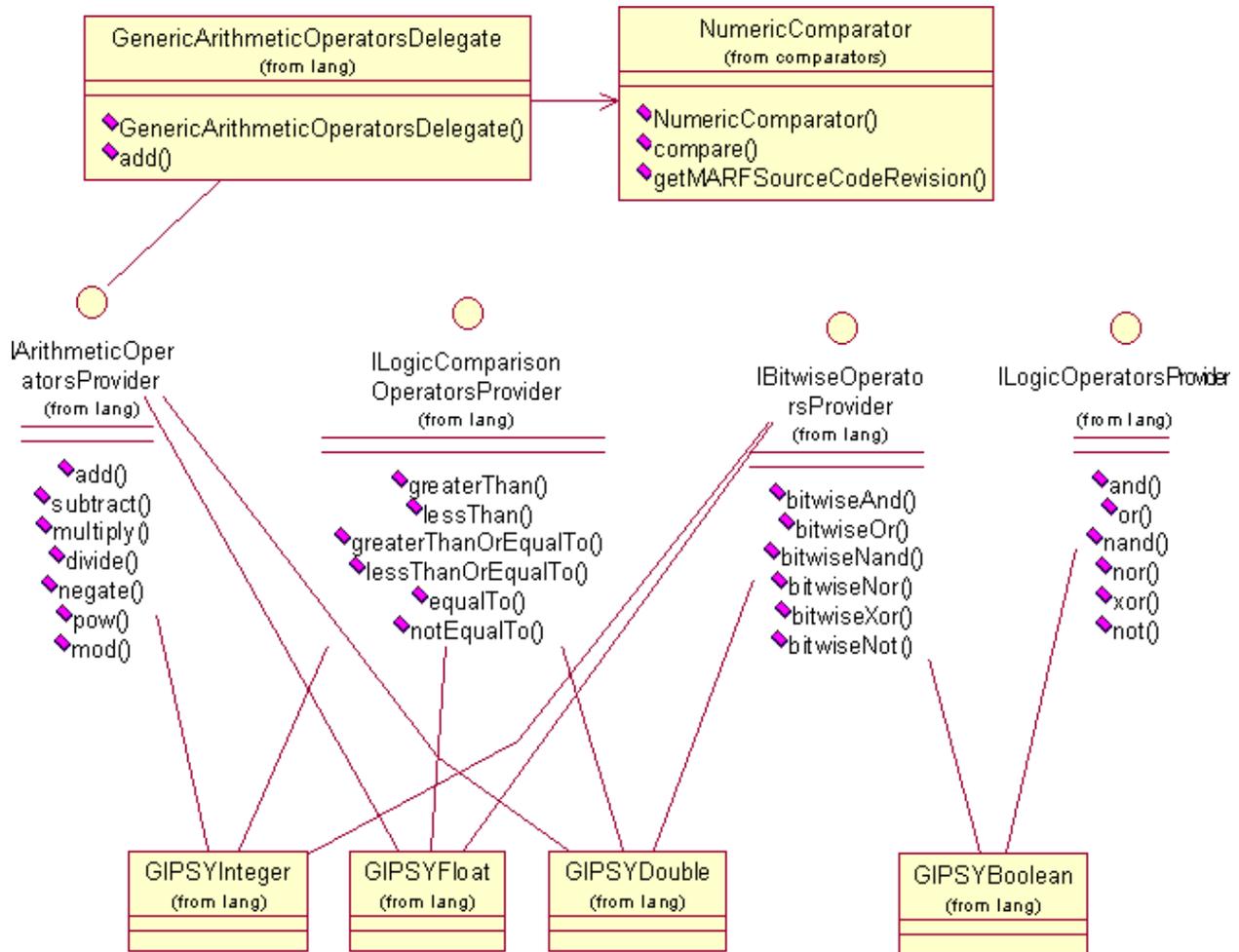

Figure 76: Example of provider interfaces and comparators [301]

A generalized implementation of the arithmetic operators is done by realization of the interface called `IArithmeticOperatorsProvider` and its concrete implementation developed in the general delegate class `GenericArithmeticOperatorsDelegate`. This design and implementation allow not only further exposure of kinds-as-first-class values later on after several iterations of refinement, but also will allow operator and type overloading or replacement of type handling implementation altogether if some researchers wish to do so. The implementation of the engine, GEE, is thus changed, to only refer to the interface type implementation when dealing with these operators. Equivalently for the logic comparison operators we have

`ILogicComparisonOperatorsProvider`

and the

`GenericLogicComparisonOperatorsDelegate`

classes. The latter relies on the comparator implemented for the numerical kind, such as `NumericComparator`. Using comparators (that is classes that implement the standard JAVA `Comparator` interface) allows JAVA to use and to optimize built-in sorting and searching algorithms for collections of user-defined types. In our case,

`GenericLogicComparisonOperatorsDelegate`

is the implementation of the delegate class that also relies on it. The example for the numeric types for the described design is in Figure 76 [301].

It is important to mention, that grouping of the numeric kind of integers and floating-point numbers does not violate the IEEE 754 standard [181], as these kinds implementation-wise wrap the corresponding JAVA's types (which are also grouped under numeric kind) and their semantics including the implementation of IEEE 754 by JAVA in accordance with the *Java Language Specification* [141, 301].

### B.5.1.2 Logic Kind

Similarly to numeric types, the primitive type `GIPSYBoolean` fits the category of the types that can be used in Boolean expressions. The operations the type provides expect the arguments to be of the same type—Boolean. The following set of operators on the logic type $B$ we provide in the GIPSY type system [301]:

1. $B_{and} : B \bigwedge B \to B$
2. $B_{or} : B \bigvee B \to B$
3. $B_{not} : \neg B \to B$
4. $B_{xor} : B \bigoplus B \to B$
5. $B_{nand} : \neg(B \bigwedge B) \to B$
6. $B_{nor} : \neg(B \bigvee B) \to B$

Note that the *logical XOR* operator (denoted as $\bigoplus$) is distinct from the corresponding *bitwise XOR* operator in Section B.5.1.3 in a way similar to the logical vs. bitwise AND and OR are respectively distinct. Again, similarly to the generalized implementation of arithmetic operators, logic operator providers implement the `ILogicOperatorsProvider` interface, with the most general implementation of it in `GenericLogicOperatorsDelegate` [301].

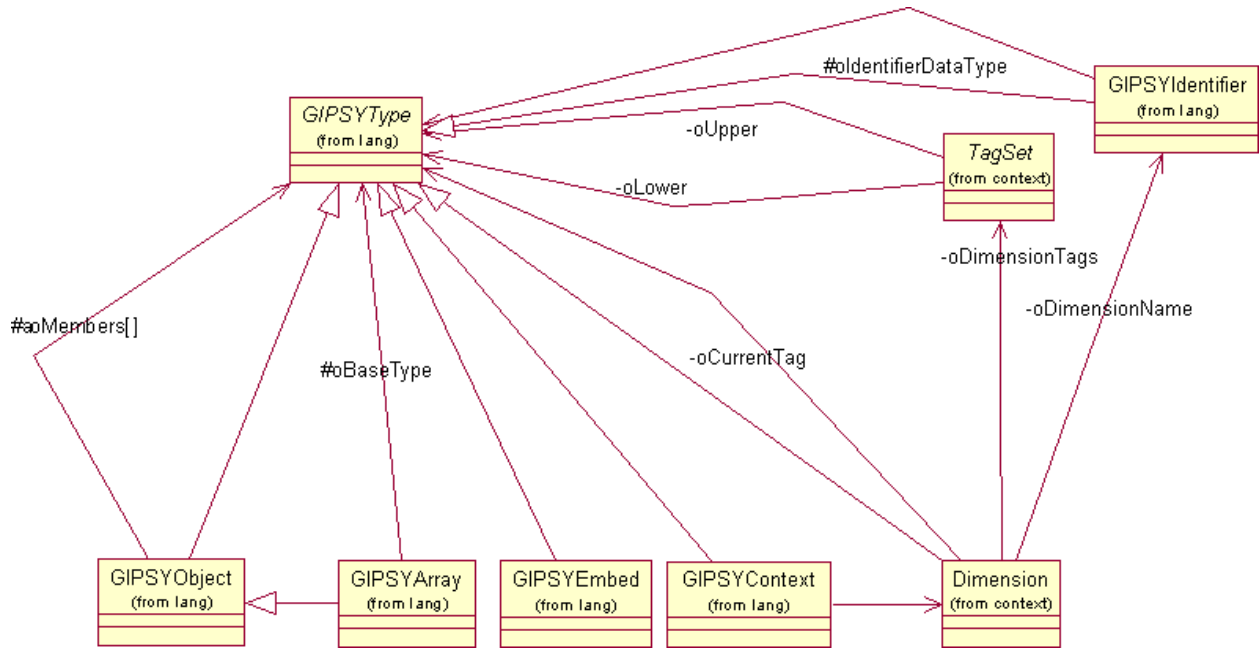
Figure 77: Composite types [301]

### B.5.1.3 Bitwise Kind

Bitwise kind of types covers all the types that can support bitwise operations on the entire bit length of a particular type $T$. Types in this category include the numerical and logic kinds described earlier in Section B.5.1.2 and Section B.5.1.1. The parameters on both sides of the operators and the resulting type are always the same. There are no implicit compatible type casts performed unlike for the numeric kind [301].

1. $T_{bit-and} : T \& T \rightarrow T$
2. $T_{bit-or} : T|T \rightarrow T$
3. $T_{bit-not} : !T \rightarrow T$
4. $T_{bit-xor} : T \bigotimes T \rightarrow T$
5. $T_{bit-nand} : !(T \& T) \rightarrow T$
6. $T_{bit-nor} : !(T|T) \rightarrow T$

The implementation of this kind's operators is done through the interface

$$\texttt{IBitwiseOperatorsProvider},$$

which is in turn generically implemented in the

$$\texttt{GenericBitwiseOperatorsDelegate}$$

class [301].

### B.5.1.4 Composite Kind

As the name suggests, the composite kind types consist of compositions of other types, possibly basic types. The examples of this kind are arrays, structs, and abstract data types and their realization such as objects and collections. In the studied type system these are `GIPSYObject`, `GIPSYArray`, and `GIPSYEmbed`. This kind is characterized by the constructors, dot operator to decide membership as well as to invoke member methods and define equality. The design of these types, just like the entire type system, adheres to the Composite design pattern [121, 128, 224]. The most prominent examples of this kind are in Figure 77, including `GIPSYContext` which is composed of `Dimension`s and indirectly of the `TagSet`s [301].

### B.5.1.5 Intensional Kind

The intentional types kind primarily has to do with encapsulation dimensionality, context information, and their operators. These are represented by the `GIPSYContext`, `Dimension`, and `TagSet`[1] types. The common operators on these types include the context switching and querying operators `@` and `#` as well as context calculus operators. Additional operators can be included depending on the intensional dialect used, but the mentioned operators are said to be the baseline operators that any intensional language can be translated to use. Implementation-wise there is a `IContextSetOperatorsProvider` interface and its general implementation in `GenericContextSetOperatorsDelegate`. The context calculus operators on simple contexts include standard set operators, such as \union, \difference, \intersection, \in, and LUCX-specific, such as \isSubContext, \projection, \hiding, and \override [301, 365, 473, 513].

### B.5.1.6 Function Kind

The function types represent either "functional" functions, imperative procedures, binary and unary operators that, themselves, can be passed as parameters or returned as results. In our type system these are represented by `GIPSYFunction`, `GIPSYOperator`, and `GIPSYEmbed`. The common operators on these include equality and evaluation [301].

### B.5.2 Existential Types

All of the presented GIPSY types are existential types because they represent and encapsulate modules and abstract data types and separate their implementation from the public interface specified by the abstract `GIPSYType`. These are defined as shown in Figure 78. They are implemented concretely as shown in Figure 79 [301].

$$\boxed{T = \exists \texttt{GIPSYType}\{\texttt{Object}\ a;\ \texttt{Object getEnclosedTypeObject}();\ \texttt{Object getValue}();\,\}}$$

Figure 78: Abstract GIPSY existential types [301].

---

[1]Not mentioned here; please refer to [365, 473].

```
boolT {Boolean a; Object getEnclosedTypeObject(); Boolean getValue(); }
intT {Long a; Object getEnclosedTypeObject(); Long getValue(); }
doubleT {Double a; Object getEnclosedTypeObject(); Double getValue(); }
functionT {FunctionItem a; Object getEnclosedTypeObject(); FunctionItem getValue(); }
...
```

Figure 79: Concrete GIPSY existential types [301].

All these correspond to be subtypes of the more abstract and general existential type $T$. Assuming the value $t \in T$, then $t$.getEnclosedTypeObject() and $t$.getValue() are well typed regardless the what the actual GIPSYType may be [301].

### B.5.3 Union Types

A union of two types produces another type with the valid range of values that is valid in *either* of the two; however, the operators defined on the union types must be those that are valid for *both* of the types to remain type-safe. A classical example of that is in C or C++ the range for the signed char is $-128\ldots 127$ and the range of the unsigned char is $0\ldots 255$, thus:

$$\text{signed char} \cup \text{unsigned char} = -128\ldots 255$$

In C and C++ there is a union type that roughly corresponds to this notion, but it does not enforce the operations that are possible on the union type that must be possible in both left and right types of the uniting operator. In the class hierarchy, such as in GIPSY, the union type among the type and its parent is the parent class; thus, in our specific type system the following holds [301]:

$$\forall T \in G : T \cup \text{GIPSYType} = \text{GIPSYType}$$
$$\text{GIPSYArray} \cup \text{GIPSYObject} = \text{GIPSYObject}$$
$$\text{GIPSYVoid} \cup \text{GIPSYBoolean} = \text{GIPSYBoolean}$$
$$\text{GIPSYOperator} \cup \text{GIPSYFunction} = \text{GIPSYFunction}$$

where $T$ is any concrete GIPSY type and $G$ is a collection of types in the GIPSY type system we are describing. Equivalently, the union of the two sibling types is their common parent class in the inheritance hierarchy. Interestingly enough, while we do not explicitly expose kinds of types, we still are able to have union type relationships defined based on the kind of operators they provide as siblings under a common interface, as shown in Figure 80, where $T$ is any concrete GIPSY type and $A$ is a collection of types that provide arithmetic operators, $L$—logic operators providers, $B$—bitwise operators providers, $C$—context operators providers, $D$—composite operator providers, and $F$—function operator providers. Thus, resulting in types shown in Figure 81 [301]. Another particularity of the GIPSY type system is the union of the string and integer types under dimension:

$$\text{GIPSYInteger} \cup \text{GIPSYString} = \text{Dimension}$$

$$\forall T \in A : T \cup \texttt{IArithmeticOperatorProvider} = \texttt{IArithmeticOperatorProvider}$$
$$\forall T \in L : T \cup \texttt{ILogicOperatorProvider} = \texttt{ILogicOperatorProvider}$$
$$\forall T \in B : T \cup \texttt{IBitwiseOperatorProvider} = \texttt{IBitwiseOperatorProvider}$$
$$\forall T \in C : T \cup \texttt{IContextOperatorProvider} = \texttt{IContextOperatorProvider}$$
$$\forall T \in D : T \cup \texttt{ICompositeOperatorProvider} = \texttt{ICompositeOperatorProvider}$$
$$\forall T \in F : T \cup \texttt{IFunctionOperatorProvider} = \texttt{IFunctionOperatorProvider}$$

Figure 80: Concrete GIPSY union types (providers) [301]

$$\{\texttt{GIPSYInteger}, \texttt{GIPSYFloat}, \texttt{GIPSYDouble}\} \in A$$
$$\{\texttt{GIPSYBoolean}\} \in L$$
$$\{\texttt{GIPSYInteger}, \texttt{GIPSYFloat}, \texttt{GIPSYDouble}, \texttt{GIPSYBoolean}\} \in B$$
$$\{\texttt{GIPSYContext}, \texttt{Dimension}\} \in C$$
$$\{\texttt{GIPSYObject}, \texttt{GIPSYArray}, \texttt{GIPSYEmbed}\}, \texttt{GIPSYString}\} \in D$$
$$\{\texttt{GIPSYFunction}, \texttt{GIPSYOperator}, \texttt{GIPSYEmbed}\} \in F$$

Figure 81: Concrete GIPSY union types [301]

and this is because in our dimension tag values we allow them to be either integers or strings. While not a very common union in the majority of type system, they do share a common set of tag set operators defined in [365, 473] for ordered finite tag sets (e.g., `next()`, etc.) [301].

### B.5.4 Intersection Types

An intersection type of given two types is a range where the sets of valid values overlap. Such types are safe to pass to methods and functions that expect either of the types as the intersection types are more restrictive and compatible in both original types. A classical example of an intersection type if it were implemented in C or C++ would be:

$$\texttt{signed char} \cap \texttt{unsigned char} = 0\ldots 127$$

The intersection types are also useful in describing the overloaded functions. Sometimes they are called as refinement types. In a class hierarchy, the intersection between the parent and child classes is the most derived type, and the intersection of the sibling classes is empty. While the functionality offered by the intersection types is promising, it is not currently explicitly or implicitly considered in the GIPSY type system, but planned for the future work [301].

## B.5.5 Linear Types

Linear (or "uniqueness") types are based on linear logic [134]. The main idea of these types is that values assigned to them have one and only one reference to them throughout. These types are useful to describe immutable values like strings or hybrid intensional-imperative objects (see [526] for details). These are useful because most operations on such an object "destroy" it and create a similar object with the new values, and, therefore, can be optimized in the implementation for the in-place mutation. Implicit examples of such types in the GIPSY type system are `GIPSYString` that internally relies on Java's `StringBuffer` that does something very similar as well as the immutable `GIPSYObject` is in JOOIP [526] and immutable `GIPSYFunction`. Since we either copy or retain the values in the warehouse (DST, see Section 6.2.2), and, therefore, one does not violate referential transparency or create side effects in, and at the same time be more efficient as there is no need to worry about synchronization overhead [301].

## B.6 Summary

Through a series of discussions, specification, design, and implementation details we presented a type system and its theory used by the GIPSY project for static and dynamic type checking and evaluation of intensional and hybrid intensional-imperative HOIL expressions potentially written in multiple languages. We highlighted the particularities of the system that does not attribute to a particular specific language as traditionally most type systems do, but to an entire set of languages and hybrid paradigms that are linked through the type system [315]. This is a necessary contribution to GIPSY-like systems to have a homogeneous environment to statically and dynamically type-check and evaluate intensional and hybrid programs that are based on a sound theory [301, 315].

The type system described in this chapter has been implemented to the large extent in the GIPSY. However, our implementation still needs thorough testing using complex program examples testing the limits of the type system. Additional endeavours, noted in the previous sections, include [301, 315]:

- Exposing *kinds* as first class entities, allowing programs to have a more explicit manipulation of types.

- Allowing custom user-defined types and extension of the existing operators and operator overloading.

- Exposing more operators for composite types to Lucid code segments.

- Adding intersection types for more flexibility in the future development of the type system, allowing more type casting possibilities at the programming level.

# Appendix C

# MARFL

This edited chapter is primarily based on the corresponding article [272] presented at the first SECASA workshop at COMPSAC 2008. We focus on defining context expressions in terms of initial syntax and semantics for an intensional MARF (Modular Audio Recognition Framework, see Section 5.2) Language, MARFL. It is there to allow scripting MARF-based applications as context-aware, where the notion of context represents coarse-grained and fine-grained configuration details of a given MARF instance and a set of overloaded context operators borrowed from the Generic Intensional Programming Language (GIPL) `@` and `#` to help with the task. This is a preliminary research on MARFL that has considerable practical implications on the usability of MARF's resources and beyond. In this chapter we focus exclusively on the context specification for multimedia pattern recognition tasks and available MARF resources for its applications [272].

## C.1 Overview

### C.1.1 Motivation

The Modular Audio Recognition Framework (MARF, introduced in Section 5.2) has a good potential for multimedia pattern recognition research and comparison of various pattern-recognition algorithms. It can also serve as a re-usable library in applications that have to do with audio, natural language text, image processing due to its generality. The existing example applications include speaker, gender, accent, emotion, writer, language detection and identification; authentication and others. While the designers of MARF made every effort and example applications to be simpler and more accessible to a wider audience of users, they still require relatively skilled JAVA developers and relatively good understanding of the MARF's design architecture. Often scientists who just need to run their experiments and simulations do not fall into this category. Thus, to make it possible we state a requirement to be able to "script" in a convenient manner applications like `SpeakerIdentApp` [317], `FileTypeIdentApp` [290], `WriterIdentApp` [273, 318], `MARFCATApp` [285], etc., by specifying parts or all of the configuration context parameters. The syntax of the scripting language should be simpler and more natural than that of JAVA [272].

## C.1.2 Approach

We introduce syntactical and semantic constructs of context definitions for the configuration into this new MARF language that inherits some properties for the intensional programming languages, e.g., GIPL and LUCX (see Chapter 4) as well as their context operators, such as `@` and `#` to switch and query the current context. Intensionality and LUCID-compatible or near-similar syntax and semantics makes it more available to the scientific community and can be easily implemented in the intensional evaluation platforms like GIPSY (see Chapter 6). Through the examples of the syntactical expressions in MARFL, we build the initial syntax, semantics, and a brief type system to express the MARF configuration-as-context in the new language. We show how we deal with the resulting context-hierarchies occurring in our solution from the level-of-detail (LOD) point of view. We further re-write one of the MARF applications as how it would look in MARFL. We discuss the more general implications of our design. We propose to either interpret the MARFL scripts directly or "compile" them and generate an equivalent JAVA code (or even other language) of a MARF-based application. This can well integrate with plug-in frameworks for application development and verification like the Java Plug-in Framework (JPF), Eclipse [99], and existing cyberforensic toolkits (see Section 2.1) [272].

## C.2 Theoretical Framework

This is a theoretical framework with practical considerations for the existing system. The said framework focuses very specifically on the initial syntactical and semantic specification of the context in the MARFL language, the issues that come up with it, and how to address them, followed by implications of the work on the future work in this and other areas, such as common media pattern recognition applications and cyberforensics. We leave out other aspects (such as storage, statistics gathering, input/output in general) of MARFL unless they are absolutely necessary for illustrative purposes [272].

Dealing with MARF-specific contexts needs to be put into a perspective of the inner workings of MARF and its applications (cf., Section 5.2). This gives the needed requirements specification, syntax, and semantics of the contextual expressions. The rich parameterization of the MARF contexts necessitates a *context hierarchy*, i.e., contexts, sub-contexts, and sub-sub-contexts and so-on for the different levels of detail in the configuration settings. This *contextuality* translates into a specific configuration instance with the needed settings. If some dimensions are omitted in the scripted MARFL specification, they are filled in with the values that would come from the specific module instance defaults backing the system. The modules represent higher-order dimensions and can be added, updated, and removed as the system evolves to support change [380]. We do not support user-defined variables and identifiers in this case nor functions, so the corresponding rules have been excluded from the syntax and semantics. All our present dimension identifiers are preset dictated by the available modules of MARF as a precompiled dictionary. They make up reserved words of MARFL [272] as a result.

We borrow two prime operators of GIPL [361, 365], `@` to switch the context and `#` to query for the current context [361]. We treat them as functions, and we overload their definitions to accept different types of arguments of their expressions and as a result return different types of dimension values. We add the dot operator for the dimensions allowing us to navigate into the *depth* of the higher-order dimension objects [272].

We build our theory by examples to capture the desired properties of MARFL. A comprehensive example of a context for processing a WAV file in the MARF's pipeline can be modeled as shown in Figure 82. In that example we illustrate a complex hierarchical context expression where several nested dimensions are explicitly specified. In general, MARFL's context follows the definition of a *simple context* in Lucx [365, 513], where it is defined as a collection of $\langle dimension : tag \rangle$ pairs. What we do differently from Lucx (Section 4.3.1, page 92) is that a single pair may not necessarily be an atomic *micro context* in Lucx terminology, but may contain sub-contextual elements. The inner-most context is always simple and atomic as in Lucx and typically has dimensions of primitive types, such as integer, IEEE 754 floating point value [181], or a string. The outer layers of context hierarchy are composite objects. Thus, a `[sample loader:WAV]` denotes a dimension of type `sample loader` with its higher-order tag value `WAV`. The `WAV` dimension value can further be decomposed to an atomic simple context if needed, that contains three dimensions of primitive types [272].

In away, the described theory of higher-order context so far is similar to equivalent definitions described by Swoboda, Wadge, *et al.* in [452, 453, 454, 455, 511], where a tree of contexts is defined for languages like ɪHTML [510] (with nested tags) and the depth for LOD, functional intensional databases annotated with XML, and so on. While very similar in nature, the authors there suggest that they traverse their context tree all the way to the "leaf" simple contexts before doing evaluation and do not an actual evaluation at the higher-order contextual entities, which we allow here. We are also being very detailed of the specification and its operators including semantics in this work [272].

```
[
  sample loader      : WAV [ channels: 2, bitrate: 16, encoding: PCM, f : 8000 ],
  preprocessing      : LOW-PASS-FFT-FILTER [ cutoff: 2024, windowsize: 1024 ],
  feature extraction : LPC [ poles: 20, windowsize: 1024 ],
  classification     : MINKOWSKI-DISTANCE [ r : 5 ]
]
```

Figure 82: Example of hierarchical context specification for an evaluation configuration of MARF [272]

The `sample loader` dimension can vary over tags `WAV`, `SINE`, `MP3`, `MIDI`, `OGG`, `TXT`, `JPG`, `TIFF`, etc., each of which in turn can vary over `channels`, `bitrate`, sampling frequency (it can even apply to the text (natural or programming) and images depending on the context of interpretation). The `channels` dimension usually varies over 1 (mono) and 2 (stereo), but theoretically can have other valid tags. Similarly, the `bitrate` is commonly 8 or 16 data storage bits per frequency sample, and the frequency sampling rate `f` is typically 4, 8, 16, 44 kHz, but is not restricted to those values. The latter three are a finer-grained context examples while `WAV` and others are coarse-grained. A MARFL-written application should be able to alter any of these or switch between them. We should also be able to easily express all that [272].

We need to extend the semantics of `@` and `#` from GIPL to deal with the hierarchy of contexts, i.e., being able to switch sub-contexts and query them arbitrarily. This implies the argument types and the resulting type may not be necessarily the same. This is where we define the context operators overloading. In Table 18 is an incomplete collection of operator overloading rule examples. The meaning of the mappings in that table is as follows: a file type $F$ or a data vector type $V$ can be evaluated at the higher-order dimension of type sample loader $LD$ producing a "loaded sample" (that maps to an internal data structure). Evaluating a directory with sample files $D$ produces a collection of samples $S$. $S$ can be evaluated at dimension types preprocessing $PR$ and feature extraction $FE$, per MARF's specification, though the latter usually means $S$ has been preprocessed that is why the return types are internally different in their content. The feature data vector $V$ can be used for training or classification resulting in either a new or updated training set $TR$ or a result set $RS$ respectively. The subdivision of dimension types can go on further until the primitive tag values are reached, e.g., the $RS$ contains a collection of an integer $ID$ and a double *outcome* pairs and their index (not illustrated in the table). The unary operator `#` doesn't have a left argument, and the evaluation of the right argument is similar to that of `@`. The highest-order dimension we define is $MARF$ itself that if queried upon, returns the higher order collection of the main processing dimensions, and we can go all the way down the configuration context hierarchy. The dot operator provides a syntactical OO-like way of traversing the depth of the higher-level dimensions to the lowest ones. Though this may look like we make up OO dimensions, it is not presently so as we do not provide any true direct member access, including method objects, plus behind a single dimension may not necessarily be a single implementing class [272].

A few small examples: `#MARF.sl` would return `WAV`. Then `#MARF.WAV.channels` would return 2. `WAV @ [channels:1]`—switches WAV loader's configuration dimension `channels` to mono, essentially returning a loader with the new configuration setting. `WAV @ [channels:1, bitrate:8, f:4000]`—would conceptually return (configure) a `WAV` sample loader object ready and configured to load mono WAV samples, 8 bits per sample with the sampling frequency 4000 Hz. `''sample1.wav'' @ WAV @ [channels:1, bitrate:8, f:4000]` – would load the specified sample given the configured loader object producing an interpreted (loaded and decoded) double-stream of amplitude values, an array, encoded in the sample object [272].

The `preprocessing` dimension's higher-order valid tag values can range over tags like `RAW-PREPROCESSING`, `DUMMY`, `FFT-LOW-PASS-FILTER`, `ENDPOINTING`, and others, each of which may have common sub-dimensions and algorithm-specific sub-dimensions. For example, most FFT-based filters have a `cutoff` frequency dimension but endpointing doesn't, so trying to refer to the sub-context dimension of `cutoff` in endpointing would return an error as it is not a part of its sub-context. The `feature extraction` dimension is currently allowed to vary along the values `FFT`, `LPC`, `MINMAX`, `RANDOM-FEATURE-EXTRACTION`, `RAW` and others. The `classification` dimension is allowed to vary along the abstracted classification modules similarly to the preprocessing and others (see MARF pipeline in Section 5.2, page 102). By labeling the dimension identifiers as being a part of the reserved keywords necessitates making them a part of the syntax, which is illustrated in part in Figure 83. If the finer details of the higher-level dimensions are omitted, they assume the defaults, defined by the semantics of the implemented modules. For example, the default for `LPC`'s `poles` dimension is 20 and `windowsize` is 128. Here we do not describe how the defaults are picked and what they mean for every single module; for such details please refer to the related work [268, 270] and Section 5.2 [272].

Valid dimension tag values are defined by the MARF's currently implemented modules

$$
\begin{array}{rcll}
E & ::= & \#E & (1)\\
 & | & [E:E,...,E:E] & (2)\\
 & | & \{E,...,E\} & (3)\\
 & | & E \text{ where } Q & (4)\\
 & | & E@E & (5)\\
 & | & E.did & (6)\\
 & | & E \text{ cxtop } E & (7)\\
Q & ::= & \texttt{sample loader } SLid & (8)\\
 & | & \texttt{preprocessing } PRid & (9)\\
 & | & \texttt{feature extraction } FEid & (10)\\
 & | & \texttt{classification } CLid & (11)\\
 & | & did = E & (12)\\
 & | & QQ & (13)\\
SLid & ::= & \texttt{WAV} & (14)\\
 & | & \texttt{MP3} & (15)\\
 & | & \texttt{SINE} & (16)\\
 & | & \texttt{MIDI} & (17)\\
 & | & \ldots & (18)\\
 & | & \texttt{TEXT} & (19)\\
PRid & ::= & \texttt{FFT-LOW-PASS-FILTER} & (20)\\
 & | & \texttt{FFT-HIGH-PASS-FILTER} & (21)\\
 & | & \texttt{FFT-BAND-PASS-FILTER} & (22)\\
 & | & \texttt{FFT-BAND-STOP-FILTER} & (23)\\
 & | & \texttt{CFE-LOW-PASS-FILTER} & (24)\\
 & | & \ldots & (25)\\
 & | & \texttt{DUMMY-PREPROCESSING} & (26)\\
 & | & \texttt{RAW-PREPROCESSING} & (27)\\
FEid & ::= & \texttt{FFT} & (28)\\
 & | & \texttt{LPC} & (29)\\
 & | & \texttt{MINMAX} & (30)\\
 & | & \ldots & (31)\\
 & | & \texttt{RANDOM-FEATURE-EXTRACTION} & (32)\\
CLid & ::= & \texttt{CHEBYSHEV-DISTANCE} & (33)\\
 & | & \texttt{EUCLIDEAN-DISTANCE} & (34)\\
 & | & \texttt{MINKOWSKI-DISTANCE} & (35)\\
 & | & \texttt{DIFF-DISTAMCE} & (36)\\
 & | & \texttt{HAMMING-DISTAMCE} & (37)\\
 & | & \texttt{COSINE-SIMILARITY-MEASURE} & (38)\\
 & | & \ldots & (39)\\
 & | & \texttt{RANDOM-CLASSIFICAION} & (40)\\
cxtop & ::= & \textbf{train} & (41)\\
 & | & \textbf{classify} & (42)\\
\end{array}
$$

Figure 83: MARFL context syntax [272]

and the list can grow. Consequently, we need to be able to automatically accommodate the growth if possible making MARFL more adaptable and sustainable. All modules in MARF are plug-ins. When a new plug-in is registered, it can "deposit" the valid operational characteristics of itself into the "pool" ($\mathcal{D}_0$) of valid dimension identifiers and their range. The identifiers are largely predefined in this work that symbolize reserved words from MARF's resource collection and are not user-defined variables at this time [272].

Syntactically and semantically, we need to separate the operators for training and classification for the last stage as we need to know how to evaluate a feature vector $V$ at the classification module $CL$. $V@CL$ is ambiguous as we do not know whether we want to train $CL$ on $V$ or identify $V$. As a result, we create the `train` and `classify` operators to resolve the @ ambiguity. The dimension type system is thus becoming more apparent: we have object-like higher-level dimension types as well as primitive numerical types [272].

Table 18: Common example types of context operators' arguments and resulting types [272]

| Left Type | operator | Right Type | → | Resulting Type in $\mathcal{D}$ |
|---|---|---|---|---|
| $F$ | @ | $LD$ | → | $S$ |
| $V$ | @ | $LD$ | → | $S$ |
| $D$ | @ | $LD$ | → | $[S, \ldots, S]$ |
| $S$ | @ | $PR$ | → | $S$ |
| $S$ | @ | $FE$ | → | $V$ |
| $V$ | train(@) | $CL$ | → | $TS$ |
| $V$ | classify(@) | $CL$ | → | $RS$ |
|  | # | $MARF$ | → | $[LD:TAG, PR:TAG, FE:TAG, CL:TAG]$ |
|  | # | $LD$ | → | $[channels:TAG, bitrate:TAG, f:TAG]$ |
|  | # | $channels$ | → | $INTEGER$ |
|  | # | $LD.channels$ | → | $INTEGER$ |
|  | # | $LD.f$ | → | $FLOAT$ |
| $MARF$ | . | $LD$ | → | $LD$ |
| $MARF$ | . | $PR$ | → | $PR$ |
| $MARF$ | . | $FE$ | → | $FE$ |
| $MARF$ | . | $CL$ | → | $CL$ |
| $LD$ | . | $channels$ | → | $INTEGER$ |
| $CL$ | . | $TS$ | → | $TS$ |
| $CL$ | . | $RS$ | → | $RS$ |
| $RS$ | . | $ID$ | → | $INTEGER$ |
| $RS$ | . | $outcome$ | → | $FLOAT$ |

Semantically speaking for our context definitions we draw a lot from Lucx and GIPL (Chapter 4), as shown in Figure 84. The definition environment, $\mathcal{D}$, internally implemented by the `marf.Configuration` and `marf.MARF` classes that encapsulate all the higher- and lower-level configuration's contextual information $\mathcal{P}$ in their data members. The semantic rules other than (C.2.4) come from GIPL and Lucx and depict identifier definitions of constants (C.2.1), operators (C.2.2), regular Lucid dimensions (C.2.3), then the operators themselves (C.2.5), dimension tag values (C.2.6), the classical operator @ (C.2.7), the language `where` clause that we use for declarative configuration customization (C.2.8), the current context query with # (C.2.9), a simple context (collection of $\langle dimension : tag \rangle$ pairs) construction rule (C.2.10) and its navigation with @ (C.2.11), and re-definition of any dimension type in the syntactic `where` clause in (C.2.12). The rule (C.2.4) is introduced by MARFL to allow object-like dot-notation for dimension identifiers. This is a preliminary research semantics specification of MARFL. Notice also, we entirely omit the issues of storage management of the intermediate training sets and the returning of the results other than the sample loading aspect (and not even recording). MARF internally maintains a local database or intermediate cache of the processed utterances stored as training sets and returns the results (usually a some sort of measure and rank) as a result set. The training sets internally can be just plain serializable Java objects, CSV text data, XML data, or relational SQL data to a various degree depending on the options. None of this we cover in this work at the present and defer it to the future work [272].

$$\mathbf{E_{cid}} \quad : \quad \frac{\mathcal{D}(id) = (\mathtt{const}, c)}{\mathcal{D}, \mathcal{P} \vdash id : c} \tag{C.2.1}$$

$$\mathbf{E_{opid}} \quad : \quad \frac{\mathcal{D}(id) = (\mathtt{op}, f)}{\mathcal{D}, \mathcal{P} \vdash id : id} \tag{C.2.2}$$

$$\mathbf{E_{did}} \quad : \quad \frac{\mathcal{D}(id) = (\mathtt{dim})}{\mathcal{D}, \mathcal{P} \vdash id : id} \tag{C.2.3}$$

$$\mathbf{E_{E.did}} \quad : \quad \frac{\mathcal{D}(E.id) = (\mathtt{dim})}{\mathcal{D}, \mathcal{P} \vdash E.id : id.id} \tag{C.2.4}$$

$$\mathbf{E_{op}} \quad : \quad \frac{\mathcal{D}, \mathcal{P} \vdash E : id \quad \mathcal{D}(id) = (\mathtt{op}, f) \quad \mathcal{D}, \mathcal{P} \vdash E_i : v_i}{\mathcal{D}, \mathcal{P} \vdash E(E_1, \ldots, E_n) : f(v_1, \ldots, v_n)} \tag{C.2.5}$$

$$\mathbf{E_{tag}} \quad : \quad \frac{\mathcal{D}, \mathcal{P} \vdash E : id \quad \mathcal{D}(id) = (\mathtt{dim})}{\mathcal{D}, \mathcal{P} \vdash \#E : \mathcal{P}(id)} \tag{C.2.6}$$

$$\mathbf{E_{at}} \quad : \quad \frac{\mathcal{D}, \mathcal{P} \vdash E' : id \quad \mathcal{D}(id) = (\mathtt{dim}) \quad \mathcal{D}, \mathcal{P} \vdash E'' : v'' \quad \mathcal{D}, \mathcal{P}\dagger[id \mapsto v''] \vdash E : v}{\mathcal{D}, \mathcal{P} \vdash E \ @E' \ E'' : v} \tag{C.2.7}$$

$$\mathbf{E_w} \quad : \quad \frac{\mathcal{D}, \mathcal{P} \vdash Q : \mathcal{D}', \mathcal{P}' \quad \mathcal{D}', \mathcal{P}' \vdash E : v}{\mathcal{D}, \mathcal{P} \vdash E \ \mathtt{where} \ Q : v} \tag{C.2.8}$$

$$\mathbf{E_{\#(cxt)}} \quad : \quad \frac{}{\mathcal{D}, \mathcal{P} \vdash \# : \mathcal{P}} \tag{C.2.9}$$

$$\mathbf{E_{construction(cxt)}} \quad : \quad \frac{\mathcal{D}, \mathcal{P} \vdash E_{d_j} : id_j \quad \mathcal{D}(id_j) = (\mathtt{dim}) \quad \mathcal{D}, \mathcal{P} \vdash E_{i_j} : v_j \quad \mathcal{P}' = \mathcal{P}_0\dagger[id_1 \mapsto v_1]\dagger\ldots\dagger[id_n \mapsto v_n]}{\mathcal{D}, \mathcal{P} \vdash [E_{d_1} : E_{i_1}, E_{d_2} : E_{i_2}, \ldots, E_{d_n} : E_{i_n}] : \mathcal{P}'} \tag{C.2.10}$$

$$\mathbf{E_{at(cxt)}} \quad : \quad \frac{\mathcal{D}, \mathcal{P} \vdash E' : \mathcal{P}' \quad \mathcal{D}, \mathcal{P}\dagger\mathcal{P}' \vdash E : v}{\mathcal{D}, \mathcal{P} \vdash E \ @ \ E' : v} \tag{C.2.11}$$

$$\mathbf{Q_{dim}} \quad : \quad \frac{}{\mathcal{D}, \mathcal{P} \vdash \mathtt{dimension} \ id \ : \ \mathcal{D}\dagger[id \mapsto (\mathtt{dim})], \mathcal{P}\dagger[id \mapsto 0]} \tag{C.2.12}$$

Figure 84: MARFL context semantic rules [272]

## C.3 Applications

Generalization of MARFL and context-oriented intensional programming can be extended to other similar pattern-recognition and multimedia processing systems, including mobile applications. A possible PoC example of `SpeakerIdentApp` [317] rewritten in MARFL is in Figure 85. It is much more concise and clear than its raw JAVA equivalent, which is approximately 500 times longer (15 lines in MARFL vs. 1700 in JAVA modulo all the omitted functionality and comments). Other possible applications include the aforementioned cyberforensic analysis and others where nested context definitions are a case [272].

```
[ train /var/training-samples && classify /var/testing-samples ] @ MARF [sl, pr, fe, cl]
where
  // Our dimension types with the tag values we are interested to evaluate over
  sample loader      sl = WAV;
  preprocessing      pr = { FFT-LOW-PASS, FFT-BAND-STOP };
  feature extraction fe = { FFT, LPC, MINMAX, RANDOM };
  classification     cl = { CHEBYSHEV-DISTANCE, MINKOWSKI-DISTANCE, NEURAL-NETWORK };

  // Custom configuration for LPC and Minkowski distance
  // other than the defaults
  where
    LPC = { poles:40, windowsize:256 };
    MINKOWSKY-DISTANCE = { r:5 };
  end;
end
```

Figure 85: Initial `SpeakerIdentApp` re-written in MARFL [272]

## C.4 Summary

We presented initial context definition for the syntax and semantics of MARFL, the *MARF Language*, which is bound around practical concerns to script MARF-based applications that are context-aware. The context in MARF is a hierarchical configuration that we manipulate using two intensional operators. We override the definitions of @ and # from GIPL and introduce the object membership operator (dot-notation) to accept hierarchies of contexts. This brings a lot more understandability to the MARF applications allowing non-programmers, e.g., scientists, to benefit from the MARF-provided resources and make the application significantly more re-usable and maintainable from the software engineering point of view [272].

In the presented application, all contexts are always finite. While the theoretical framework could support potentially infinite contexts (e.g., very long continuous audio streams coming from a recorder or microphone which is normally never turned off), it is generally not very practical [272].

# Appendix D

# Self-Forensics

This chapter introduces the concept of *self-forensics* in addition to the standard autonomic *self-\** properties (also known as *self-CHOP*: self-configuration, self-healing, self-optimization, and self-protection in autonomic computing [327, 371]) of self-managed systems, specified in the FORENSIC LUCID language. We argue that self-forensics, with the digital forensics aspect taken out of the cybercrime domain, is applicable to "self-dissection" of autonomous software and hardware systems for automated incident and anomaly analysis and event reconstruction for the benefit of the engineering teams in a variety of incident scenarios [277, 308].

We focus on the core concepts and fundamentals of self-forensics for ICT using FORENSIC LUCID to specify the evidence. Self-forensics can be roughly understood as being the application of autonomic or semi-autonomic self-diagnostics combined with reasoning. This includes forensic logging by the software/hardware sensors, evidence encoding, and case modeling. The process can be automated or interactive, "after-the-fact", with the investigator present. Thus, we take out the FORENSIC LUCID language from the cybercrime context to apply to any autonomic software or hardware systems (e.g., vehicles or distributed software systems) as an example of self-forensic case specification [277, 308, 321]. The earlier presented properties in Chapter 7 make FORENSIC LUCID an interesting and appropriate choice for self-forensic computing in self-managed systems to complement the existing self-\* properties [277, 308, 321].

This proposition in this chapter in large part is a synthesis of published and unpublished works and proposals of the author of various applications [269, 276, 277, 308, 321, 322] for the self-forensics concept as well as a work towards improving the safety and incident investigation of autonomous systems.

## D.1 Overview

In this work we introduce a new concept and aggregate our previous findings in requirements specification for autonomous systems, flight-critical integrated software and hardware systems, etc. to analyze themselves forensically on demand as well as keeping forensics data for further automated analysis in cases of reports of anomalies, failures, and crashes. We insist this should be a part of the protocol for each such a system, (even not only space missions or flight systems, but any large and/or critical self-managed system [277, 308]).

We review some of the related work that these ideas are built upon prior describing the requirements for self-forensics-enabled components. We subsequently describe the general requirements as well as limitations and advantages [277].

The property of self-forensics [276, 277] was introduced to encompass and formally apply to or be included in the design of not only autonomic hardware and/or software systems, which are inherently complex and hard to analyze in general when incidents happen, but also as an optional requirement for smaller-scale projects [322].

Self-forensics in a nutshell includes a dedicated module or modules observing the regular modules in some way, logging the observations and events that led to them in a predefined format suitable for automatic forensic processing and deduction, and event reconstruction in case of incidents. The modules can optionally have a capability of automatically diagnose themselves based on the collected evidence and make more complex and complete decisions after the analysis than ad-hoc binary decisions. In a sense, the self-forensic modules can be seen as smart "blackboxes" like in aircraft, but can be included in spacecraft, road vehicles, large and small software systems, etc. that assist with the incident analysis. Human experts can be also trained in investigation techniques based on the forensic data sets collected during different incidents. In a traditional sense, one argues that any self-diagnostics, as well as traditional logging, hardware or software, are necessary parts of self-forensics that have been around for a long time [322].

We compile examples of FORENSIC LUCID specifications for self-forensics of a few software projects as case studies. The specifications are there to be built into the systems for the purpose of tracing and understanding complex relationships and events within some components of the said systems, especially the distributed ones. In this work, we first reasonably narrowly focus on "exporting" the states of the systems as data structures in chronological encoded order as FORENSIC LUCID contexts using its syntax in accordance with the grammar for compiling the cases [322] and try to generalize the concept to a wider application spectrum.

We further proceed to describe the background and the related work, followed by the example specifications of the core data structures and dataflows in FORENSIC LUCID of the case studies, such as the *Distributed Modular Audio Recognition Framework* (DMARF, Section 5.2, page 102), *General Intensional Programming System* (GIPSY, Chapter 6), *Java Data Security Framework* (JDSF, Section D.3.3, page 363), and *Cryptolysis*—an automated cryptanalysis framework for classical ciphers. These are primarily academic/open-source systems with the first two being distributed and a lot more complex than the last two. After their specification, we generalize to other domains, such as cloud forensics, and we conclude and list a few near future work items [322].

Additionally, this preliminary conceptual work discusses a notion of self-forensics as an autonomic property to augment the Autonomic System Specification Language (ASSL) framework of formal specification tools for autonomic systems. The core of the proposed methodology leverages existing designs, theoretical results, and implementing systems to enable rapid completion of and validation of the experiments and their the results initiated in this work. Specifically, we leverage the ASSL toolkit to add the self-forensics autonomic property (SFAP) to enable generation of the JAVA-based Object-Oriented Intensional Programming (JOOIP, Section 4.3.2.3, page 94) language code laced with traces of FORENSIC LUCID to encode contextual forensic evidence and other expressions [322].

The notion and definition of self-forensics was introduced by the author Mokhov in circa 2009 to encompass software and hardware capabilities for autonomic and other systems to record their own states, events, and others encoded in a forensic form suitable for (potentially automated) forensic analysis, evidence modeling and specification, and event reconstruction for various system components. For self-forensics, "self-dissection" is possible for analysis using a standard language and decision making if the system includes such a self-forensic subsystem. The self-forensic evidence is encoded in a cyberforensic investigation case and event reconstruction language, FORENSIC LUCID (Chapter 7). The encoding of the stories depicted by the evidence comprise a context as a first-class value of a FORENSIC LUCID "program", after which an investigator models the case describing relationships between various events and pieces of information. It is important to get the context right for the case to have a meaning and the proper meaning computation. As PoC case studies of some small-to-medium (distributed or not) primarily academic open-source software systems are examined. In this work, for the purpose of implementation of the small self-forensic modules for the data structures and event flow, we specify the requirements of what the context should be for those systems. The systems share in common the base programming language—JAVA, so our self-forensic logging of the JAVA data structures and events as FORENSIC LUCID context specification expressions is laid out ready for an investigator to examine and model the case [322].

## D.2 Motivation

We motivate this work through examples that include enhancement and help with the blackboxes like in aircraft, or where self-forensics would have been helpful to analyze anomalies say in the spacecraft, when *Mars Exploration Rover*s behaved strangely [335], or even with one is doing a hard disk recovery, such as from the shuttle *Columbia* [116], or automatically as well as interactively reasoning about events, possibly speeding up the analysis of the anomalies in subsystems following sound methodology. Another example is when the *Hubble Space Telescope* was switched from its "side A" of its instruments to the redundant "side B". The self-forensics units would have helped Hubble to analyze the problem and self-heal later. Of course, the cost of such self-forensic units would not be negligible; however, the cost of self-forensics units maybe well under than the costs of postponing missions, as, e.g., it was happening with the *Hubble Space Telescope Servicing Mission 4 (SM4)* and the corresponding shuttle processing delay and costs of moving shuttles around [331, 332, 333, 334, 415, 475] and others [277, 308].

Furthermore, the concept of self-forensics would be even a greater enhancement and help with flight-critical systems, (e.g., included with the blackboxes in aircraft) to help with crash investigations [72, 277]. Similar examples can made for a bus crash in California in 2009 [71] or the *Kepler* spacecraft entering safe mode in 2010 [336]. Any large-scale software systems, distributed or not, web services, operating systems, embedded systems are a natural environment for self-forensics analysis, prevention, and reaction just as well. Thus, we insist that self-forensics, if included earlier in the design and development of the spacecraft, craft or vehicles, systems, not only helps during the validation and verification, but also *a posteriori*, during the day-to-day operations incident investigation and instrumenting corrections by the system itself when it is unreachable by personnel or human investigators after the incident if the system is not recoverable [277].

### D.2.1 Self-Forensic Computing

Many ideas in this work come from computer forensics and forensic computing. Computer forensics has traditionally been associated with computer crime investigations (see Section 2.1, page 24). We show the approach is useful in autonomic systems [308]. Additionally, the approach is useful as an aid in for validation and verification during design, testing, and simulations of such systems as well as during the actual operations and any kind of incident investigations. We earlier argued [276, 308] if the new technologies are built with the self-forensics components and toolkits from the start, it would help a wide spectrum of various autonomous and autonomic software and hardware industries [277].

Existing self-diagnostics, computer BIOS/UEFI-like reports with *name:value* attributes, Cisco IOS states/log reports (e.g., see Section 9.5), and S.M.A.R.T. [12, 67, 519] reporting for hard drives as well as many other devices are a good source for such data computing. The idea is to be more forensics-friendly and provide appropriate interfaces for self-forensic analysis and investigation as well as allowing engineering teams extracting, analyzing, and reconstructing events using such data [277, 308, 322].

### D.2.2 Problem and Proposed Solution

The emerging concept of self-forensics and the idea of its proposed realization within ASSL and GIPSY are described through their core founding related work. These preliminary findings and discussions are currently at the conceptual level, but the author provides the complete requirements, design, and implementation of the concept described here by leveraging the resources provided by the previous research work. To the author's knowledge there is no preceding work other than the author's own that does attempt something similar [322].

First, we give a glimpse overview of the founding background work on ASSL and self-forensics in Section D.3.6 and Section D.4.1. Then, we describe the core principles and ideas of the methodology of realization of the self-forensics autonomic property (SFAP) within the ASSL framework in Section D.4.5. We provide a quick notion of the syntactical notation of SFAP and where it fits within the generating toolset of ASSL. We conclude in Section D.5 for the merits and the future endeavors for the developments in this direction [322].

## D.3 Related Work

There are a number of inter-related works of relevance listed below.

### D.3.1 Available Related Technologies and Tools

Some of these have been previously mentioned:

- S.M.A.R.T. technologies [12, 67, 519]

- Recent Cisco netflow uplink modules running at the 10G speeds allowing collection of the pcap netflow data. FORENSIC LUCID rules like ACLs in Cisco equipment would run at line speed. Local data with SD storage until transmitted later to the control and maintenance centers for analysis.

- Gigamon span-port splitters for sorting out the pcap netflow traffic.

- Linux state tracing kernel [150].

- ASPECTJ [29] (Section 8.2.1) can be used to create forensics-relevant aspects for JAVA programs.

- Various software based logging mechanisms provided by operating systems and services (syslog, Event Log, etc.).

### D.3.2 Expert Systems

The AT&T early diagnosing expert system for troubleshooting their phone line networks with a corresponding knowledge base is a good early example [186] of relevance to us. There are many of such examples [186] where self-forensics would be useful.

### D.3.3 Java Data Security Framework

JDSF [275, 294, 295, 299, 316] has been proposed to allow security researches working with several types of data storage instances or databases in JAVA to evaluate different security algorithms and methodologies in a consistent environment. The JDSF design aims at the following aspects of data storage security: confidentiality (data are private), integrity (data are not altered in transit), origin authentication (data are coming from a trusted source), and SQL randomization (for relational databases only, not discussed here any further) [271, 322].

JDSF also provides an abstraction of the common essential cryptographic primitives. The abstraction exposes a common API proxied to the common JAVA open-source implementations of the cryptographic algorithms for encryption, hashing, digital signatures, etc. The higher-level JDSF design summary is illustrated in several UML specification diagrams documented in the related works [295, 299, 316]. The design presented in those works illustrates all the necessary main subpackages of JDSF and its configuration, the design of security-enhanced storage, the authentication subframework, privacy subframework, integrity subframework, and the abstraction API over the concrete cryptographic primitives [271, 322].

JDSF is convenient to use in the scope of the presented research, and, like GIPSY and DMARF, it is implemented in JAVA, and is open-source [271, 322].

### D.3.4 Cryptolysis

Cryptolysis [298] is a small framework that includes a collection of automated attacks on the classical ciphers using a set heuristics algorithm implementation in JAVA. The algorithms primarily come from the related work on the same subject [69]. Cryptolysis also features additional algorithms that are wrappers around classification and signal processing tasks of MARF [465] for additional imprecise and spectral reasoning. Cryptolysis, among other things, also performs some NLP parsing that segments the deciphered whole-block text and inserts spaces in-between the word boundaries automatically for the ease of readability. Cryptolysis, like the others, is an open-source project [322].

### D.3.5 Autonomic Computing and Self-Management Properties

The common aspects of self-managing systems, such as self-healing, self-protection, self-optimization, and the like (self-CHOP) are now fairly well understood in the literature and R&D [167, 178, 207, 327, 371, 476, 491, 500]. We formally augment that list with the *self-forensics* autonomic property that we would like to be a part of the standard list of autonomic systems requirements specification and design specification [277, 308].

The self-forensics property is meant to embody and formalize all existing and future aspects of self-analysis, self-diagnostics, related data collection and storage, software and hardware components ("sensors") and automated decision making that were not formalized as such and define a well-established category in the industry and academia [277, 308]. In that view, self-forensics encompasses self-diagnostics, blackbox-like recording, traditional logs, web services, (Self-Monitoring, Analysis, and Reporting Technology) S.M.A.R.T. reporting [12, 67, 519], and encoding this information in analyzable standard form of FORENSIC LUCID [304, 310] for later automated analysis and event reconstruction using the corresponding expert system tool [277, 308]. Optional parallel logging is done of the forensics events during the normal operation of the autonomous and semi-autonomous systems, especially during the "blackout" periods (when a given remote system is unreachable by operators) will further enhance the durability of the live forensics data logged from the system to a nearby *loghost* (a remote logging facility), some sort of receiving station, etc. [277, 308].

### D.3.6 ASSL Formal Specification Toolset

The ASSL (Autonomic System Specification Language) framework [486, 493, 502] takes as an input a specification of properties of autonomic systems [6, 173, 178, 179, 207, 327, 371], does formal syntax and semantics checks of the specifications, and if the checks pass, it generates a collection of JAVA classes and interfaces corresponding to the specification. Subsequently, a developer has to fill in some overridden interface methods corresponding to the desired autonomic policies in a proxy implementation within the generated JAVA skeleton application or map them to the existing legacy application [322, 486, 493, 502].

Similarly to GIPSY (Section 6.2, page 136), the ASSL framework [502] includes the autonomic multi-tier system architecture (AS) including formal language constructs to specify service-level objectives (SLOs), core self-CHOP (i.e., self-configuration, self-healing, self-optimization, and self-protection) autonomic properties, corresponding architecture, allowed actions, events, and metrics to aid the self-management aspect of the system. It also specifies the interaction protocols between the AS's managed autonomic elements, including specification of messages exchanged and how they are communicated. Finally, it provides for specification of the autonomic element (AE) architecture, like for the whole system, each element is a subject to the SLOs, self-CHOP policies, behavior, actions, metrics, and interaction protocols, the summary of all of which is enumerated in Figure 86, page 365 [322].

```
I. Autonomic System (AS)
 * AS Service-level Objectives
 * AS Self-managing Policies
 * AS Architecture
 * AS Actions
 * AS Events
 * AS Metrics
II. AS Interaction Protocol (ASIP)
 * AS Messages
 * AS Communication Channels
 * AS Communication Functions
III. Autonomic Element (AE)
 * AE Service-level Objectives
 * AE Self-managing Policies
 * AE Friends
 * AE Interaction Protocol (AEIP)
   - AE Messages
   - AE Communication Channels
   - AE Communication Functions
   - AE Managed Elements
 * AE Recovery Protocol
 * AE Behavior Models
 * AE Outcomes
 * AE Actions
 * AE Events
 * AE Metrics
```

Figure 86: Vassev's ASSL multi-tier model [322]

ASSL formal modeling, specification, and model checking [489, 490] has been applied to a number open-source, academic, and research software system specifications, e.g., such as Voyager imagery processing [488]; the Distributed Modular Audio Recognition Framework (DMARF) [320, 494, 495]; the General Intensional Programming System (GIPSY) [500]; reliability of self-assessment, distributed, and other autonomic aspects of the autonomic system-time reactive model (AS-TRM) [492, 497]; self-adapting properties of NASA swarm missions [167, 476, 491], and others [322, 487].

# D.4 Self-Forensics Methodology Overview

## D.4.1 Self-Forensics Concept

The study of self-forensics [276, 277, 321] is an additional property the author is investigating throughout this topic with the contextual forensic logging with FORENSIC LUCID and case specification [269, 300, 304, 306, 307]. As previously mentioned, FORENSIC LUCID is an intensional context-oriented forensic case specification, modeling, and evaluation language (see Chapter 7 for details). FORENSIC LUCID was initially proposed for specification and automatic deduction and event reconstruction in the cybercrime domain of digital forensics [357]. It has been proposed to extend its use onto other domains such as investigation of incidents in various vehicle crash investigations, and autonomous software and hardware systems. Its primary feature inherited from the LUCID family of languages is to be able to specify and work with context [365, 473, 513] as a first-class value, and the context represents the evidence and witness accounts [322].

FORENSIC LUCID's primary experimental platform for compilation (FORENSIC LUCID compiler is a member of the General Intensional Programming Compiler (GIPC) framework, Section 6.2.1, page 139) and evaluation is the General Intensional Programming System (GIPSY, Chapter 6) [161, 302, 362]. GIPSY's run-time system, the General Eduction Engine (GEE, Section 6.2.2, page 142), is designed to be flexible to allow various modes of execution, including the use of the evaluation by the PRISM- [467] and ASPECTJ-based [29] backends as illustrated in Figure 32, page 137 [315] and Figure 58, page 219 [322] (Chapter 6 and Chapter 1).

## D.4.2 Generalities

As we encode the observation sequences for our case study systems, we observe some general tendencies. It is possible to have gaps in the evidence stories when some of the expected data structures were not instantiated or some events did not occur. To recap, we model them as *no-observation*s [136, 137], \$. A *no-observation* is an observation $\$ = (C_T, 0, \mathit{infinitum})$ that puts no restrictions on computations (Section 2.2, page 29). The *infinitum* is an integer constant that is greater than the length of any finite computation that may have happened and $C$ is a set of all possible computations in the system $T$ [137, 321] (Chapter 2).

All systems' high-level evidence follows a similar traditional blackbox evidential pattern: *given an instance of a configuration followed by the initial "input" data followed by one or more computations on the data followed by a result or a result set.* This contextual pattern can be exploded at the context level to the very fine-grained details to the lowest level of core data structures and variables. The computations can be arbitrarily complex and nesting of the context expressions, as evidential logs, can be arbitrarily deep, depending on the desired level-of-detail (LOD) [321].

Inherently, a parallel evaluation can also be captured as variable-length generic observations [136, 137] (Section 2.2.4.5, page 39, Section 7.5.5, page 208), forming a multidimensional *box* [513], where each element can be evaluated concurrently. All elements should then complete the evaluation prior transiting to the final result or result set observation [321].

Subsequently we describe some possible example applications of self-forensics and its requirements that must be made formal, in the industry, if the property to be used in a wider context and scope and included in the design of the various systems early on [277, 308, 321].

Given this initial discussion we further proceed with some of the contextual FORENSIC LUCID specifications of our studied systems, specifically DMARF in Section D.4.6.1, GIPSY in Section D.4.6.2, JDSF in Section D.4.6.3, and Cryptolysis in Section D.4.6.4 [321].

### D.4.3 Requirements

Based on the previously cited work on different aspects of self-forensics we define the requirements scope for the autonomic self-forensics property. This is a mix of mostly high-level functional and some non-functional requirements, aggregated as the initial "check-list" to be used in the real system requirements specification in a much more elaborate detail and expected to be carried over into the design specification.

Here we define very rough high-level the requirements scope for the autonomic self-forensics property:

- Should be optional if affected by the severe budget constraints for less critical system components, but easily "pluggable". Must not be optional for mission critical and safety-critical systems and blackboxes [277, 308].

- Must be included in the design specification at all times using some formal method of specification (e.g., FORENSIC LUCID, ASSL, and Isabelle) [277, 308].

- Must cover all types of the self-diagnostics events (e.g., S.M.A.R.T.-like capabilities and others.) [277, 308].

- Must have a formal specification (that what it makes it different from various typical self-diagnostics and otherwise various aspects of self-monitoring) [277, 308].

- Must have tools for automated reasoning and reporting about incident analysis matching the specification, real-time or *a posteriori* during investigations. The tools may accompany the system in question if resources allow, and also be remotely available for independent parallel analysis and investigation [277, 308].

- Context should be specified in the terms of system specification involving the incidents, e.g., parts and the software and hardware engineering design specification should be formally encoded (e.g., in FORENSIC LUCID) in advance, during the design and manufacturing. These are the static forensic data. The dynamic forensic-relevant data are recorded in real-time during the system operation [277, 308].

- Preservation of forensic evidence must be atomic, reliable, robust, and durable [277, 308].

- The forensic data must be able to include any or all related not-necessarily-forensic data for analysis when needed in case it can be a source of a problem (incident). Depending on the system, e.g., the last image or application-specific data passing through reconnaissance or science images for military, exploration, and scientific aircraft taken by a camera, measurements done around the incident by an instrument and/or protocol states around the time of the incident. The observing modules may include the entire trace of a lifetime of a system or system components logged *somewhere* for automated analysis and event reconstruction [277, 308].

- Levels of forensic logging and detail should be optionally configurable in collaboration with other design requirements in order not to hog other activities' resources, create significant overhead, or fill in the network bandwidth of downlinks or to preserve power [277, 308].

- Self-forensics components should optionally be duplicated in case they themselves also fail [277].

- Event co-relation mechanisms optionally should be specified, e.g., as in time-reactive models and systems [277, 308], or combined within FORENSIC LUCID.

- Probabilistic trustworthiness and credibility factors need be specified if corrupt and/or unreliable (evidential) data is detected, so its credibility weight is reduced during self-investigation to reduce the risk of "wrong" decisions.

- Some forensic analysis can be automatically done by the autonomous system itself (provided having enough resources to do so), e.g., when it cannot communicate with system administrators, or with flight controllers, or the engineering crew (e.g., during solar conjunction for a week or two the robotic spacecraft are left unattended) [277, 308].

### D.4.4 Example Scenario

Below is a high-level description of the self-forensic process described as a scenario that is to become a refined algorithm. Terminology: an *autonomous system* is any software and/or hardware system exhibiting self-management properties, e.g., distributed systems, operating systems, cloud computing platforms, road-side networks (RSNETs), any time of autonomous craft or vehicle. A *sensor* is a software or hardware component observing events of interest and doing forensic logging. A *functional unit* is a software/hardware component or instrument under observation.

1. Self-forensic "sensors" observe functional units (e.g., instruments and subsystems of spacecraft, aircraft in flight, systems of IRSNET, subsystems of a cloud, or software systems) [277, 308].

2. Every event of interest (engineering or scientific) is logged using FORENSIC LUCID [277].

3. Each forensic sensor observes one or several components or instruments (hardware or software).

4. Each sensor composes a witness testimony in the form of FORENSIC LUCID observational sequence *os* about a component, a subsystem or instrument it observes [277].

5. A collection of witness accounts from multiple sensors, properly encoded, represent the evidential statement, $es$ forming the local knowledge base either during the verification or actual operation [277].

6. If an incident (simulated or real) happens, systems and engineers define and encode theories (or hypotheses) in FORENSIC LUCID about what happened. The theories are likewise encoded as observation sequences $os$. When evaluated against the collected forensic evidence in $es$, they are added to $es$. Then any appropriate evaluating expert system (e.g., based on GIPSY with extensions in our case studies) can automatically verify their theory against the context of the evidential statement $es$: if the theory $T$ agrees with the evidence, meaning this theory has an explanation within the given evidential context (and the amount of evidence can be significantly large for "eyeballing" it by humans), then likely the theory is a possible explanation of what has happened. It is possible to have multiple explanations and multiple theories agreeing with the evidence. In the latter case usually the "longer" (in the amount of reconstructed events and observations involved) theory is preferred or the one that has a higher cumulative credibility weight $w$. Given the previously collected and accumulated knowledge base of FORENSIC LUCID facts, some of the analysis and events reconstructed can be done automatically. The transition function and its inverse in this case are assumed to have already been incorporated and available from the time the system was designed. Its FORENSIC LUCID specification is simply processed with the $es$ that was received [277, 308].

7. The incident is also processed in a somewhat similar fashion by the autonomous system in question if resources allow. It may have actions defined on what to do based on the results of the self-investigation. The specification of what the potential claims may be in this case can be pre-defined in the hardware or software for expected incidents. The unexpected incidents may be more difficult to deal with, but over time those can be either machine-learned or added by the engineering team over time after simulations with the test systems.

### D.4.5 Self-Forensics Autonomic Property (SFAP) in ASSL

In the PoC ASSL specification of self-forensics we add a notion of a SELF_FORENSICS policy specification for the AS tier and AE, just like it is done for the self-CHOP properties. The proposed property introduction consists of two major parts: (1) adding the syntax and semantic support to the lexical analyzer, parser, and semantic checker of ASSL as well as (2) adding the appropriate code generator for JOOIP and FORENSIC LUCID to translate forensic events. The JOOIP code is mostly JAVA with embedded fragments of FORENSIC LUCID-encoded evidence [269, 321, 322].

We use ASSL's managed-element (ME) specification of AE to encode any module or subsystem of any software system under study to increase or reduce the amount of forensic evidence logged as FORENSIC LUCID events depending on the criticality of faults (that can be expressed as ASSL metrics) [322].

A very high-level example of the generic self-forensic specification is in Listing D.1, page 371. Many details are presently omitted due to the preliminary work on this new concept and will be provided in our subsequent future publications [322].

Wu and the GIPSY R&D team came up with a hybrid intensional OO language, JOOIP ([526, 528], Section 4.3.2.3, page 94), to allow mixing JAVA and LUCID code by placing LUCID fragments nearly anywhere within JAVA classes (as data members or within methods). As a part of this conceptual research work, we propose that the ASSL toolset in this instance be augmented with a code-generation plug-in that generates JOOIP [526, 528] code laced with FORENSIC LUCID contextual expressions for forensic analysis. The evaluation of the JOOIP+FORENSIC LUCID code further is to be performed by the GIPSY's general eduction engine (GEE), described in detail in [161, 302, 322, 362] (see also Section 6.2.2, page 142).

Furthermore, in this proposed prototype the `EVENTS` members are the basic building blocks of the contextual specification of the FORENSIC LUCID observation sequences. The `INITIATED_BY` and `TERMINATED_BY` clauses correspond to the beginning and end-of-data-stream LUCID operators `bod` and `eod`. ASSL fluents map to the LUCID streams of the observation sequences where each stream is a witness account of systems behavior. All fluents constitute an evidential statement. The mapping and actions correspond to the handling of the anomalous states within the JOOIP's JAVA code [322].

In the proposed design, once JOOIP code with FORENSIC LUCID fragments is generated by the ASSL toolset, it is passed on to the hybrid compiler of GIPSY, the GIPC to properly compile the JOOIP and FORENSIC LUCID specifications, link them together in a executable code inside the GEE engine resources (GEER), which then would have three choices of evaluation of it—the traditional eduction model of GEE, ASPECTJ-based eduction model, and probabilistic model checking with the PRISM backend [322] (see Chapter 8).

### D.4.6 Example Applications

We illustrate some simple examples of how a FORENSIC LUCID specification of self-forensics context will begin in a form of a scripting template. Then, we follow with some more concrete examples in the subsequent subsections. Having FORENSIC LUCID helps scripting the forensics events in the autonomous systems. The real blackboxes can contain the forensic data encoded anyhow including forensics expressions, XML, or just compressed binary and using external tool to convert it to a FORENSIC LUCID specification at a later time. Below is a high-level preamble of any typical specification of the hardware or software system in question.

```
invtrans(T @ system_es)
where
  evidential statement system_es = { ... };
  // T is a theory or hypothesis of what has transpired
  observation sequence T = { ... };
end
```

Default binary FORENSIC LUCID specification corresponds to a serialized AST (as in earlier presented GEER) for more optimal evaluation and retrieval. The evidential statement *es* is a dictionary of the evidential observation sequences. It may contain definitions that are not necessarily used by the modeled case program. The following template can be remodeled for

```
AS ADMARF {
    TYPES {      MonitoredElement }
    ASSELF_MANAGEMENT {
        SELF_FORENSICS {
            FLUENT inIntensiveForensicLogging {
                INITIATED_BY { EVENTS.anomalyDetected }
                TERMINATED_BY {
                    EVENTS.anomalyResolved ,
                    EVENTS.anomalyFailedToResolve
                }
            }

            MAPPING {
                CONDITIONS { inIntensiveForensicLogging }
                DO_ACTIONS { ACTIONS.startForensicLogging }
            }
        }
    }
    ACTIONS {
        ACTION startForensicLogging {
            GUARDS { ASSELF_MANAGEMENT.SELF_FORENSICS.inIntensiveForensicLogging }
              VARS { Boolean ... }
              DOES {
                  ...
                  FOREACH member in AES {
                      ...
                  };
              }
              ONERR_DOES {
                  // if error then log it too
                  ...
              }
        }
    } // ACTIONS
    EVENTS { // these events are used in the fluents specification
        EVENT anomalyDetected {
            ACTIVATION { SENT { ASIP.MESSAGES.... } }
        }
        ...
    } // EVENTS
    METRICS {
        METRIC thereIsInsecurePublicMessage {
            METRIC_TYPE { CREDIBILITY }
            DESCRIPTION { "sets event's trustworthiness/credibility AE" }
            VALUE { ... }
            ...
        }
    }
} // AS ADMARF
// ...
MANAGED_ELEMENTS
{
   MANAGED_ELEMENT STAGE_ME
   {
       INTERFACE_FUNCTION logForensicEvent
       {
           PARAMETERS { ForensicLucidEvent poEvent }
           RETURNS { Boolean }
       }
   }
}
```

Listing D.1: The prototype syntactical specification of the SELF_FORENSICS in ASSL for ADMARF [322]

example to fit network equipment, such as a switch or a router, based on a lot more level of detail exemplified in the MAC spoofer investigations (see Section 9.5, page 257).

```
invtrans(T @ switch_es)
where
 evidential statement switch_es = { ... };
 // T is a theory of what has happened, e.g., with a switch port
 observation sequence T = { ... };
end
```

Such functions can be precompiled as "stored procedures" (using the DBMS terminology) for very well known manufacturer specifications of components as well as validated software, possibly with Petri nets, or the PRISM state machines alike [136, 137].

As the application examples, the following subsections illustrate some context encoding examples in FORENSIC LUCID for small-to-medium software systems followed by the IRSNETs and cloud computing systems as larger scale project proposals where it is sensible to implement self-forensics.

### D.4.6.1 DMARF

Distributed aspect of DMARF (Section 5.2, page 102), as opposed to the classical MARF, makes gathering of more forensic data due to intrinsically more complex communication between modules, and potentially remote ones. In addition to the earlier specification recited in Figure 68, page 235, we need to capture the configuration data related to the connection settings, protocols, and any other properties related to the distributed systems [321].

The main difference is that now given the same configuration it is possible to have multiple distributed/parallel training sets to be computed as well as multiple results produced on different nodes and the configuration object extension is designed to include the communication protocols and the related information. In Equation D.4.1 is a complete high-level specification of a DMARF run of observations. $dconfo$ is an observation of the initial DMARF configuration, $tseto_i$ is the observation of the $X \times Y$ training sets, where $X$ and $Y$ form a *box* of all possible training sets that can be available in that run, and the $resulto_i$ is an observation of possible corresponding results of the pipelines' runs. There could be multiple pipelines formed through the lifetime of a DMARF network of the computing stages and nodes [321].

$$(dconfo, 1, 0), (tseto_i, X, Y), \ldots, (resulto_i, X, Y) \qquad (D.4.1)$$

A system instance that has not produced a result (e.g., due to a crash at the classification stage) would still have a 3-observation sequence, with the last one being a *no-observation*, as shown in Equation D.4.2.

$$(dconfo, 1, 0), (tseto_i, X, Y), \ldots, \$ \qquad (D.4.2)$$

It is theoretically possible to have no-observations for the configuration or training set data components say in case such real evidential data were lost or are not trustworthy due to poor handling or improper chain of custody. In such a case, only partial reasoning can be performed on the evidence [321].

Note, at this point we do not cover the typical advanced distributed systems features such as the replication, write-ahead-logging, and load balancing and their related data structures, communication, etc. that deserve a separate large treatment and for now we focus only on the normal business operation of the system [321].

### D.4.6.2 GIPSY

In GIPSY (Chapter 6), the core data structure, connecting the compiler GIPC (Section 6.2.1, page 139) and the execution engine GEE (Section 6.2.2, page 142) is the GEER, represented by the `GIPSYProgram` class. This is the initial observation as far as GEE concerned. It causes the engine to produce a number of demands in the order of the abstract syntax tree (AST) traversal, abstracted by the `Demand` class and the corresponding `IDemand` interface. Demands live and die inside a some sort of a store (that acts like a cache) and are transported by the transport agents (TAs) using various communication technologies mentioned earlier in the previous section. `IDemand` is used for both demands and their results. When the result of computation reaches the top of the evaluation tree, the final result of the program is computed—it is an instance of some subclass of the `GIPSYType` from the GIPSY Type System [315] (Appendix B). Thus, the high-level observation sequence is encoded as [321]:

$$(p, 1, 0), (d_i, X, Y), \ldots, (r, 1, 0) \tag{D.4.3}$$

where `GIPSYProgram` $p$—is the initial program (GEER), `IDemand` $d_i$—is a cross product $X \times Y$ of all possible demands that may be observed, `GIPSYType` $r$—is the result of computation. $X$ and $Y$ form a cross product of all possible demands that can be made throughout the program execution, of whether they need to be computed or retrieved from a cache, an intensional data warehouse. An example of the observation sequence expression is in Figure 87, page 374. The `AST` corresponds to the abstract syntax tree of a compiled GIPSY program, followed by a dictionary of identifiers and the format of the compiled language. Then, at run-time demands are generated for the two identifiers at the context $d:1$ as well as their sum with the + operator yielding a resulting value $r$ [321].

The high-level encoding does not constitute the compilation data structures used in the GIPC compiler family (Section 6.2.1, page 139) as we are more concerned with the run-time execution at this point; however, one has to plan for certified compilation eventually, which is becoming more important these days, and that's what we will recommend to the GIPSY designers and developers for the future work. Once certified compilation is in place, the corresponding compilation evidence can be encoded for further offline or run-time analysis. Likewise, as in DMARF, we avoid talking about typical distributed system properties that ensure availability of the services, replication, etc., but in the final self-forensics modules design all the distributed middleware also needs to be covered [321].

### D.4.6.3 JDSF

JDSF's core data structures comprise secure data beans amended with security information layered as needed [275, 294, 295, 299, 316] (Section D.3.3, page 363). Thus, most of the

```
observation sequence GIPSYos = { p, d1, d2, di, r } =
{
  (
    [
      AST        : ..., // abstract syntax tree
      dictionary : [id1:'x', id2:'y'],
      formattag  : COMPILED_GEER
    ], 1, 0
  ),

  ([identifier:id1, tag:d, tagval:1], 1, 0),
  ([identifier:id2, tag:d, tagval:1], 1, 0),
  ([identifier:+, tag:d, tagval:1], 1, 0),

  ([result:X], 1, 0)
}
```

Figure 87: Example of observations in GIPSY [321]

observations comprise these data structures (the beans), and the observation sequence, using strict ordering gives away the sequence, in which a given security information was applied, e.g., signing happened before encryption and so on. Another observation is the configuration object instance that provides the details of all the module configuration parameters. The final observation is the secure object after application of all the security levels. As a result, the high-level sequence of events captures the observation context of given data and configuration that result in a secured bean. Therefore, the data structures covered by the FORENSIC LUCID context include the base `SecureBean` object type, and its derivatives `AuthenticatedObject`, `EncryptedObject`, `IntegrityAddedObject`. The basic forensic context structure is, therefore [321]:

```
bean: { value:X, security_info:Y }
```

The dimensions `value` (i.e., payload content) and `security_info` comprise the concrete security-enhanced payload data $X$ and the details about the said security technique applied $Y$. On top of the actual data beans, we have the corresponding configuration object that is a part of the system's context. The security `Configuration` object's context varies in three basic dimensions [321]:

```
configuration:
[
  confidentiality:C,
  integrity:I,
  authentication:A
]
```

```
observation sequence JDSFos = { config, data, bean } =
{
  (
    ordered [
      confidentiality : RSA  [ key:12345, keylength:1024 ],
      integrity       : MD5,
      authentication  : DSA  [ key:23456, keylength:256 ]
    ], 1, 0
  ),

  ([1,2,3,4], 1, 0),

  ([value:[3,1,4,2], security_info:Y], 1, 0)
}
```

Figure 88: Example of observations in JDSF [321]

These dimension's tag values evaluate to integers enumerating concrete algorithms used for each of these stages. The context also determines in which order they were applied, which may be an important factor, e.g., in the public key encryption schemes where the order of signing and encryption is important. In Figure 88, page 375 is an example of the FORENSIC LUCID context instance of a particular configuration [321].

### D.4.6.4 Cryptolysis

Cryptolysis [280, 298] being a small system does not have a lot of deep contextual complexity. While it has no explicit configuration object, there are implicit configuration state that impacts the results of the runs of the system. There there are key several operational evidential context pieces: the cryptographic Key, the plain-text and cipher-text data, cipher algorithm type, cryptanalysis algorithm type, statistics, and a result. The Key $k$ in the forward encryption process is a part of input configuration, along with the cipher algorithm to use $r$, and the input plain text data $t$, and the result is the cipher text $c$. In the reverse process, which is the main mode of operation for Cryptolysis, the configuration consists of the cryptanalysis algorithm type $a$, followed by the input cipher text data $c$, and the primary result is the guessed encryption Key $k$ and the secondary result is the associated accuracy statistics $s$ built along, and the plain text $t$, as shown in Equation D.4.4 and Equation D.4.5. The two equations denote these two scenarios; the latter is exemplified in Figure 89, page 376 [321].

$$(k, 1, 0), (r, 1, 0), (t, 1, 0), (c, 1, 0) \tag{D.4.4}$$

$$(a, 1, 0), (c, 1, 0), (k, 1, 0), (s, 1, 0), (t, 1, 0) \tag{D.4.5}$$

There is also a natural language processing (NLP) aspect of Cryptolysis that we do cover in [280] that deals with natural language word boundary detection in the deciphered text

```
observation sequence Cryptolysis_os = { a, c, k, s, t } =
{
  ([alogrithm:GENETIC_ALGORITHM], 1, 0),
  (['k','h','o','o','r'], 1, 0)
  ([key:[xyzabc...], length:26], 1, 0),
  ([accuracy:0.95], 1, 0),
  (['h','e','l','l','o'], 1, 0)
}
```

Figure 89: Example of observations for the second equation in Cryptolysis [321]

that has no punctuation or spaces [321].

### D.4.6.5  Hardware/Software Systems

Various types of craft (space or air) usually have many instruments and sensors on board, including high-end computers. All of these, or most important of these, "normal" operational units can also have their design augmented (hardware and software) to include additional functional units to observe the instruments and components, both hardware and software, for anomalies and log them appropriately for forensics purposes [277, 308]. Given enough on-board processing resources for the forensic log data, the system can automatically perform self-investigation of any related anomalies and institute corrective measures whenever possible while a remote operator access in unavailable.

For example, in the event when a spacecraft is "safing" itself due to, say, a cosmic ray hit, is a critical and appropriate moment to have some forensic data logged before and after (as well as during if possible) to help achieving a quicker analysis of what happened to bring out the spacecraft out of the safe mode sooner because presently the engineering teams spend a lot of effort to do the investigation after getting the telemetry data (which is would be, e.g., impossible when a craft is behind a planetary body or the Sun and is unreachable for a period of time) [308].

Finally, self-forensic *investigation engines* could run continuously in the aircraft and automotive vehicles, employing remote secure logging while doing in-situ self-investigation for problems and reporting to the pilots or drivers or taking a corrective action.

### D.4.6.6  Complex, Distributed Software Systems and Services

This topic was explored in some detail earlier [321] covering example distributed software systems as case studies, such as the *Autonomic Distributed Modular Audio Recognition Framework* (ADMARF) [496] and the *Autonomic General Intensional Programming System* (AGIPSY) [500] among others. The principles and concepts are very similar to the ones described in the preceding section except applied to the software components only and the corresponding middleware technologies used for communications. The *MAC Spoofer Investigation* presented in Section 9.5, page 257 also fits this criteria.

### D.4.6.7 Self-Forensics for Intelligent Roadside Networks

In [276] we have already discussed the idea of self-forensics for road vehicles (and similar ideas proposed in response to NASA's call in a white paper for spacecraft and aircraft safety and autonomous operation of distributed software systems [277, 321]), but there was almost next to no mention of the backbone and other support on the roadside networks (RSNETs), which was argued to aid the road safety and incident investigations. Self-forensics is important for diagnosing in remote areas where human monitors and operators may not be available on the highways. We'd like to include the term of self-forensics into the vocabulary of the roadside networks. RSNETs are also the receiving end of the vehicle undergoing an accident, it is therefore important to capture all the relevant data for the investigation by expert systems and human investigators with the intent of event reconstruction later on. Roadside networks have to be autonomous (e.g., for self-protection) in the modern age to sustain and self-manage to the maximally possible extent. As a result the corresponding equipment and intelligent software support must be designed accordingly.

**Main Objectives.**  Main objectives of this topic include:

- Standardize the terminology and requirements of roadside networks for intelligent self-management and self-forensic support for roadside backbones specified for VCNs.

- Define the R&D activities in roadside networks for the future test fields for self-forensics.

- Define and formalize self-monitoring and specification aspects for the design and development of the roadside networks to monitor and react in case of intrusions, power failures, incidents, and other faults with a blueprint of autonomous recovery strategies.

- Formalize context specification for the internetworking for VANETs and roadside networks, mobility, ubiquity, etc. reducing the management traffic overhead and improve incident response.

- Relevant QoS and security aspects and related protocols as solutions to performance and safety aspects of self-forensics in IRSNETs.

### D.4.6.8 Self-Forensics for Cloud Computing

This represents another project topic for the use of the self-forensics techniques previously proposed for both manual and autonomous security monitoring, event reconstruction, incident response, and investigation in distributed software systems or information systems in general onto cloud computing and its platforms. We argue that self-forensics, which is applicable to "self-dissection" of autonomous software in clouds for automated incident and anomaly analysis and event reconstruction including after-the-fact analysis by the investigators in a variety of incident scenarios. The idea, however, is the same, to construct and analyze the relevant formalizations, terminology, and standards to be included into the design of cloud computing platforms and tools to facilitate their management and digital investigations due to such systems being large and complex.

The concepts presented in the preceding sections of self-forensics for autonomic operation of distributed software systems [321]), multimedia information systems forensic specification and ASSL-based specification [322] include the notion of self-protection and other self-management aspects as well as forensics and incident investigations that we intend to include into the vocabulary cloud computing as well.

This project is set to cover the following topics to a certain degree in the context of cloud computing with self-forensics as a part of the self-management properties that cross-cover a number of aspects and topics.

- Similarly for the other scenarios, standardize the terminology and requirements of cloud computing platforms for intelligent self-management and self-forensic support, including self-forensic sensors deployed in the cloud servers logging forensic contexts in FORENSIC LUCID format.

- Like for IRSNETs, define and formalize self-monitoring and specification aspects for the design and development of clouds to monitor and react in case of intrusions, attack surface, other kinds of cybercrime incidents, and other faults with a blueprint of autonomous recovery strategies.

## D.5 Summary

We introduced a new concept of self-forensics with FORENSIC LUCID to aid validation and verification for critical systems to be used during the design, manufacturing, and testing, as well as its continuous use during the actual operation [277].

We outlined some of the background work, including the forensic case specification language, FORENSIC LUCID, that we adapted from the cybercrime investigations domain to aid the validation and verification of the autonomic subsystems design by proposing logging the forensics data in the FORENSIC LUCID context format available for manual/interactive analysis on the ground as well as real-time by a corresponding expert system [277].

We drafted some of the requirements for such a property to be included into the design as well as its potential most prolific limitations today [277]. Additionally, we laid out some preliminary groundwork of requirements to implement formally the self-forensics autonomic property within the ASSL toolset in order to allow any implementation of the self-forensics property added to the legacy small-to-medium open-source and academic software systems [322]. We note that all studied software systems are remarkably similar in their evidence structure suggesting that a reusable model can be built to observe a large number of similar systems for self-forensic evidence [321]. Using an ASPECTJ-based implementation for these software-based systems and component- or even method-level external observations can give very fine-grained as well as coarse-grained observation sequences and allow to model not only the contexts in the form of observation sequences of data structures and dataflows, but also *automatically* build the state transition function $\psi$ in FORENSIC LUCID to be ready to be exported for the use by investigators to allow modeling the $\Psi^{-1}$ for event reconstruction. This approach can be very well used for automated debugging of complex systems allowing the developers and quality assurance teams trace the problem and recreate the sequence of events back to what may have been the cause better than a mere stack trace and the resulting core dump from individual modules. Moreover, mobile devices these days increasingly use JAVA, so the said forensic data can be created on-the-fly with ASPECTJ observation of mobile device OS components [321].

We argued that a property of self-forensics formally encompassing self-monitoring, self-analyzing, self-diagnosing systems along with a decision engine and a common forensics logging data format can standardize the design and development of hardware and software of airborne vehicles, road vehicles, spacecraft, marine and submarine vehicles, to improve safety and autonomicity of the ever-increasing complexity of the software and hardware systems in such vehicles and further their analysis when incidents happen [277].

After having specified the high-level evidential observation sequences, each self-forensic module designer needs specify the complete details, as required and deemed important. Then, following the normal operation of the systems the self-forensic modules are turned on and logging data in the FORENSIC LUCID format continuously (notice, the forensic logs of course themselves must of high integrity and confidentiality, and there should be enough external storage to accommodate them, but these problems are not specific to this research, are addressed for other needs as well elsewhere like any type of logging or continuous data collection). The format, assuming being a standard, would be processable by FORENSIC LUCID-enabled expert systems, fully automatically (appropriate for autonomic systems) or interactively with the investigators making their case and analyzing the gathered evidence [321].

In summary, such a self-forensic specification is not only useful for cybercrime investigations, incident analysis, and response, but also useful to train new (software and other) engineers on a team, flight controllers, and others involved, in data analysis, to avoid potentially overlooking facts and relationships in the data and making incorrect ad-hoc decisions when performing an investigation.

In an FORENSIC LUCID-based expert system (see Section D.3.2, page 363) on the "human-side" of investigation (that's what was the original purpose of FORENSIC LUCID in the first place in cybercrime investigations) one can accumulate a number of contextual facts from the self-forensic evidence to form the knowledge base and the trainees can formally construct their theories of what happened and see of their theories agree with the evidential data. Over time, (unlike in most cybercrime investigations) it is possible to accumulate a general enough contextual knowledge base of encoded facts that can be analyzed across cases, deployed systems, flights, or missions, networks via sharing and firmware updates locally or globally and on the deep web when multiple agencies and aircraft manufactures collaborate for scalability [277, 308]. Such data can be shared across law enforcement, military, and research institutions [321] similarly to the Security Information Exchange (SIE) project of ISC (isc.org).

For the purposes of portability or model-checking, depending on the evaluating engine, FORENSIC LUCID constructs comprising the context of the encoded self-forensic evidence can be translated into XML specifications (that can be unsatisfiable [169] and less compact, but maybe more portable) of a specification for embedded systems and the like or translated into the PRISM probabilistic model checker's language [467] for model validation [321].

### D.5.1 Advantages FORENSIC LUCID for Self-Forensics

FORENSIC LUCID (Chapter 7) is context-enabled, built upon intensional logic and the LUCID language family that existed in the literature and mathematics for more than 35 years and served the initial purpose of program verification [25, 26]. FORENSIC LUCID and its evaluation engine provide an ability for event reconstruction upon validating claims against the available evidence [308] including when evidence is incomplete or imprecise. It is, therefore, a natural choice for self-forensics.

### D.5.2 Limitations of Self-Forensics

The self-forensics autonomic property is essential to have for automated analysis of simulated (testing and verification) and real incidents the autonomic or semi-autonomic hardware and software systems, but it can not be mandated as absolutely required due to a number of limitations it creates. However, whenever the monetary, time, and otherwise resources allow, it should be included in the design and development of the autonomous spacecraft, military equipment, or software systems [277, 308]. What follows is a summary of the limitations.

1. The cost of the overall autonomic systems will increase [277, 308]. The cost can be offset if the data are logged centrally and processed by central controllers.

2. If built into software, the design and development requires functional long-term storage and CPU power [277, 308].

3. Required communications will likely increase the bandwidth requirements (e.g., for scientific and exploratory aircraft if the science data are doubled in the forensic stream. If the forensic data are mirrored into the scientific one, more than twice bandwidth and storage used.) [277, 308].

4. An overall overhead if collecting forensics data continuously (e.g., can be offloaded along with the usual science and control data down to the flight control towers or centers periodically.) [277, 308].

5. The self-forensics logging and analyzing software ideally should be in ROM or similar durable flash type of memory or an SSD; but should allow firmware and software upgrades [277, 308].

6. We do not tackle other autonomic requirements of the system assuming their complete coverage and presence in the system from the earlier developments and literature, such as self-healing, protection, etc. [277, 308].

7. Transition function modeling the hardware and software specification has to be produced by the qualified engineering team throughout the design phase and encoded in FORENSIC LUCID. To simplify the task, the data-flow graph (DFG) [89, 311] interactive development environment (IDE), like a CAD application is to be available [308].

### D.5.3 Future Directions

As a part of the future work in this direction we plan to complete the implementation of the self-forensics modules to generate evidence according to the specifications presented earlier. Specifically, our priority is to implement the notion of self-forensics in the GIPSY and DMARF systems. Then, we plan to gather performance and storage overhead statistics when the self-forensics modules are turned on. We also need to improve the granularity of the event collection, correlation, and encoding with ASPECTJ-based tracing on the level of the method calls with the *before* and *after* triggers. Further, we want to explore events with probabilities, credibility, trustworthiness factors and the like using a probabilistic approach when exact reasoning is not possible [321].

We also need to explore and specify on how malicious and determined attackers would be dealt with by the self-forensics methods and techniques to retain valuable trustworthy forensic evidence. (In the nutshell, the issue is similar to remove logging in the specified format, perhaps offsite, maintaining a secure channel, but this deserves a separate complete discussion; additionally an investigated concept, like self-protection for DMARF [320] can be employed as one of the assurance techniques for self-forensics) [321].

#### D.5.3.1 Expected Outputs

Summarizing the expected output of this self-forensics project: a formal model, for validation and verification, of self-forensics, mathematically sound and complete to be finalized. It is to be followed by the concrete implementation of it first in the two mentioned research distributed agent systems such as GIPSY and DMARF. Then we move on to larger scale commercial and open-source software systems with the companies listed, extending it to the actual networked devices further. This would allow gather overhead and other metrics and refine the model and its scalability as well as user tools for analysis and investigation by the human investigators (expert system with updatable forensic knowledge base). Then the outcomes would be robotic and research craft prototypes—based on whatever we get access to or even build a mini research nano-satellite (e.g., a part of the `ServerSky.org` project), have self-forensic modules and sensors in a car and a plane.

#### D.5.3.2 Expected Outcomes

Standardized self-forensic hardware and software components in robotic devices, road vehicles, aircraft, spacecraft, marine vessels, operating systems and distributed autonomous software systems alike. The corresponding industries are the audience as well as law enforcement agencies and any type of human investigators involved in incident analysis, even reconstruction, and response.

# Appendix E

# Graph-based Representation and Visualization

This chapter is formed from the previously published and unpublished discussions on the need and ideas of a variety of visualization aspects, techniques, and tools, pertaining to this research. The purpose is to determine how the findings can be applied to FORENSIC LUCID and investigation case management. It is also natural to want a convenient and usable evidence visualization, its semantic linkage and the reasoning machinery for it. We present some of these deliberations as a future work item with the detailed related work review.

Visualization requirements in FORENSIC LUCID have to do with different levels of case knowledge abstraction, representation, aggregation, as well as the operational aspects as the final long-term goal of this proposal. It encompasses anything from the finer detailed representation of hierarchical contexts (Section 7.2.2, page 158) to FORENSIC LUCID programs, to the documented evidence and its management, its linkage to programs, to evaluation, and to management of GIPSY software networks. This includes an ability to arbitrarily switch between those views combined with usable multimodal interaction.

## E.1 Related Work

There is a number of items and proposals in graph-based visualization and the corresponding languages. In 1982 Faustini proved that any INDEXICAL LUCID program can be represented as a DFG [108]. In 1995, Jagannathan defined various graphical intensional and extensional models for GLU programming—arguably one of the first practical graph-based visualization proposals for LUCID programs [187]. Paquet subsequently in 1999 expanded on this for multidimensional intensional programs as exemplified in Figure 90 [361]. Stankovic, Orgun, *et al.* proposed the idea of visual parallel programming in 2002 [441].

The GIPSY R&D team's own work in the area (Chapter 6) includes the theoretical foundation and initial practical implementation of the DFGs by Ding [89, 361]. Ding provided the first implementation of Paquet's notion within the GIPSY project in 2004 [89] using Graphviz's `lefty`'s GUI and `dot` languages [32, 33] along with bi-directional translation between GIPL's or INDEXICAL LUCID's abstract syntax trees (ASTs) and `dot`'s [311]. Additionally, an idea was proposed for visualization and control of communication patterns and load balancing by having a "3D editor" within RIPE [282]. The concept behind such an editor is to render graphs in 3D space the current communication patterns of a GIPSY program-in-execution or to replay it back and allow the user to visually redistribute demands if the workload between workers becomes unbalanced. It is designed as a kind of a "virtual 3D remote control" with a mini-expert system, an input from which can be used to teach the planning, caching, and load-balancing algorithms to perform more efficiently the next time a similar GIPSY application is run. Related research work on visualization of load balancing, configuration, formal systems for diagrammatic modeling and visual languages and the corresponding graph systems was also presented by several authors in [13, 38, 258, 505, 540]. They defined a number of key concepts that are relevant to our visualization mechanisms within GIPSY [311] and its corresponding General Manager Tier (GMT) [191], for which Rabah provided the initial PoC visualization and basic configuration management of the GIPSY nodes and tiers with the graphical GMT [393] (e.g., Figure 38, page 151).

More recently (2012), another very interesting work of relevance was proposed by Tao *et al.* on visual representation of event sequences, reasoning and visualization [458] (in their case of the EHR data) and around the same time Wang *et al.* proposed a temporal search algorithm for personal history event visualization [517]. Monore *et al.* note the challenges of specifying intervals and absences in temporal queries and approach those with the use of a graphical language [323]. This could be particular useful for no-observations (Section 7.5.1, page 204) in our case. A recent novel HCI concept of documentary knowledge visual representation and gesture and speech-based interaction in the *Illimitable Space System* (ISS) was introduced by Song [437] in 2012. A multimodal case management interaction system was proposed for the German police called *Vispol Tangible Interface: An Interactive Scenario Visualization*[1].

We propose to build upon those works to represent the nested evidence, crime scene as a 2D or even 3D DFG, and the reconstructed event flow upon evaluation. Such a feature is designed to support the work on intensional forensic computing, evidence modeling and encoding, and FORENSIC LUCID [304, 310, 312, 321, 322] (Chapter 7) and MARFL ([269, 272], Appendix C)  in order to aid investigator's tasks to build and evaluate digital forensic cases [311].

## E.2  Visualization Example Prototypes

From the related work, a conceptual example of a 2D DFG corresponding to a simple LUCID program is in Figure 90 produced by Paquet [361]. The actual rendering of such graphs currently in the GIPSY environment is exemplified in Figure 91 by Ding [89] in 2004 [311].

In Figure 40, page 158 is the conceptual model of hierarchical nesting of the evidential statement *es* context elements, such as observation sequences *os*, their individual observations *o* (consisting of the properties being observed $(P, min, max, w, t)$, details of which are discussed in the referenced related works and in Chapter 7). These 2D conceptual visualizations are proposed to be renderable at least in 2D or in 3D via an interactive interface to allow modeling complex crime scenes and multidimensional evidence on demand. The end result is envsioned to look like either expanding or "cutting out" nodes or complex-type

---

[1]`http://www.youtube.com/watch?v=_2DywsIPNDQ`

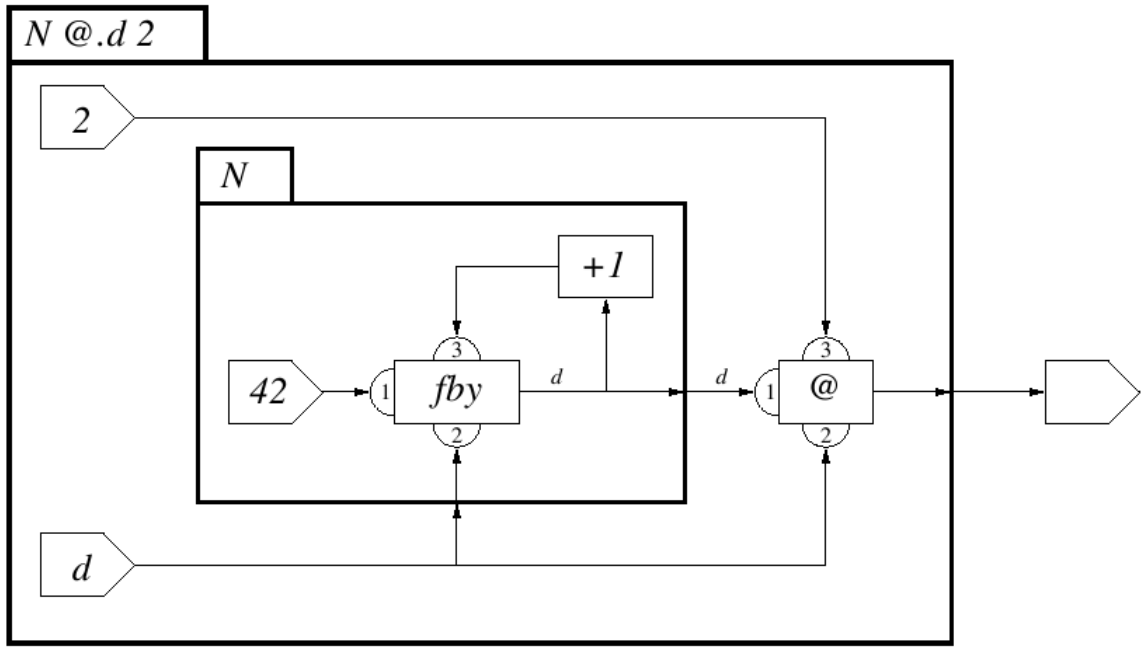

Figure 90: Canonical example of a 2D dataflow graph-based program [361]

results as exemplified in Figure 92[2] [311].

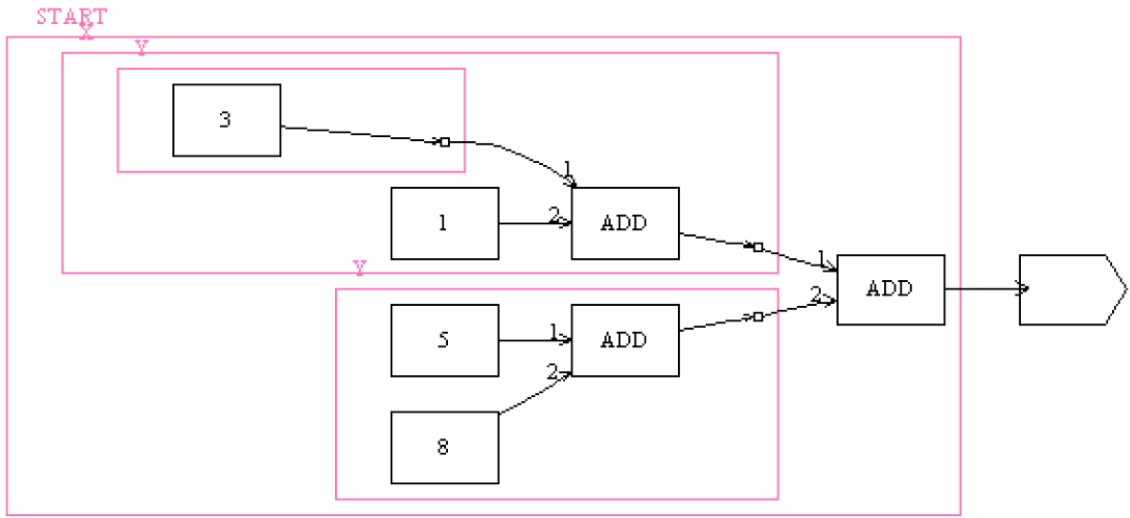

Figure 91: Example of an actual rendered 2D DFG-based program with Graphviz [89]

## E.3 Visualization of FORENSIC LUCID

To aid investigators to model the scene and evaluate it, we propose to expand the design and implementation of the LUCID DFG programming onto FORENSIC LUCID case modeling and specification to enhance the usability of the language and the system and its behavior in 3D. With the ongoing advances the visualization project was proposed to further enhance usability of the discussed language and system and the related tools following the good interaction design practices [404].

---

[2]cutout image credit is that of Europa found on Wikipedia http://en.wikipedia.org/wiki/File:PIA01130_Interior_of_Europa.jpg from NASA

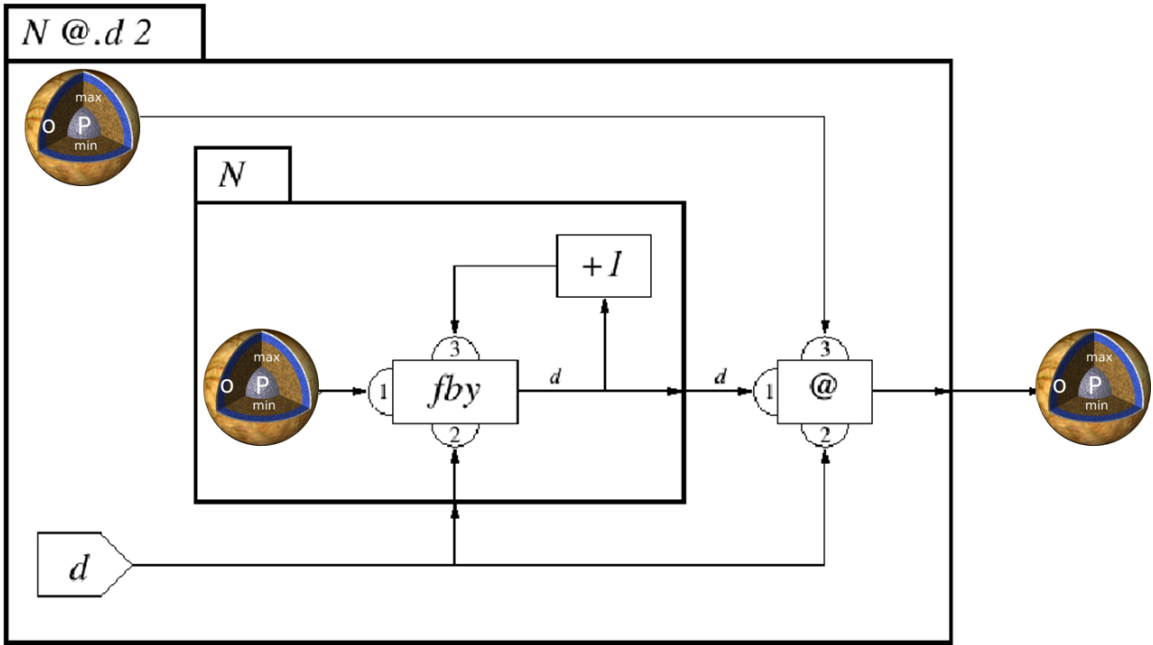

Figure 92: Modified conceptual example of a 2D DFG with 3D elements [311]

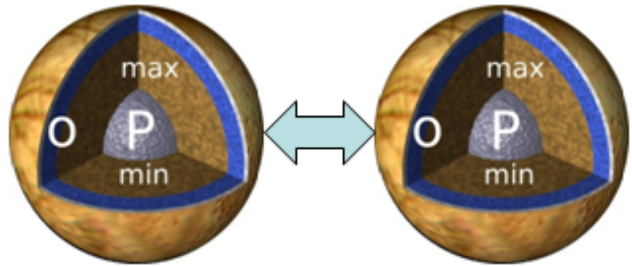

Figure 93: Conceptual example of linked 3D observation nodes [311]

### E.3.1 3 Dimensions

The need to represent visually forensic cases, evidence, and other specification components is obvious for usability and other issues. Placing it in 3D helps to structure the "program" (specification) and the case in 3D space can help arrange and structure the case in a virtual environment better with the evidence items encapsulated in 3D spheres akin to Russian dolls, and can be navigated in depth to any level of detail, e.g., via clicking (see Figure 93) [311].

The depth and complexity of operational semantics and demand-driven (eductive) execution model are better represented and comprehended visually in 3D especially when doing event reconstruction. Ding's implementation allows navigation from a graph to a subgraph by expanding more complex nodes to their definitions, e.g., more elaborate operators such *whenever* (`wvr`) or *advances upon* (`upon`), their reverse operators, forensic operators, and others [311] found in Section 7.3, page 166.

## E.3.2 Requirements Summary

Based on the preceding discussion, some immediate requirements to realize the envisioned DFG visualization of FORENSIC LUCID programs and their evaluation are summarized below [311]:

- Visualization of the hierarchical evidential statements (potentially deeply nested contexts), see Figure 40, page 158 [311].

- Placement of hybrid intensional-imperative nodes into the DFGs such as mixing JAVA and LUCID program fragments [311]. The GIPSY Research and Development group's previous research did not deal with aspects on how to augment the `DFGAnalyzer` and `DFGGenerator` of Ding to support hybrid GIPSY programs. This can be addressed by adding an "unexpandable" (one cannot click their way through its depth) imperative DFG node to the graph. To make it more useful, i.e., expandable, and so it's possible to generate the GIPSY code off the DFG or reverse it back, it should possible to leverage recent additions to Graphviz and GIPSY [311]. The newer versions of Graphviz support additional features that are more usable for our needs at the present. Moreover, with the advent of JOOIP ([528], Section 4.3.2.3, page 94), the JAVA 5 ASTs are made available along with embedded LUCID fragments that can be tapped into when generating the `dot` code's AST [311].

- There has to be a JAVA-based wrapper for the DFG Editor of Ding [89] to enable its native use within the JAVA-based GIPSY and plug-in IDE environments like Eclipse or NetBeans [311].

- Leveraging visualization and control of communication patterns and load balancing for the task in Euclidean space with the GGMT of Rabah [393].

- Ability to switch between different views (control, evidence, DFG, etc.).

## E.3.3 Selection of the Visualization Languages and Tools

One of the goals of this work is to find the most optimal technique, with soundness and completeness and formal specifications along with the ease of implementation and better usability. We began by gathering insights and requirements in selecting a technique or a combination of techniques with the most plausible outcome [311]. The current design allows any of the implementation paths to be chosen [311].

### E.3.3.1 Graphviz

First, the most obvious is Ding's [89] basic DFG implementation within GIPSY as it is already a part of the project and done for the two predecessor LUCID dialects: GIPL and INDEXICAL LUCID. Additionally, the modern version of Graphviz now has some integration done with Eclipse [99], so GIPSY's IDE—RIPE (Run-time Interactive Programming Environment)— may very well be an Eclipse-based plug-in [311].

#### E.3.3.2 PureData

Puckette came up with the PureData [388] language and its commercial offshoots (Jitter/Max/MSP [74]), which also employ DFG-like programming with boxes and inlets and outlets of any data types graphically placed and connected as "patches" and allowing for sub-graphs and external implementations of inlets in procedural languages. Puckette's original design was targetting signal processing for electronic music and video processing and production for interactive artistic and performative processes that has since outgrown that notion. The PureData externals allow deeper media visualizations in OpenGL, video, etc. thereby potentially enhancing the whole aspect of the process significantly [311]. Curiously, LUCID as a dataflow language, is said to have influenced PureData[3].

#### E.3.3.3 BPEL

The BPEL (Business Process Execution Language) and its visual realization within NetBeans [338, 450] for SOA (service-oriented architectures) and web services is another good model for an inspiration [177, 210, 349] that has recently undergone a lot of research and development, including flows, picking structures, faults, and parallel/asynchronous and sequential activities. More importantly, BPEL notations have a backing formalism modeled based upon Petri nets. (See, e.g., visual BPEL graph in BPEL Designer (first came with the NetBeans IDE) that illustrates two flows and three parallel activities in each flow as well asynchrony and timeouts modeling.) BPEL specifications actually translate to executable JAVA web services composition code [311].

#### E.3.3.4 Illimitable Space System

We explore an idea of a scalable management, visualization, and evaluation of digital evidence with modifications to the interactive 3D documentary subsystem of the *Illimitable Space System* (ISS) [437] to represent, semantically link, and provide a usable interface to digital investigators. As we stated in Chapter 1, that work may scale when properly re-engineered and enhanced to act as an interactive "3D window" into the evidential knowledge base grouped into the semantically linked "bubbles" visually representing the documented evidence. By moving such a contextual window, or rather, navigating within the theoretically illimitable space an investigator can sort out and re-organize the knowledge items as needed prior launching the reasoning computation. The interaction design aspect would be of a particular usefulness to open up the documented case knowledge and link the relevant witness accounts and group the related knowledge together. This is a proposed solution to the large scale visualization problem of large volumes of "scrollable" evidence that does not need to be all visualized at once, but be like a snapshot of a storage depot.

Stills from the actual ISS installation hosting multimedia data (documentary videos) users can call out by voice or gestures to examine the contents are in Figure 94[4]. We propose to re-organize the latter into more structured spaces linked together by the investigators grouping the relevant evidence semantically instead of the data containing bubbles floating around randomly.

---

[3]http://en.wikipedia.org/wiki/Lucid_(programming_language)
[4]http://vimeo.com/51329588

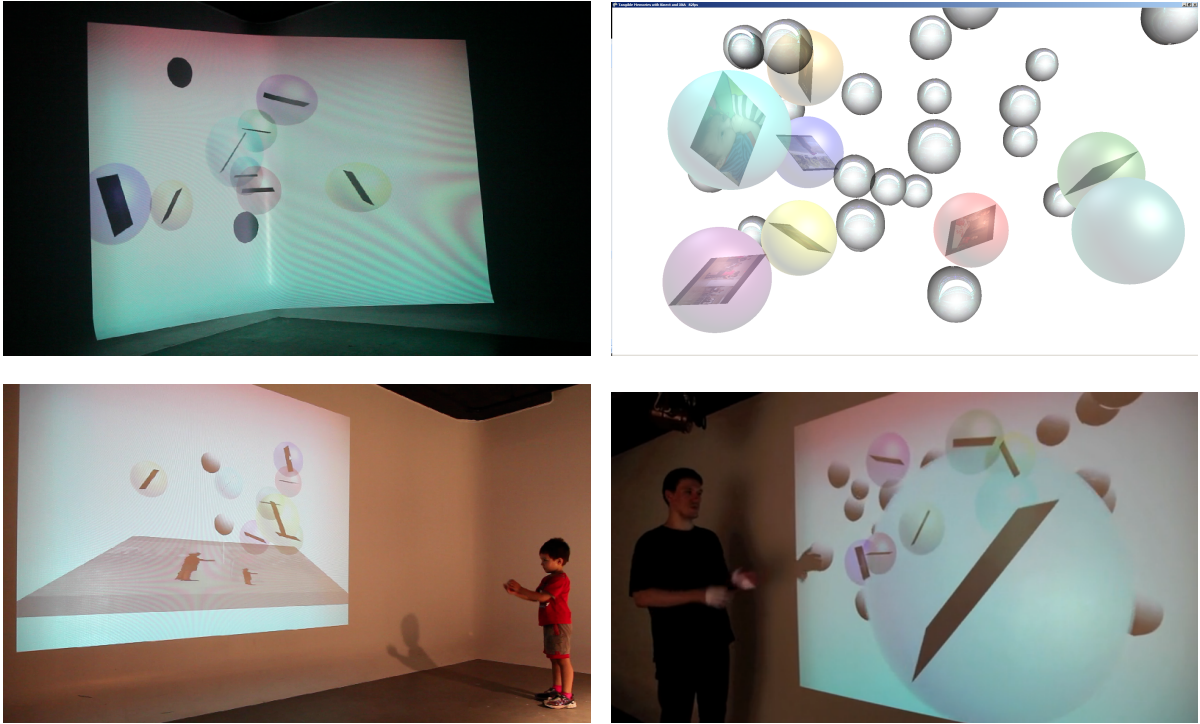

Figure 94: Interactive documentary illimitable space visualization and management [437]

# E.4 Summary

With the goal to have a visual DFG-based tool to model FORENSIC LUCID case specification, we deliberate on the possible choice of the visualization languages and paradigms within today's technologies and their practicality and attempt to build upon previous sound work in this area. Main choices so far identified include Ding-derived Graphviz-based implementation, PureData-based, BPEL-like, or the ISS. All languages are more or less industry standards and have some formal backing. The ones that are not, may require additional work to formally specify their semantics and prove correctness and soundness of the translation to and from FORENSIC LUCID [311].

The main problem with PureData and Graphviz'es `dot` is that their languages do not have formal semantics specified only some semantic notes and lexical and grammatical structures (e.g., see `dot`'s [32]). If we use any and all of these, we will have to provide translation rules and their semantics and equivalence to the original FORENSIC LUCID specification similarly as it is (e.g., was done by Jarraya between the UML 2.0/SysML state/activity diagrams and probabilities in [190] when translating to PRISM) [311]. ISS seems the most scalable approach that can aggregate all the others, but requires significant number of modifications.

Thus, this aspect of the work is at the active research stage. Given author's some familiarity with the languages and systems mentioned, the final choice may result being an intermediate form or a combination of inter-translatable forms [311] of those languages and systems.

# Index